\pdfoutput=1 
\documentclass[cernpreprint,english,USenglish]{na61doc}

\usepackage{amsmath}
\usepackage{enumerate}
\usepackage{placeins}
\usepackage{appendix}
\usepackage[LY1]{fontenc}
\usepackage[utf8]{inputenc}
\usepackage{chngcntr}
\usepackage{microtype}
\usepackage{lineno}
\usepackage{color}

\usepackage{colortbl}
\definecolor{darkred}{rgb}{0.5,0,0}
\definecolor{darkblue}{rgb}{0,0,0.5}
\definecolor{firebrick}{rgb}{0.75,0.125,0.125}
\definecolor{darkgreen}{rgb}{0,0.5,0}
\graphicspath{{Figures/}}

\newcommand{\eV}{\ensuremath{\mbox{e\kern-0.1em V}}\xspace}
\newcommand{\GeV}{\ensuremath{\mbox{Ge\kern-0.1em V}}\xspace}
\newcommand{\MeV}{\ensuremath{\mbox{Me\kern-0.1em V}}\xspace}
\newcommand{\GeVc}{\ensuremath{\mbox{Ge\kern-0.1em V}\!/\!c}\xspace}
\newcommand{\GeVcc}{\ensuremath{\mbox{Ge\kern-0.1em V}\!/\!c^2}\xspace}
\newcommand{\AGeV}{\ensuremath{A\,\mbox{Ge\kern-0.1em V}}\xspace}
\newcommand{\AGeVc}{\ensuremath{A\,\mbox{Ge\kern-0.1em V}\!/\!c}\xspace}
\newcommand{\MeVc}{\ensuremath{\mbox{Me\kern-0.1em V}/c}\xspace}

\newcommand{\cm}{\ensuremath{\mbox{cm}}\xspace}

\newcommand{\dd}{\ensuremath{{\mathrm{d}}}\xspace}
\newcommand{\dedx}{\ensuremath{\dd E\!/\!\dd x}\xspace}

\newcommand{\pt}{\ensuremath{p_{\mathrm{T}}}\xspace}

\newcommand{\mt}{\ensuremath{m_{\mathrm{T}}}\xspace}
\newcommand{\atanh}{\ensuremath{atanh}\xspace}

\newcommand{\Urqmd}{{\scshape U}r{\scshape QMD}\xspace}

\newcommand{\GeantThree}{{\scshape Geant3}\xspace}

\newcommand{\Epos}{{\scshape EPOS}\xspace}

\newcommand{\CernVM}{\textsc{Cern\-\kern-0.05emVM}\xspace}

\ShineTitle{Measurements of $\pi^\pm$, K$^\pm$, p and $\bar{\textrm{p}}$ spectra in proton-proton interactions at 20, 31, 40, 80 and 158~\GeVc with the \NASixtyOne spectrometer at the CERN SPS}

\PreprintIdNumber{CERN-EP-2017-066}

\ShineJournal{Eur. Phys. J. C}
\ShineAbstract{ 
Measurements of inclusive spectra and mean multiplicities of $\pi^\pm$, K$^\pm$, p and $\bar{\textrm{p}}$ produced in inelastic p+p interactions at incident projectile momenta of 20, 31, 40, 80 and 158~\GeVc ($\sqrt{s} = $ 6.3, 7.7, 8.8, 12.3 and 17.3~\GeV, respectively) were performed at the CERN Super Proton Synchrotron using the large acceptance \NASixtyOne hadron spectrometer. Spectra are presented as function of rapidity and transverse momentum and are compared to predictions of current models. The measurements serve as the baseline in the \NASixtyOne study of the properties of the onset of deconfinement and search for the critical point of strongly interacting matter. 
}
\begin{document}
\maketitle

\section{Introduction}
\label{sec:Intro} 

\begin{sloppypar}
This paper presents experimental results on
inclusive spectra and mean multiplicities of
 $\pi^\pm$, K$^\pm$, p and $\bar{\textrm{p}}$  produced in inelastic p+p interactions
at 20, 31, 40, 80 and 158~\GeVc.
The measurements were performed by the multi-purpose
\NASixtyOne{} experiment~\cite{Abgrall:2014fa,Antoniou:2006mh}
at the CERN Super Proton Synchrotron (SPS).
The new measurements complement previously published results from the same datasets
on $\pi^-$ production~\cite{Abgrall:2013pp_pim} obtained without particle identification
as well as on fluctuations of charged particles~\cite{Aduszkiewicz:2015jna}.
These studies form part of the \NASixtyOne strong interaction programme investigating
the properties of the onset of deconfinement and searching for the
critical point of strongly interacting matter. The programme is mainly motivated by
the observation of rapid changes of hadron production properties in central Pb+Pb 
collisions at about 30\AGeVc by the NA49 experiment~\cite{Afanasiev:2002mx,Alt:2007aa}
which were interpreted as the onset of deconfinement. 
These findings were recently confirmed by the RHIC beam
energy programme~\cite{Adamczyk:2017iwn} and the interpretation is supported by the LHC
results (see Ref.~\cite{Rustamov:2012np} and references therein). Clearly, 
a two dimensional scan in collision energy and size of colliding nuclei
is required to explore systematically the phase diagram of strongly interacting matter~\cite{Gazdzicki:2014sva}. 
\end{sloppypar}

\begin{sloppypar}
Pursuing this programme \NASixtyOne already recorded data on p+p, Be+Be, Ar+Sc and  p+Pb collisions 
and data taking on Xe+La collisions is scheduled for 2017. Moreover, measurements of
Pb+Pb interactions are planned for the coming years~\cite{PbAddendum}.
\end{sloppypar}


An interpretation of the rich experimental results on nucleus--nucleus
collisions relies to a large extent on a comparison to the
corresponding data on p+p and p+A interactions.
However, the published measurements 
mainly refer to basic features of unidentified charged hadron production
and are sparse. Results on identified hadron spectra, fluctuations and
correlations are mostly missing.
Detailed measurements of hadron spectra in a large acceptance
in the beam momentum range covered by the data in this
paper exist only from the NA49 experiment for inelastic p+p
interactions at 158~\GeVc~\cite{Alt:2005zq,Anticic:2009wd,Anticic:2010yg}.
Thus new high precision measurements of hadron production properties
in p+p and p+A interactions are essential. They are performed by \NASixtyOne
in parallel with the corresponding measurements on nucleus--nucleus
collisions using the same detector and thus covering the same acceptance. 
Precise data on pion, kaon and proton production properties are crucial for
constraining basic properties of models of strong interactions.

This publication presents two-dimensional spectra of positively and negatively charged pions, 
kaons, protons and antiprotons produced in p+p interactions in the SPS momentum range (20, 31, 40, 80, 158~\GeVc).
The paper is organized as follows: after this introduction the experiment is briefly described
in Sec.~2. The analysis procedure is discussed in Sec.~3. Section~4 presents the results of the
analysis. In Sec.~5 model calculations are compared to the new measurements. A summary in Sec.~6 closes the paper.     
 
The following variables and definitions are used in this paper.
The particle rapidity is calculated in the collision center of
mass system (cms), $y = \atanh(\beta_\mathrm{L})$,
where $\beta_\mathrm{L} = cp_\mathrm{L}/E$ is the longitudinal component of the
velocity and $p_\mathrm{L}$ and $E$ are the longitudinal momentum and energy given in the cms.
The transverse component of the momentum is denoted as $\pt$ and
the transverse mass $\mt$ is defined as $\mt = \sqrt{m^2 + (c\pt)^2}$,
where $m$ is the particle mass in GeV. The momentum in the laboratory frame is denoted $p_{\mathrm{lab}}$
and electric charge in units of the electron charge as $q$.
The collision energy per nucleon pair in the center of mass
system is denoted as $\sqrt{s_\mathrm{NN}}$.

\section{\NASixtyOne experiment}
\label{sec:na61} 

\begin{figure}
\centering
\includegraphics[width=0.8\textwidth]{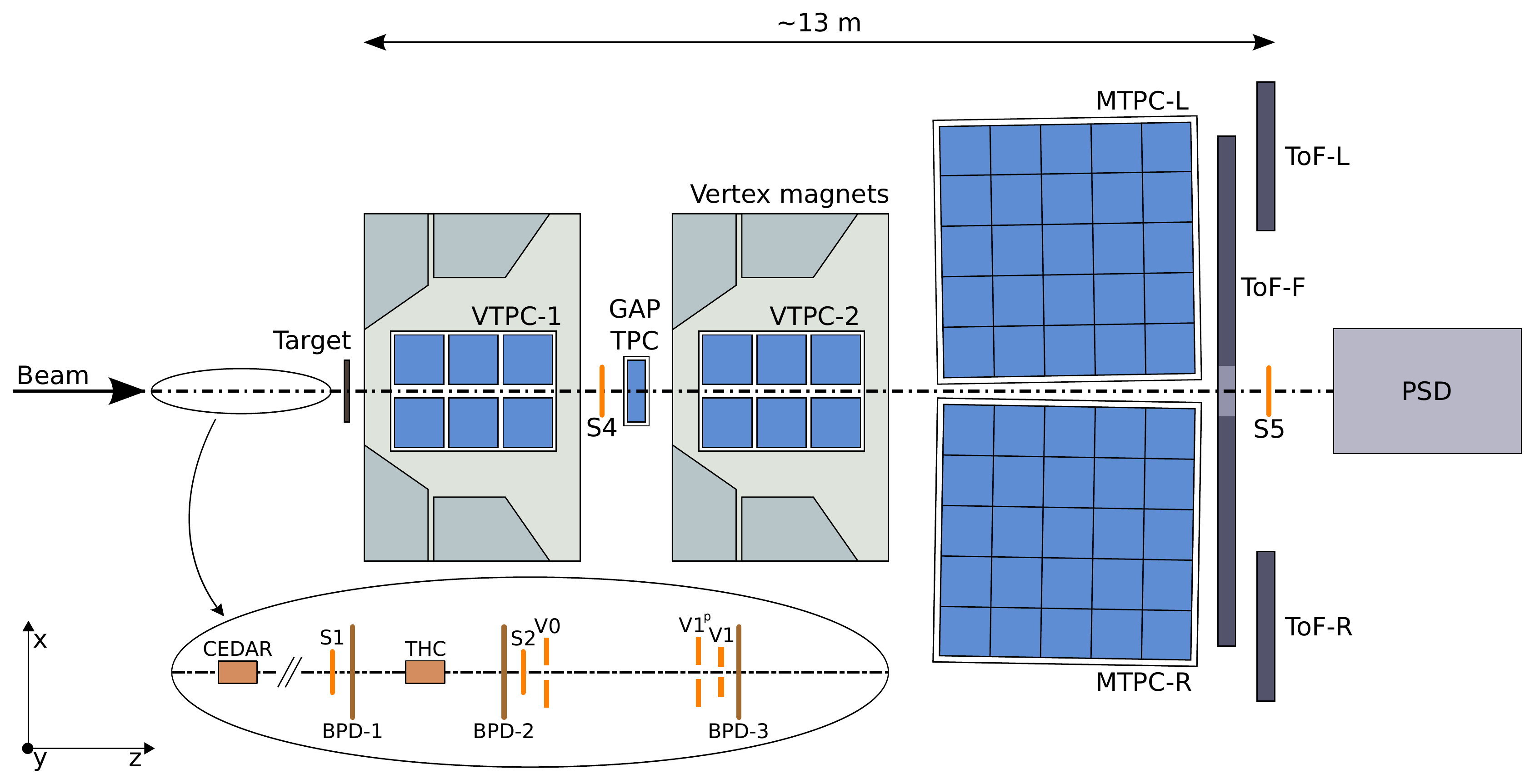}
\caption{(Color online) The schematic layout of the NA61/SHINE experiment
at the CERN SPS (horizontal cut, not to scale).
Alignment of the \NASixtyOne coordinate system is shown on the plot. 
The nominal beam direction is along the $z$ axis. The magnetic field bends charged particle
trajectories in the $x$-$z$ plane. The electron drift direction in the TPCs is along  the $y$ (vertical) axis. 
}
\label{fig:detector-setup}
\end{figure}

\NASixtyOne is a fixed target experiment employing a large acceptance hadron spectrometer situated in the North Area H2 beam-line of the CERN SPS~\cite{Abgrall:2014fa}. A schematic layout is shown in Fig.~\ref{fig:detector-setup}. 
The main components of the detection system used in the analysis are four large volume Time Projection Chambers (TPC). Two of them, called Vertex TPCs (VTPC), are located downstream of the target inside superconducting magnets with maximum combined bending power of 9~Tm. The TPCs are filled with Ar:CO$_{2}$ gas mixtures in proportions 90:10 for the VTPCs and 95:5 for the Main TPCs.
The MTPCs and two walls of pixel Time-of-Flight (ToF-L/R) detectors are placed symmetrically to the beamline downstream of the magnets. A GAP-TPC (GTPC) between VTPC-1 and VTPC-2 improves the acceptance for high-momentum forward-going tracks.

Individual beam particles are identified and precisely measured by a set of scintillation and Cherenkov counters, 
as well as three beam position detectors (BPDs) placed upstream of the target~\cite{Abgrall:2014fa}. 
Secondary beams of positively charged hadrons at momenta of 20, 31, 40, 80 and 158~\GeVc were used to collect the data for the analysis presented in this paper. 
These beams were produced from 400~\GeVc protons extracted from the SPS in a slow extraction mode with a flat-top of 10 seconds. Protons from the secondary hadron beam are identified by two Cherenkov counters, a CEDAR~\cite{CEDAR} (either CEDAR-W or CEDAR-N) and a threshold counter (THC). A selection based on signals from the Cherenkov
counters allowed to identify beam protons with a purity of about 99\%~\cite{Abgrall:2013qoa}. The proton contamination in the secondary beam The beam momentum and intensity was adjusted by proper setting of the H2 beamline magnets and collimators. The precision of the setting of the beam magnet currents is approximately 0.5\%. 
The beam momentum was verified by a direct measurement at 31~\GeVc by bending the incoming beam particles into the TPCs with the maximum magnetic field. 
The properties of the beams used for obtaining the analysed data are summarized in Table~\ref{tab:beams}. 


\begin{table}
   \caption{
Basic beam properties and number of events recorded for p+p interactions at incident proton momentum of 20, 31, 40, 80 and 158~\GeVc. 
}
\centering
\begin{tabular}{| c | c | c | c | c |}
  \hline\hline
    $p_{beam}\ [\GeVc]$ & $\sqrt{s}$ $[\GeV]$ & Particles                        & Proton  & Number of  \\
    					&				   & per spill $\times10^3$			  & fraction		& recorded events \\
    \hline 
    20  & 6.2 & 1000 &   12\% & $1.3\cdot10^{6}$    \\
    31  & 7.7 & 1000 &   14\% & $3.1\cdot10^{6}$     \\
    40  & 8.8 & 1200 &   14\% & $5.2\cdot10^{6}$     \\
    80  & 12.3& 460  &   28\% & $4.3\cdot10^{6}$     \\
    158 & 17.3& 250  &   58\% & $3.5\cdot10^{6}$     \\\hline\hline
  \end{tabular}

  \label{tab:beams}
\end{table}

A Liquid Hydrogen Target (LHT) of 20.29~\cm length (2.8\% interaction length) and 3~\cm diameter was placed 88.4~\cm upstream of VTPC-1. 
Data were taken with full (denoted as target inserted, I) and empty (denoted as target removed, R) LHT. The event statistics collected in the two configurations are summarised in Table~\ref{tab:statistic}.

Interactions in the target are selected with the trigger system
by requiring an incoming beam proton and no signal from S4,
a small 2~\cm diameter scintillation counter placed on the beam
trajectory
between the two vertex magnets. This
minimum bias trigger is based on the disappearance of the beam
proton downstream of the target.


\FloatBarrier
\section{Analysis procedure}\label{sec:analysis}

This section starts with a brief overview of the data analysis procedure and the applied corrections.
It also defines to which class of particles the final results correspond.
A description of the calibration and the track and vertex reconstruction procedure can be found in 
Ref.~\cite{Abgrall:2013pp_pim}.  

The analysis procedure consists of the following steps:
\begin{enumerate}[(i)]
  \item application of event and track selection criteria,
  \item determination of spectra of identified hadrons
        using the selected events and tracks,
  \item evaluation of corrections to the spectra based on
        experimental data and simulations,
  \item calculation of the corrected spectra and mean multiplicities,
  \item calculation of statistical and systematic uncertainties.
\end{enumerate}  

Corrections for the following biases were evaluated and applied:
\begin{enumerate}[(i)]
 \item geometrical acceptance,
 \item contribution from off-target interactions,
 \item contribution of particles other than \emph{primary} (see below)
       hadrons produced  in inelastic p+p interactions,
 \item losses of inelastic p+p interactions due to the trigger and the
       event and track selection criteria employed in the analysis 
       as well as losses of produced hadrons in accepted interactions
       due to their decays and secondary interactions.
\end{enumerate}

The final results refer to identified hadrons 
produced in inelastic p+p interactions 
by strong interaction processes and in  electromagnetic
decays of produced hadrons. 
Such hadrons are referred to as \emph{primary} hadrons.

The analysis was performed independently in ($y$, \pt) bins.
The bin sizes were selected taking into account the statistical uncertainties 
and the resolution of the momentum reconstruction~\cite{Abgrall:2013pp_pim}.
Corrections as well as statistical and systematic uncertainties
were calculated for each bin.
 
\FloatBarrier 
   \subsection{Event and track selection}\label{sec:cuts} 

\subsubsection{Event selection}

Inelastic p+p events were selected using the following criteria:
\begin{enumerate}[(i)]
\item no off-time beam particle detected within a time window of $\pm$2$~\mu$s around the trigger particle,
\item beam particle trajectory measured in at least three planes out of four of BPD-1 and BPD-2 and in both planes of BPD-3,
\item at least one track reconstructed in the TPCs and fitted to the interaction vertex,
\item $z$ position of the interaction vertex (fitted using the beam trajectory and TPC tracks) not farther away than 20~\cm from the center of the LHT,
\item events with a single, positively charged track with absolute momentum close to the beam momentum (see \cite{Abgrall:2013pp_pim}) are removed in order to eliminate elastic scattering reactions.
\end{enumerate}

\subsubsection{Track selection}

In order to select tracks of primary charged hadrons and to reduce the contamination
of tracks from secondary interactions, weak decays and off-time interactions, the following track selection criteria
were applied:
\begin {enumerate}[(i)]
\item track momentum fit at the interaction vertex should have converged,
\item fitted $x$ component of particle rigidity $\left(p_{lab,x}/q\right)$ is positive.
This selection minimizes the angle between the track
trajectory and the TPC pad direction for the chosen
magnetic field direction, reducing uncertainties of the reconstructed cluster position, energy
deposition and track parameters,
\item total number of reconstructed points on the track should be greater than 30,

\item sum of the number of reconstructed points in VTPC-1 and VTPC-2 should be greater than 15 or the number of reconstructed points in the GAP-TPC should be greater than 4,
\item the distance between the track extrapolated to the interaction plane and the interaction point (impact parameter)
should be smaller than 4~\cm in the horizontal (bending) plane and 2~\cm in the vertical (drift) plane,
%

\item the total number of reconstructed \dedx points on the track should be greater than 30,
\item in case of $tof$-\dedx identification, three additional selection criteria were used:
	\begin{enumerate}[(i)]
		\item the hit in the ToF pixel should be matched only with one TPC track,
		\item proper measurement of the hit  in Charge to Digital Converter (QDC) and Time to Digital Converter (TDC)
		 \item the last point of the track should be in the last 2 padrows of the MTPC to ensure good matching with the ToF hit.
	\end{enumerate}
\end {enumerate}


The event and track statistics after applying the selection criteria
are summarized in Table~\ref{tab:statistic}. 

{
\small
\begin{table}[!ht]
\caption{Statistics of events and tracks used in \dedx and $tof$-\dedx identification methods for target inserted and removed configurations. Events with removed target were used to
evaluate corrections for off-target interactions.}
\begin{center}
\begin{tabular}{|c|c|c|c|}
\hline\hline
 \multicolumn{4}{|c|}{Target inserted}\\
		\hline\hline
Momentum	&	Number of events	& Number of tracks			& Number of tracks \\
$[$\GeVc$]$ &						& (\dedx method) &  ($tof$-\dedx method) \\
\hline
20  & {\color{white}0}234758  & {\color{white}0}244813  & {\color{white}0}17023  \\
31& {\color{white}0}832608  & {\color{white}0}859573  & {\color{white}0}44228  \\
40  & 1604483 & 1625595 & 199775 \\
80  & 1591076 & 1592538 & 214316 \\
158 & 1625578 & 4464269 & 158520 \\
\hline\hline
 \multicolumn{4}{|c|}{Target removed}\\
		\hline\hline
Momentum	&	Number of events	& Number of tracks	& Number of tracks \\
$[$\GeVc$]$		&						& (\dedx method)  &($tof$-\dedx method)  \\
\hline
20  & {\color{white}0}3184  & {\color{white}0}2175  & {\color{white}0}402  \\
31& 12618 & 10080 & {\color{white}0}691  \\
40  & 42115 & 39893 & 4745 \\
80  & 51588 & 38132 & 8003 \\
158 & 26837  & 41234 & 3373 \\
\hline\hline
\end{tabular}
\end{center}
\label{tab:statistic}
\end{table}
}


\FloatBarrier
   \subsection{Identification techniques}
\label{sec:identification}

\begin{figure}[!ht]
	\begin{center}
	\includegraphics[width=0.3\textwidth]{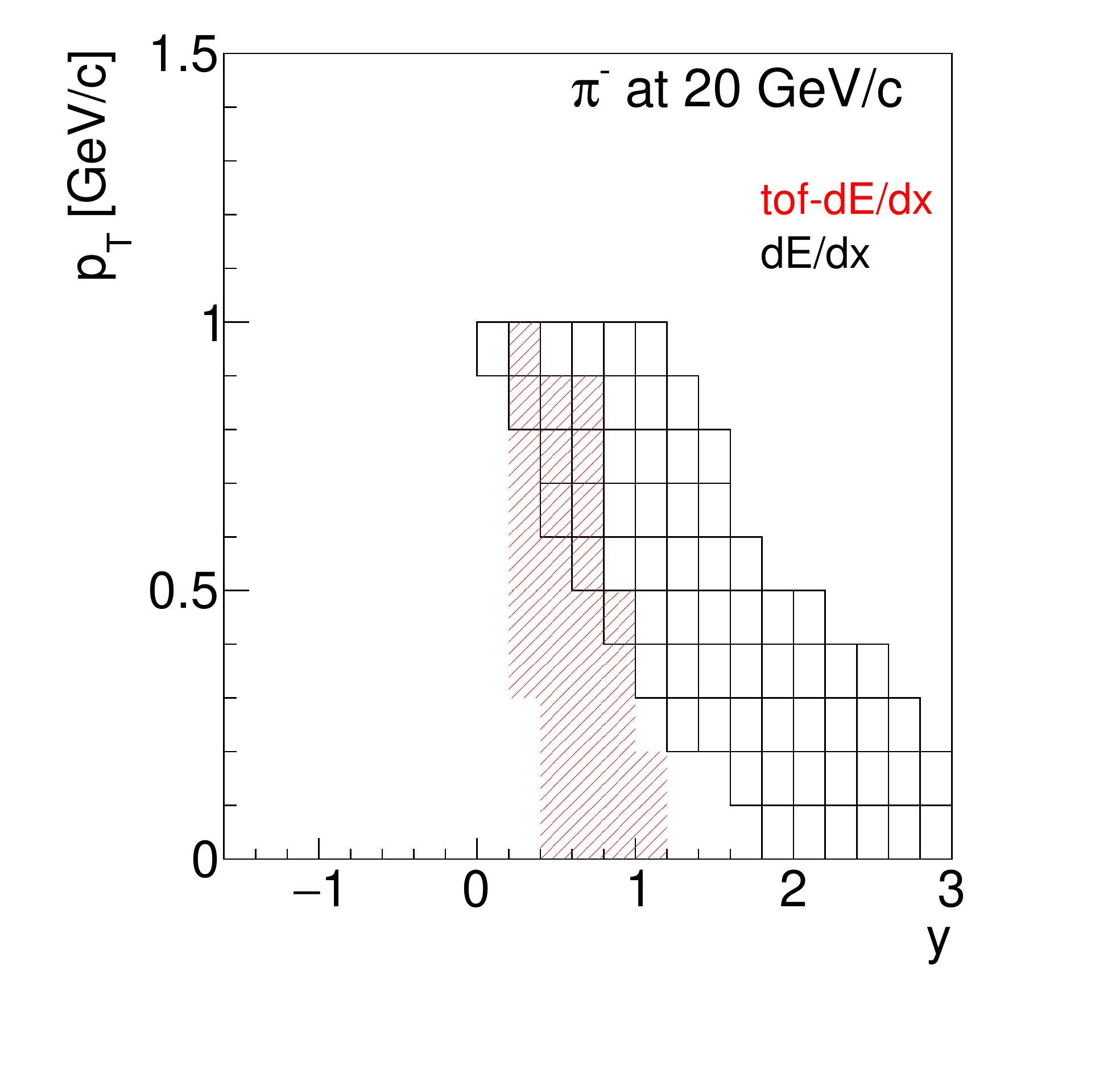}
	\includegraphics[width=0.3\textwidth]{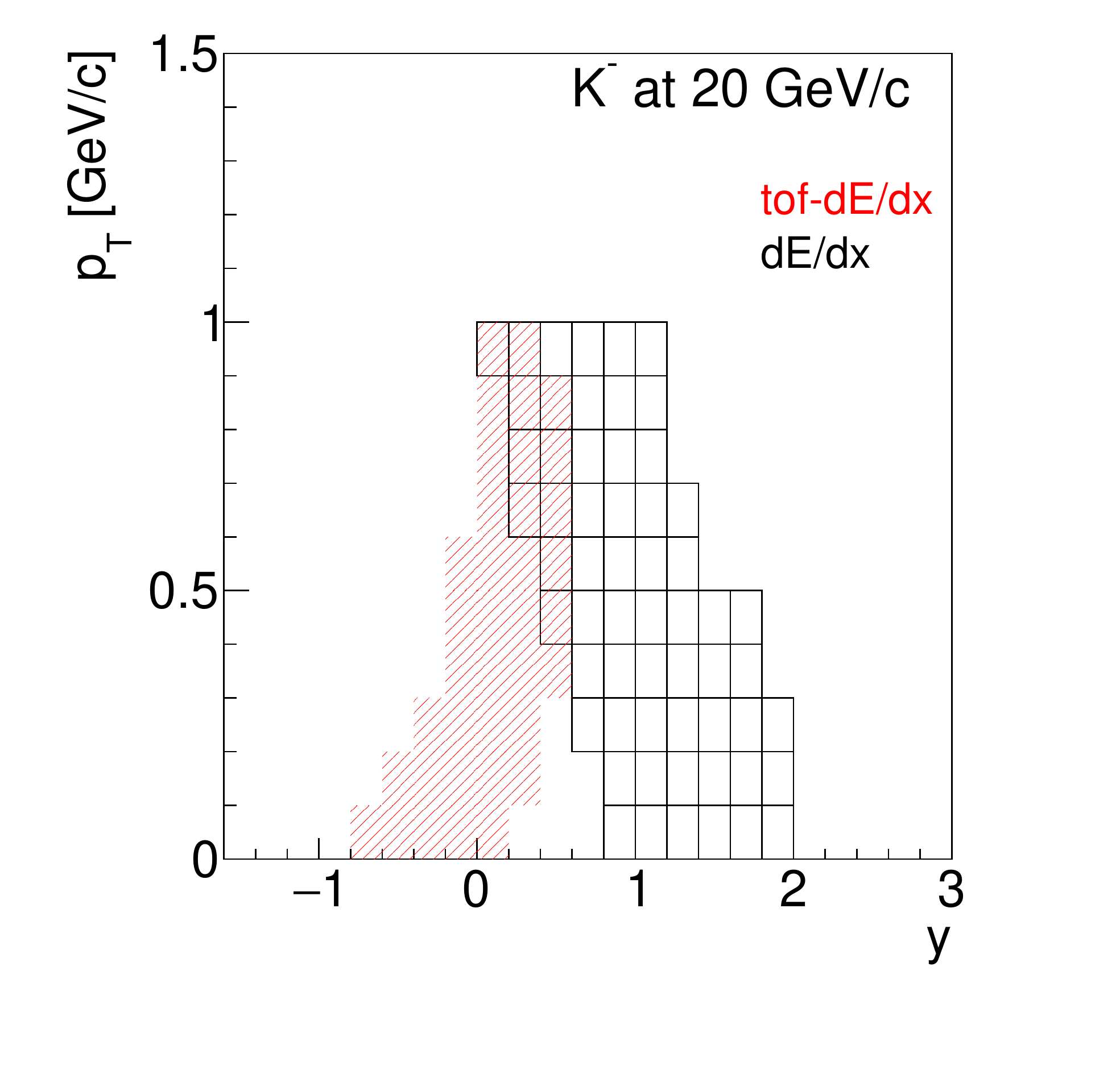}\\
	\includegraphics[width=0.3\textwidth]{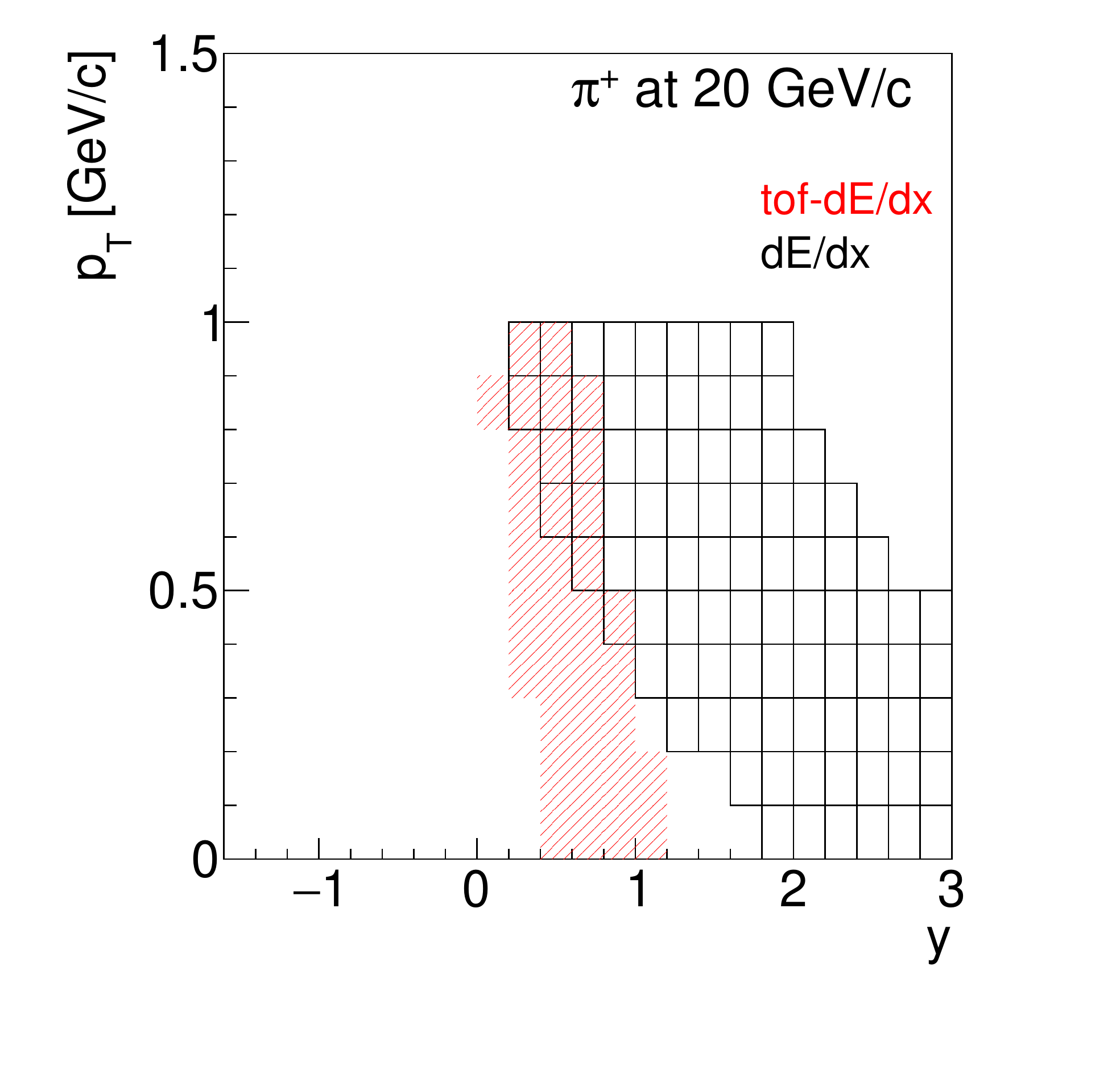}
	\includegraphics[width=0.3\textwidth]{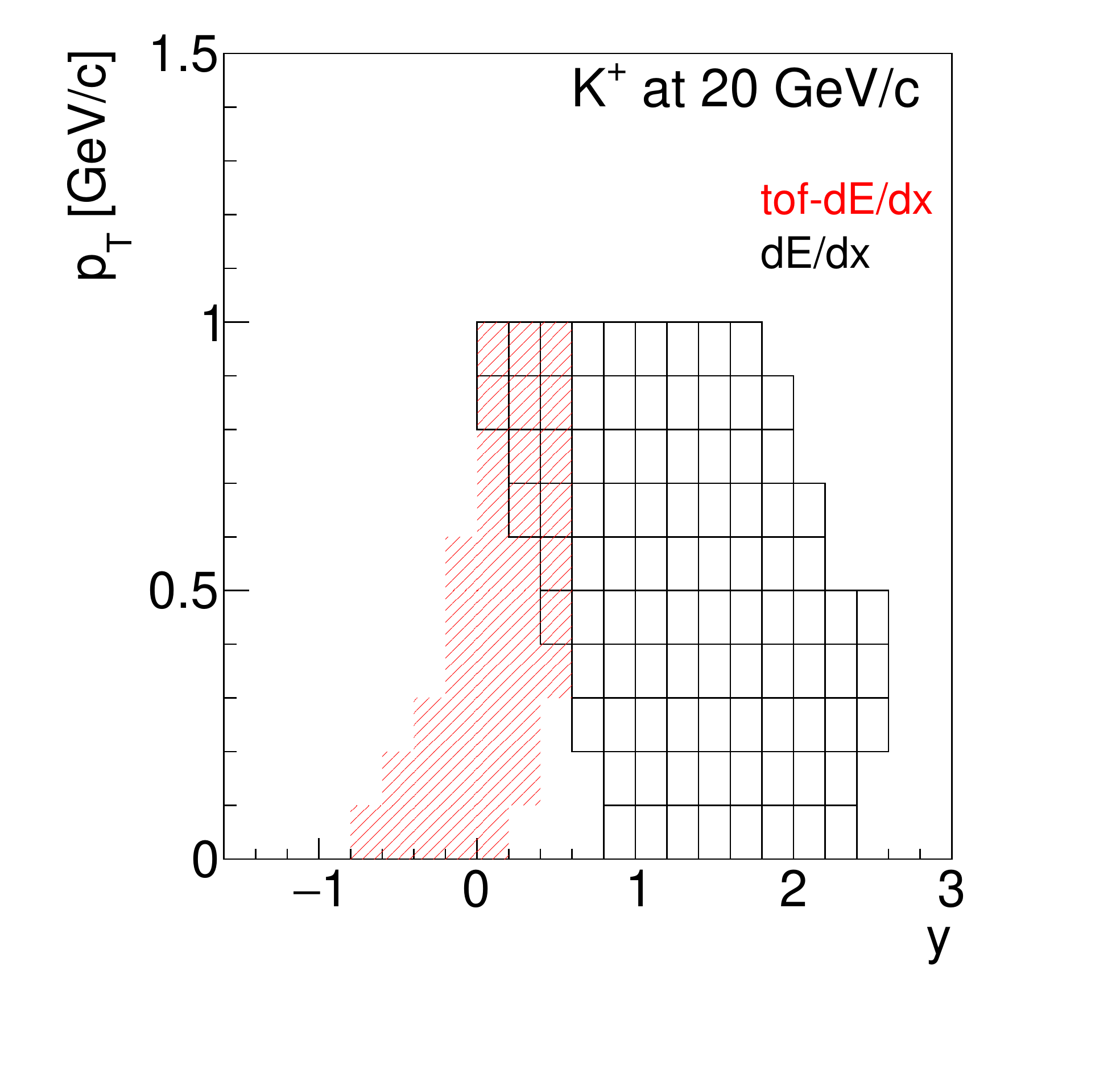}
	\includegraphics[width=0.3\textwidth]{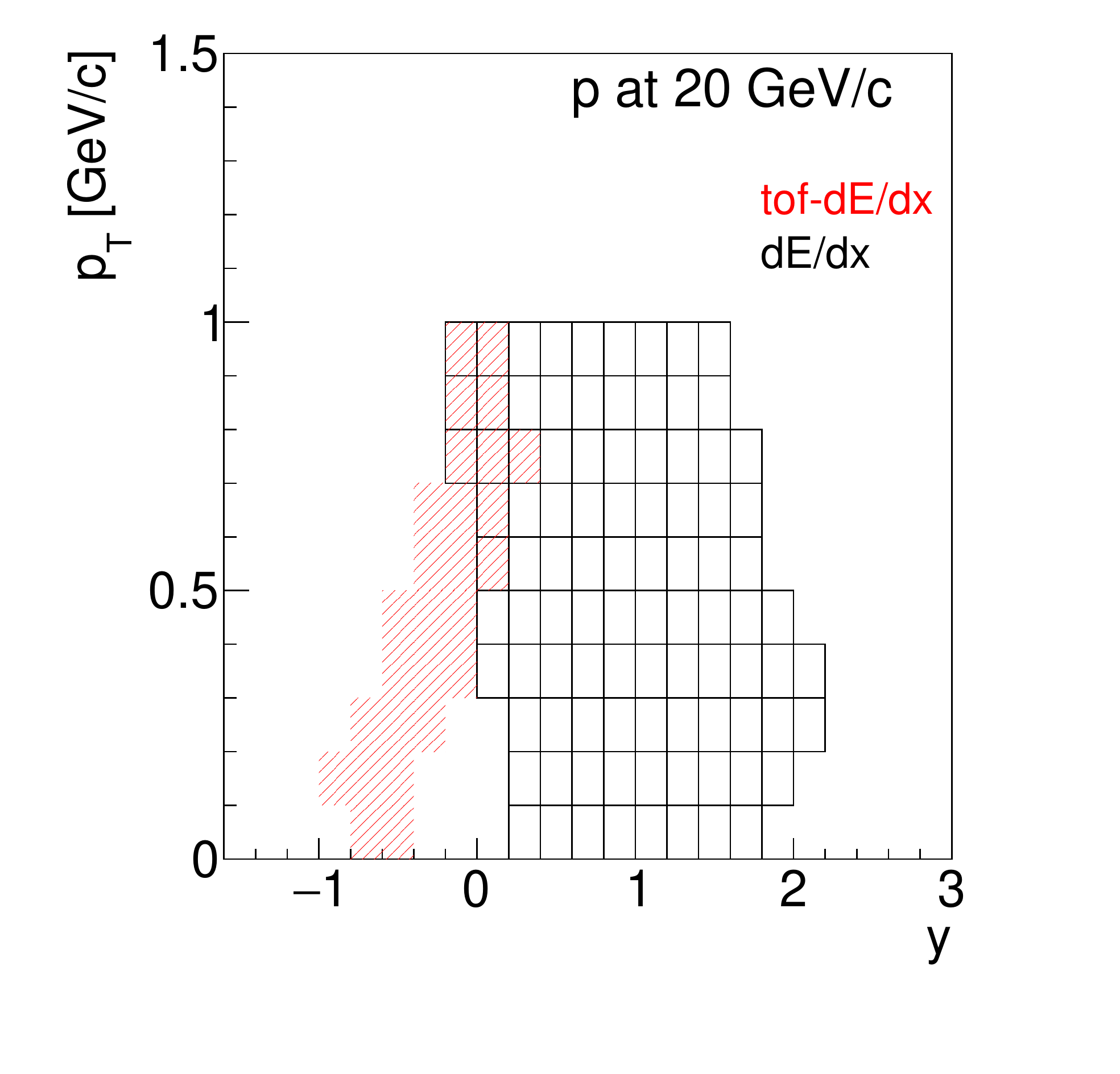}
	\end{center}
	\caption{(Color online) 
    Acceptance of the $tof$-\dedx and \dedx methods for identification of pions, kaons and protons in p+p interactions at 20~\GeVc.
}
	\label{fig:methodacc20}
\end{figure}

\begin{figure}[!ht]
	\begin{center}
	\includegraphics[width=0.3\textwidth]{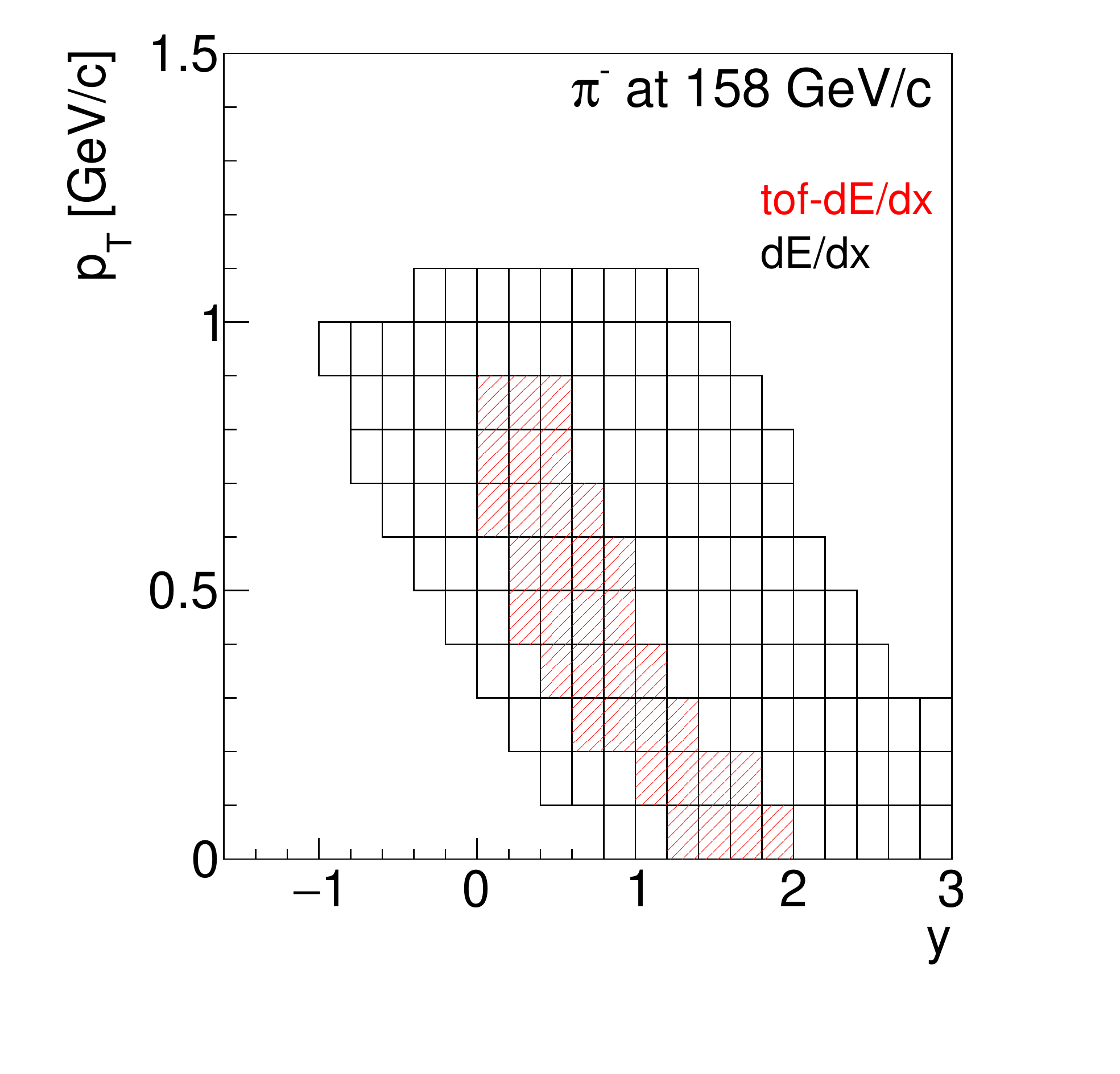}
	\includegraphics[width=0.3\textwidth]{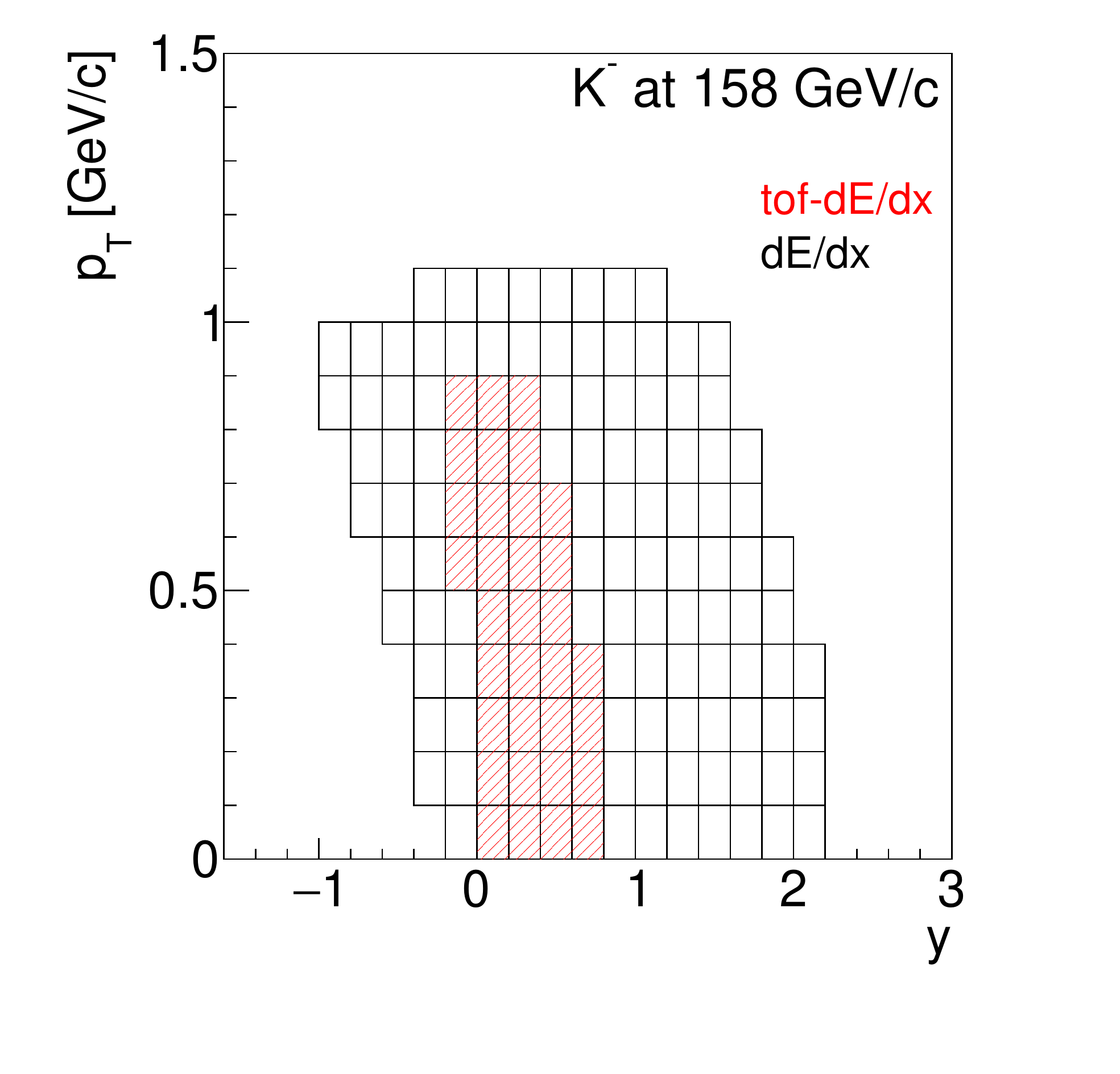}
	\includegraphics[width=0.3\textwidth]{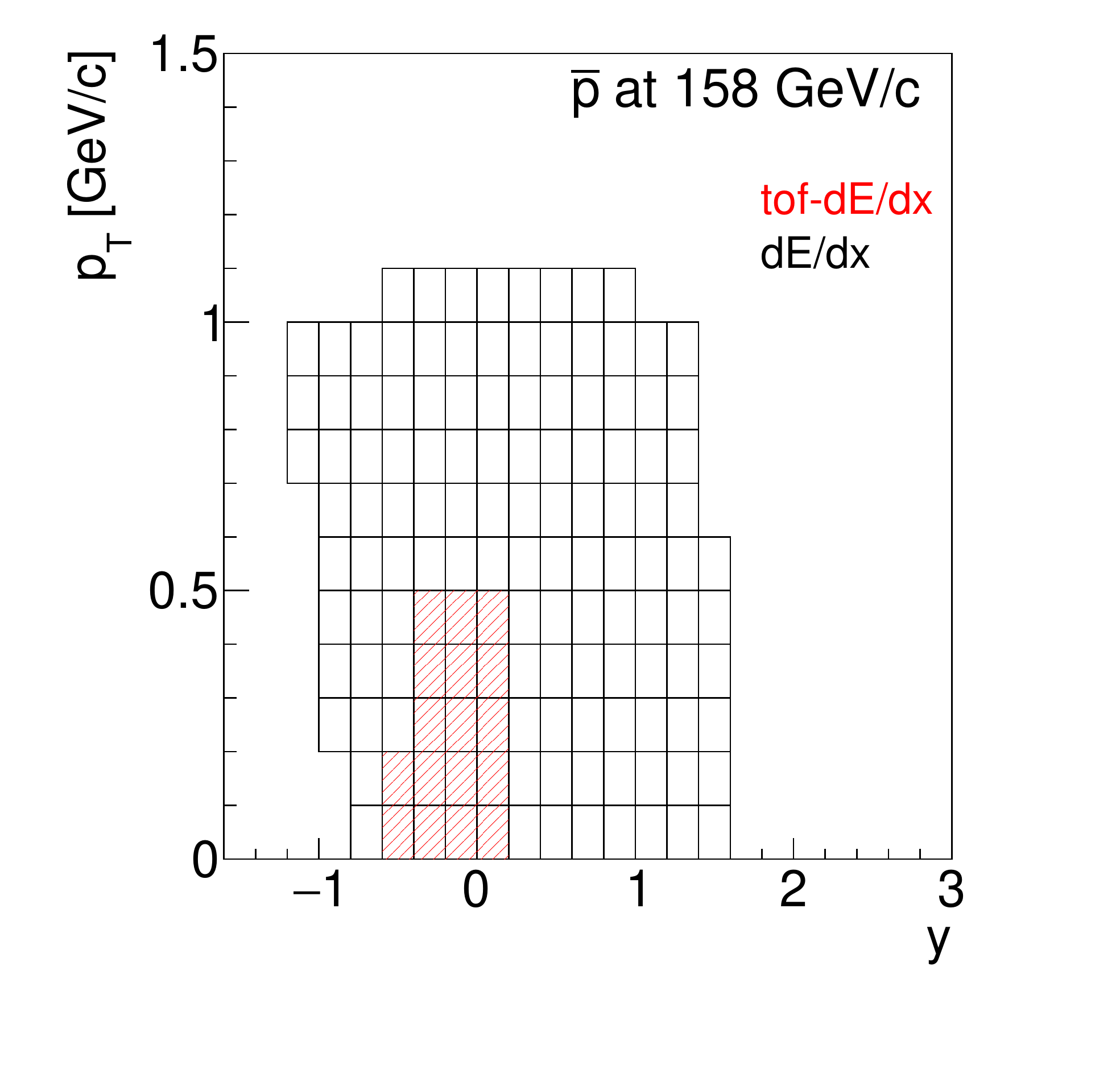}\\
	\includegraphics[width=0.3\textwidth]{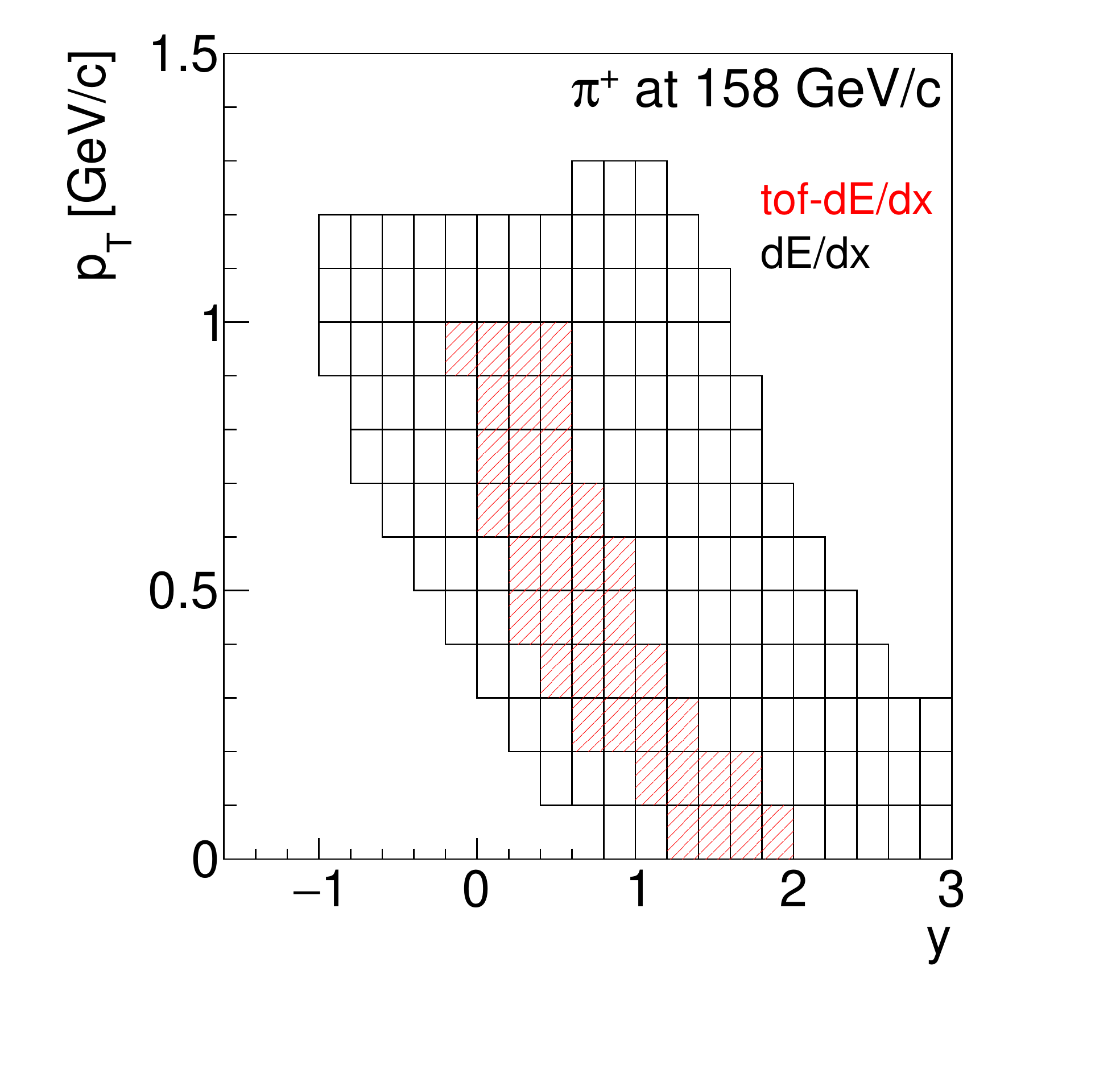}
	\includegraphics[width=0.3\textwidth]{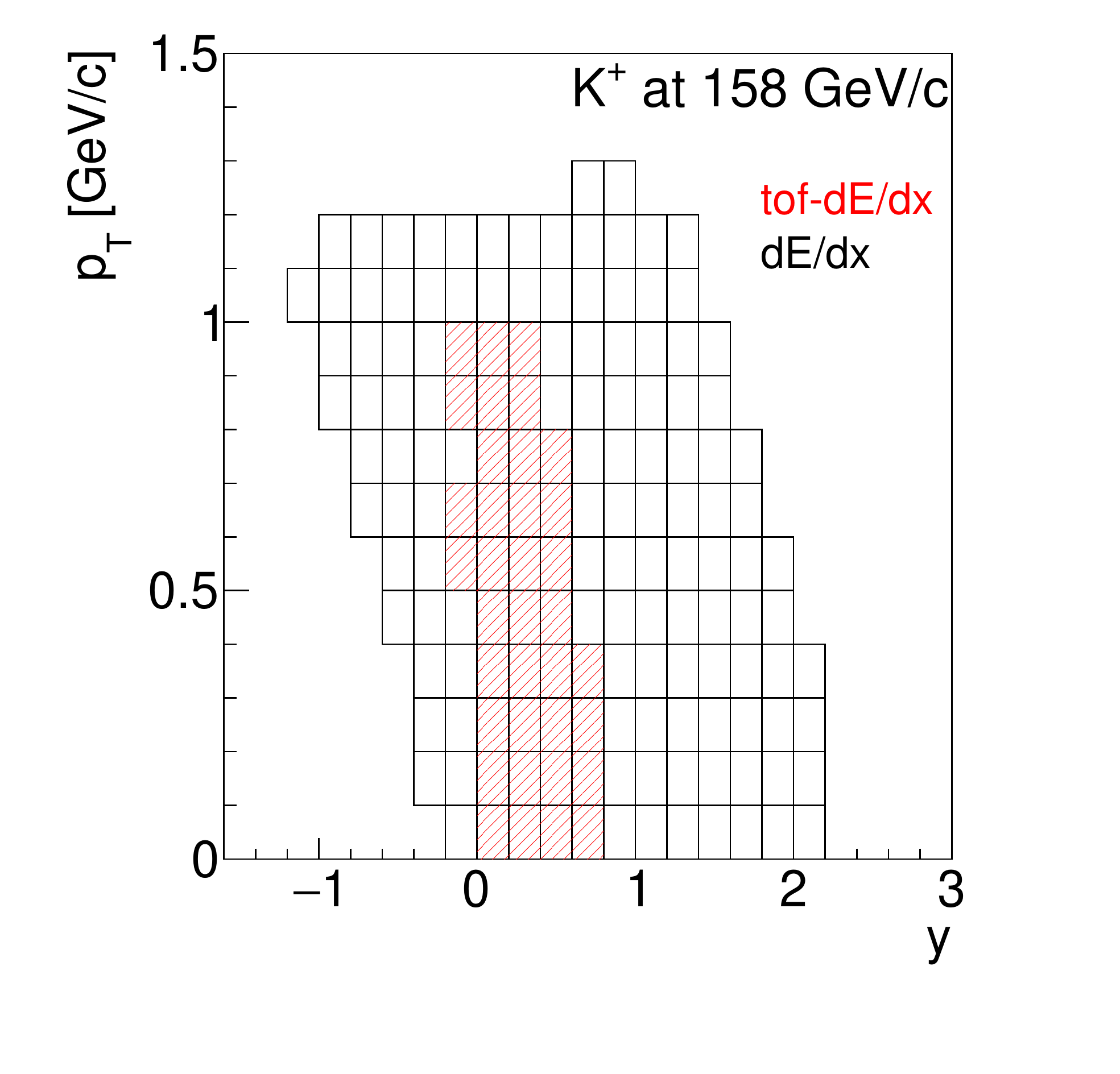}
	\includegraphics[width=0.3\textwidth]{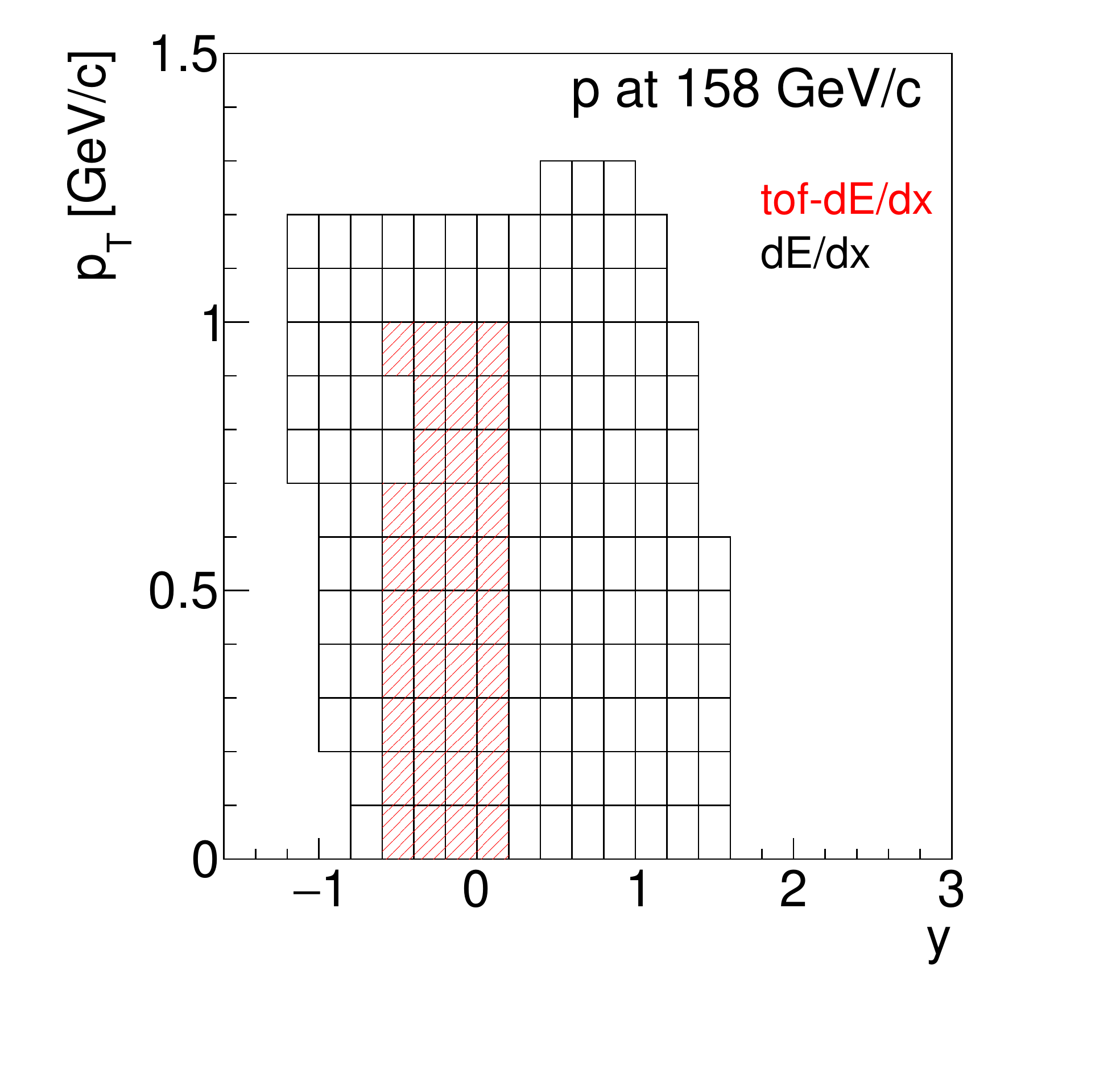}
	\end{center}
	\caption{(Color online) 
    Acceptance of the $tof$-\dedx and \dedx methods for identification of pions, kaons and protons in p+p interactions at 158~\GeVc.
}
	\label{fig:methodacc158}
\end{figure}

Charged particle identification in the \NASixtyOne experiment is based on the measurement
of the ionization energy loss \dedx in the gas of the TPCs and of the time of flight $tof$ obtained
from the ToF-L and ToF-R walls. In the region of the relativistic rise of the ionization at large momenta the measurement of \dedx alone allows identification. At lower momenta the
\dedx bands for different particle species overlap and additional measurement of $tof$ is required
to remove the ambiguity.
These two methods allow to cover most of the phase space in rapidity and transverse momentum which is of interest for the 
strong interaction program of \NASixtyOne. The acceptance of the two methods is shown in 
Figs.~\ref{fig:methodacc20} and~\ref{fig:methodacc158} for p+p interactions at 20 and 158~\GeVc, respectively. 
At low beam energies the $tof$-\dedx method extends the identification acceptance, while at top SPS energy it overlaps with the \dedx method.

\normalsize
\subsubsection{Identification based on energy loss measurement~(\dedx)}
\label{sec:dedx_id}

\begin{figure}[!ht]
	\begin{center}
	\includegraphics[width=0.9\textwidth]{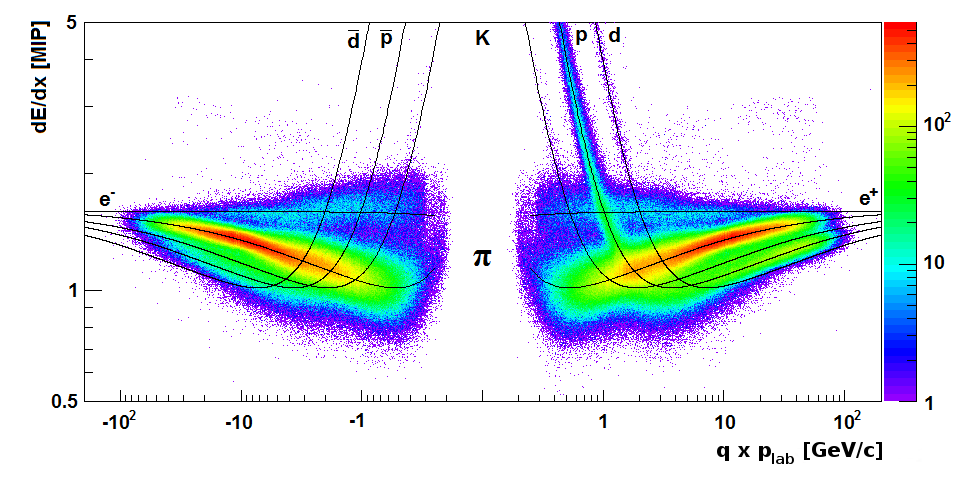}
	\end{center}
	\caption{(Color online) 
    Distribution of charged particles in the \dedx - $q \times p_{\mathrm{lab}}$ plane.	
	The energy loss in the TPCs for different charged particles for events and tracks
	selected for the analysis of p+p interactions at 80~\GeVc 
	(the target inserted configuration). 
	Expectations for the dependence of the mean \dedx on $p_{\mathrm{lab}}$ for the considered particle 
	types are shown by the curves calculated based on the Bethe-Bloch function.
}
	\label{fig:dedx}
\end{figure}

Time projection chambers can provide measurements of energy loss \dedx of charged particles in the chamber gas
along their trajectories. 
Simultaneous measurements of \dedx and $p_{\mathrm{lab}}$ allow to extract information on
particle mass. 
Here \dedx is calculated as the truncated mean (smallest 50\%) 
of cluster charges measured along a track trajectory. As an example, \dedx measured in p+p interactions at 80~\GeVc, for positively and negatively charged particles, as a function of $q \times p_{\mathrm{lab}}$  is presented in Fig.~\ref{fig:dedx}. The expected values of \dedx are shown by the Bethe-Bloch curves. 

The contributions of e$^{+}$, e$^{-}$, $\pi^{+}$, $\pi^{-}$, K$^{+}$, K$^{-}$, p and $\bar{\textrm{p}}$ are 
obtained by fitting the \dedx distributions separately for positively and negatively charged particles in bins of  $p_{\mathrm{lab}}$ and $\pt$
with a sum of four functions~\cite{vanLeeuwen:2003ru,note_MvL} each corresponding to the expected \dedx distribution for a given particle type.

In order to ensure similar particle multiplicities in each bin, 20 logarithmic bins 
are chosen in $p_{\mathrm{lab}}$ in the  range $1-100$~\GeVc to cover the full detector acceptance. 
Furthermore the data are binned in 20 equal $\pt$ intervals in the range 0-2~\GeVc.

The \dedx fits consider four
particle types ($i=$ p, K, $\pi$, e). The signal shape 
for a given particle type is parametrised as the sum of asymmetric Gaussians with widths $\sigma_{i,l}$ 
depending on the particle type $i$ and the number of points $l$ measured in the TPCs.
Simplifying the notation in the fit formulae, 
the peak position of the \dedx distribution 
for particle type $i$ is denoted as $x_{i}$. 
The contribution of a reconstructed particle track to the fit function reads:
\begin{equation}
\rho(x)=\sum_{i}\rho_{i}(x)=\sum\limits_{i=\pi,\textrm{p},\textrm{K},\textrm{e}} A_{i} \frac{1}{\sum\limits_{l} n_{l}} \sum\limits_{l} \frac{n_{l}}{\sqrt{2\pi}\sigma_{l}}exp \left [-\frac{1}{2}\left (\frac{x-x_{i}}{(1\pm\delta)\sigma_{l}} \right ) ^{2} \right ]~,
\label{Eq:AsymGaus}
\end{equation} 
where $x$ is the \dedx of the particle, $n_{l}$ is the number of tracks with number of points $l$ 
in the sample and $A_{i}$ is the amplitude of the contribution of particles of type $i$. 
The second sum is the weighted average of the line-shapes from the different numbers of measured points 
(proportional to track-length) in the sample. The quantity $\sigma_{l}$ is written as:
\begin{equation}
\sigma_{l}=\sigma_{0}\left ( \frac{x_{i}}{x_{\pi}}\right )^{0.625} / \sqrt{n_{l}},
\label{Eq:sigma}
\end{equation}
where the width parameter $\sigma_{0}$ is assumed to be common for all particle types and bins. 
A $1/\sqrt{l}$ dependence on number of points is assumed. The Gaussian peaks are allowed to 
be asymmetric (parameter $\delta$ added/subtracted above/below the peak $x_{i}$) to describe 
the tail of the Landau distribution which may still be present after truncation.

\begin{figure}[!ht]
	\begin{center}
	\includegraphics[width=0.45\textwidth]{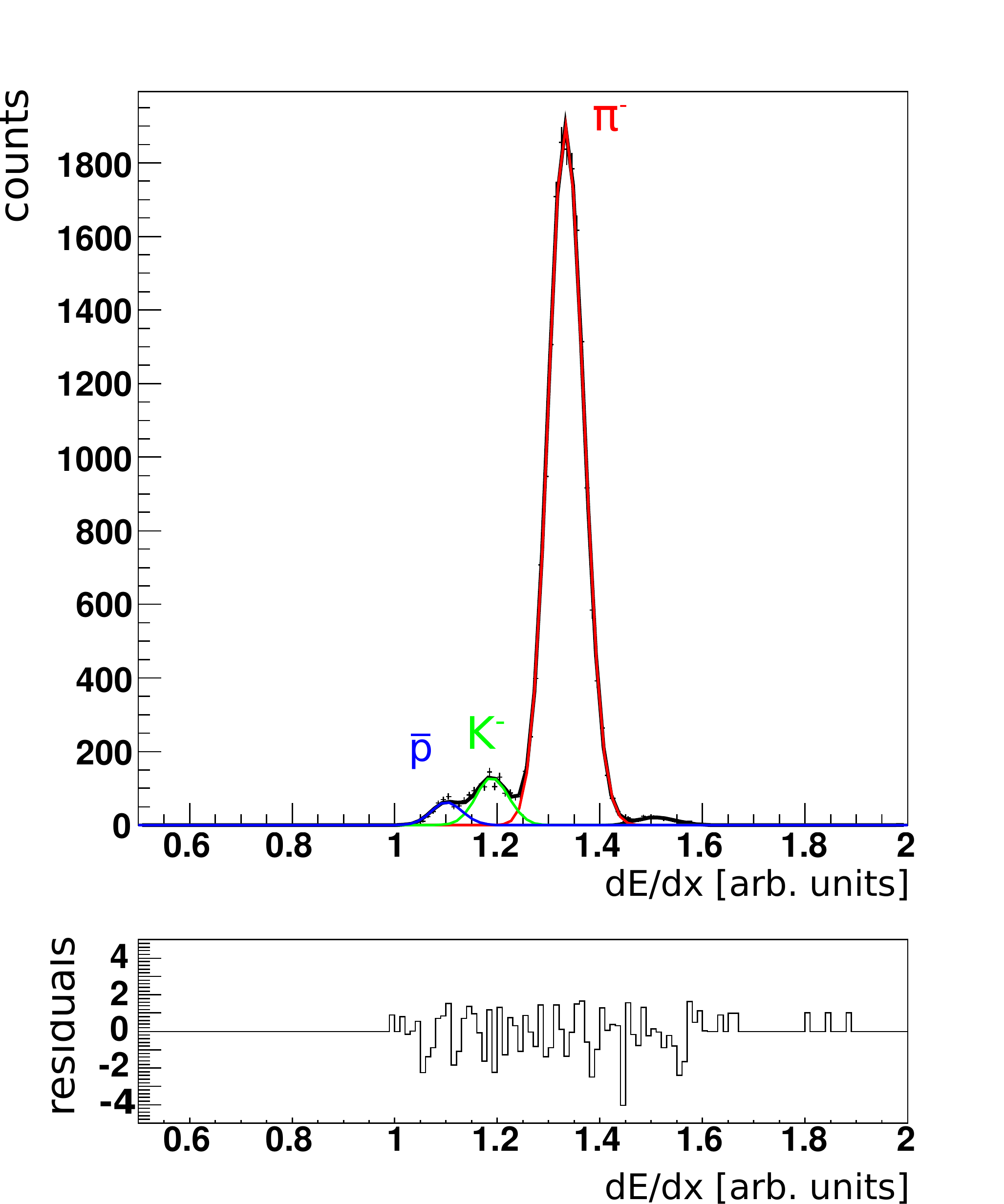}
	\includegraphics[width=0.45\textwidth]{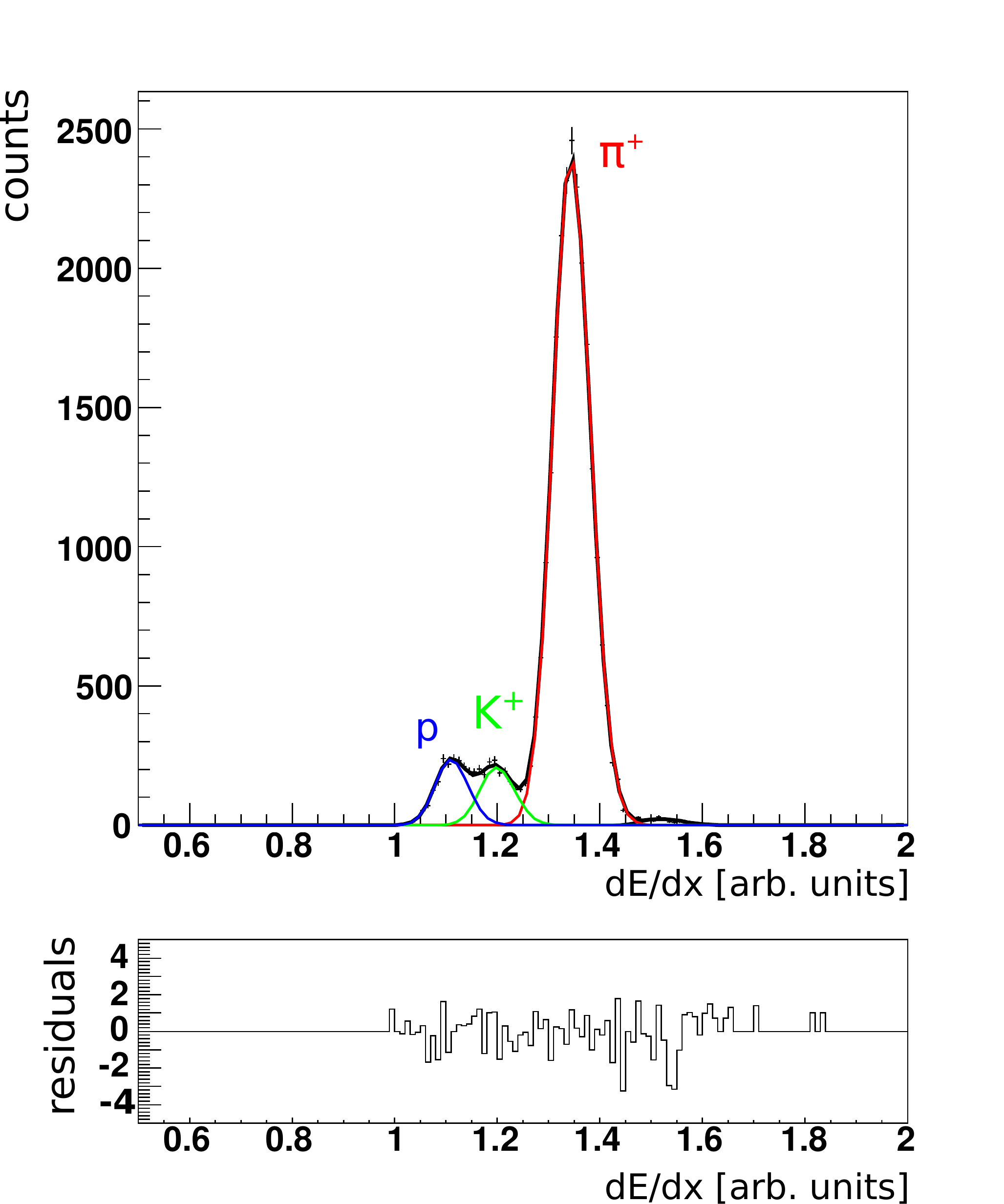}
	\end{center}
	\caption{(Color online) 
    {The \dedx distributions for negatively ($top-left$)} 
    and positively ($top-right$) charged
    particles in the bin 12.6 $< p_{\mathrm{lab}} \leq$ 15.8~\GeVc and 0.2 $< \pt \leq$ 0.3~\GeVc
    produced in p+p interactions at 158~\GeVc.
    The fit by a sum of contributions from different particle types is shown by solid lines.
    The corresponding residuals (the difference between the data and fit divided 
    by the statistical uncertainty of the data) is shown in the bottom plots.  	}
	\label{fig:exfit}
\end{figure}

The fit function has 10 parameters (4 amplitudes, 4 peak positions, width and asymmetry) which are very difficult to fit in each bin independently. Therefore the following simplifications were adopted:
\begin{enumerate}[(i)]
\item relative positions of electrons, kaons and protons to pions were assumed to be $\pt$-independent,
\item in the analysed data, the asymmetry parameter $\delta$ is smaller than 0.001 and thus was fixed to zero,
\item the fitted amplitudes were required to be greater than or equal to 0,
\item the electron amplitude was set to zero for total momentum $p_{\mathrm{lab}}>$23.4~\GeVc (i.e. starting from the 13$^{th}$ bin), as the electron contribution vanishes at high $p_{\mathrm{lab}}$,
\item if possible, the relative position of the positive kaon peak was taken to be the same as that of negative kaons determined from the negatively charged particles in the bin of the same $p_{\mathrm{lab}}$ and $\pt$. This procedure helps to overcome the problem
of the large overlap between K$^+$ and protons in the \dedx distributions. 
\end{enumerate}
The simplifications reduce the number of independently fitted parameters in each bin from 10 to 6, i.e. the amplitudes of the four particle types, the pion peak position and the width parameter $\sigma_{0}$.

\begin{figure}[!ht]
	\begin{center}
	\includegraphics[width=0.45\textwidth]{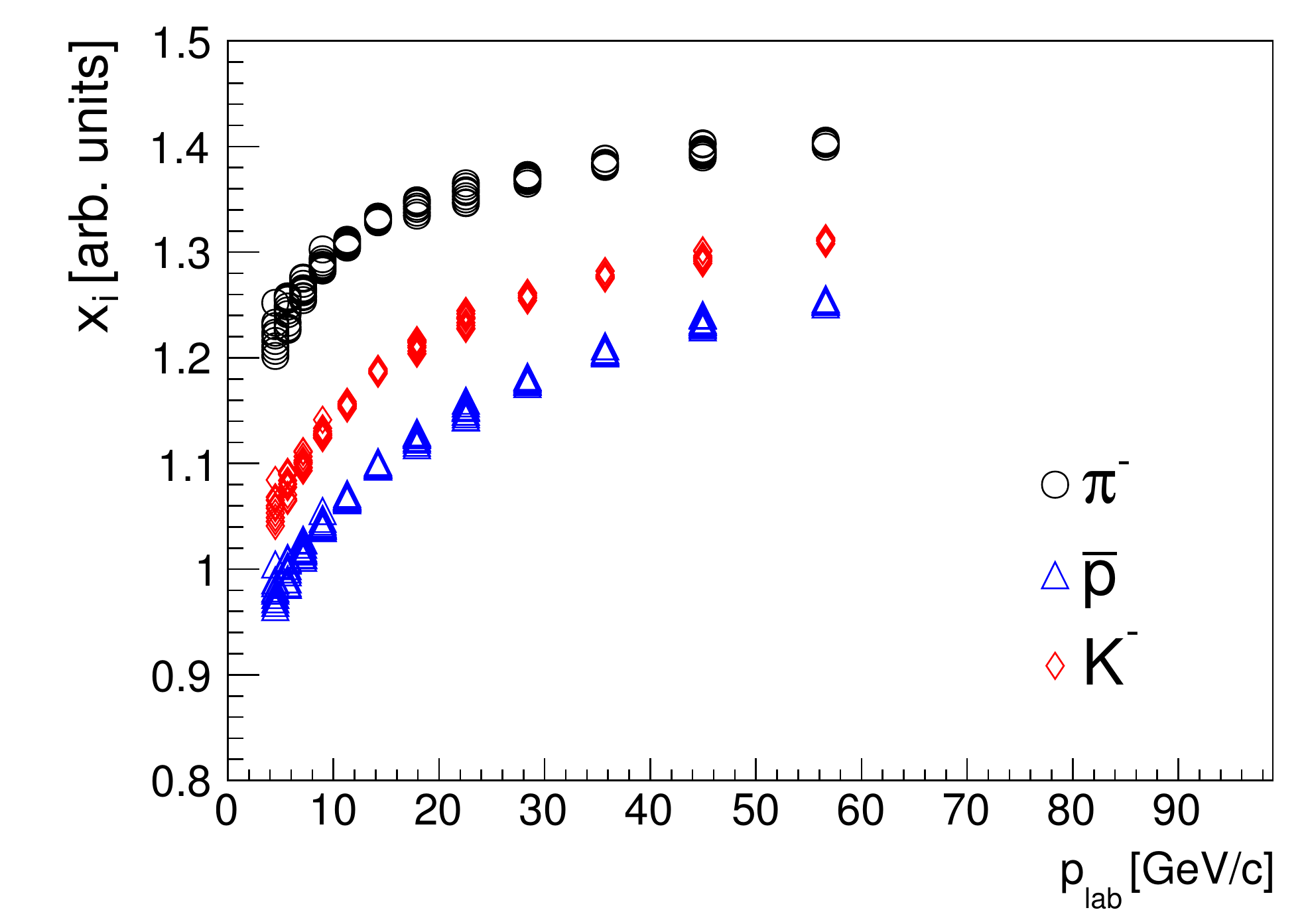}
	\includegraphics[width=0.45\textwidth]{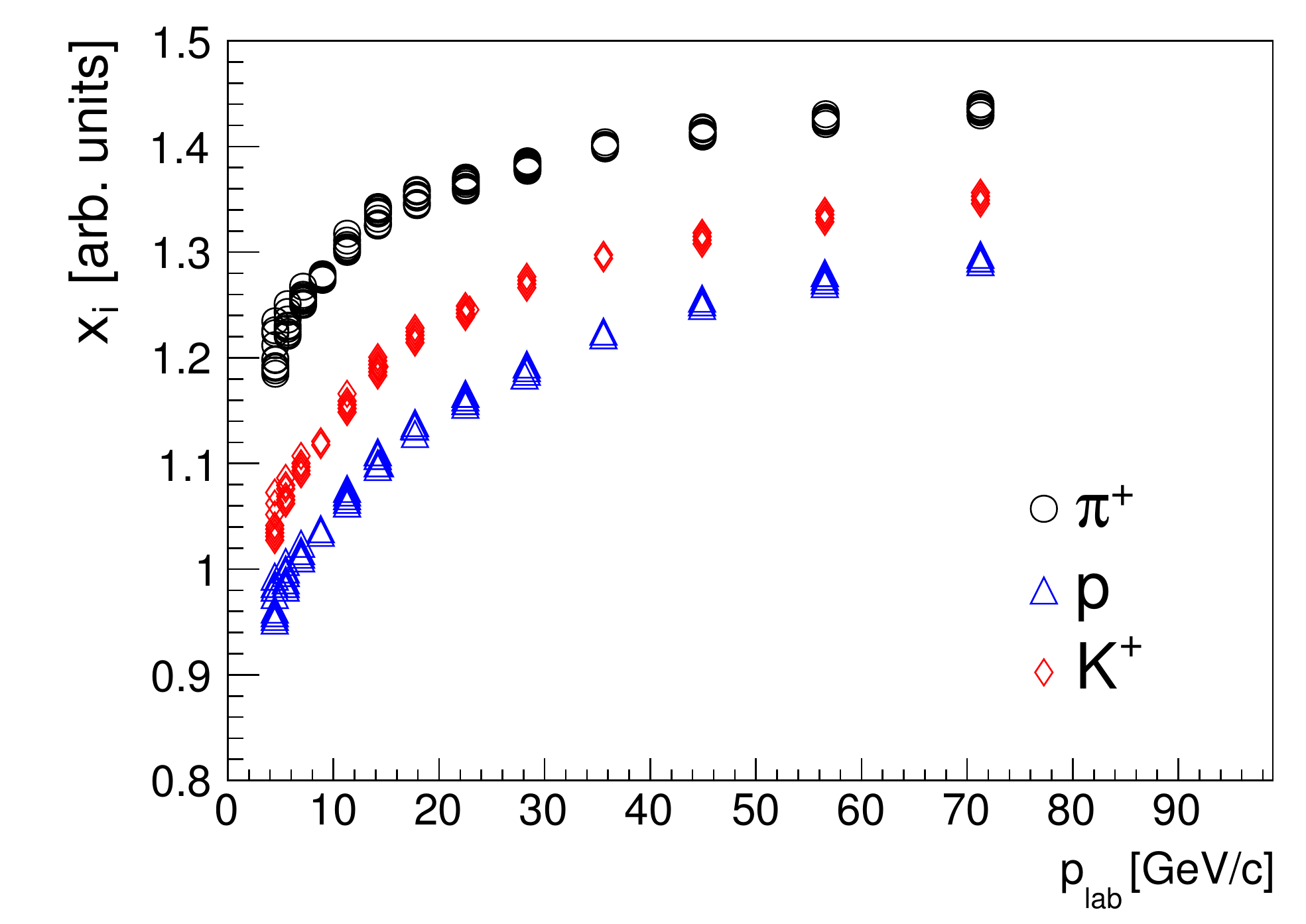}
	\end{center}
	\caption{(Color online) 
	Fitted peak positions in p+p interactions at 158~\GeVc for different particles as a function of $p_{\mathrm{lab}}$. Different points at the same value of $p_{\mathrm{lab}}$ correspond to different transverse momentum bins.}
	\label{fig:fitpos}
\end{figure}

Examples of fits are shown in Fig.~\ref{fig:exfit} and the values of the fitted peak positions $x_i$ are plotted in
Fig.~\ref{fig:fitpos} versus momentum for different particle types $i$ in p+p interactions at 158~\GeVc. As expected,
the values of $x_i$ increase with $p_{\mathrm{lab}}$ but do not depend on $\pt$.   


In order to ensure good fit quality, only bins with number of tracks grater than 300 are used for further analysis. 
The Bethe-Bloch curves for different particle types cross each other at low values of the total momentum. 
Thus, the proposed technique is not sufficient for particle identification at low $p_{\mathrm{lab}}$ and
bins with $p_{\mathrm{lab}}<$3.98~\GeVc (bins 1-5) are excluded from the analysis based solely on \dedx. 

\subsubsection{Identification based on time of flight and energy loss measurements 
($tof$-\dedx)}
\label{sec:tof_id}
Identification of $\pi^{+}$, $\pi^{-}$, K$^{+}$, K$^{-}$, p and $\bar{\textrm{p}}$
 at low momenta (from 2-8~\GeVc) 
is possible when measurement of \dedx  is combined with time of flight ($tof$) information.
Signals from the constant-fraction discriminators TDC and signal amplitude information from ADCs are
recorded for each tile of the ToF-L/R walls. 
Only $tof$ hits which satisfy quality criteria~(see Ref.~\cite{Anticic:2011ny})
are selected for the analysis. 
Tracks reconstructed in the TPCs  are
extrapolated to the front face of ToF-L/R where they are matched to the selected hits. 
The position of
the extrapolation point on the scintillator tile is used to correct the measured value of $tof$ for
the propagation time of the light signal. The distribution of the difference between the corrected
$tof$ measurement and the value predicted from the track momentum and the trajectory length can be 
well described by a Gaussian with standard deviation of 80 ps for ToF-R and 95 ps for ToF-L. 
These values represent the $tof$ resolution including all detector effects.

\begin{figure}[!ht]
        \begin{center}
        \includegraphics[width=0.45\textwidth]{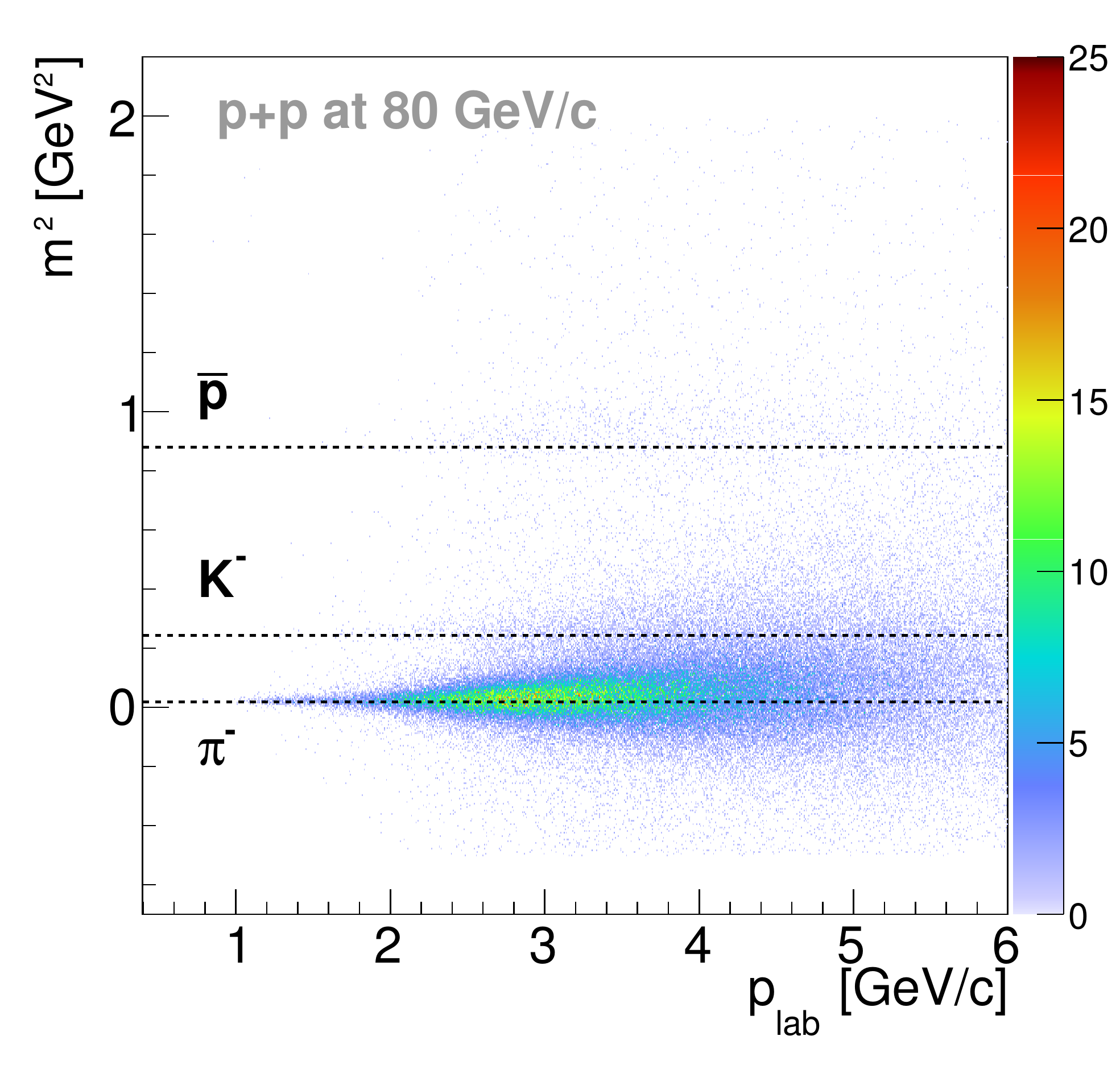}
        \includegraphics[width=0.45\textwidth]{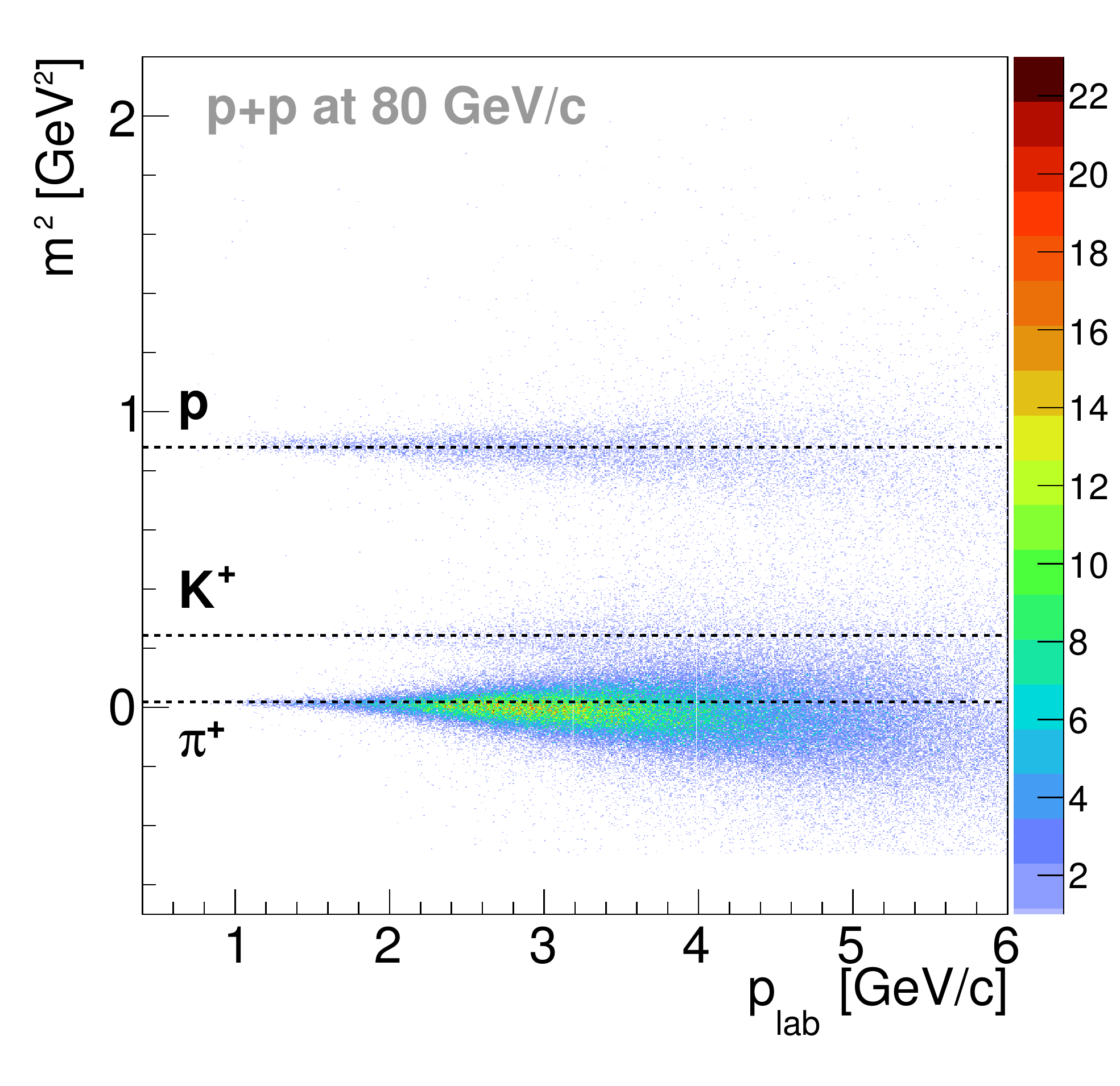}
        \end{center}
        \caption{(Color online) Mass squared versus momentum measured by ToF-R ({\em left}) and
             ToF-L ({\em right}) detectors for particles produced in p+p
             interactions at 80~\GeVc. The lines show the expected mass squared values
             for different hadrons. 
}
        \label{fig:m2vsp}
\end{figure}
The square of the particle mass $m^2$ is obtained from $tof$, the momentum $p$ and
the fitted trajectory length $l$: 
\begin{equation} 
m^2=(cp)^2
\left(\frac{c^2~tof^2}{l^2}-1 \right)~.
\label{eq:m2}
\end{equation}
For illustration distributions of $m^2$ versus $p_{\mathrm{lab}}$ are plotted in Fig.~\ref{fig:m2vsp}
for negatively ({\em left}) and positively ({\em right}) charged hadrons produced in p+p interactions at 80~\GeVc.
Bands which correspond to different particle types are visible.
Separation between pions and kaons is possible up to momenta of about 5~\GeVc, between pions and protons up to about 8~\GeVc.

\begin{figure}[!ht]
	\begin{center}
	\includegraphics[width=0.45\textwidth]{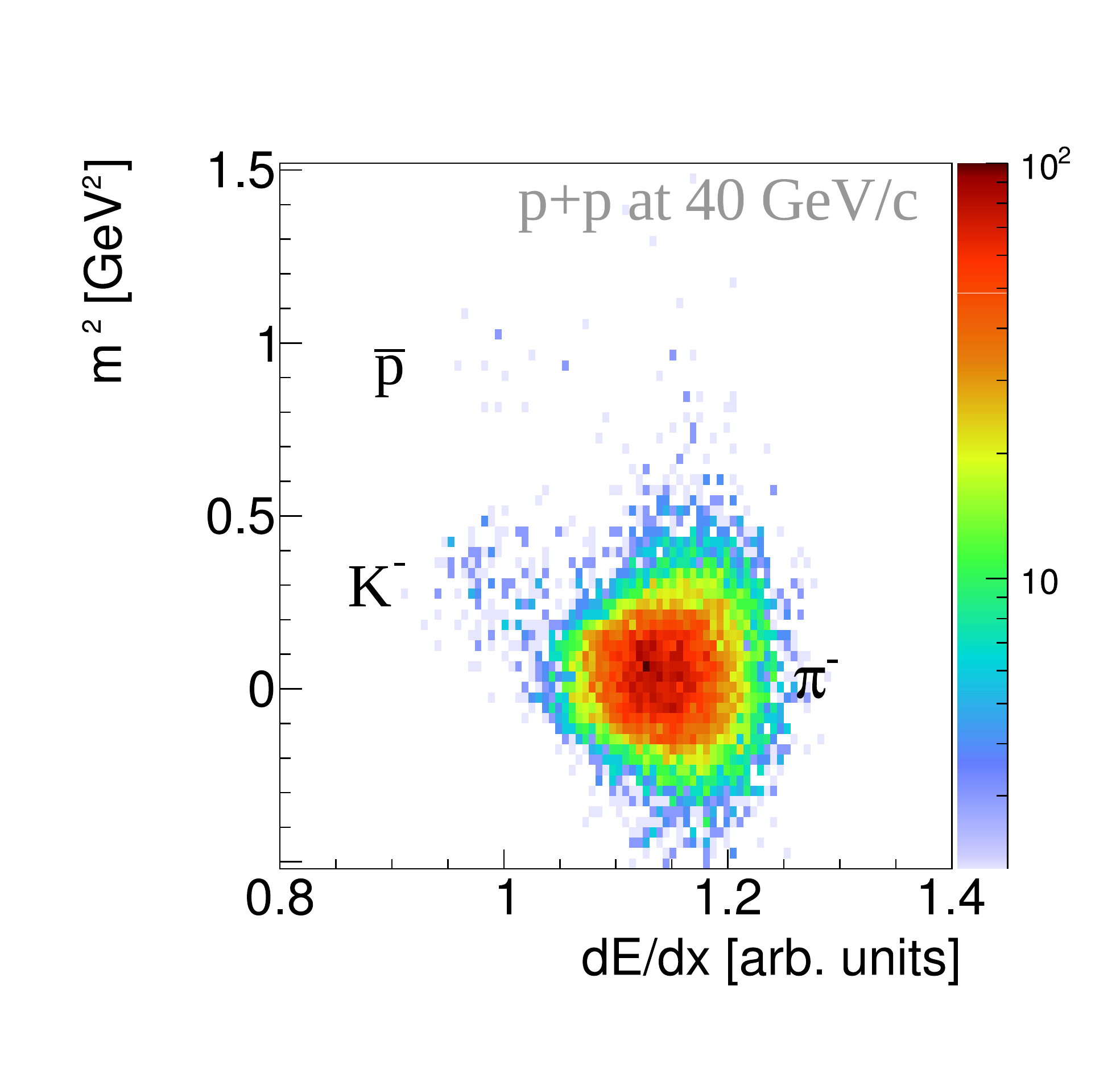}
	\includegraphics[width=0.45\textwidth]{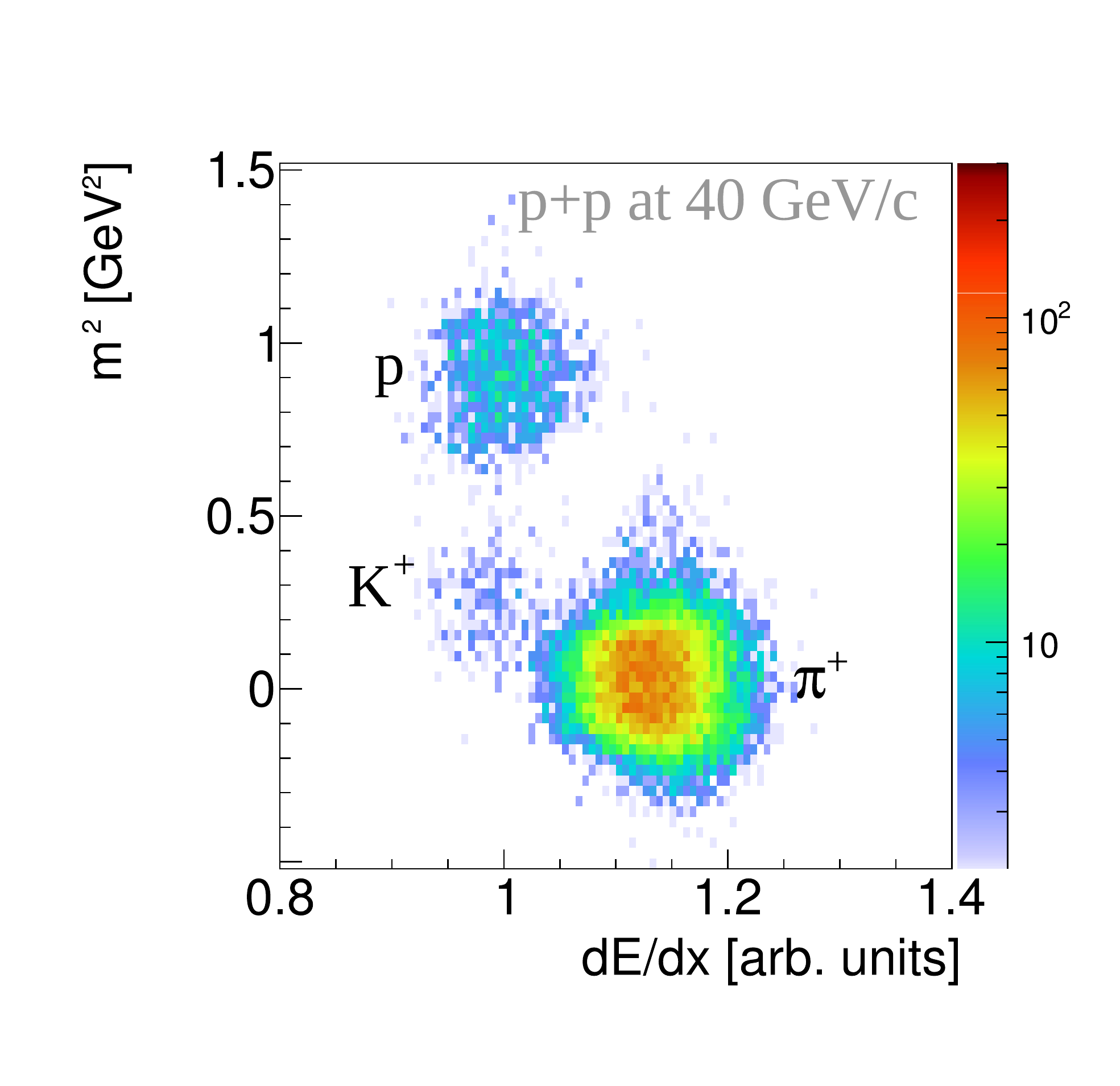}
	\end{center}
	\caption{(Color online) 
Particle number distribution in the $m^2$-\dedx plane for negatively ({\em left}) and positively ({\em right}) hadrons with momenta close to 4~\GeVc for p+p interactions 
at 40~\GeVc. Electrons are not visible since their \dedx values are beyond the \dedx plot range.
}
	\label{fig:tofdedx}
\end{figure}

Example distributions of particles in the $m^2$-\dedx plane for p+p interactions at 40~\GeVc are presented in Fig.~\ref{fig:tofdedx}. 
Simultaneous \dedx and $tof$ measurements lead to improved separation between different hadron types. 
In this case a simple Gaussian parametrization of the \dedx distribution for a given hadron type can be used.

The $tof$-\dedx identification method proceeds by fitting the 2-dimensional distribution of particles in the \dedx-$m^2$ plane.
Fits were performed in 7 equal momentum bins from 1-8~\GeVc and 20 equal bins in transverse
momentum in the range 0-2~\GeVc. For positively charged particles the fit function
included contributions of p, $K^+$, $\pi^+$ and $e^+$, whereas for negatively charged particles the corresponding anti-particles were considered.
The fit function for a given particle type was assumed to be a product of a Gauss function in \dedx and the sum of two Gauss functions in $m^2$.
Then the full fitted function (for simplicity of notation \dedx is denoted by $x$ and $m^{2}$ by $y$) reads:

\begin{equation}
\rho(x,y) = \sum_{i}\rho_{i}(x,y)
  = \sum_{i = \pi,\textrm{p},\textrm{K}} A_{i} e^{-\frac{(x-x_{i})^2}{2\sigma_{x}^2}} 
     ( f e^{-\frac{(y-y_{i})^2}{2\sigma_{y1}^2}} 
    + ( 1 - f ) e^{-\frac{(y-y_{i})^2}{2\sigma^{2}_{y2}}} )~,
\label{eq:2dgaus}
\end{equation} 
where $A_{i}$  and $f$ are the amplitude parameters, $x_i$, $\sigma_{x}$
and $y_i$, $\sigma_{y1}$, $\sigma_{y2}$ are mean and width of the \dedx and
$m^2$ Gaussians, respectively. 
The total number of parameters in Eq.~\ref{eq:2dgaus} is 16.
The fits were performed imposing the following constraints:
\begin{enumerate}[(i)]
\item $y_i = m_i^2$, where $m_i$ is a particle mass~\cite{Agashe:2014kda},
\item relative \dedx positions of electrons, kaons and protons to pions were assumed to be $\pt$-independent,
\item the fitted amplitudes were required to be greater than or equal to 0,
\item if possible, the relative \dedx position of the positive kaon peak was taken to be the same as that of negative kaons determined from the negatively charged particles in the bin of the same $p_{\mathrm{lab}}$ and $\pt$. This procedure helps to overcome the problem
of the large overlap between K$^+$ and protons in the \dedx distributions,
\item $\sigma_{y1} < \sigma_{y2} $ and $f > 0.7 $, the ''core'' distribution dominates the
$m^2$ fit.
\end{enumerate}
The total number of fitted parameters is then reduced to 5.

\begin{figure}[!ht]
	\begin{center}
	\includegraphics[width=0.7\textwidth]{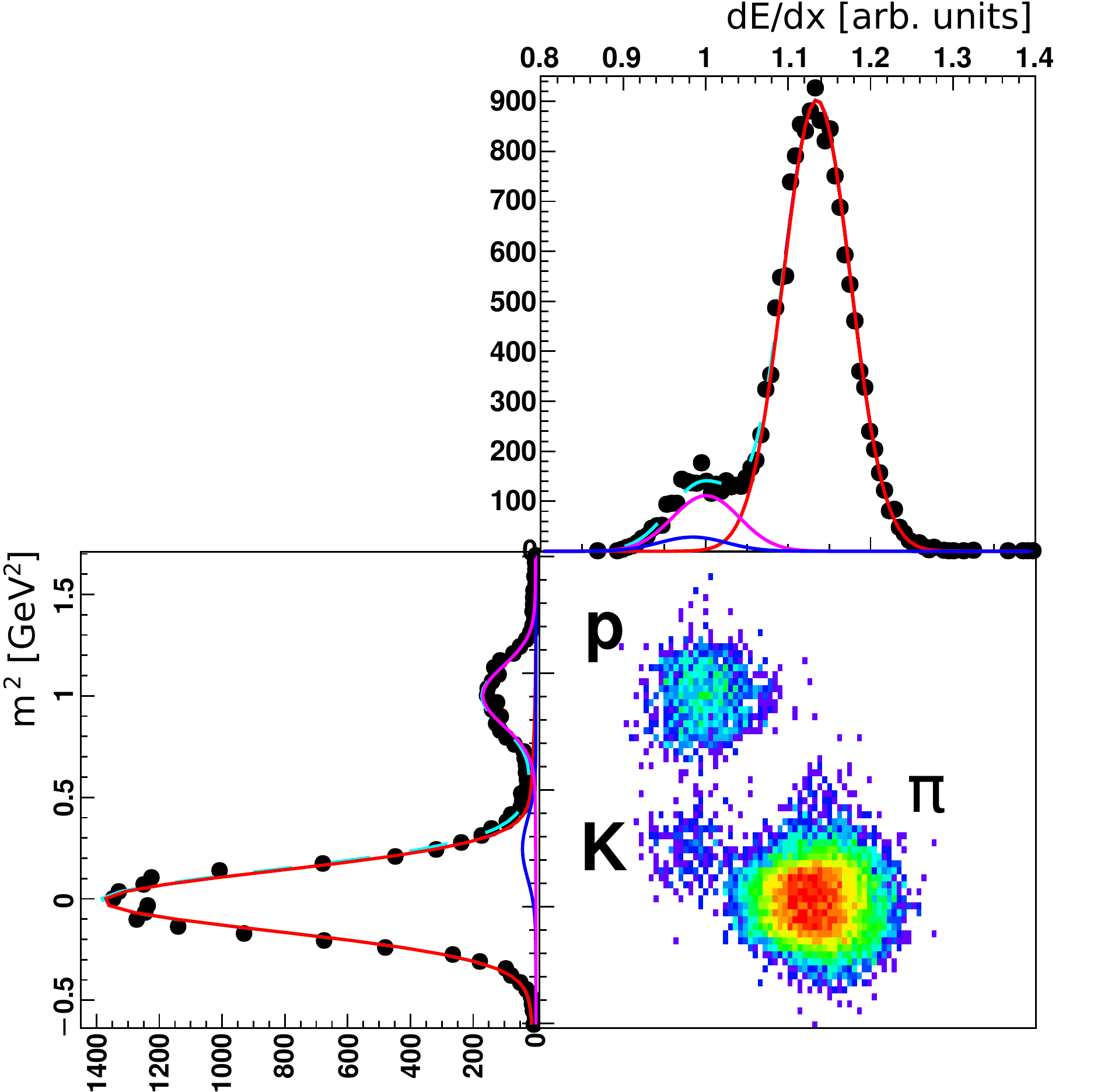}
	\end{center}
	\caption{(Color online) Example of the $tof$-\dedx fit (Eq.~\protect{\ref{eq:2dgaus}}) obtained in a single phase-space bin (3 $< p_{\mathrm{lab}} <$ 4~\GeVc and 0.2 $< \pt <$ 0.3~\GeVc) for positively charged particles in p+p interactions at 158~\GeVc. Lines show projections of obtained fits for pions (red), kaons (blue) and protons (magenta).}
	\label{fig:exmtof}
\end{figure}

An example of the $tof$-\dedx fit obtained in a single phase-space bin for positively charged particles in p+p interactions at 158~\GeVc is shown in Fig.~\ref{fig:exmtof}. 

The $tof$-\dedx method allows to fit the kaon yield close to mid-rapidity. This 
is not possible using the \dedx method. Moreover, the kinematic domain in which pion and proton yields can be fitted is enlarged. The results from both methods partly overlap 
at the highest beam momenta. In these regions the results from the \dedx method were selected since 
they have smaller uncertainties.

\subsubsection{Probability method}
\label{sec:propability}

The probability method allows to transform fit results performed in
($p_{\mathrm{lab}}$, $\pt$) bins to results in ($y$, $\pt$) bins.
The fit results allow to calculate a probability $P_i$ that a measured particle is 
of a given type $i = \pi, K, \textrm{p}, e$, namely for the \dedx fits 
(see Eq.~\ref{Eq:AsymGaus}) one gets:
\begin{equation}
P_{i}(dE/dx)_{p_{\mathrm{lab}}, \pt}=\frac{\rho_{i}(dE/dx)_{p_{\mathrm{lab}}, \pt}}
{\sum\limits_{i=\pi,\textrm{K},\textrm{p},\textrm{e}}^{} 
\rho_{i}(dE/dx)_{p_{\mathrm{lab}},p_{T}}},
\label{eq:propdedx}
\end{equation}
where $\rho_{i}$ is the value of the fitted function in a given ($p_{\mathrm{lab}}$, $\pt$) bin
calculated for \dedx  of the particle. 

\begin{figure}[!ht]
	\begin{center}
	\includegraphics[width=0.45\textwidth]{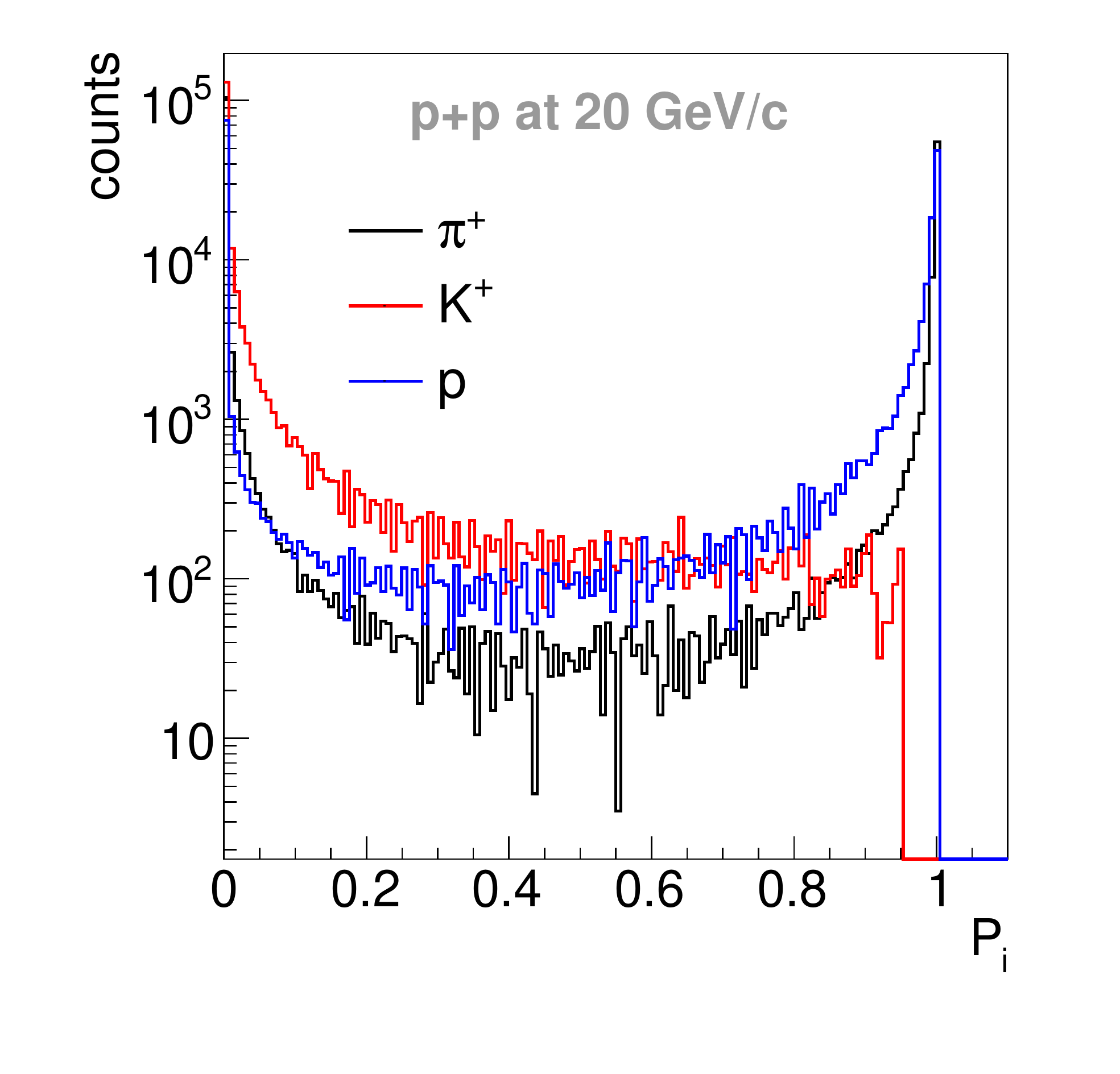}
	\includegraphics[width=0.45\textwidth]{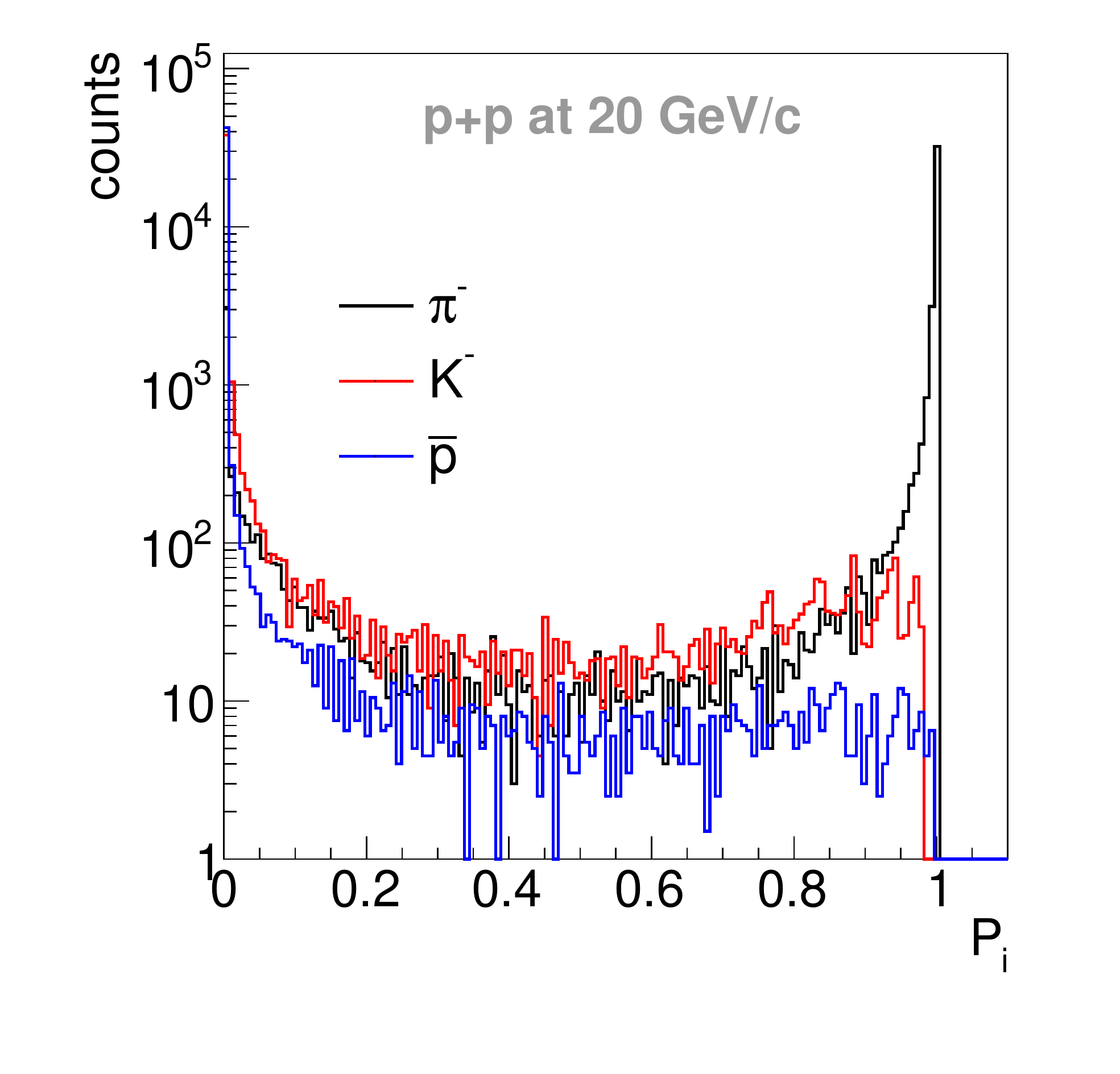}\\
	\includegraphics[width=0.45\textwidth]{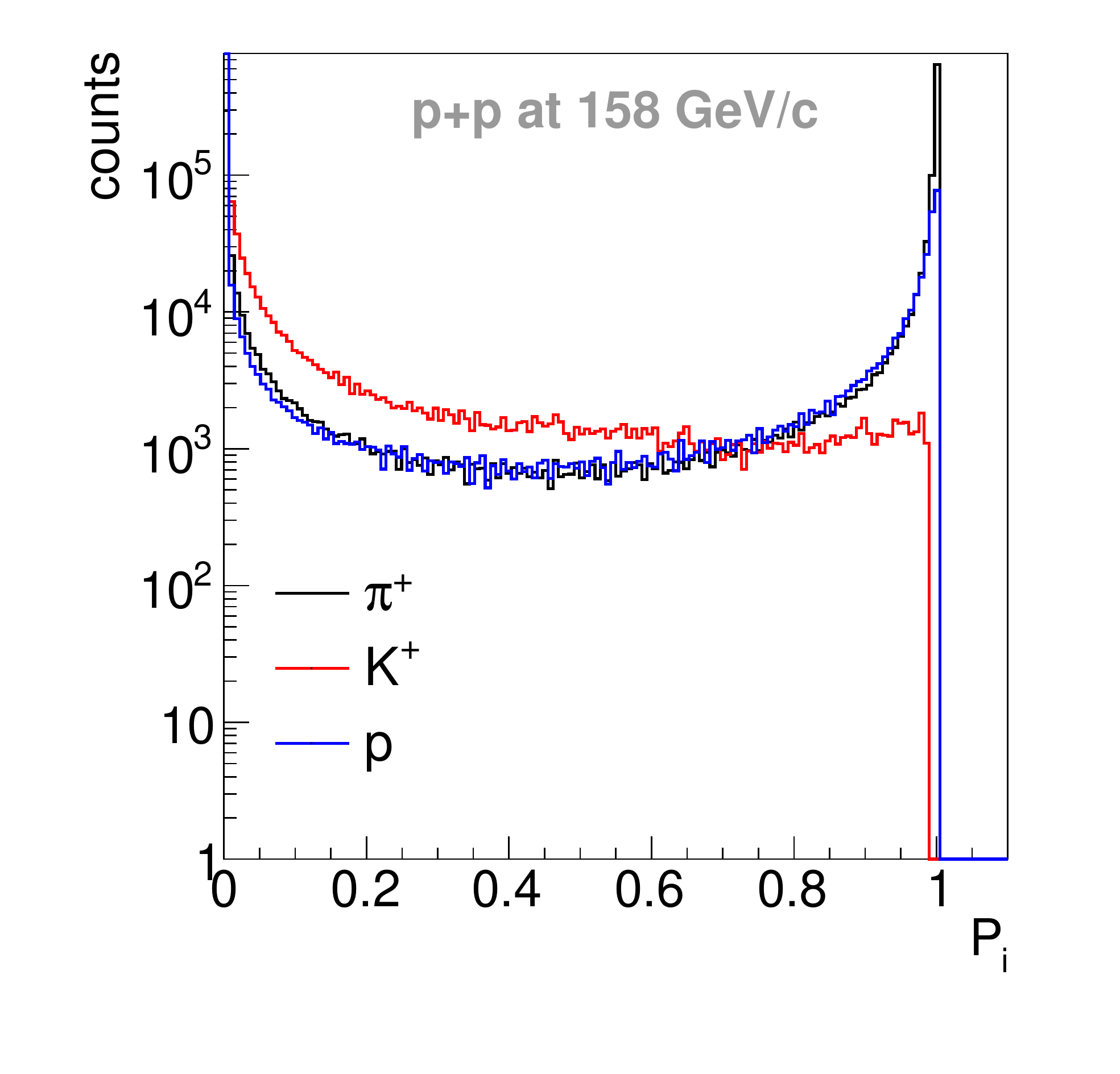}
	\includegraphics[width=0.45\textwidth]{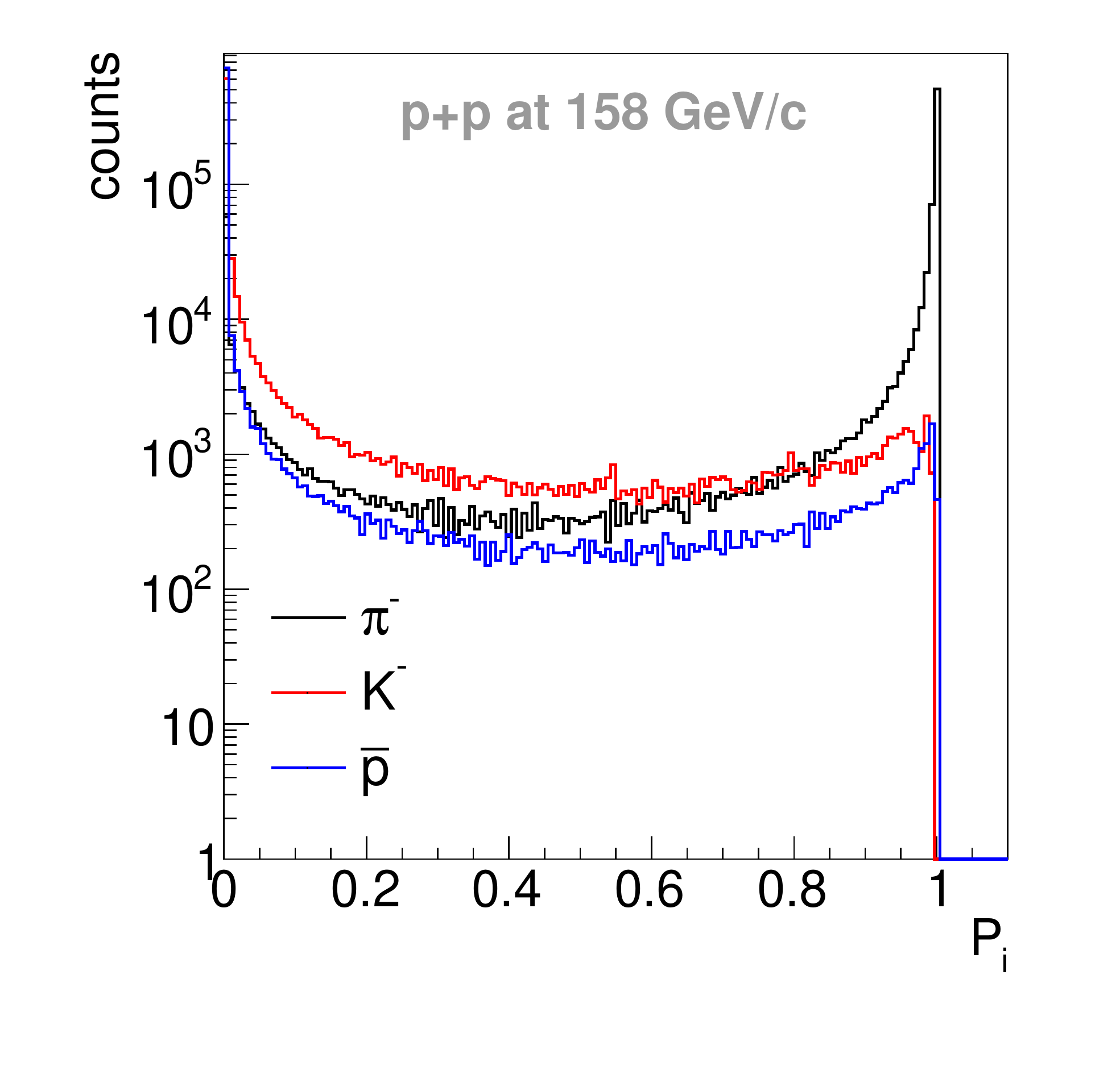}
	\end{center}
	\caption{(Color online) Probability of a track being a pion, kaon, proton for positively (left panels) and negatively (right panels) charged tracks from \dedx measurements in p+p interactions at 20 and 158~\GeVc.}
	\label{fig:propdedx}
\end{figure}

Similarly the $tof$-\dedx fits (see Eq.~\ref{eq:2dgaus})
give a particle type probability as 
\begin{equation}
P_{i}(dE/dx, m^{2})_{p_{\mathrm{lab}}, \pt}=\frac{\rho_{i}(dE/dx,m^{2})_{p_{\mathrm{lab}},\pt}}{\sum\limits_{i=\pi,\textrm{K},\textrm{p},\textrm{e}}^{} \rho_{i}(dE/dx,m^{2})_{p_{\mathrm{lab}},\pt}}.
\label{eq:proptof}
\end{equation}

\begin{figure}[!ht]
	\begin{center}
	\includegraphics[width=0.45\textwidth]{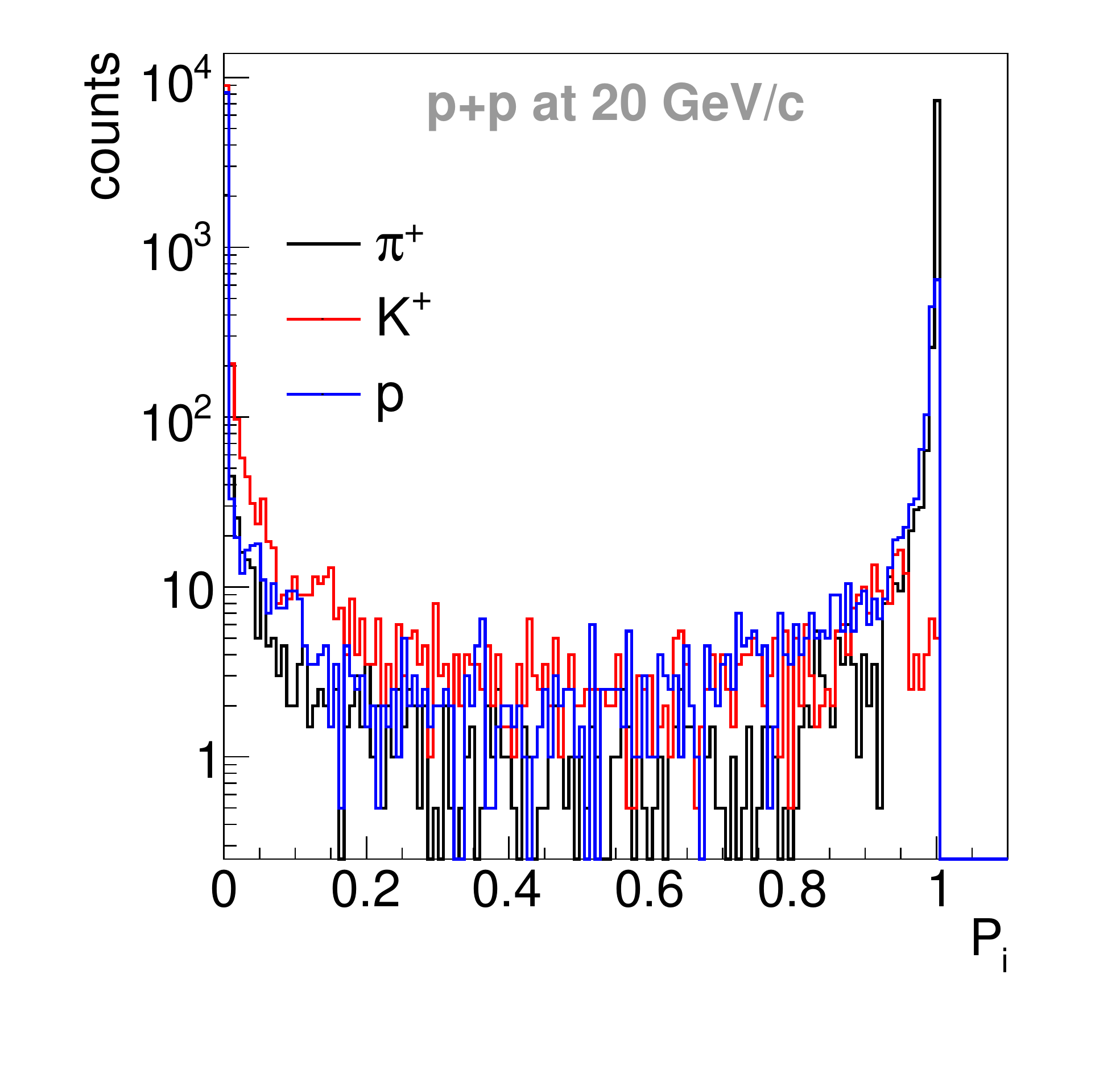}
	\includegraphics[width=0.45\textwidth]{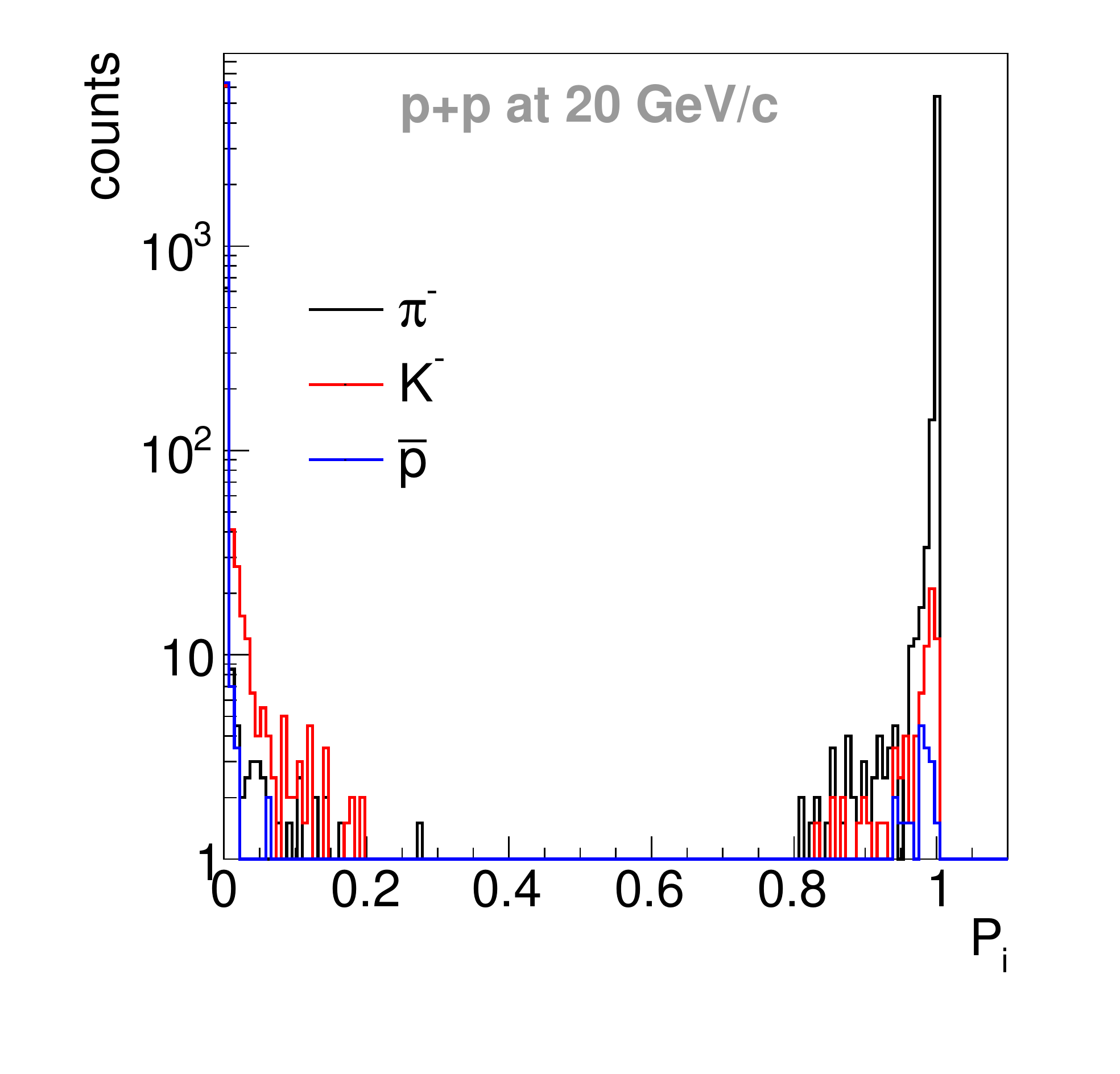}\\
	\includegraphics[width=0.45\textwidth]{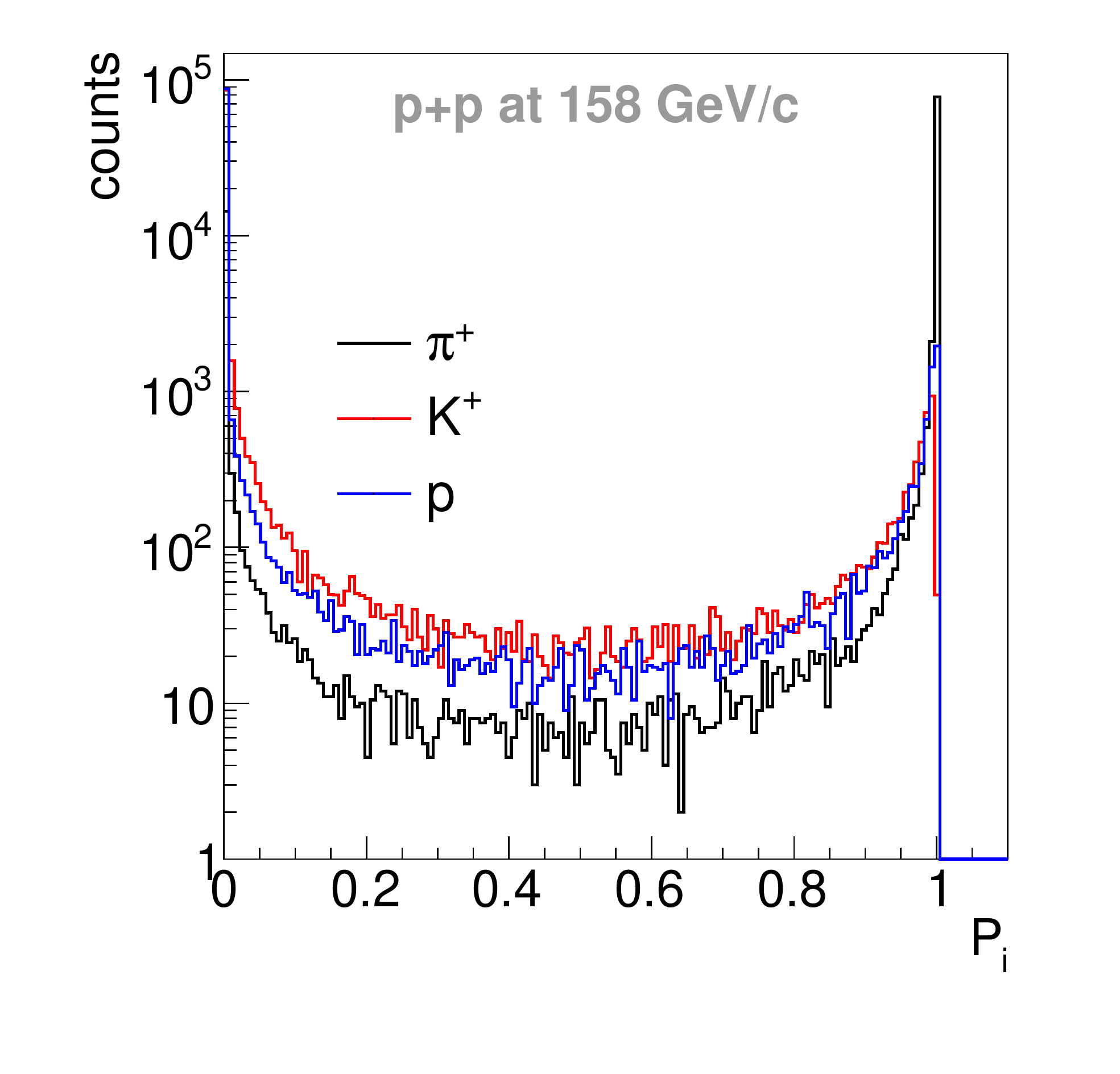}
	\includegraphics[width=0.45\textwidth]{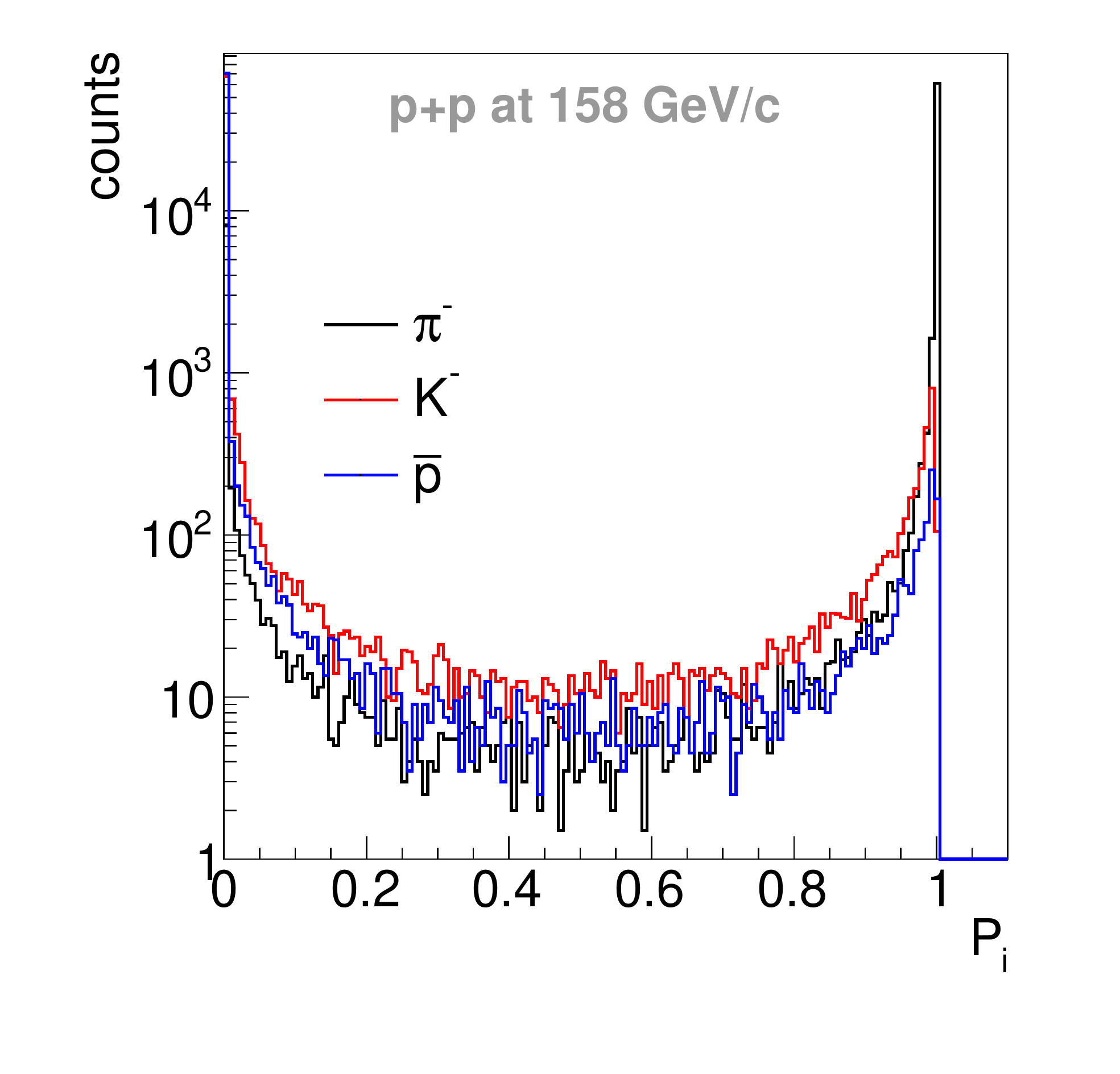}
	\end{center}
	\caption{(Color online) Probability of a track being a pion, kaon, proton for positively (left panels) and negatively (right panels) charged tracks from $tof$-\dedx measurements in p+p interactions at 20 and 158~\GeVc.}
	\label{fig:proptof}
\end{figure}

For illustration particle type probability distributions for positively and negatively charged particles
produced in p+p interactions at 20 and 158~\GeVc
are presented in Fig.~\ref{fig:propdedx}
for the \dedx fits and in Fig.~\ref{fig:proptof} for the $tof$-\dedx fits. Only in the case of perfect particle type discrimination the probability distributions in Figs.~\ref{fig:propdedx} and \ref{fig:proptof} will show peaks at 0 or 1. In the case of non-ideal discrimination (overlapping \dedx or $tof$-\dedx distributions) values between these extremes will be populated.

The numbers of identified particles in a given kinematical bin 
(e.g.,($p_{\mathrm{lab}}$, $\pt$)) are given by~\cite{Rustamov:2012bx}:
\begin{equation}
n_{i=\pi,\textrm{K},\textrm{p}}=\sum_{j=1}^{n}P_{i},
\label{eq:ntracks}
\end{equation}
where $P_i$ is the probability of particle type $i$ given by 
Eqs.~\ref{eq:propdedx} and~\ref{eq:proptof} and $n$ is the number of particles in the
given kinematical bin.

Additionally the probability method allows to implement the efficiency correction particle by particle. The corrected number of identified particles in the given kinematical bin is:
\begin{equation}
n_{i=\pi,\textrm{K},\textrm{p}}=\sum_{j=1}^{n}P_{i}\epsilon,
\label{eq:ntracksepilon}
\end{equation}
where as in Eq.~\ref{eq:ntracks} $P_i$ is the probability of particle type $i$ given by 
Eqs.~\ref{eq:propdedx} and~\ref{eq:proptof} and $n$ is the number of particles in the
given kinematical bin and $\epsilon$ is the efficiency. Usually the efficiency is calculated
with a different binning scheme. Thus the value of $\epsilon$ has to be taken from the
bin which corresponds to the kinematic quantities of particle $i$. 

The probability method has some subtleties when the coverage of phase-space differs between identification and final variable. For example particles in bins selected for the identification ($p_{\mathrm{lab}}$) may not fully populate the edges of bins in the final variable (y) due to the non-orthogonal transformation between them. However, only the edges of the spectra are affected. For the \dedx analysis method bins could be selected in such a way that edges almost overlap. The remaining small effect was corrected by the overall Monte-Carlo correction factor. The mismatches are larger for the $tof$-\dedx method where identification had to be performed in linear, equal bins in total momentum due to the $ToF$ acceptance. Therefore a separate correction factor $\epsilon$ was calculated using the simulation based on the \Epos model (see below):
\begin{equation}
\epsilon=\frac{n_{MC}^{\mathrm{accepted}}}{n_{MC}^{\mathrm{generated}}}
\label{eq:epsilontof}
\end{equation} 
where $n_{MC}^{\mathrm{accepted}}$ are the numbers of tracks accepted in a rapidity bin within the phase space covered by the identification technique and $n_{MC}^{\mathrm{generated}}$ the number of tracks generated in the corresponding rapidity bin. Bins in which this correction exceeds 30\% are rejected and the others are corrected by this factor.


\FloatBarrier
   \subsection{Corrections}
\label{sec:Corrections}
In order to determine the true number of each type of identified particle produced in inelastic p+p interactions a 
set of corrections was applied to the extracted raw results. The main effects for which corrections were introduced 
are the following: contribution of interactions outside the liquid hydrogen of the target (off-target events), 
detector effects (acceptance, efficiency) and particles from weak decays (feed-down). Note that the manner
of application and the number of used correction factors depend on the particle identification technique 
(i.e. \dedx or $tof$-\dedx)

	
\subsubsection{Correction for off-target interactions}
\label{sec:target-off}
	
To estimate the off-target interactions about 10\% of the data were collected without the liquid hydrogen in
the target (so-called target removed data denoted as R). 	
Before the identification procedure (see section ~\ref{sec:dedx_id}) a suitably normalized target removed yield was
subtracted from target inserted data. This correction was applied for each bin of total momentum and transverse momentum.  
	

	
The normalization of the target removed data was based on the fitted vertex z distribution. The ratio of the
numbers of events with fitted vertex outside the target (in the range from -400~\cm to -200~\cm) was calculated
for target inserted and removed data and used subsequently as the normalization factor.	
The contamination of out of target events in the target inserted sample is given in Table~\ref{tab:offtarget} 
for p+p interactions at 20, 31, 40, 80 and 158~\GeVc.
	
	\begin{table}[!ht]
	\begin{center}
	\caption{Measured fraction of out-of-target events in recorded p+p interactions at SPS energies.}
	\begin{tabular}{|c|c|}
	\hline
	Beam momentum [GeV/c] & Fraction of target removed events \\
	\hline
	20   & 20.22\% \\
	31 & 26.17\% \\
	40   & 15.84\% \\
	80   & 12.53\% \\
	158  & 9.62\% \\
	\hline
	\end{tabular}
	\label{tab:offtarget}
	\end{center}
	\end{table}
	 
\subsubsection{Corrections for detector effects and particles from weak decays (feed-down)}
\label{sec:decays}	
	

A simulation of the \NASixtyOne detector is used to correct the data for
reconstruction efficiency and acceptance. Only inelastic p+p interactions on the hydrogen in the target cell
were simulated and reconstructed. The \Epos model~\cite{Werner:2008zza} was selected to generate
the primary interactions as it best describes the \NASixtyOne measurements. A \GeantThree based program chain
was used to track particles through the spectrometer, generate decays and secondary interactions and
simulate the detector response (for more detail see Ref.~\cite{Abgrall:2013pp_pim}).
The reconstructed tracks were matched to the simulated particles based on the cluster positions.
The derived corrections can be applied only for inelastic events.
The contribution of elastic events in the data was eliminated by the event and track selection cuts.
Hadrons which were not produced in the primary interaction can amount to a significant fraction of the selected
track sample. Thus a special effort was undertaken to evaluate and subtract this contribution (see above).  
As mentioned before correction factors depend on the particle identification technique and are described separately.
							
\emph{\textbf The \dedx method.}\\
The correction factor $C$ is defined as:

\begin{equation}
C_{i} = \frac{\left(n_{i}\right)^{MC}_{sel}} { \left(n_{i}\right)^{MC}_{gen}},
\label{eq:corectionfactor}
\end{equation}

where:
\begin{itemize}
	\item $\left(n_{i}\right)\ ^{MC}_{gen}$ - multiplicity of particle type $i$ $\left(i=\pi^{+/-},\ \textrm{K}^{+/-},\ \textrm{p},\ \bar{\textrm{p}} \right)$ generated by the \Epos model,
	\item $\left(n_{i}\right)\ ^{MC}_{sel}$ - multiplicity of particle type $i$ $\left(i=\pi^{+/-},\ \textrm{K}^{+/-},\ \textrm{p},\ \bar{\textrm{p}} \right)$ after applying the selection criteria described in the previous section,
\end{itemize}

\begin{figure}[!ht]
\begin{center}
\includegraphics[width=0.3\textwidth]{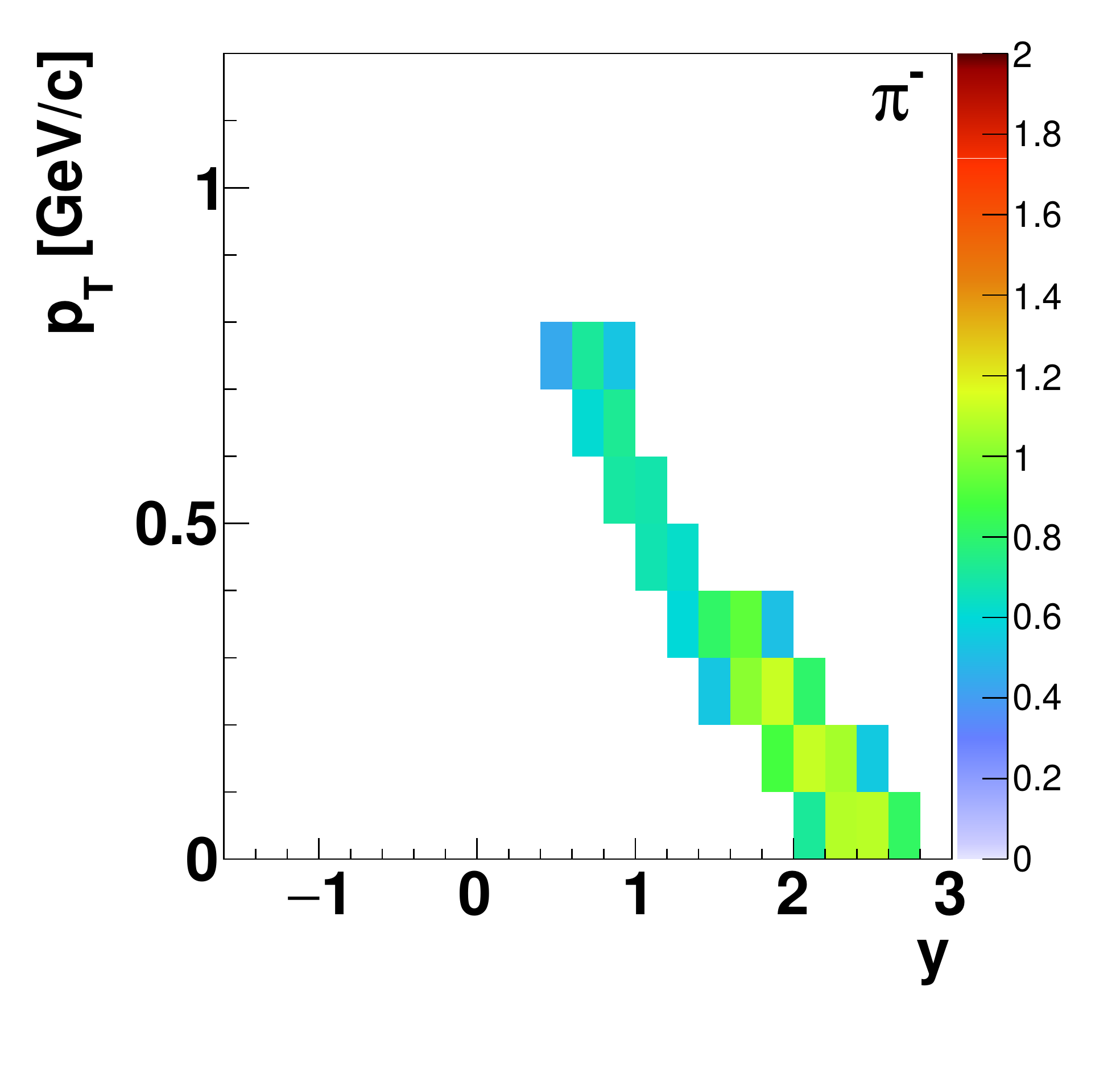}
\includegraphics[width=0.3\textwidth]{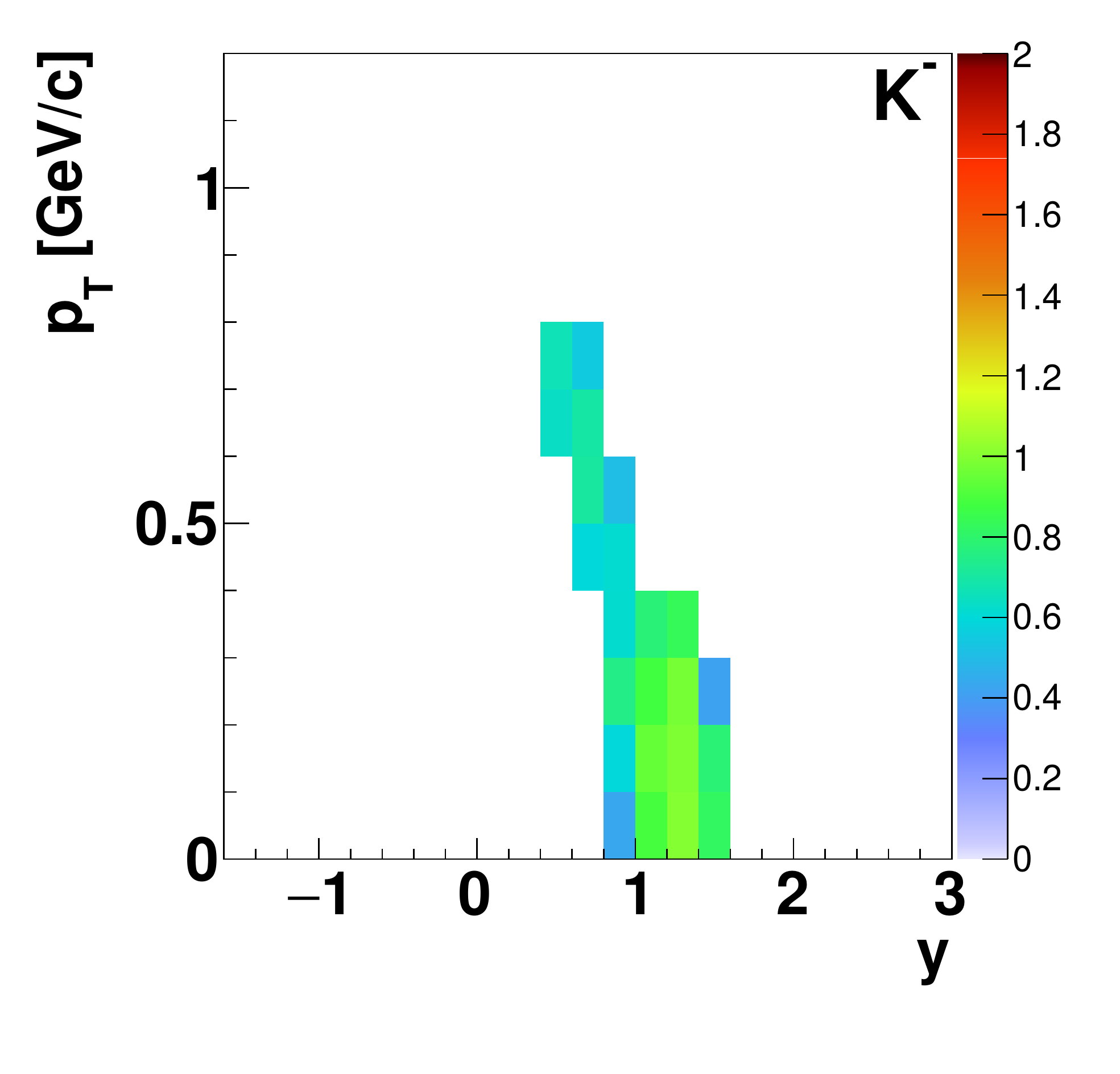}\\
\includegraphics[width=0.3\textwidth]{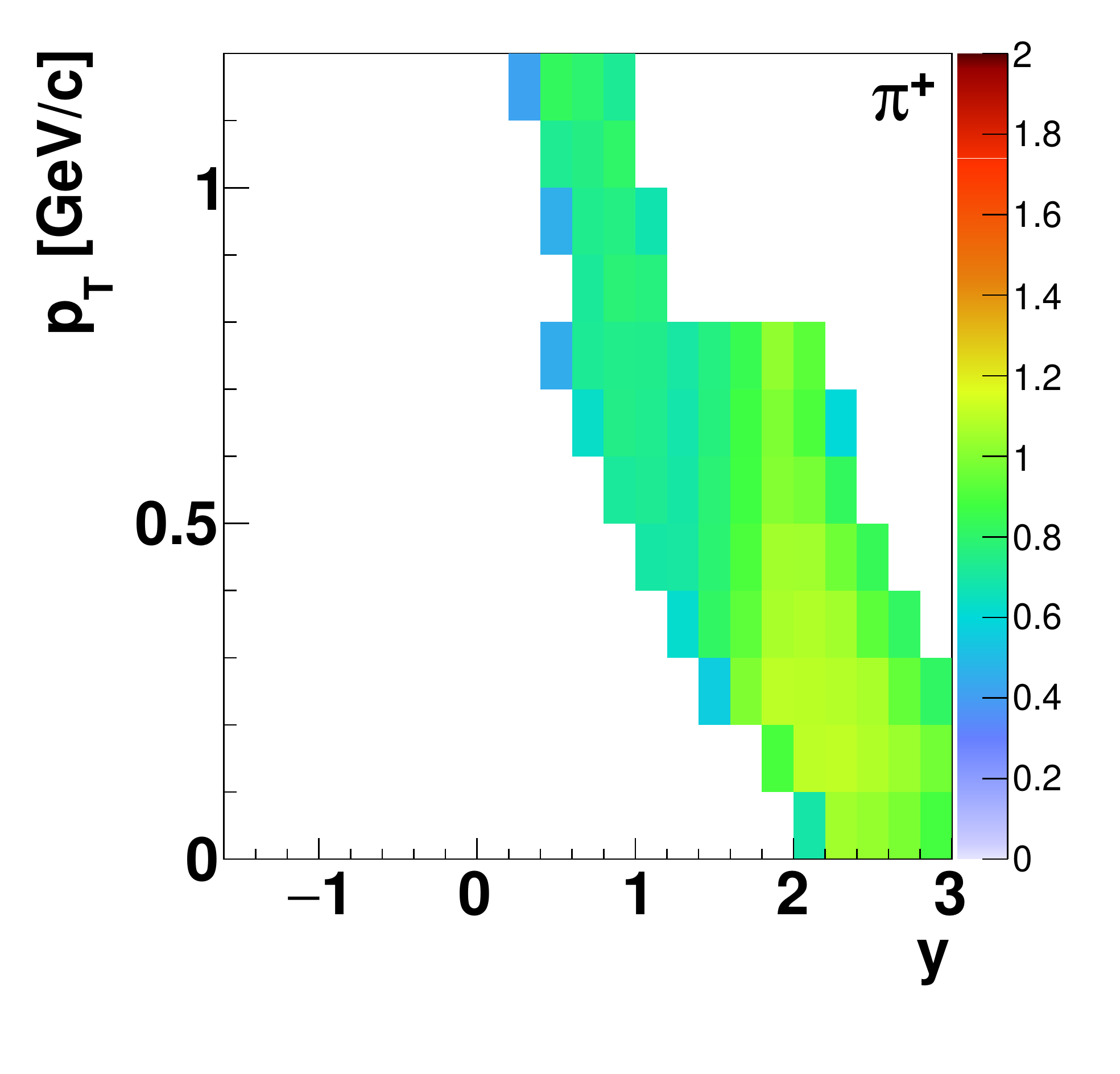}
\includegraphics[width=0.3\textwidth]{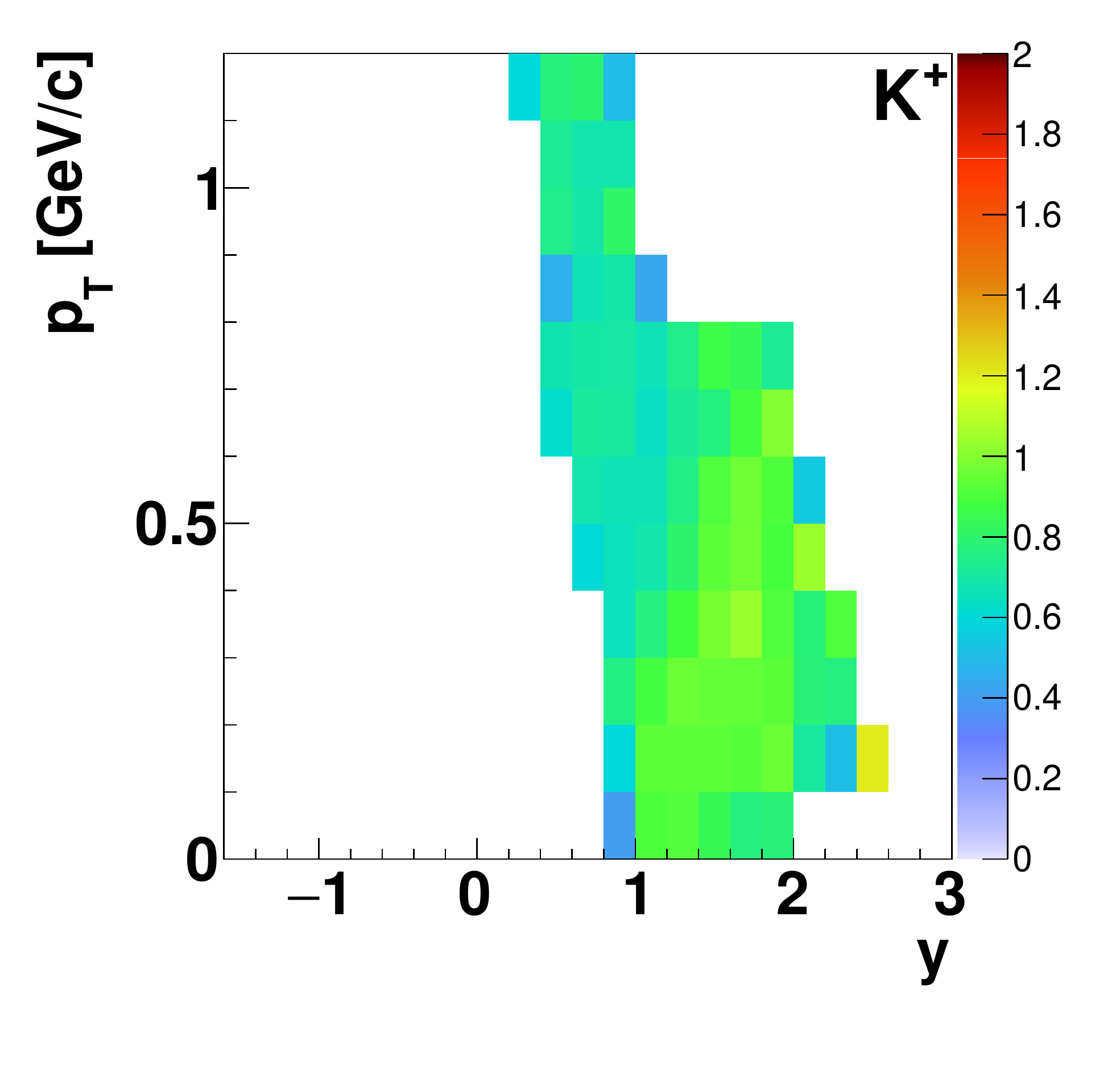}
\includegraphics[width=0.3\textwidth]{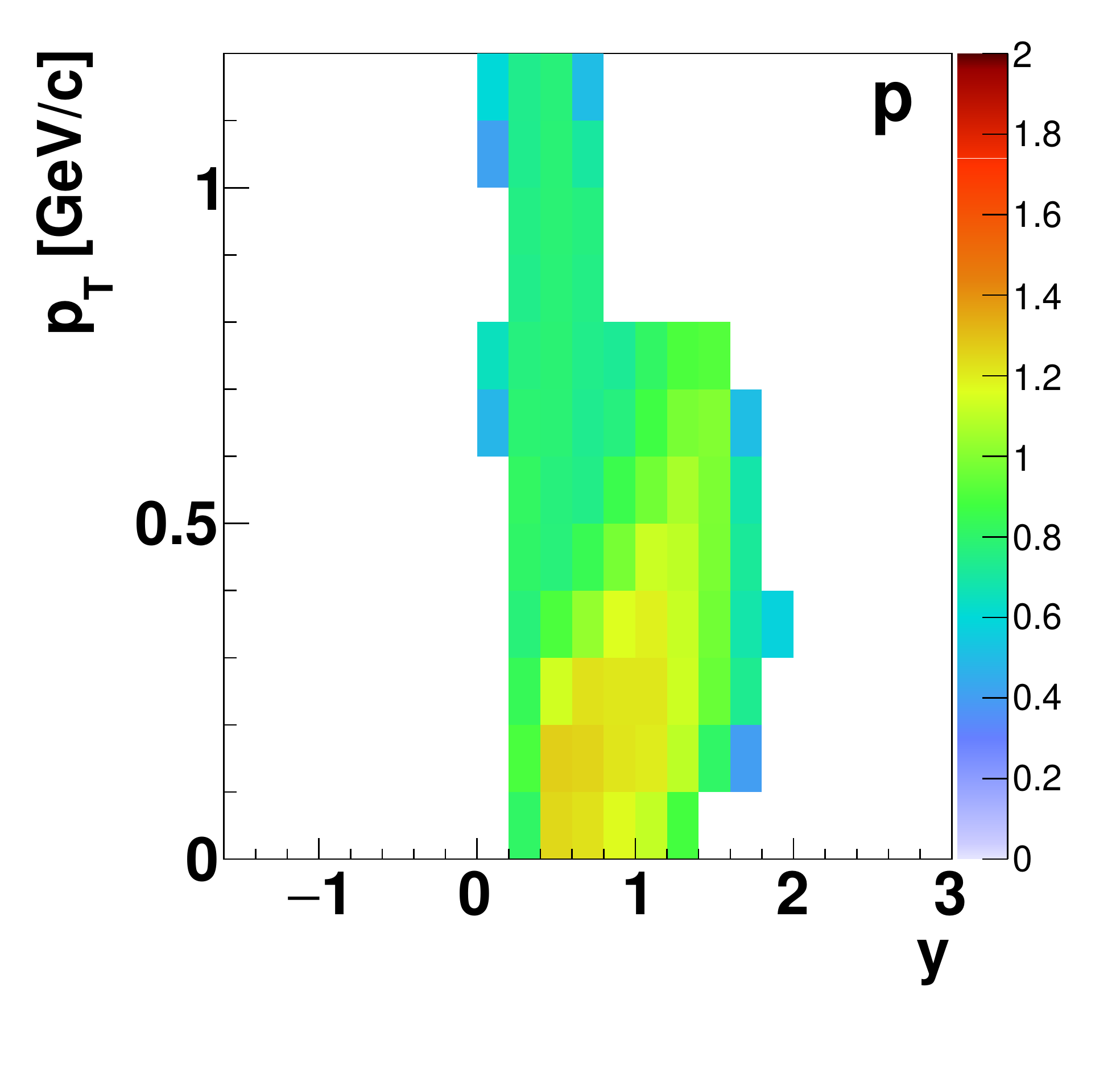}
\end{center}
\caption{Correction factors $C_{i}^{-1}$ for the \dedx identification method for positively and 
negatively charged pions, kaons and protons in p+p interactions at 20~\GeVc. The correction factor for antiprotons produced in inelastic p+p interactions is not presented due to insufficient statistics.}
\label{fig:c_dedx_20}
\end{figure}

\begin{figure}[!ht]
\begin{center}
\includegraphics[width=0.3\textwidth]{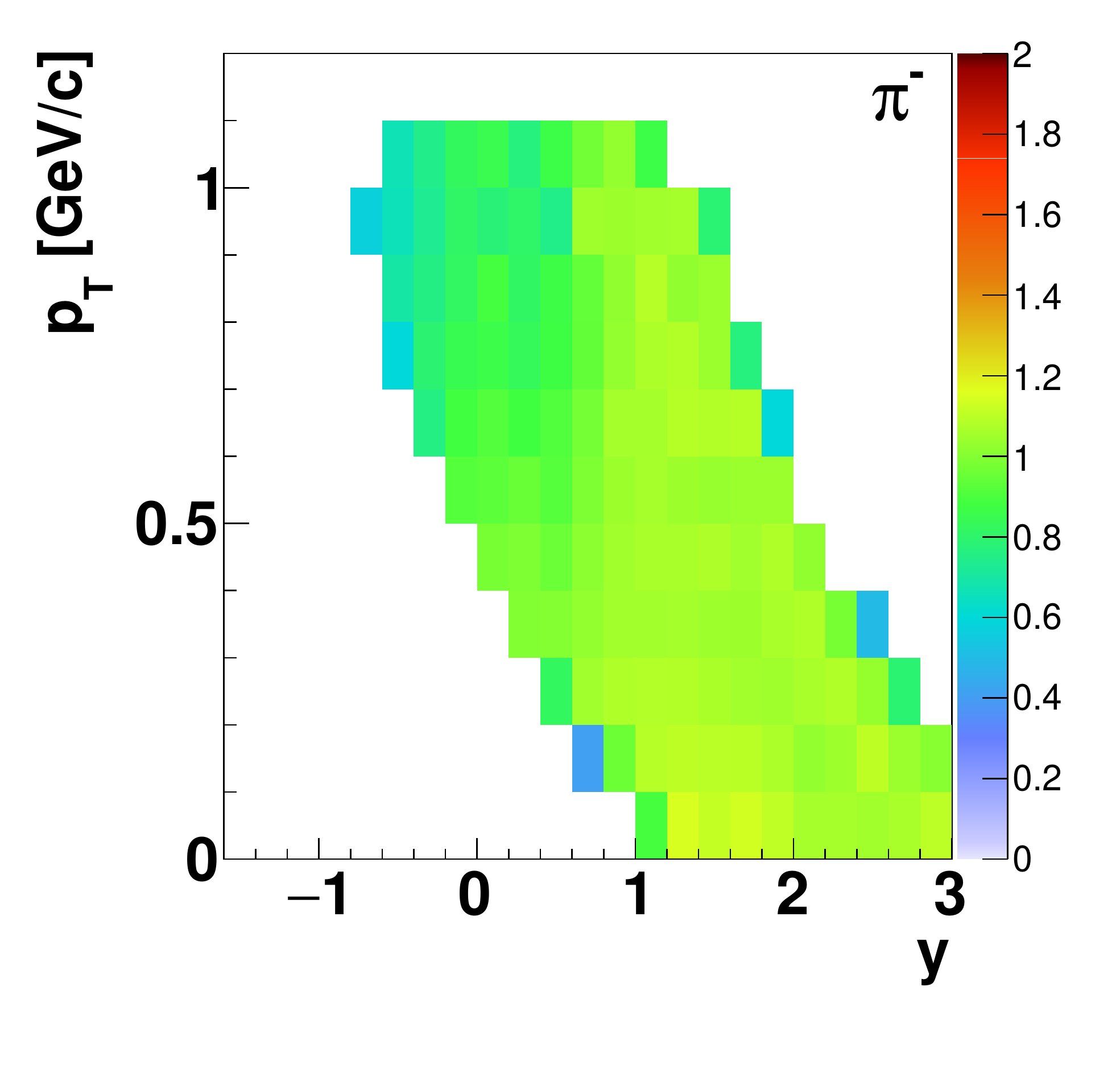}
\includegraphics[width=0.3\textwidth]{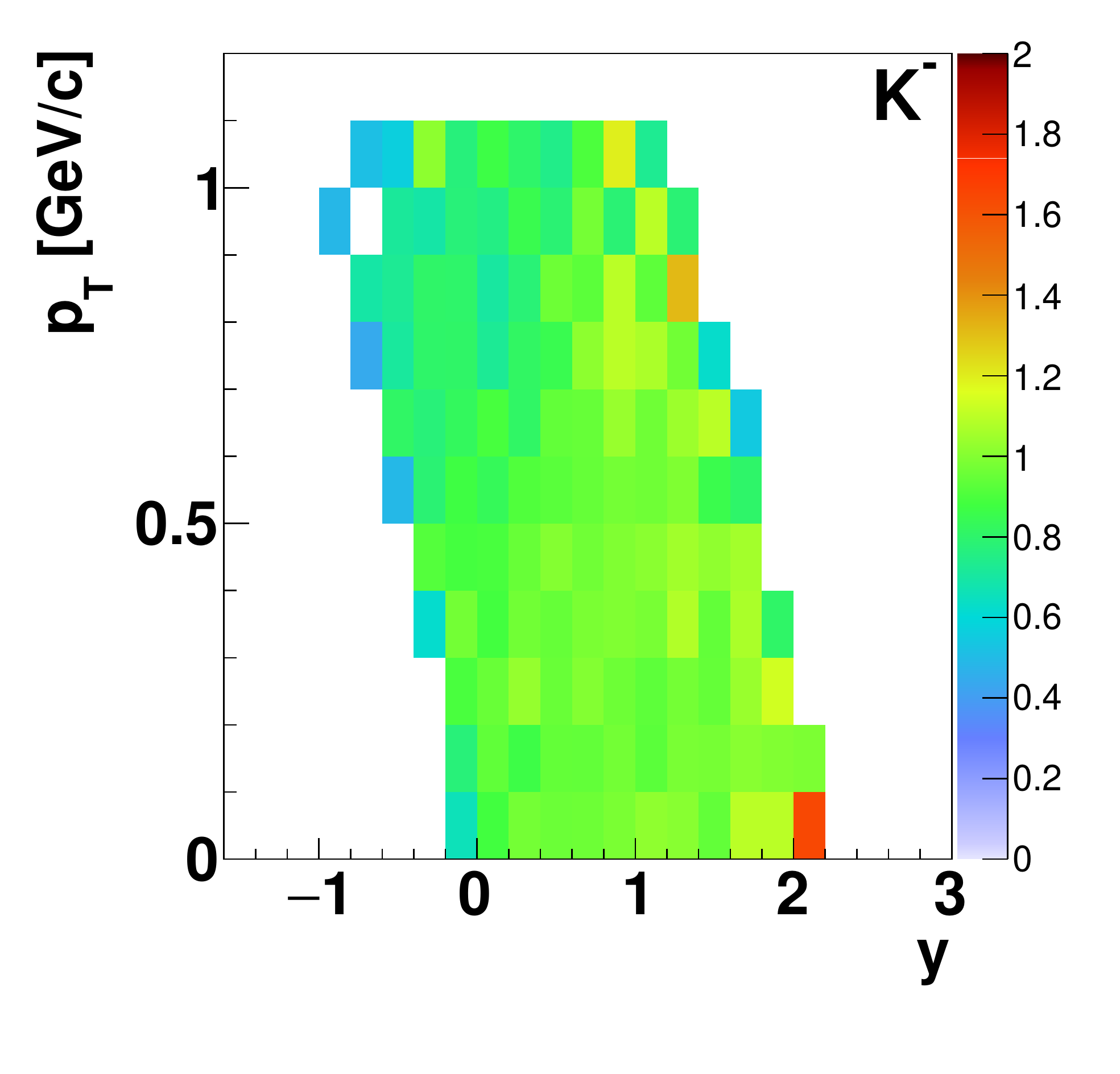}
\includegraphics[width=0.3\textwidth]{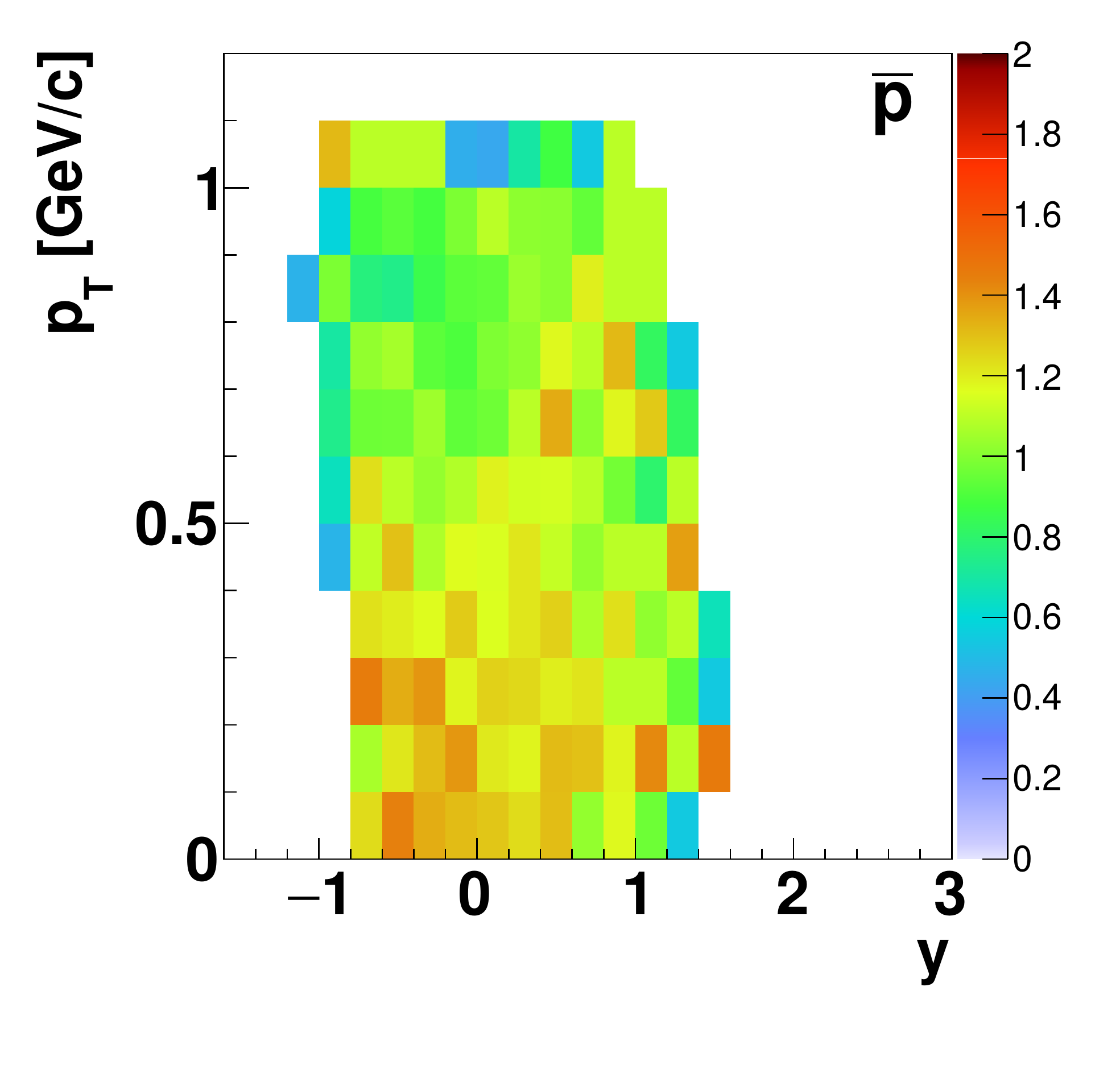}\\
\includegraphics[width=0.3\textwidth]{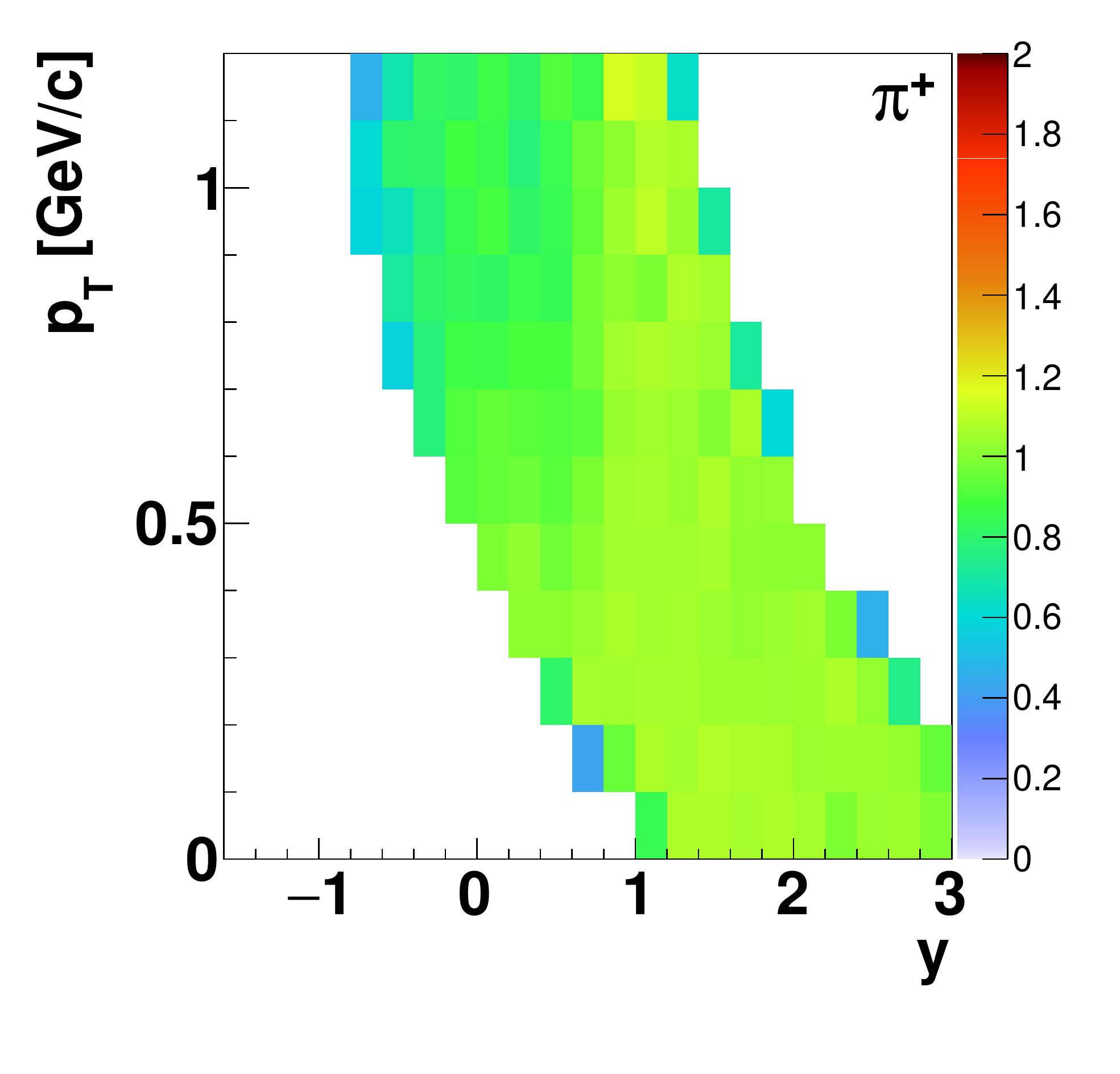}
\includegraphics[width=0.3\textwidth]{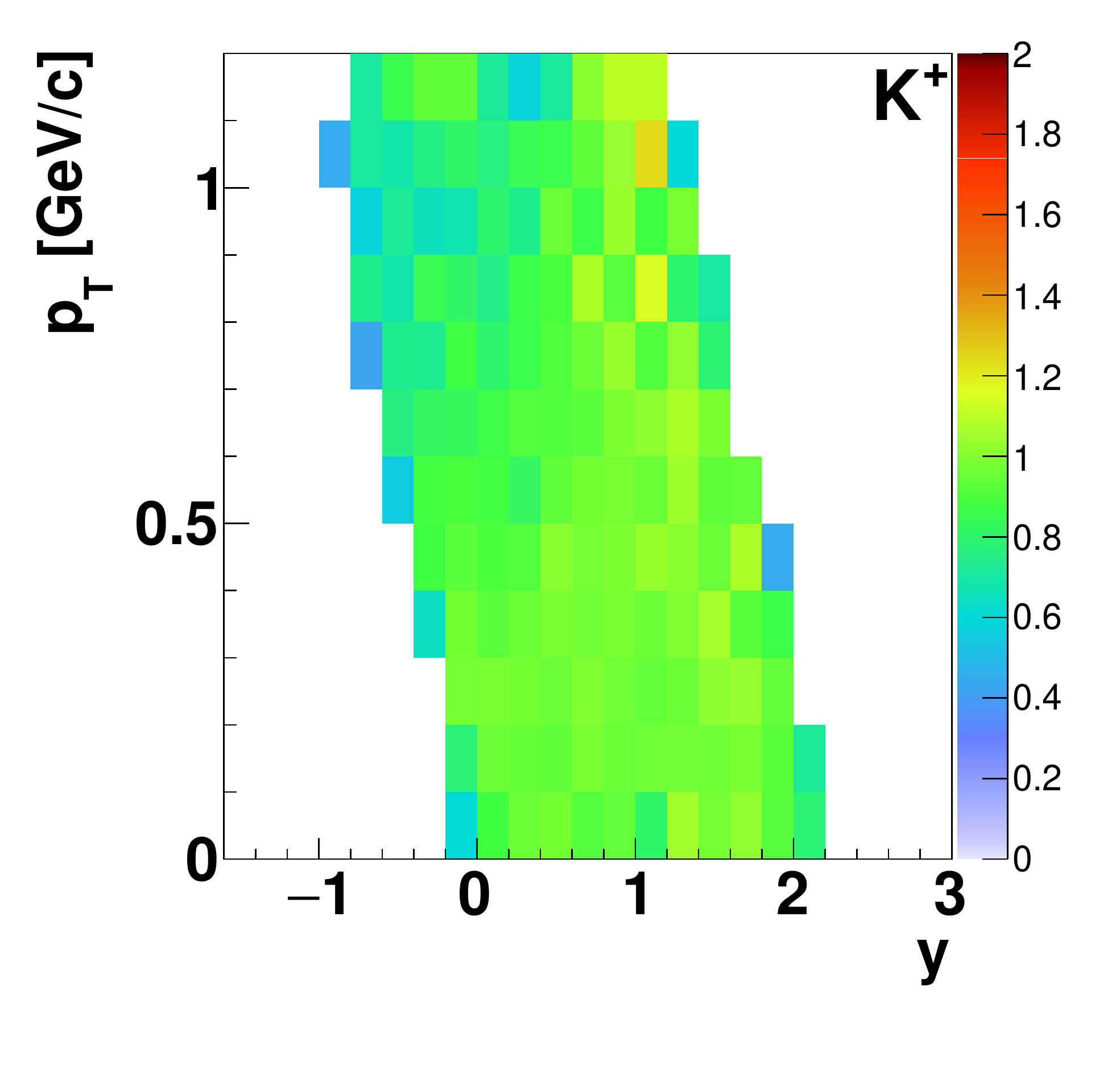}
\includegraphics[width=0.3\textwidth]{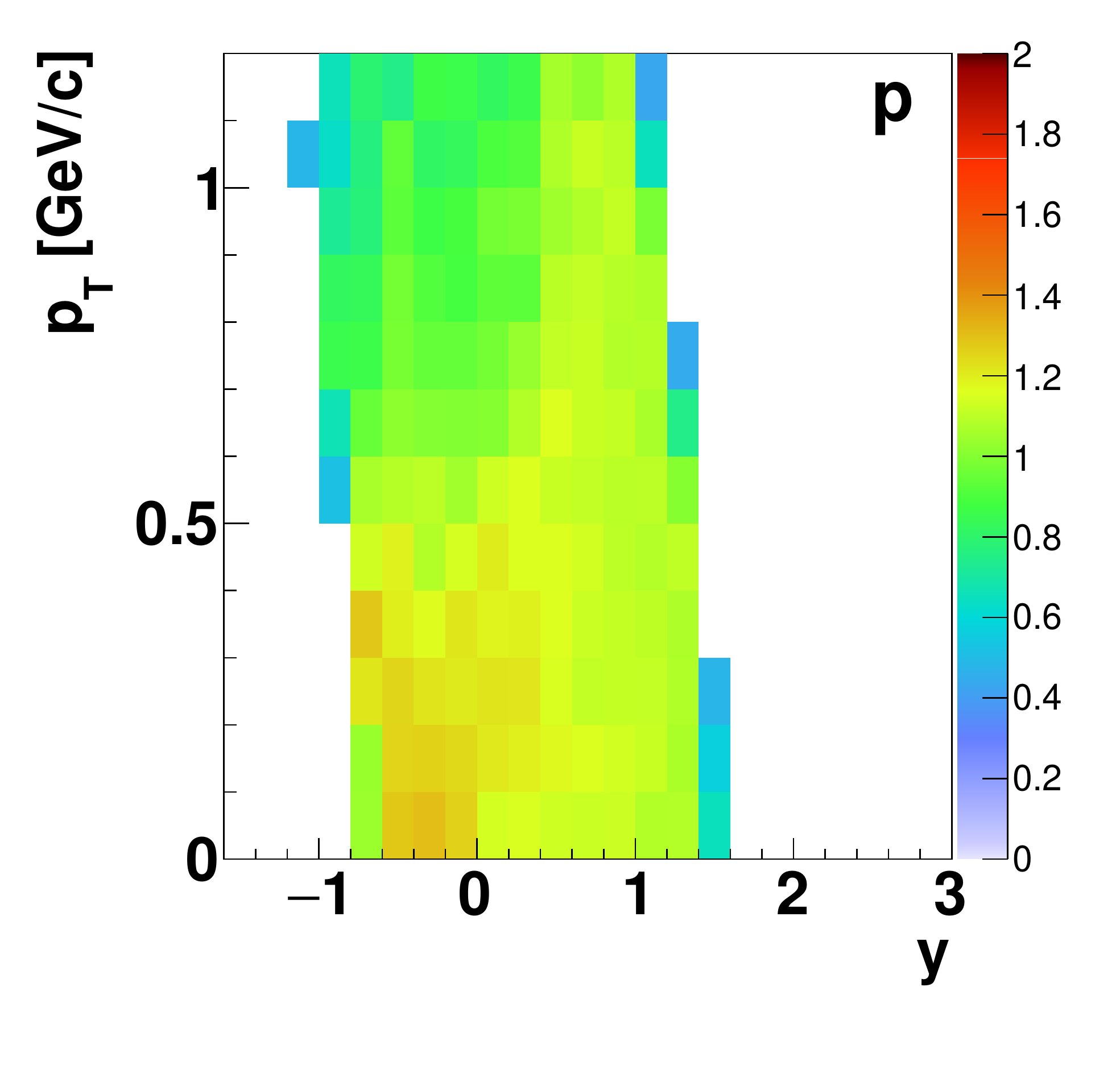}
\end{center}
\caption{Correction factors $C_{i}^{-1}$ for the \dedx identification method for positively and 
negatively charged pions, kaons and protons in p+p interactions at 158~\GeVc.}
\label{fig:c_dedx_158}
\end{figure}

The correction factor was calculated in the same bins of $y$ and $p_{T}$ as the particle spectra. 
Bins with correction factor lower than 0.5 and higher than 1.5 were rejected from the final results 
due to low acceptance or high contamination of particles from weak decays. Statistical uncertainties of the 
correction factors were calculated from the binomial distribution. The inverse correction factors for p+p interactions 
at 20 and 158~\GeVc are presented in Fig.~\ref{fig:c_dedx_20} and Fig.~\ref{fig:c_dedx_158}.

\emph{{\textbf The $tof$-\dedx method.}}\\
Due to the lack of a simulation of the ToF system the corrections for the $tof$-\dedx identification procedure 
had to be done in a different way than for the \dedx method, namely they partially employed a data based approach. 
Each simulated and reconstructed track was extrapolated to the ToF walls and if it crossed one of the ToF pixels 
it was classified as having a ToF hit. This defined the geometrical acceptance. The efficiency of the ToF pixels
was estimated from data.

The correction factor for the $tof$-\dedx identification method comprises the following components:

\begin{enumerate}[(i)]
\item Correction for the detector efficiency and geometrical acceptance \\
Based on the event and detector simulation the combined geometrical and reconstruction efficiency $C_i^{\mathrm geo}$ was calculated as:
\begin{equation}
C_i^{\mathrm geo} = \frac{\left(n_{i}\right)^{MCrec}_{\mathrm geo}}{\left(n_{i}\right)^{MCgen}},
\end{equation}
where $\left(n_{i}\right)^{MCrec}_{\mathrm geo}$ is the multiplicity of particle type 
$i=\pi^{-},\ \pi^{+}$, K$^{-}$, K$^{+}$, p,\ $\bar{\textrm{p}}$ after the TPC selection criteria and track extrapolation 
to the ToF wall resulting in a ToF hit, and $\left(n_{i}\right)^{MCgen}$ is the multiplicity of 
particle type $i$ generated by the \Epos model. 
In addition, the last point on the track was required to be located in the last two padrows of a MTPC. 
The resulting geometrical correction factors for p+p interactions at 158~\GeVc are presented in Fig.~\ref{fig:tofgeo}. 
Differences between efficiencies for pions, kaons and protons are due to the different particle lifetimes.


\item Correction for the pixel efficiency of ToF-L and ToF-R detectors\\
The pixel efficiency was calculated from data as the ratio between $\left(n_{i}\right)^{tof}_{\mathrm hit}$, the number of 
tracks with ToF hits in working pixels (pixel efficiency from data higher than 50\%) with correct TDC and QDC 
measurements and $\left(n_{i}\right)^{tof}_{\mathrm geo}$, the number of all tracks which were extrapolated 
to the particular pixel. 
\begin{equation}
C_i^{pixel} = \frac{\left(n_{i}\right)^{tof}_{\mathrm hit}}{\left(n_{i}\right)^{tof}_{\mathrm geo}},
\end{equation}
The pixel efficiency obtained for p+p interactions at 158~\GeVc is shown in Fig.~\ref{fig:tofdead}.

\item Correction for decays and interactions between the last measured point in the MTPC and the ToF detectors \\

The probability of decays and interactions between the last measured point in the MTPC and the ToF detectors was 
estimated by simulations. The survival probability is defined in the following way:
\begin{equation}
C_i^{\mathrm survive}= \frac{\left(n_{i}\right)^{MCrec}_{\mathrm survive}}{\left(n_{i}\right)^{MCrec}_{\mathrm geo}},
\end{equation}
where $\left(n_{i}\right)^{MCrec}_{\mathrm survive}$ is the number of particles which hit a working ToF pixel
and which did not decay or interact between the last measured point in the MTPC and the ToF walls.
The survival probability is lower than expected from decay only, due to interactions in the Forward Time of Flight detector. The survival probability $C_i^{\mathrm survive}$ is presented in Fig.~\ref{fig:tofdecay}.


\end{enumerate}
The combined efficiency factor $C_{i}^{tof}$ was calculated as:
\begin{equation}
\label{eq:correctiontof}
C_{i}^{tof} = C_i^{\mathrm geo} \times C_i^{pixel} \times C_i^{\mathrm decay} 
\end{equation}

\begin{figure}[!ht]
\begin{center}
\includegraphics[width=0.3\textwidth]{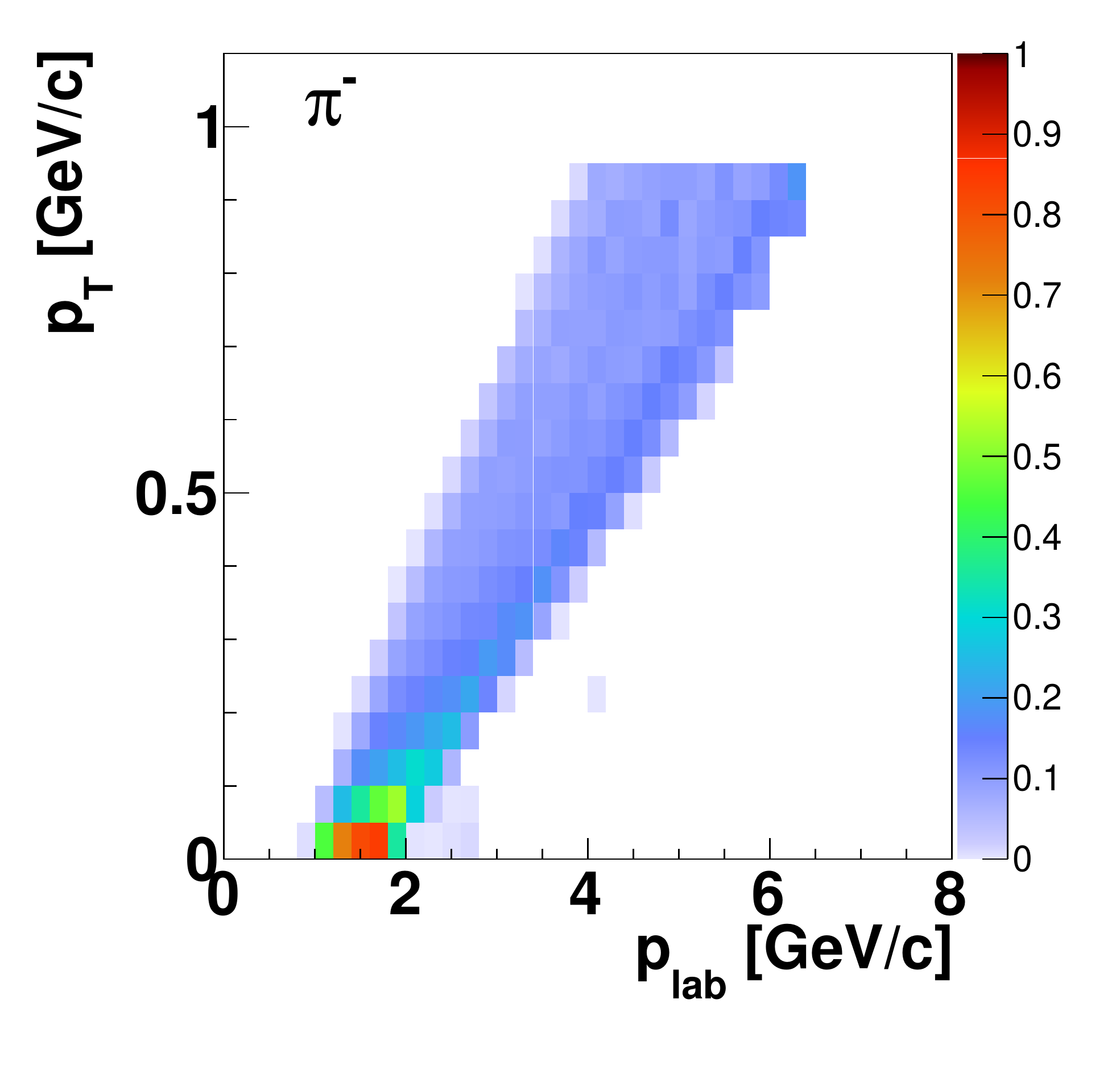}
\includegraphics[width=0.3\textwidth]{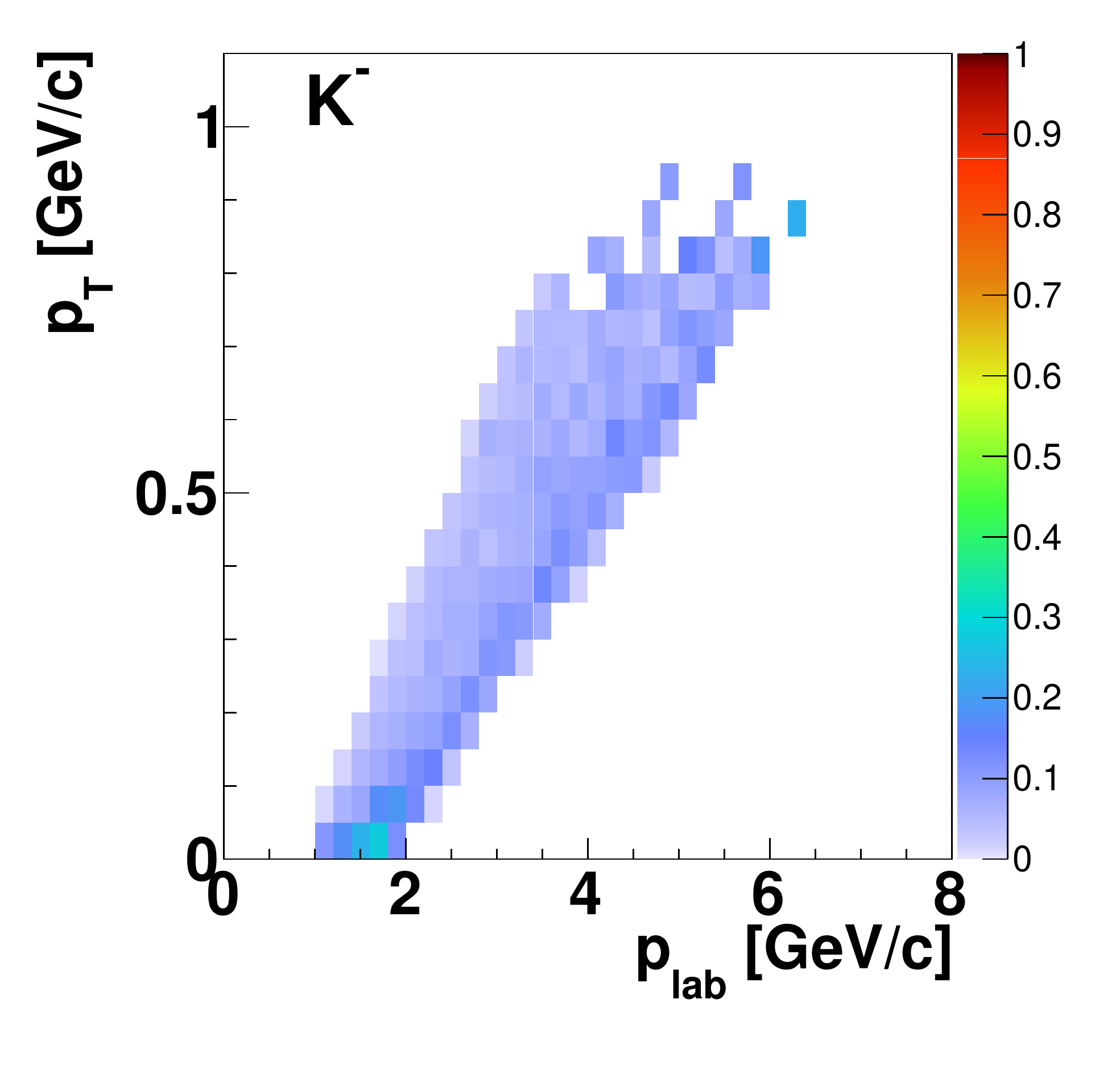}\\
\includegraphics[width=0.3\textwidth]{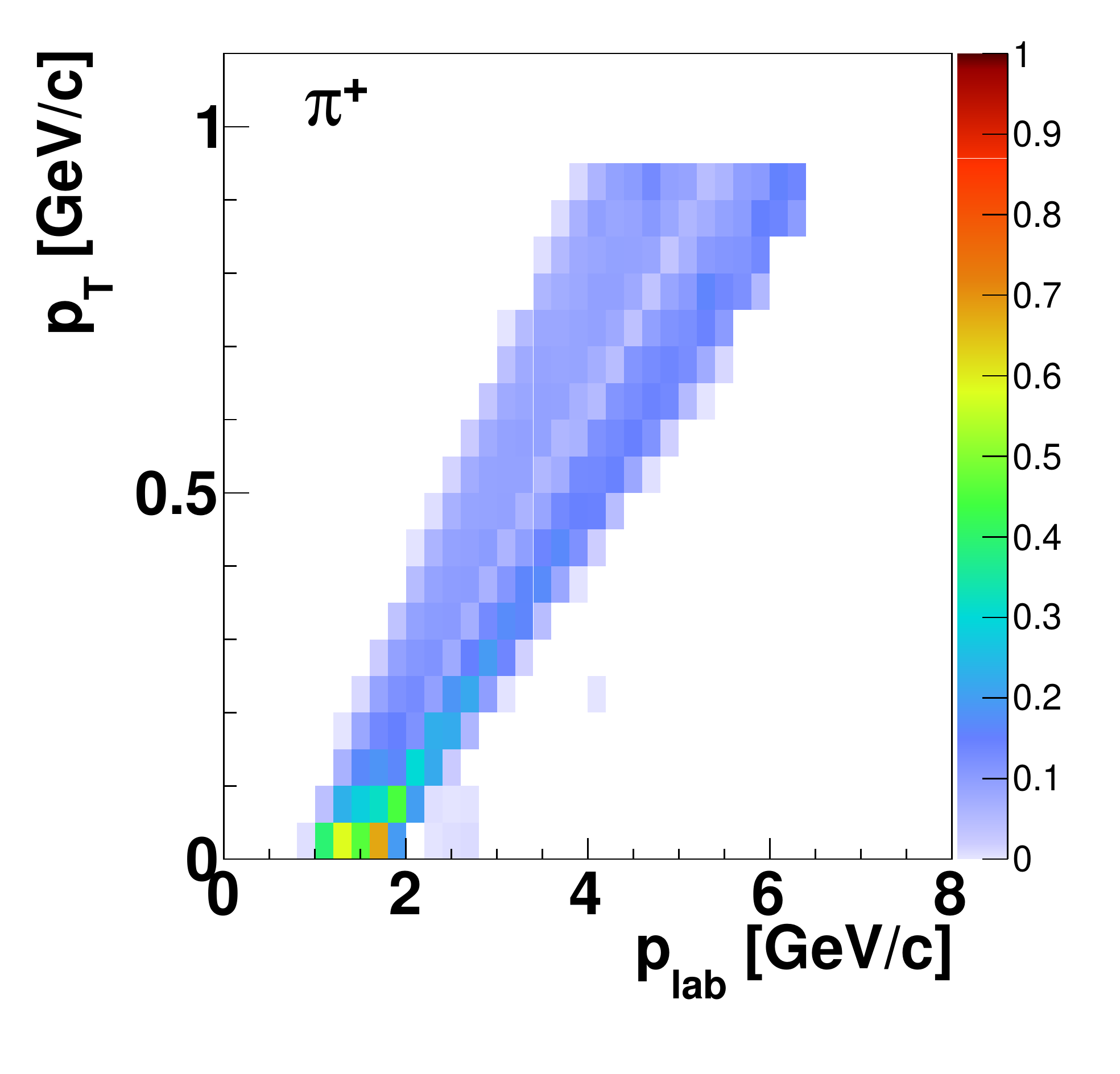}
\includegraphics[width=0.3\textwidth]{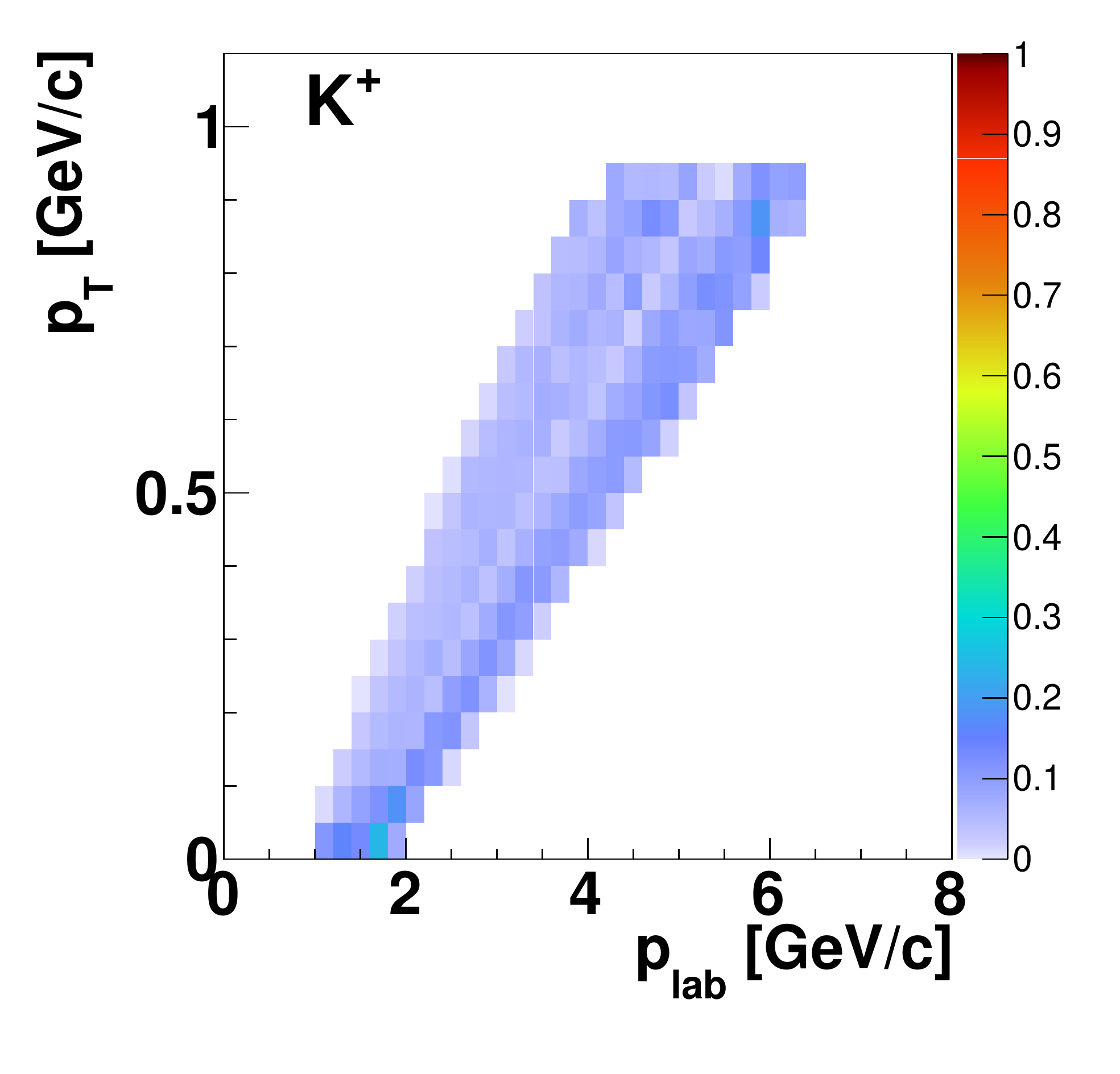}
\includegraphics[width=0.3\textwidth]{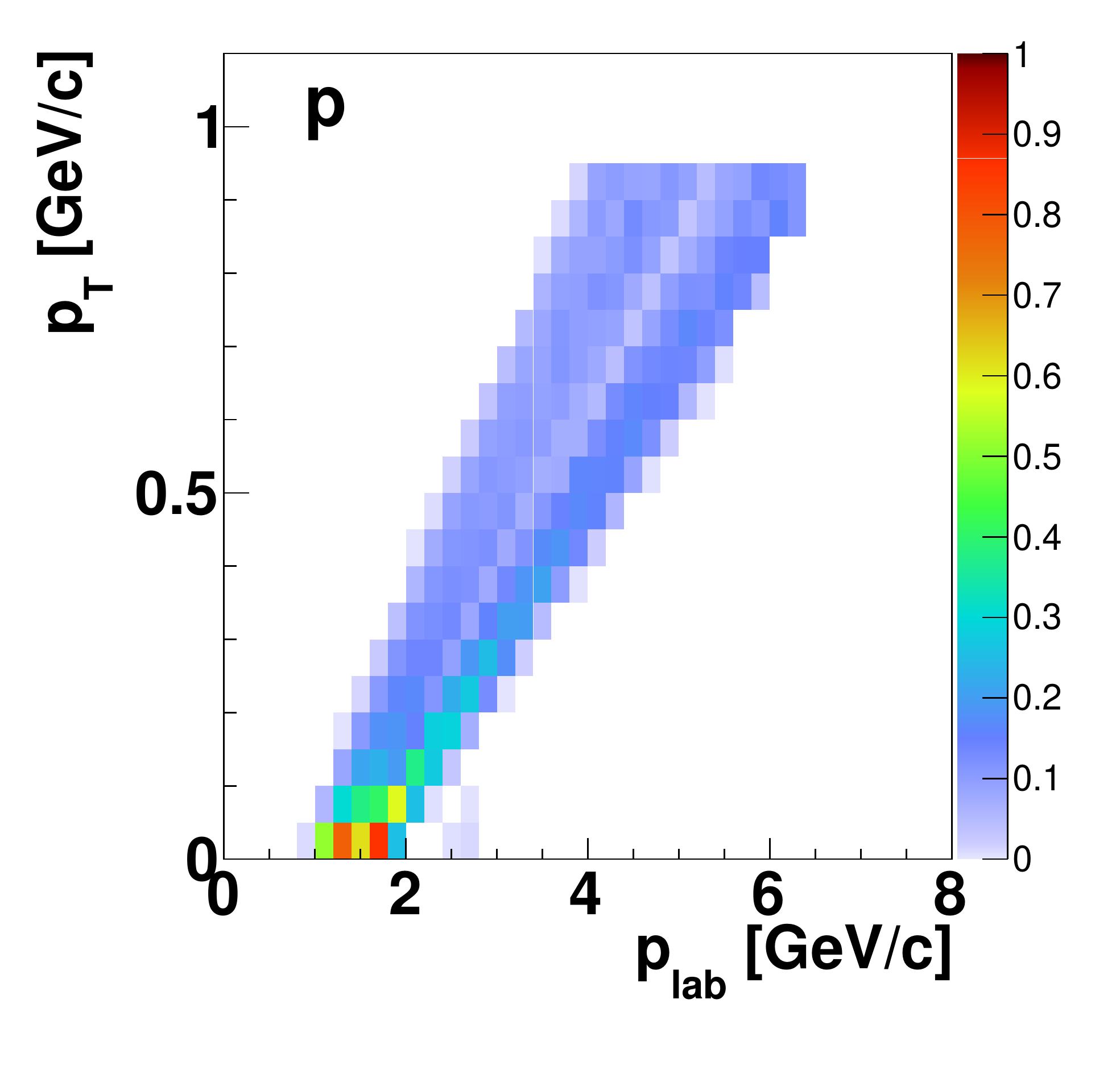}
\end{center}
\caption{Efficiencies $C_{i}^{tof}$ for the $tof$-\dedx identification method for positively and negatively charged pions, kaons and protons for p+p interactions at 20~\GeVc.}
\label{fig:c_tof_20}
\end{figure}

\begin{figure}[!ht]
\includegraphics[width=0.3\textwidth]{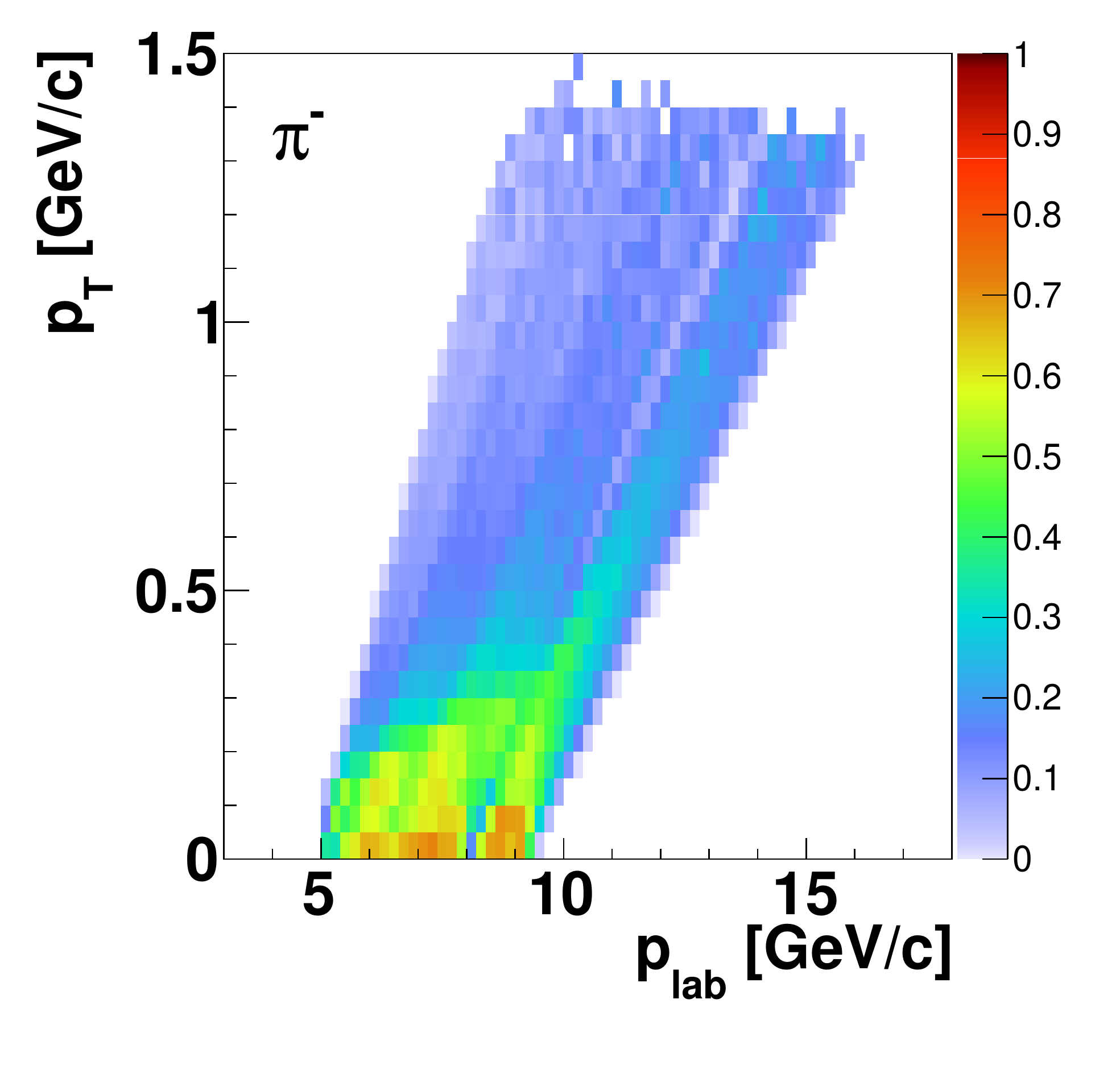}
\includegraphics[width=0.3\textwidth]{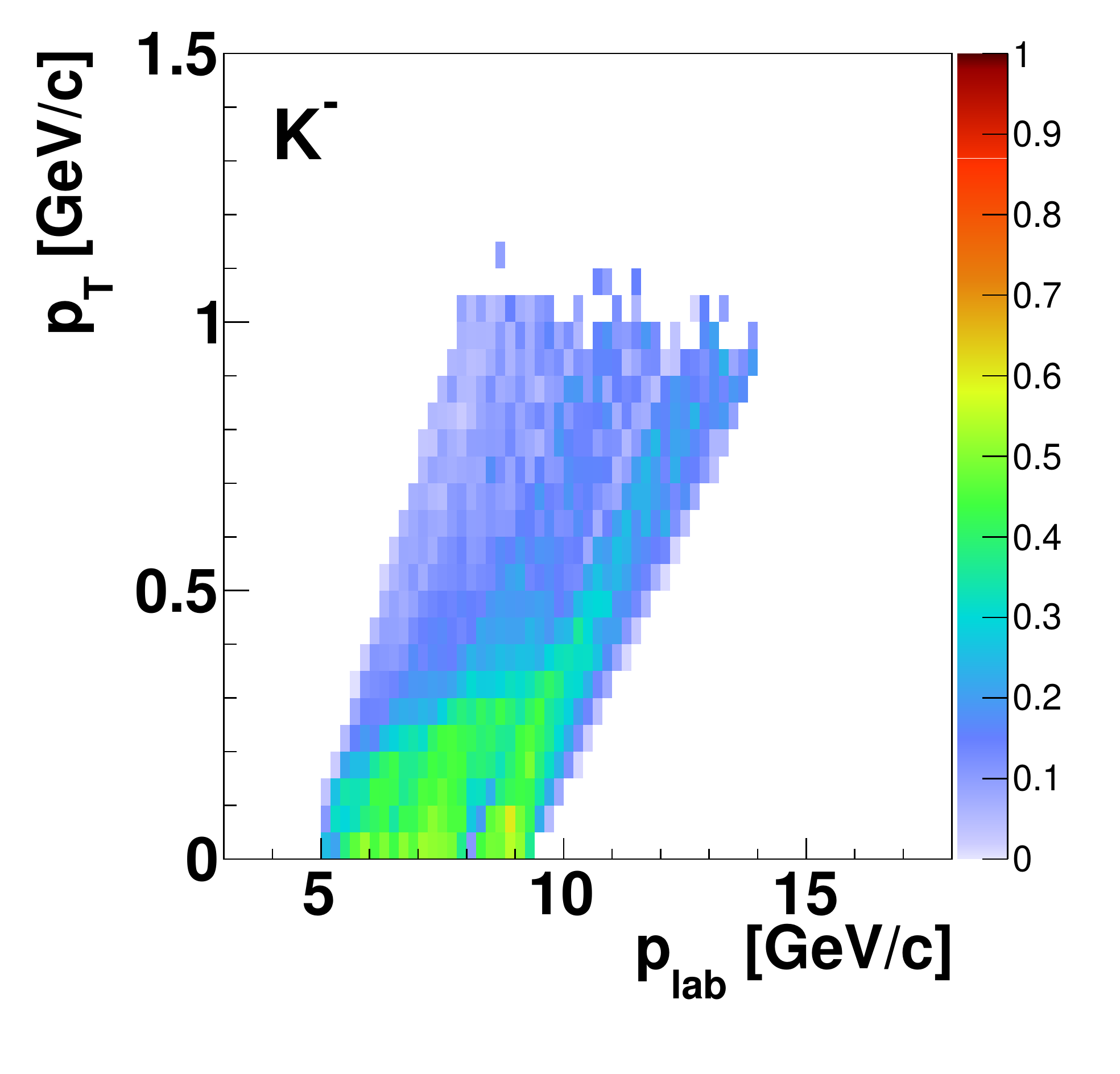}
\includegraphics[width=0.3\textwidth]{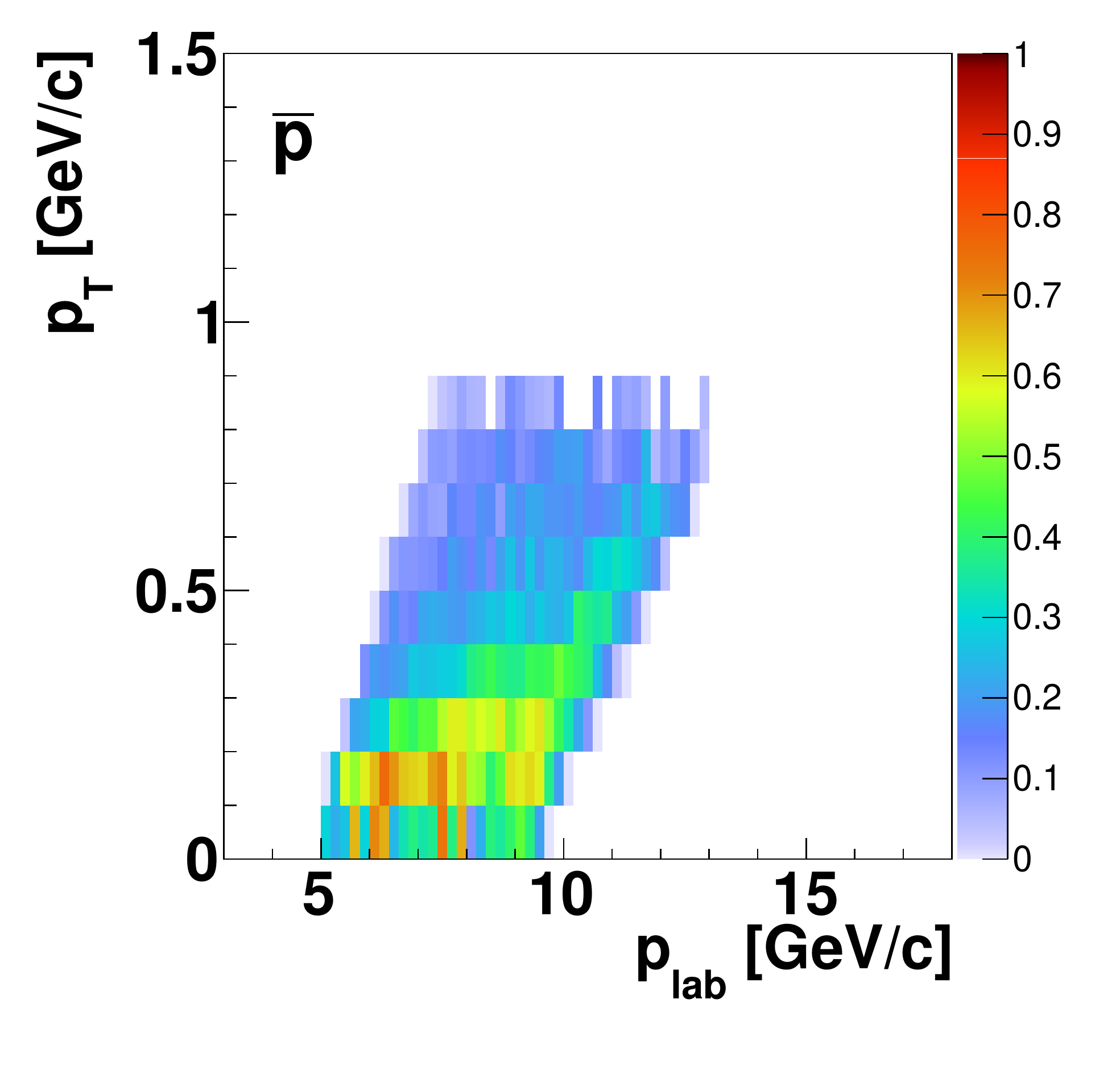}\\
\includegraphics[width=0.3\textwidth]{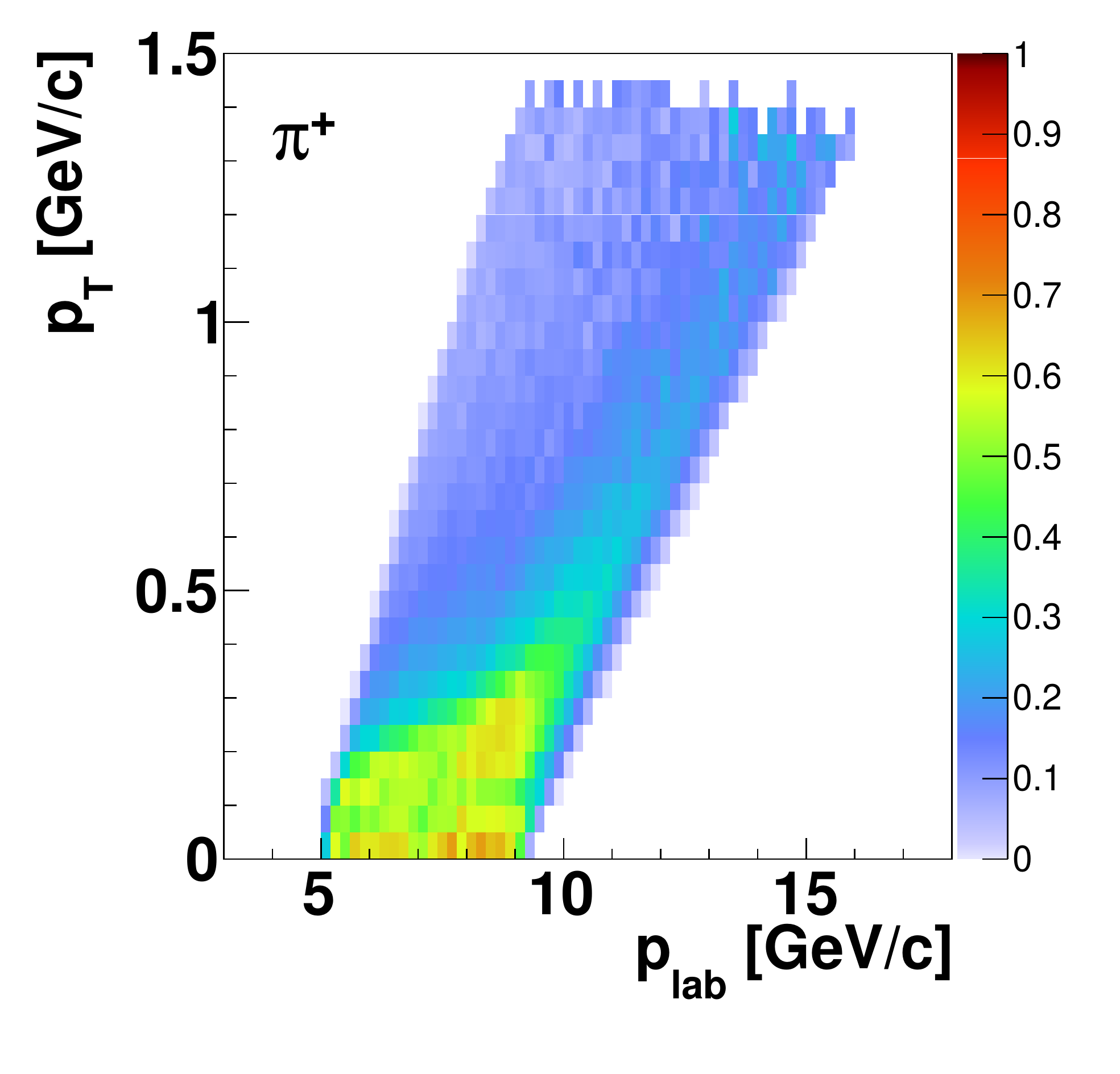}
\includegraphics[width=0.3\textwidth]{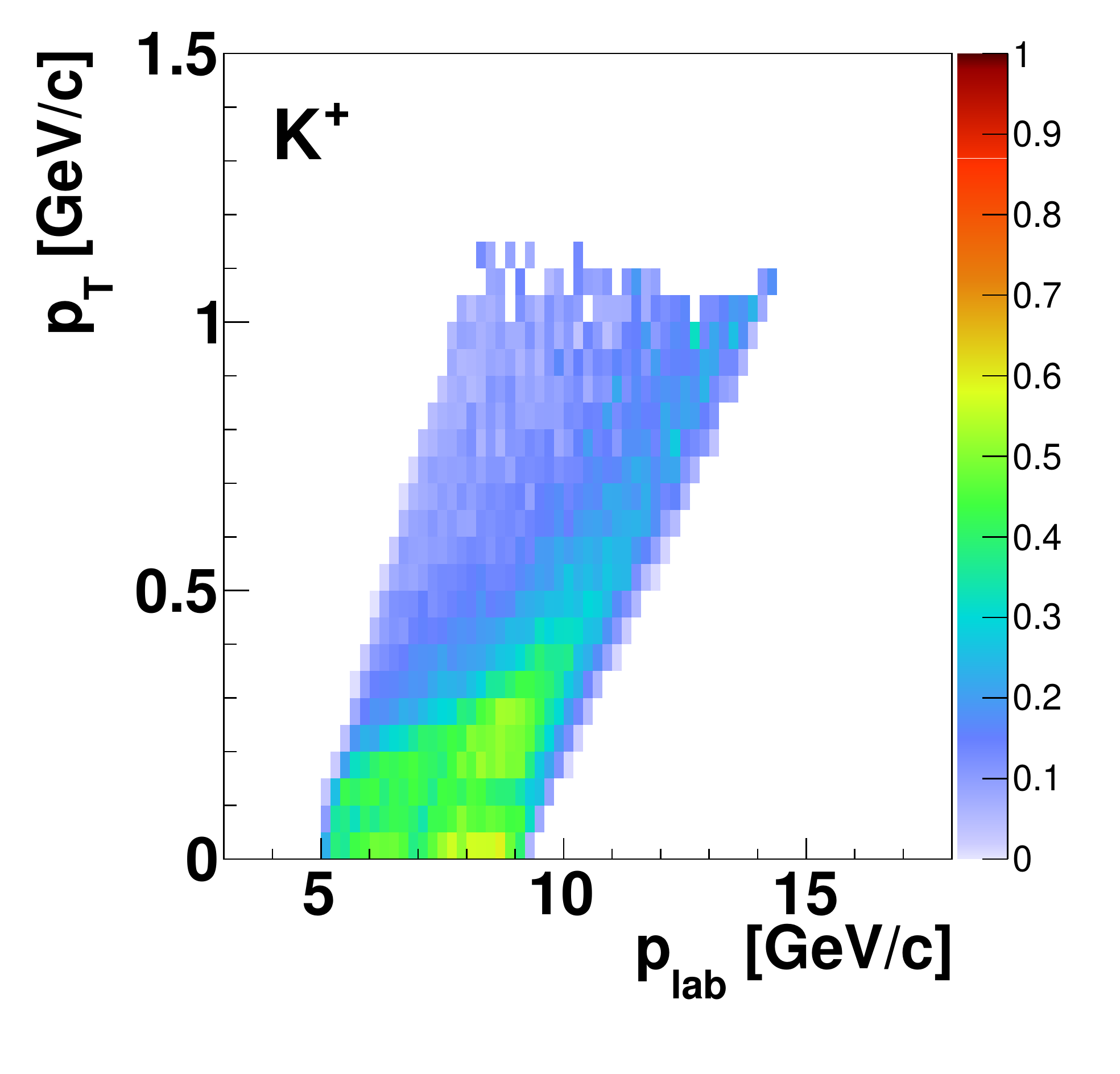}
\includegraphics[width=0.3\textwidth]{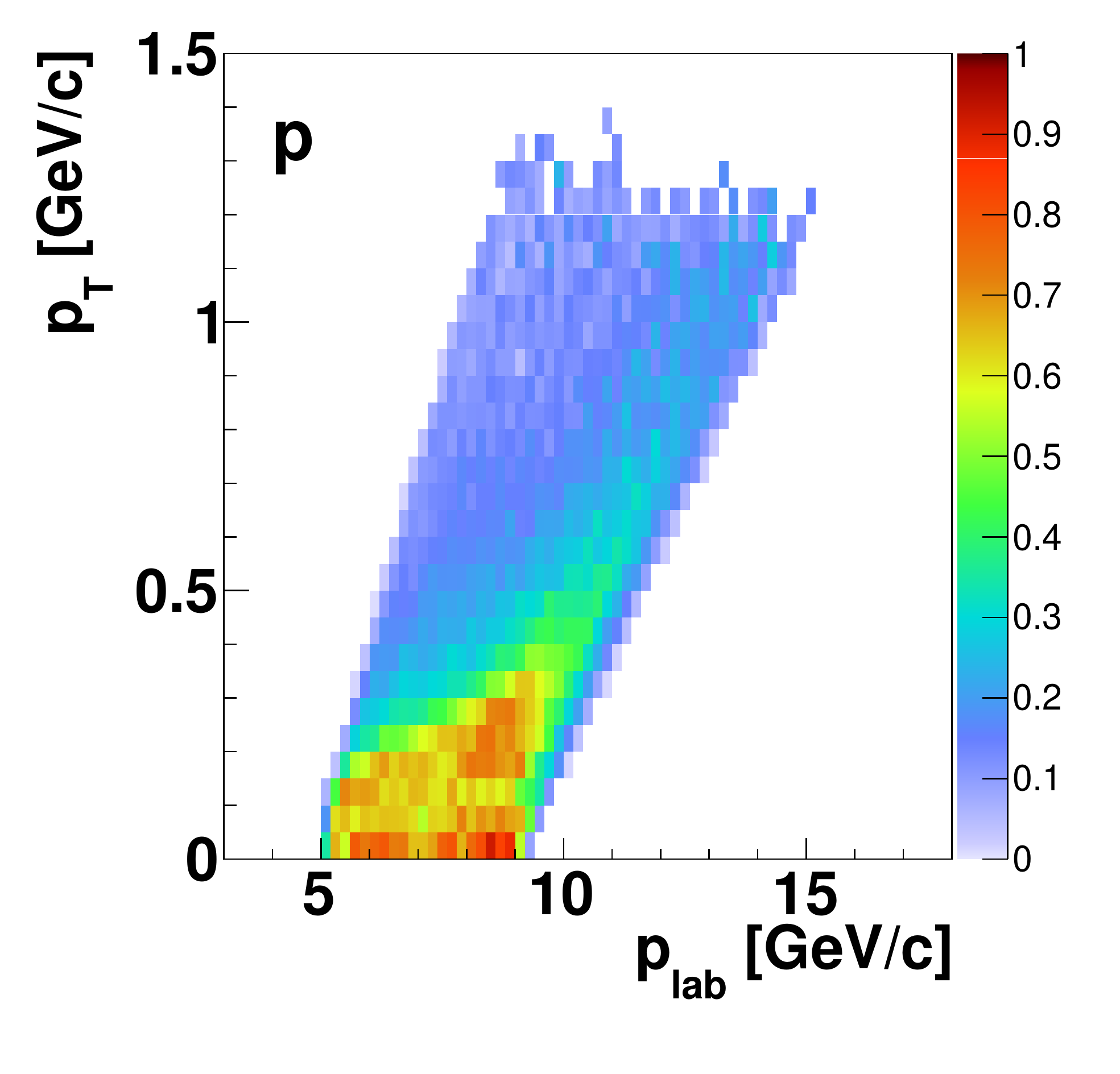}
\caption{Efficiencies $C_{i}^{tof}$ for the $tof$-\dedx identification method for positively and negatively charged pions, kaons and protons for p+p interactions at 158~\GeVc.}
\label{fig:c_tof_158}
\end{figure}

Combined efficiency factors $C_{i}^{tof}$ for the $tof$-\dedx method calculated for each (p, \pt) bin independently applied to p+p interactions 
at 20 and 158~\GeVc are presented in Fig.~\ref{fig:c_tof_20} and Fig.~\ref{fig:c_tof_158}.

\FloatBarrier
   \subsection{Corrected spectra and uncertainties}
\label{sec:results_calc}

\subsubsection{Corrected spectra}
\label{sec:Corrected_spectra}
		
The multiplicity of different types of hadrons from inelastic p+p interactions measured by the \dedx technique 
is defined as the sum of probabilities divided by the number of events corrected for detector effects, feed-down 
and contamination by target removed events. For particle type $i=\pi^{+}$, $\pi^{-}$, K$^{+}$, K$^{-}$, p, $\bar{\textrm{p}}$, 
the multiplicity is defined as:
		\begin{equation}
		\label{finalresdEdx}
		\frac{n_{i}}{N}=
		\frac{1}{C_{i}}
		\frac{\sum\limits_{I} P_{i}(dE/dx)_{p_{\mathrm lab}, \pt} - B \sum\limits_{R} P_{i}(dE/dx)_{p_{\mathrm lab}, \pt}}{N_{I} - B N_{R}},
		\end{equation}
		
		where:
		\begin{itemize}
		\item[--] $C_{i}$ - correction factor defined in Eq.~\ref{eq:corectionfactor},
		\item[--] $\sum\limits_{I} P_{i}(dE/dx)_{p_{\mathrm lab}, \pt} \equiv \sum\limits_{I} P_{i}$ - sum over probabilities $P_{i}$ defined in Eq.~\ref{eq:propdedx} for all tracks for inserted target (abbreviated as "I"), 
		\item[--] $\sum\limits_{R} P_{i}(dE/dx)_{p_{\mathrm lab}, \pt} \equiv \sum\limits_{R} P_{i}$ - sum over probabilities $P_{i}$ defined in Eq.~\ref{eq:propdedx} for all tracks for removed target (abbreviated as "R"),
		\item[--] $B$ - the normalization factor applied to target removed events,
		\item[--] $N_{I}$ and $N_{R}$ - the number of events of target inserted and removed, respectively.
		\end{itemize}

Particle multiplicities for the $tof$-\dedx technique can be calculated in a similar way:
		\begin{equation}
		\label{finalrestof}
		\frac{n_{i}}{N}=
		\frac{1}{\epsilon}
		\frac{\sum\limits_{I} \frac{P_{i}(dE/dx, m^{2})_{p_{\mathrm lab}, \pt}}{C_{i, p_{\mathrm lab}, \pt}^{tof}} - B \sum\limits_{R} \frac{P_{i}(dE/dx, m^{2})_{p_{\mathrm lab}, \pt}}{C_{i, p_{\mathrm lab}, \pt}^{tof}}}{N_{I} - B N_{R}} ,
		\end{equation}
where: $P_{i}$ are the probabilities defined in Eq.~\ref{eq:proptof}, $C_{i}^{tof}$ is the efficiency described 
by Eq.~\ref{eq:correctiontof} and $\epsilon$ is the correction for the non-orthogonal transformation between the phase spaces (momentum $\rightarrow$ rapidity) described by Eq.~\ref{eq:epsilontof}. Sum over probabilities are the same like in eq.~\ref{finalresdEdx}.
		
\subsubsection{Statistical uncertainties}
\label{sec:statistics}

Statistical uncertainties of multiplicities obtained by the \dedx method were calculated in the following way:
\begin{equation}
\begin{split}
	\left(\Delta\frac{n_{i}}{N}\right)^{2} =
	\left(\left|\frac{-1}{C_{i}^{2}} \frac{\sum\limits_{I} P_{i} - B \sum\limits_{R} P_{i}}{N_{I} - B N_{R}}\right| \Delta C_{i} \right)^{2} + \\
	\left(\left|\frac{1}{C_{i}} \frac{1}{N_{I} - B N_{R}}\right| \sqrt{\sum\limits_{I}P_{i}}\right)^{2} + \\ 
	\left(\left|\frac{1}{C_{i}} \frac{-B}{N_{I} - B N_{R}}\right| \sqrt{\sum\limits_{R}P_{i}}\right)^{2}  \ ,
\end{split}
\end{equation}
where $\Delta$ denotes the statistical uncertainty of the quantities used to calculate the particle multiplicity. 
The uncertainty of the normalization factor ($B$) as well as of the number of events ($N_{I}$ and $N_{R}$) are not 
taken into account in this calculation due to their negligible influence on the uncertainty value.

%

Calculation of statistical uncertainties of the multiplicities from the $tof$-\dedx identification technique 
is more complicated. Assuming that the statistical uncertainty of $C_{i}^{tof}$ is equal to the mean statistical 
uncertainty in the full $y-p_{T}$ bin, it can be moved in front of the sums. The total statistical 
uncertainty can then be calculated according to the following formula:
	\begin{equation}
\begin{split}
	\left(\Delta\frac{n_{i}}{N}\right)^{2} = 	\left(\left|\frac{-1}{C_{tof}^{2}} \frac{\sum\limits_{I} \frac{P_{i}}{C_{i}^{tof}} - B \sum\limits_{R} \frac{P_{i}}{\epsilon}}{N_{I} - B N_{R}}\right| \Delta \epsilon \right)^{2} + \\
	\left(\left|\frac{1}{\epsilon} \frac{\sum\limits_{I}\frac{1}{C_{i}^{tof}}}{N_{I} - B N_{R}}\right| \sqrt{\sum\limits_{I}P_{i}}\right)^{2} + \\ 
	\left(\left|\frac{1}{\epsilon} \frac{\sum\limits_{R}\frac{-B}{C_{i}^{tof}}}{N_{I} - B N_{R}}\right| \sqrt{\sum\limits_{R}P_{i}}\right)^{2} + \\
	\left(\left|\frac{-1}{\epsilon \langle C_{i}^{tof} \rangle^2} \frac{\sum\limits_{I} P_{i} - B \sum\limits_{R} P_{i}}{N_{I} - B N_{R}}\right| \langle \Delta C_{i}^{tof} \rangle\ \right)^{2} .
\end{split} 
	\end{equation}
	

	
\subsubsection{Systematic uncertainties}
\label{sec:systematics}

\begin{figure}[!ht]
	\begin{center} 
	\includegraphics[width=0.3\textwidth]{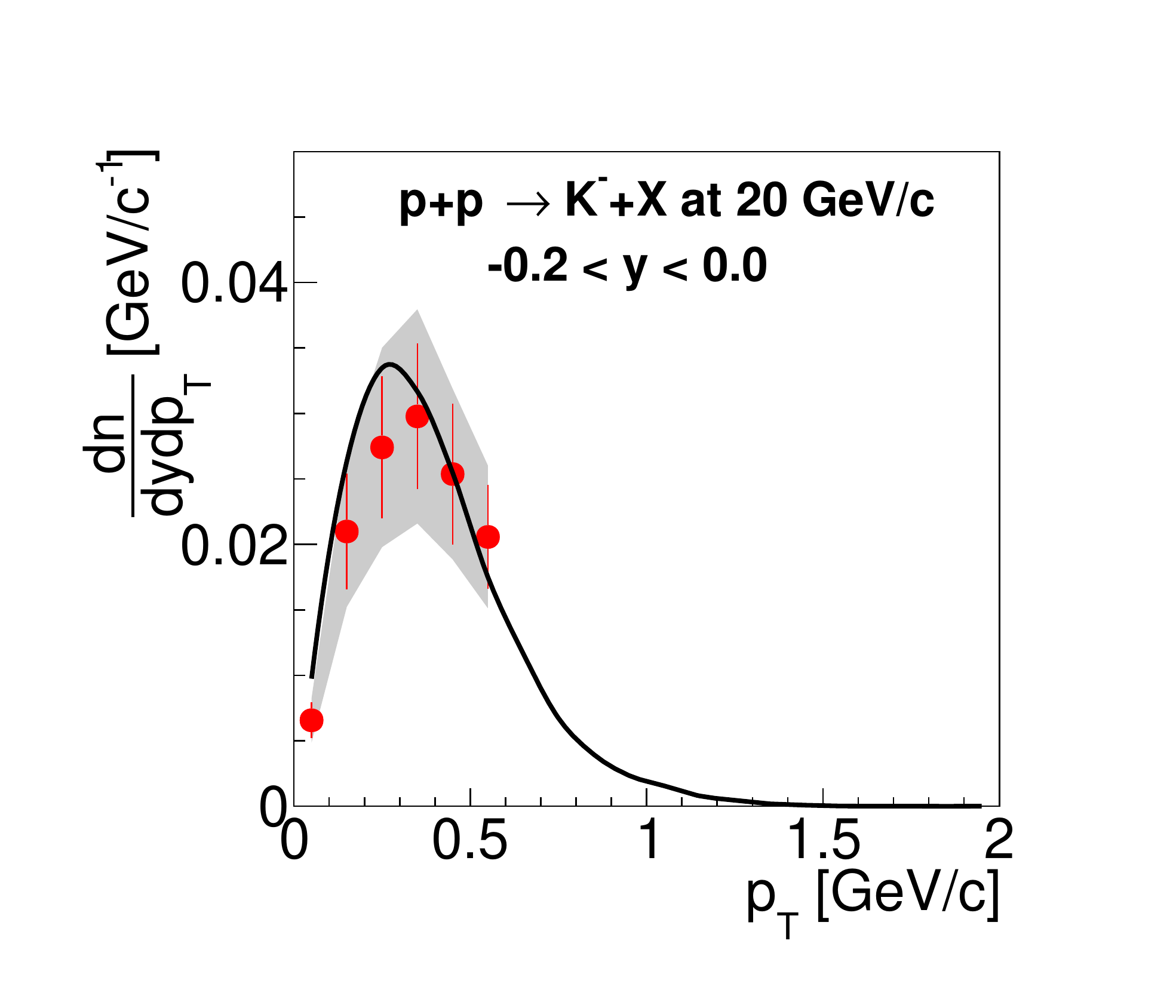}
	\includegraphics[width=0.3\textwidth]{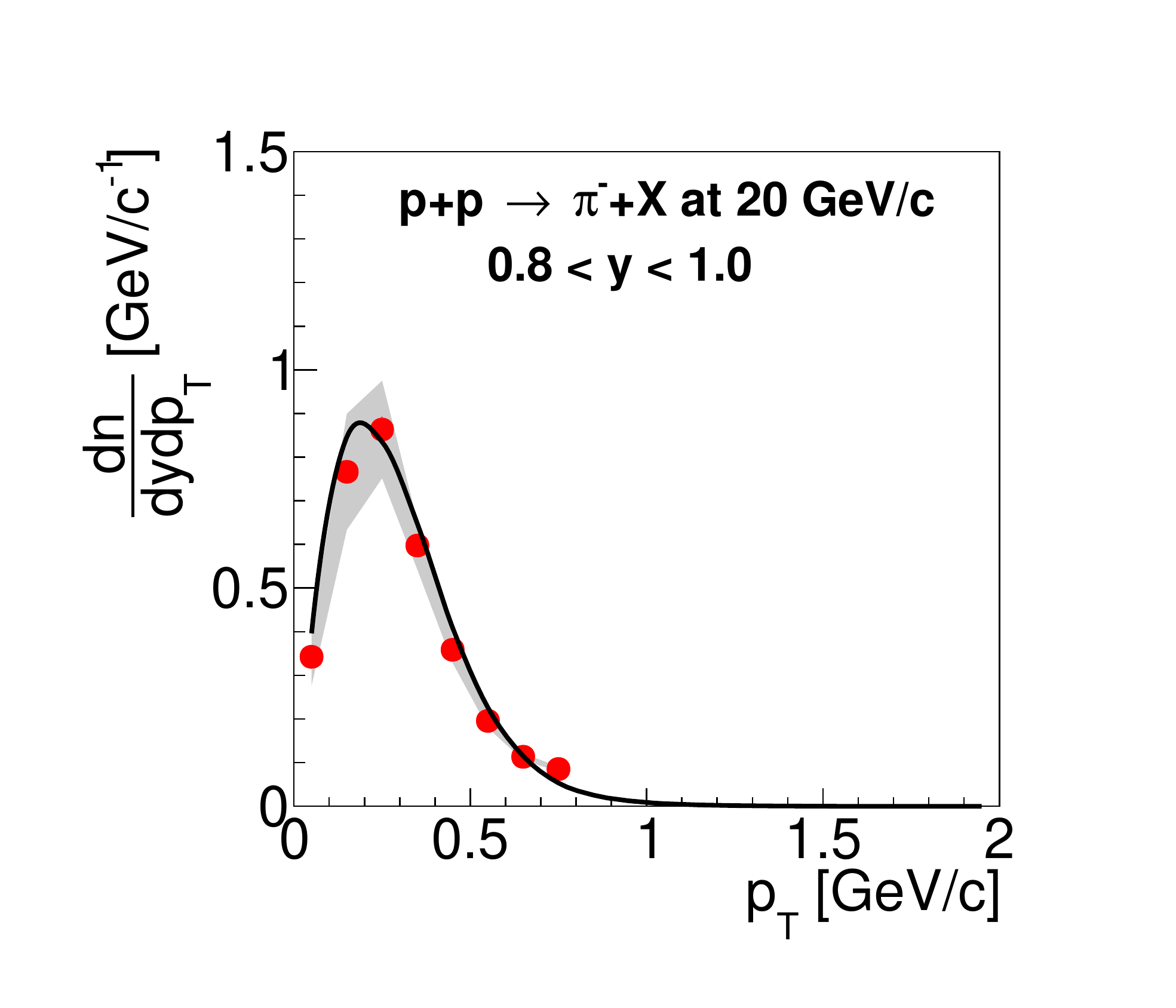}\\
	\includegraphics[width=0.3\textwidth]{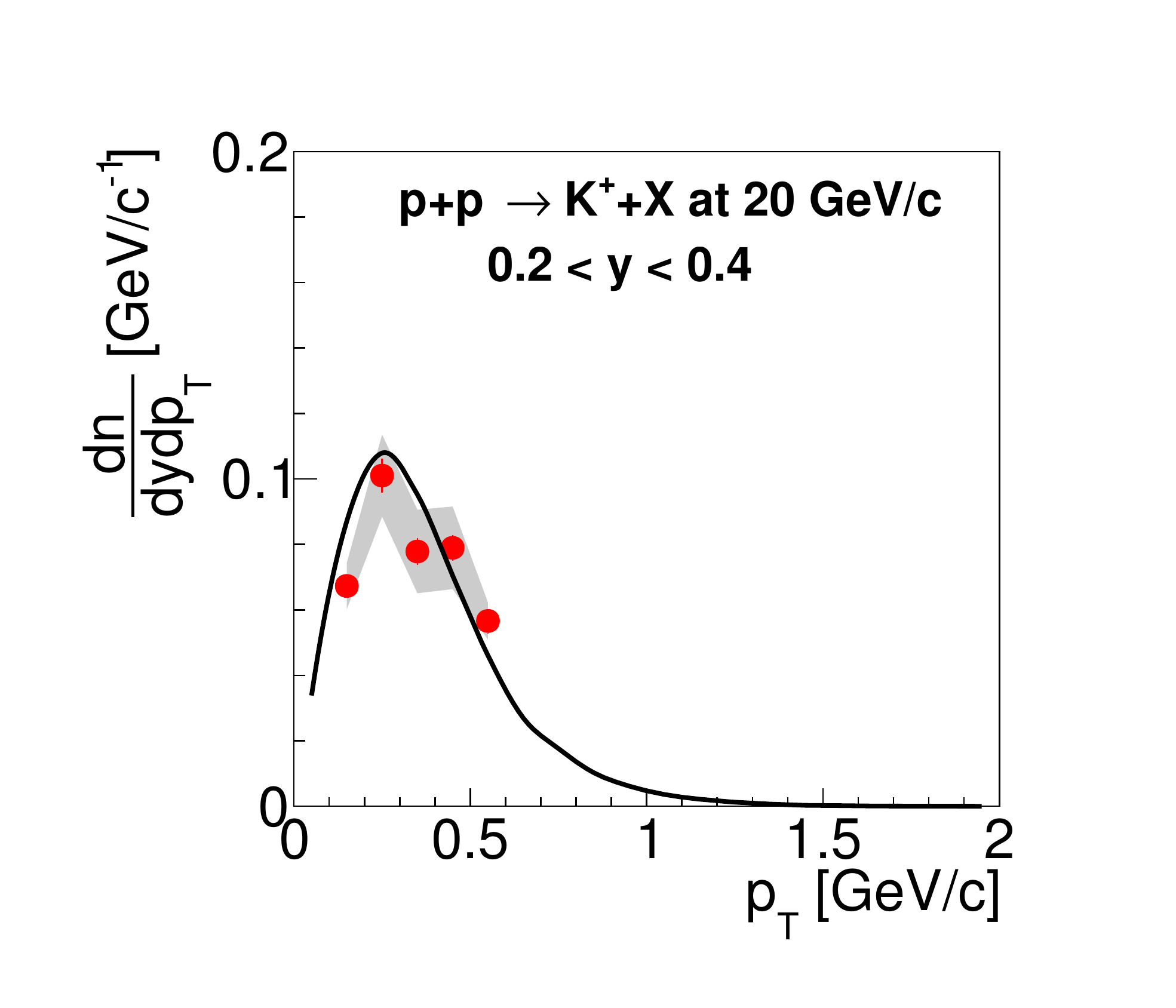}
	\includegraphics[width=0.3\textwidth]{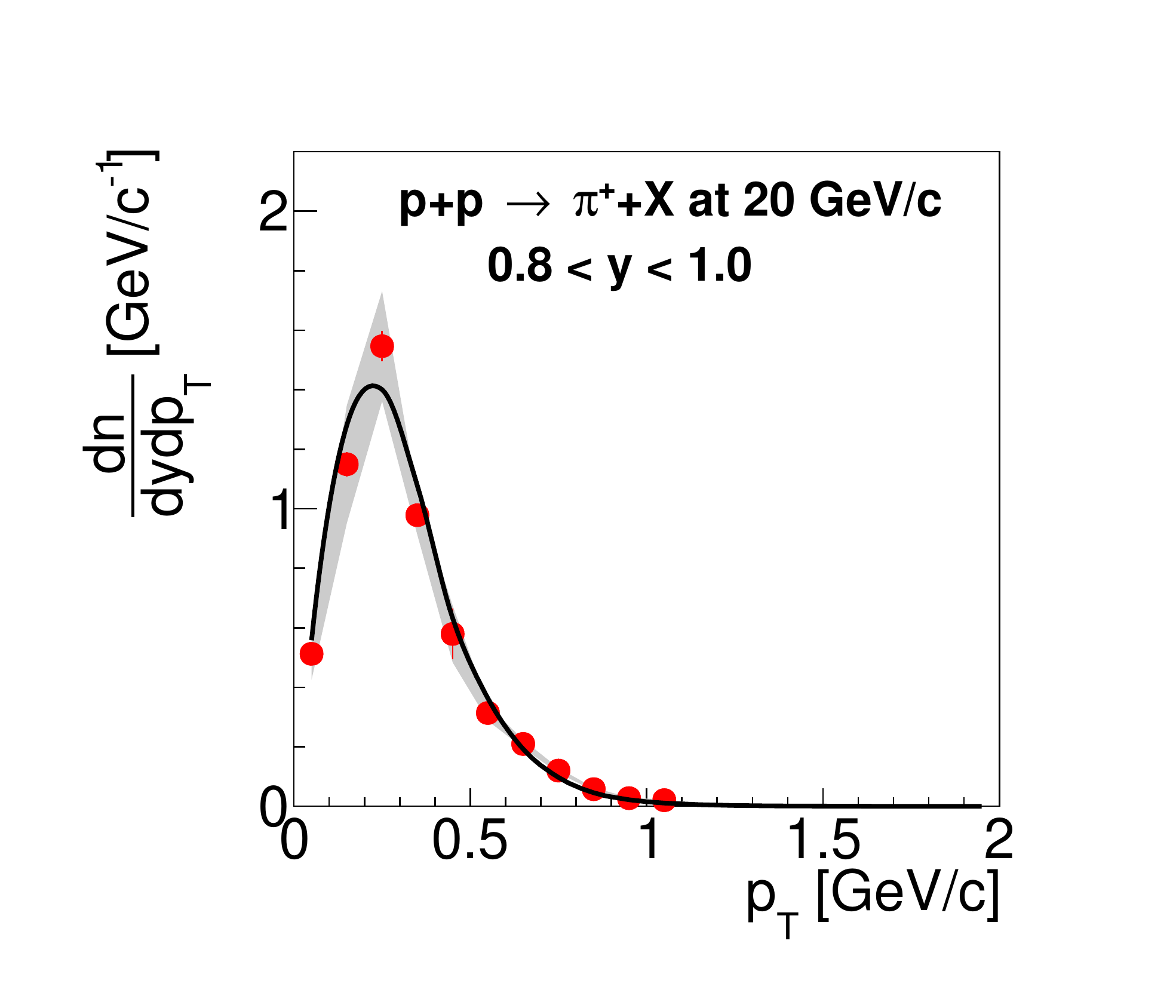}
	\includegraphics[width=0.3\textwidth]{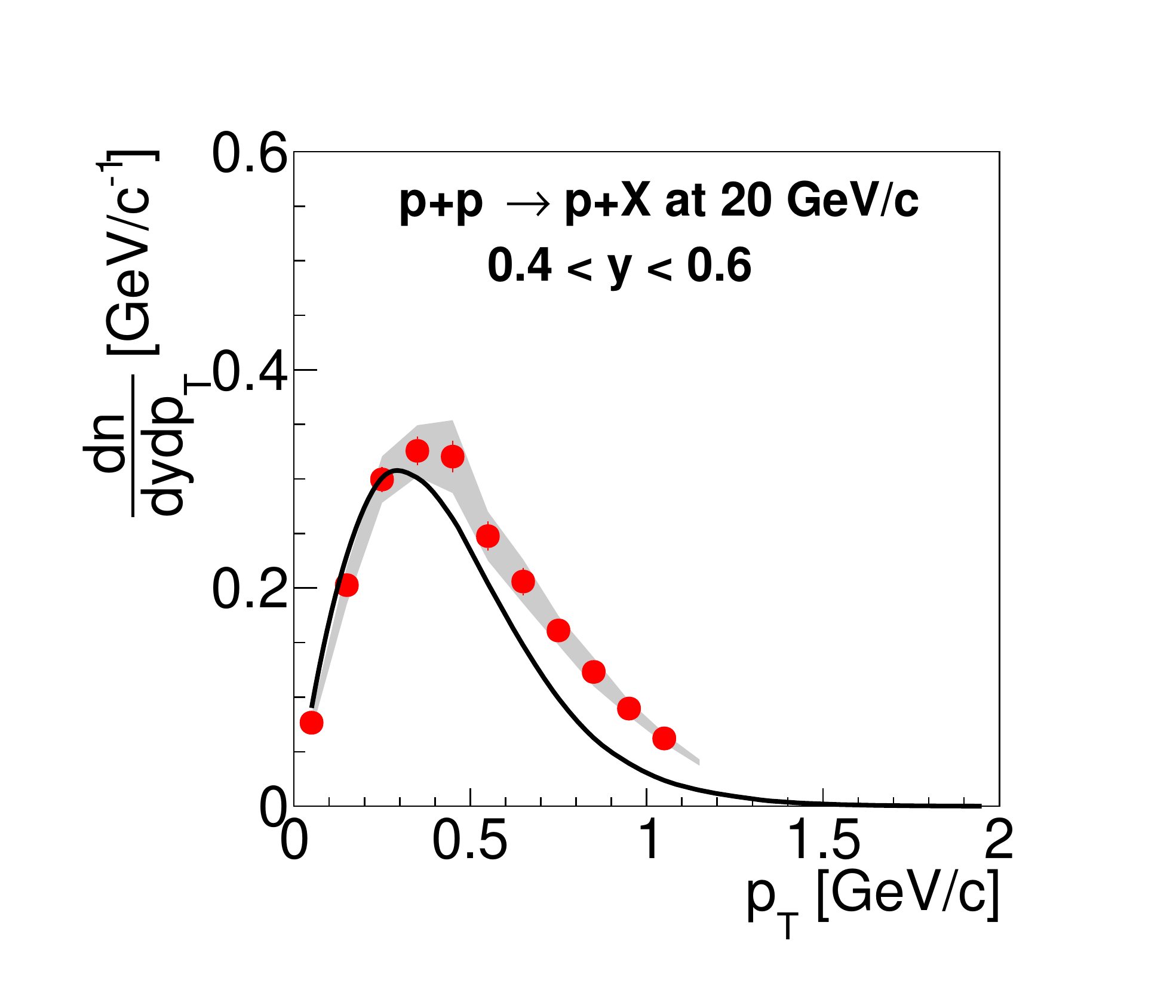}\\
	\end{center}
	\caption{(Color online) Transverse momentum spectra of identified hadrons produced in inelastic p+p interactions at 20~\GeVc in selected rapidity intervals  in comparison to the \Epos model~\cite{Werner:2008zza} (black line). Systematic uncertainties are shown as shaded bands.}
	\label{fig:epos_comps20}
\end{figure}

\begin{figure}[!ht]
	\begin{center}
	\includegraphics[width=0.3\textwidth]{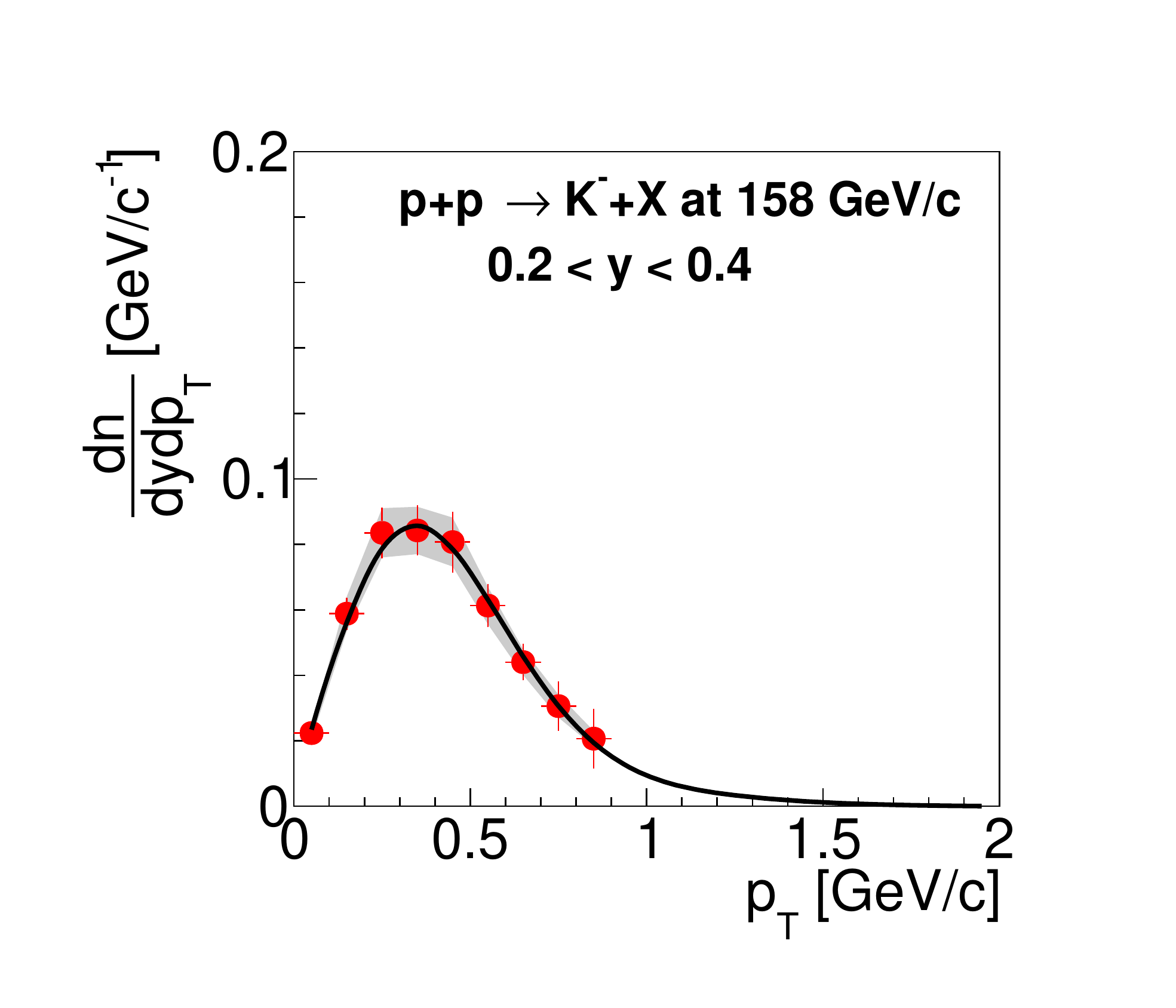}
	\includegraphics[width=0.3\textwidth]{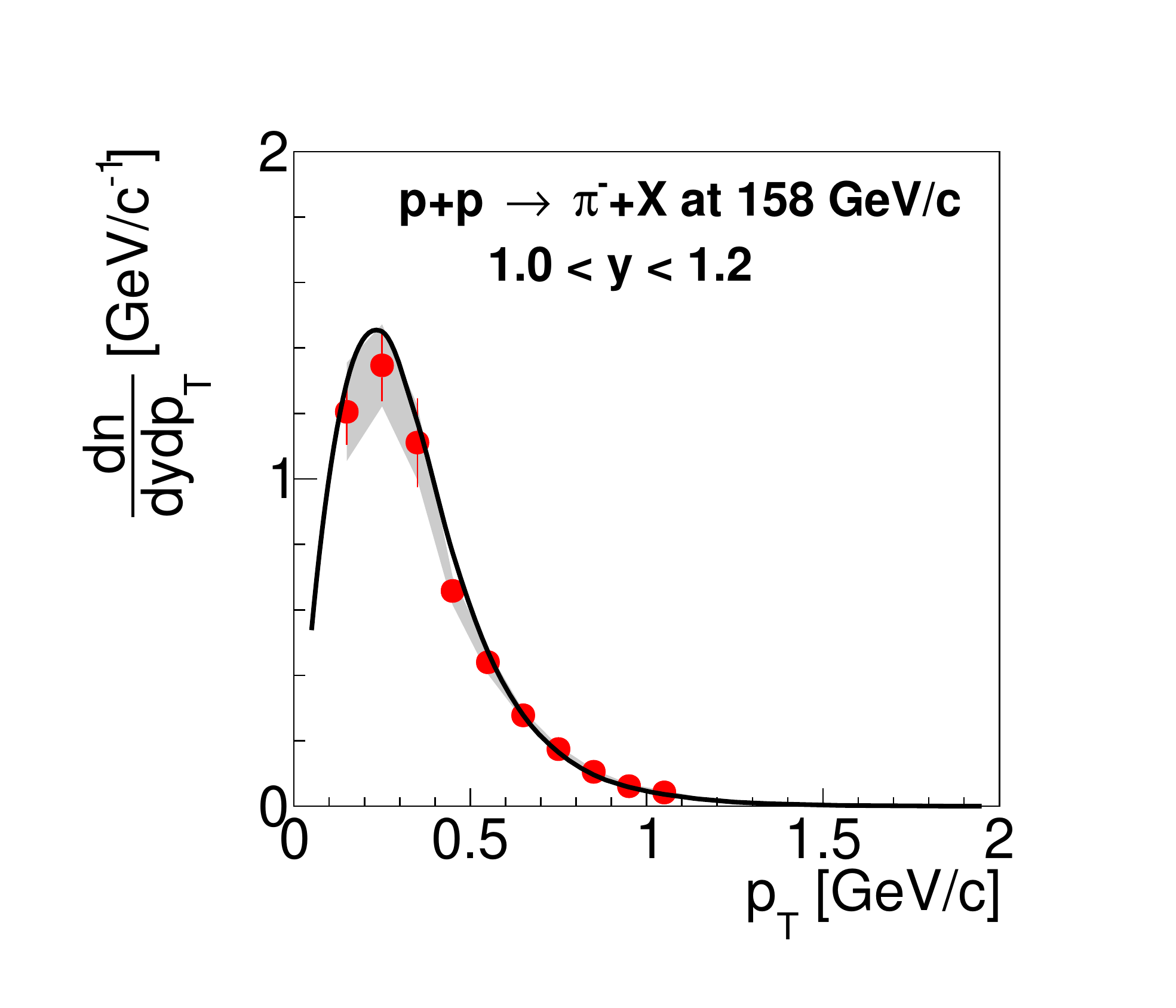}
	\includegraphics[width=0.3\textwidth]{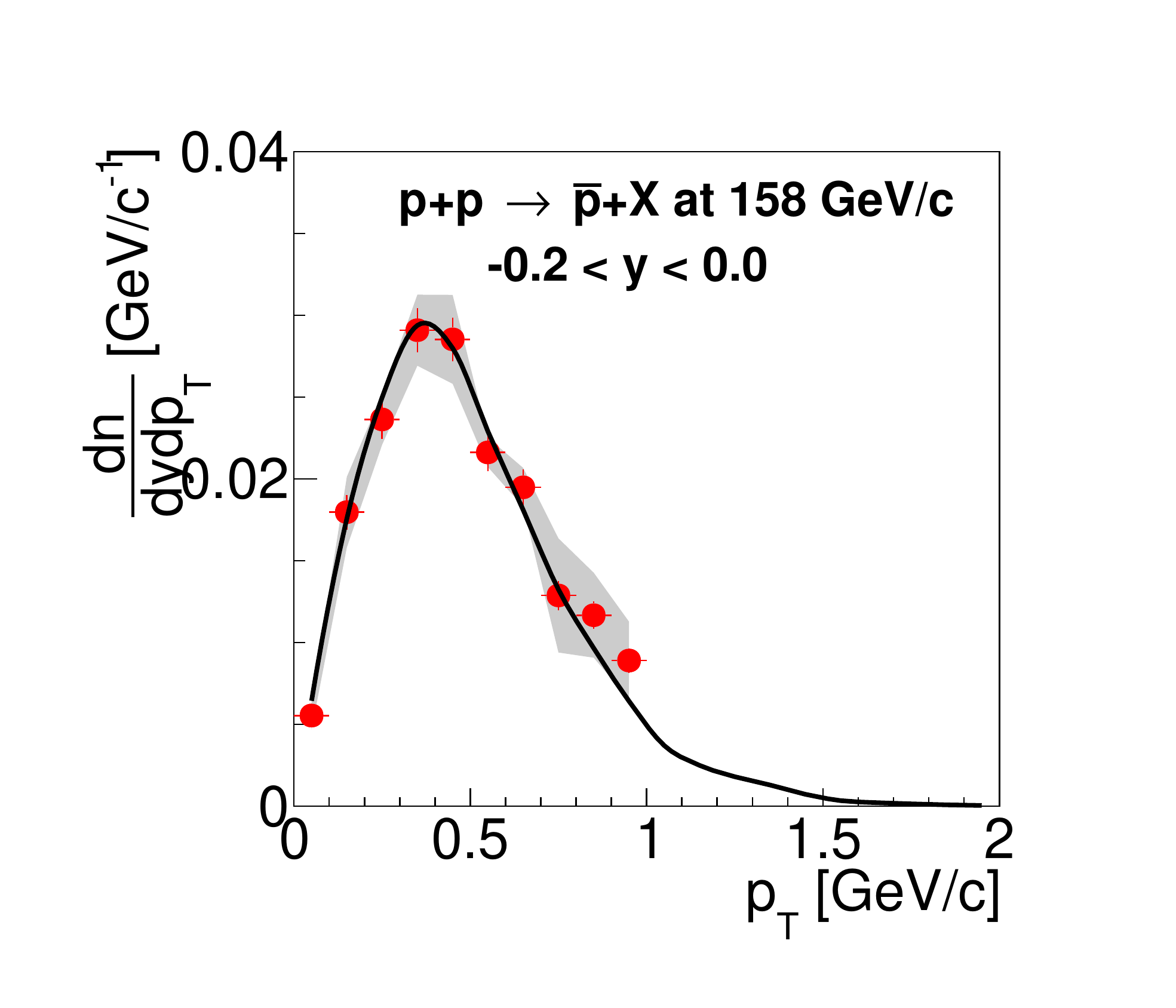}\\
	\includegraphics[width=0.3\textwidth]{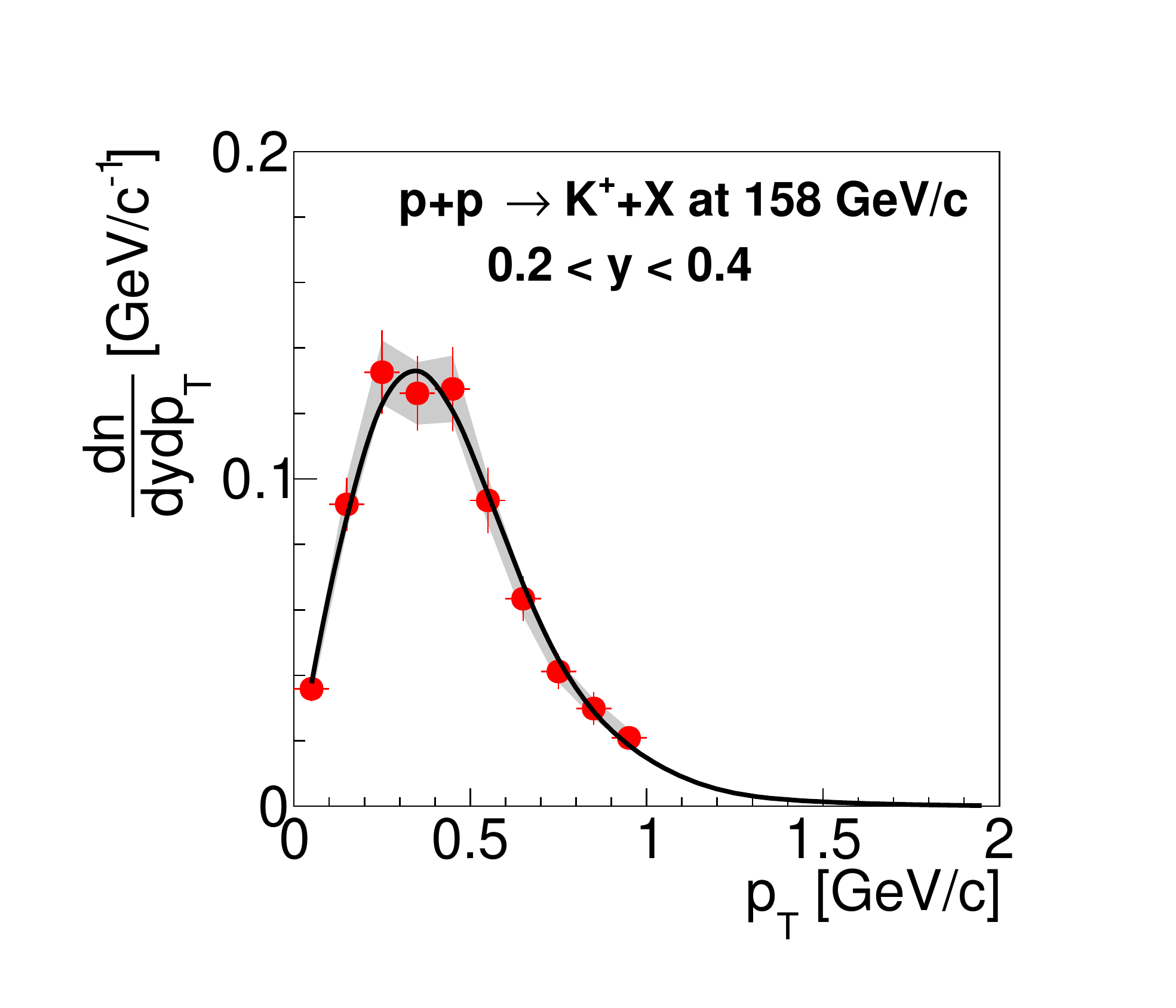}
	\includegraphics[width=0.3\textwidth]{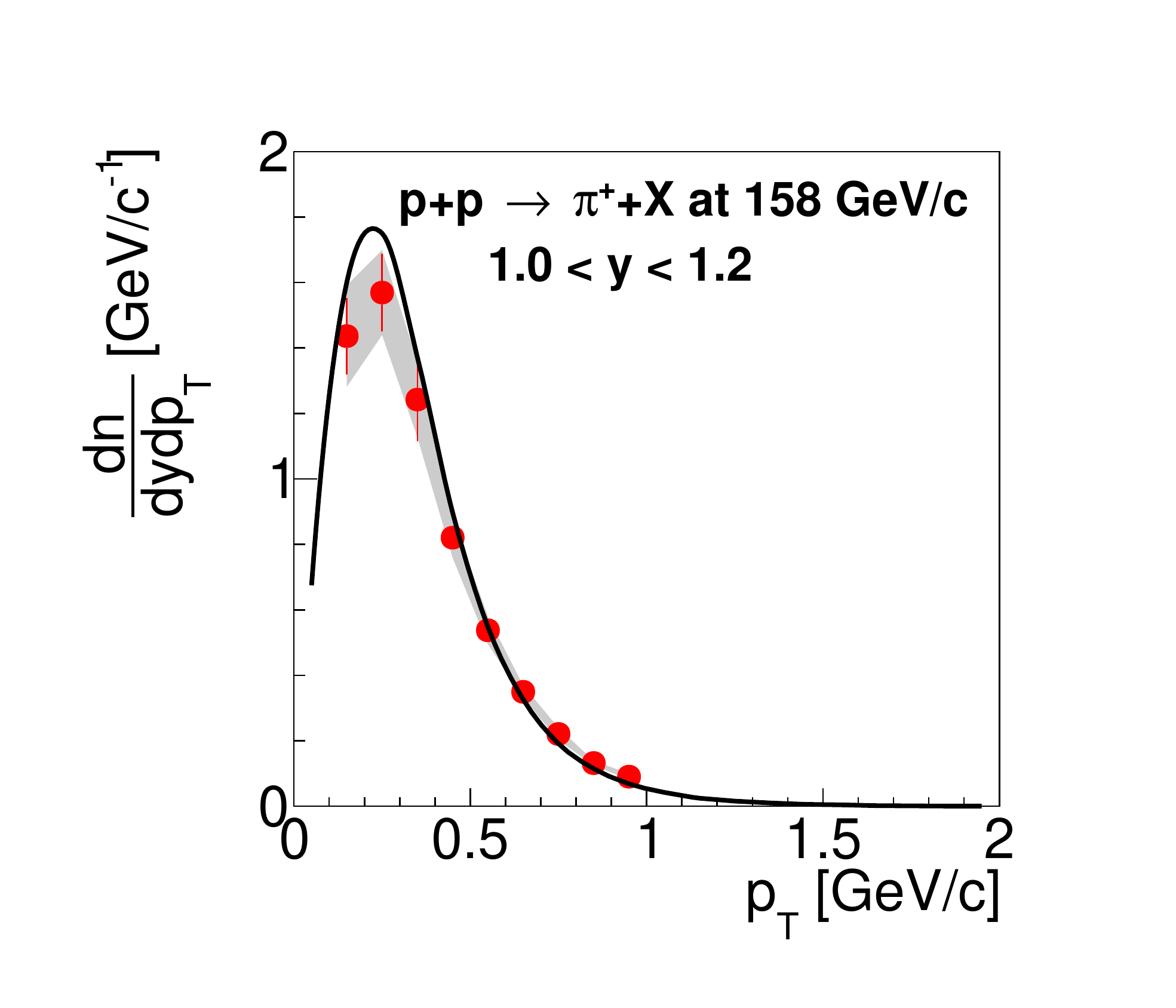}
	\includegraphics[width=0.3\textwidth]{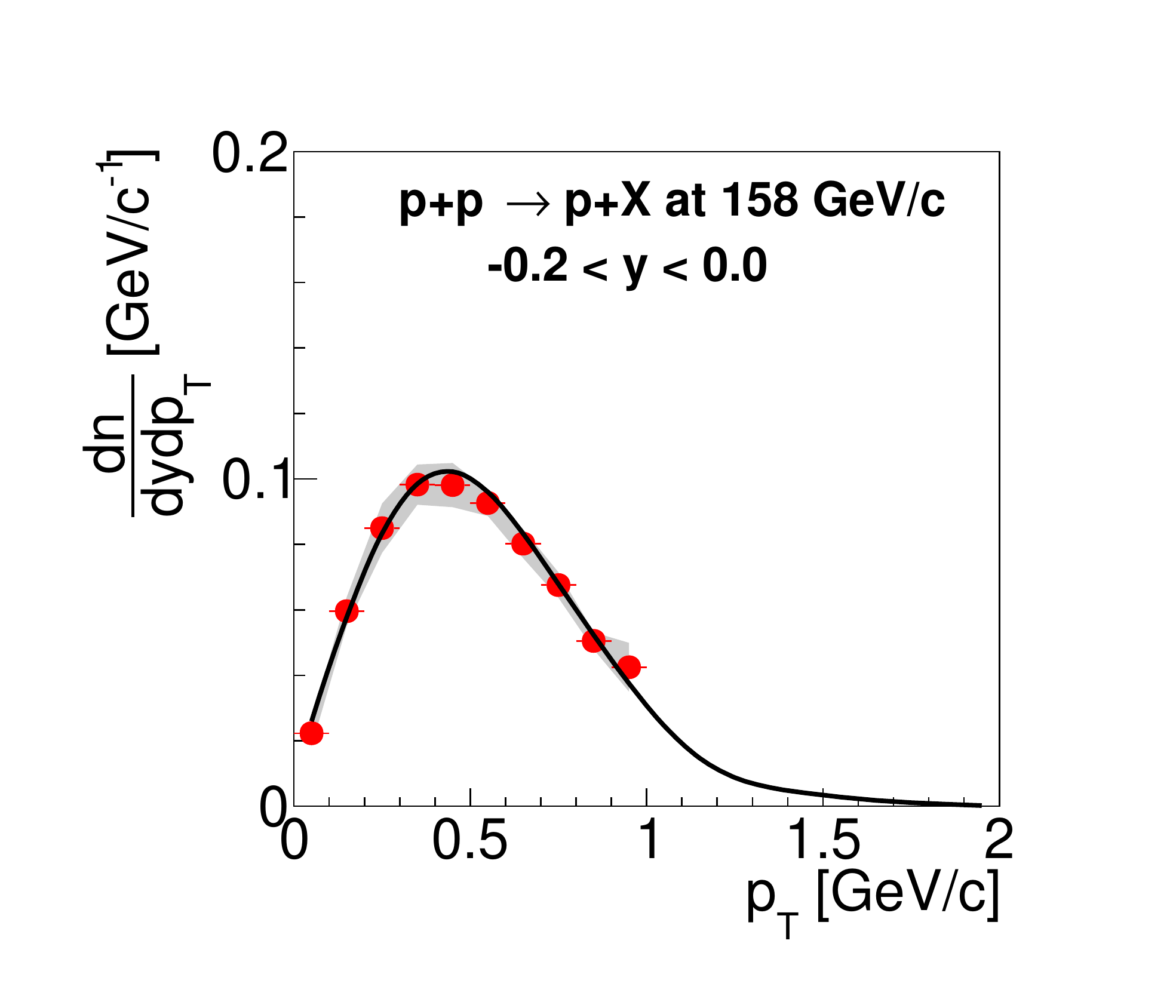}\\
	\end{center}
	\caption{(Color online) Transverse momentum spectra of identified hadrons produced in inelastic p+p interactions at 158~\GeVc in selected rapidity intervals  in comparison to the \Epos model~\cite{Werner:2008zza} (black line). Systematic uncertainties are shown as shaded bands.}
	\label{fig:epos_comps158}
\end{figure}

Systematic biases to the measurements could result mainly from:
	\begin{enumerate}[(i)]
    \item Methods of event selection.
    
    The first uncertainty is related to the acceptance of events with additional tracks from off-time particles. This systematic uncertainty was estimated by changing the width of the time window in which no second beam particle is allowed by $\pm~1~\mu s$ (variation by $\pm$50\%) with respect to the nominal value of $\pm$2$~\mu$s. The maximal difference of the results was assigned as the systematic uncertainty of the selection.
	
	The second source of possible systematic bias are losses of inelastic events due to the interaction trigger. The S4 detector veto mainly selects inelastic and removes elastic scattering events. However, it can also result in some losses of inelastic events. To estimate this effect, the analysis was done using correction factors calculated with and without applying the S4 trigger in the simulation. The difference between these two results was taken as a contribution to the systematic uncertainty.
	
	The next source of systematic uncertainty related to the normalization came from the selection window
for the $z$-position of the fitted vertex. To estimate the systematic uncertainty the selection criteria for
the data and the \Epos model were varied in the range of $\pm$10~\cm (50\%) around the nominal value.
	
	\item Methods of track selection.
	
	To estimate systematic uncertainty related to the track selection the following variations were performed independently: number of requested points on a track in all TPCs was changed by $\pm$5 (33\% of the standard selection), number of requested points on a track in the VTPCs was reduced and increased by 5 (18\% of the criterion).
		
	\item Identification techniques.
	
	Moreover, uncertainties of the \dedx identification method were studied and estimated by a 10\% variation of the parameter constraints for the function Eq.\ref{Eq:AsymGaus} used for particle identification. In case of $tof$-\dedx identification uncertainties were estimated by changing the selection criteria related to this technique - the last point of the track should be in the last 2$\pm$ 1 padrows of the MTPC, the QDC signal is or is not required. 
	
    \item Feeddown correction.
    
  The determination of the feeddown correction is based on the \Epos model which describes well the available cross section data for strange particles (see e.g. for $\Lambda$ at 158~\GeVc Ref.~\cite{Aduszkiewicz:2015dmr} and at 40~\GeVc Ref.~\cite{SQM2016_HS} and Figs.~\ref{fig:epos_comps20} and \ref{fig:epos_comps158} in this paper for K$^{+}$ and K$^{-}$). Systematic uncertainty comes from the lack of precise knowledge of the production cross section of K$^{+}$, K$^{-}$, $\Lambda$, $\Sigma^{+}$, $\Sigma^{-}$, K$^{0}_{s}$ and $\bar{\Lambda}$ in case of pions, and in addition of $\Sigma^{+}$ in case of protons, and $\bar{\Lambda}$ in case of antiprotons. Since the corrections are only at the level of a few percent in the phase space region of the measurements a small additional systematic error of 1\% was assumed in case of pions and 2\% for protons and antiprotons.
    
	\end{enumerate}

\begin{figure}[!ht]
		\begin{center}
		\includegraphics[width=0.8\textwidth]{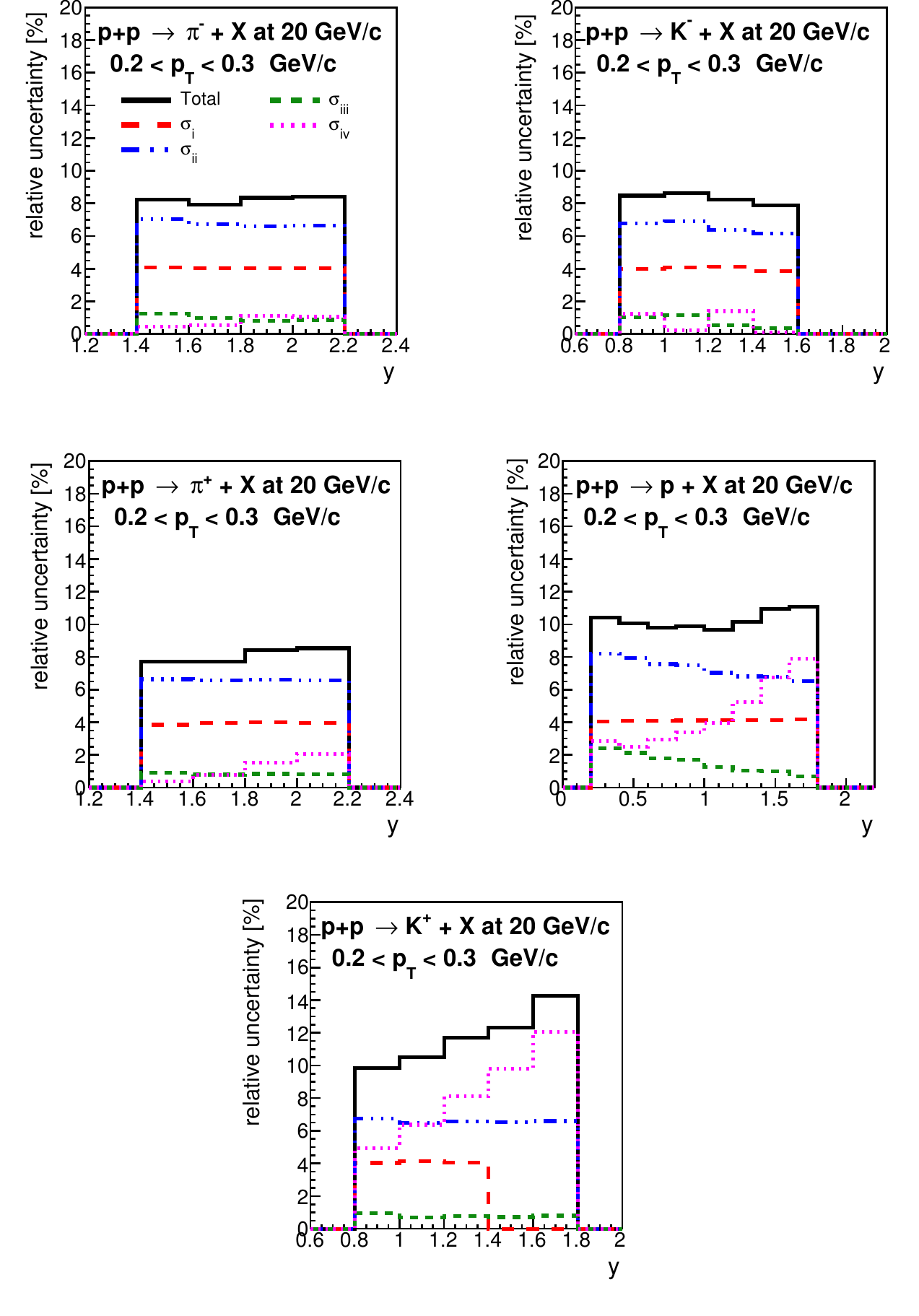}
		\end{center}
		\caption{(Color online) Components of systematic uncertainty of particle spectra 
obtained from the \dedx method in inelastic p+p interaction at 20~\GeVc as function of rapidity for the transverse momentum interval between $0.2 - 0.3$~\GeVc. $\sigma_{\textrm{i}}$ refers to acceptance of events with off-time beam tracks, $\sigma_{\textrm{ii}}$ to possible bias of the S4 trigger, $\sigma_{\textrm{iii}}$ to event vertex and track selection procedure} and $\sigma_{\textrm{iv}}$ to the identification technique. Black lines (Total) show the total systematic uncertainty calculated as the square root of the sum of squares of the components.
		\label{fig:sysexa20}
	\end{figure}

\begin{figure}[!ht]
		\begin{center}
		\includegraphics[width=0.8\textwidth]{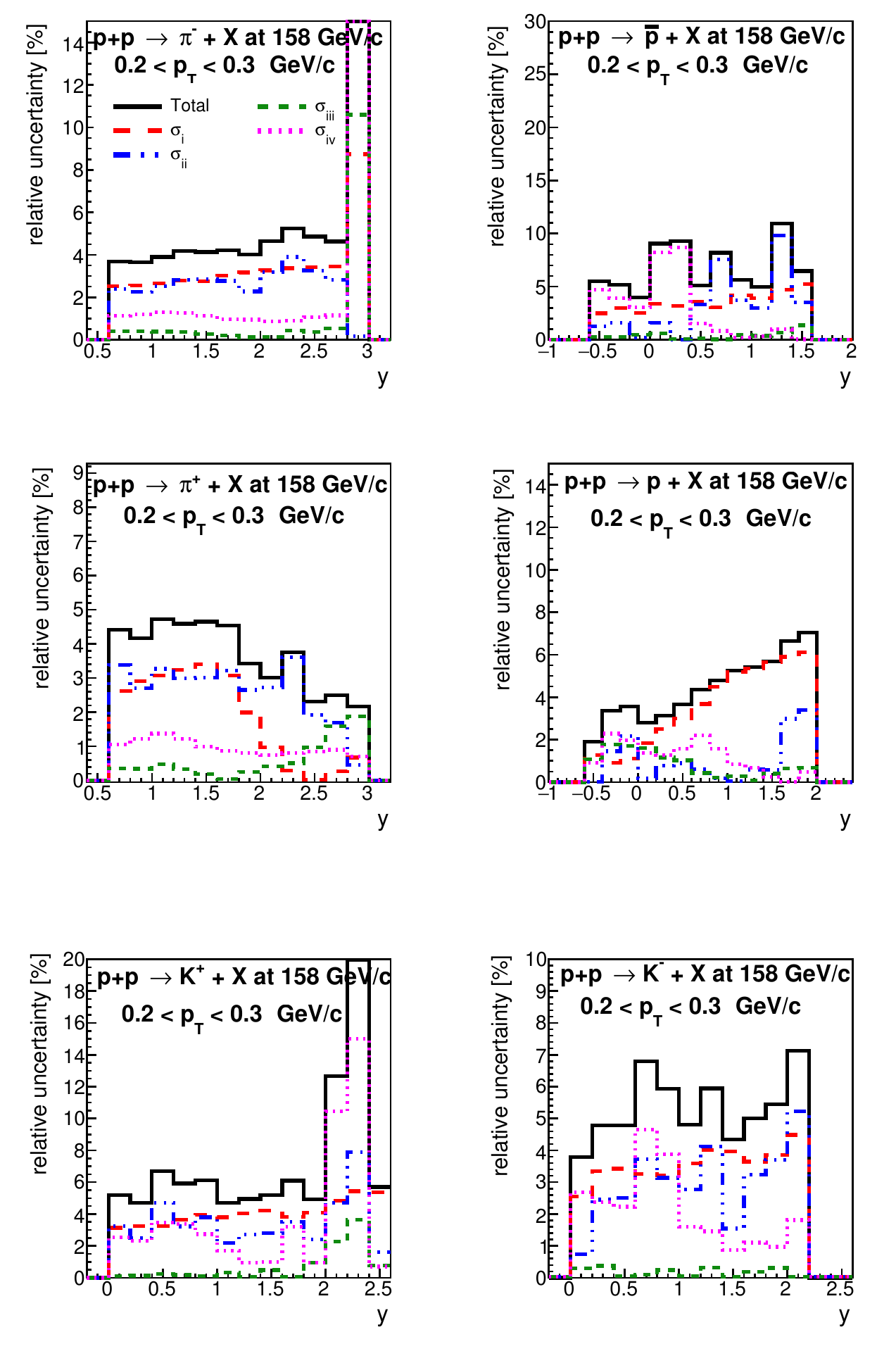}
		\end{center}
		\caption{(Color online) Components of systematic uncertainty of particle spectra 
obtained from the \dedx method in inelastic p+p interaction at 158~\GeVc as function of rapidity for the transverse momentum interval between $0.2 - 0.3$~\GeVc. $\sigma_{\textrm{i}}$ refers to removal of events with off-time beam tracks, $\sigma_{\textrm{ii}}$ to possible bias of the S4 trigger, $\sigma_{\textrm{iii}}$ to event vertex and track selection procedure} and $\sigma_{\textrm{iv}}$ to the identification technique. Black lines (Total) show the total systematic uncertainty calculated as the square root of the sum of squares of the components.
		\label{fig:sysexa158}
	\end{figure}

\begin{figure}[!ht]
		\begin{center}
		\includegraphics[width=0.8\textwidth]{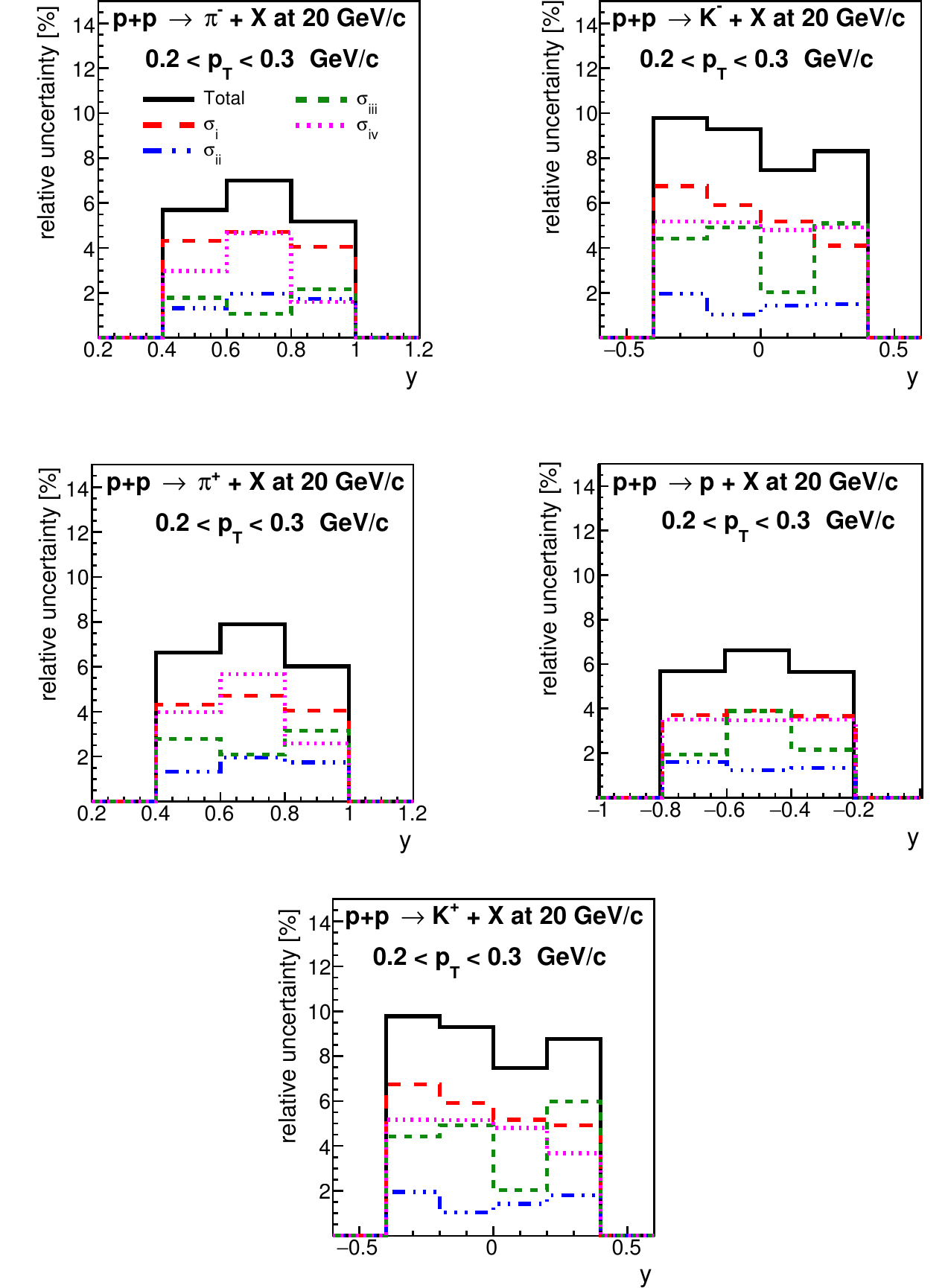}
		\end{center}
		\caption{(Color online) Components of systematic uncertainty of particle spectra 
obtained from the $tof$-\dedx method in inelastic p+p interaction at 20~\GeVc as function of rapidity for the transverse momentum interval between $0.2 - 0.3$~\GeVc. $\sigma_{\textrm{i}}$ refers to removal of events with off-time beam tracks, $\sigma_{\textrm{ii}}$ to possible bias of the S4 trigger, $\sigma_{\textrm{iii}}$ to event vertex and track selection procedure} and $\sigma_{\textrm{iv}}$ to the identification technique. Black lines (Total) show the total systematic uncertainty calculated as the square root of the sum of squares of the components.
		\label{fig:sysexa20t}
	\end{figure}

\begin{figure}[!ht]
		\begin{center}
		\includegraphics[width=0.8\textwidth]{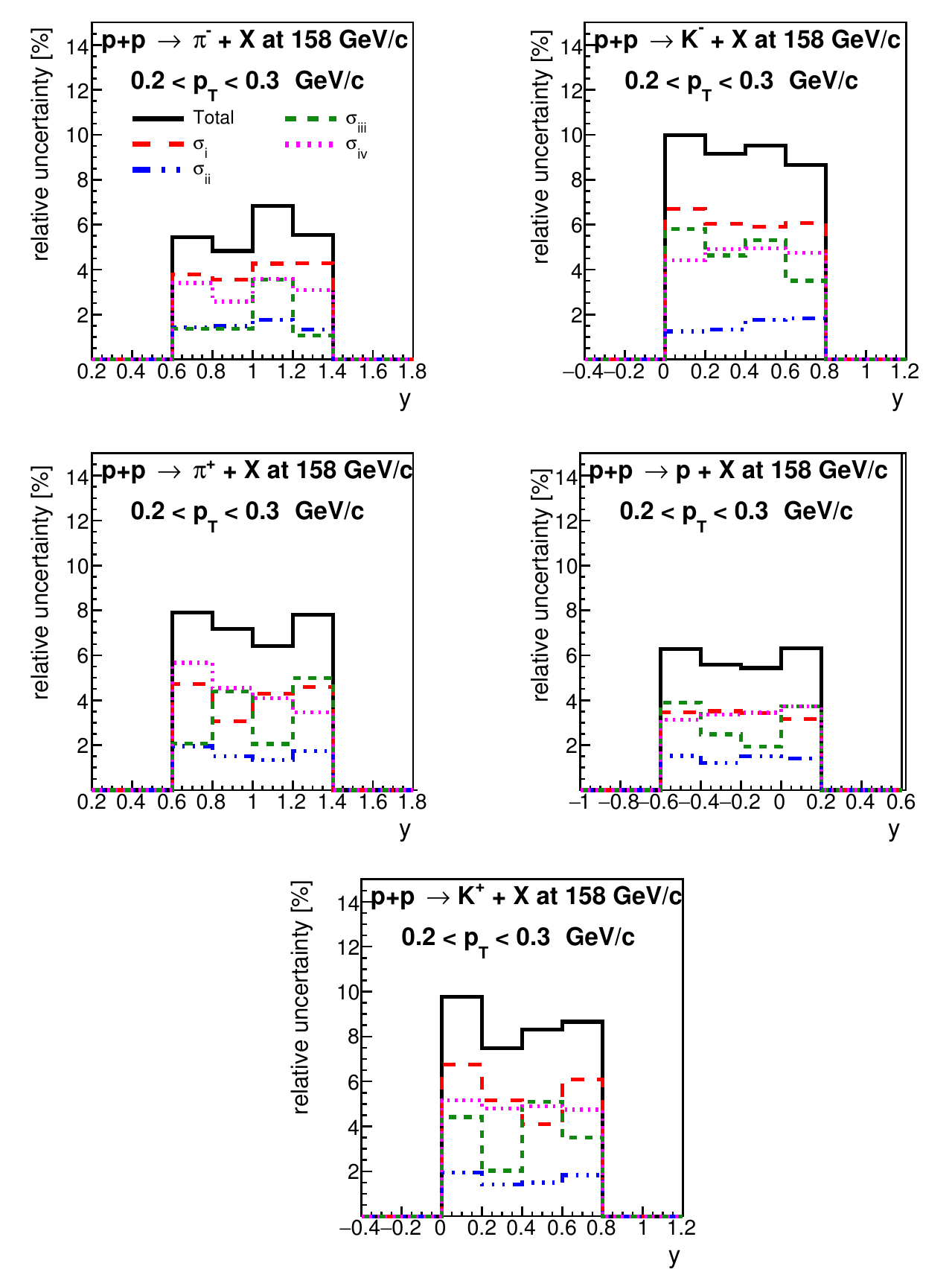}
		\end{center}
		\caption{(Color online) Components of systematic uncertainty of particle spectra
obtained from the $tof$-\dedx method in inelastic p+p interaction at 158~\GeVc as function of rapidity for the transverse momentum interval between $0.2 - 0.3$~\GeVc. $\sigma_{\textrm{i}}$ refers to removal of events with off-time beam tracks, $\sigma_{\textrm{ii}}$ to possible bias of the S4 trigger, $\sigma_{\textrm{iii}}$ to event vertex and track selection procedure} and $\sigma_{\textrm{iv}}$ to the identification technique. Black lines (Total) show the total systematic uncertainty calculated as the square root of the sum of squares of the components.
		\label{fig:sysexa158t}
	\end{figure}

	The components of the systematic uncertainty and the total for the \dedx and $tof$-\dedx methods, for a selected $p_{T}$ interval and beam momenta of 20 and 158~\GeVc are presented in Figs.~\ref{fig:sysexa20}, \ref{fig:sysexa158}, \ref{fig:sysexa20t} and \ref{fig:sysexa158t}. Assuming independence of all systematic error sources, the total systematic uncertainty was calculated as the square root of the sum of squares of the described components.
	
\begin{figure}[!ht]
	\begin{center}
	\includegraphics[width=0.3\textwidth]{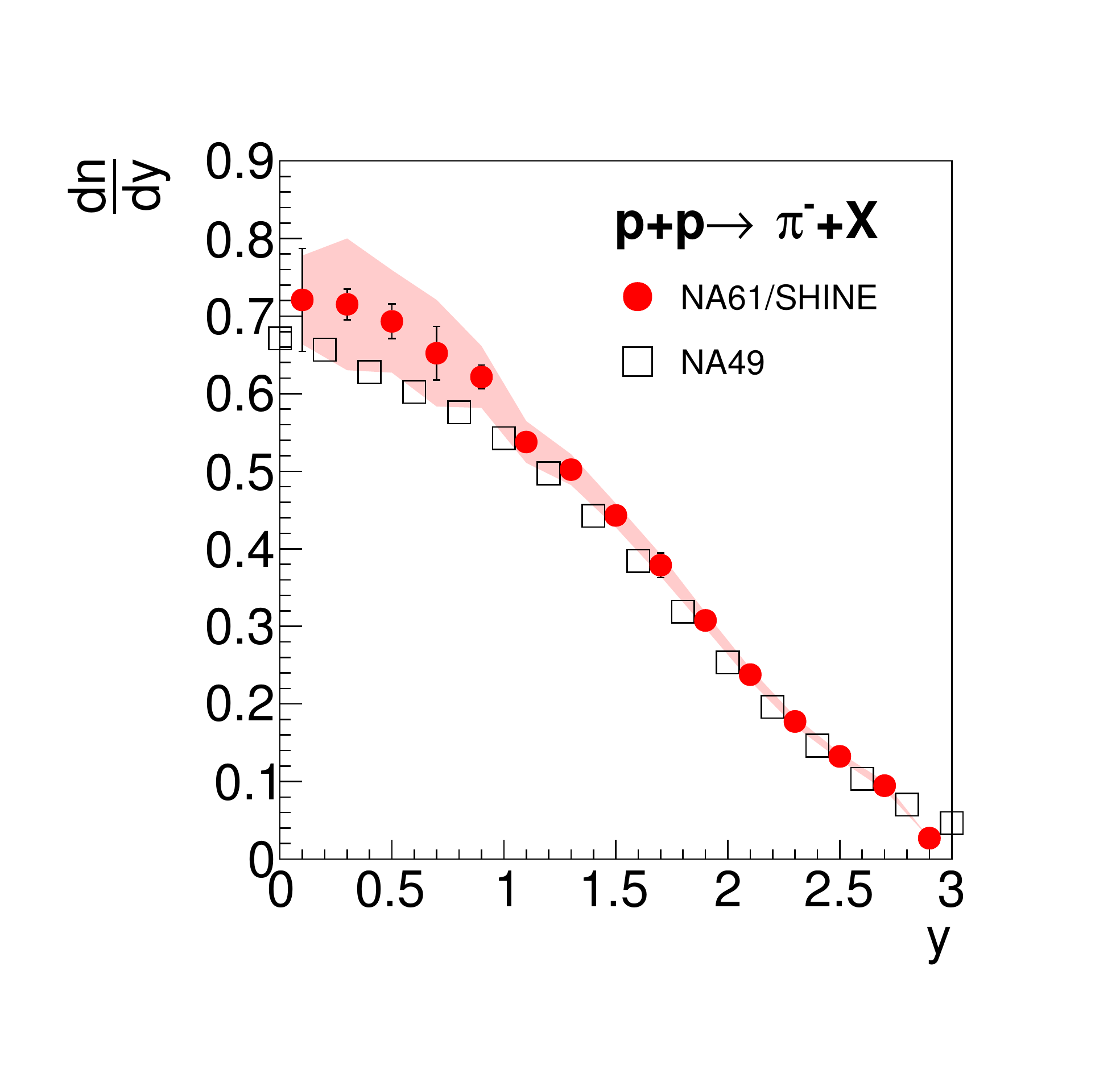}
	\includegraphics[width=0.3\textwidth]{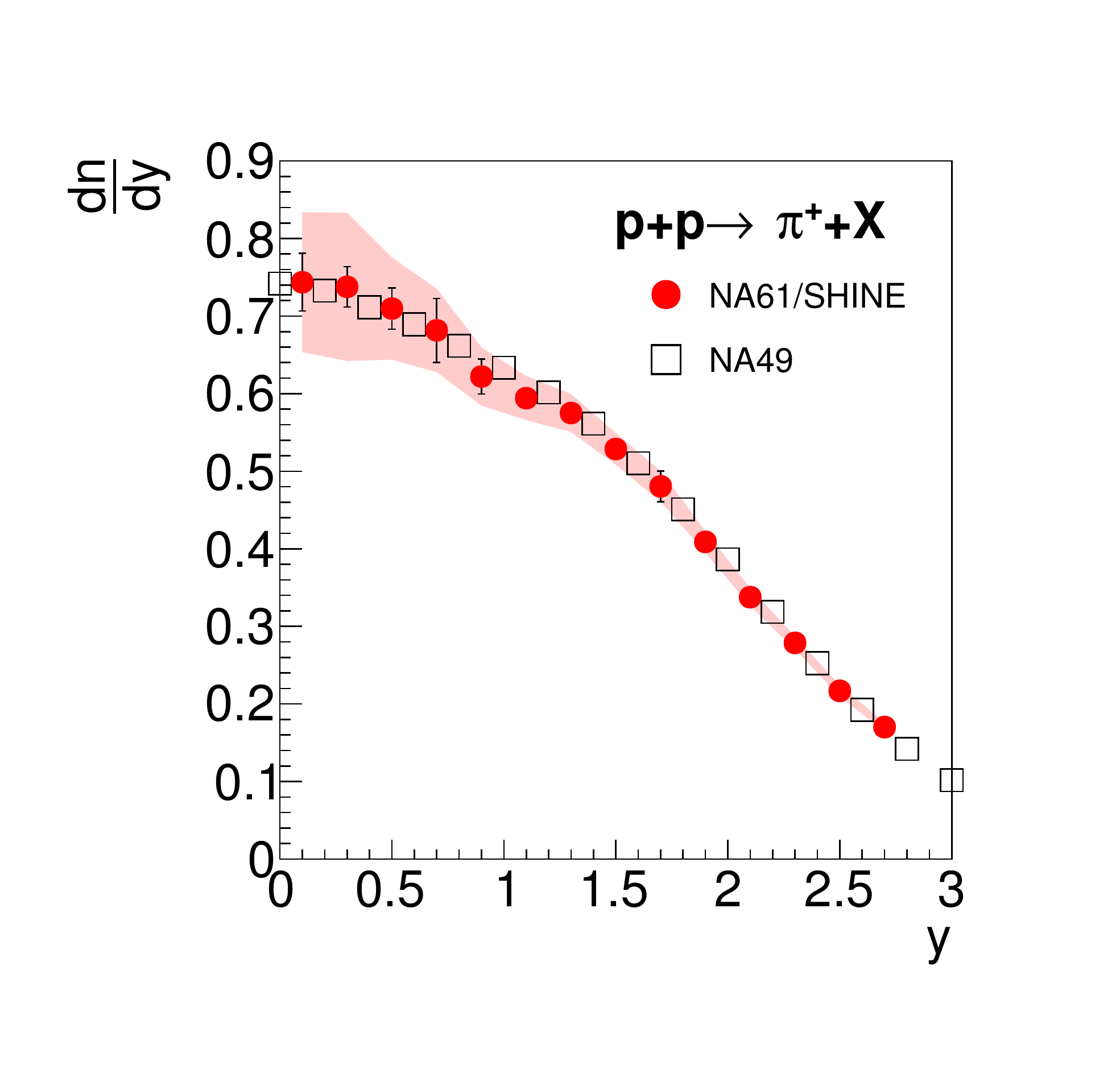}
	\includegraphics[width=0.3\textwidth]{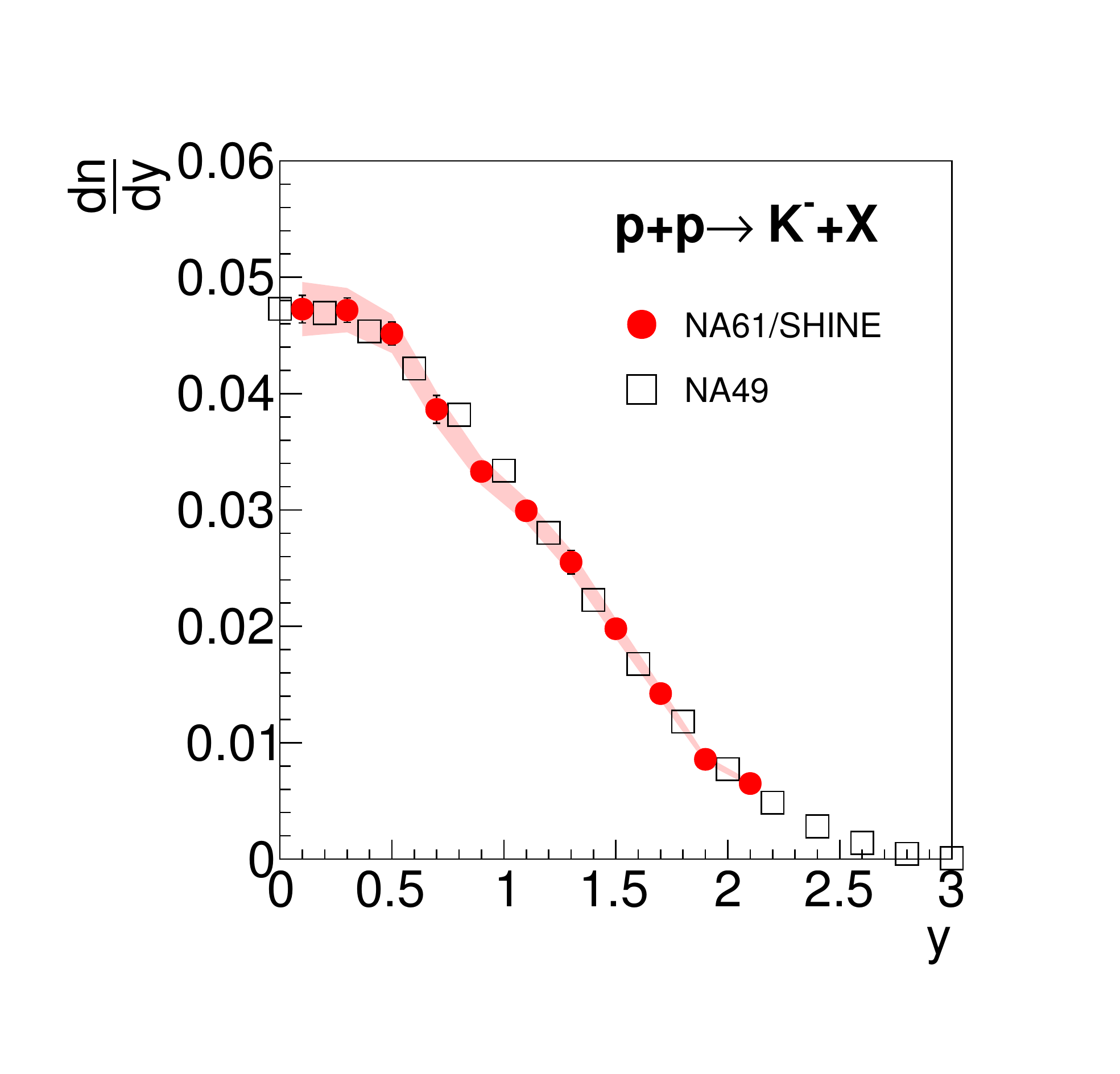}\\
	\includegraphics[width=0.3\textwidth]{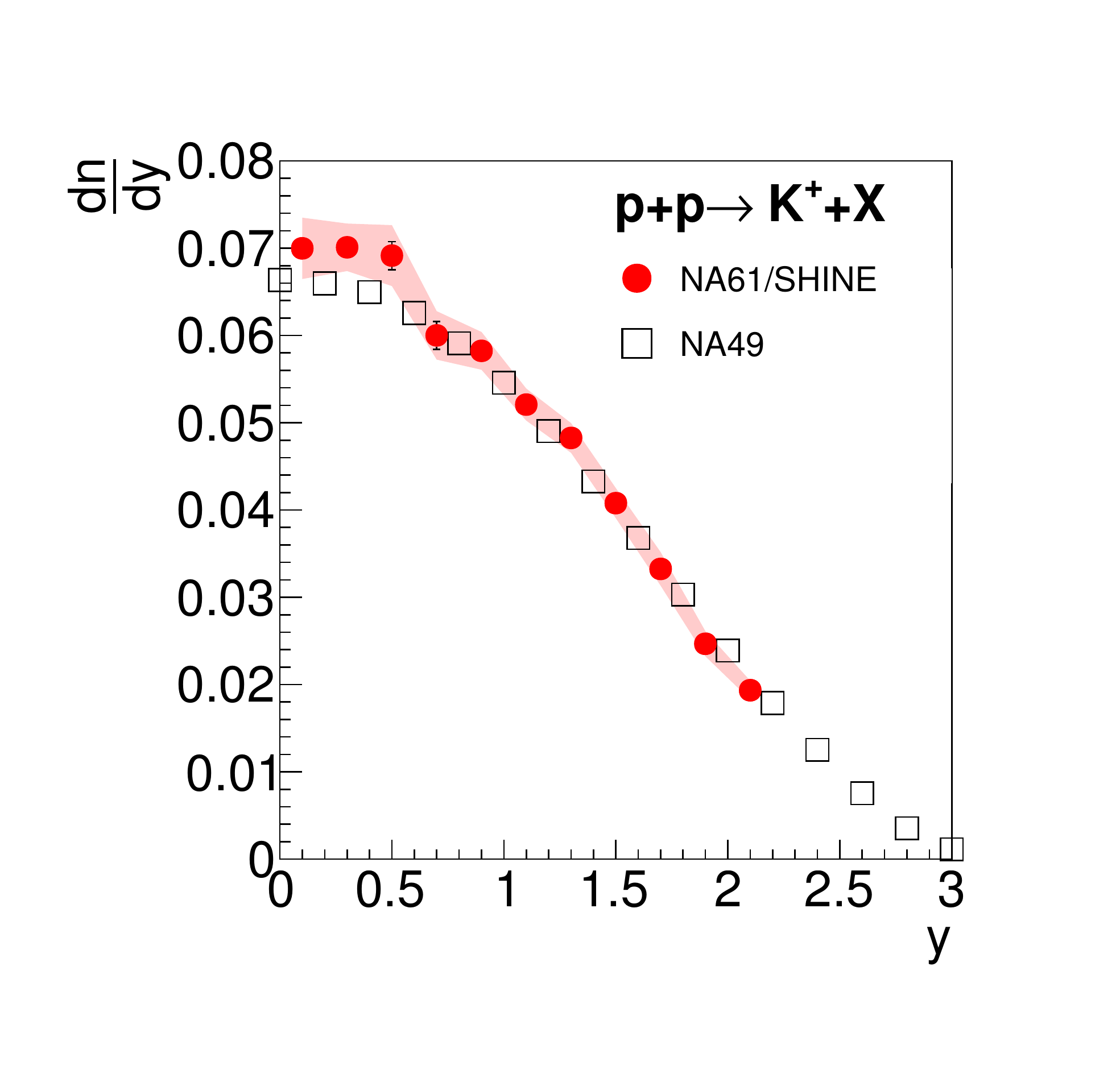}
	\includegraphics[width=0.3\textwidth]{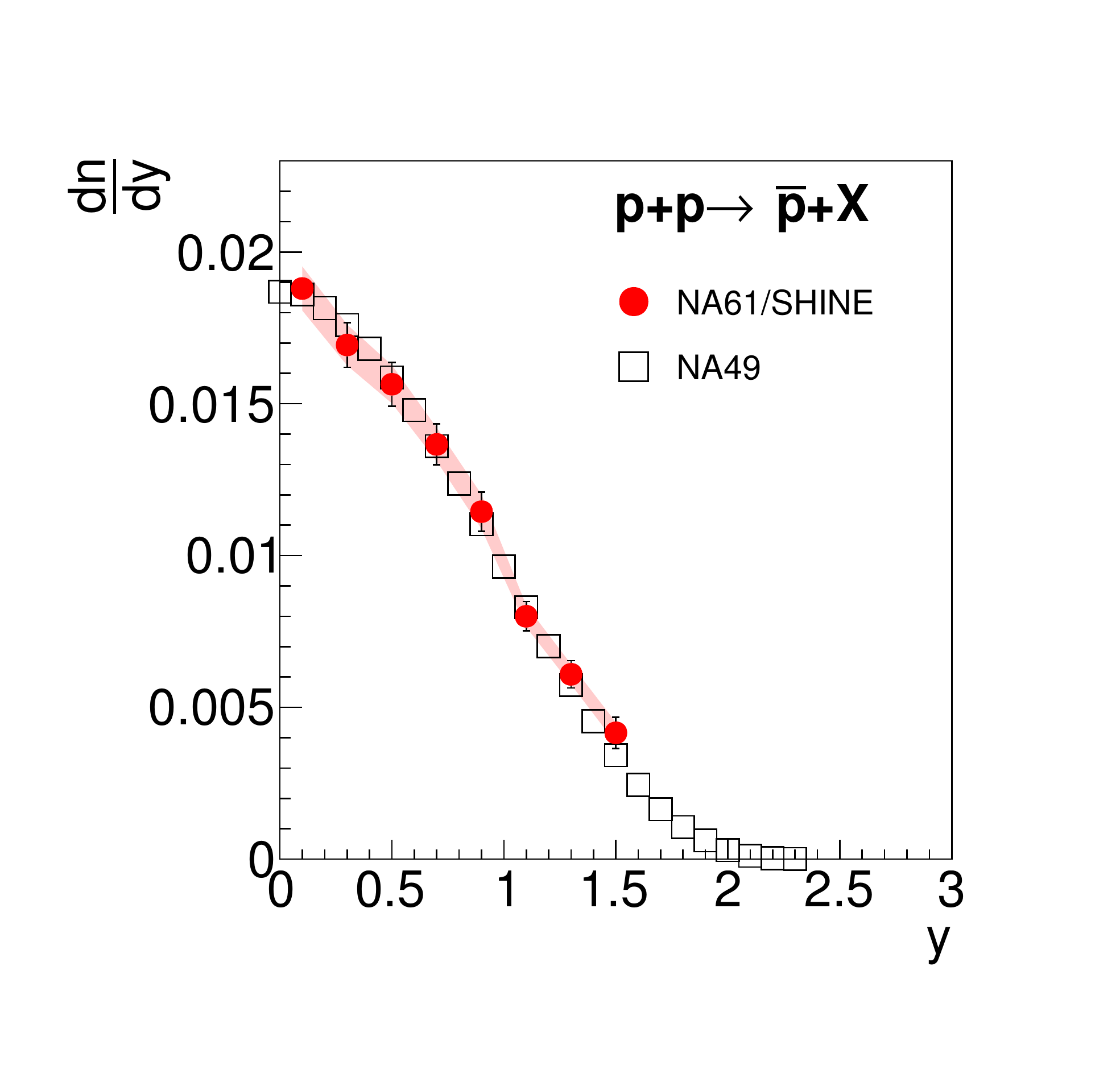}
	\includegraphics[width=0.3\textwidth]{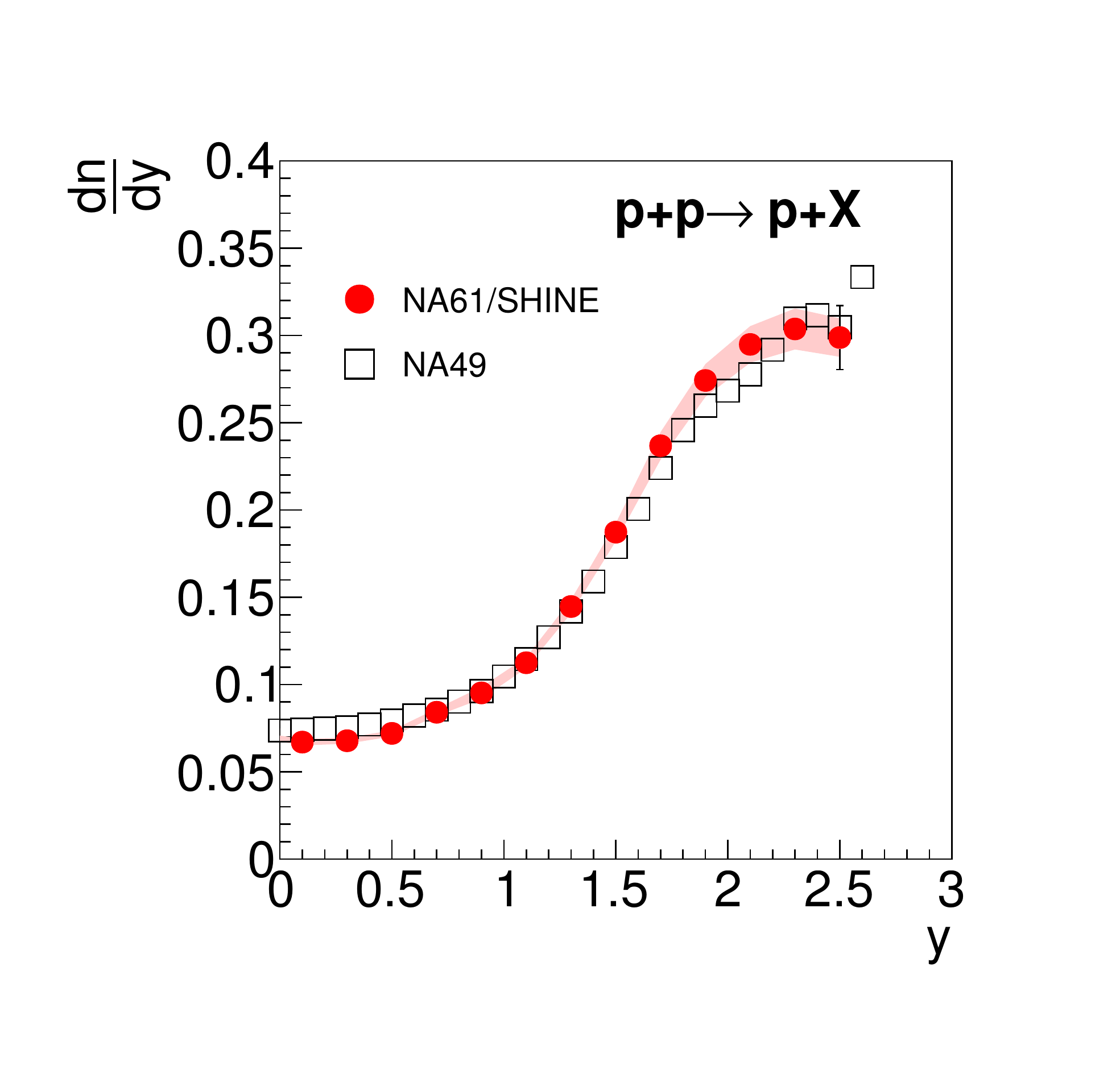}
	\end{center}
	\caption{(Color online) Comparison of rapidity distributions of pions, kaons and protons produced in inelastic p+p interactions at 158~\GeVc. The plotted NA49 results were published without uncertainties~\cite{Alt:2005zq,Anticic:2009wd,Anticic:2010yg}. \NASixtyOne systematic uncertainties are shown as shaded red bands.}
	\label{fig:withna49}
\end{figure}

Production of positive and negative pions, kaons and protons in p+p interactions at 158~\GeVc was measured before by the NA49 experiment~\cite{Alt:2005zq,Anticic:2009wd,Anticic:2010yg}. Comparison of \NASixtyOne and NA49 rapidity distributions is presented in Fig.~\ref{fig:withna49}. The results are consistent within the systematic uncertainty. 

\begin{figure}[!ht]
	\begin{center}
	\includegraphics[width=0.3\textwidth]{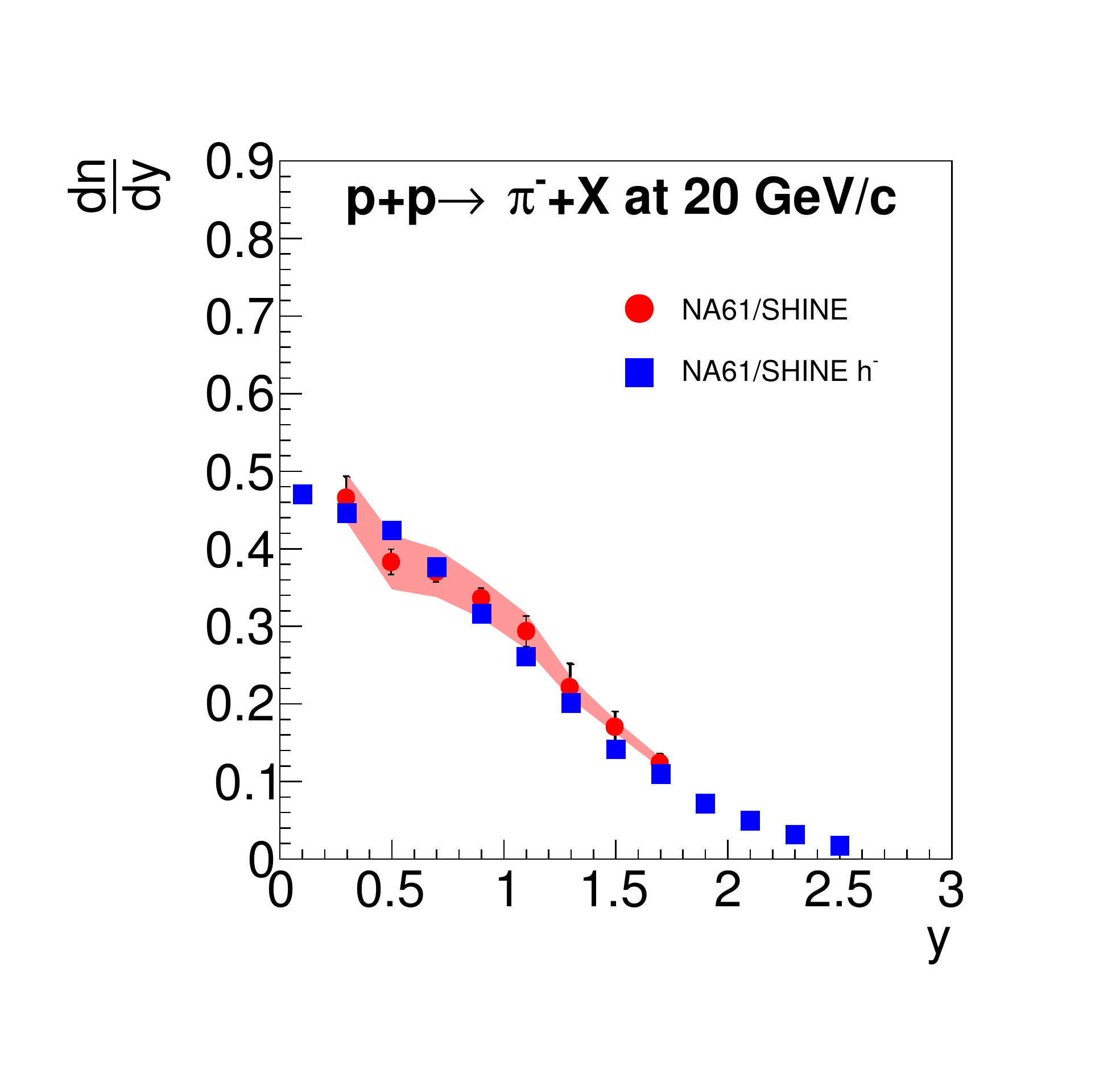}
	\includegraphics[width=0.3\textwidth]{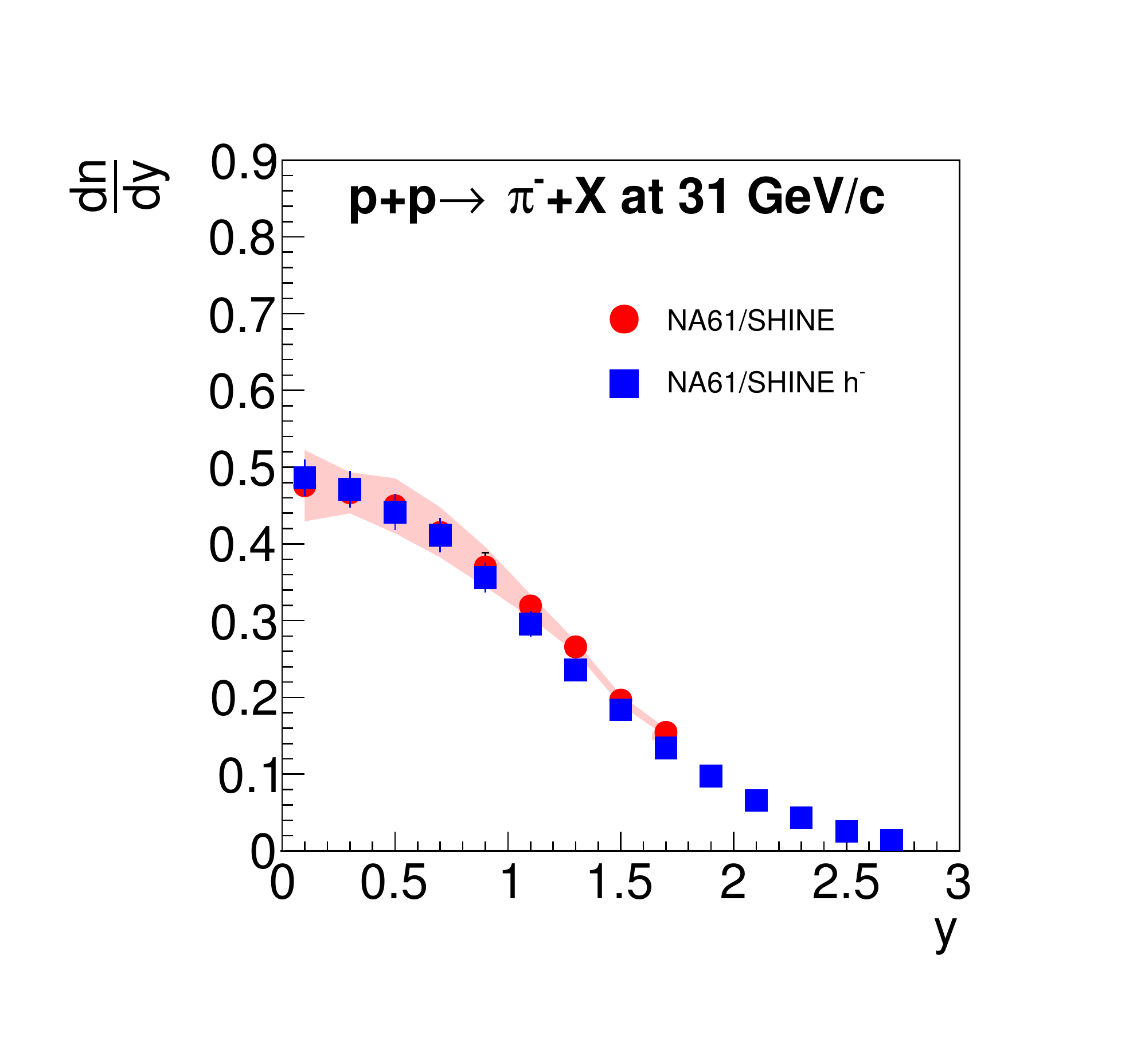}\\
	\includegraphics[width=0.3\textwidth]{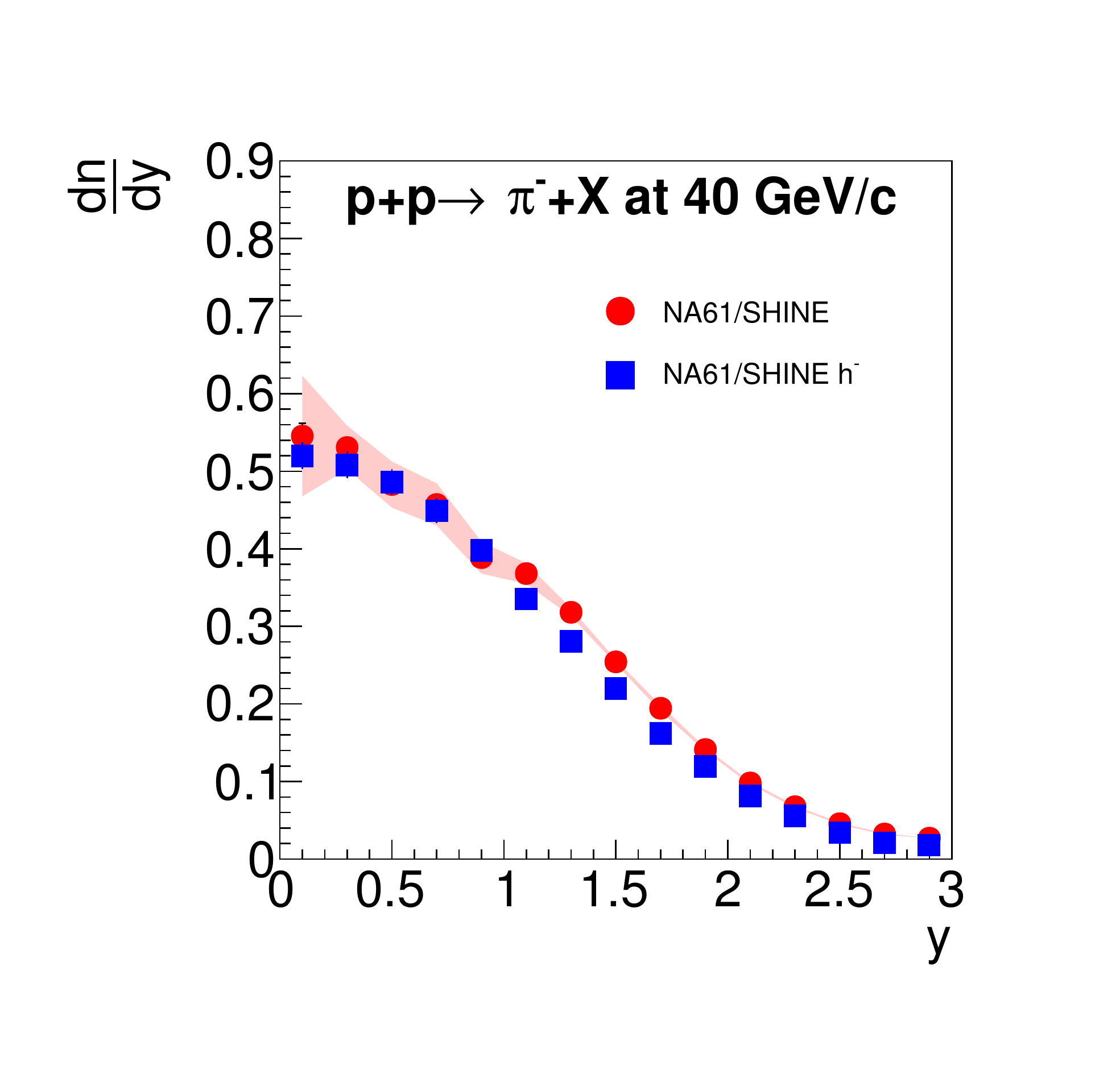}
	\includegraphics[width=0.3\textwidth]{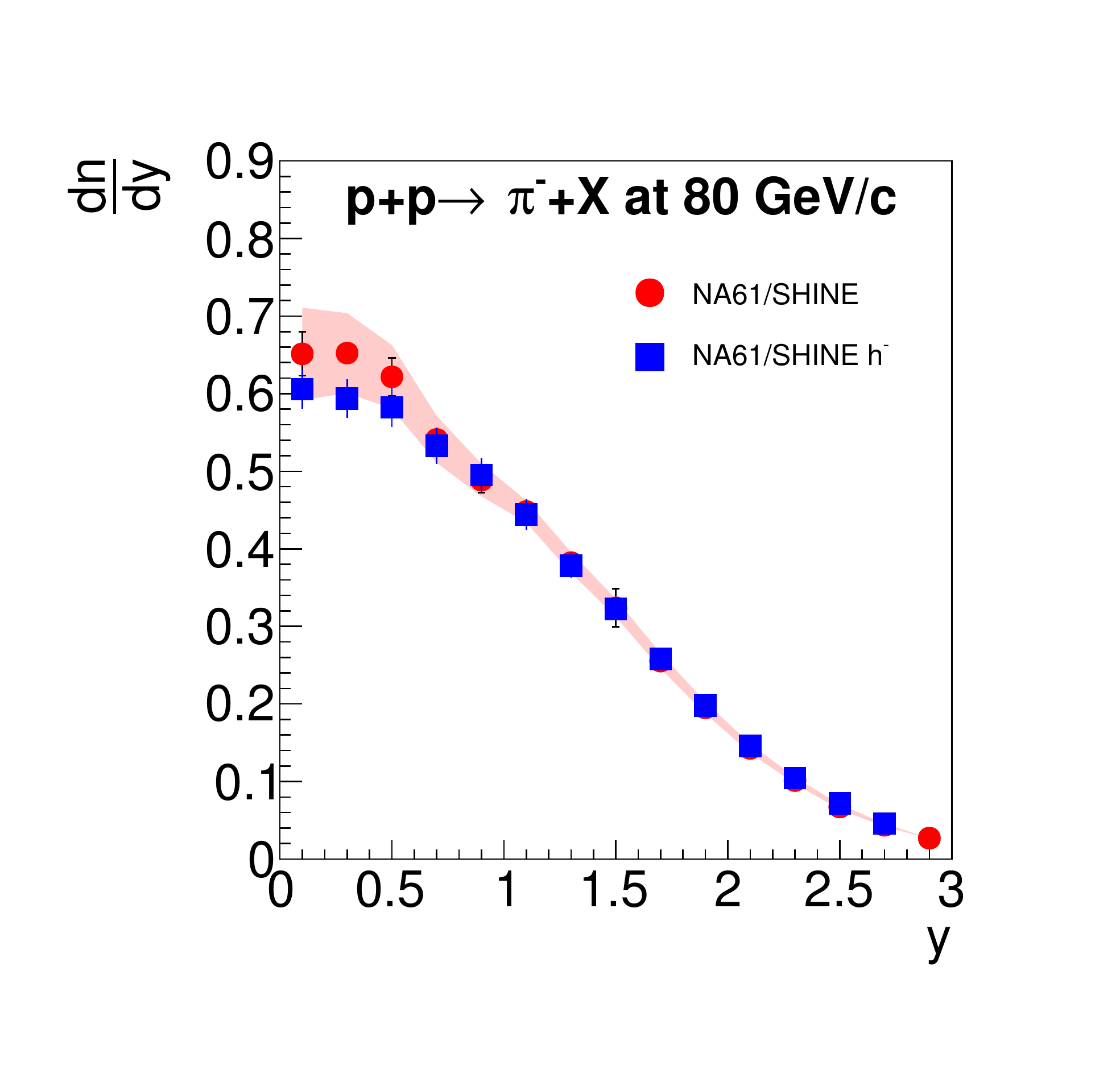}
	\includegraphics[width=0.3\textwidth]{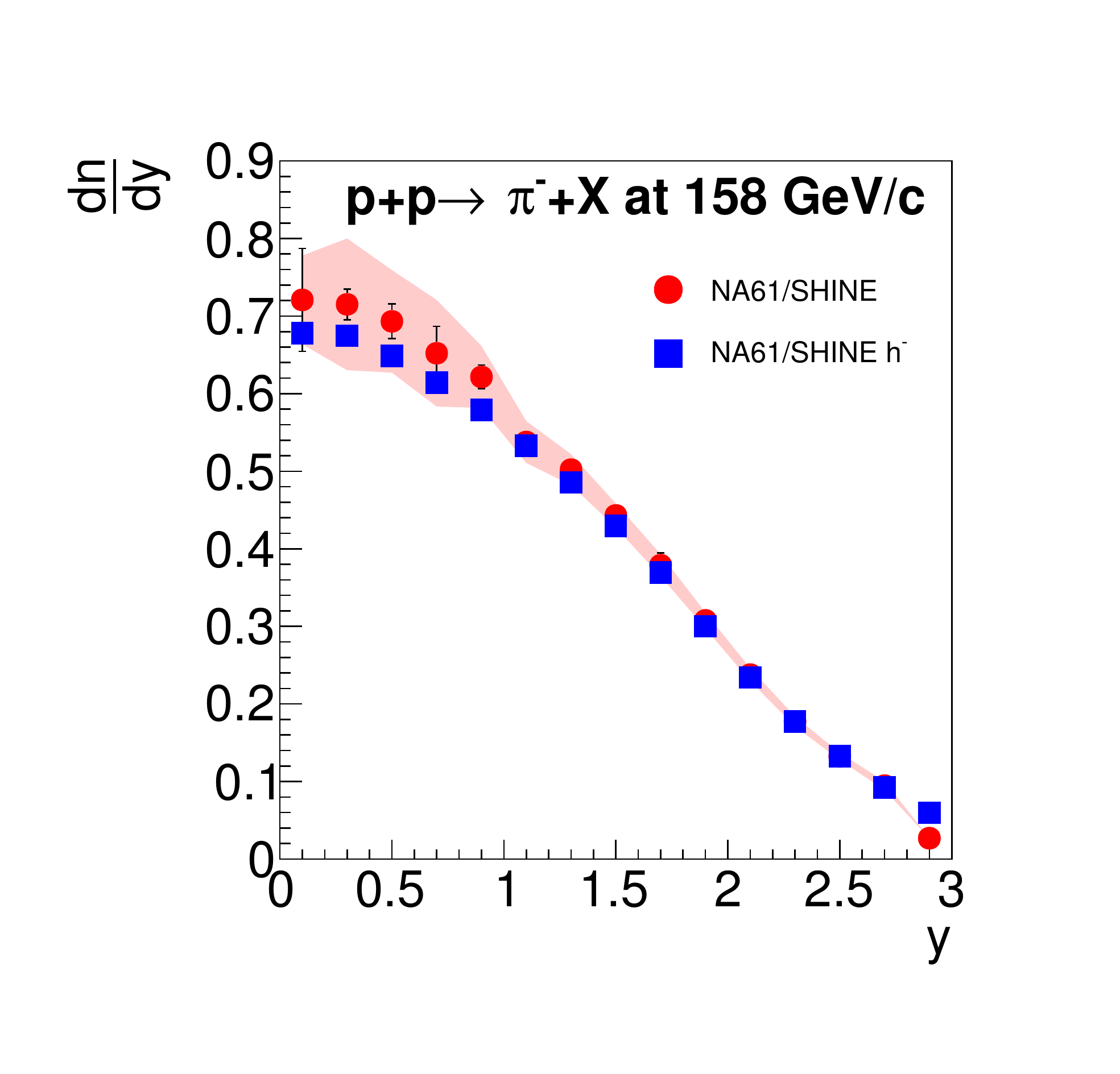}
	\end{center}
	\caption{(Color online) Comparison of $\pi^{-}$ rapidity distributions obtained by \dedx and $tof$-\dedx methods with results from the $h^{-}$ technique~\cite{Abgrall:2013pp_pim}. The $h^{-}$ results are presented with statistical uncertainty. Red shaded bands show the systematic uncertainty of particle multiplicities obtained by the $tof$-\dedx method.}
	\label{fig:withantoni}
\end{figure}

The \NASixtyOne experiment also published $\pi^{-}$ spectra obtained by the so-called $h^{-}$~\cite{Abgrall:2013pp_pim} analysis procedure. This technique is based on the fact that the majority of negatively charged particles are $\pi^{-}$ mesons. The contribution of the other particles was subtracted using predictions of the \Epos model. Comparison of rapidity spectra from the $h^{-}$ method with those obtained in this publication using particle identification is shown in Fig.~\ref{fig:withantoni}. Both methods give results for $\pi^{-}$ spectra which agree within uncertainties .

\begin{figure}[!ht]
	\begin{center}
	\includegraphics[width=0.3\textwidth]{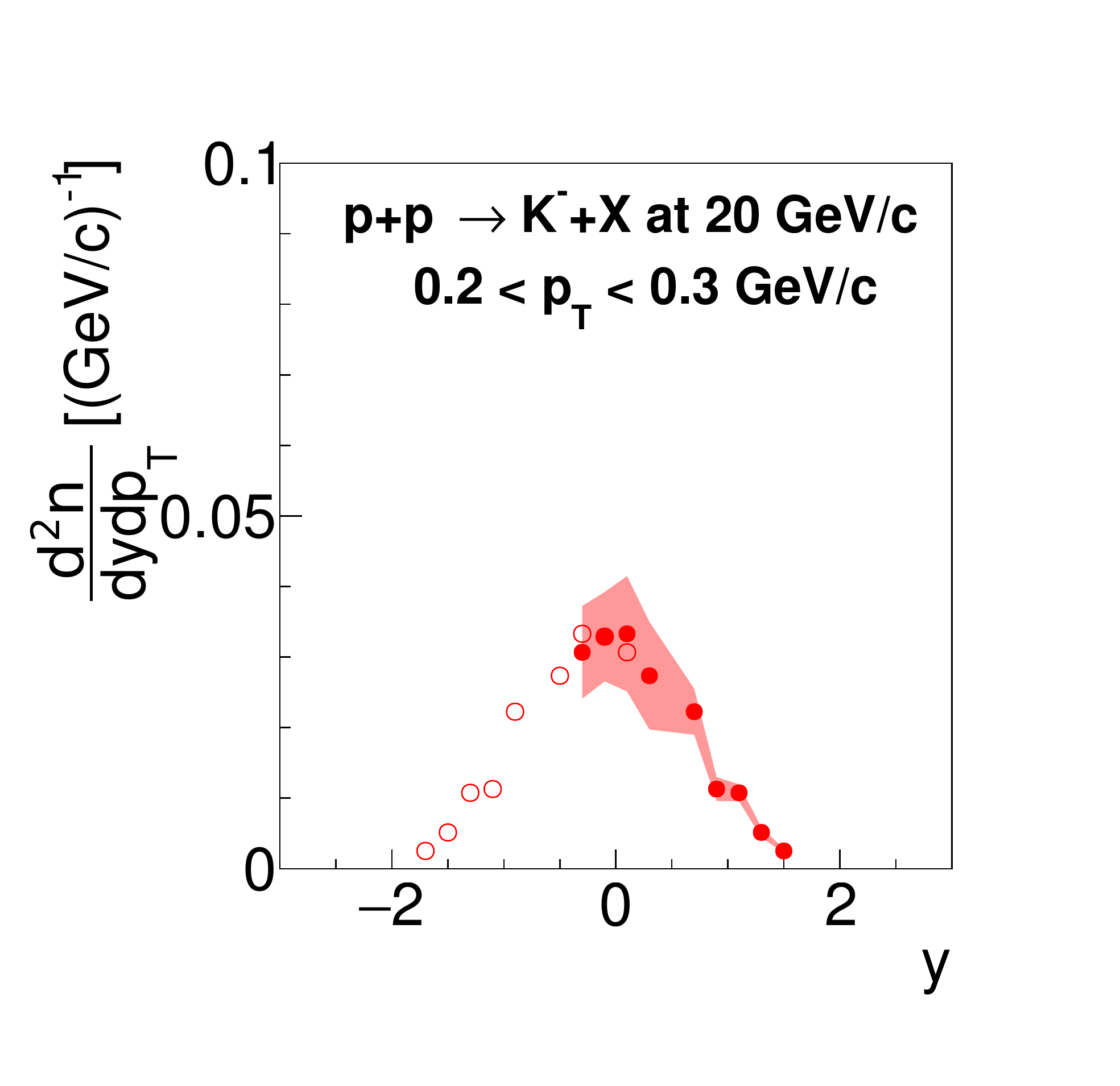}
	\includegraphics[width=0.3\textwidth]{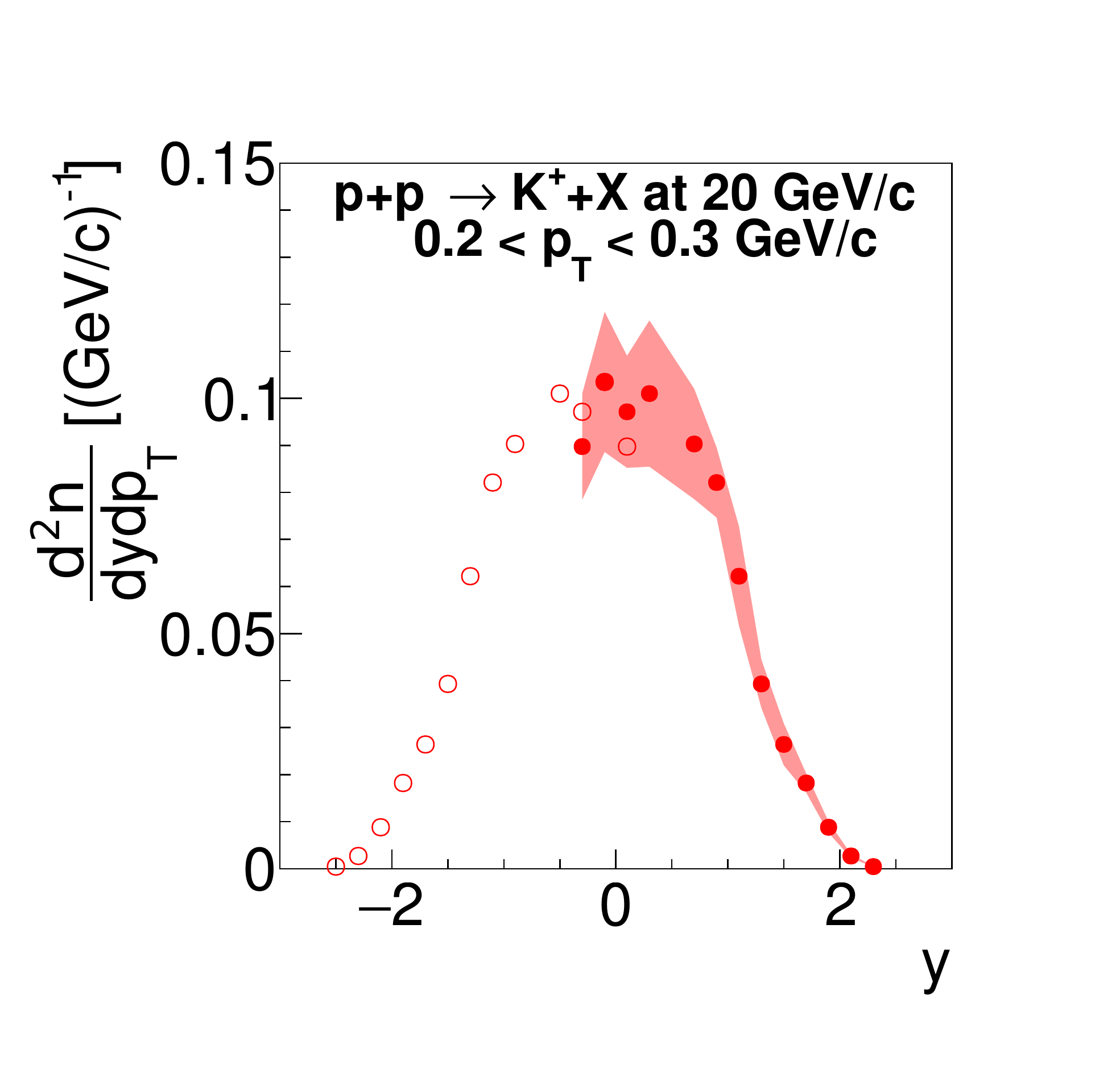}
	\includegraphics[width=0.3\textwidth]{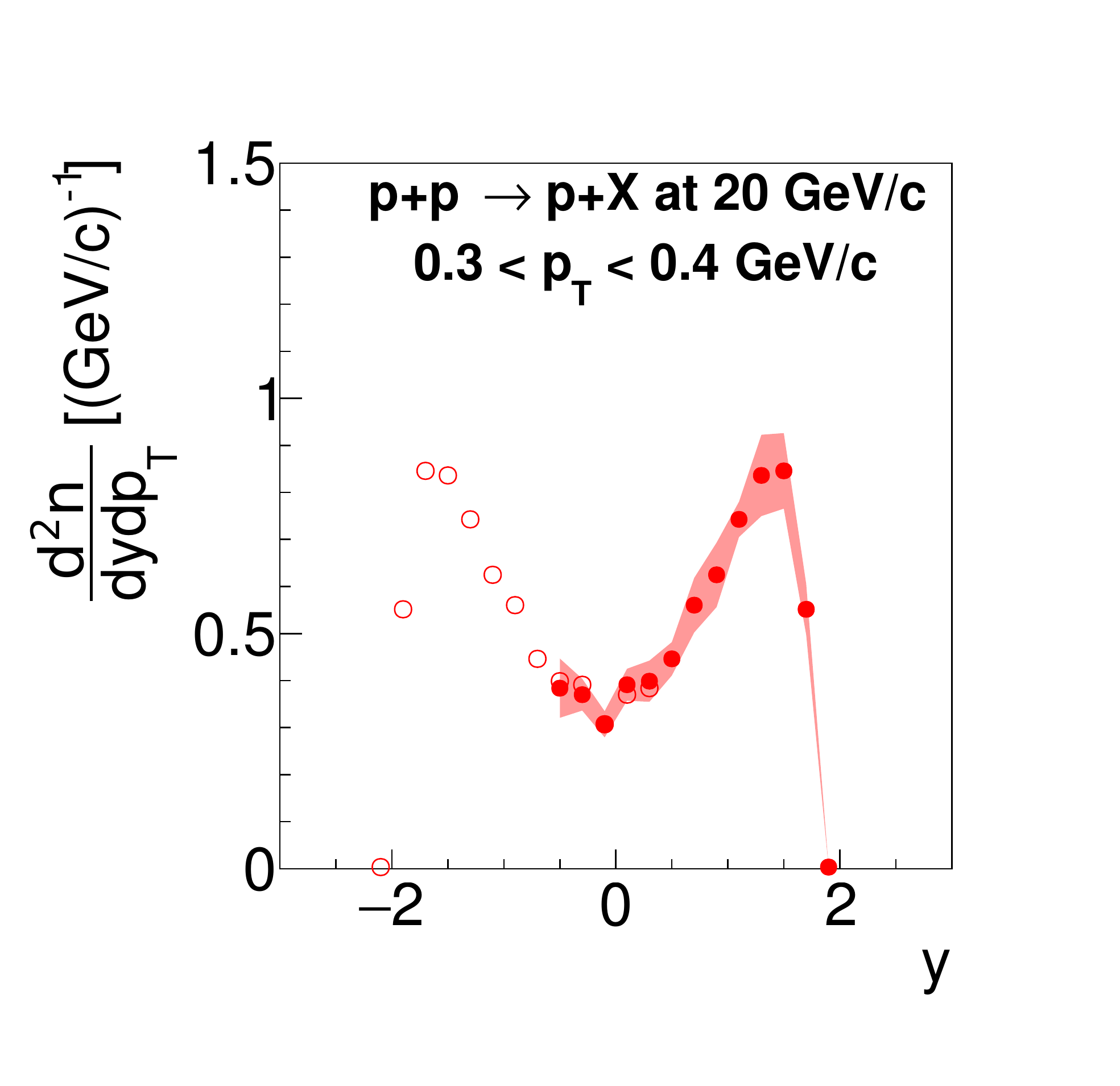}\\
	\includegraphics[width=0.3\textwidth]{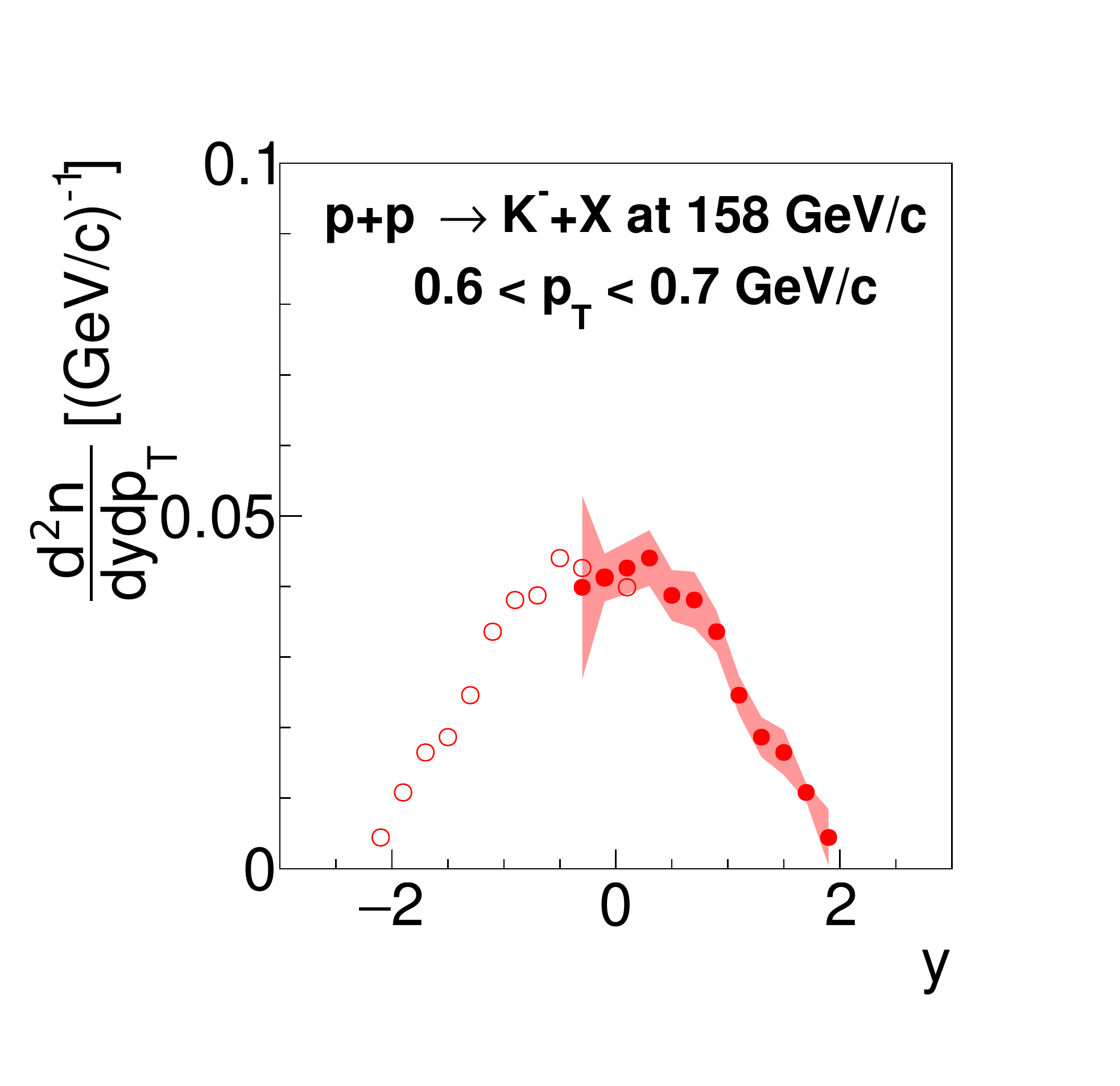}
	\includegraphics[width=0.3\textwidth]{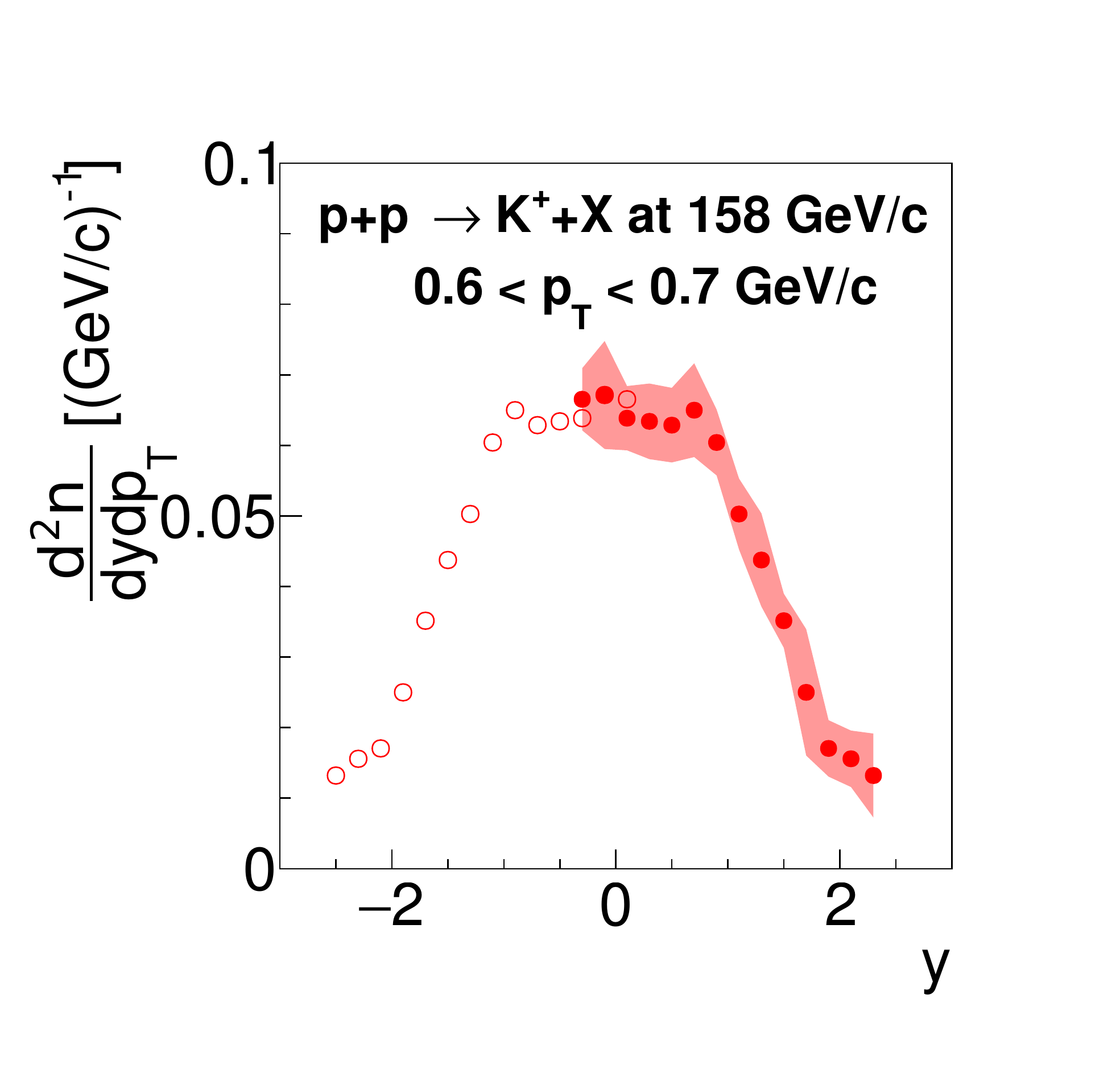}
	\includegraphics[width=0.3\textwidth]{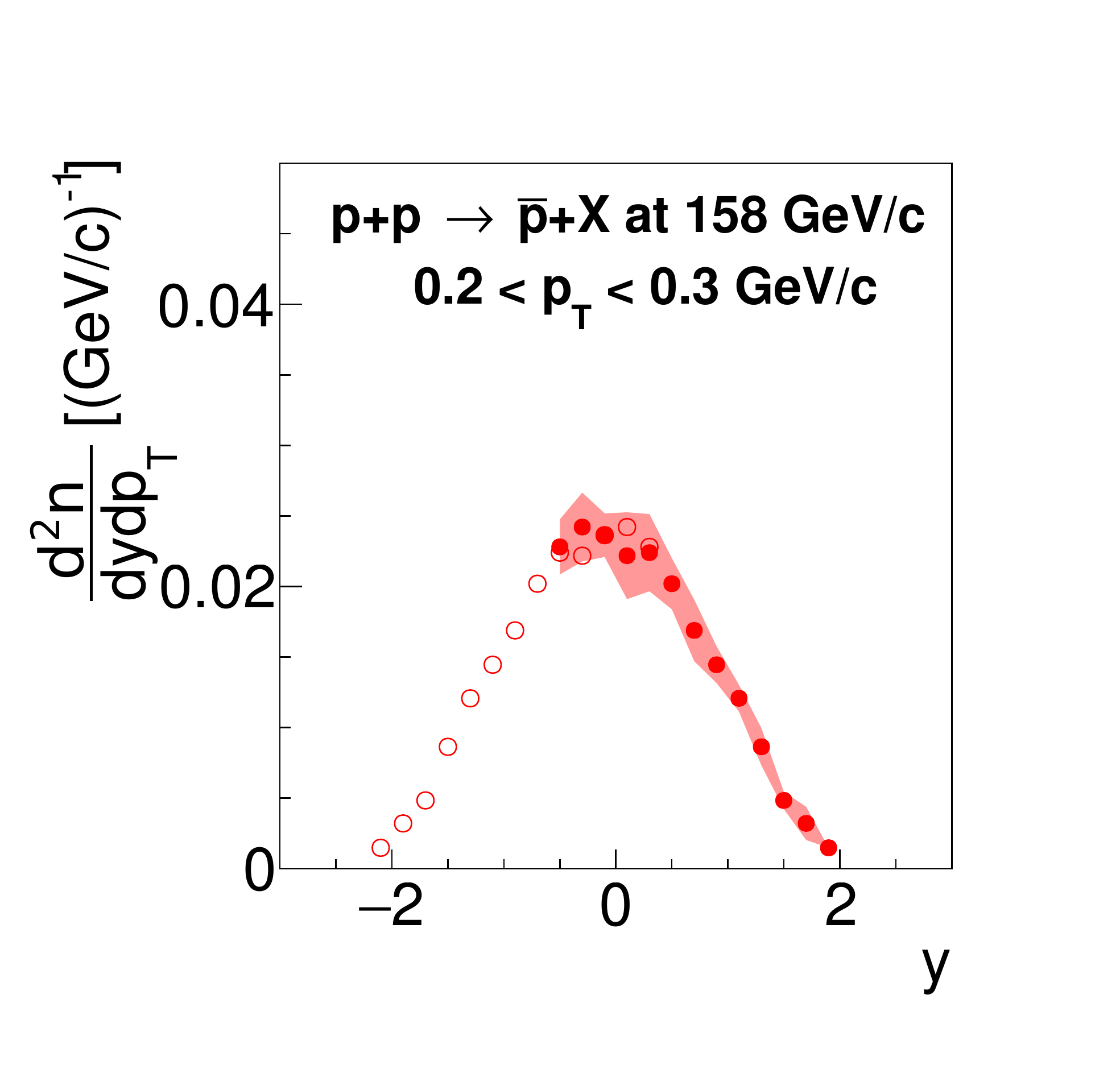}
	\end{center}
	\caption{(Color online) Rapidity spectra of K$^{-}$, K$^{+}$ and p measured (filled dots) and reflected with respect to mid-rapidity (open dots) obtained in inelastic p+p interactions in the interval $\pt = 0.2 - 0.3$~\GeVc at 20~\GeVc (upper row) and $\pt = 0.6 - 0.7$~\GeVc at 158~\GeVc (bottom row).}
	\label{fig:reflection}
\end{figure}

Spectra measured in p+p interactions should obey reflection symmetry with respect to mid-rapidity. As the
NA61/SHINE acceptance extends somewhat below mid-rapidity a test of the reflection symmetry can be performed
to check the consistency of the measurements. As examples these reflection tests are presented in Fig.~\ref{fig:reflection} in selected $p_{T}$ intervals for positively and negatively charged kaons, antiprotons and protons produced at 20 and 158~\GeVc. One observes that the yields measured for $y < 0$ agree with those measured for $y > 0$ in the reflected acceptance within uncertainties.

The \Epos model was chosen as the event generator in the simulation chain to calculate corrections. Figures \ref{fig:epos_comps20} and \ref{fig:epos_comps158} present \Epos model predictions and experimental results obtained in this analysis in selected rapidity intervals for inelastic p+p interactions at 20 and 158~\GeVc. The \Epos model describes well particle production in p+p interactions at 20~\GeVc and at 158~\GeVc. Since the corrections were applied differentially in rapidity and transverse momentum bins the effect of minor discrepancies are not expected to cause significant biases.

\FloatBarrier
\section{Results}
\label{sec:results}

\subsection{Spectra}
\label{sec:spectra}

\begin{figure*}
\begin{center}
\newcolumntype{S}{>{\centering} m{0.03\textwidth} }
\newcolumntype{A}{>{\centering} m{0.19\textwidth} }
\hspace{-10mm}
\begin{tabular}{S A A A A A A}
& \tiny20~\GeVc & \hspace{-15mm}\tiny31~\GeVc & \hspace{-30mm}\tiny40~\GeVc & \hspace{-45mm}\tiny80~\GeVc  & \hspace{-60mm}\tiny158~\GeVc 
\tabularnewline
\vspace{-5mm}$\pi^{-}$ &
\includegraphics[width=0.18\textwidth]{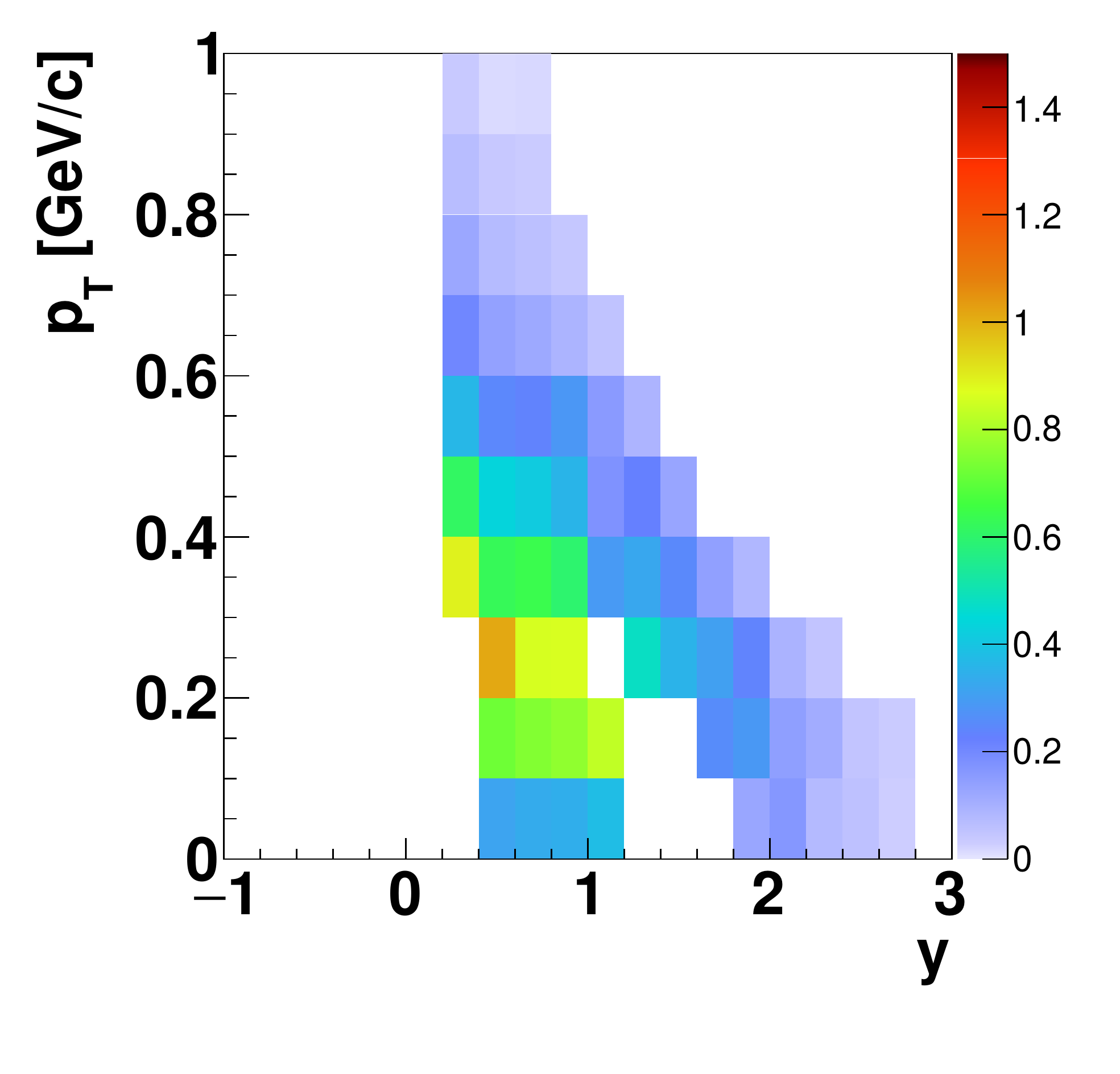} &
\hspace{-15mm}
\includegraphics[width=0.18\textwidth]{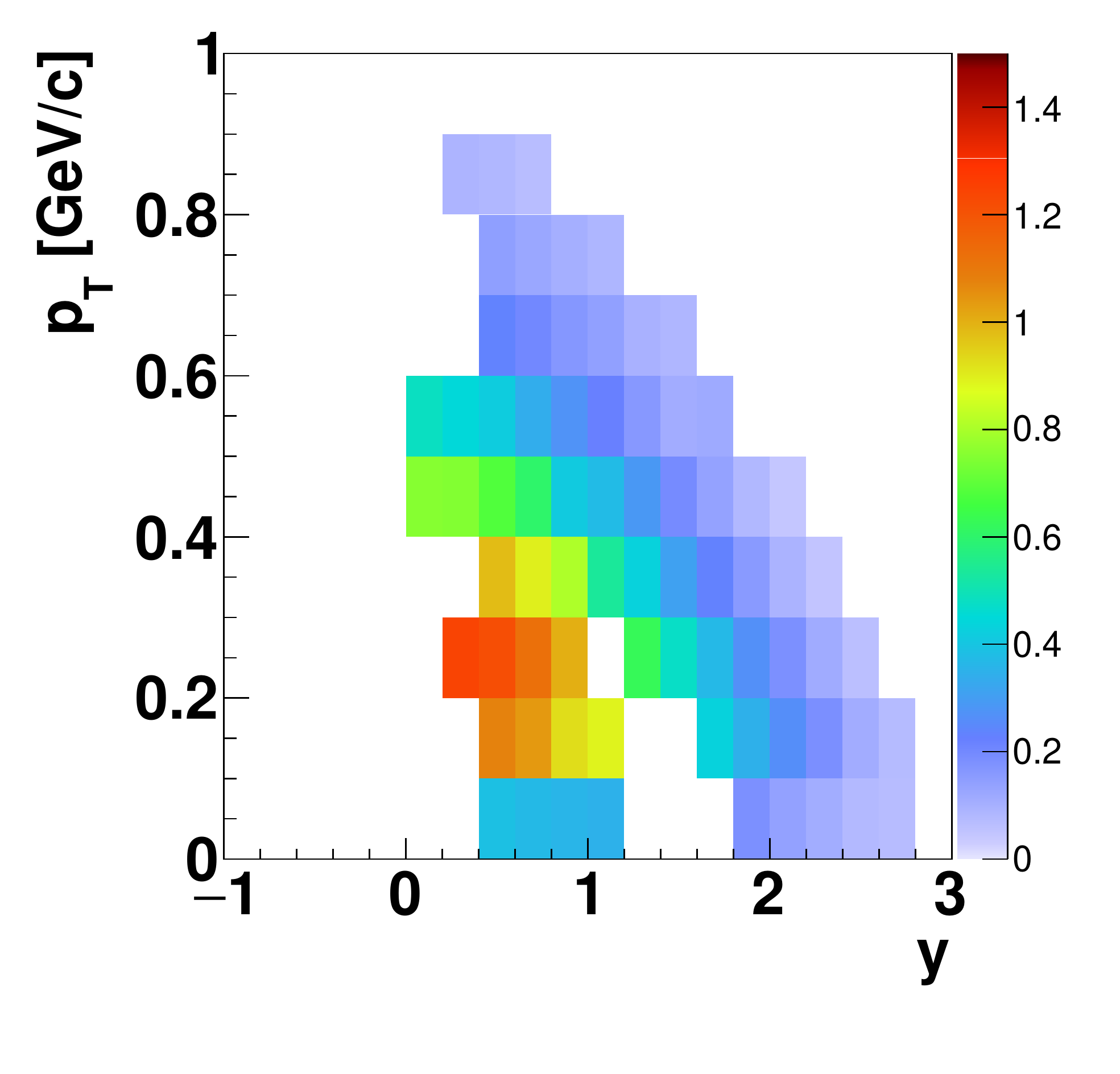} &
\hspace{-30mm}
\includegraphics[width=0.18\textwidth]{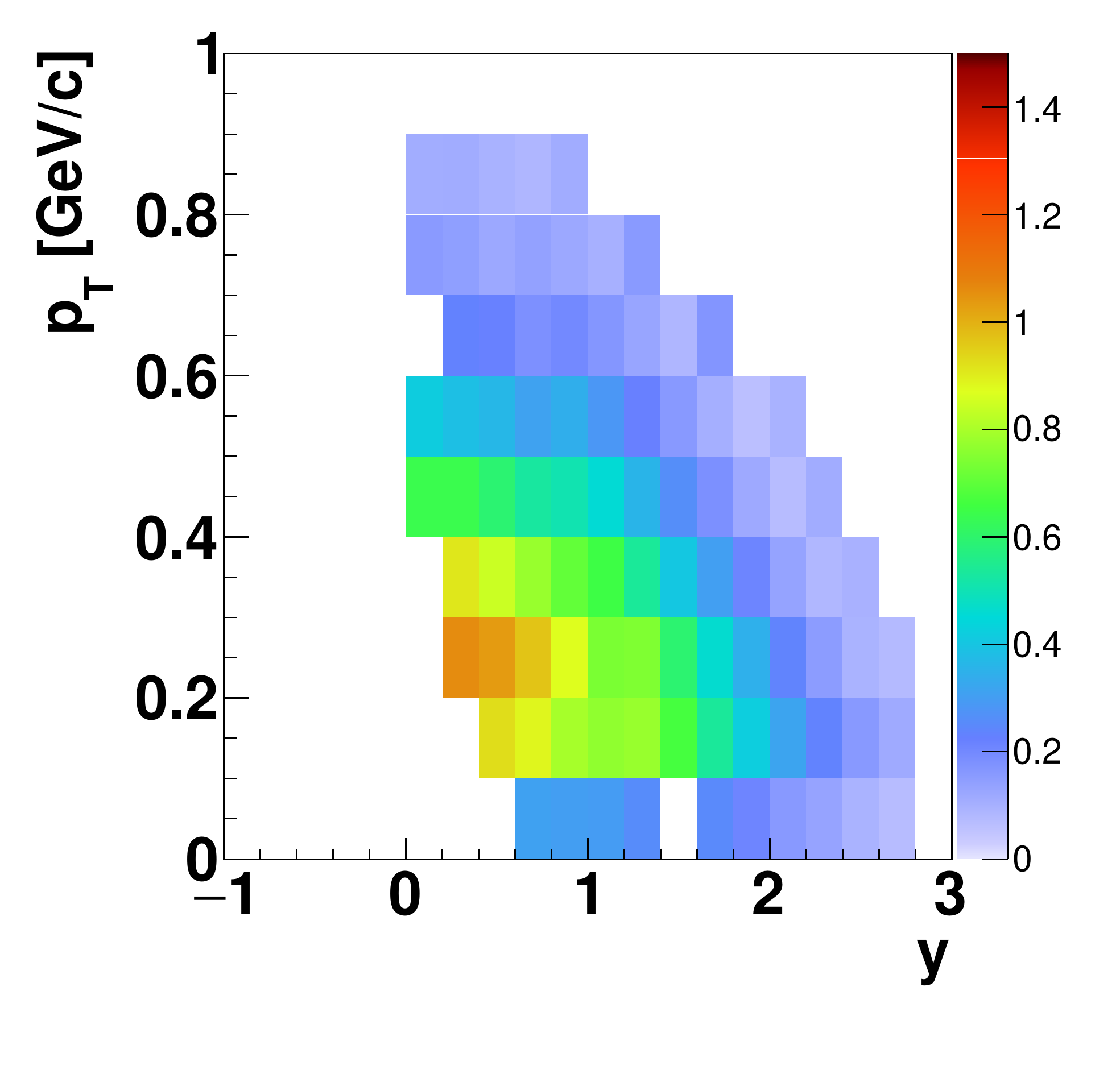} &
\hspace{-45mm}
\includegraphics[width=0.18\textwidth]{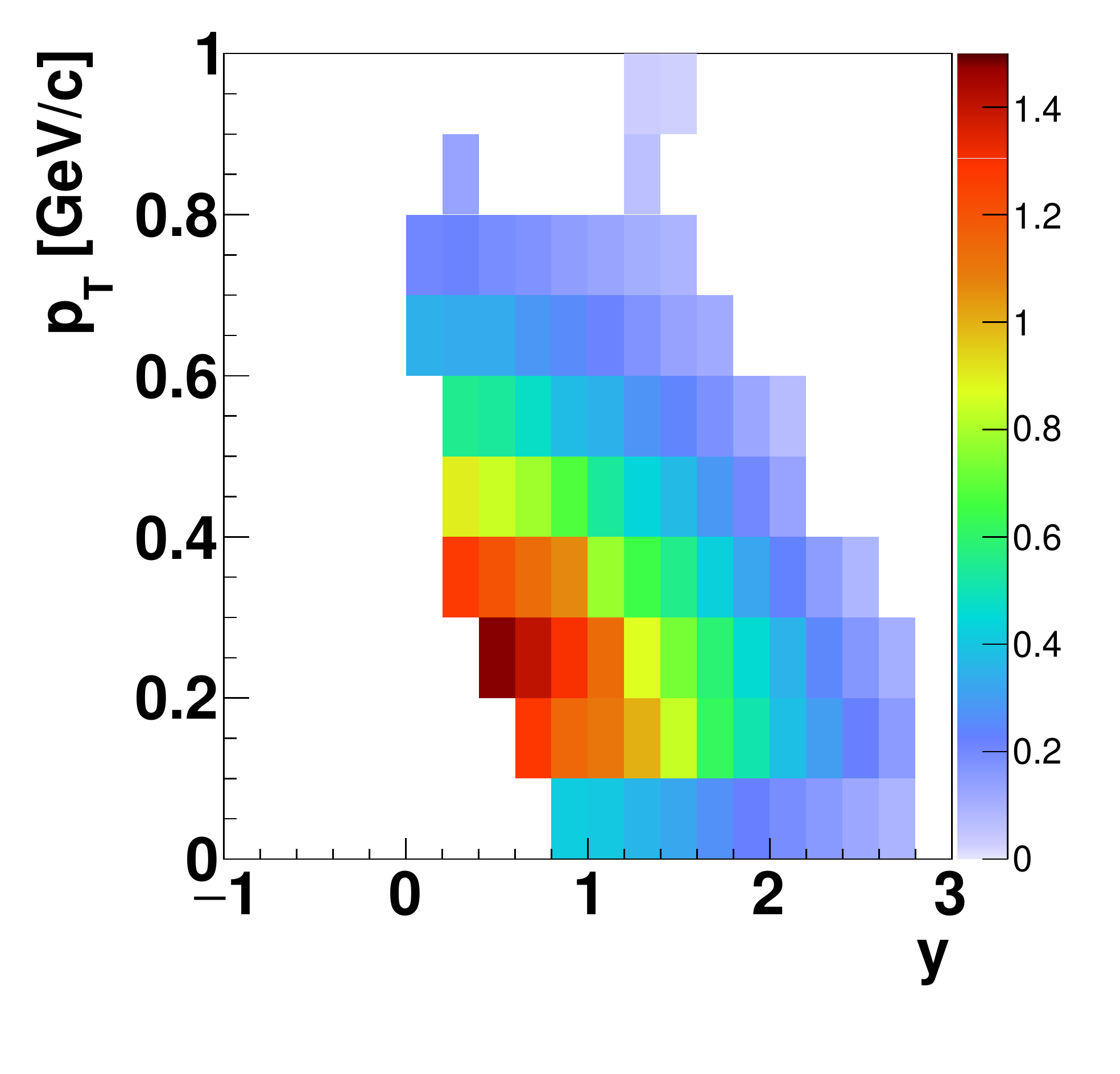} &
\hspace{-60mm}
\includegraphics[width=0.18\textwidth]{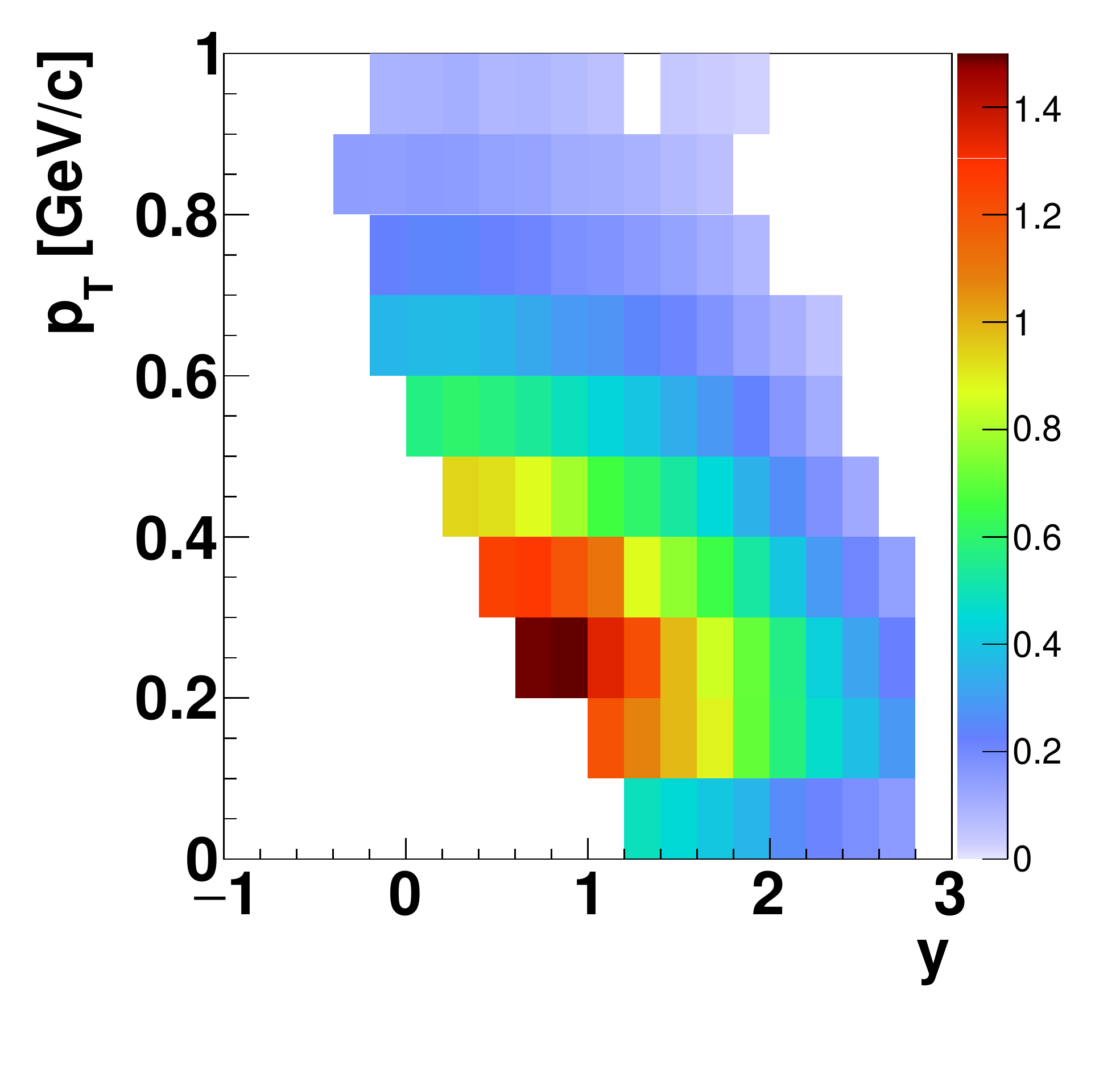} 
\tabularnewline
\vspace{-5mm}$\pi^{+}$ &
\includegraphics[width=0.18\textwidth]{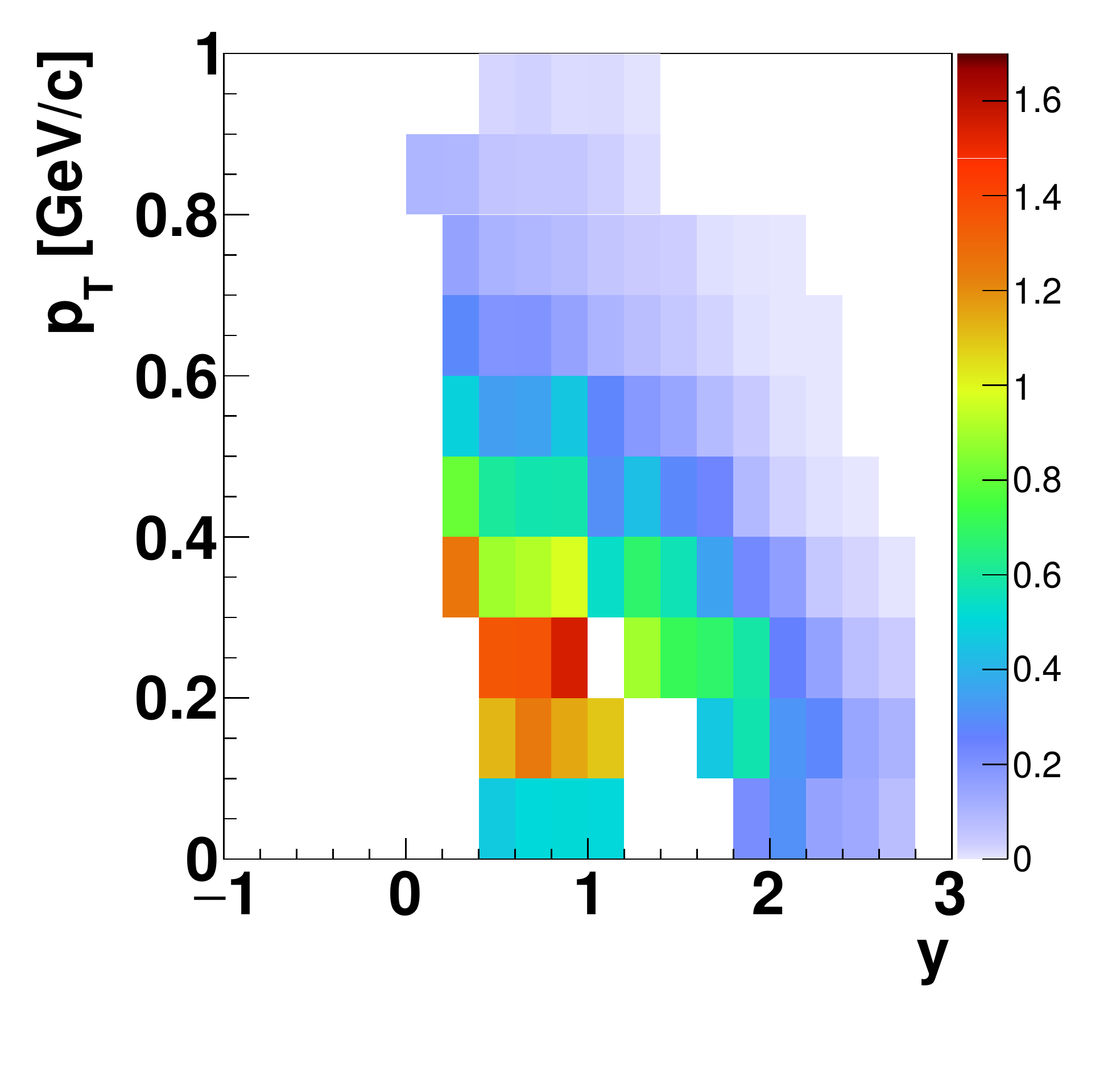} &
\hspace{-15mm}
\includegraphics[width=0.18\textwidth]{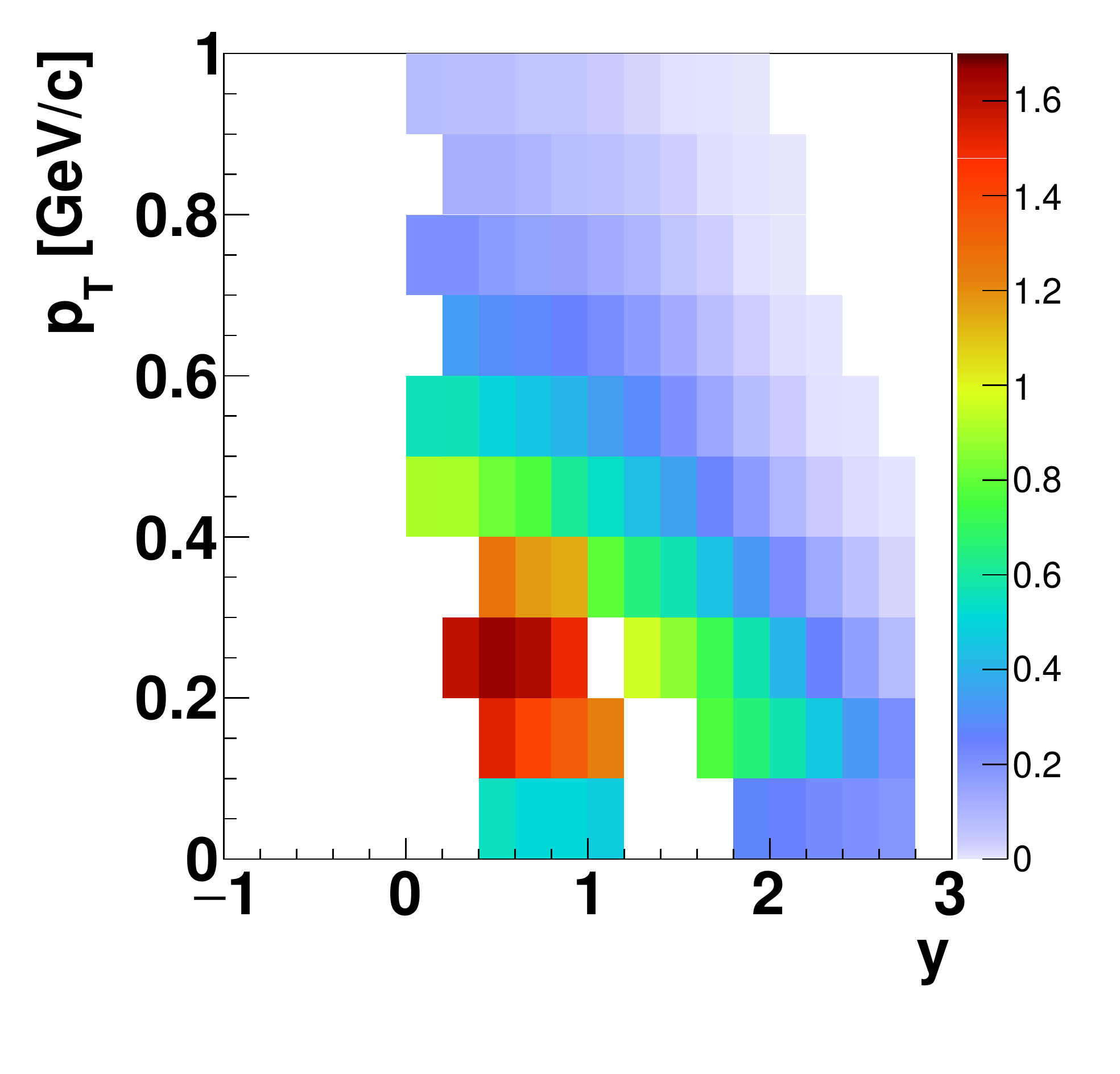} &
\hspace{-30mm}
\includegraphics[width=0.18\textwidth]{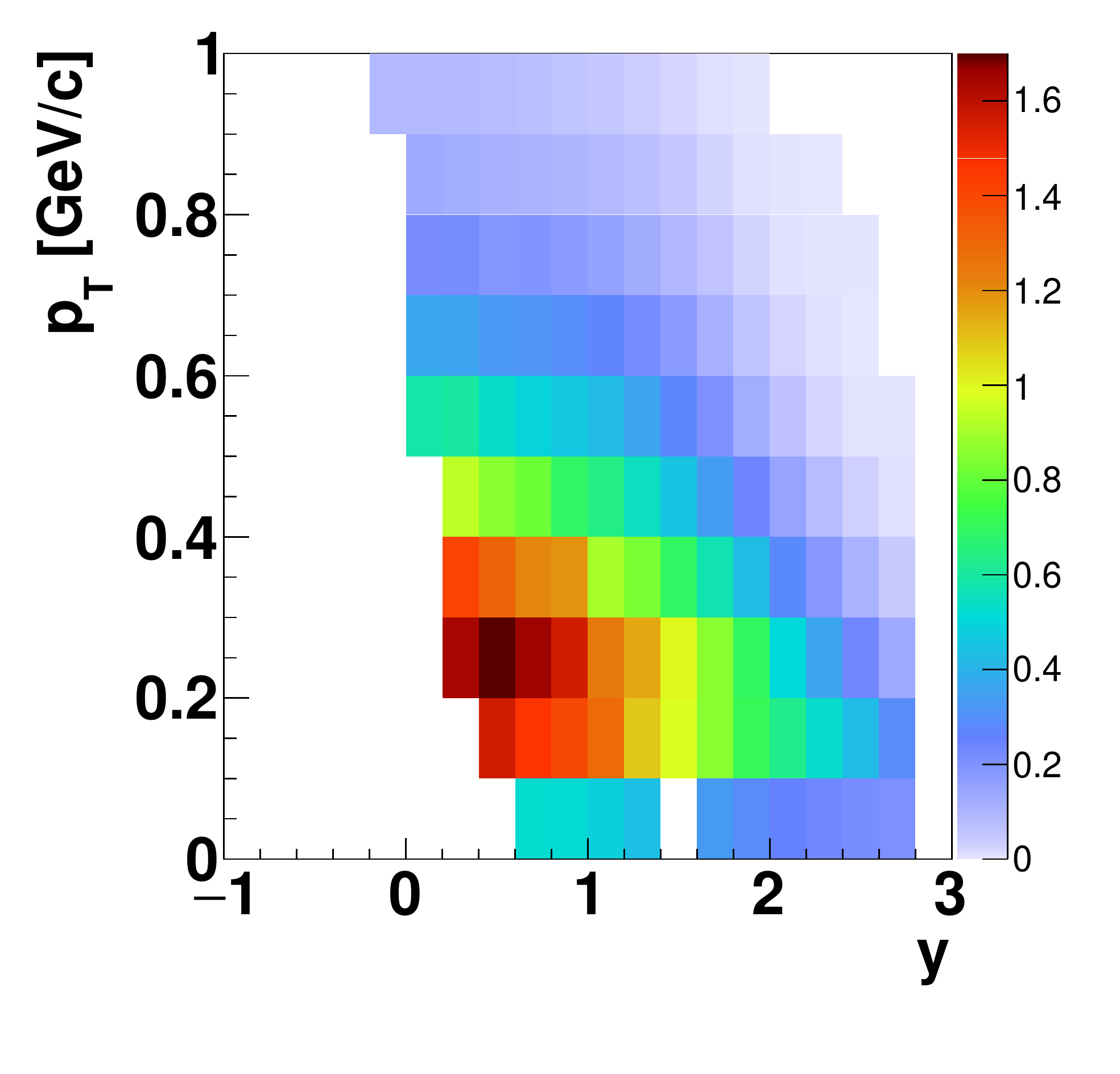} &
\hspace{-45mm}
\includegraphics[width=0.18\textwidth]{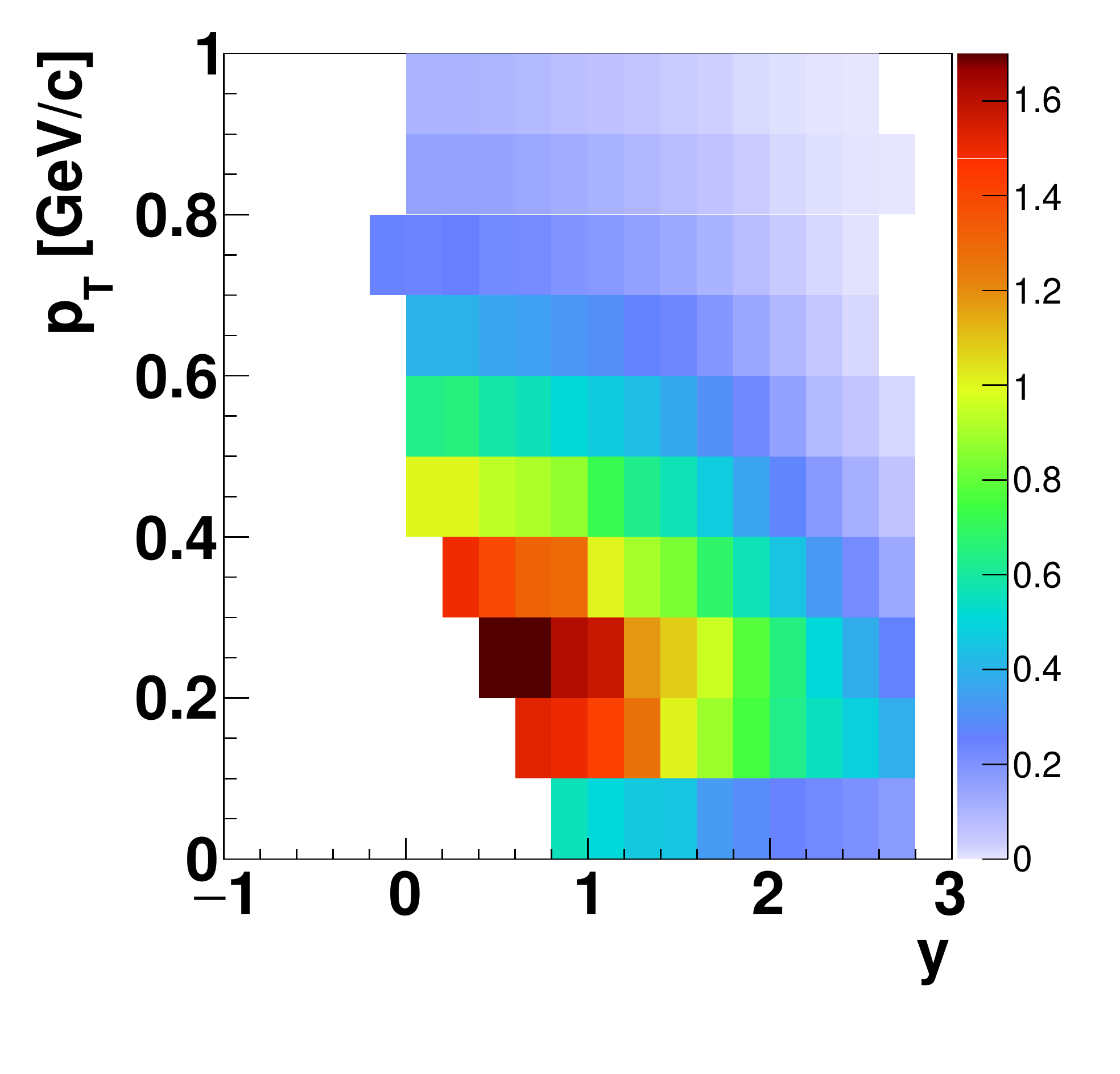} &
\hspace{-60mm}
\includegraphics[width=0.18\textwidth]{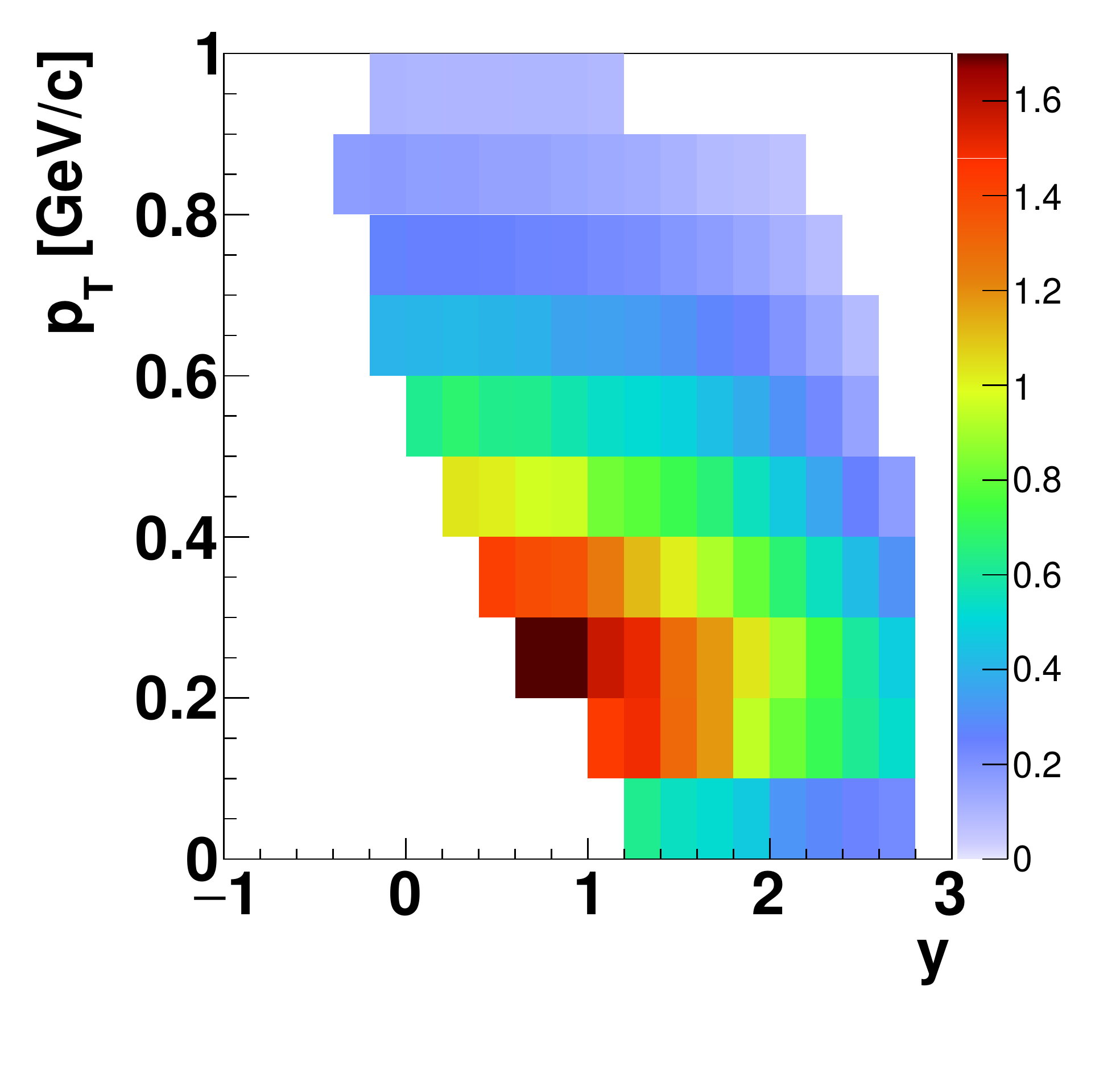} 
\tabularnewline
\vspace{-5mm}K$^{-}$ &
\includegraphics[width=0.18\textwidth]{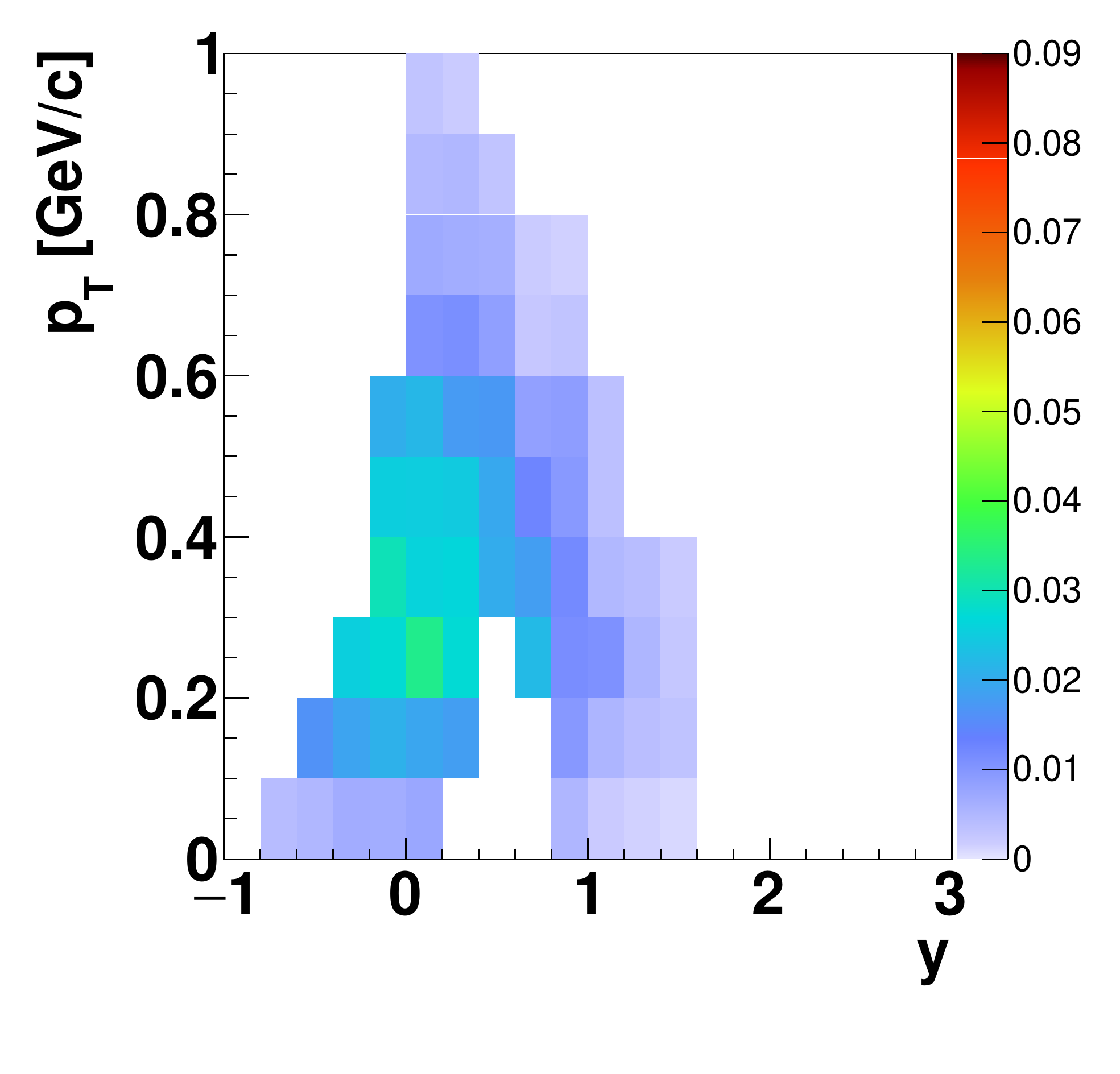} &
\hspace{-15mm}
\includegraphics[width=0.18\textwidth]{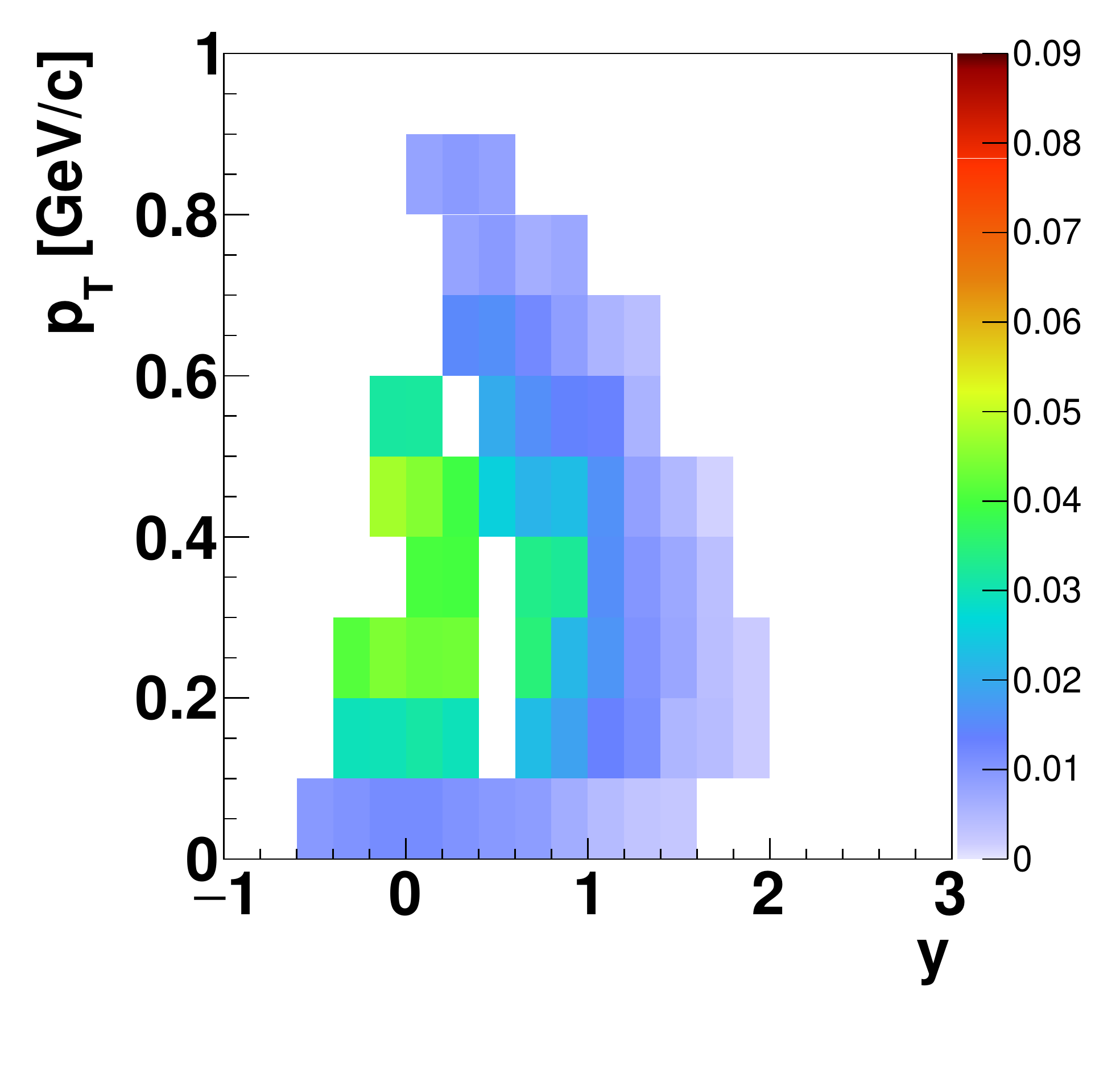} &
\hspace{-30mm}
\includegraphics[width=0.18\textwidth]{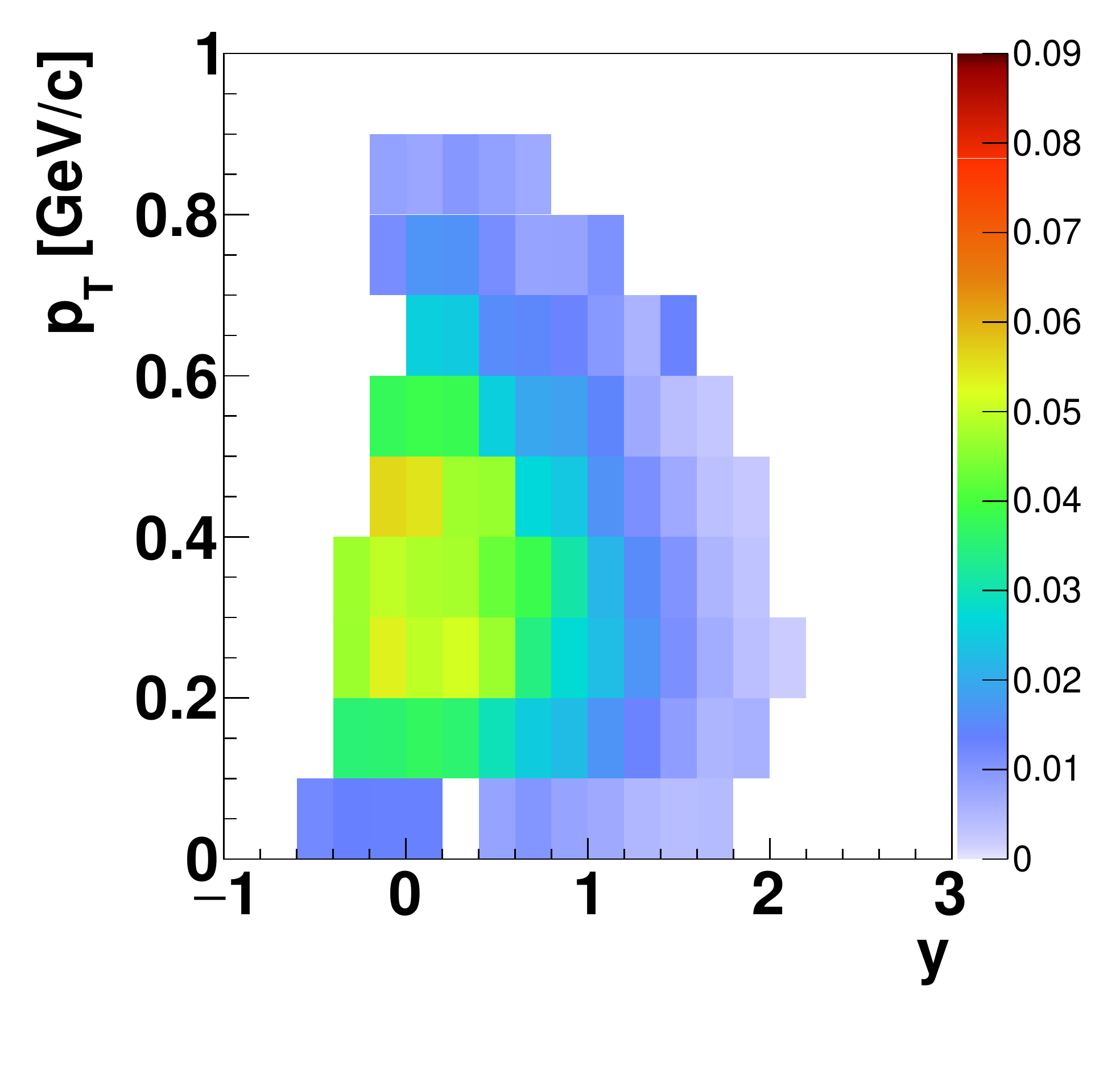} &
\hspace{-45mm}
\includegraphics[width=0.18\textwidth]{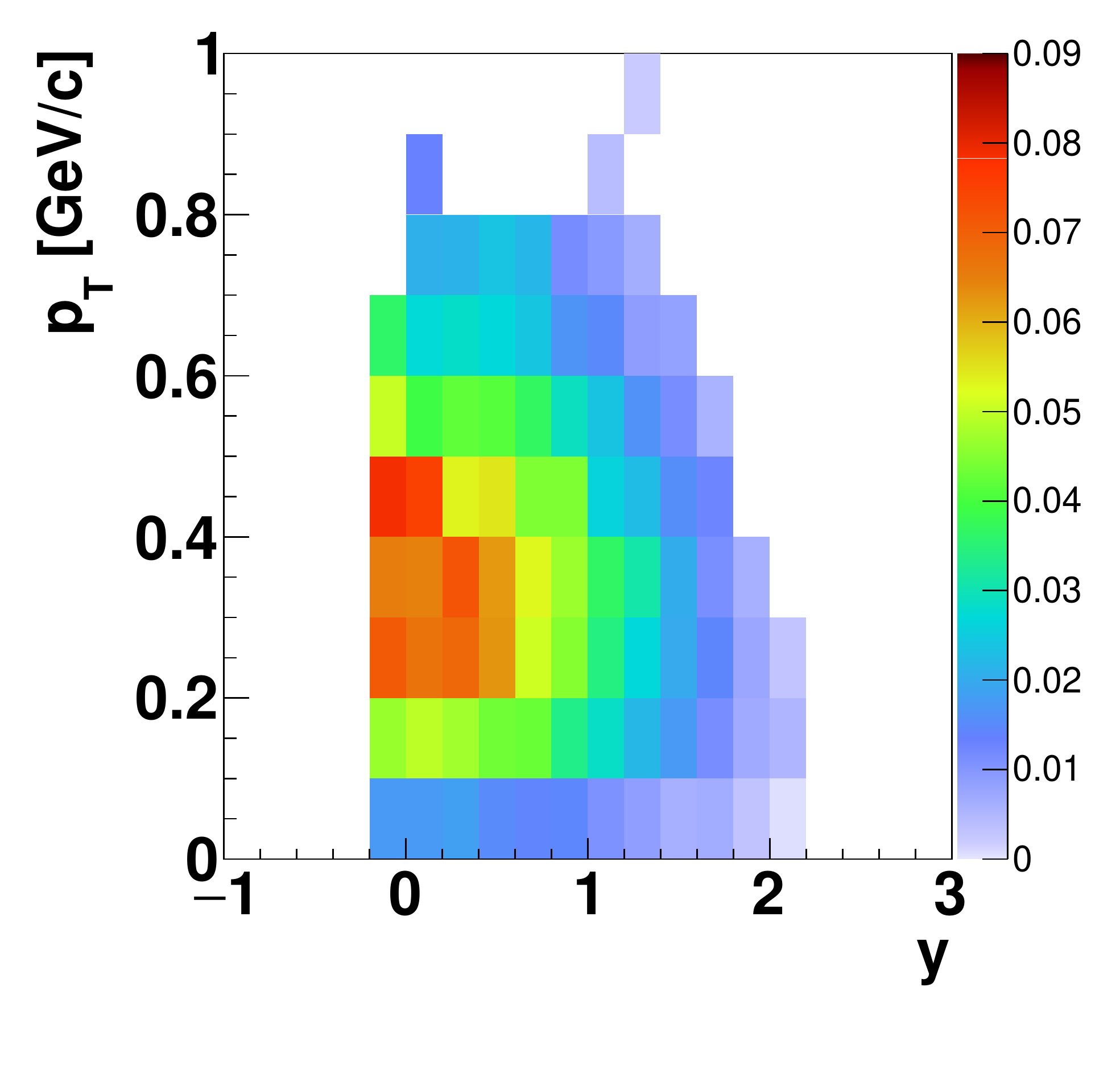} &
\hspace{-60mm}
\includegraphics[width=0.18\textwidth]{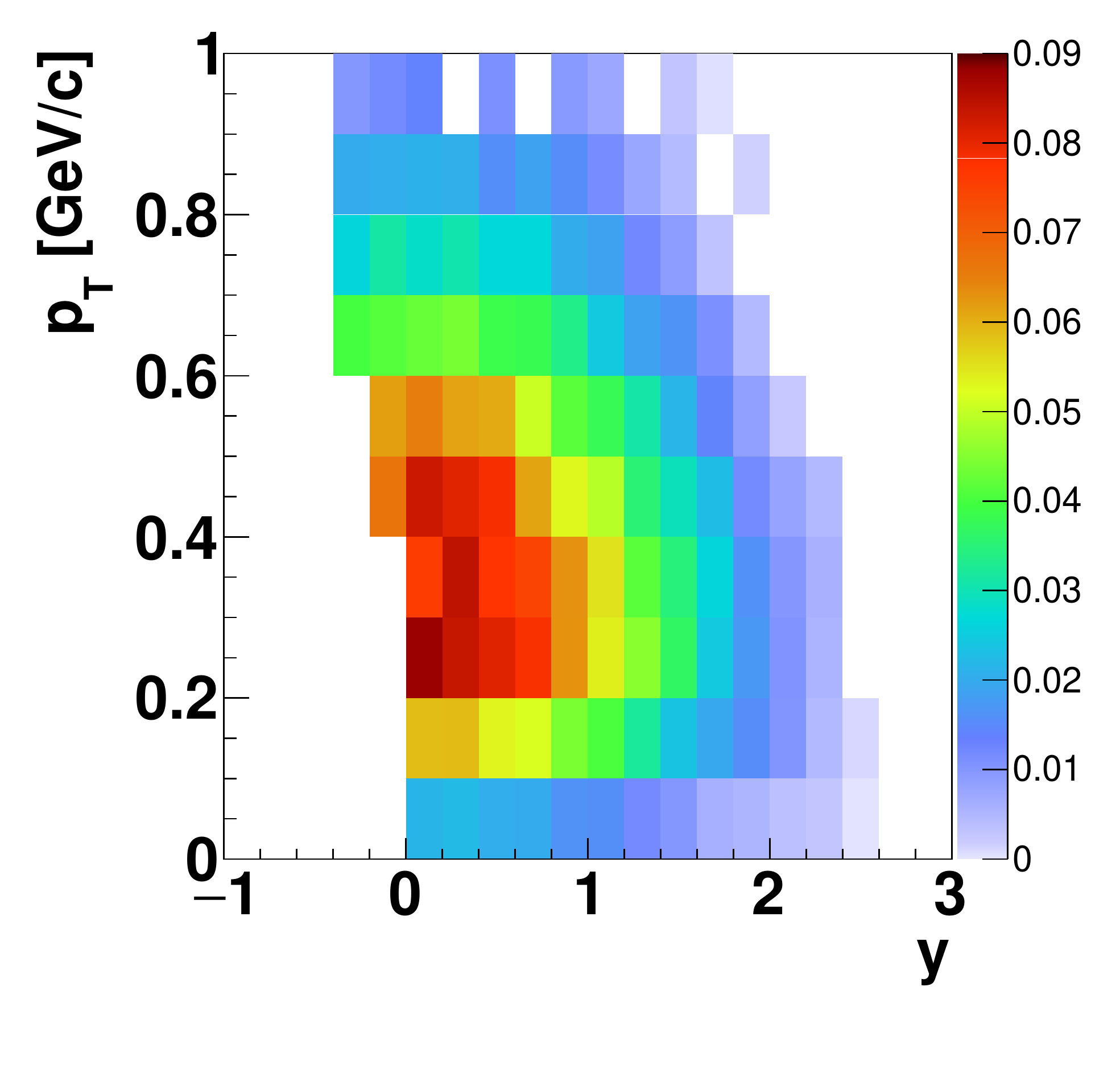} 
\tabularnewline
\vspace{-5mm}K$^{+}$ &
\includegraphics[width=0.18\textwidth]{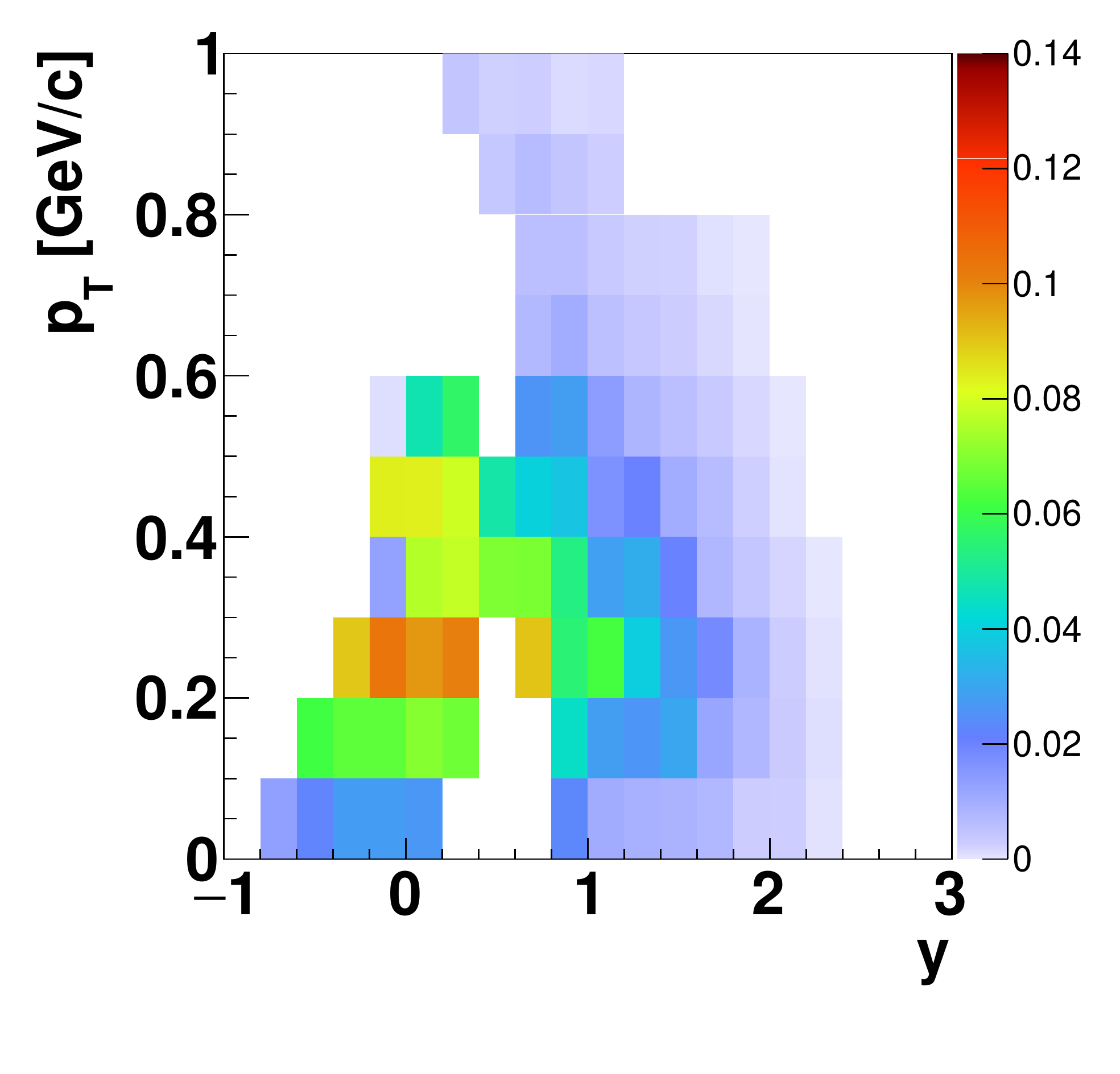} &
\hspace{-15mm}
\includegraphics[width=0.18\textwidth]{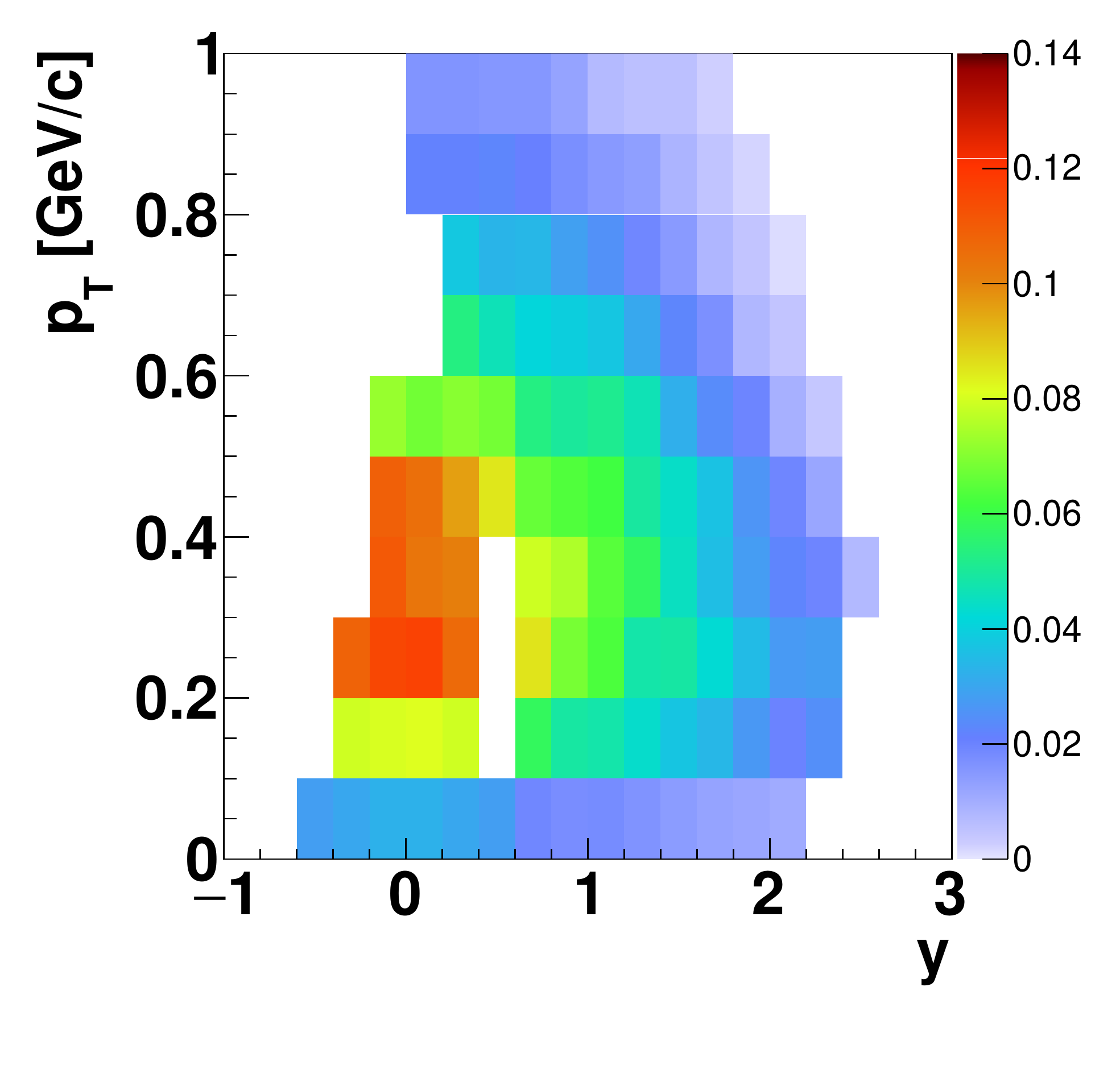} &
\hspace{-30mm}
\includegraphics[width=0.18\textwidth]{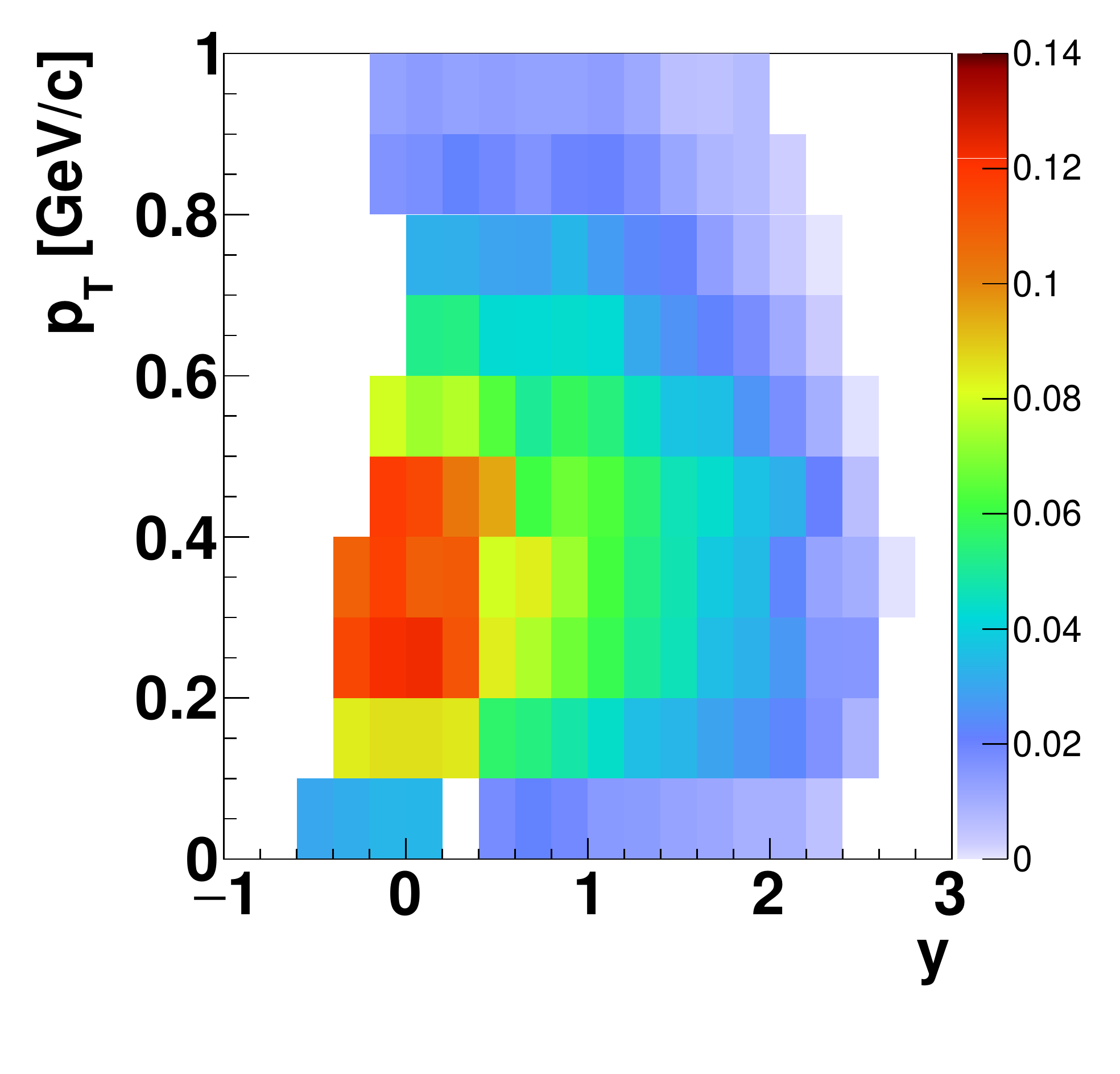} &
\hspace{-45mm}
\includegraphics[width=0.18\textwidth]{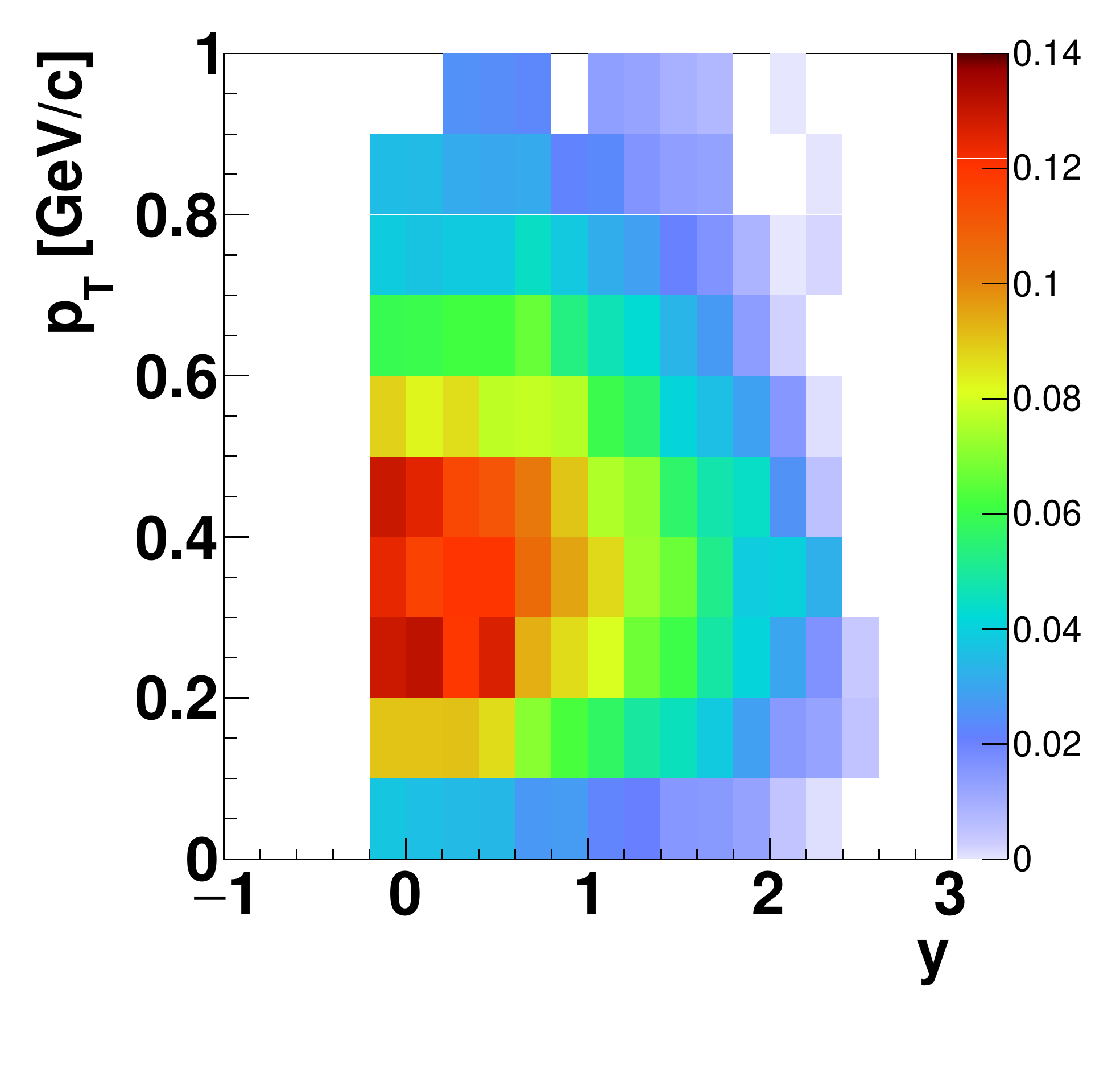} &
\hspace{-60mm}
\includegraphics[width=0.18\textwidth]{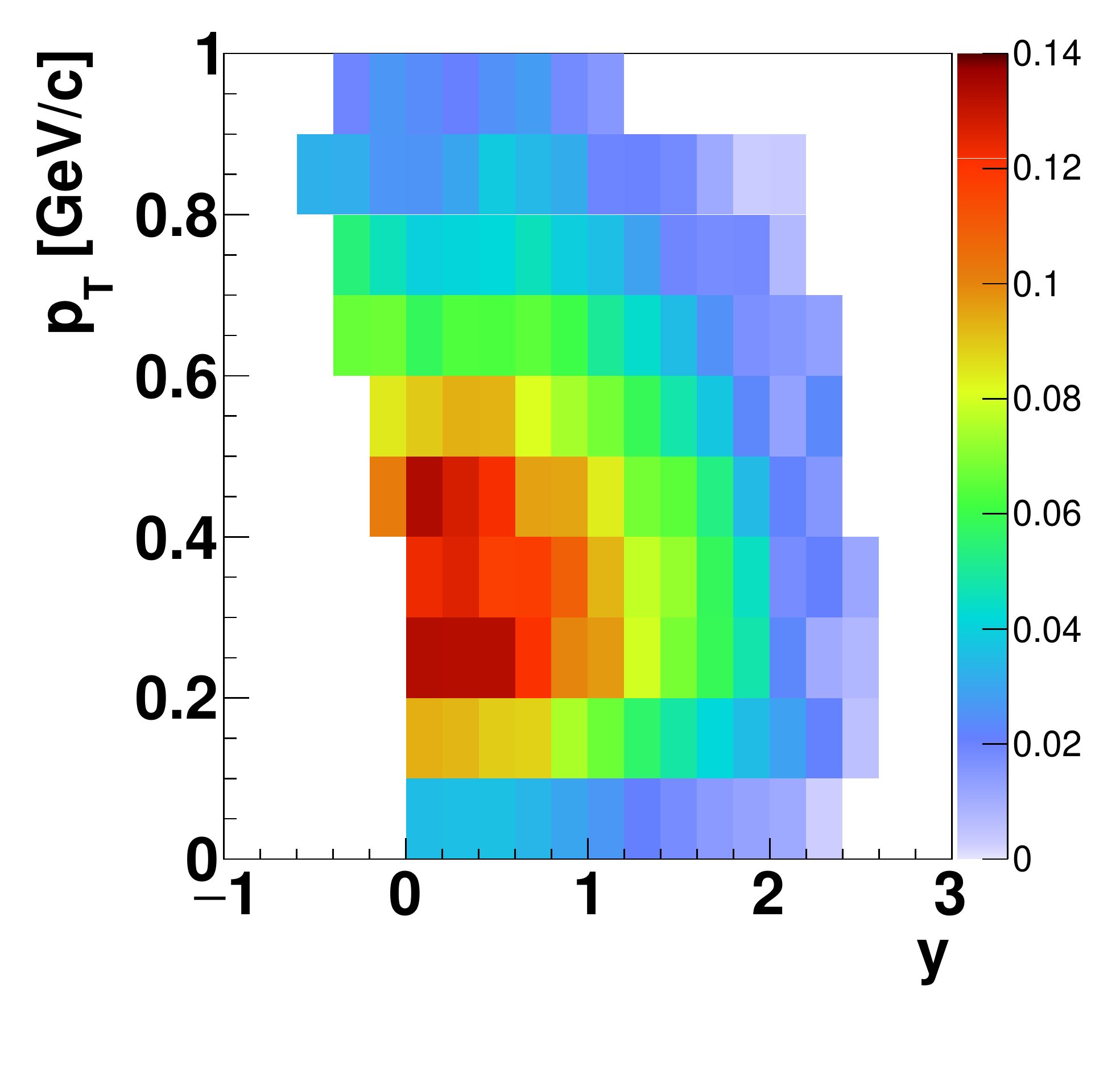} 
\tabularnewline
\vspace{-5mm}p &
\includegraphics[width=0.18\textwidth]{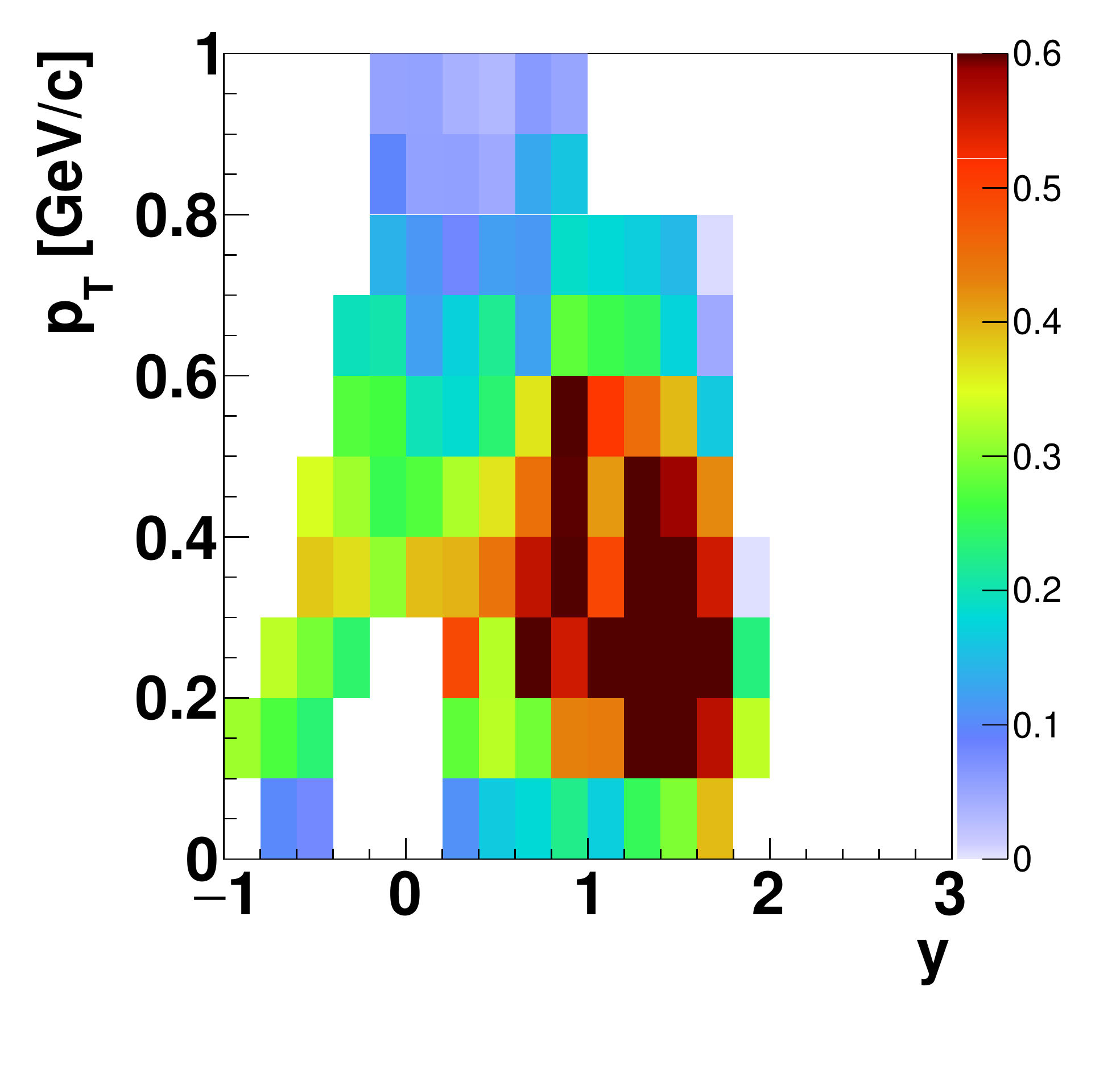} &
\hspace{-15mm}
\includegraphics[width=0.18\textwidth]{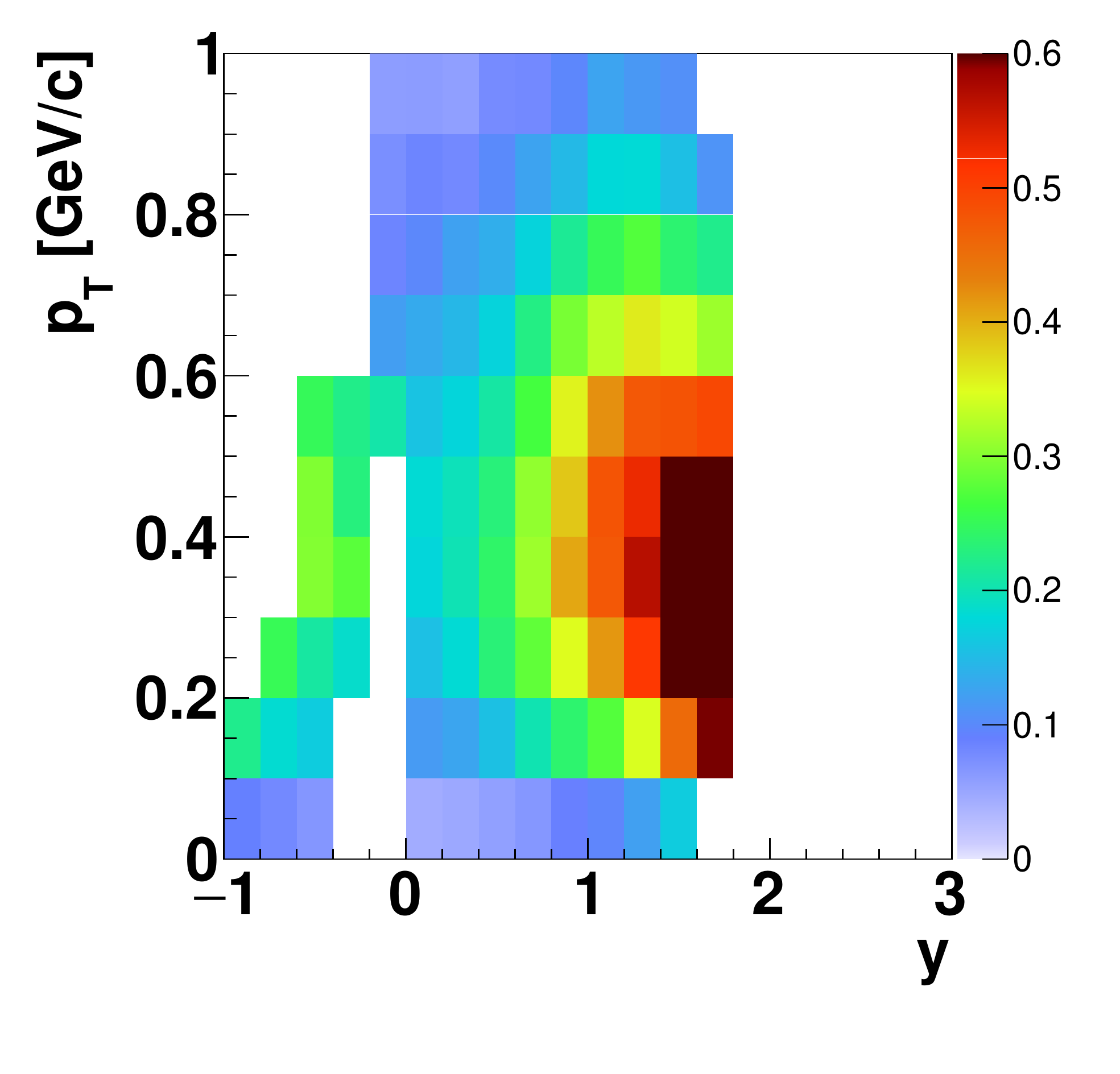} &
\hspace{-30mm}
\includegraphics[width=0.18\textwidth]{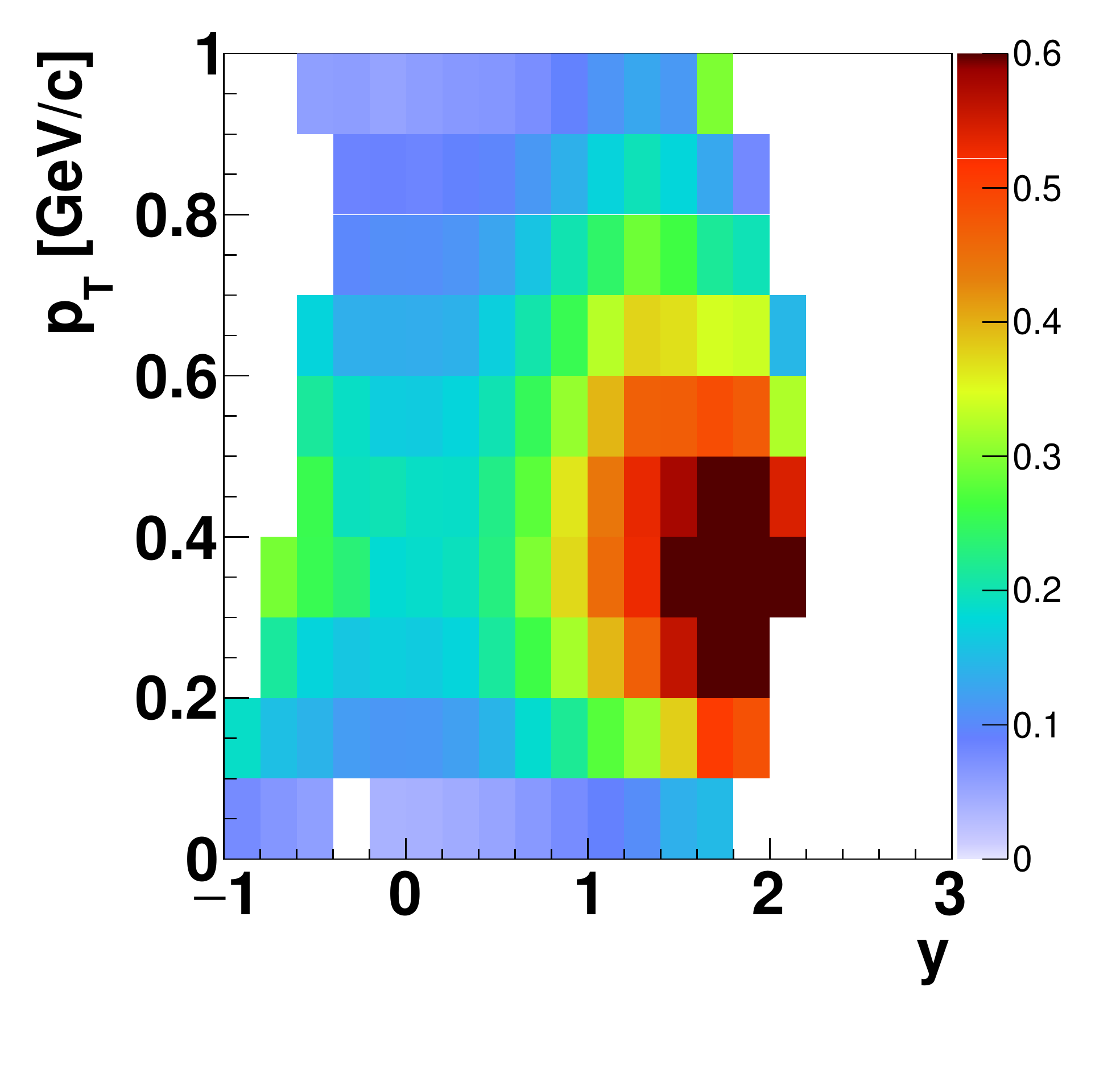} &
\hspace{-45mm}
\includegraphics[width=0.18\textwidth]{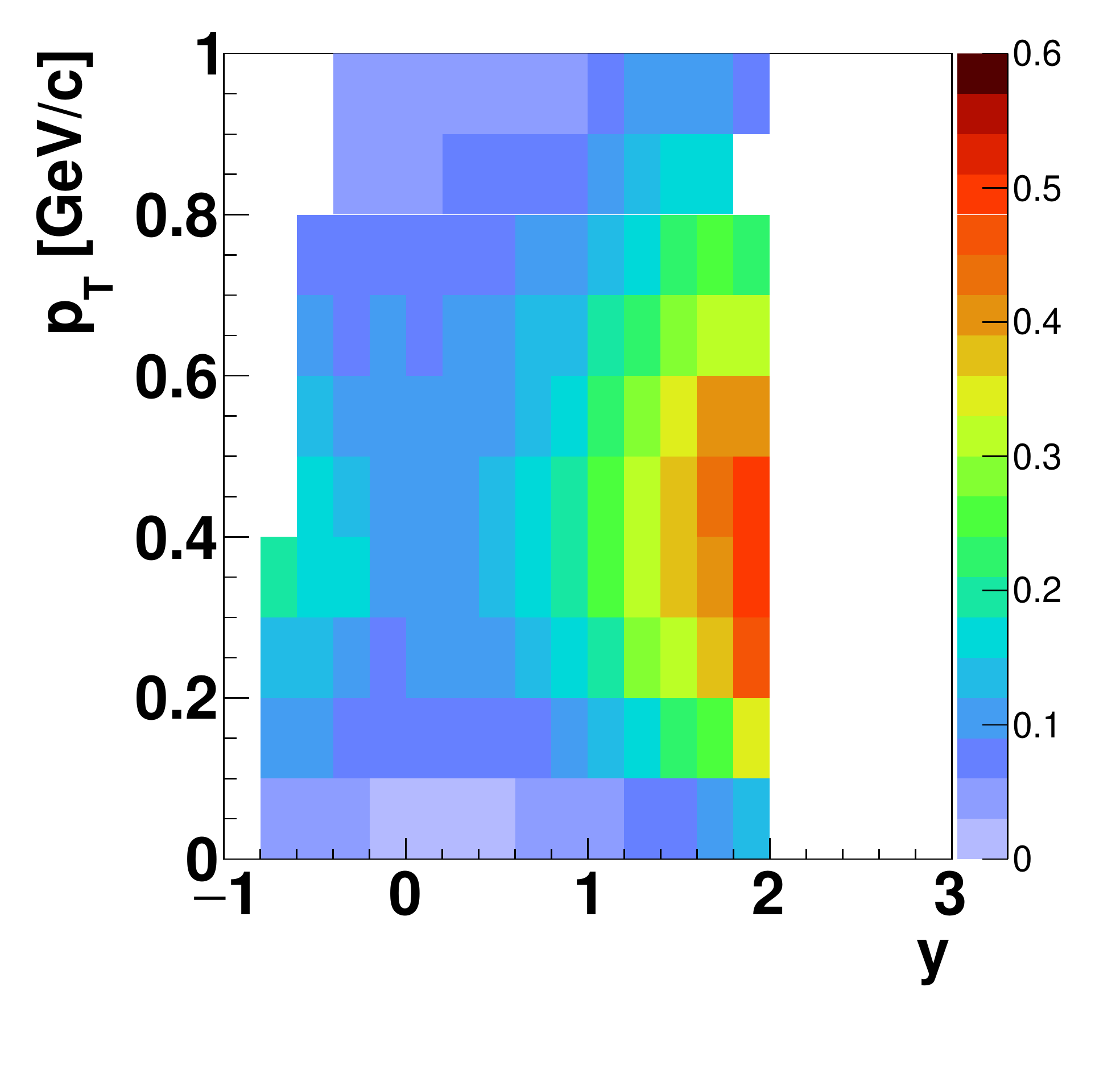} &
\hspace{-60mm}
\includegraphics[width=0.18\textwidth]{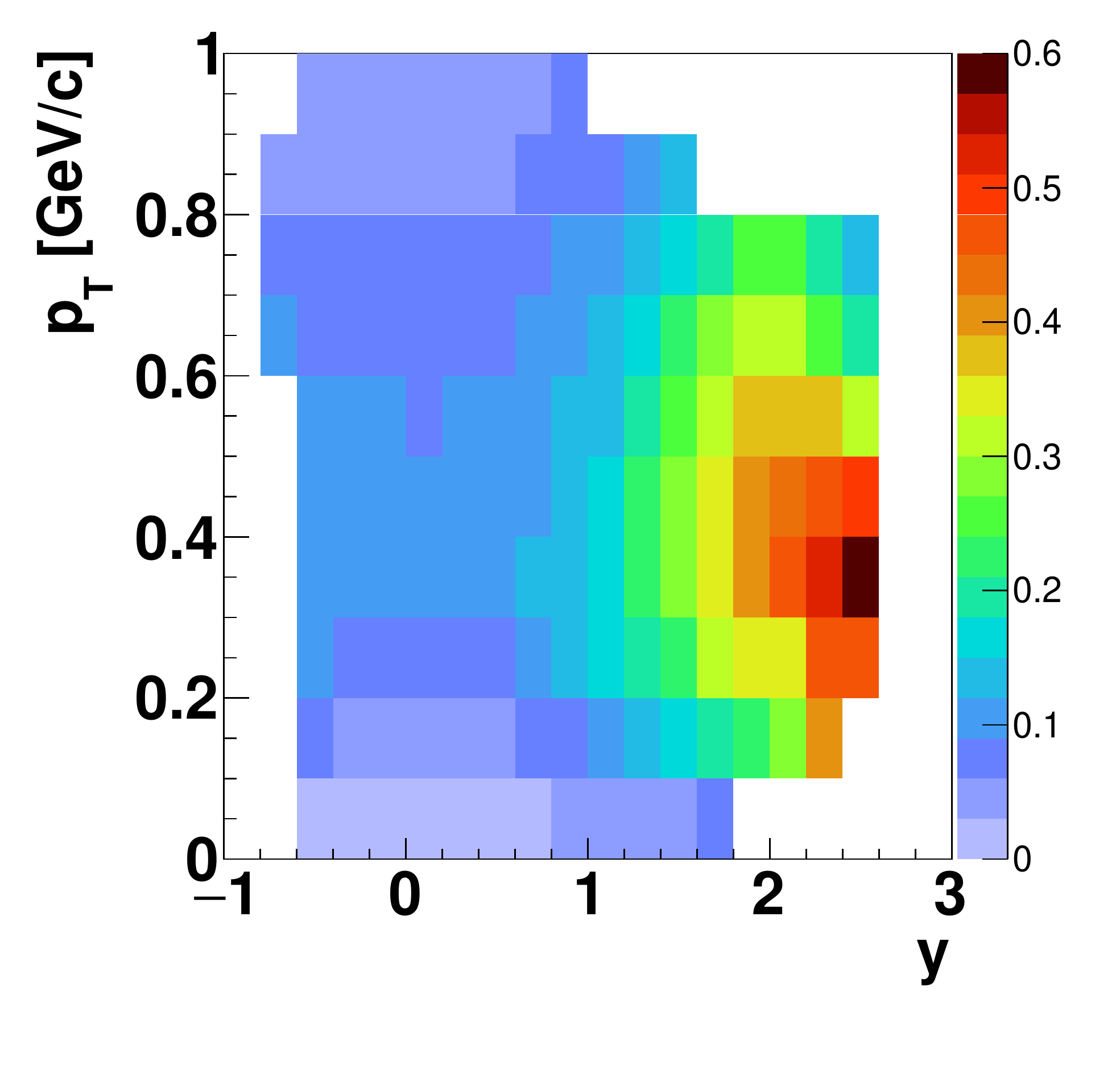} 
\tabularnewline
\vspace{-5mm}$\bar{\textrm{p}}$ &
&
\hspace{-15mm}
\includegraphics[width=0.18\textwidth]{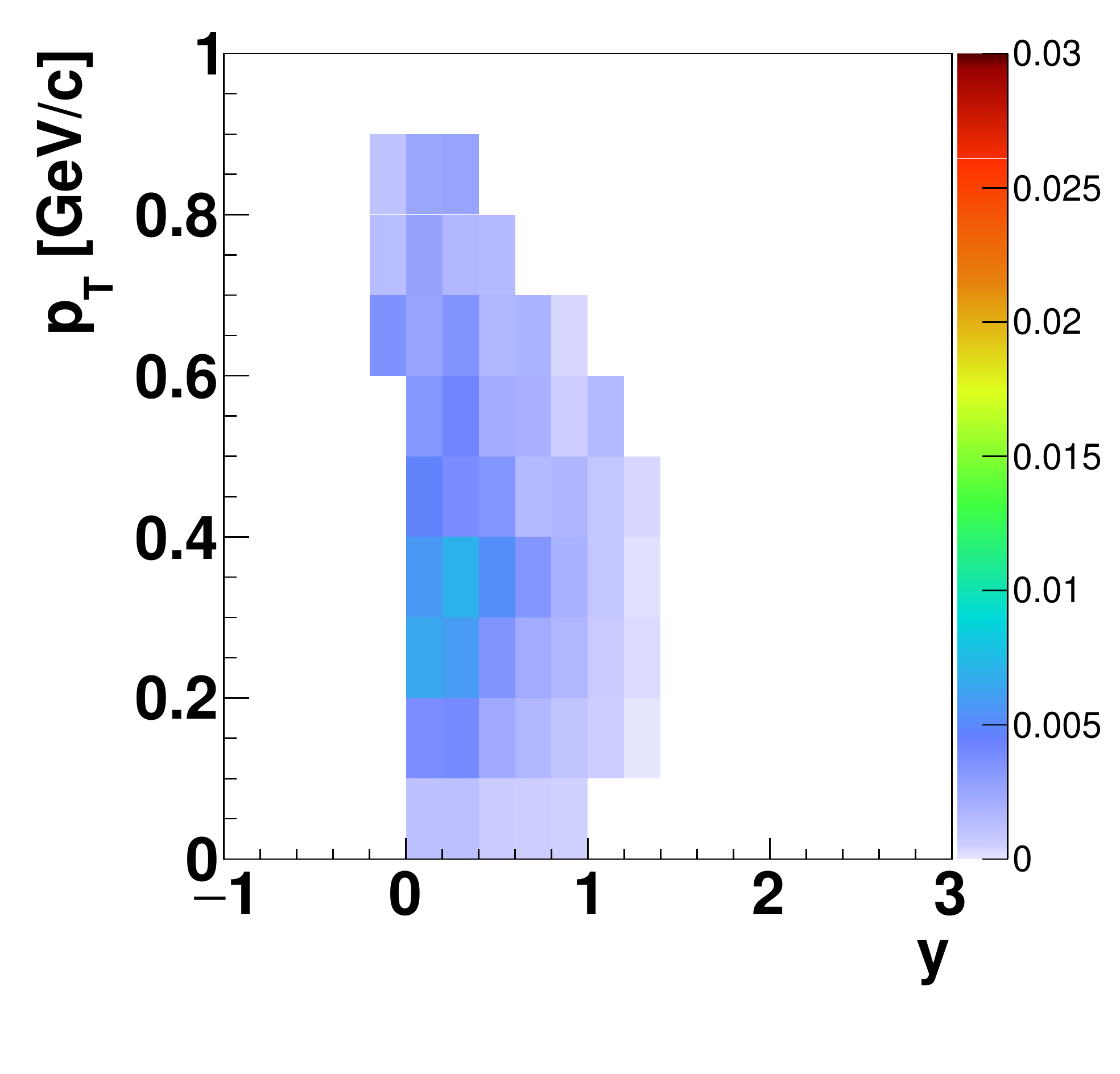} &
\hspace{-30mm}
\includegraphics[width=0.18\textwidth]{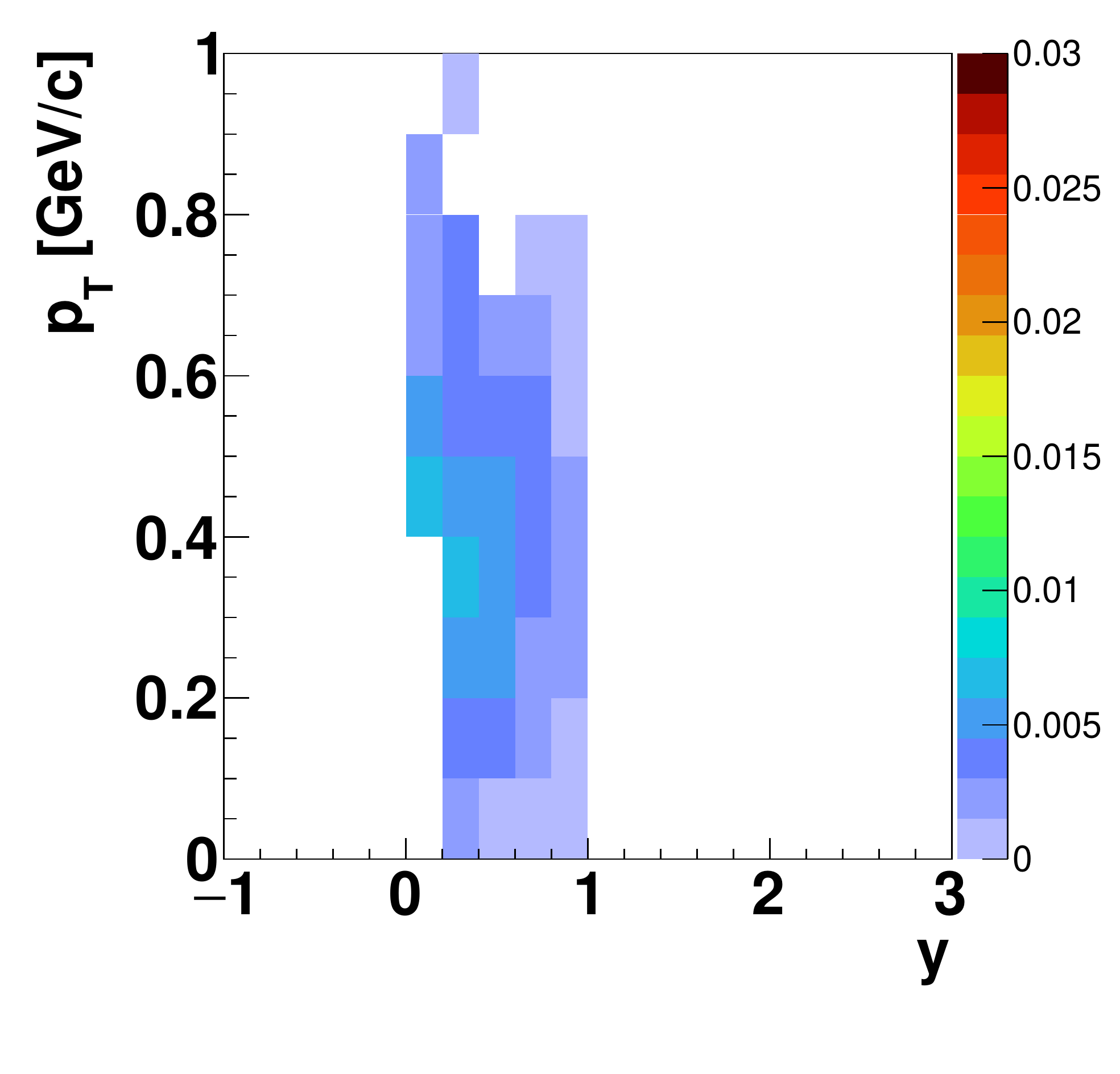} &
\hspace{-45mm}
\includegraphics[width=0.18\textwidth]{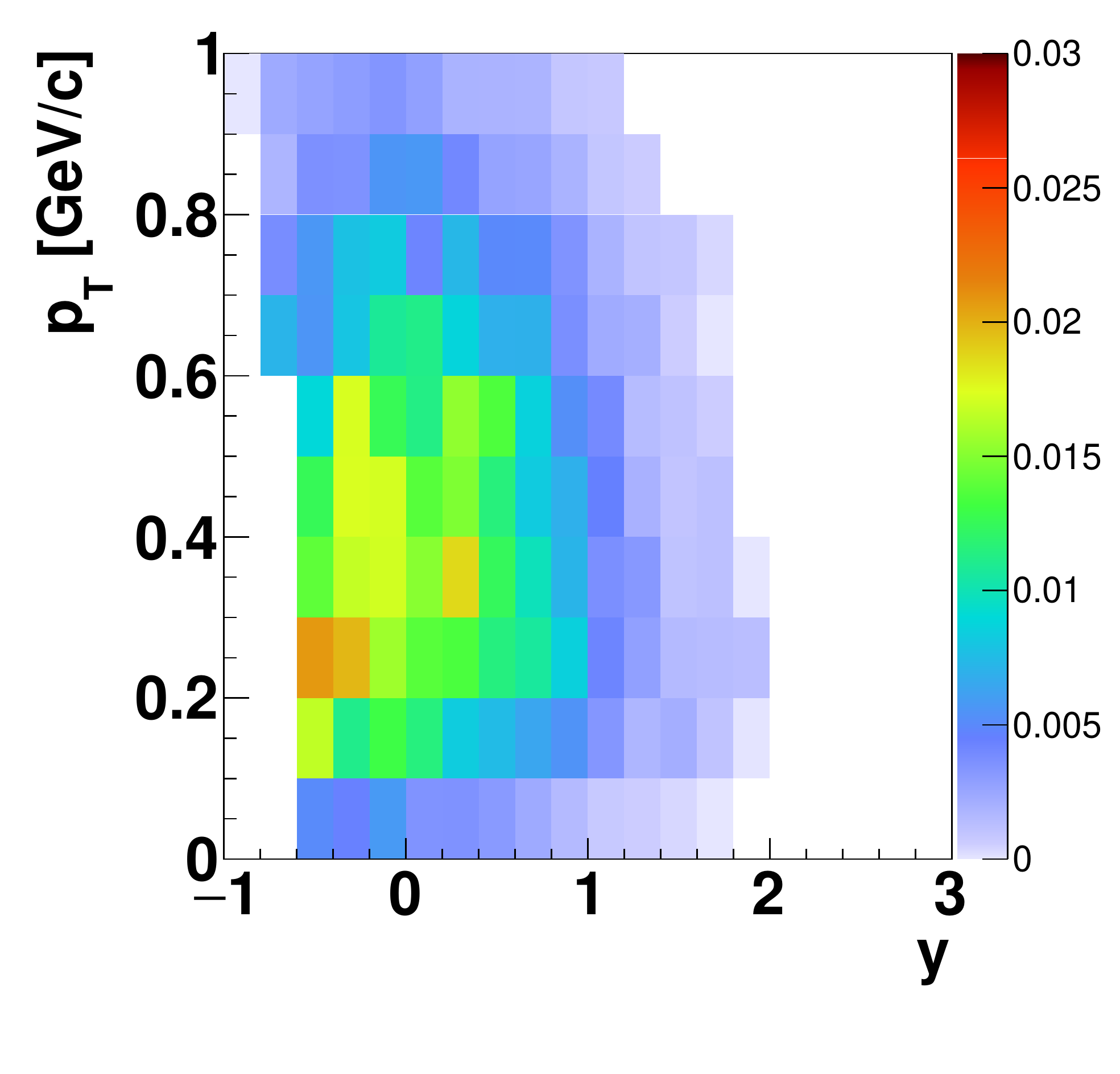} &
\hspace{-60mm}
\includegraphics[width=0.18\textwidth]{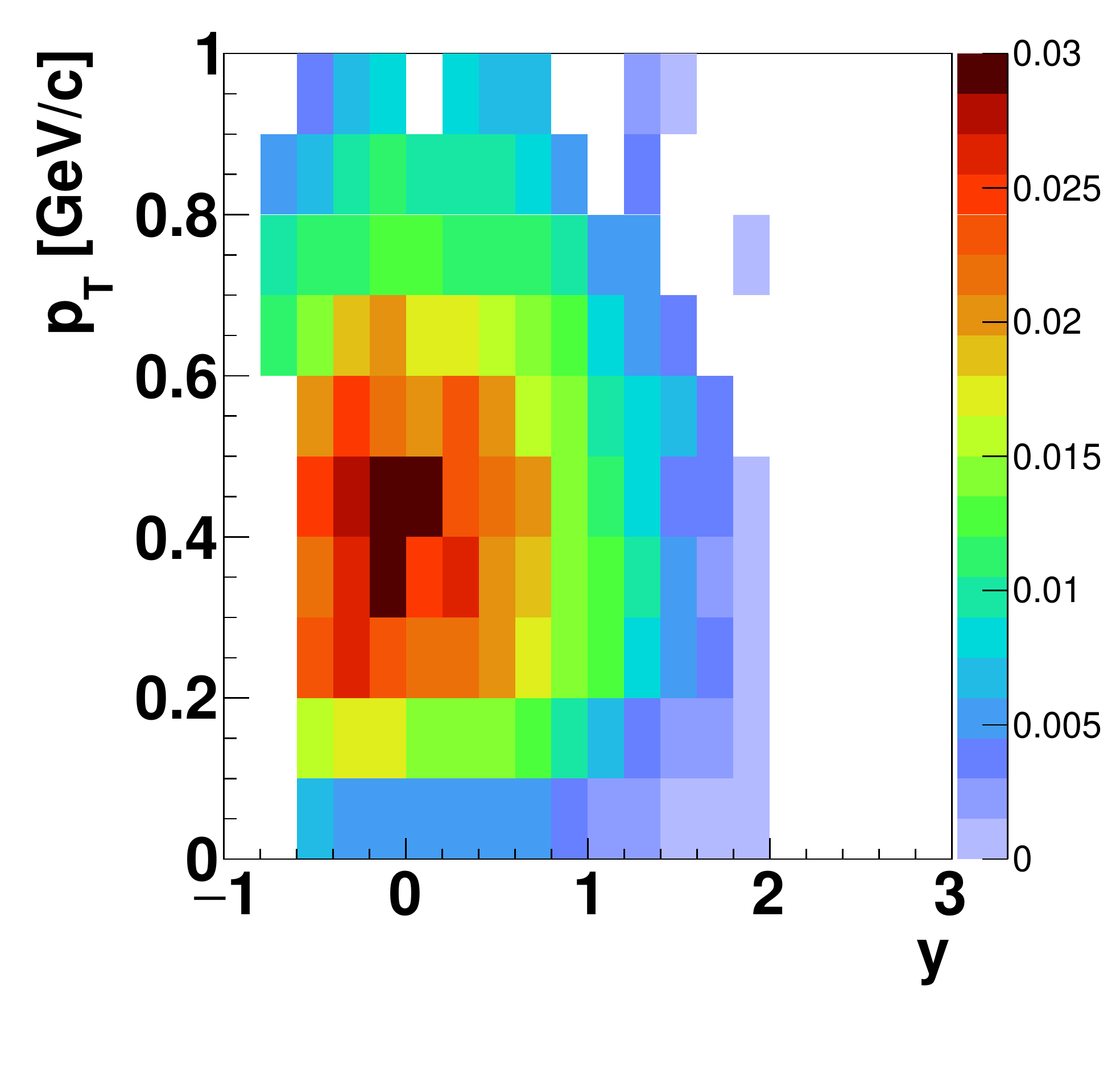} 
\tabularnewline
\end{tabular}
		\end{center}
		\caption{(Color online) Transverse momentum-rapidity spectrum of $\pi^{-}$, $\pi^{+}$, K$^{-}$, K$^{+}$, p and $\bar{\textrm{p}}$ produced in inelastic p+p interactions at 20, 31, 40, 80 and 158~\GeVc. Color scale represents particle multiplicities normalized to the phase-space bin size ($\frac{dn}{dydp_{T}}$).}
		\label{fig:final2D}
	\end{figure*}

Two dimensional distributions normalised per event ($y$ vs. $p_{T}$) of $\pi^{-}$, $\pi^{+}$, K$^{-}$, K$^{+}$, p and $\bar{\textrm{p}}$ produced in inelastic p+p interactions at different SPS energies are presented in Fig.~\ref{fig:final2D}. 
Where available, results from the \dedx methods were used because of their smaller statistical uncertainties. Results from the $tof$-\dedx method were taken to extend the phase space coverage. Anti-proton yields at 20~\GeVc could not be determined due to the insufficient statistics of the collected data. Empty bins in phase-space (mostly for lower energies) are caused by insufficient acceptance for the methods used in the analysis. Yields for all particle types except protons are seen to increase with beam momentum in the SPS energy range.
	
\begin{figure*}
		\begin{center}
		\includegraphics[width=0.3\textwidth]{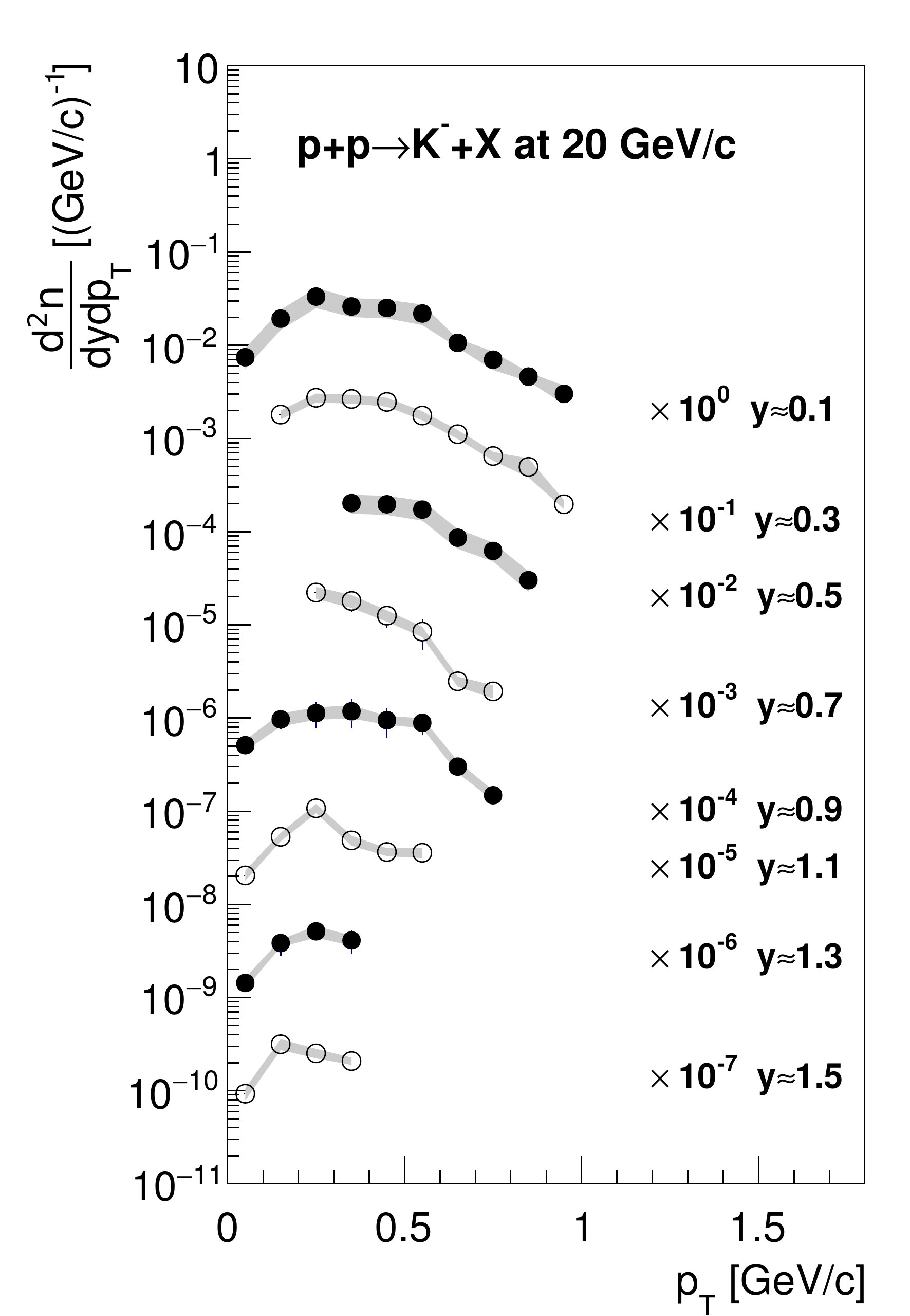}
		\includegraphics[width=0.3\textwidth]{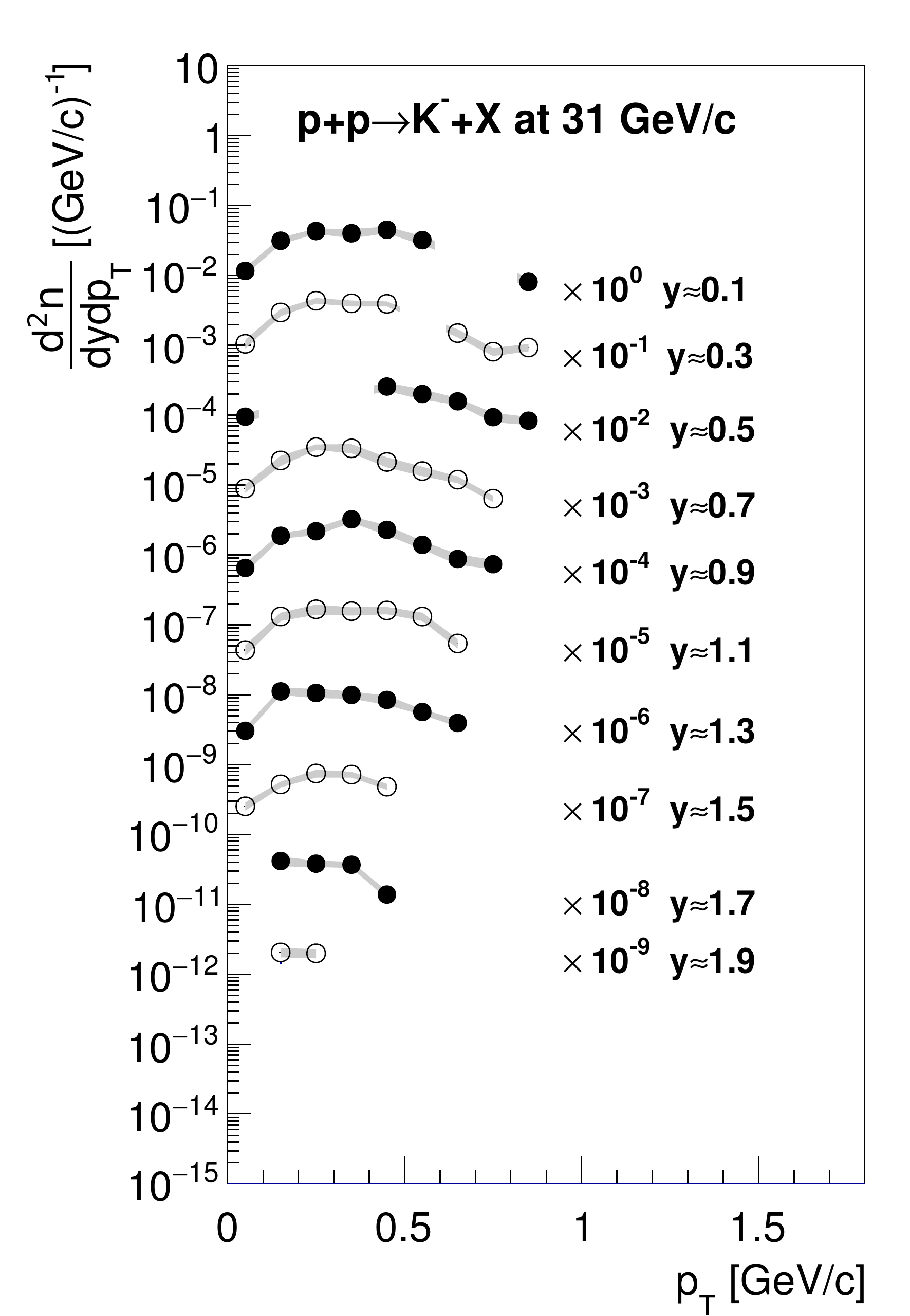}\\
		\includegraphics[width=0.3\textwidth]{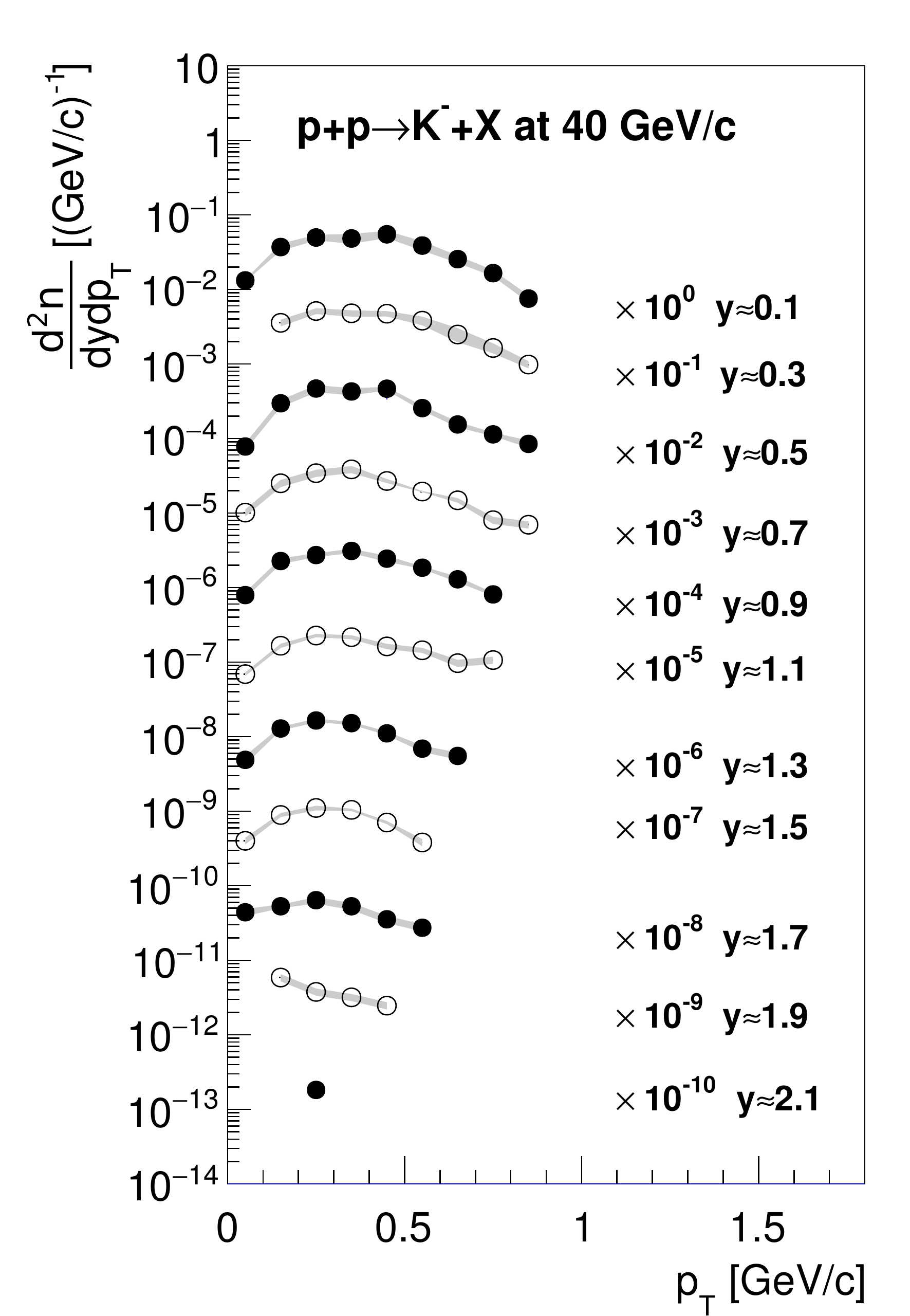}
		\includegraphics[width=0.3\textwidth]{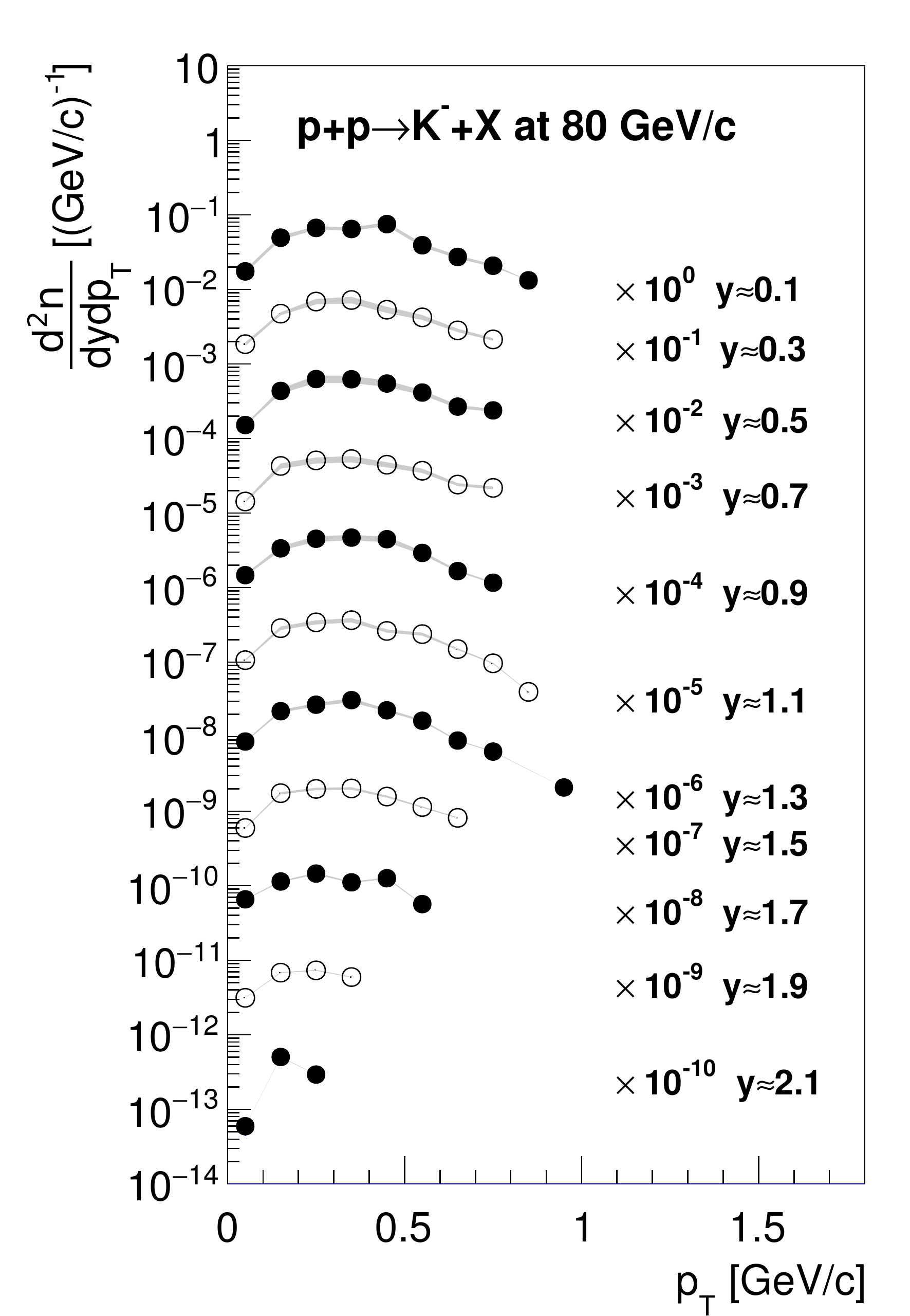}
		\includegraphics[width=0.3\textwidth]{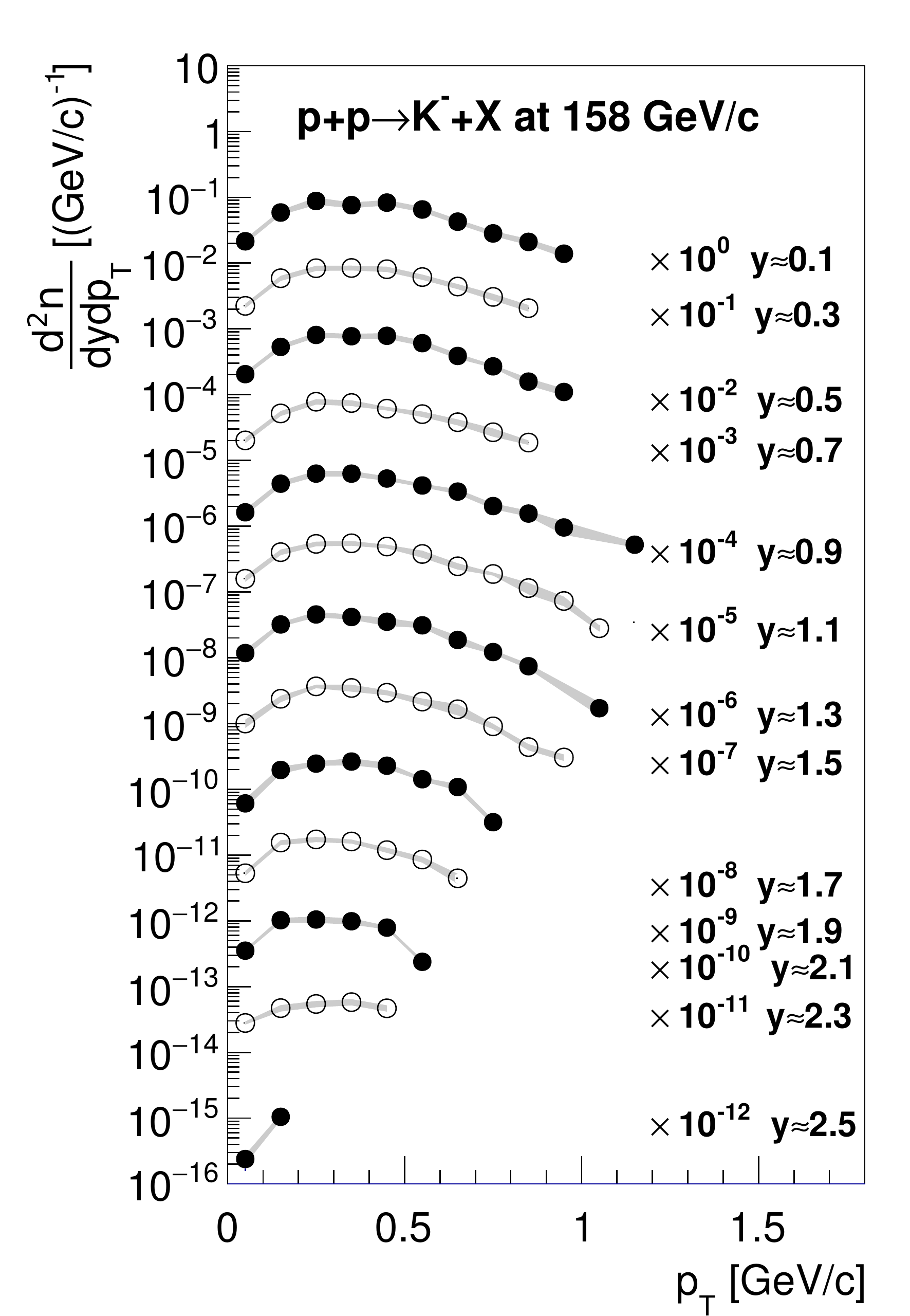}
		\end{center}
		\caption{(Color online) Transverse momentum K$^{-}$ spectra in rapidity slices produced in inelastic p+p interactions at 20, 31, 40, 80, 158~\GeVc. Rapidity values given in the legends correspond to the middle of the corresponding interval. Shaded bands show systematic uncertainties.}
		\label{fig:nptkaon}
\end{figure*}

\begin{figure*}
		\begin{center}
		\includegraphics[width=0.3\textwidth]{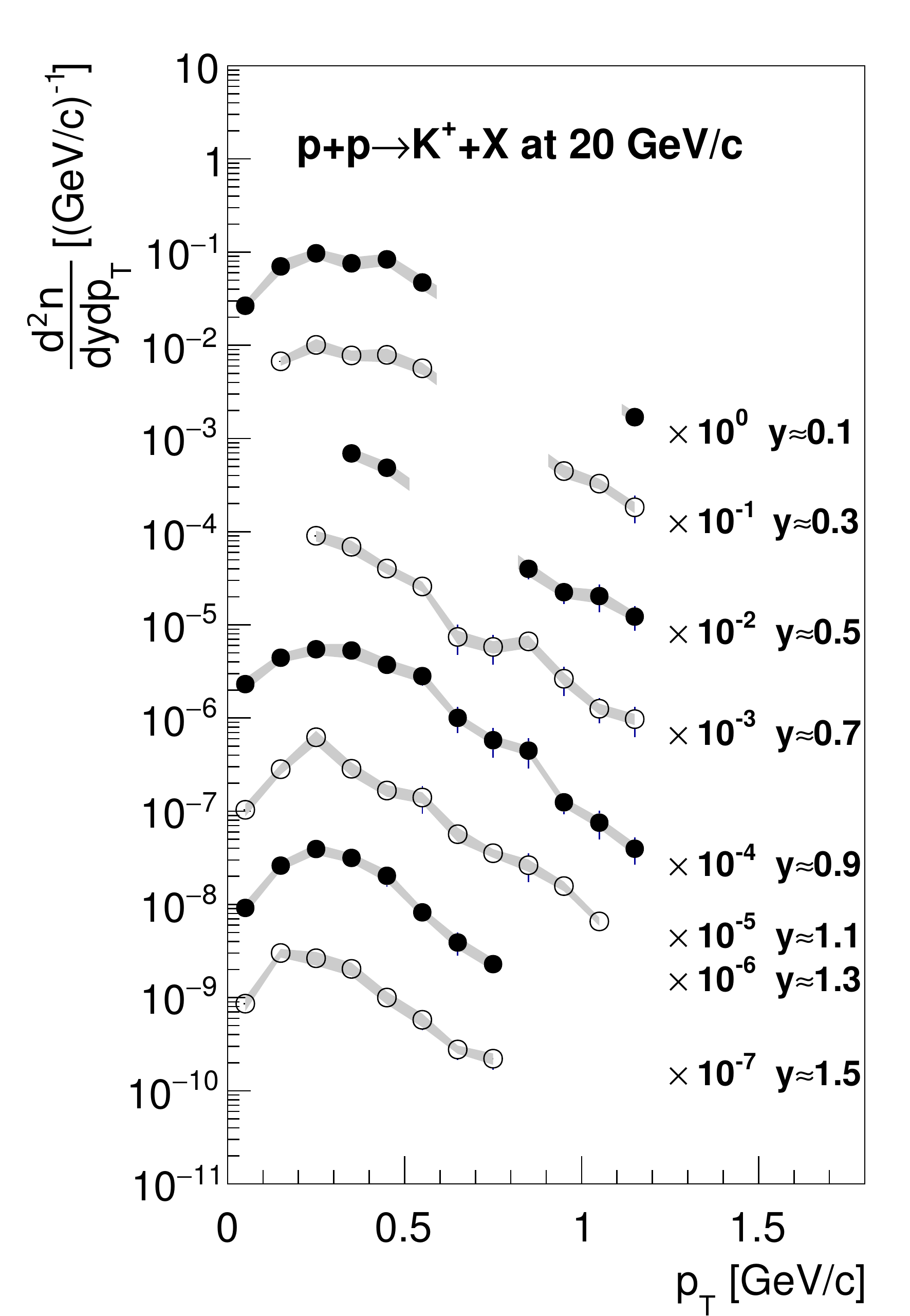}
		\includegraphics[width=0.3\textwidth]{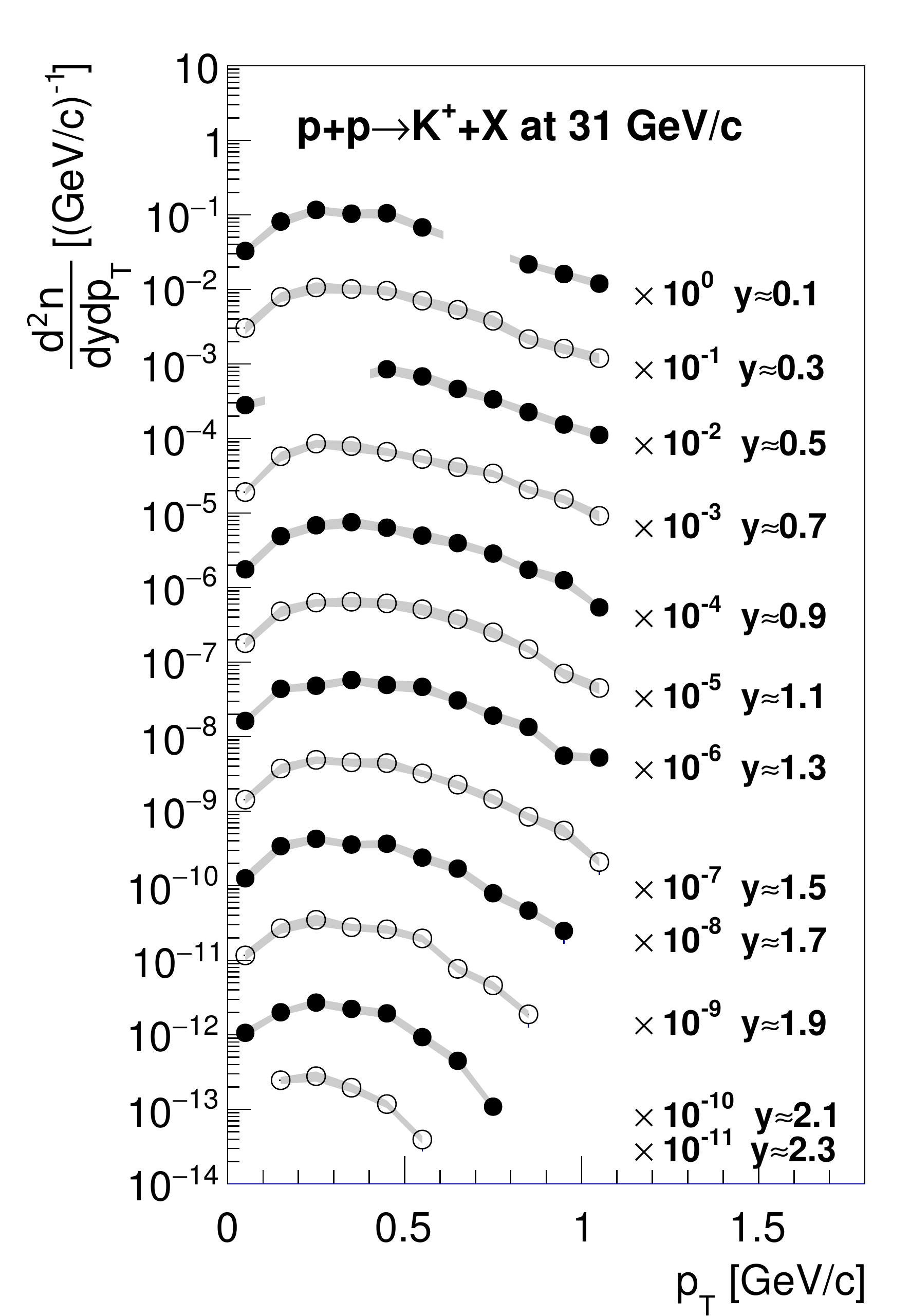}\\
		\includegraphics[width=0.3\textwidth]{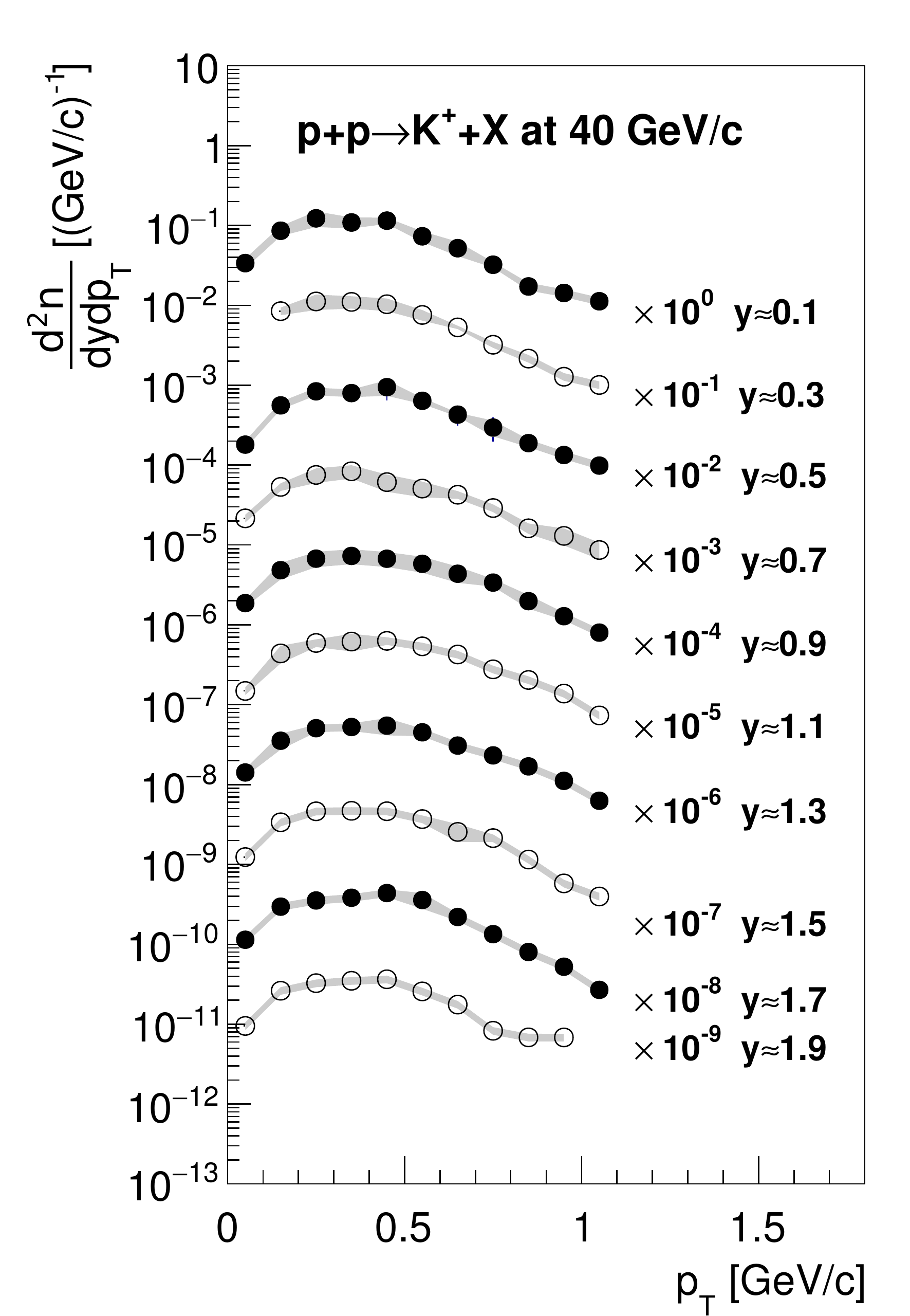}
		\includegraphics[width=0.3\textwidth]{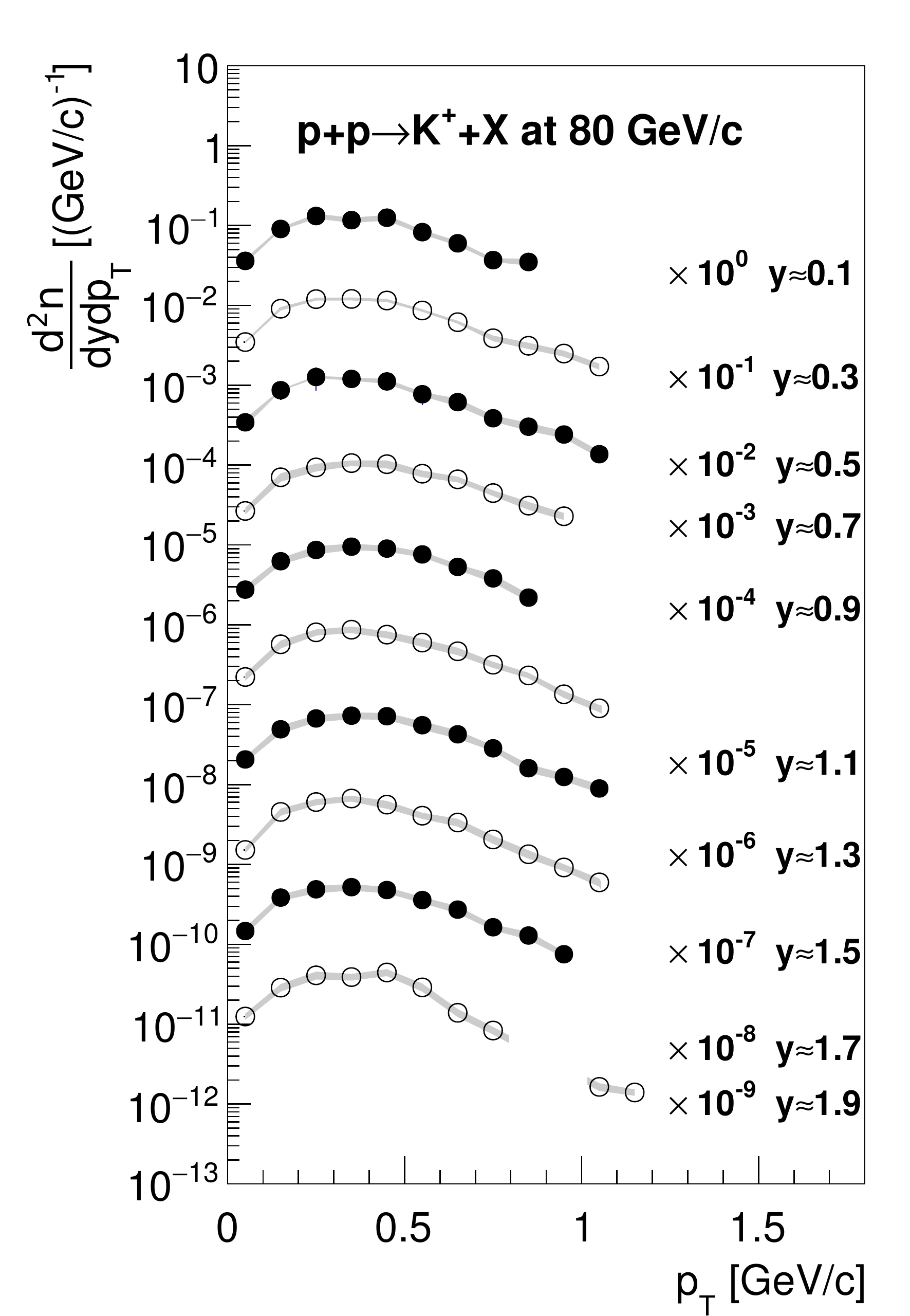}
		\includegraphics[width=0.3\textwidth]{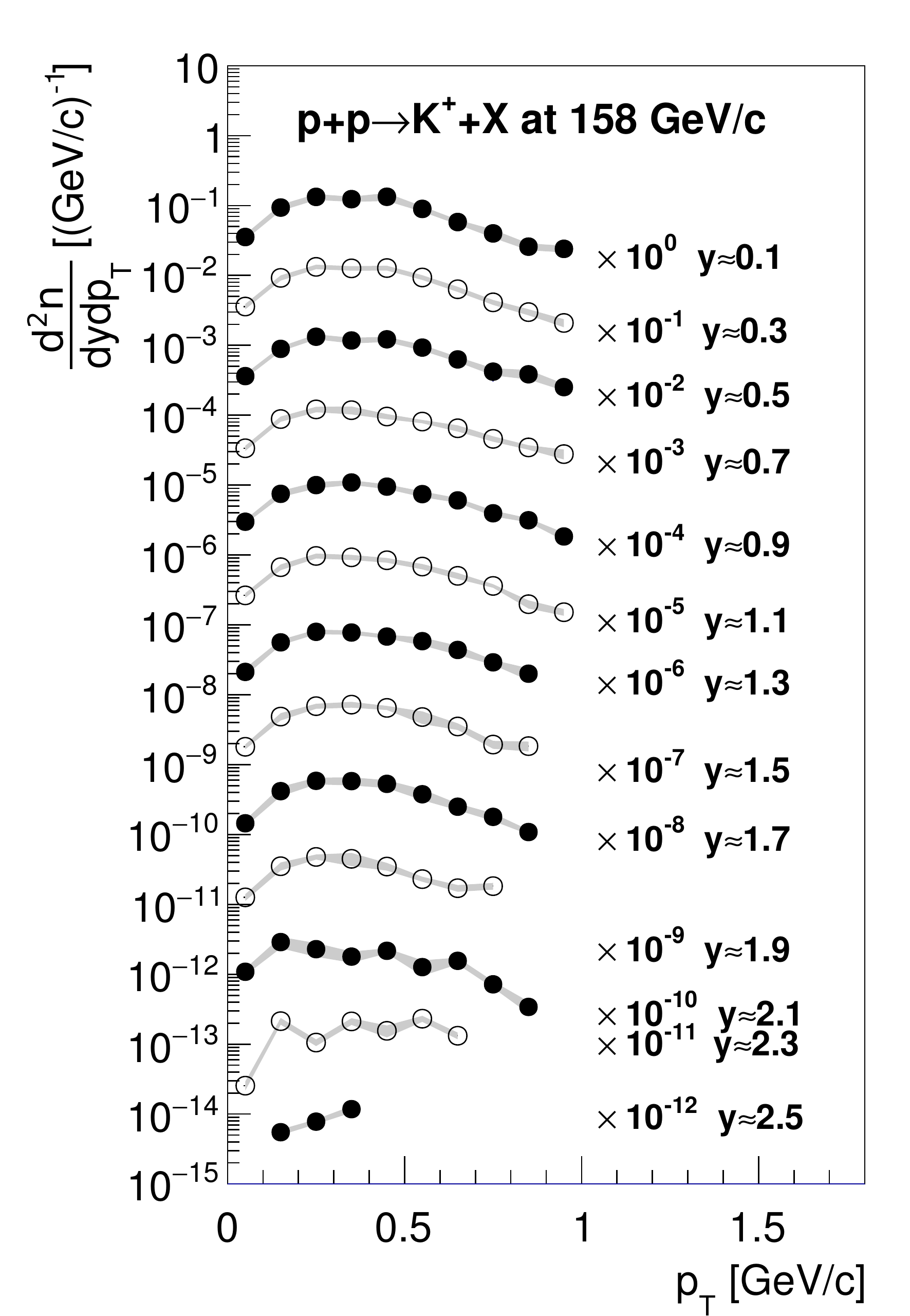}
		\end{center}
		\caption{(Color online) Transverse momentum K$^{+}$ spectra in rapidity slices produced in inelastic p+p interactions at 20, 31, 40, 80, 158~\GeVc. Rapidity values given in the legends correspond to the middle of the corresponding interval. Shaded bands show systematic uncertainties.}
		\label{fig:pptkaon}
\end{figure*}

\begin{figure*}
		\begin{center}
		\includegraphics[width=0.3\textwidth]{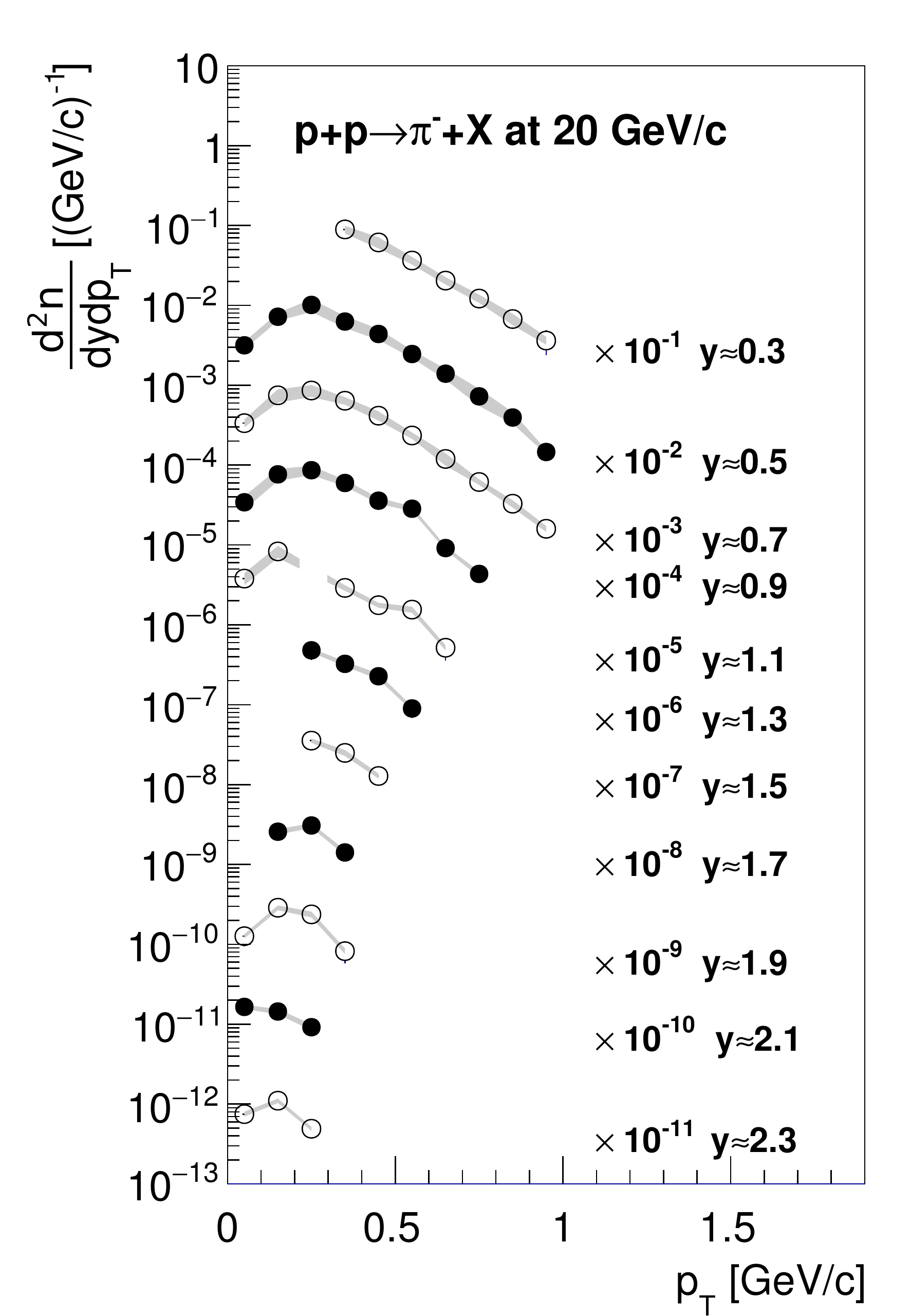}
		\includegraphics[width=0.3\textwidth]{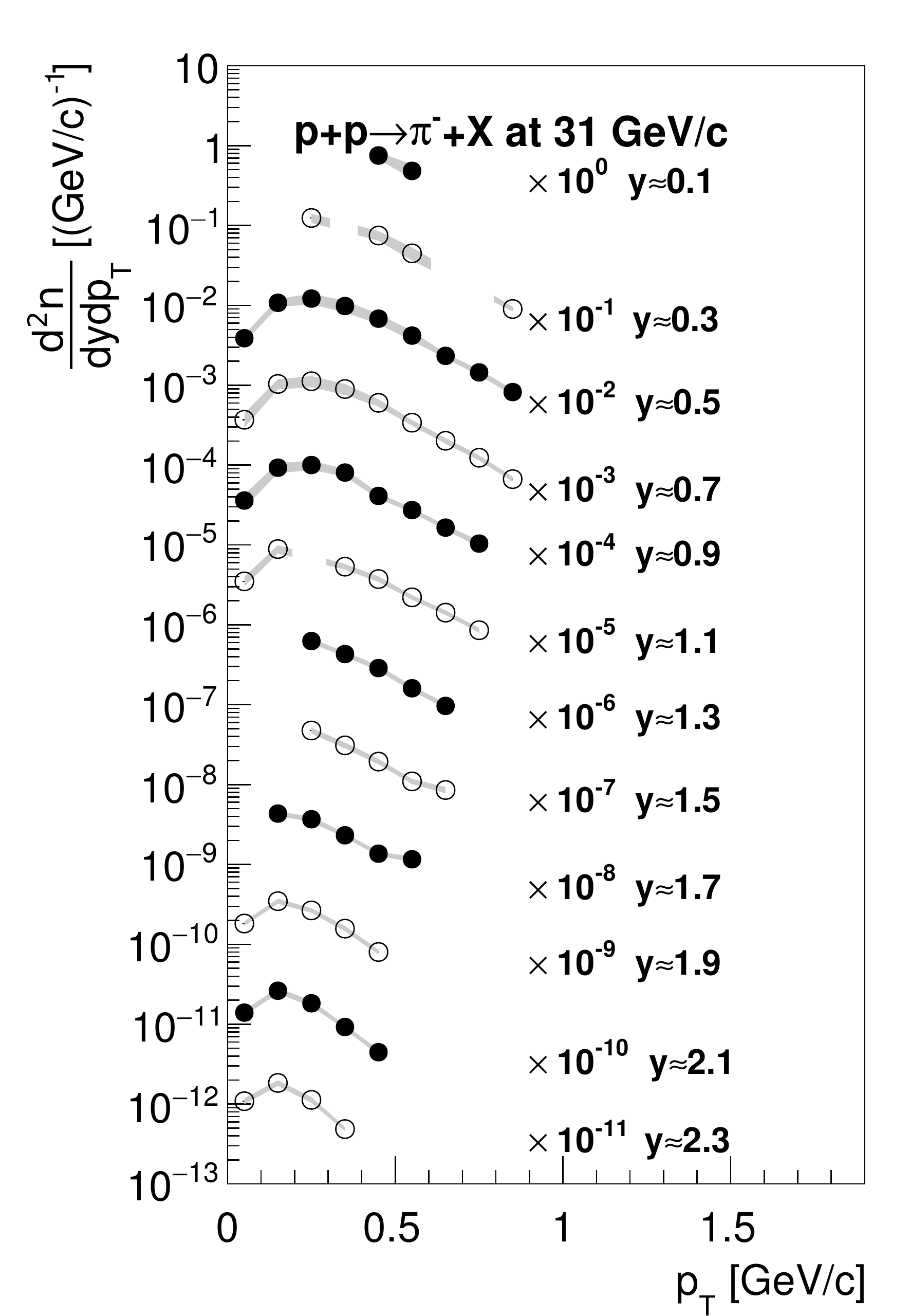}\\
		\includegraphics[width=0.3\textwidth]{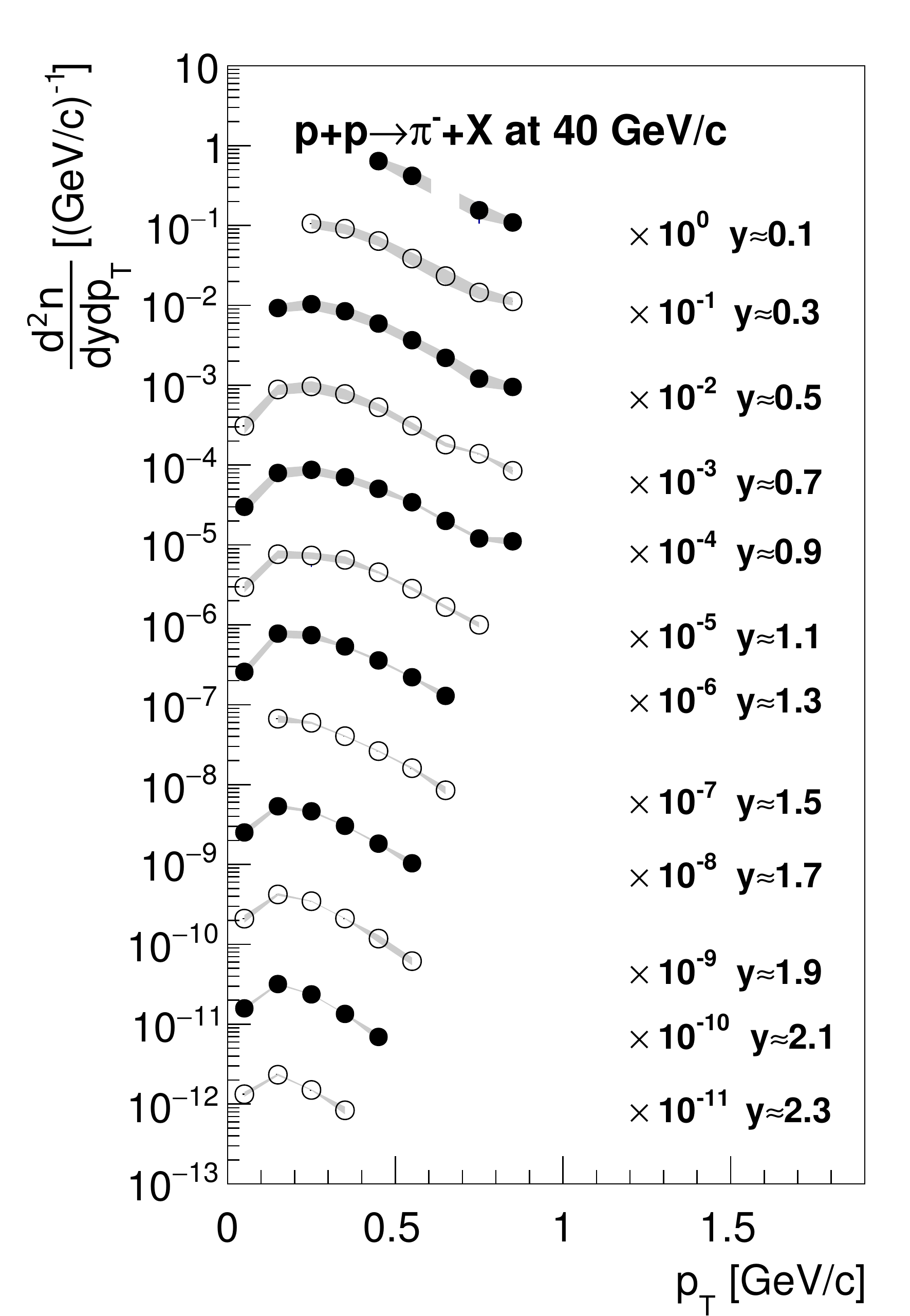}
		\includegraphics[width=0.3\textwidth]{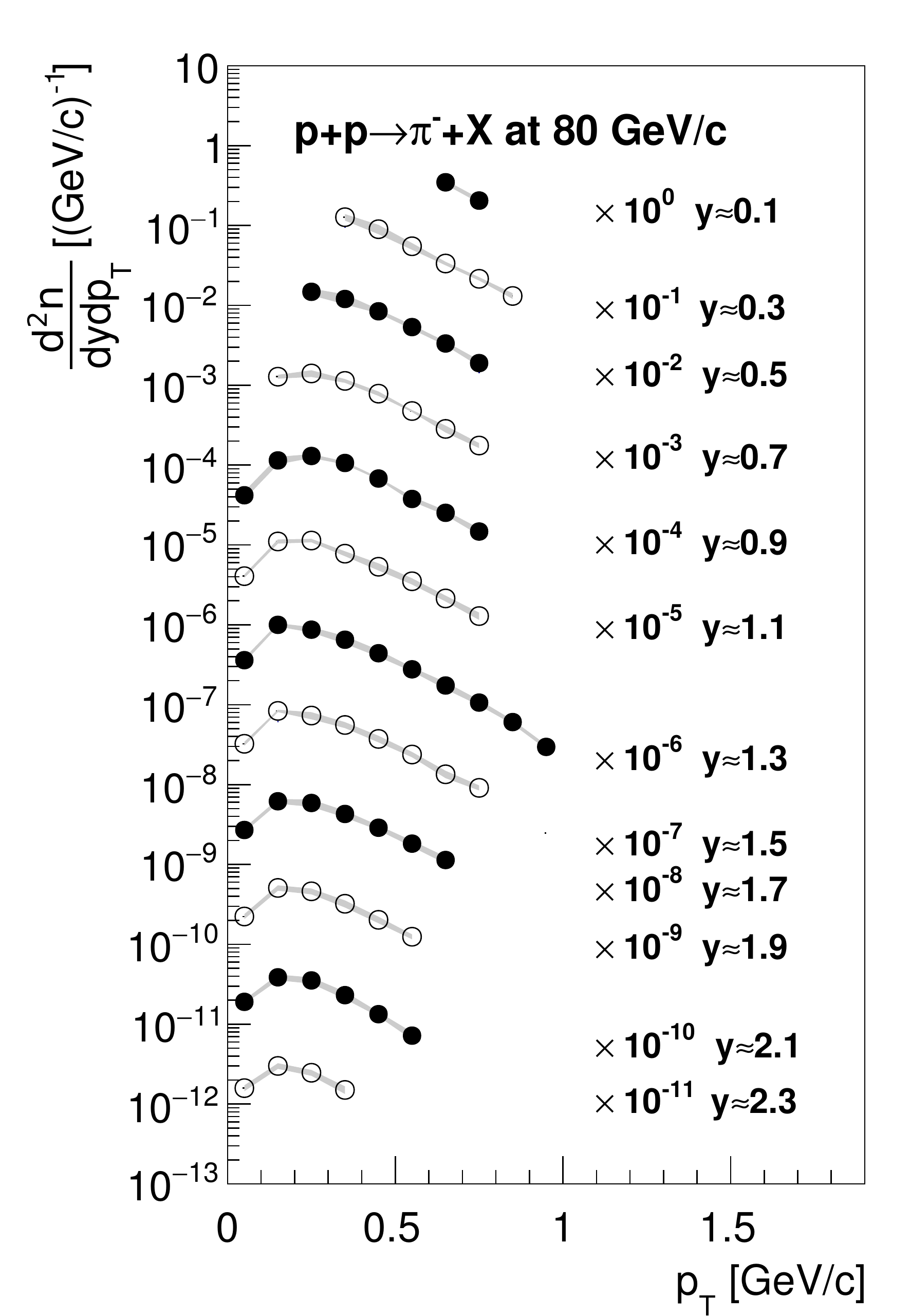}
		\includegraphics[width=0.3\textwidth]{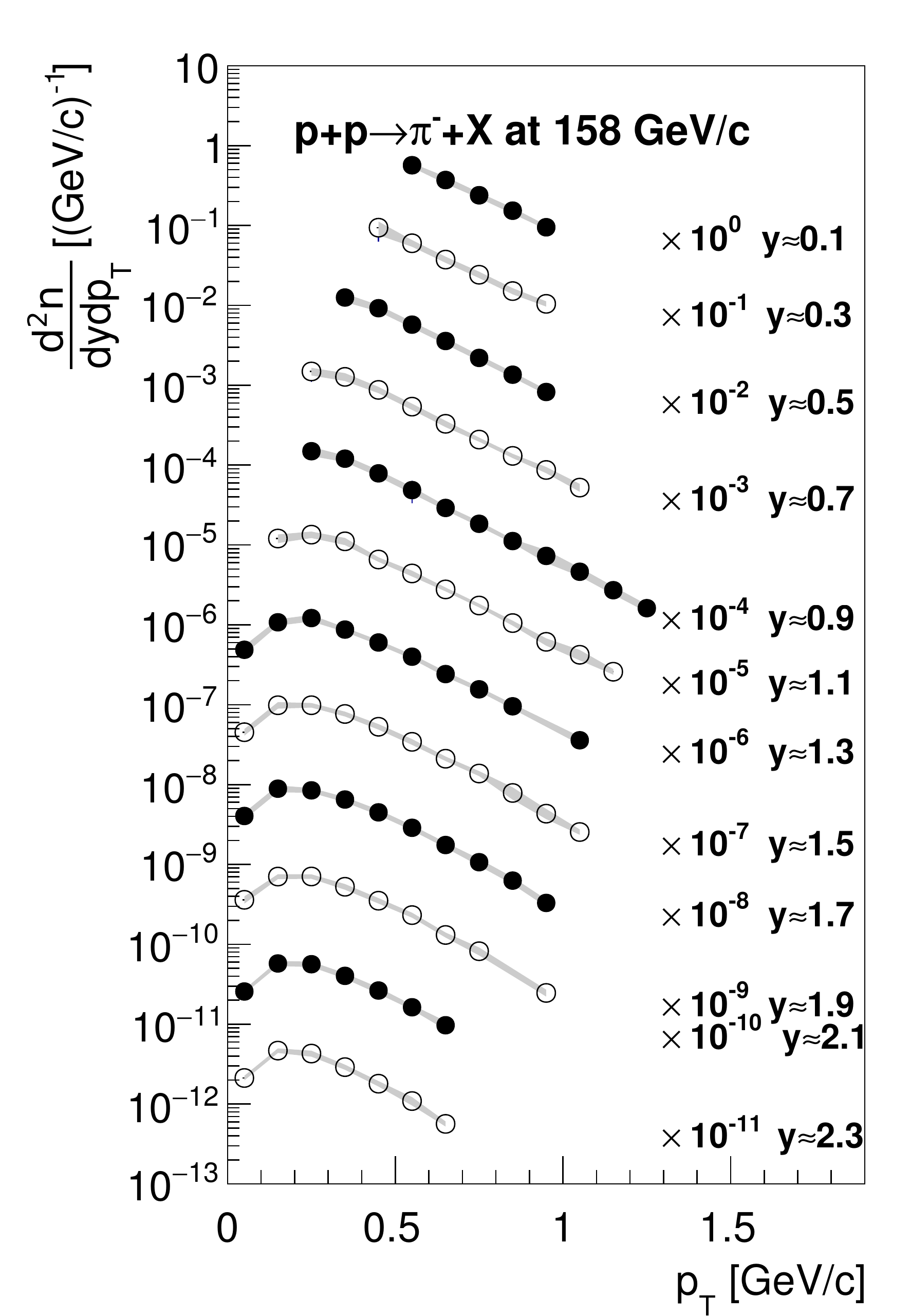}
		\end{center}
		\caption{(Color online) Transverse momentum $\pi^{-}$ spectra in rapidity slices produced in inelastic p+p interactions at 20, 31, 40, 80, 158~\GeVc. Rapidity values given in the legends correspond to the middle of the corresponding interval. Shaded bands show systematic uncertainties.}
		\label{fig:nptpion}
\end{figure*}

\begin{figure*}
		\begin{center}
		\includegraphics[width=0.3\textwidth]{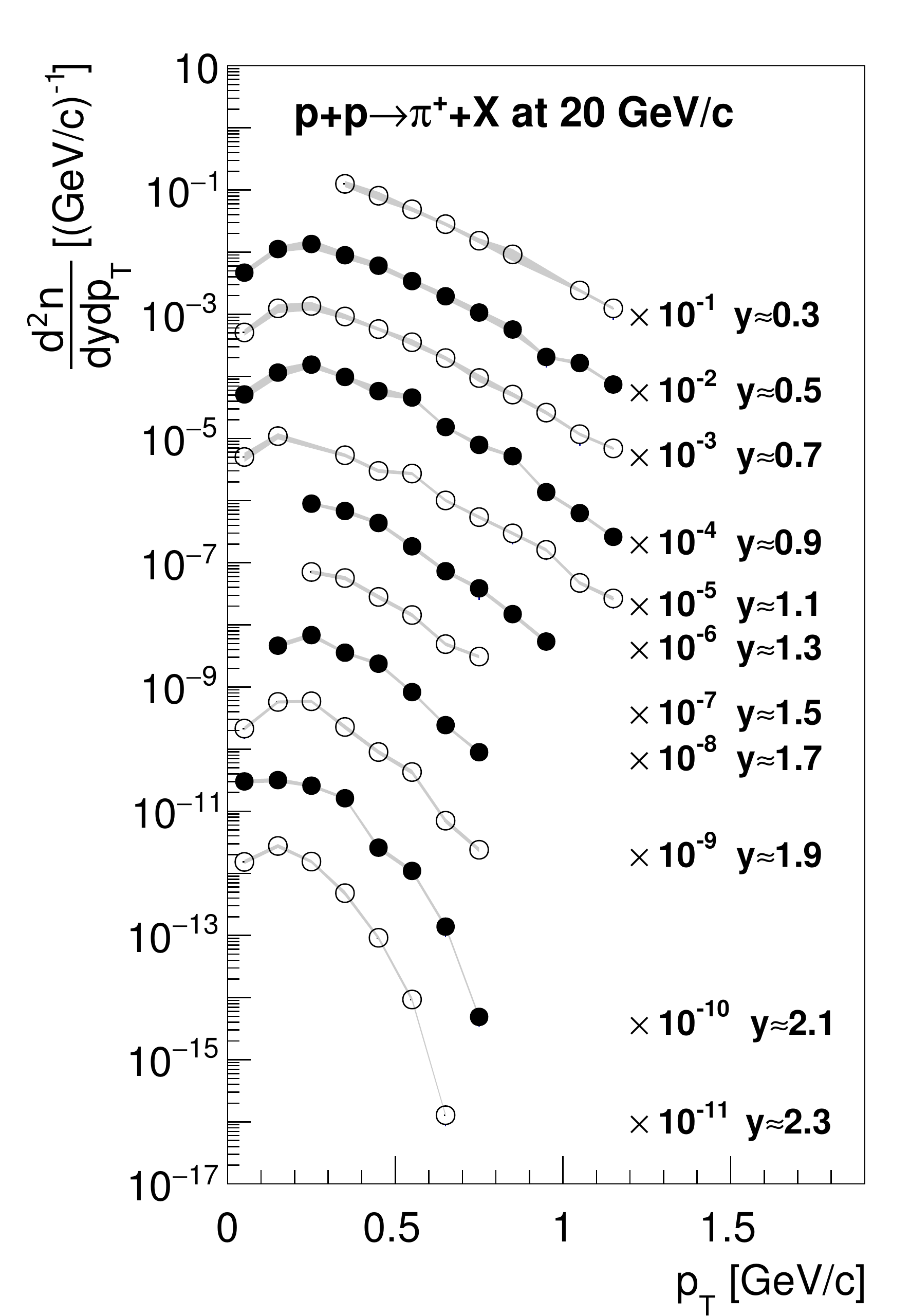}
		\includegraphics[width=0.3\textwidth]{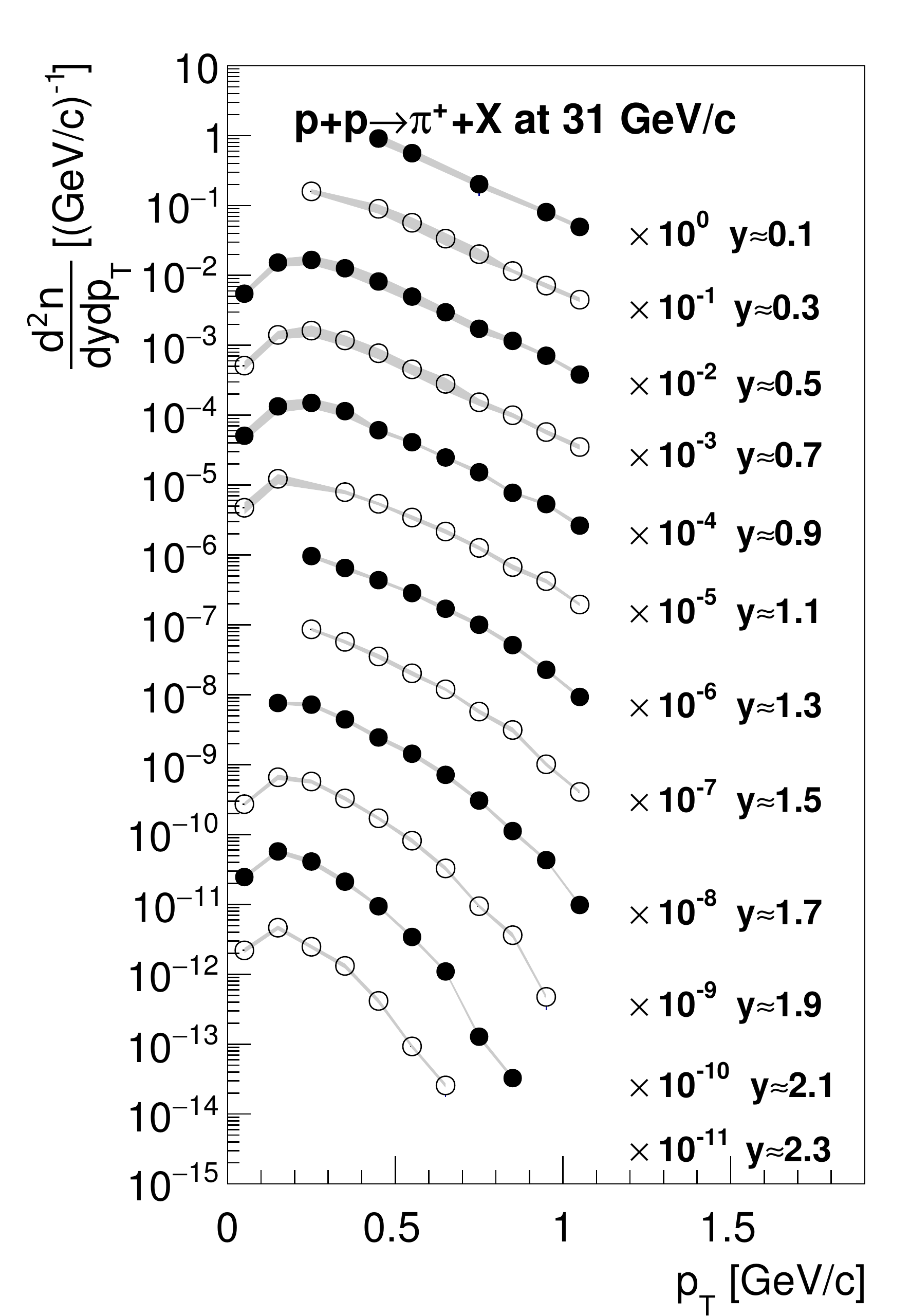}\\
		\includegraphics[width=0.3\textwidth]{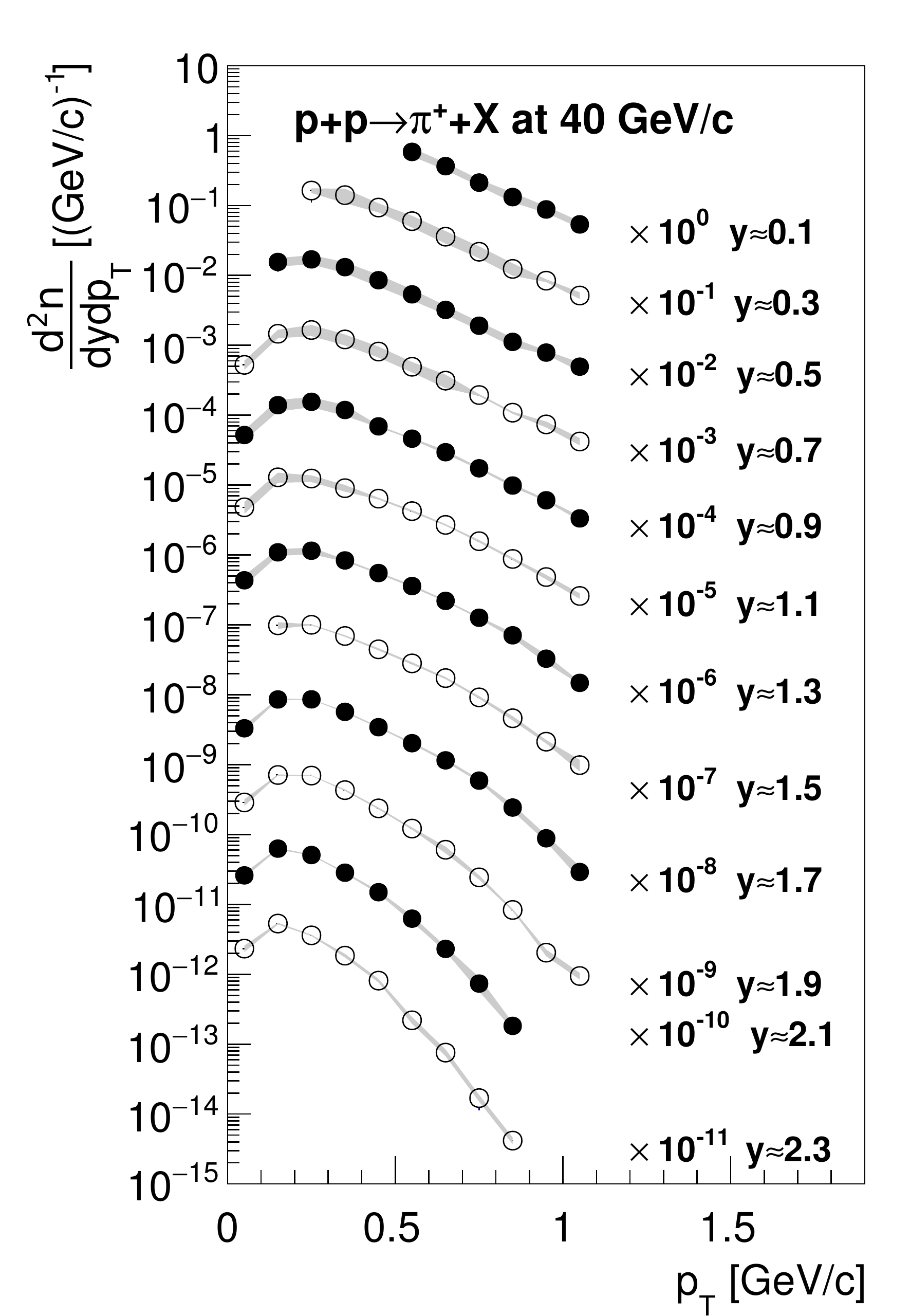}
		\includegraphics[width=0.3\textwidth]{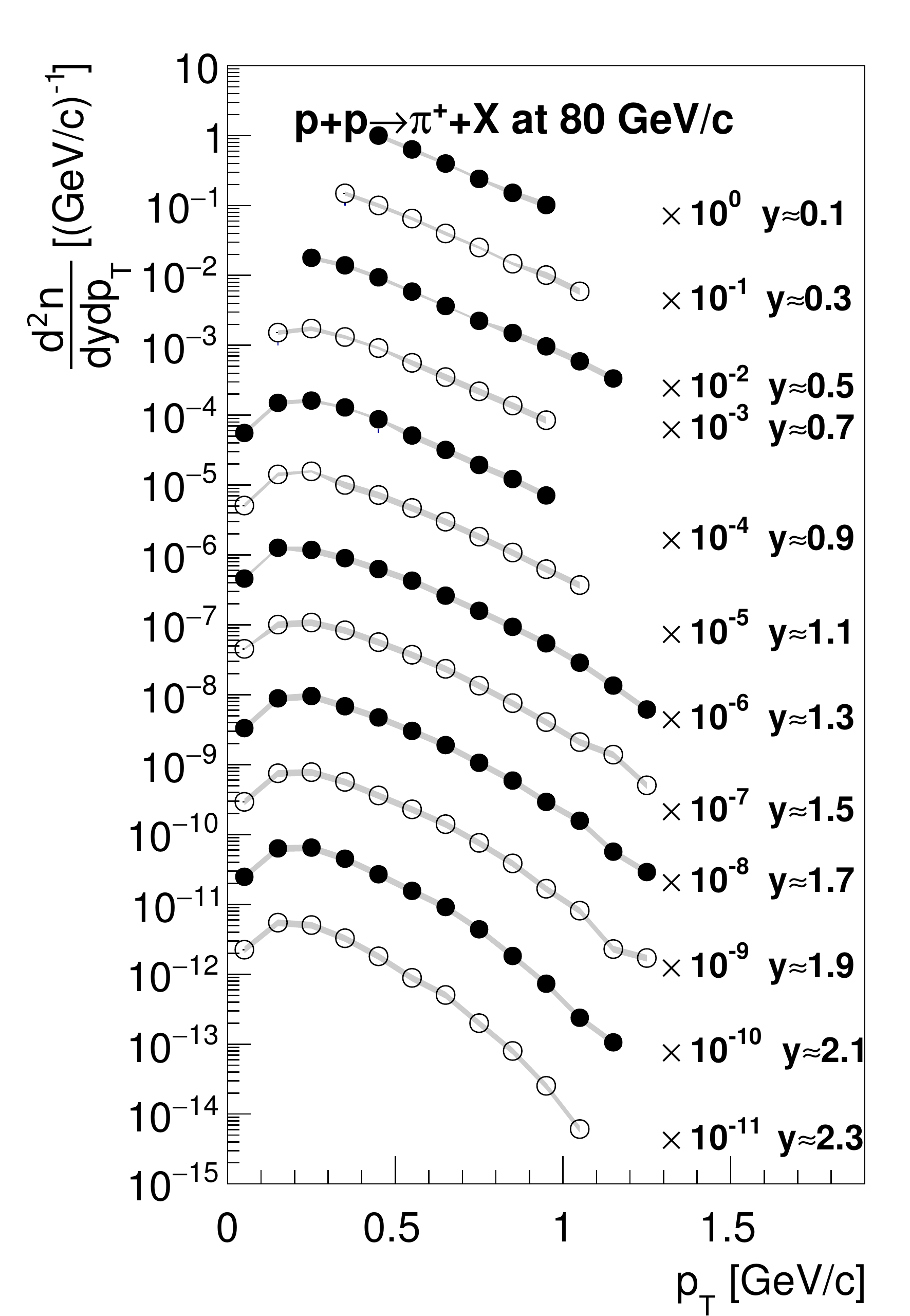}
		\includegraphics[width=0.3\textwidth]{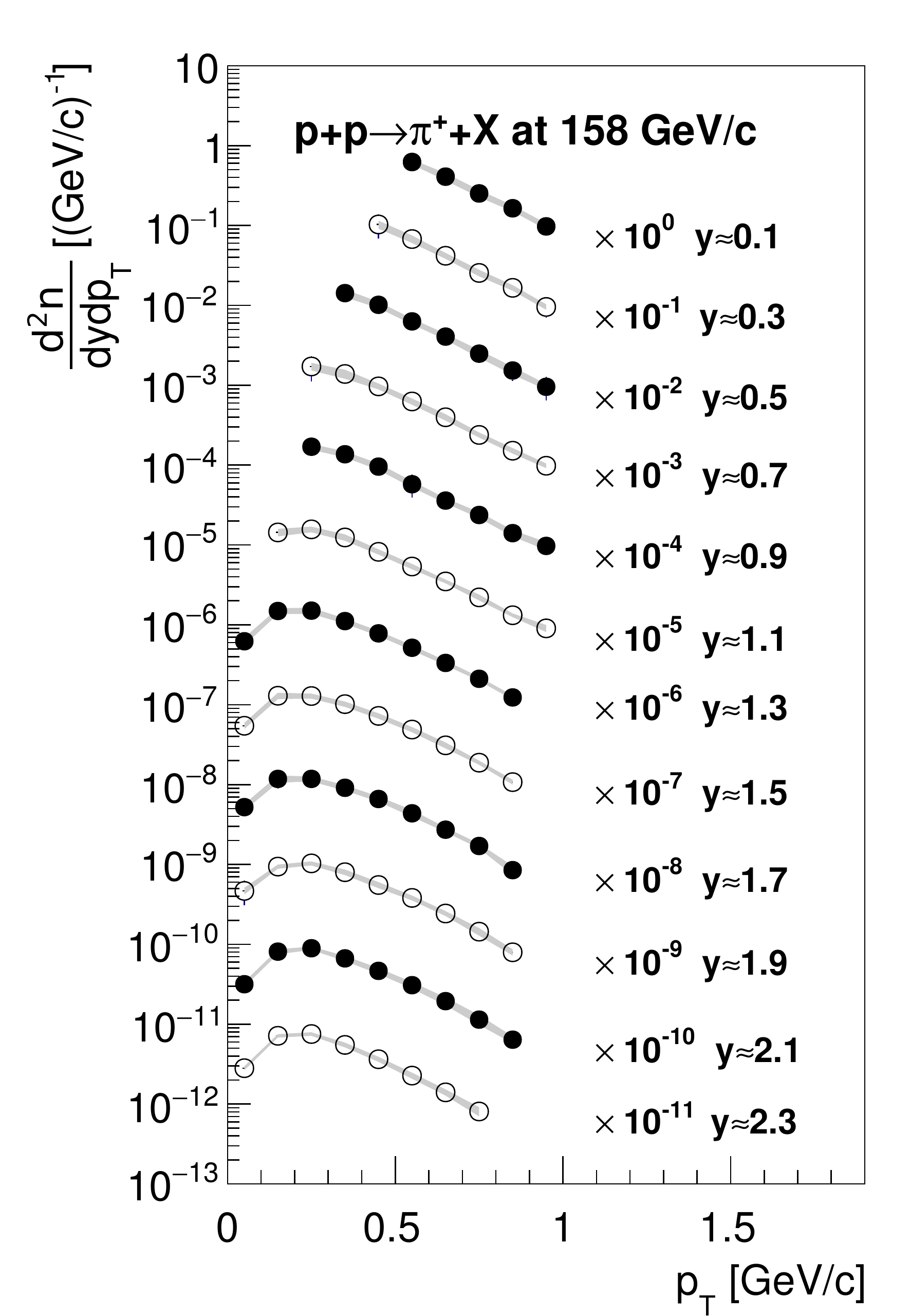}
		\end{center}
		\caption{(Color online) Transverse momentum $\pi^{+}$ spectra in rapidity slices produced in inelastic p+p interactions at 20, 31, 40, 80, 158~\GeVc. Rapidity values given in the legends correspond to the middle of the corresponding interval. Shaded bands show systematic uncertainties.}
		\label{fig:pptpion}
\end{figure*}

\begin{figure*}
	 	\begin{center}
	 	\includegraphics[width=0.3\textwidth]{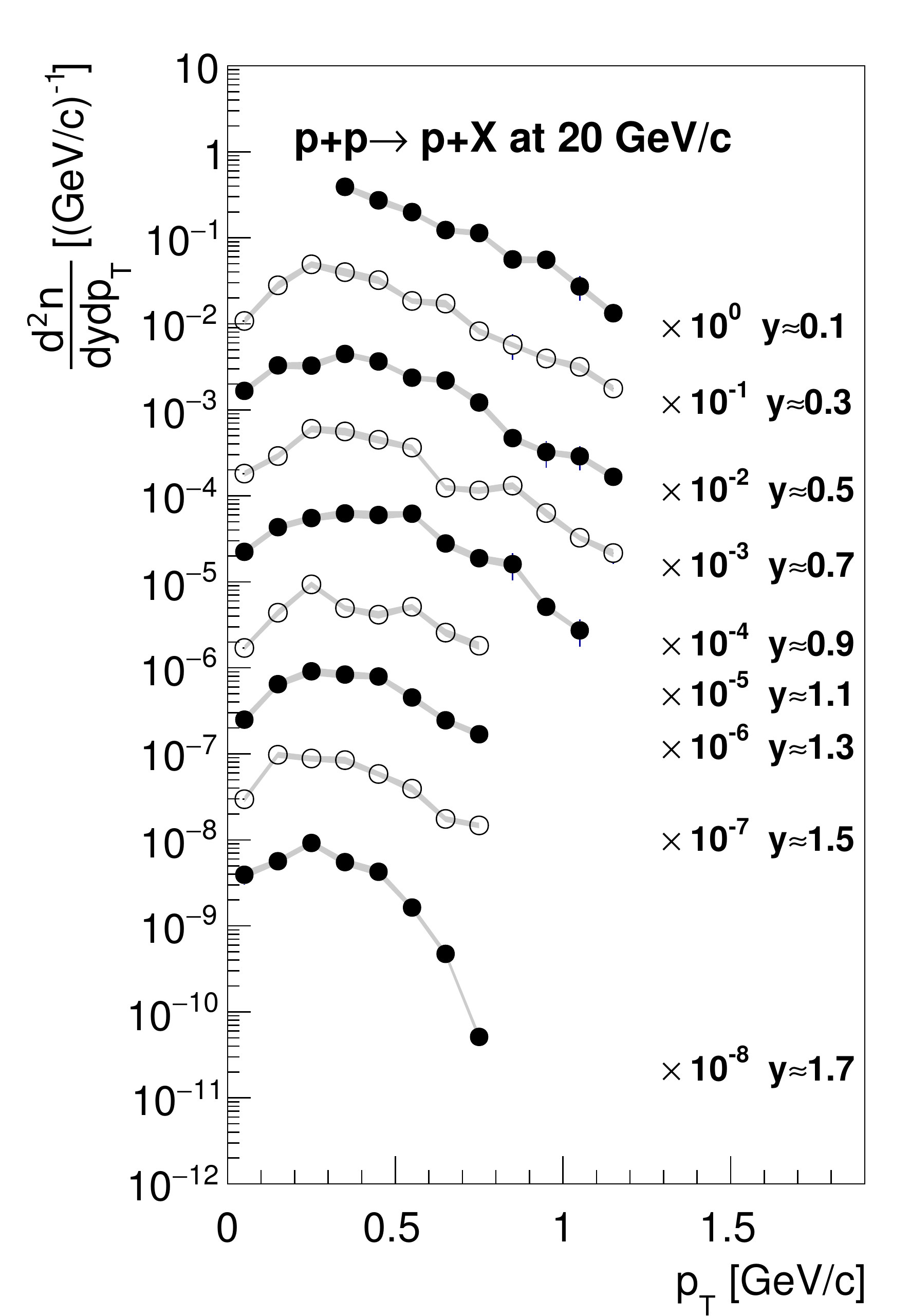}
	 	\includegraphics[width=0.3\textwidth]{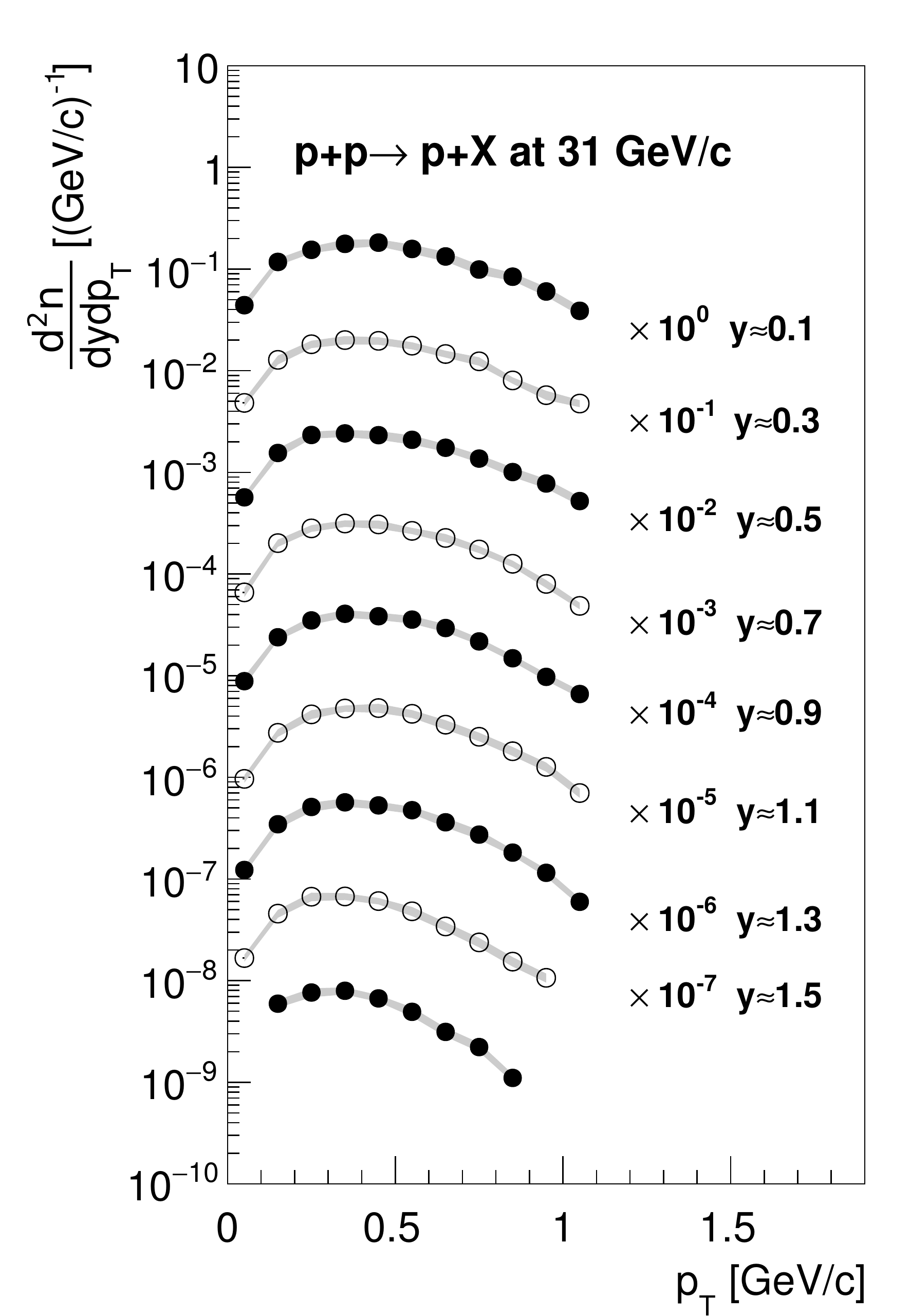}\\
	 	\includegraphics[width=0.3\textwidth]{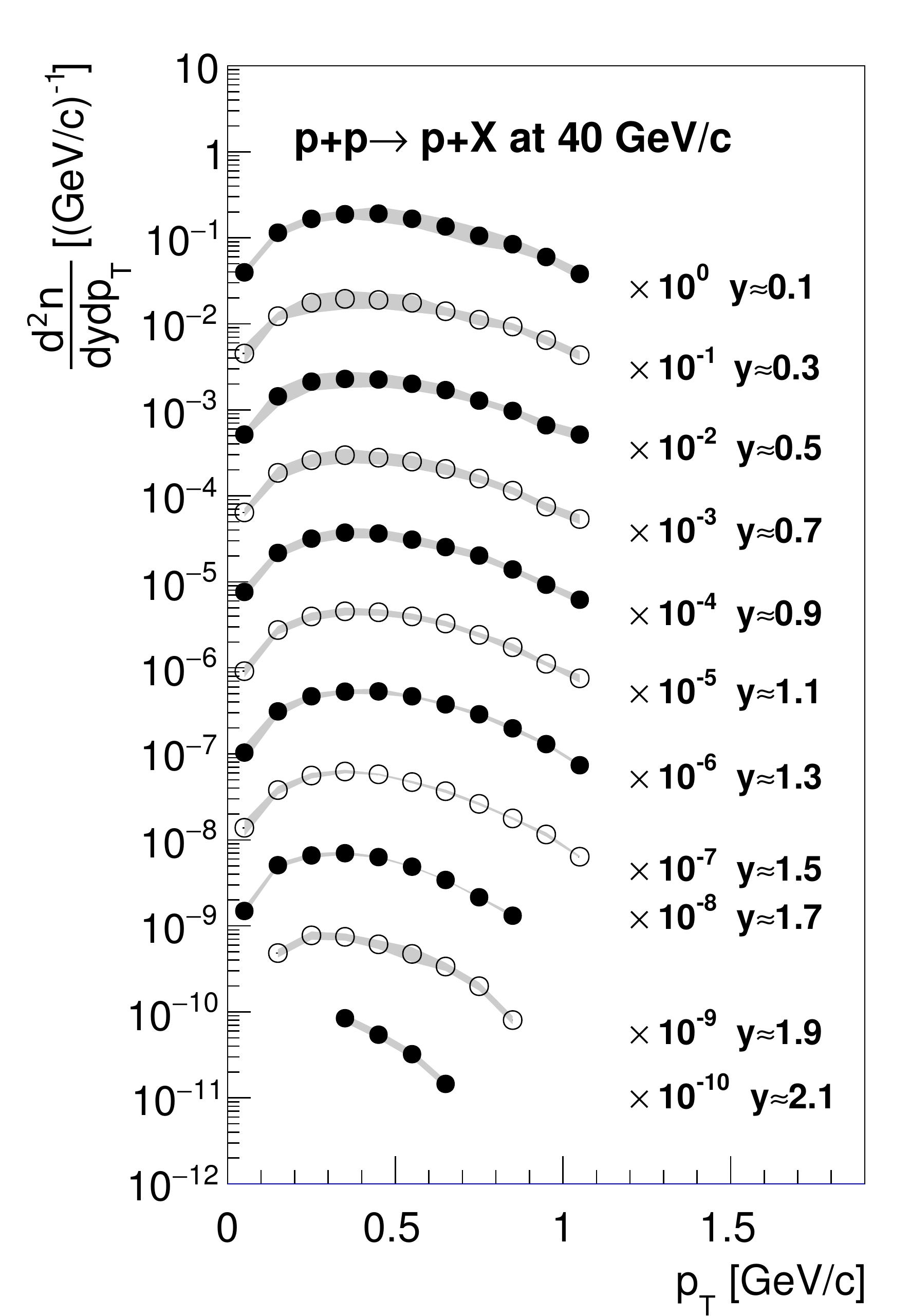}
	 	\includegraphics[width=0.3\textwidth]{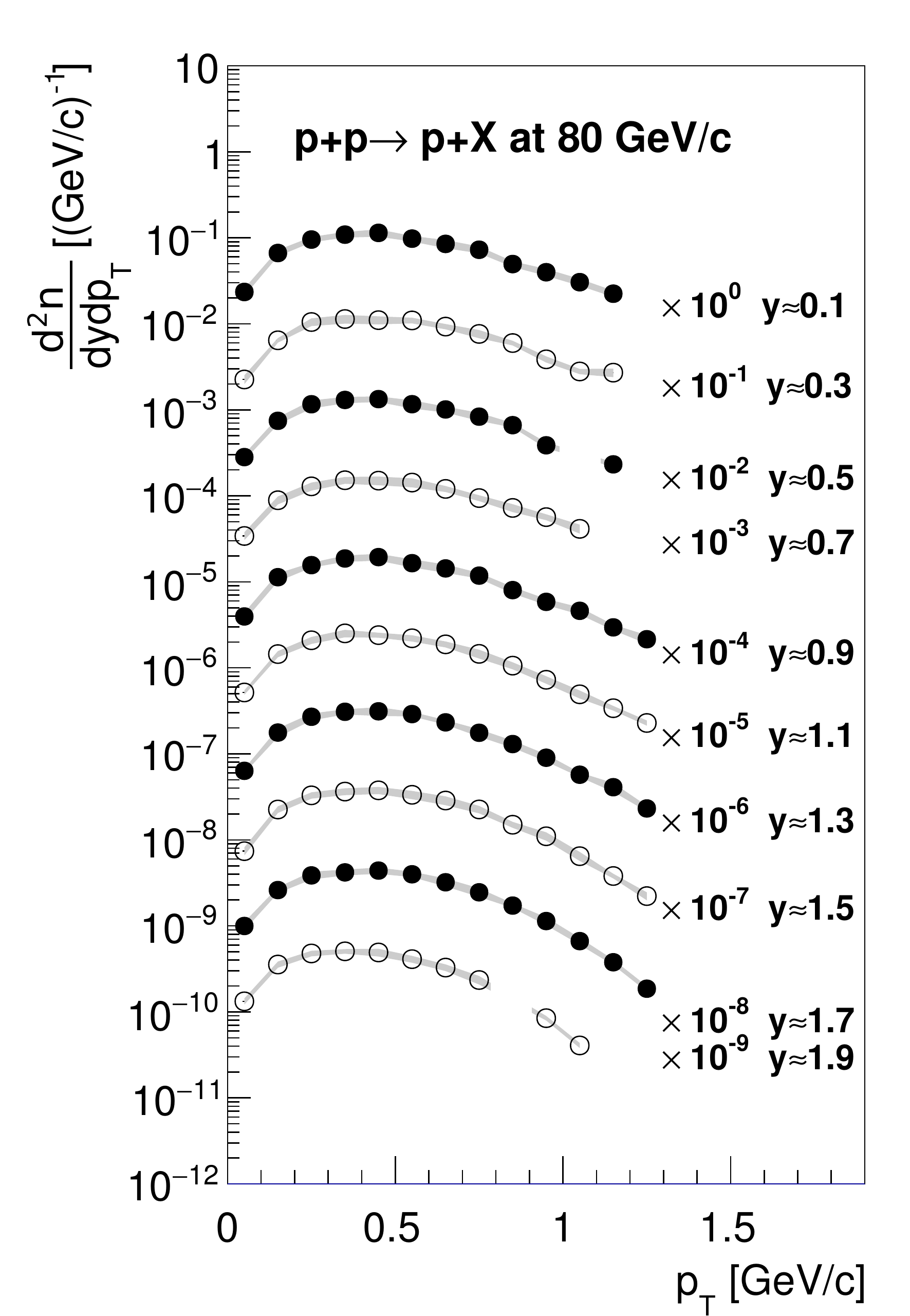}
	 	\includegraphics[width=0.3\textwidth]{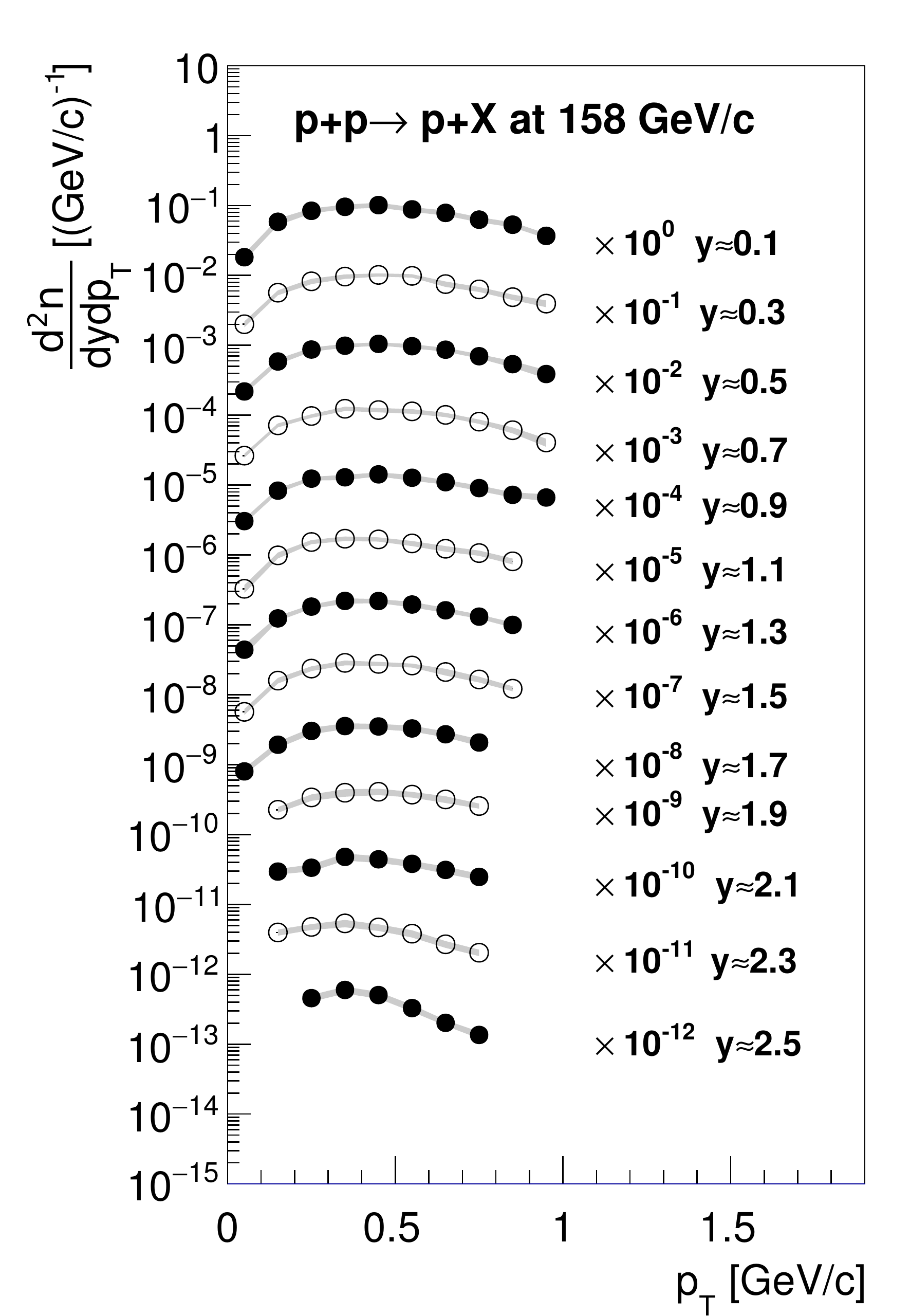}
	 	\end{center}
	 	\caption{(Color online) Transverse momentum p spectra in rapidity slices produced in inelastic p+p interactions at 20, 31, 40, 80, 158~\GeVc. Rapidity values given in the legends correspond to the middle of the corresponding interval. Shaded bands show systematic uncertainties.}
	 	\label{fig:ppy}
	 \end{figure*}

\begin{figure*}
	 	\begin{center}
	 	\includegraphics[width=0.3\textwidth]{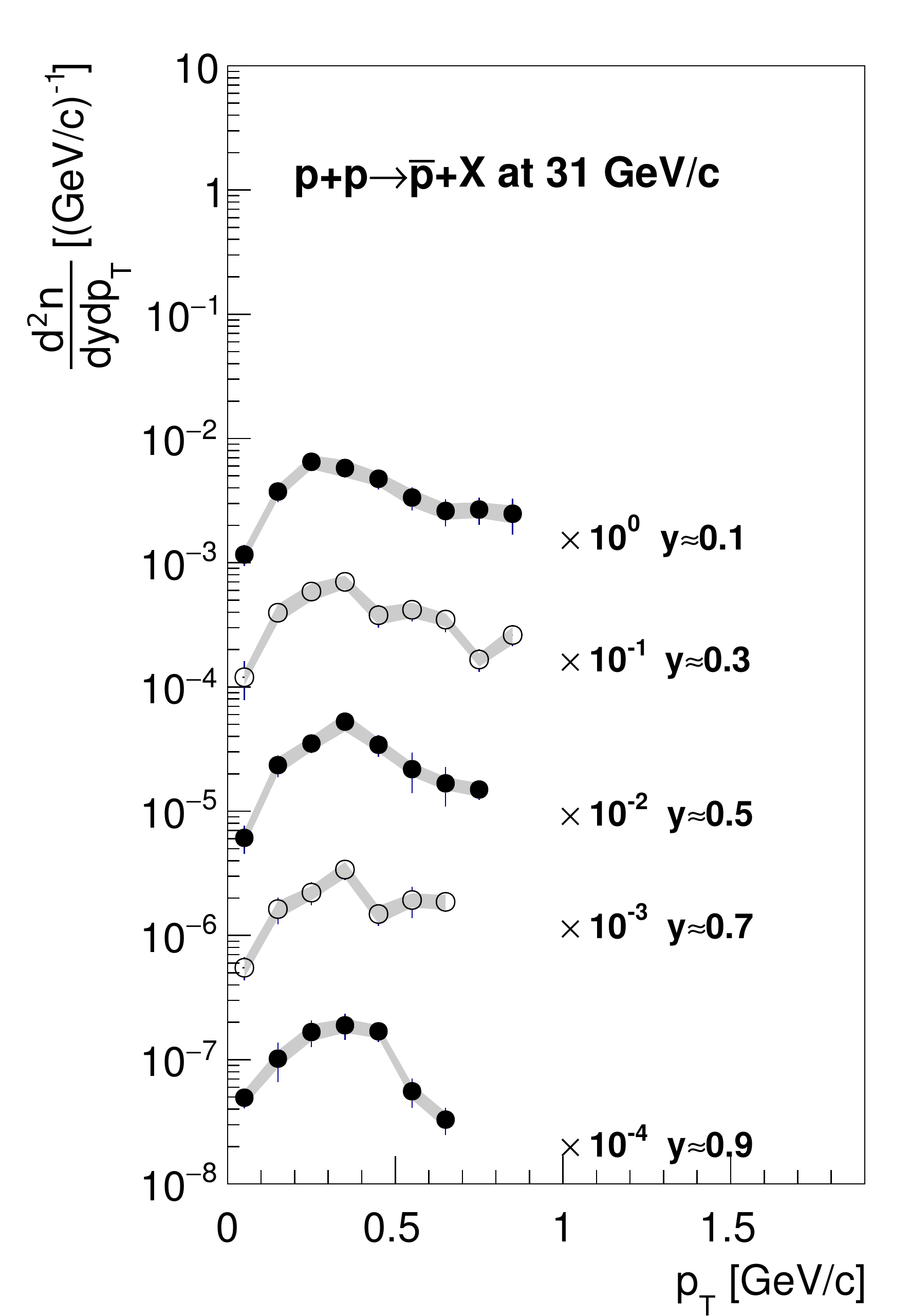}
	 	\includegraphics[width=0.3\textwidth]{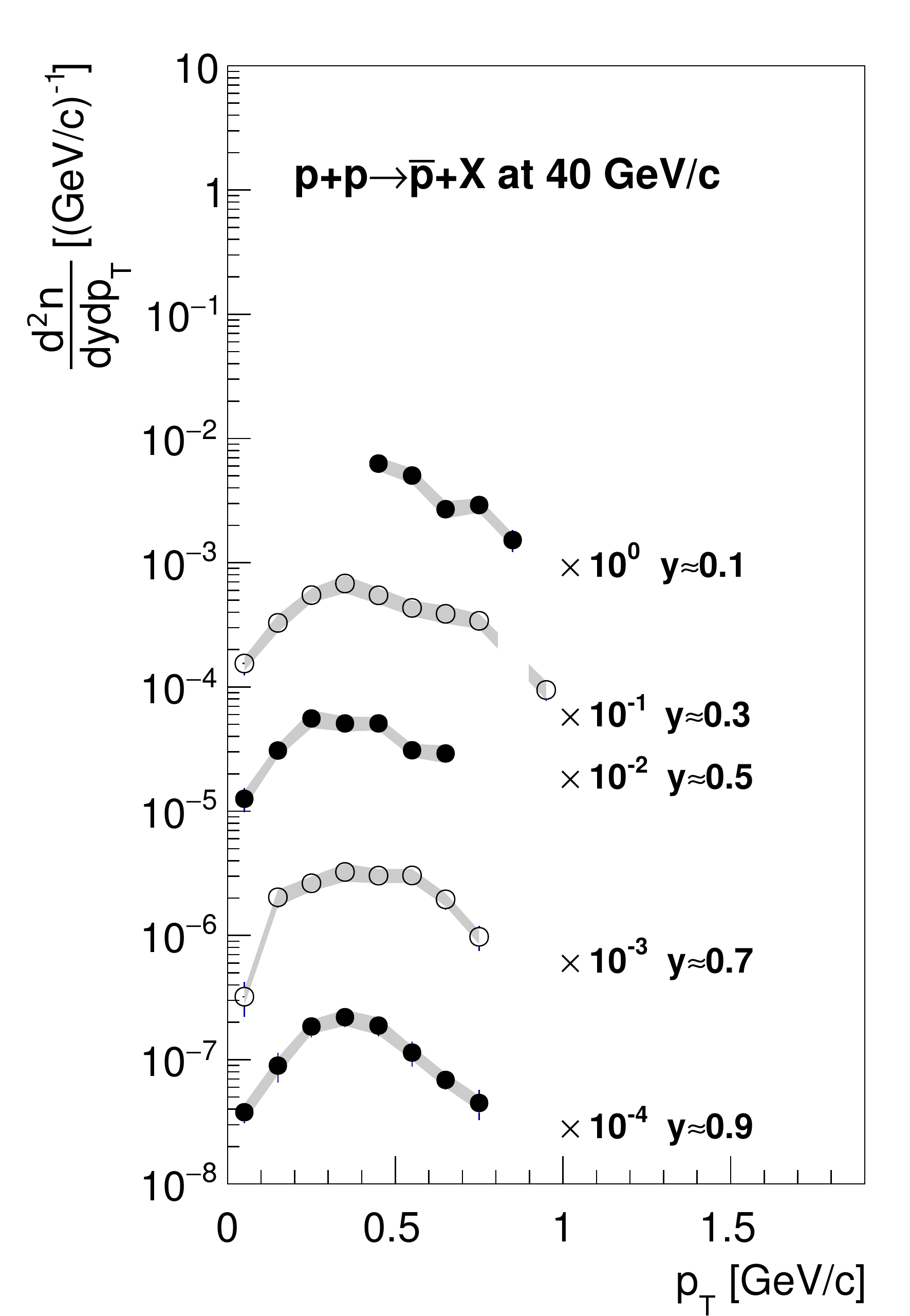}\\
	 	\includegraphics[width=0.3\textwidth]{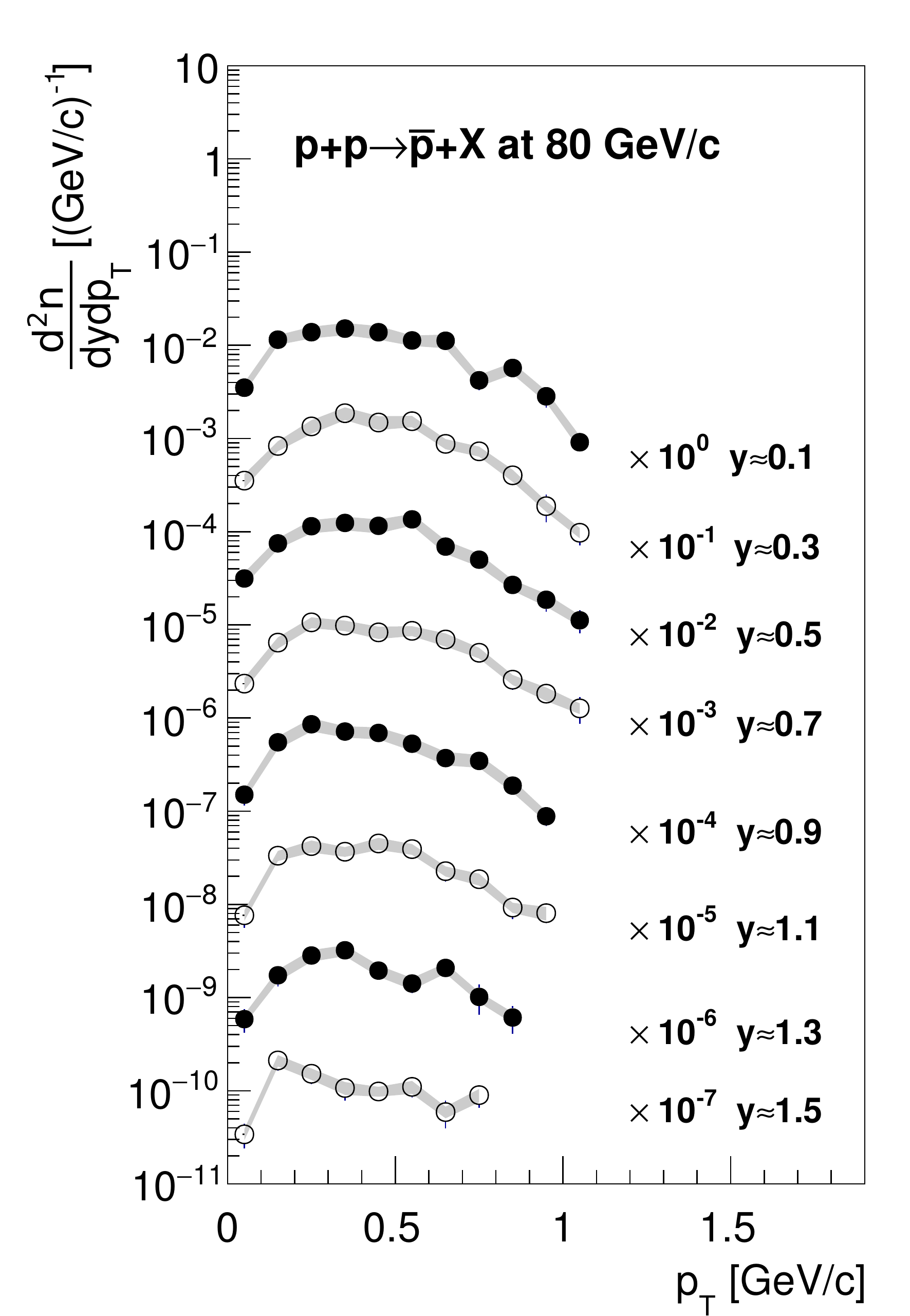}
	 	\includegraphics[width=0.3\textwidth]{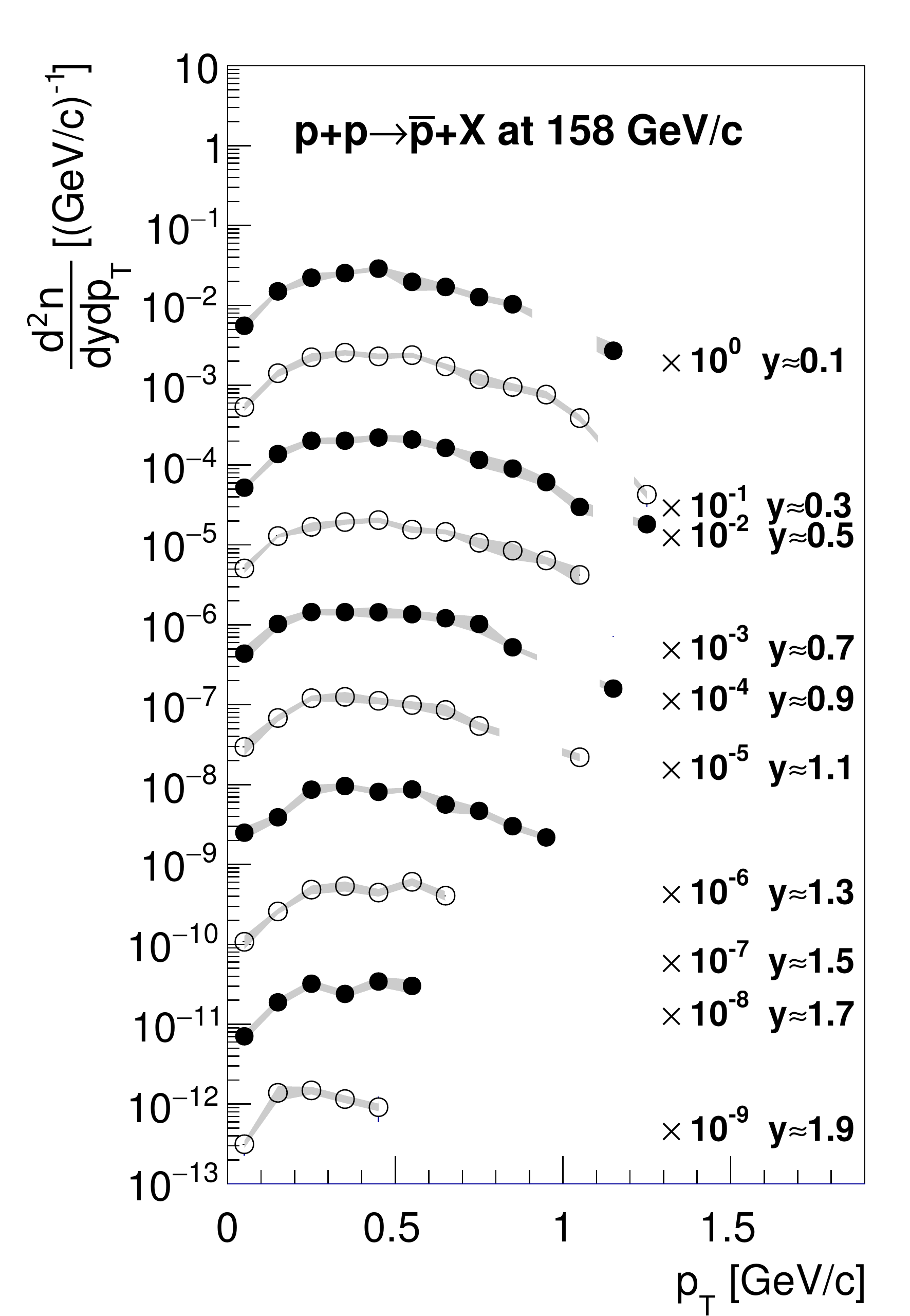}
	 	\end{center}
	 	\caption{(Color online) Transverse momentum $\bar{\textrm{p}}$ spectra in rapidity slices produced in inelastic p+p interactions at 20, 31, 40, 80, 158~\GeVc. Rapidity values given in the legends correspond to the middle of the corresponding interval. Shaded bands show systematic uncertainties.}
	 	\label{fig:appy}
	 \end{figure*}

	The measurements shown in Fig.~\ref{fig:final2D} were studied as a function of transverse momentum ($p_{T}$) in intervals of rapidity ($y$). Resulting double differential spectra of K$^{-}$, K$^{+}$, $\pi^{-}$, $\pi^{+}$, p and $\bar{\textrm{p}}$ produced in p+p interactions at 20, 31, 40, 80, 158~\GeVc are plotted in Figs.~\ref{fig:nptkaon}, \ref{fig:pptkaon}, \ref{fig:nptpion}, \ref{fig:pptpion},  \ref{fig:ppy} and \ref{fig:appy}, respectively. Spectra in successive rapidity intervals were scaled by appropriate factors for better visibility. Vertical bars on data points correspond to statistical, shaded bands to systematic uncertainties.
	
\clearpage

\begin{figure}[!ht]
		\begin{center}
		\includegraphics[width=0.4\textwidth]{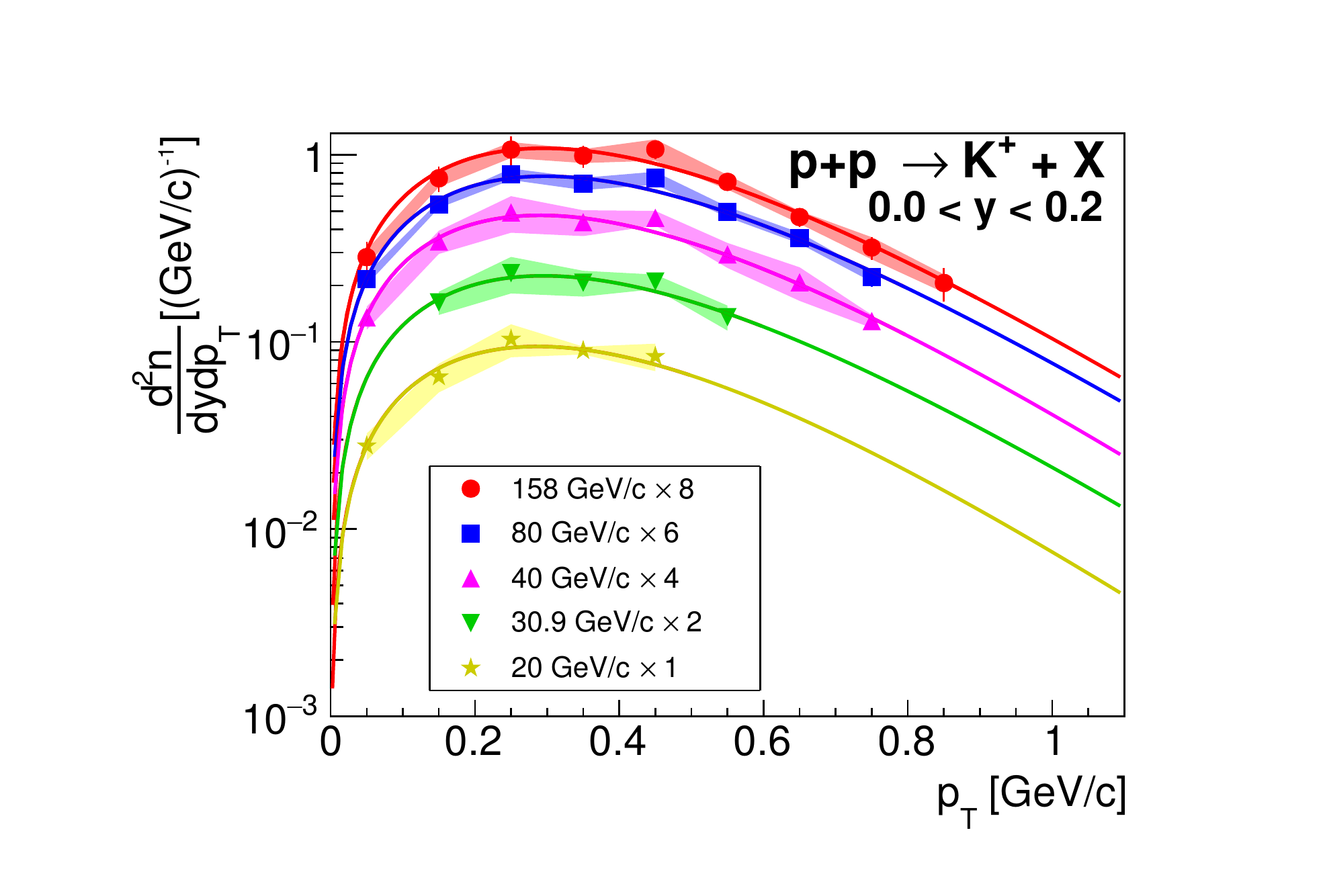}
		\includegraphics[width=0.4\textwidth]{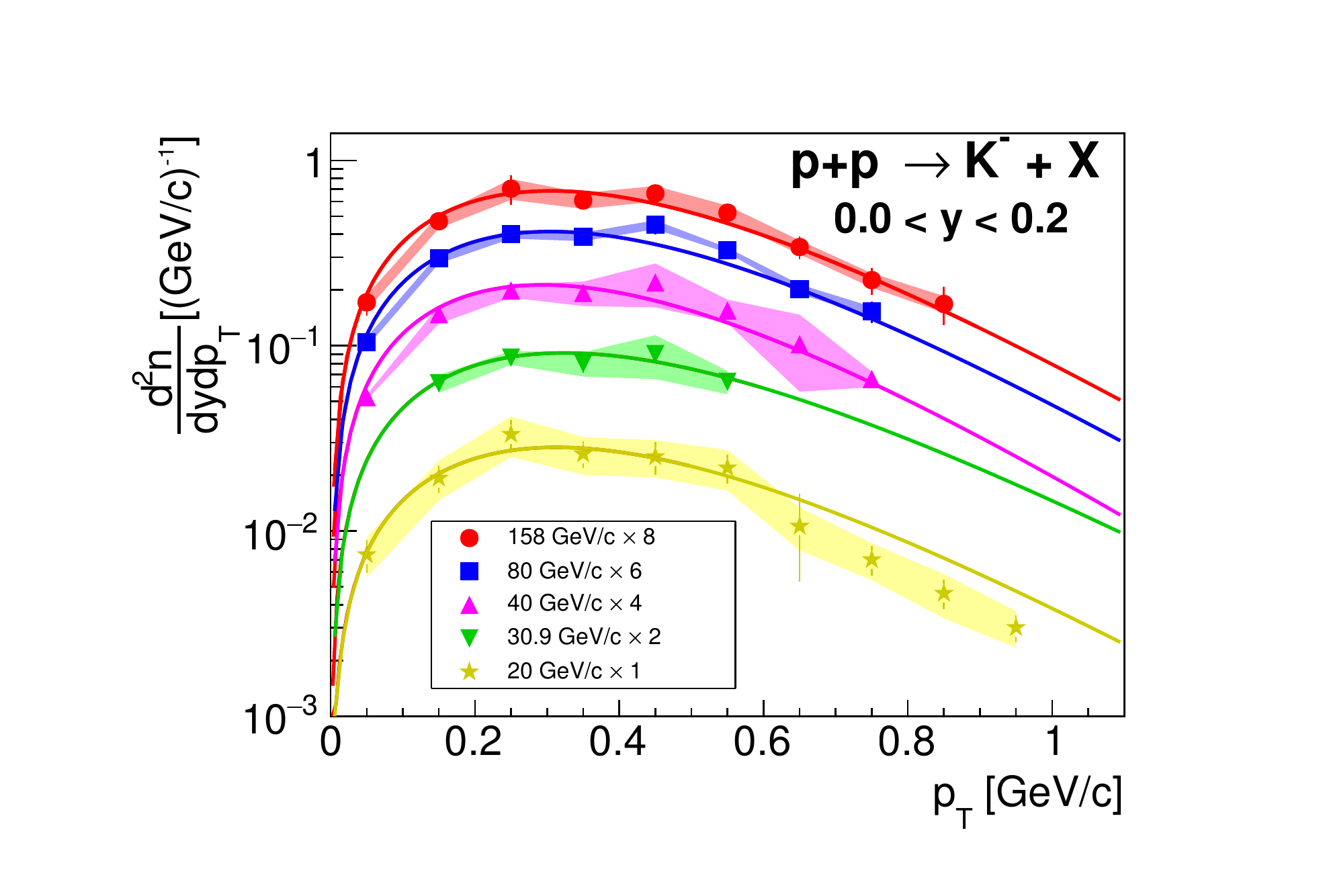}
		\end{center}
		\caption{(Color online) Transverse momentum spectra of $K^{+}$ and $K^{-}$ mesons produced at $y\approx 0$ in inelastic $p+p$ interactions. Colored lines represent the fitted function (Eq.~{\ref{eq:inverse}}).}
		\label{fig:kaonsptfinal}
	\end{figure}

\begin{figure}[!ht]
		\begin{center}
\newcolumntype{S}{>{\centering} m{0.03\textwidth} }
\newcolumntype{A}{>{\centering} m{0.22\textwidth} }
\includegraphics[width=0.91\textwidth]{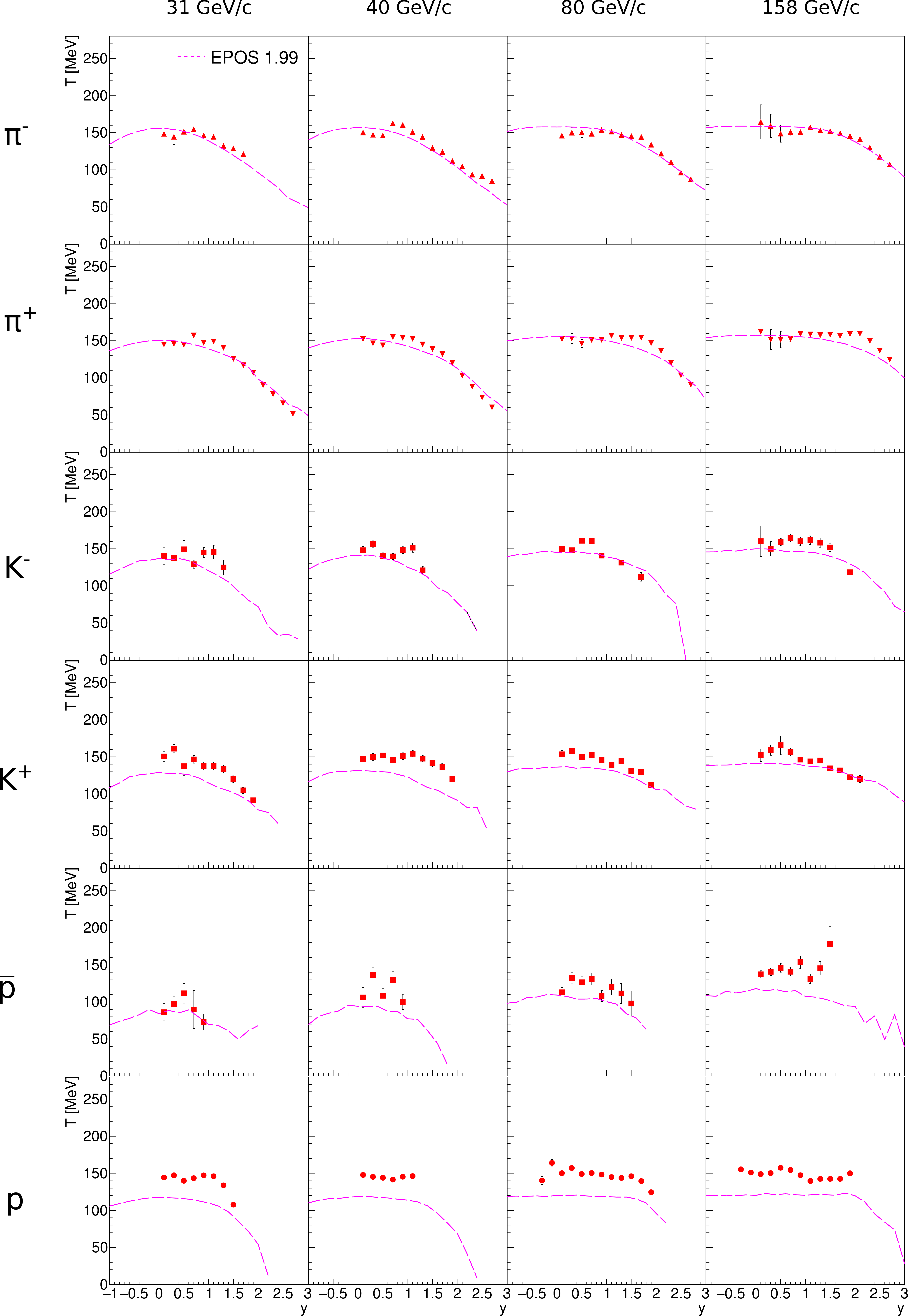} 

		\end{center}
		\caption{(Color online) Inverse slope parameter T obtained from fits with Eq.~\ref{eq:inverse} to the $p_{T}$ distributions of K$^{+}$, K$^{-}$, $\pi^{+}$, $\pi^{-}$, p and $\bar{\textrm{p}}$ as function of rapidity in inelastic p+p collisions at 30, 40, 80 and 158~\GeVc. Results are compared with \Epos model~\cite{Werner:2008zza} predictions fitted in the same range. Bars show statistical uncertainties. Systematic uncertainties of about 10\% are not shown.}
		\label{fig:Tasy}
	\end{figure}

	Results presented in Figs.~\ref{fig:nptkaon}, \ref{fig:pptkaon}, \ref{fig:nptpion}, \ref{fig:pptpion},  \ref{fig:ppy} and \ref{fig:appy} were parametrized by the exponential function~\cite{Hagedorn:1968jf,Broniowski:2004yh}:
\begin{equation}
 \frac{d^{2}n}{dp_{T}dy}=\frac{S~c^{2}p_{T}}{T^{2} + m~T}\exp(-(\mt - m)/T),
\label{eq:inverse}
\end{equation}
where $m$ is the particle mass and  $S$ and $T$ are the yield integral and  the inverse slope parameter, respectively. Examples of this parametrization fitted to K$^{+}$ and K$^{-}$ spectra at mid-rapidity are presented in Fig.~\ref{fig:kaonsptfinal}. The obtained values of the inverse slope parameter for all spectra shown in Figs.~\ref{fig:nptkaon}-\ref{fig:appy} are plotted in Fig.~\ref{fig:Tasy} as function of the rapidity for four beam momenta 31, 40, 80 and 158~\GeVc. Note that results are only plotted for those rapidity intervals for which there were more than 6 data points in the $p_{\textrm{T}}$-distribution. 

\begin{figure}[!ht]
                \begin{center}
                \includegraphics[width=0.4\textwidth]{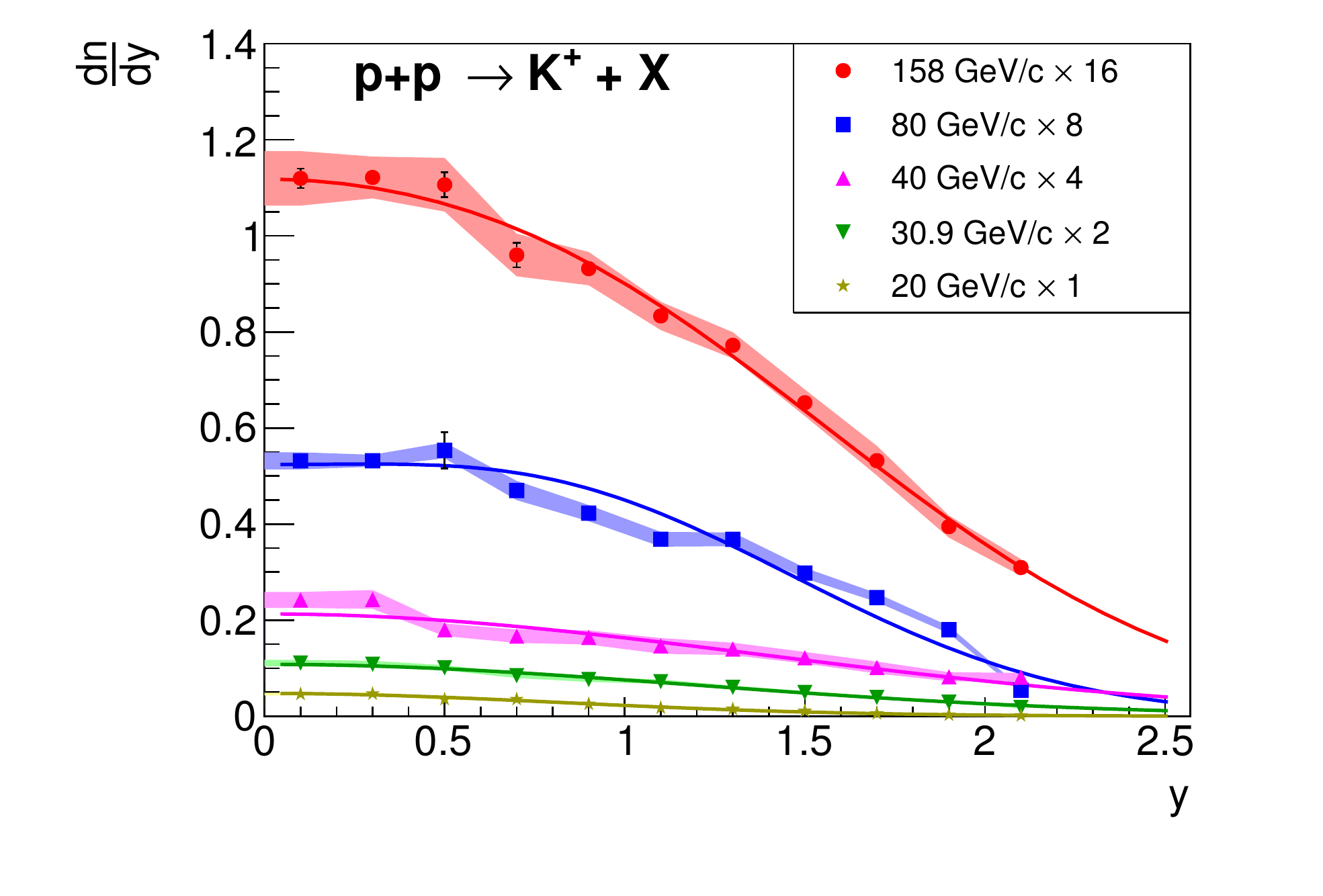}
                \includegraphics[width=0.4\textwidth]{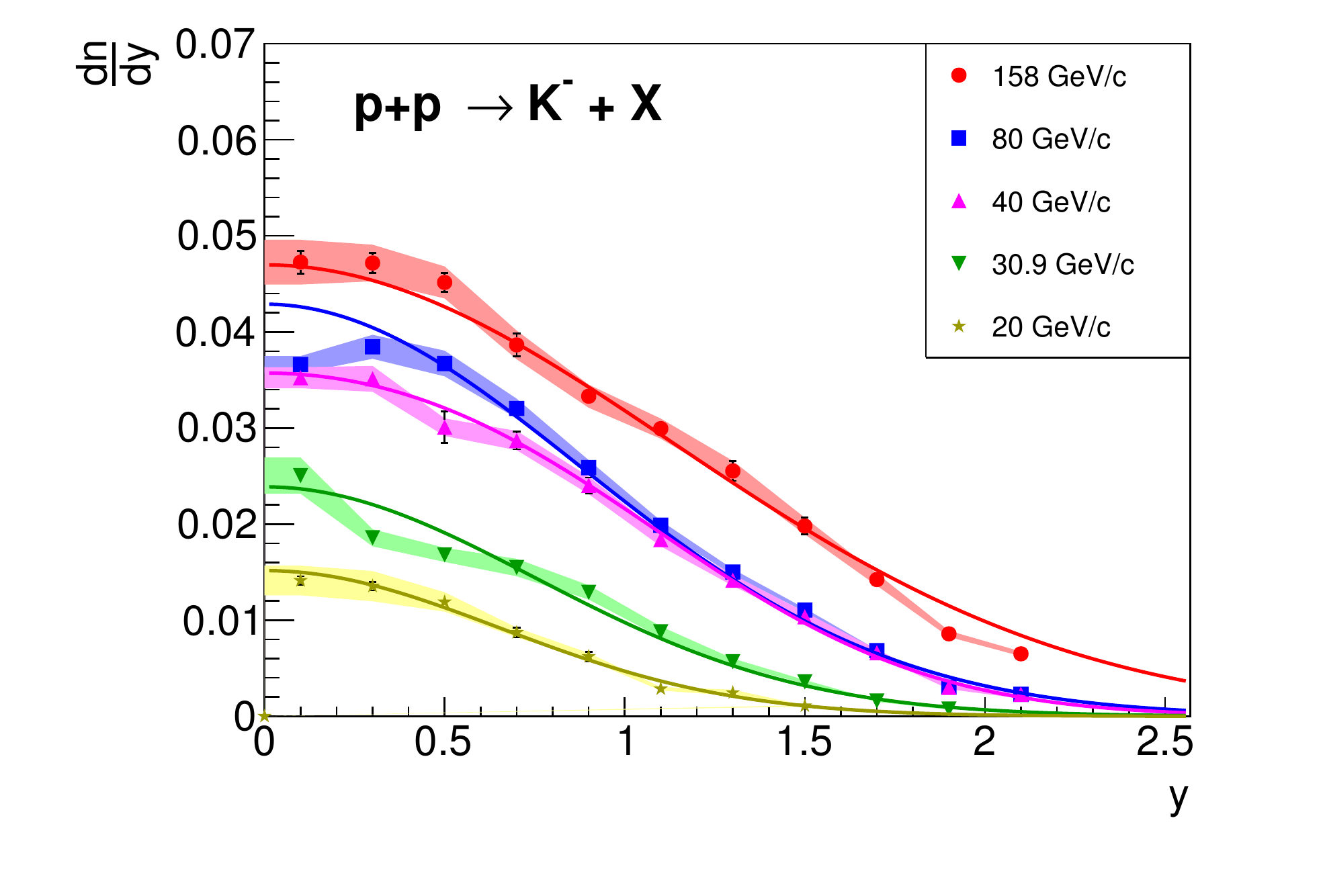}\\
                \includegraphics[width=0.4\textwidth]{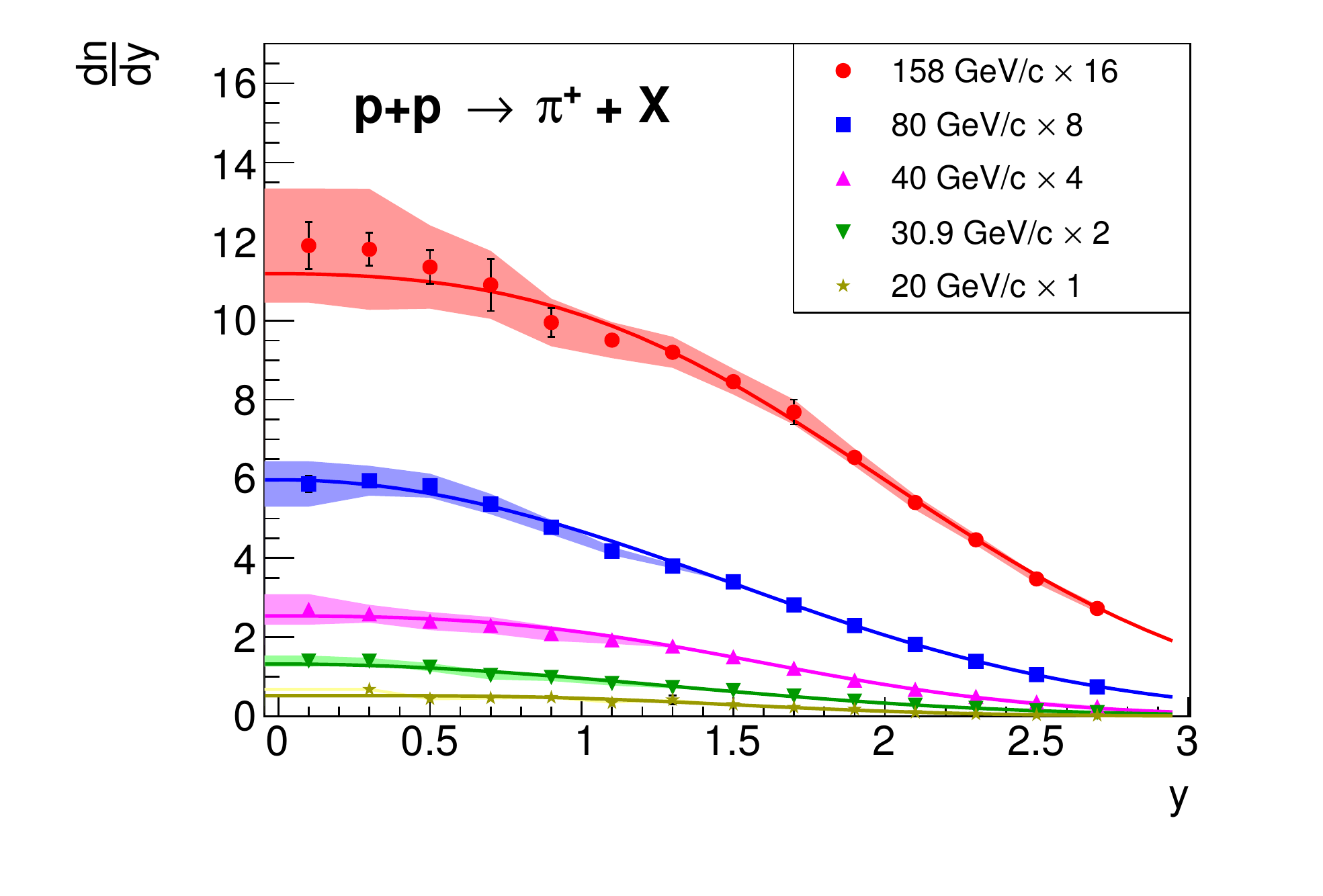}
                \includegraphics[width=0.4\textwidth]{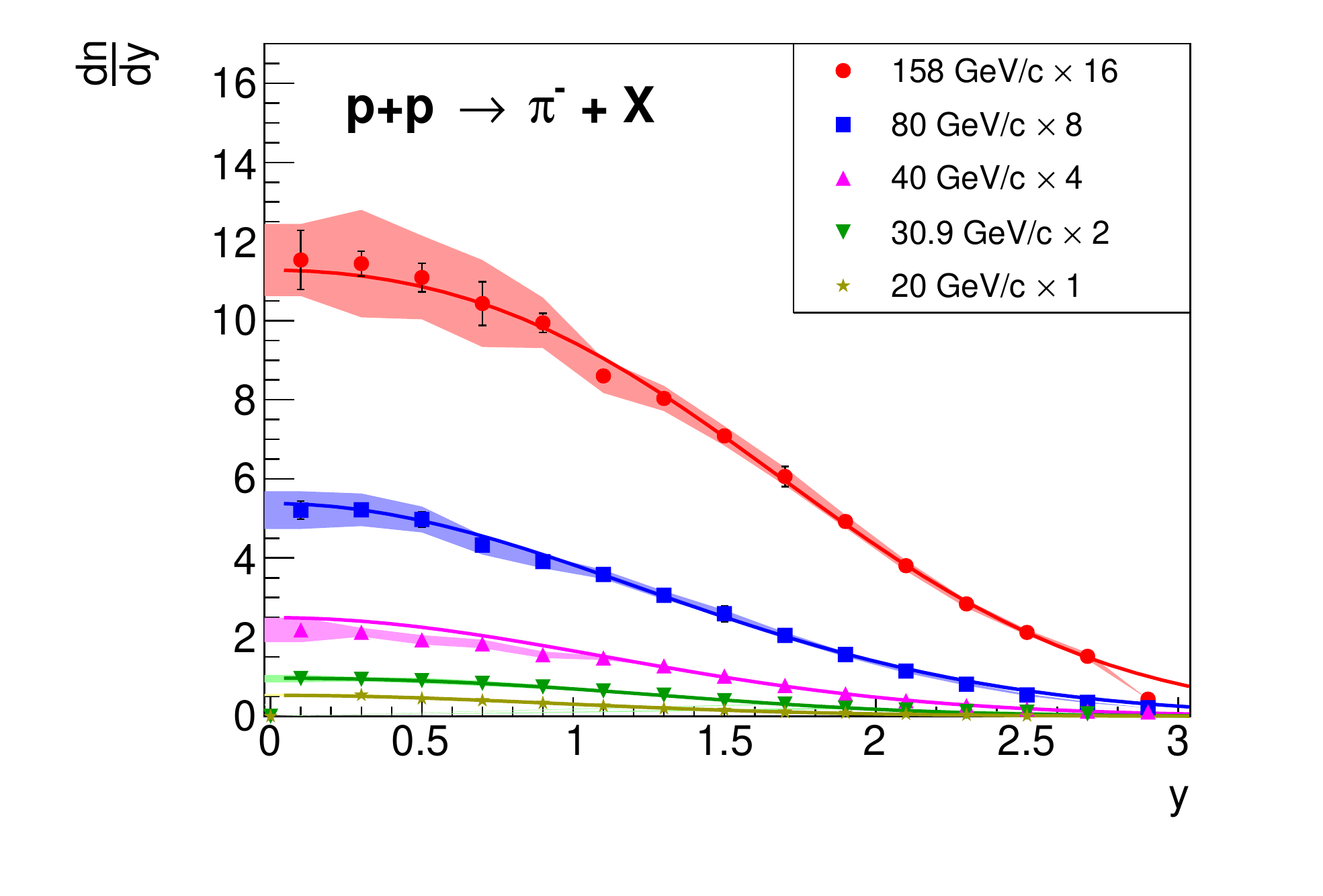}\\
                \includegraphics[width=0.4\textwidth]{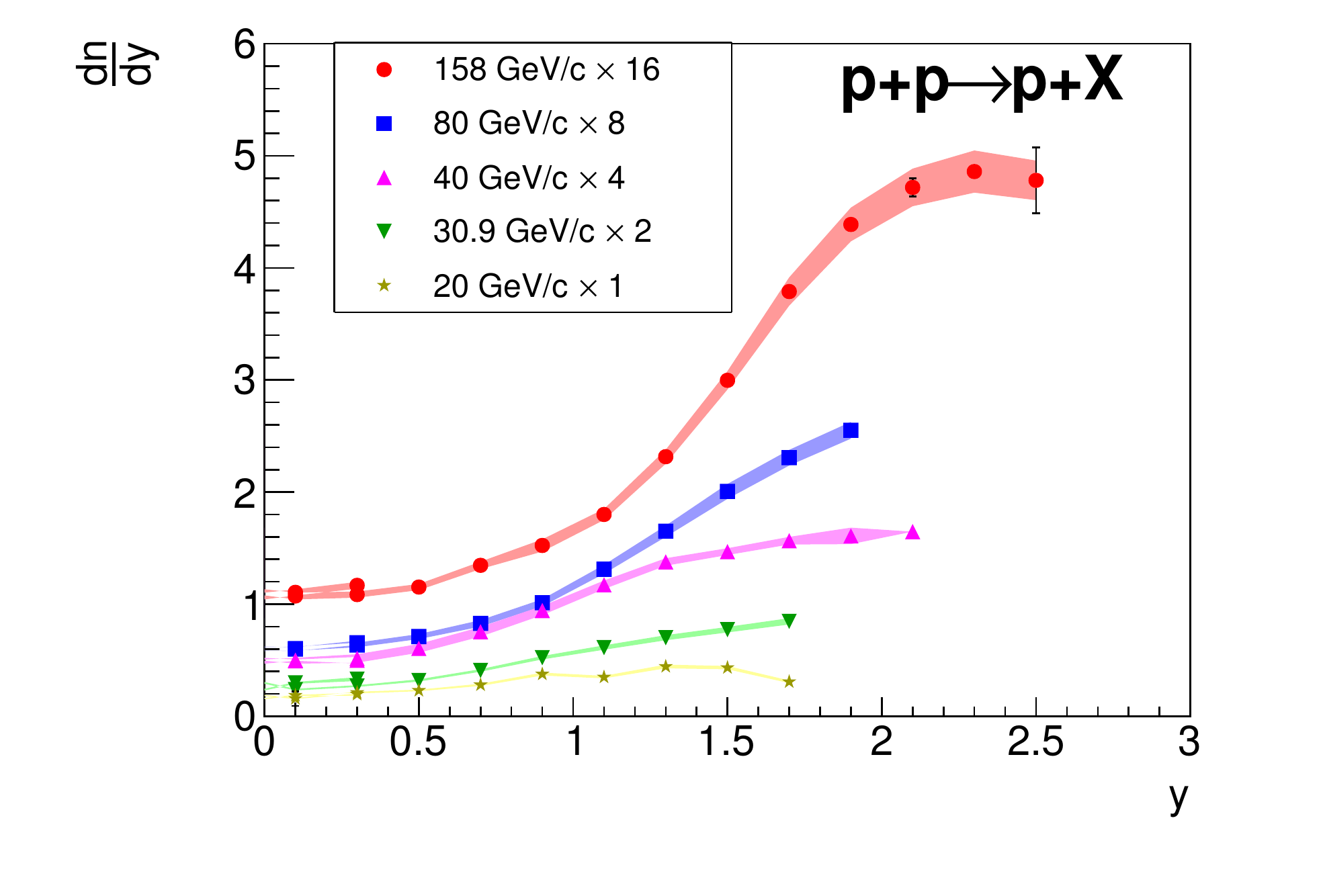}
                \includegraphics[width=0.4\textwidth]{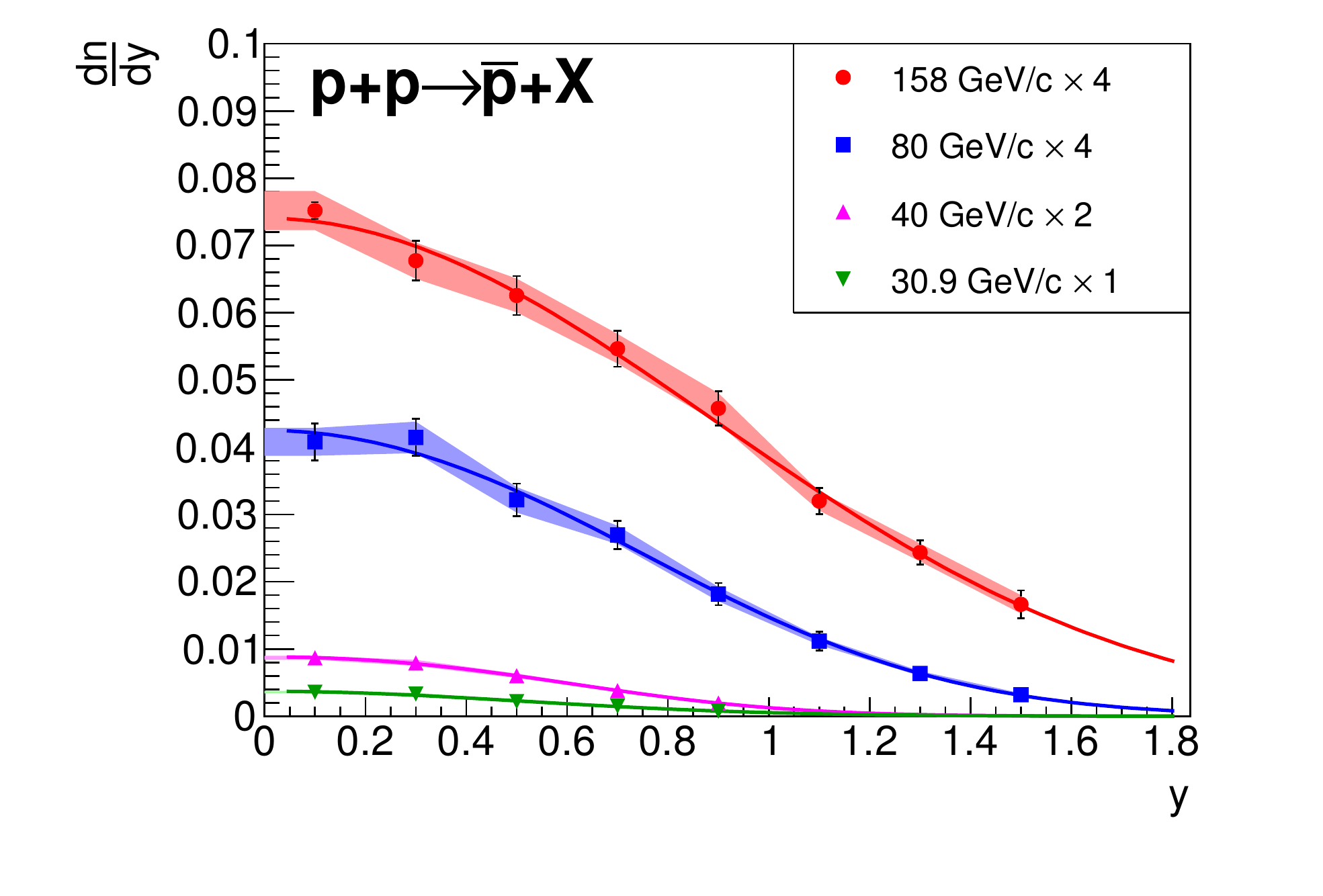}
                \end{center}
                \caption{(Color online) Rapidity spectra of K$^{+}$, K$^{-}$, $\pi^{+}$, $\pi^{-}$, p and $\bar{\textrm{p}}$ produced in inelastic p+p interactions at SPS energies, scaled by appropriate factors for better visibility. Vertical bars indicate statistical, shaded bands systematic uncertainties of the measurements. Curves depict Gaussian fits used to determine total multiplicities.}
                \label{fig:finaldndy}
        \end{figure}

	Rapidity distributions were then obtained by integrating the transverse momentum spectra also using reflection symmetry around mid-rapidity ($y=0$). Extrapolation to unmeasured regions in $\pt$ was performed in rapidity intervals using fits with function Eq.~\ref{eq:inverse} where results were reliable (number of measured \pt points higher than 6). In the other rapidity intervals the yield was obtained by summing the measured values and multiplying the result by an extrapolation factor calculated from the \Epos model. This factor was taken as the ratio of the integrated yield in the rapdity interval to that in the region of the measurements. 
The results are shown in Fig.~\ref{fig:finaldndy}. Statistical uncertainties, indicated by vertical bars, were calculated as the square root of the sum of the squares of statistical errors of the summed bins. Systematic uncertainties (shaded bands) were calculated by varying the uncertainty sources described in section~\ref{sec:systematics} and adding half of the extrapolated yield.

\FloatBarrier
   \subsection{Mean multiplicities}
\label{sec:multiplicities}		

Next, mean multiplicities produced in the forward region $y > 0$ were calculated by integrating the rapidity distributions shown in Fig.~\ref{fig:finaldndy}. The distributions are seen to be nearly Gaussian at all energies except for protons. In order to obtain a good description of the data points, fits were performed with a sum of two identical Gaussian functions with mean position symmetrically displaced around mid-rapidity. The integrated result was taken as the sum of measured values plus a contribution from the unmeasured region obtained from the fit function. The statistical uncertainty is obtained as the square root of the sum of squares of statistical uncertainties of the measured data points and the square of the statistical error of the extrapolation. The systematic uncertainty was estimated by repeating the complete analysis procedure by varying the components described in section~\ref{sec:systematics}.
Doubling of the results gives the $4\pi$ multiplicities which are listed in Table~\ref{tab:meanmultp} for $\pi^+$, K$^+$ and p and in Table~\ref{tab:meanmultn} for $\pi^-$, K$^-$ and $\bar{\textrm{p}}$.

\begin{figure}
		\begin{center}
		\includegraphics[width=0.4\textwidth]{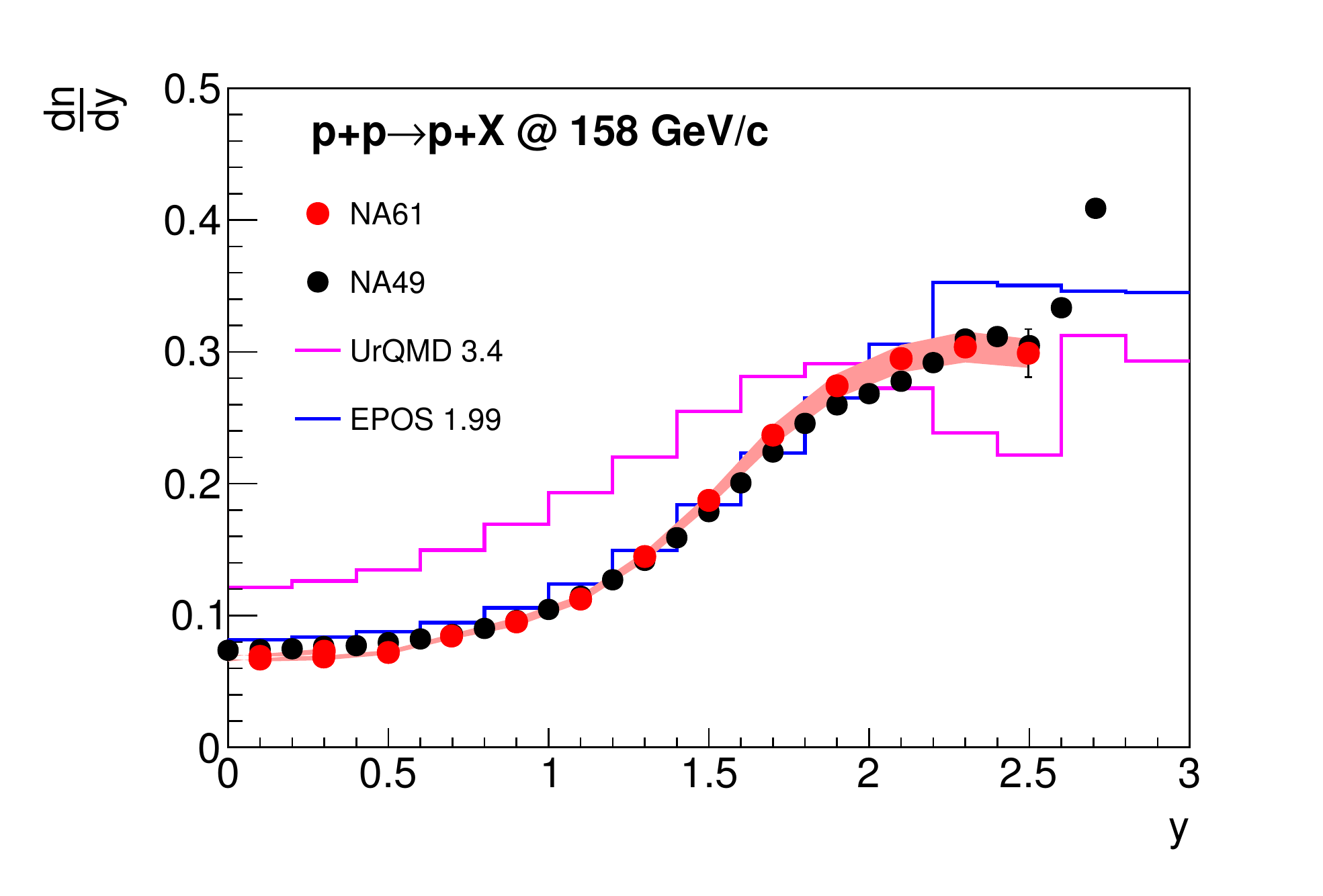}
		\end{center}
		\caption{(Color online) Proton rapidity distribution in inelastic p+p interactions at 158~\GeVc compared with NA49 measurements~\cite{Baatar:2012fua} and predictions of the  \Urqmd~\cite{Bass:1998ca,Bleicher:1999xi} and \Epos~\cite{Werner:2008zza} models.}
		\label{fig:protonmodel}
	\end{figure}

The determination of $4\pi$ multiplicity is more complicated for protons due to the rapid rise of the yield towards beam rapidity and the lack of measurments in this region. A comparison of the rapidity distributions obtained in this analysis with measurements of the NA49 experiment~\cite{Baatar:2012fua} and calculations using the  \Urqmd~\cite{Bass:1998ca,Bleicher:1999xi} and \Epos~\cite{Werner:2008zza} models at 158~\GeVc beam momentum is shown in Fig.~\ref{fig:protonmodel}. One observes that the models do not describe well the measurements at highest rapidities at this beam energy. Nevertheless, the models were used to extrapolate the proton yields into the unmeasured region since they provide predictions at all beam energies. Extrapolation factors were calculated from the models as the ratio of total multiplicity to that in the region covered by the measurements. The average of the two model results was then used for extrapolating the measured yield to $4\pi$ phase space. The difference between results obtained using the \Urqmd and \Epos models was added to the
systematic uncertainty. Final results for protons are given in Table~\ref{tab:meanmultp}.

	\begin{table}
	\caption{Mean multiplicity of positively charged  particles at SPS energies with statistical and systematic uncertainties.}
	\begin{center}
	\begin{tabular}{ c || c | c | c}
	
    & K$^{+}$ & $\pi^{+}$ & p\\
	\hline
	\hline
	\tiny20~\GeVc   & $ 0.097\pm 0.014\pm 0.006$ &  $ 1.884\pm 0.012\pm 0.20$ & $1.069\pm 0.010\pm 0.13$\\
	\tiny31~\GeVc   & $ 0.157\pm 0.010\pm 0.015$  & $2.082\pm 0.021\pm 0.20$  & $0.977\pm 0.003\pm 0.14$\\
	\tiny40~\GeVc   & $ 0.170\pm 0.009\pm 0.023$  & $ 2.390\pm 0.022\pm 0.16$ & $1.095\pm 0.003\pm 0.09$\\
	\tiny80~\GeVc   & $ 0.201\pm 0.010\pm 0.010$  & $ 2.671\pm 0.022\pm 0.14$ & $1.093\pm 0.004\pm 0.07$\\
	\tiny158~\GeVc  & $ 0.234\pm 0.014\pm 0.017$  & $ 3.110\pm 0.030\pm 0.26$ & $1.154\pm 0.010\pm 0.04$\\
	\hline
	\end{tabular}
	\end{center}
	\label{tab:meanmultp}
	\end{table}	

	\begin{table}
	\caption{Mean multiplicity of negatively charged particles at SPS energies with statistical and systematic uncertainties.}
	\begin{center}
	\begin{tabular}{ c || c | c | c}
    & K$^{-}$ & $\pi^{-}$& $\bar{\textrm{p}}$\\
	\hline
	\hline
	\tiny20~\GeVc   & $ 0.024\pm 0.006\pm 0.002$ &$ 1.082\pm 0.021\pm 0.20$ &   --- \\
	\tiny31~\GeVc   & $ 0.045\pm 0.004\pm 0.003$ & $ 1.474\pm 0.031\pm 0.19$ & $ 0.0047\pm 0.0007\pm 0.0003$  \\
	\tiny40~\GeVc   & $ 0.084\pm 0.006\pm 0.003$ & $ 1.711\pm 0.028\pm 0.17$ & $ 0.0059\pm 0.0006\pm 0.0004$  \\
	\tiny80~\GeVc   & $ 0.095\pm 0.004\pm 0.005$ & $ 2.030\pm 0.031\pm 0.17$ & $ 0.0183\pm 0.0015\pm 0.0010$  \\
	\tiny158~\GeVc  & $ 0.132\pm 0.011\pm 0.009$ & $ 2.404\pm 0.034\pm 0.18$ & $ 0.0402\pm 0.0020\pm 0.0030$  \\
	\hline
	\end{tabular}
	\end{center}
	\label{tab:meanmultn}
	\end{table}

\subsubsection{Energy dependence of mean multiplicities}

\begin{figure}[!ht]
	\begin{center}
	\includegraphics[width=0.46\textwidth]{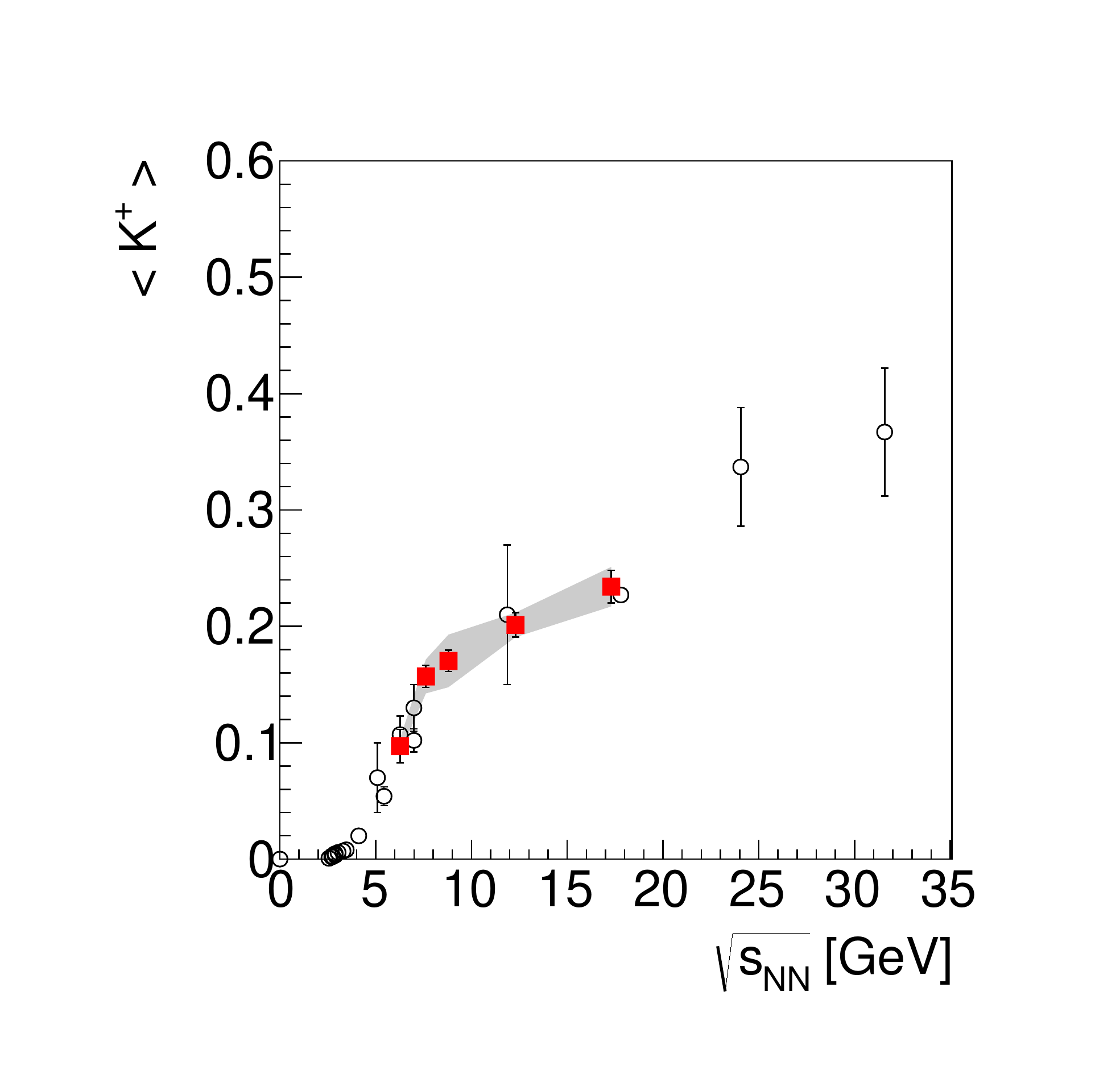}
	\includegraphics[width=0.46\textwidth]{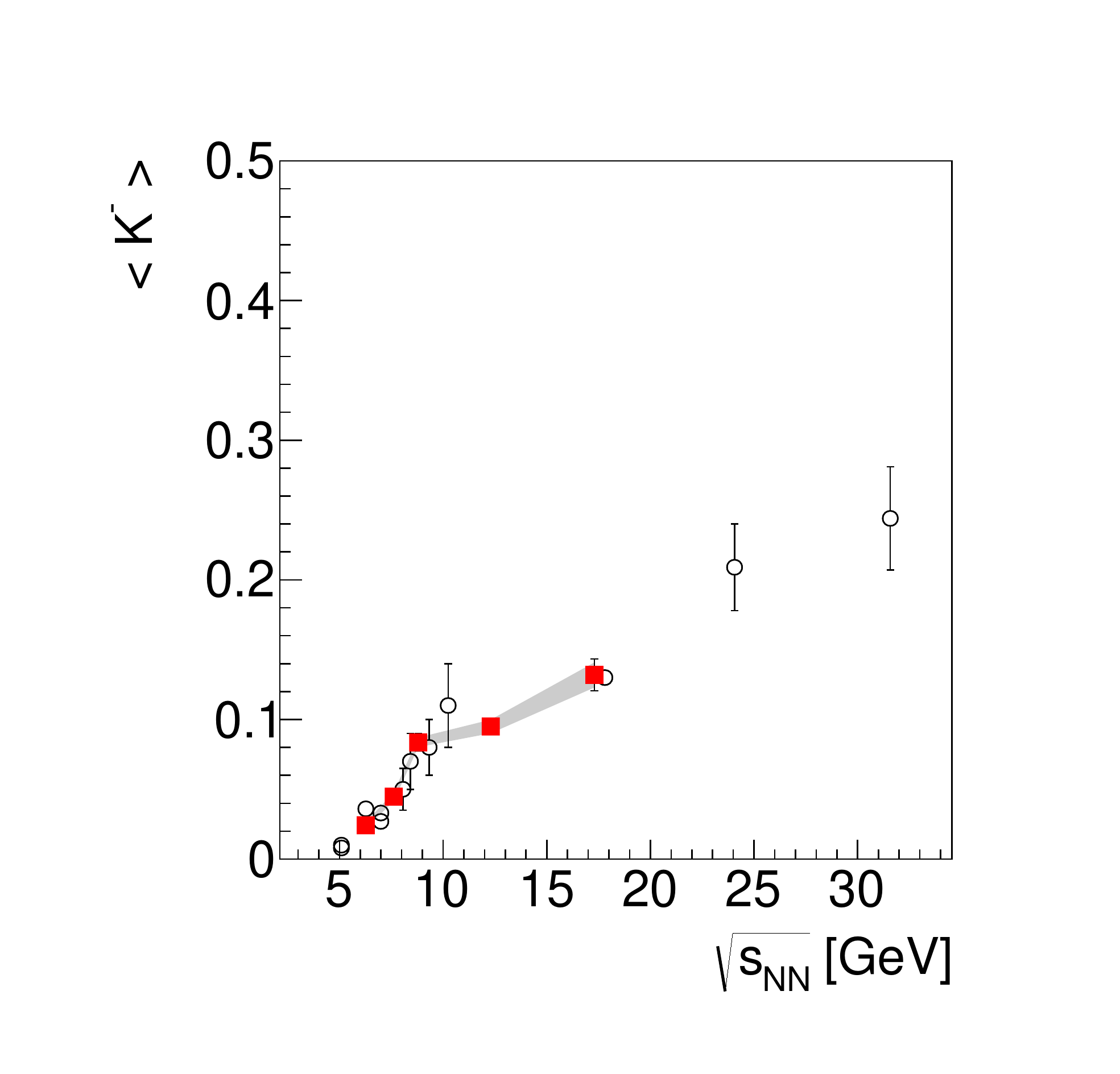}
	\end{center}
	\caption{(Color online) Mean K$^+$ (left) and K$^-$ (right) multiplicity produced in inelastic p+p interactions as function of collision energy. Measurements of \NASixtyOne are plotted as filled red squares, world data~\cite{Anticic:2010yg,Gazdzicki:1996pk,Gazdzicki:1991ih,Bachler:1999hu,Verbeure:1991tv,Rossi:1974if,Golokhvastov:2001ei,Becattini:1997rv,Back:2003xk} are shown by open circles. Statistical uncertainties are indicated by bars, systematic uncertainties of the \NASixtyOne measurements by shaded bands.}
	\label{fig:Kaonmeanmult}
\end{figure}

The energy dependence of total charged kaon multiplicities produced in inelastic p+p collisions is presented in Fig.~\ref{fig:Kaonmeanmult}. One observes a steep threshold rise and a clear flattening above about 8 GeV for both K$^{+}$ and K$^{-}$. The new \NASixtyOne measurements (filled red squares) agree with the world data (open circles) and locate the change of slope due to the high precision of the actual measurement.   

%
 
\begin{figure}[!ht]
	\begin{center}
	\includegraphics[width=0.37\textwidth]{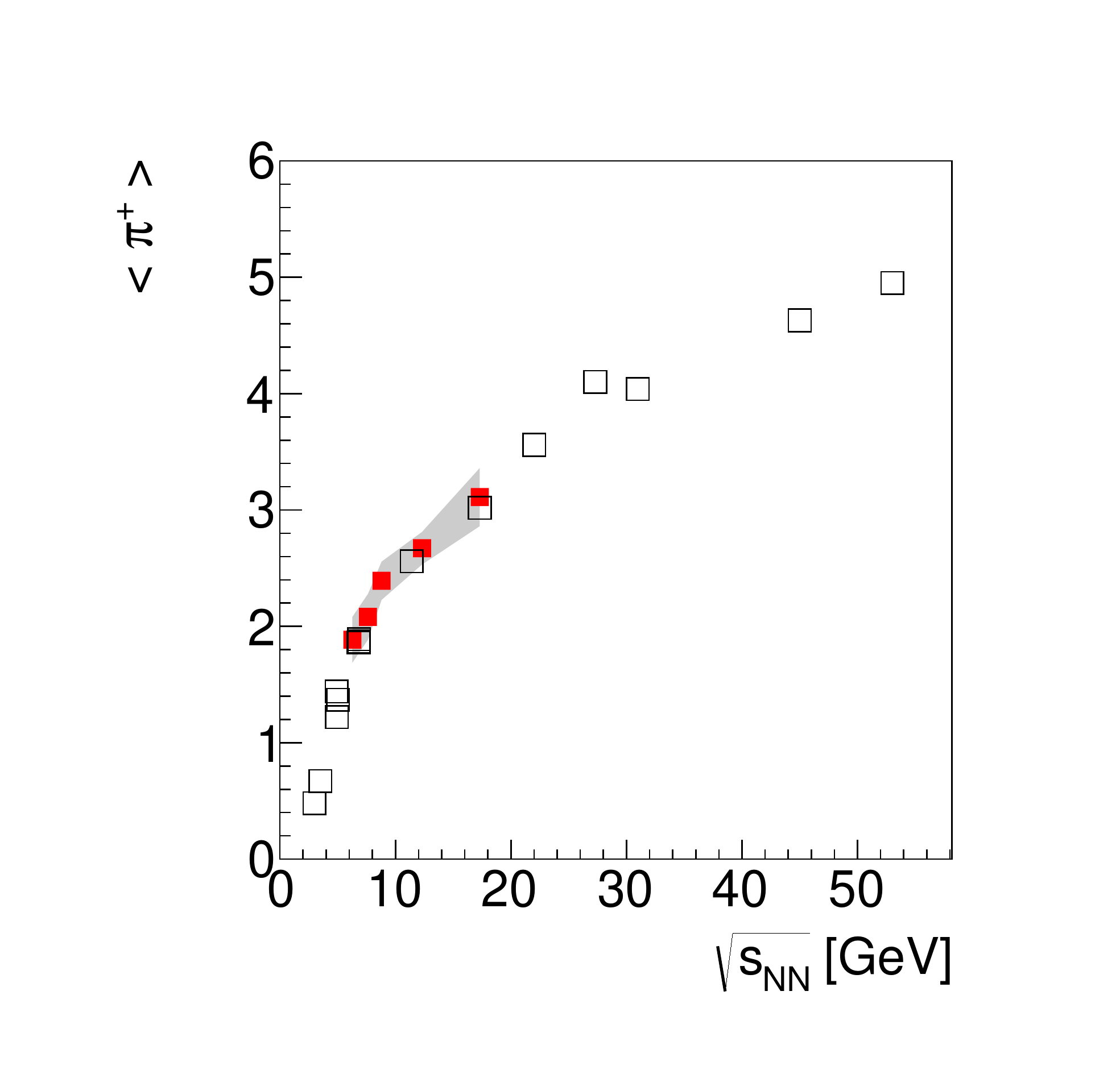}
	\includegraphics[width=0.37\textwidth]{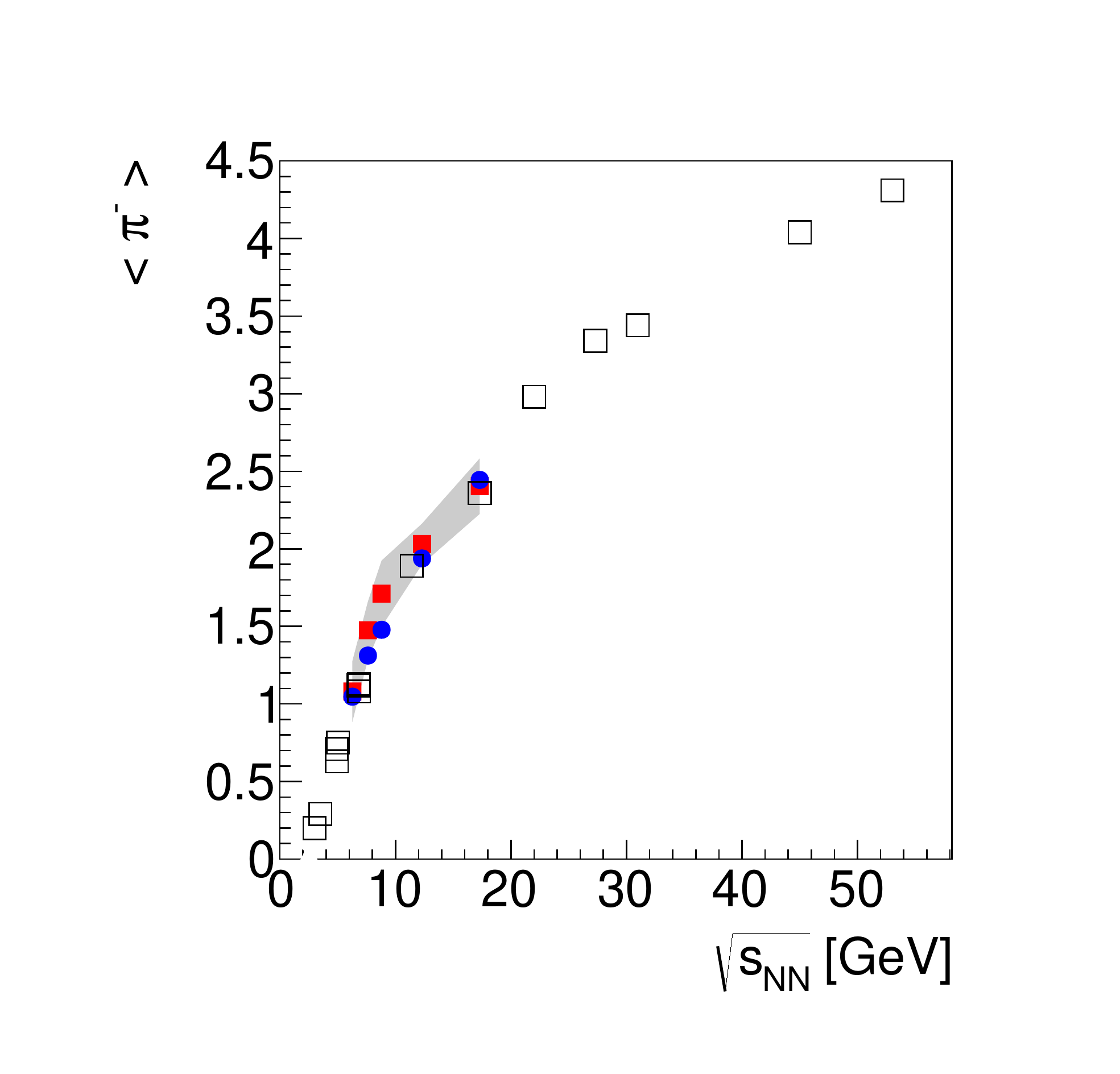}
	\end{center}
	\caption{(Color online) Mean $\pi^{+}$ (left) and $\pi^{-}$ (right) multiplicity produced in inelastic p+p interactions as function of collision energy. Measurements of \NASixtyOne for identified pions are plotted as filled red squares, whereas those obtained by the $h^{-}$ method~\cite{Abgrall:2013pp_pim} by blue dots. Black open squares represent the world data~\cite{Alt:2005zq,Blobel:1975ka,AguilarBenitez:1991yy,Melissinos:1962zz,Alexander:1967zz,Mueck:1972qz,Akerlof:1971mz,Antinucci:1972ib,Ammosov:1976zk}. Statistical uncertainties are indicated by bars, systematic uncertainties of the \NASixtyOne measurements by shaded bands.}
	\label{fig:pionmeanmult}
\end{figure}

The energy dependence of the \NASixtyOne measurements of charged pion multiplicities $\left\langle\pi^{+}\right\rangle$ and $\left\langle\pi^{-}\right\rangle$ produced in inelastic p+p collisions is compared in Fig.~\ref{fig:pionmeanmult} to that of the world data~\cite{Alt:2005zq,Blobel:1975ka,AguilarBenitez:1991yy,Melissinos:1962zz,Alexander:1967zz,Mueck:1972qz,Akerlof:1971mz,Antinucci:1972ib}. Clearly the \NASixtyOne results (filled red squares, filled blue dots) are consistent with measurements published in the literature (open squares). Also the \NASixtyOne measurements for $\left\langle\pi^{-}\right\rangle$ obtained with the identification method (filled red squares) and the $h^{-}$ method~\cite{Abgrall:2013pp_pim} (blue dots) agree. It can be seen, that $\left\langle\pi^{+}\right\rangle$ and $\left\langle\pi^{-}\right\rangle$ increase smoothly with the collision energy. The rate of the increase gradually diminishes with increasing energy. The energy dependence of the charged pion multiplicity is qualitatively similar to that of charged kaons.

%
 
\begin{figure}[!ht]
 	\begin{center}
 	\includegraphics[width=0.37\textwidth]{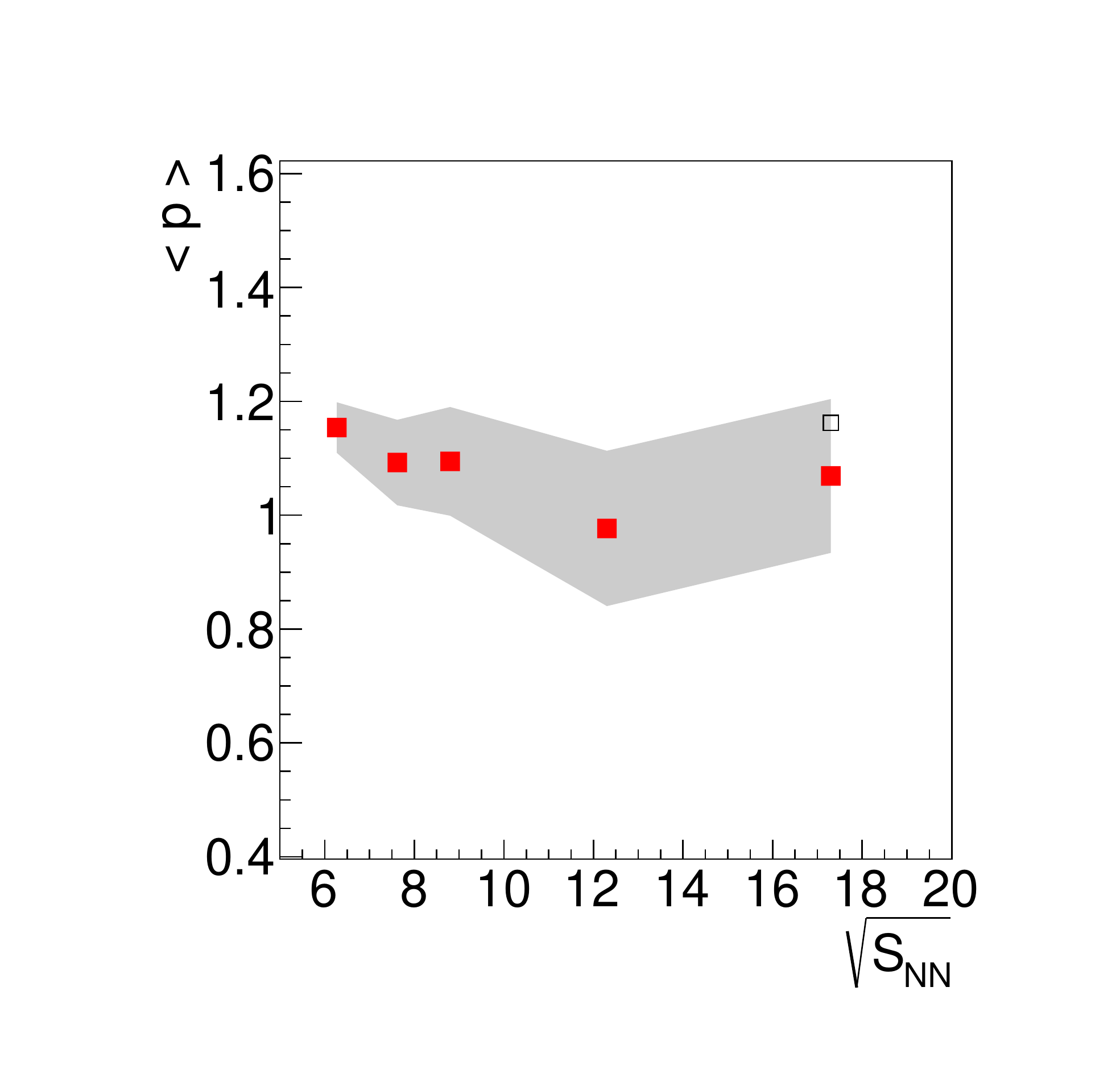}
 	\end{center}
 	\caption{(Color online) Mean proton multiplicities produced in inelastic p+p interactions as a function of collision energy. Filled red squares depict results from this analysis, the open black sqare shows the measurement of NA49~\cite{Anticic:2009wd}. The dominating uncertainty of the \NASixtyOne measurements is systematic and is indicated by the shaded band.}
 	\label{fig:meanproton}
 \end{figure}

For the determination of total proton multiplicities large acceptance over a full hemisphere of the reaction is particularly important. Due to this feature \NASixtyOne and NA49 were able to provide such measurements in the SPS energy range. The total proton multiplicity in inelastic p+p collisions is presented in Fig.~\ref{fig:meanproton} as function of collision energy. One observes that the proton yield is almost constant at SPS energies.
 
\FloatBarrier
\section{Comparison with hadron production models}
\label{sec:models}

This section compares the \NASixtyOne measurements with predictions from the publicly available codes of the
microscopic models \Epos 1.99~\cite{Werner:2008zza} and \Urqmd 3.4~\cite{Bass:1998ca,Bleicher:1999xi}. In \Epos the
reaction proceeds via excitation of strings according to Gribov-Regge theory which subsequently fragment 
into hadrons. \Urqmd generates a hadron cascade using elementary cross sections and supplements this
process by string production and fragmentation at higher energies.

\begin{figure}[!ht]
\begin{center}
\includegraphics[width=0.3\textwidth]{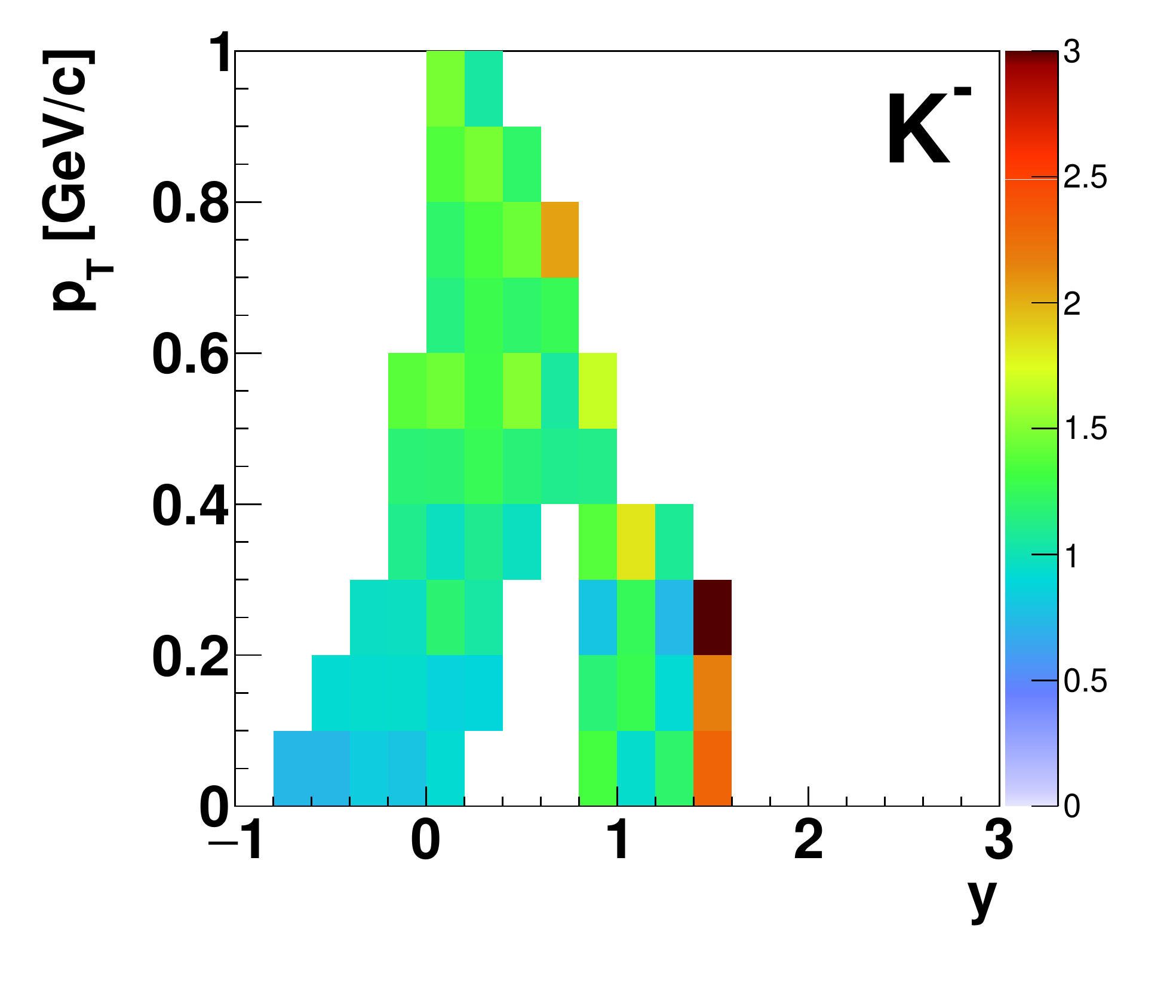}
\includegraphics[width=0.3\textwidth]{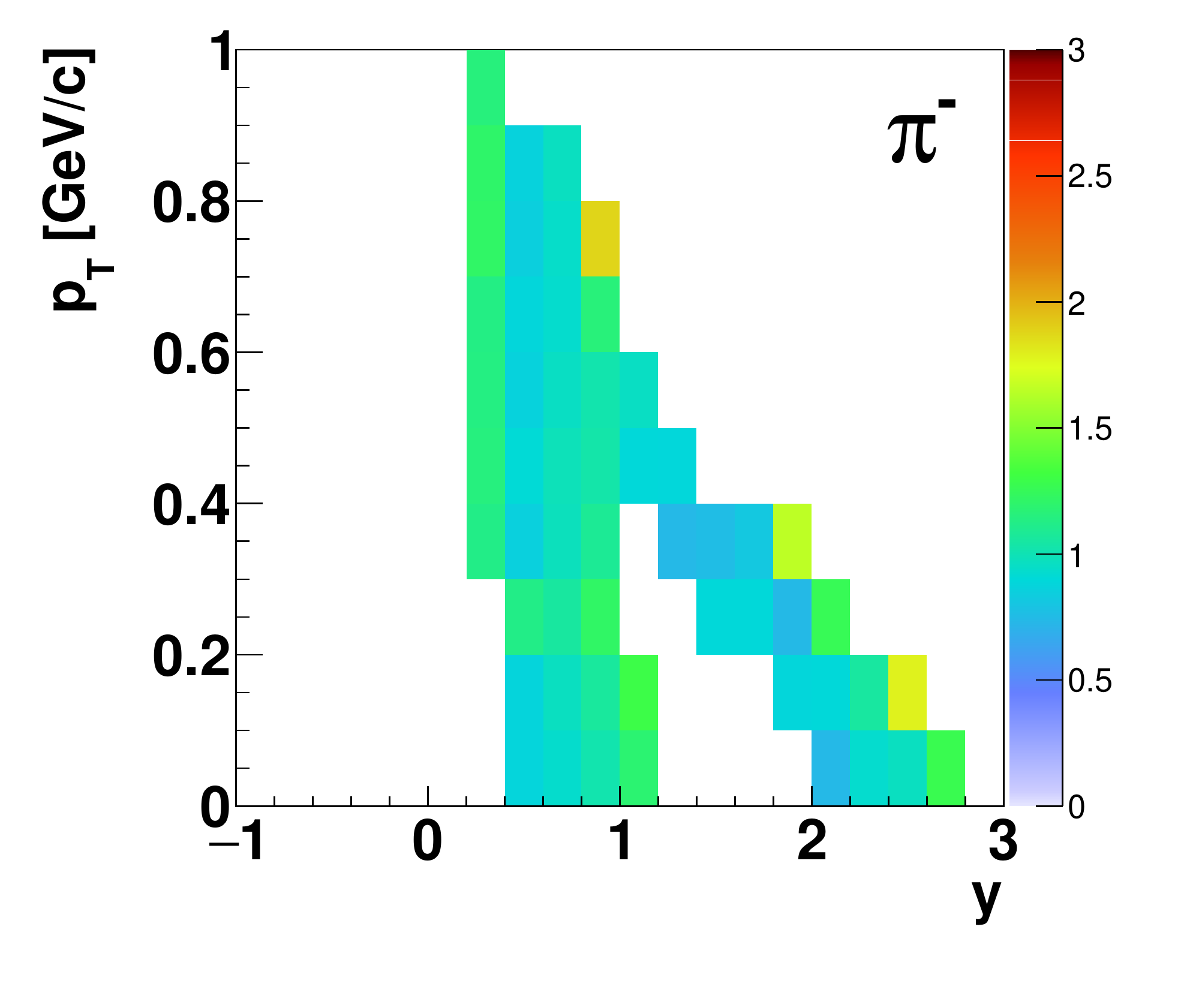}\\
\includegraphics[width=0.3\textwidth]{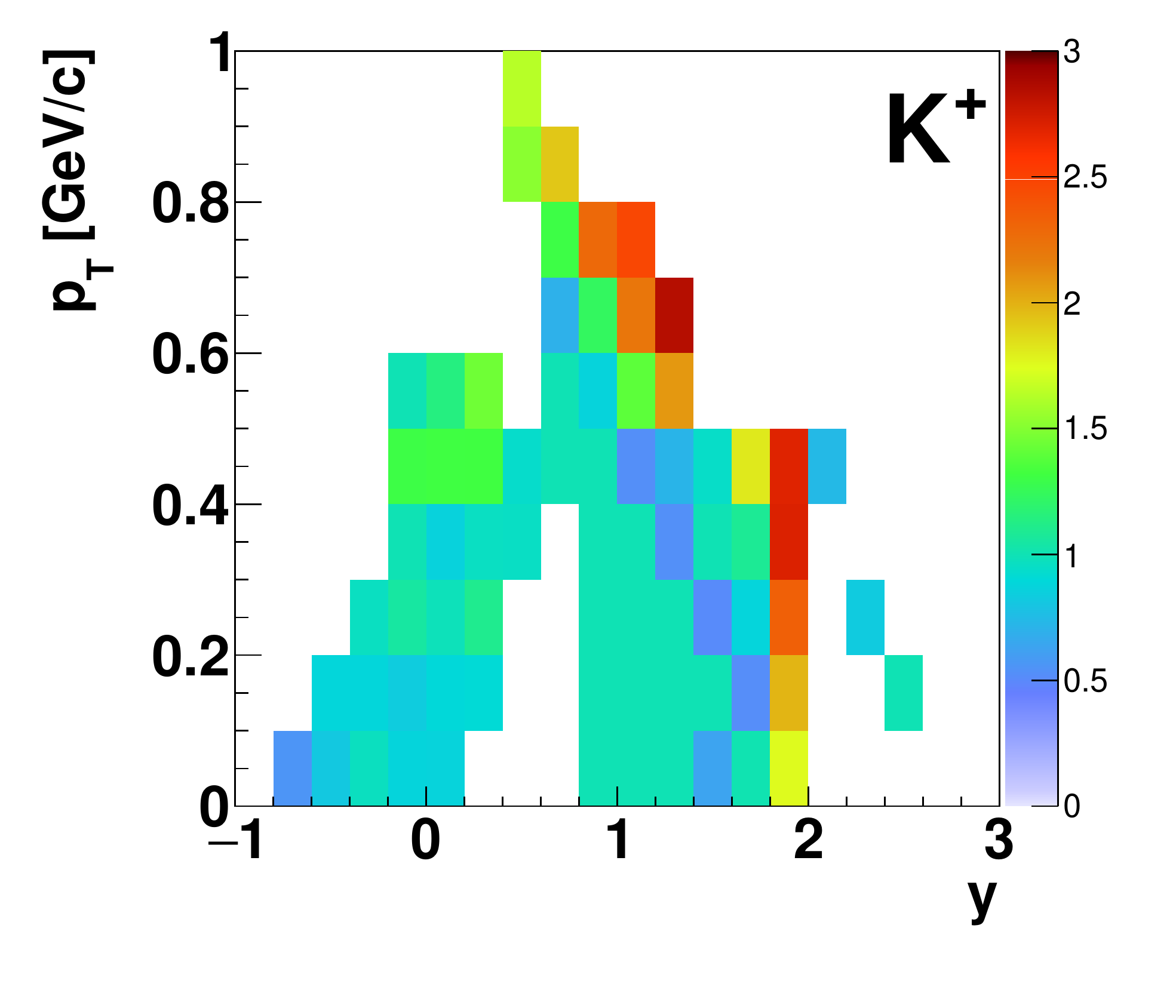}
\includegraphics[width=0.3\textwidth]{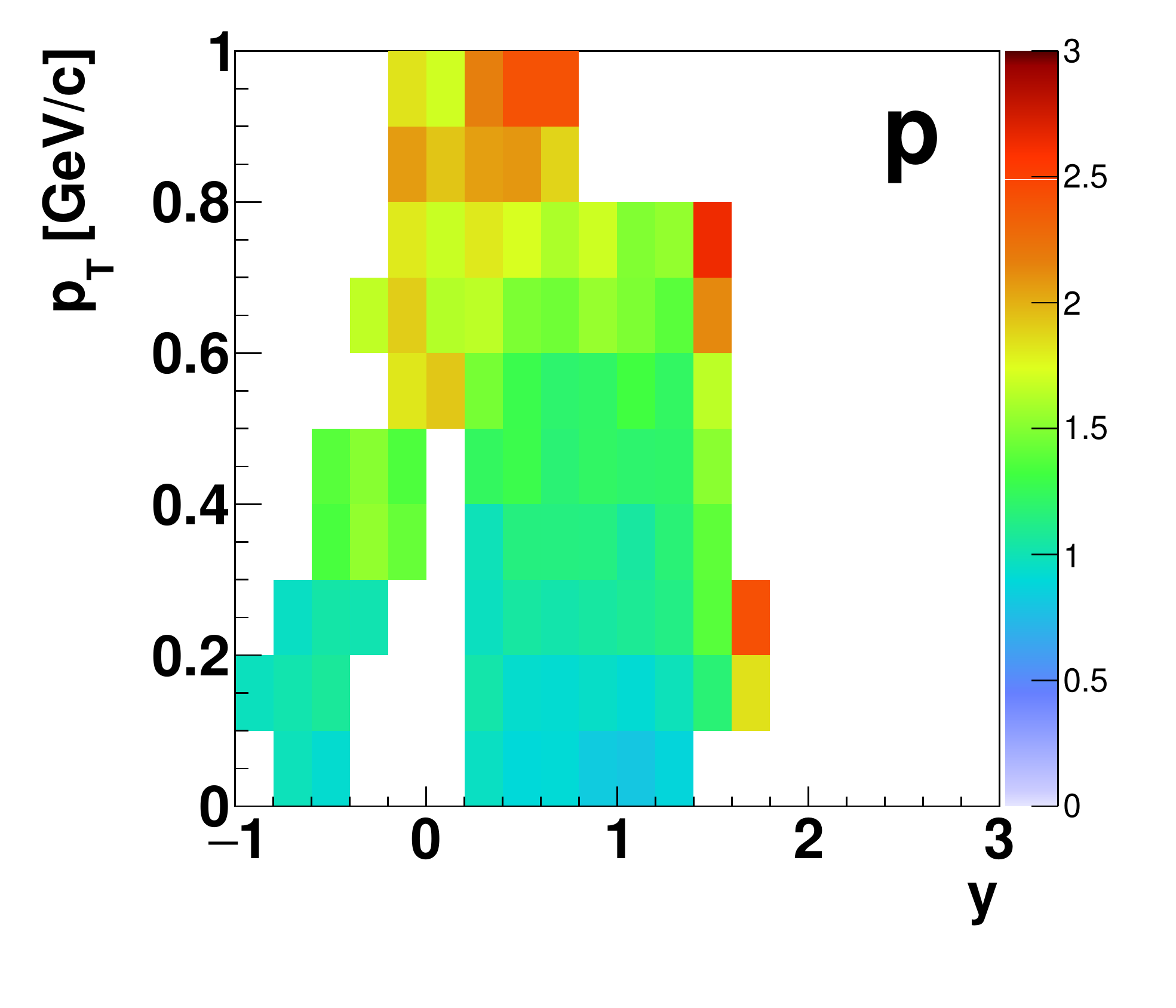}
\includegraphics[width=0.3\textwidth]{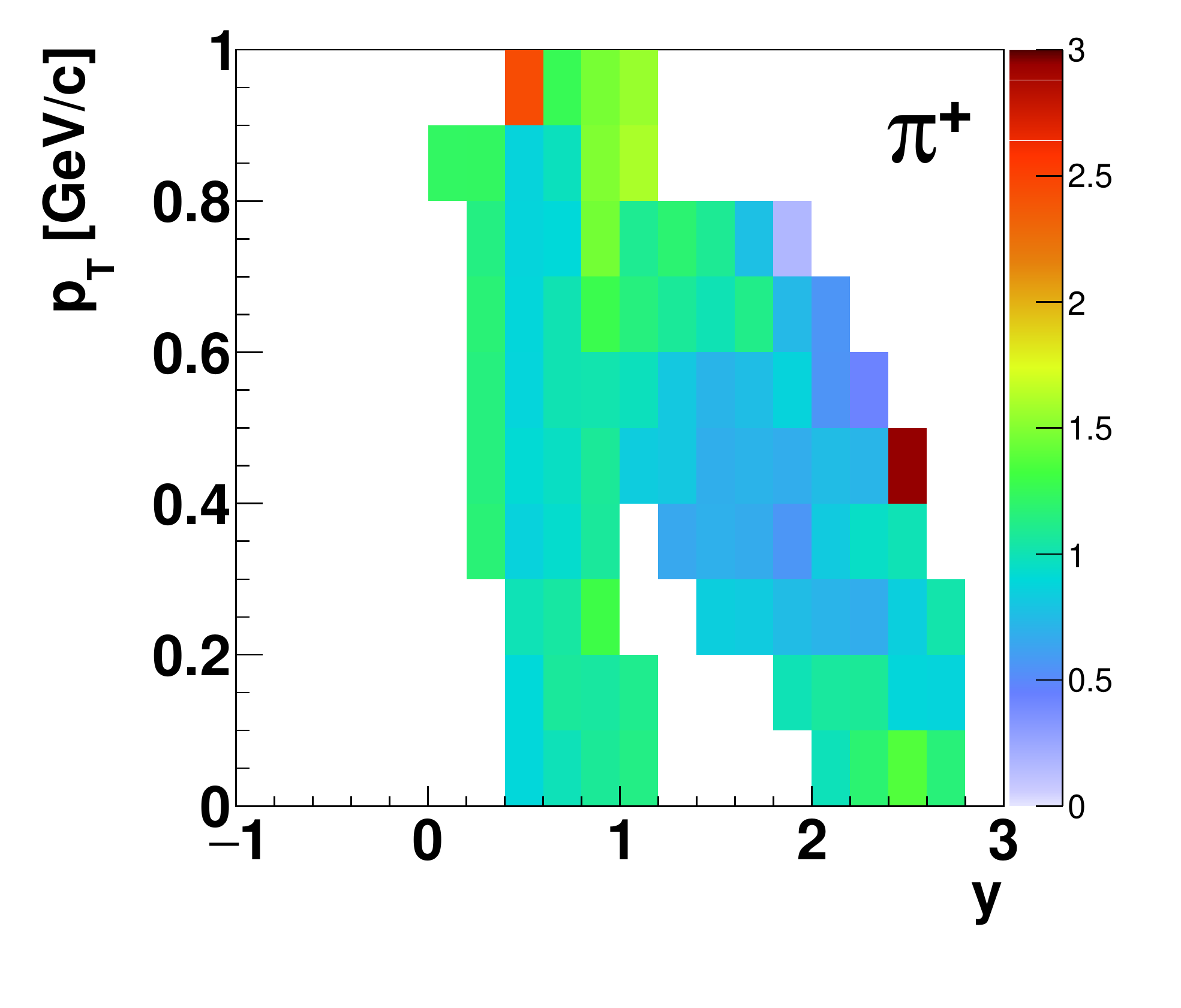}
\end{center}
\caption{(Color online) Two dimensional distributions $d^{2}n/(dp_{T}dy)$ of $\pi^{-}$, $\pi^{+}$, K$^{-}$, K$^{+}$ and p produced in inelastic p+p interactions at 20~\GeVc divided by the \Epos model~\cite{Werner:2008zza} calculations.}
		\label{fig:2DEPOS20}
\end{figure}

\begin{figure}[!ht]
\begin{center}
\includegraphics[width=0.3\textwidth]{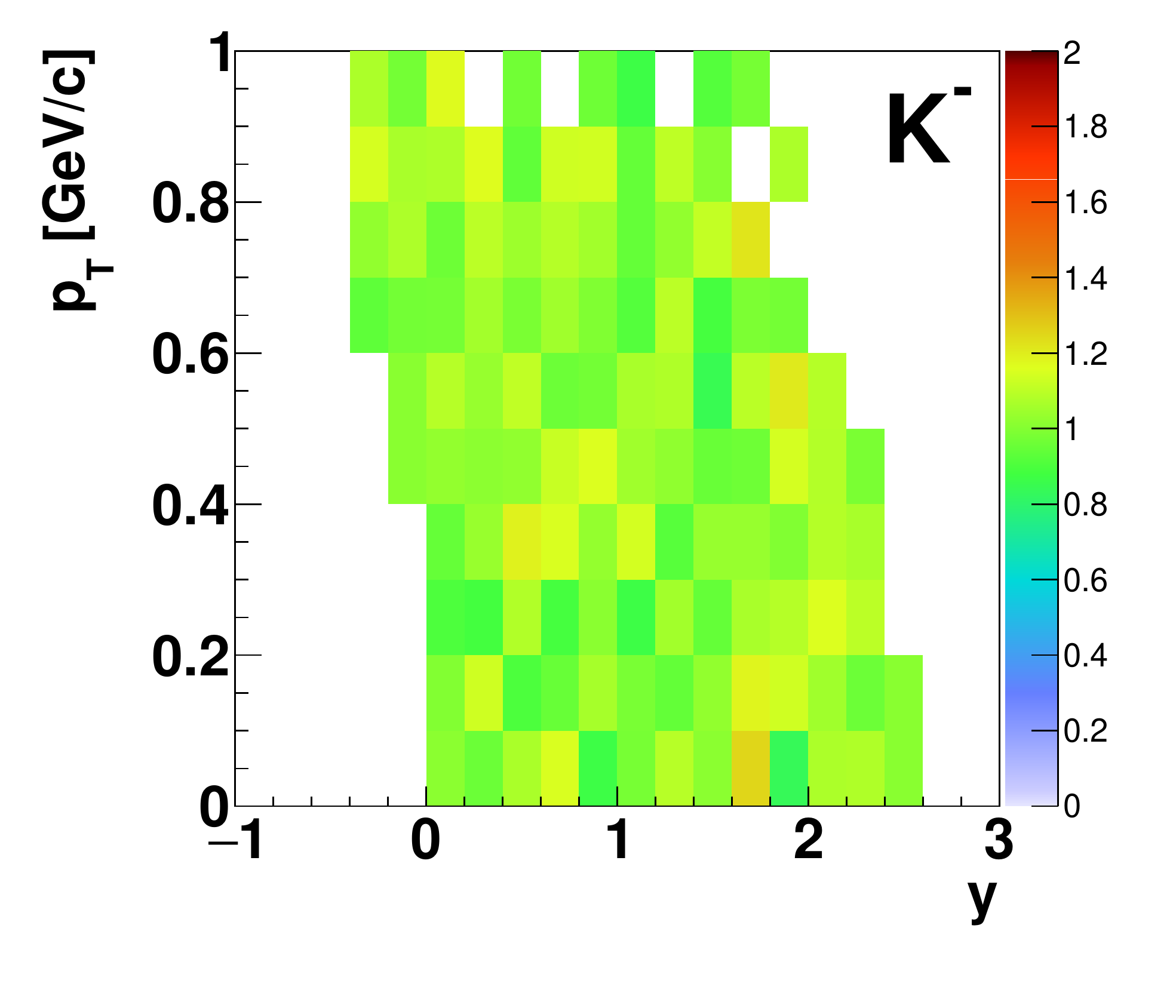}
\includegraphics[width=0.3\textwidth]{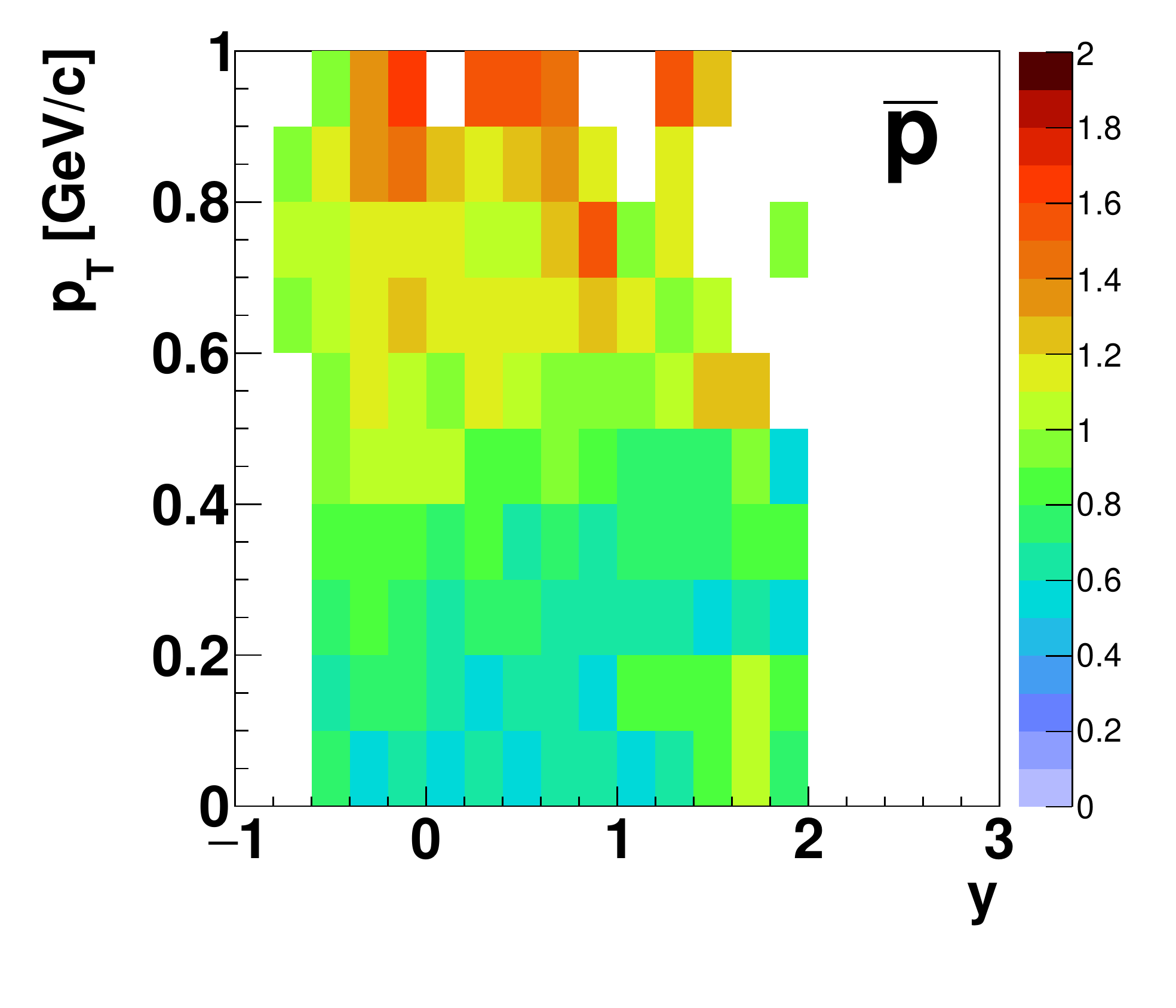}
\includegraphics[width=0.3\textwidth]{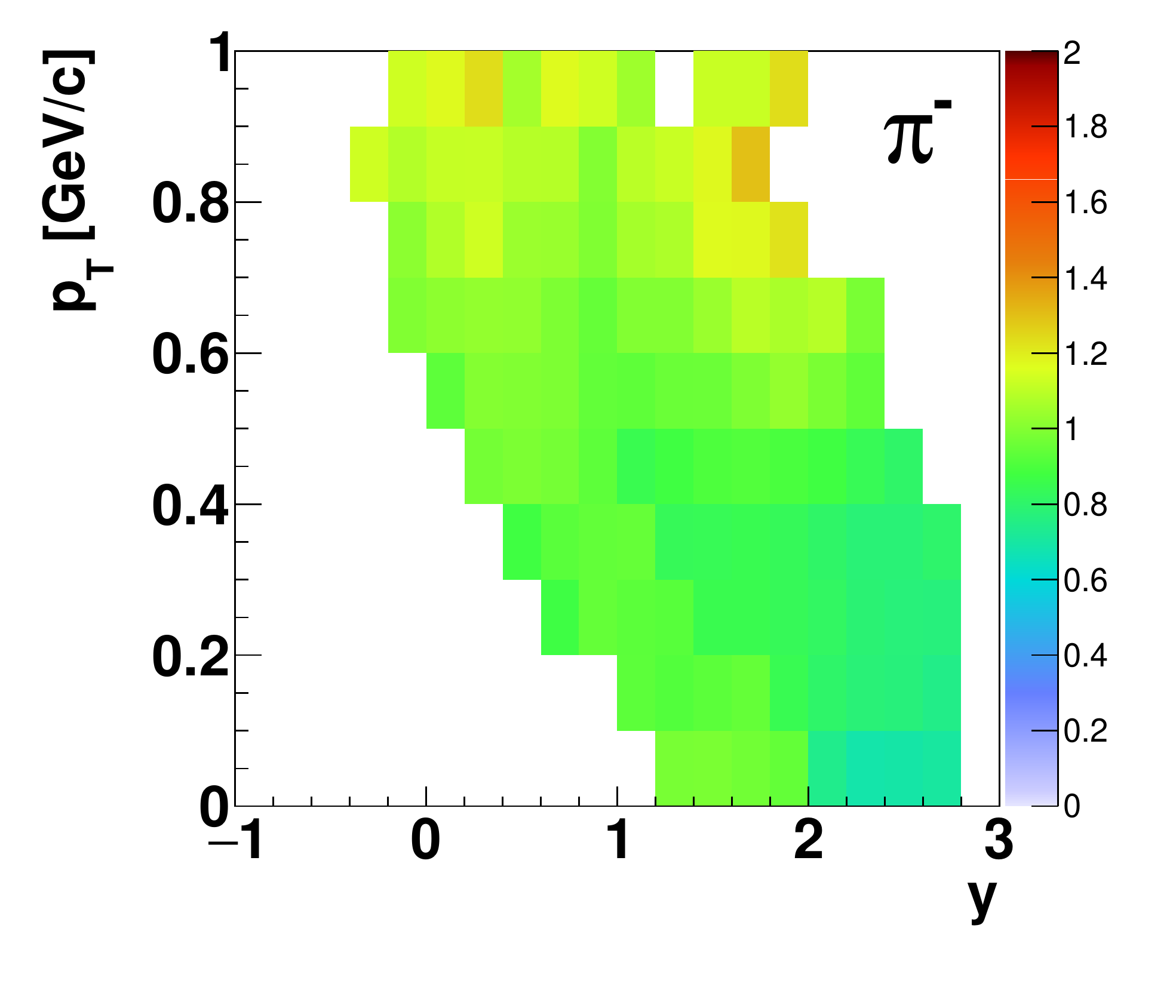}\\
\includegraphics[width=0.3\textwidth]{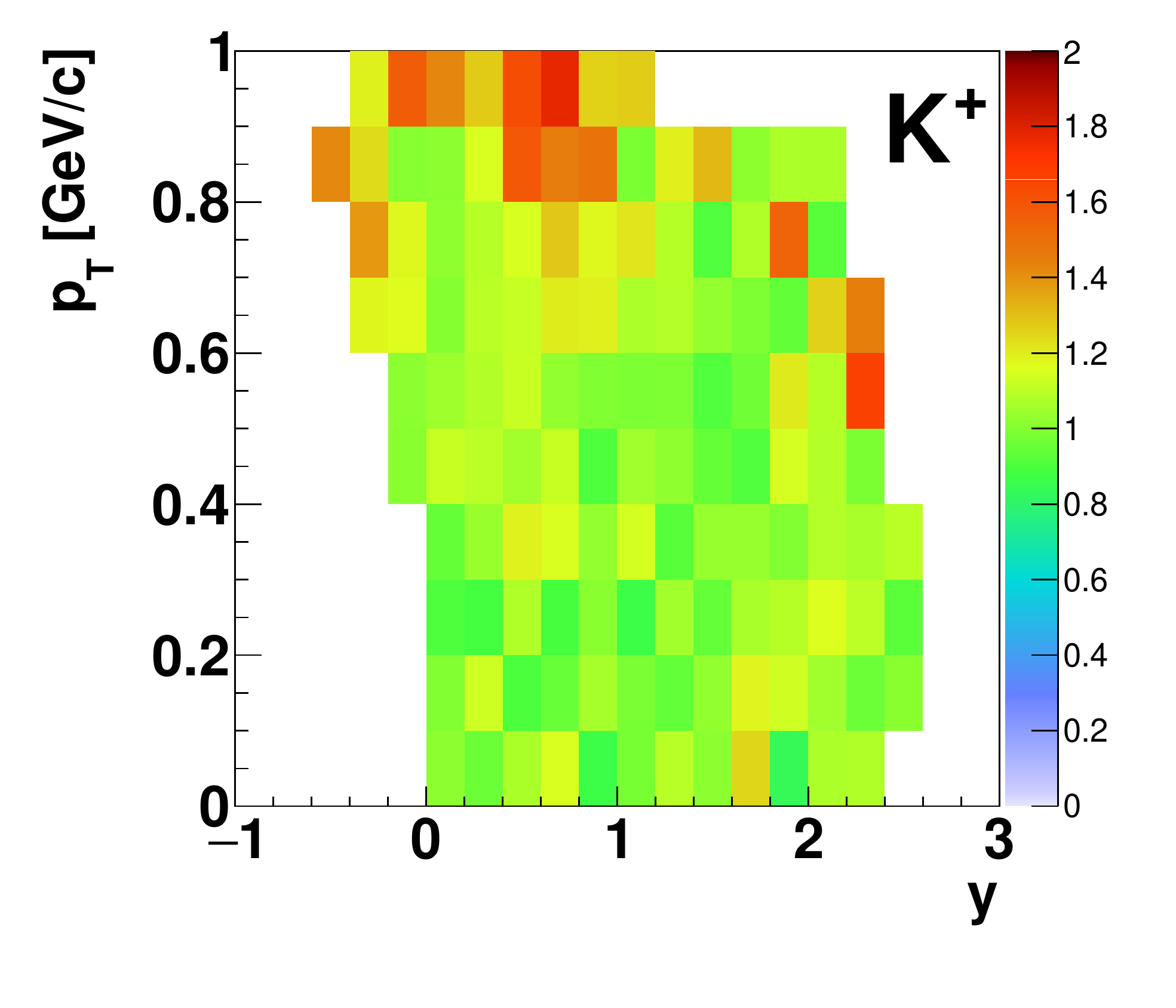}
\includegraphics[width=0.3\textwidth]{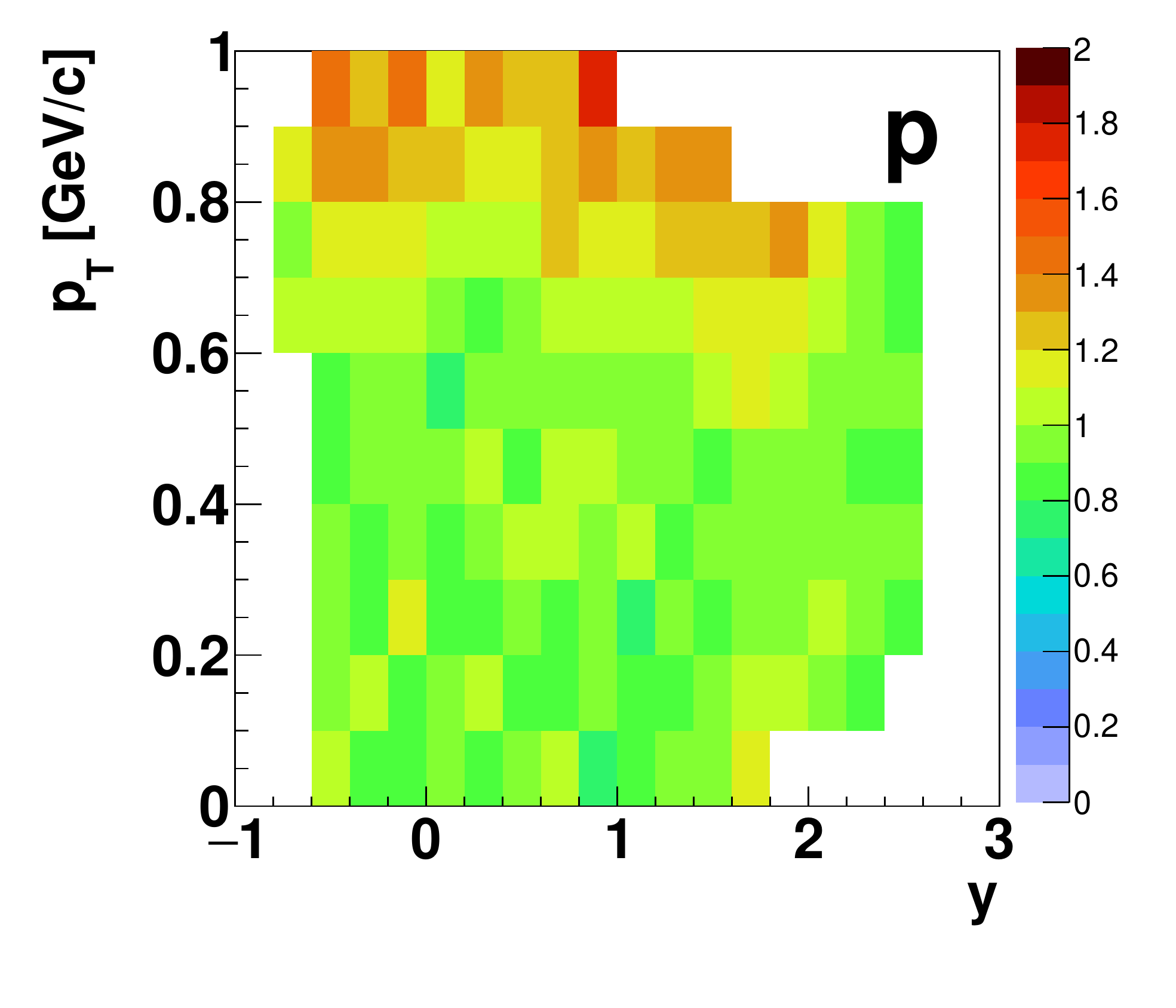}
\includegraphics[width=0.3\textwidth]{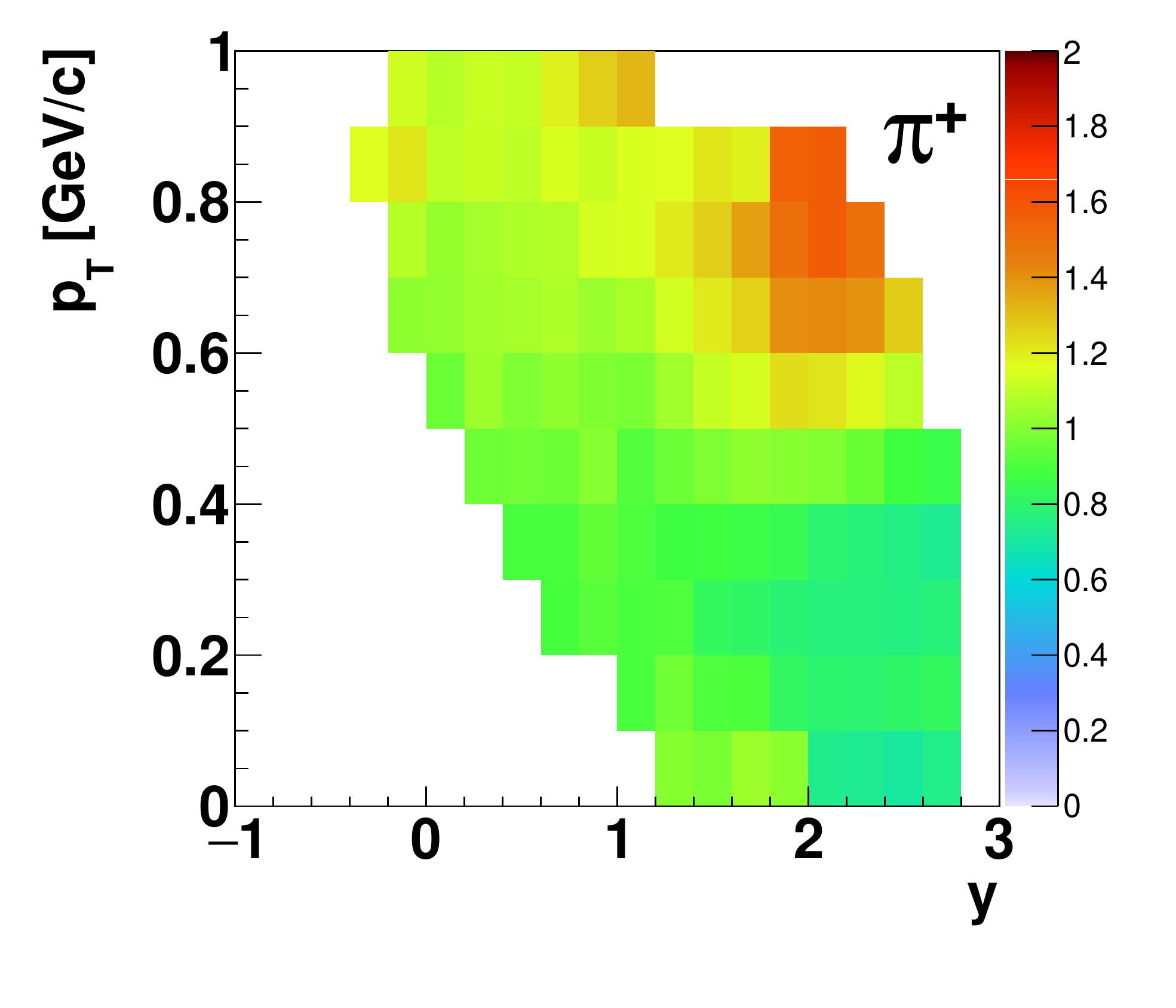}
\end{center}
\caption{(Color online) Two dimensional distributions $d^{2}n/(dp_{T}dy)$ of $\pi^{-}$, $\pi^{+}$, K$^{-}$, K$^{+}$, p and $\bar{\textrm{p}}$ produced in inelastic p+p interactions at 158~\GeVc divided by the \Epos model~\cite{Werner:2008zza} calculations.}
		\label{fig:2DEPOS158}
\end{figure}

\begin{figure}[!ht]
\begin{center}
\includegraphics[width=0.3\textwidth]{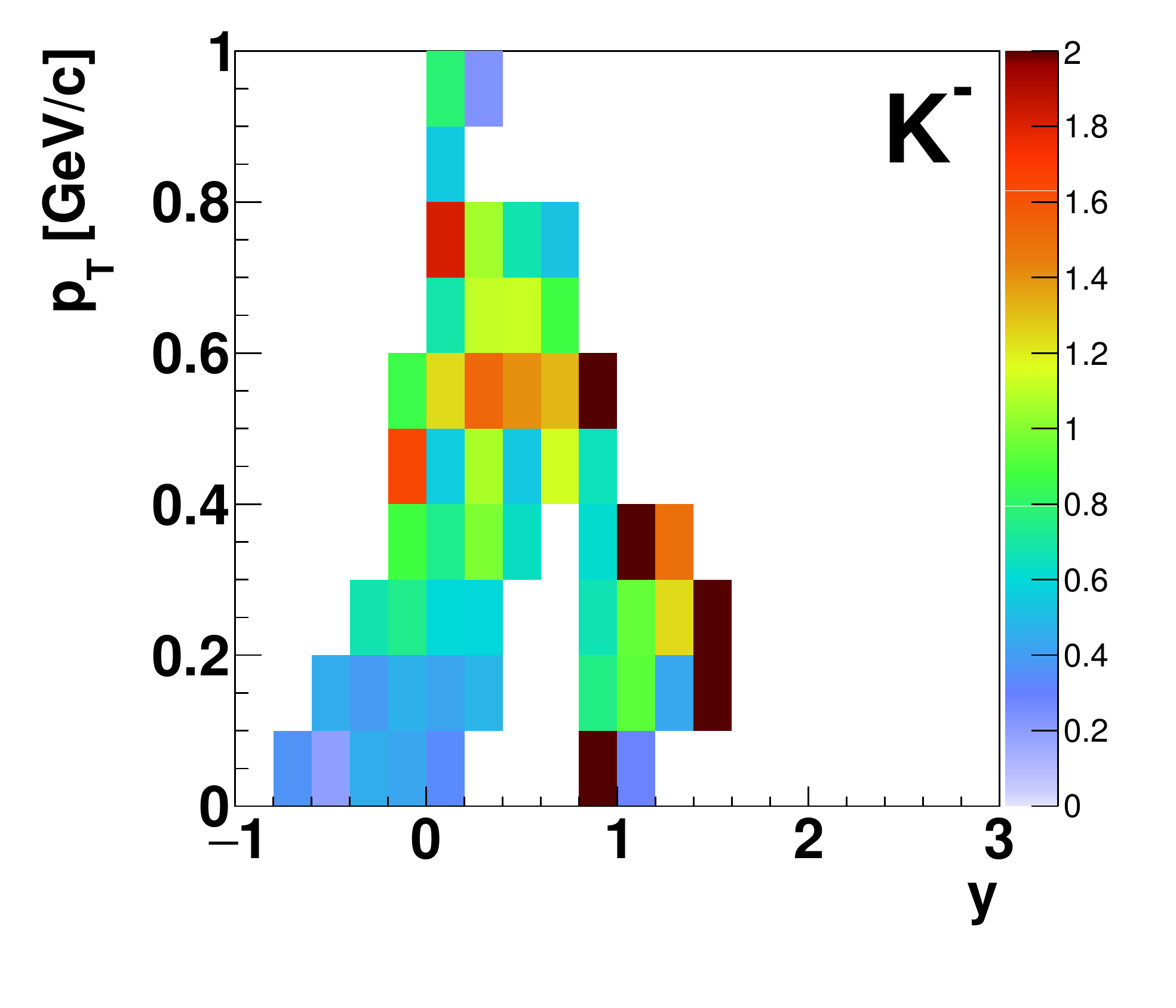}
\includegraphics[width=0.3\textwidth]{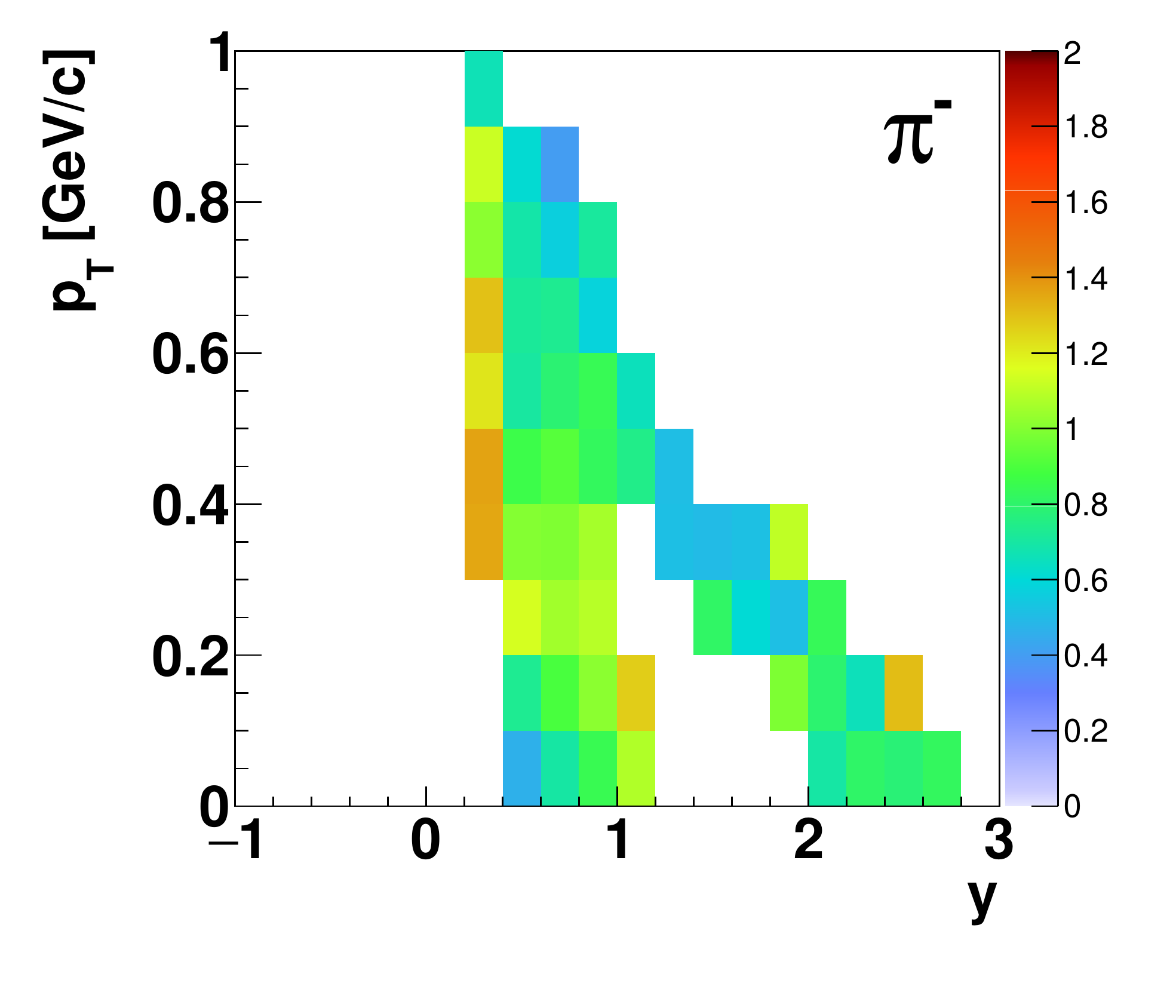}\\
\includegraphics[width=0.3\textwidth]{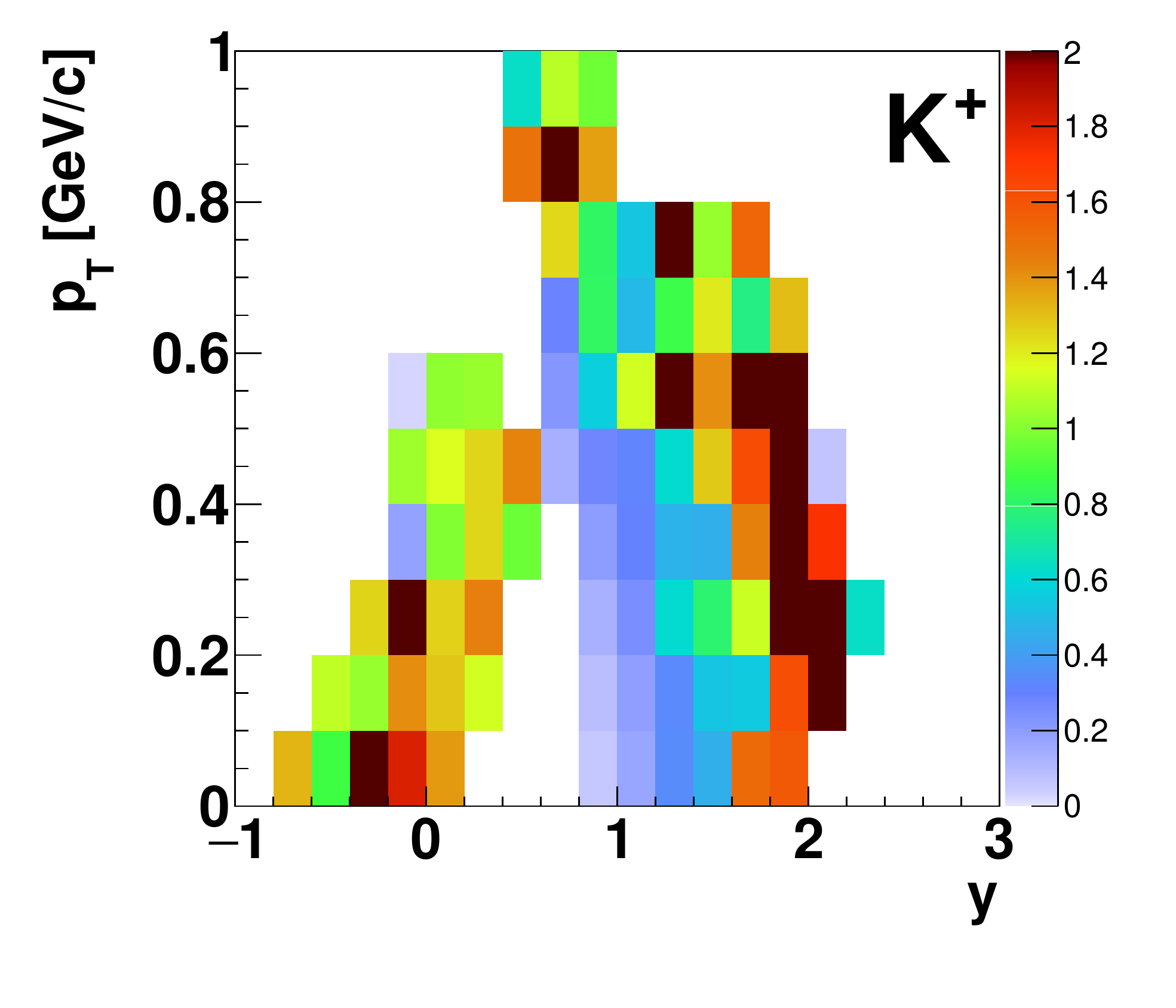}
\includegraphics[width=0.3\textwidth]{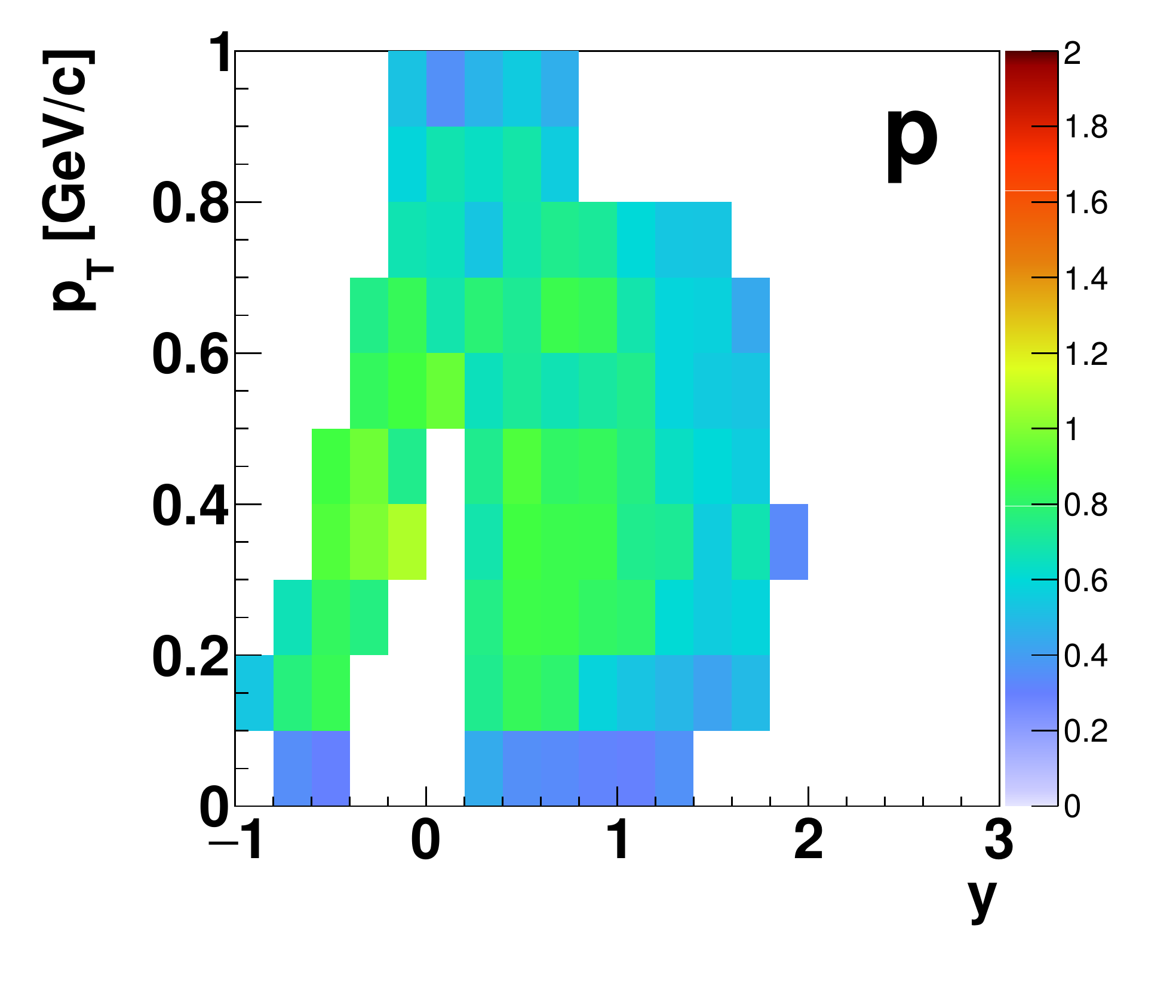}
\includegraphics[width=0.3\textwidth]{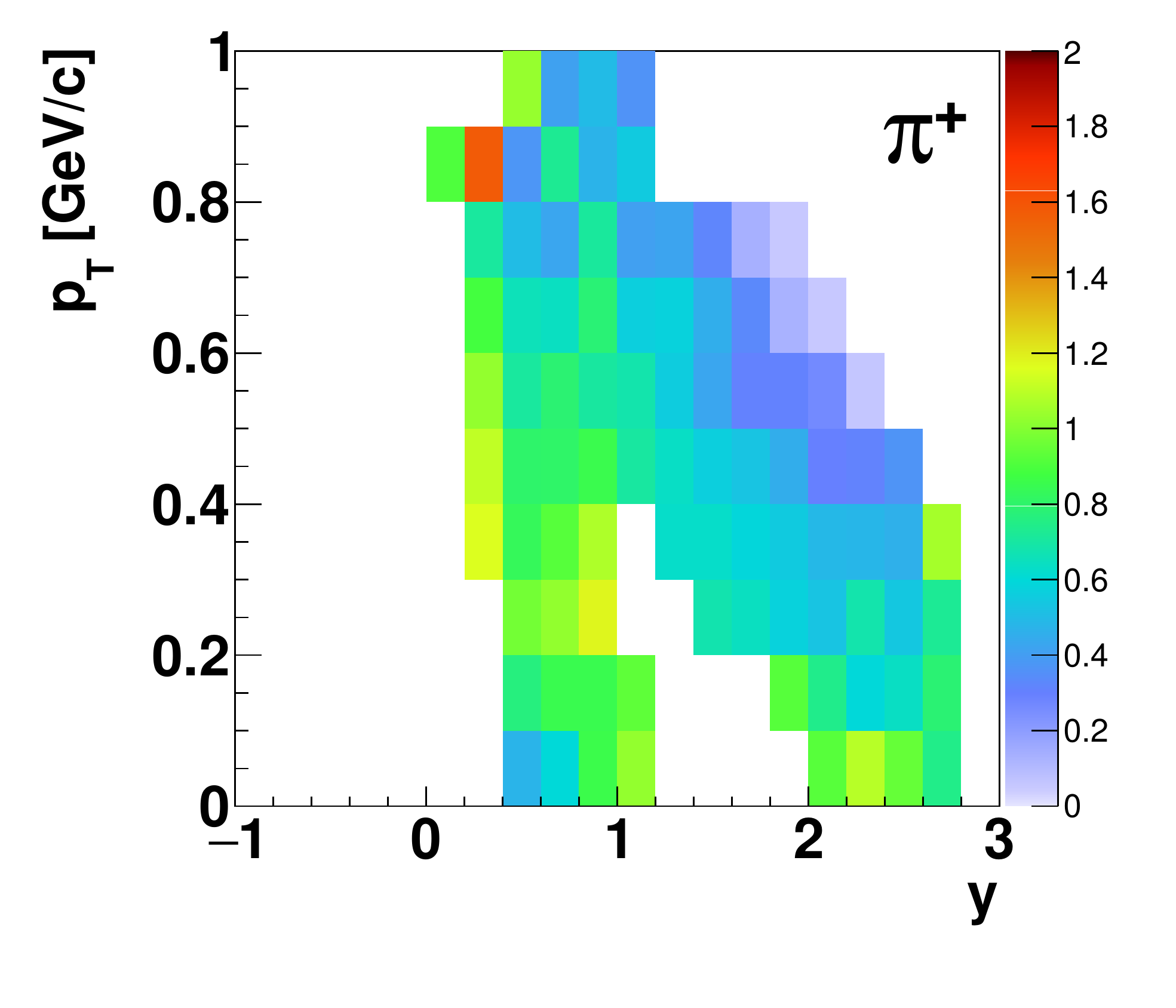}
\end{center}
\caption{(Color online) Two dimensional distributions $d^{2}n/(dp_{T}dy)$ of $\pi^{-}$, $\pi^{+}$, K$^{-}$, K$^{+}$ and p produced in inelastic p+p interactions at 20~\GeVc divided by the \Urqmd 3.4 model~\cite{Bass:1998ca,Bleicher:1999xi} calculations.}
		\label{fig:2DUR20}
\end{figure}

\begin{figure}[!ht]
\includegraphics[width=0.3\textwidth]{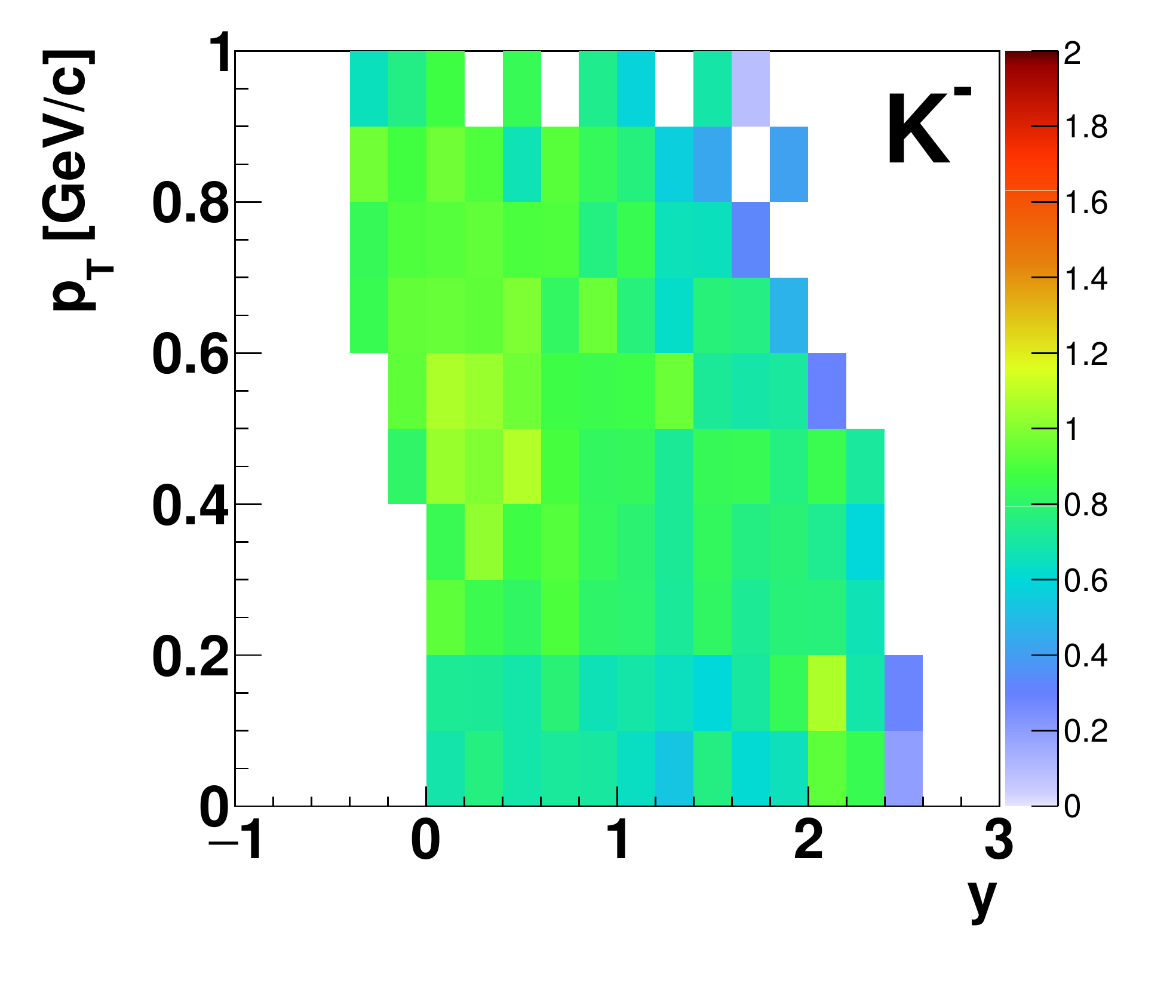}
\includegraphics[width=0.3\textwidth]{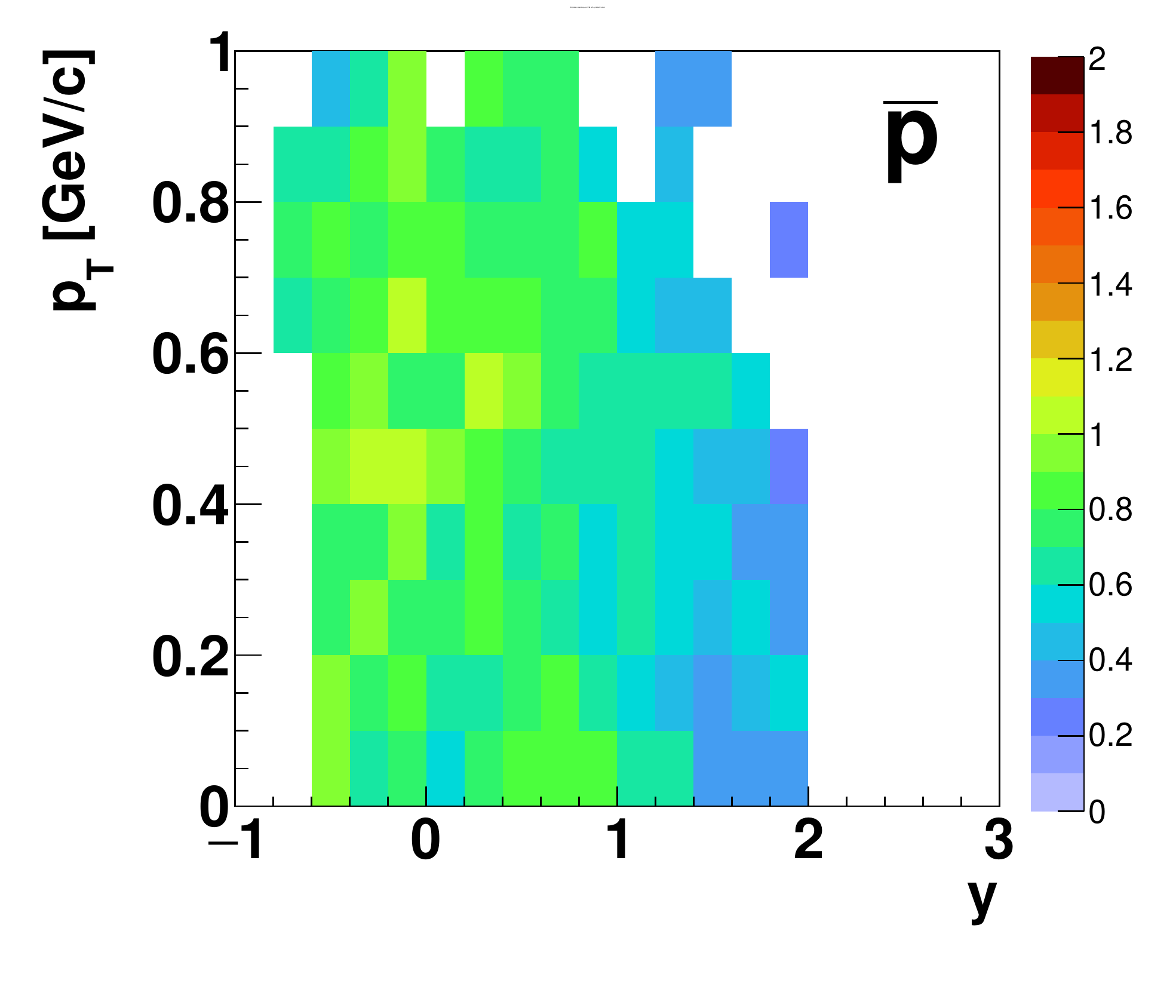}
\includegraphics[width=0.3\textwidth]{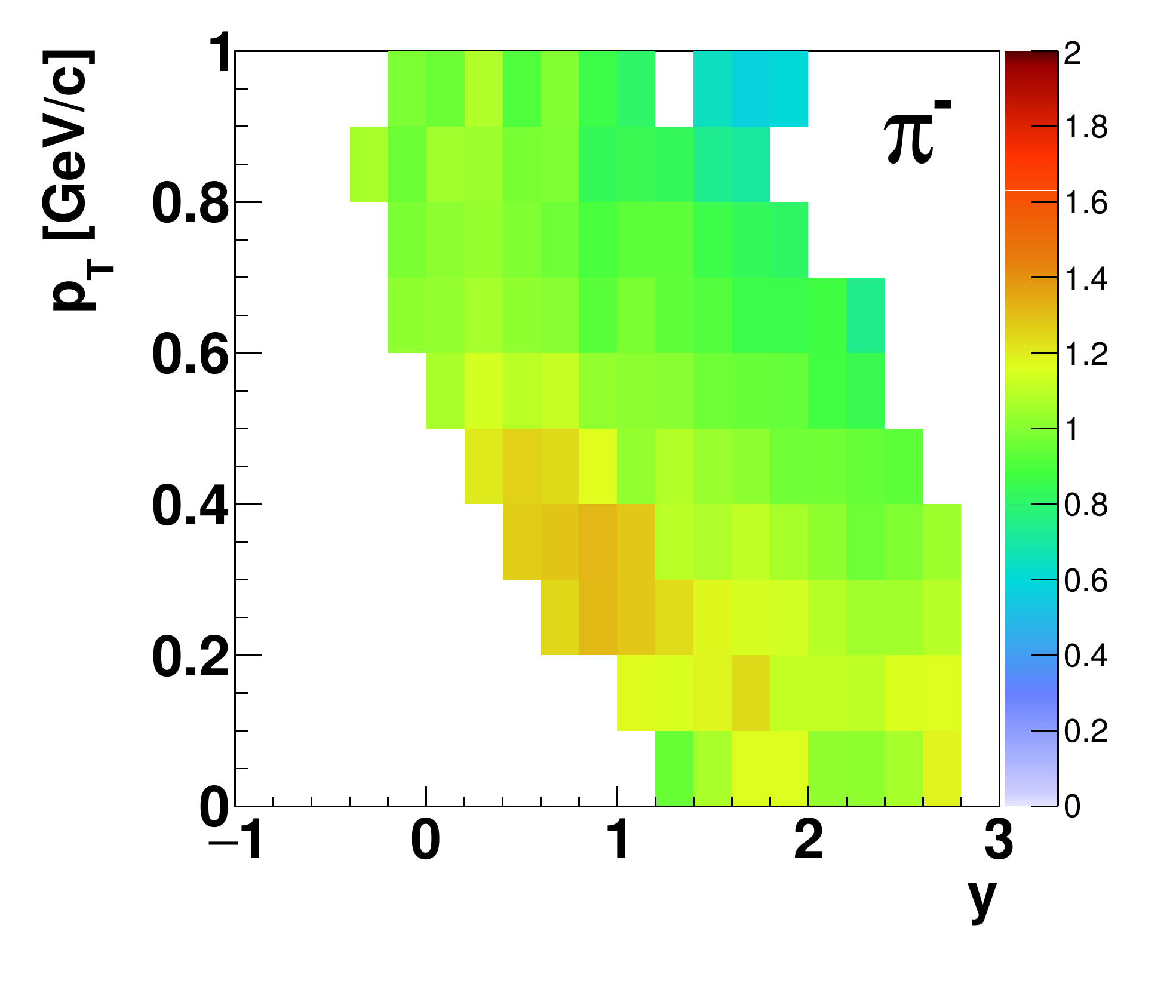}\\
\includegraphics[width=0.3\textwidth]{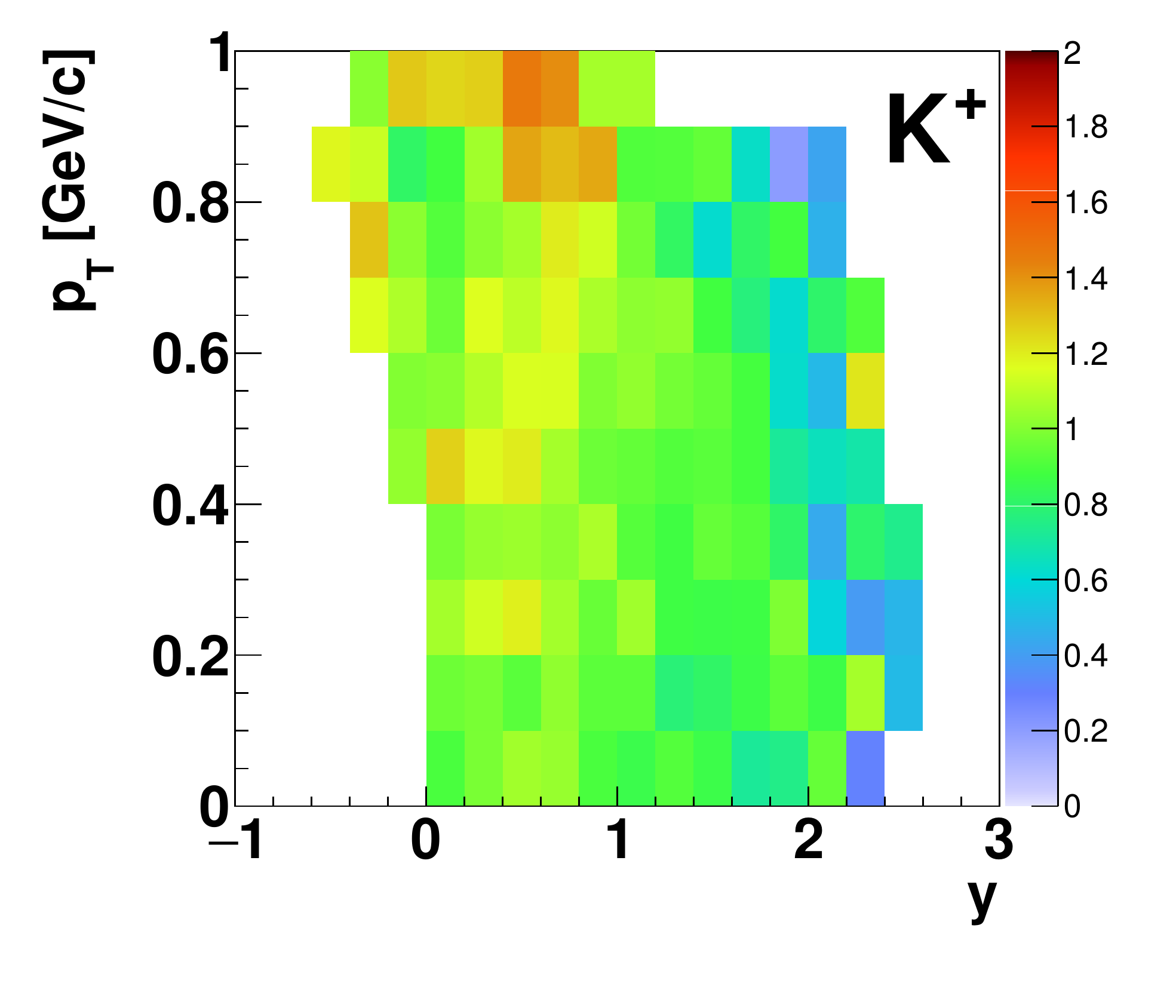}
\includegraphics[width=0.3\textwidth]{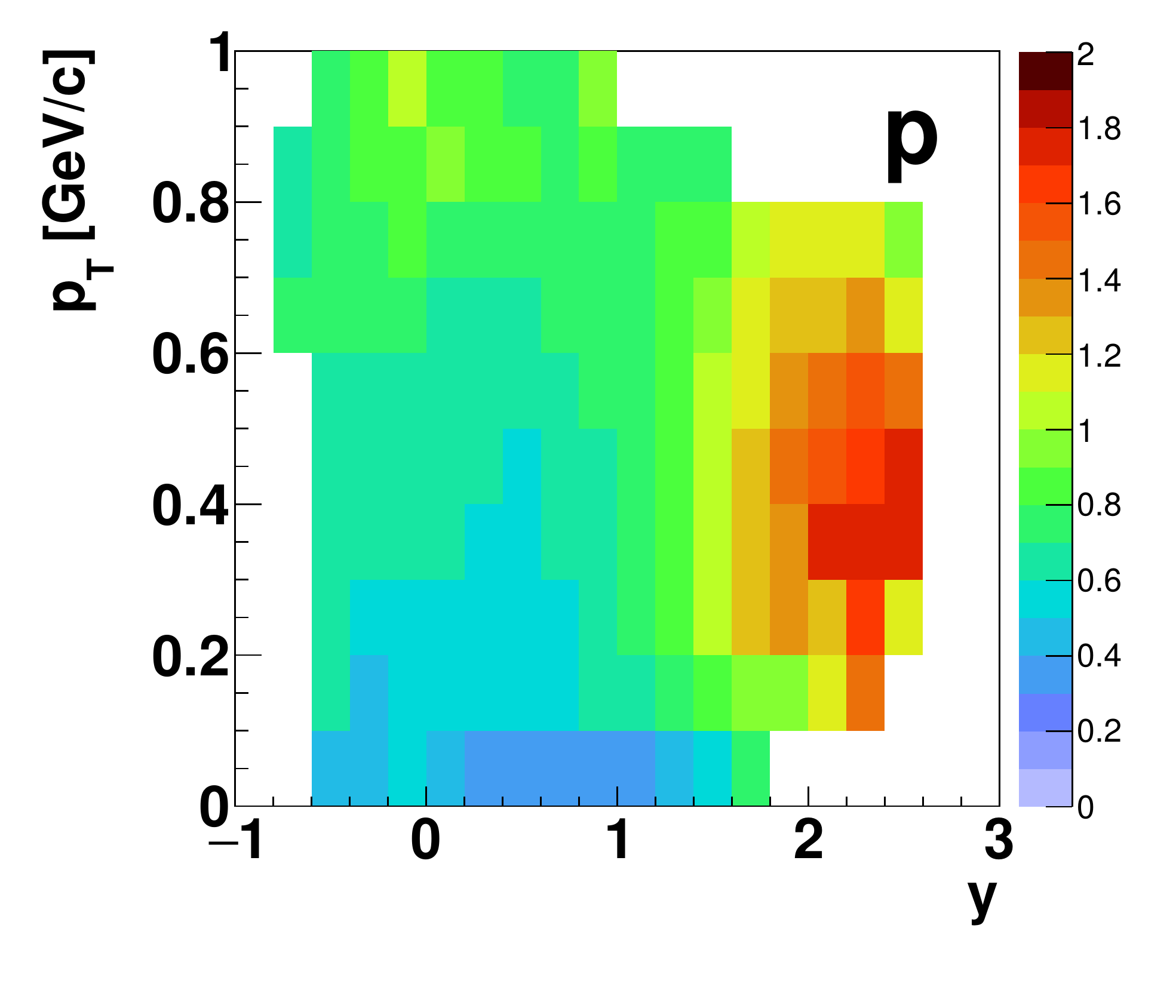}
\includegraphics[width=0.3\textwidth]{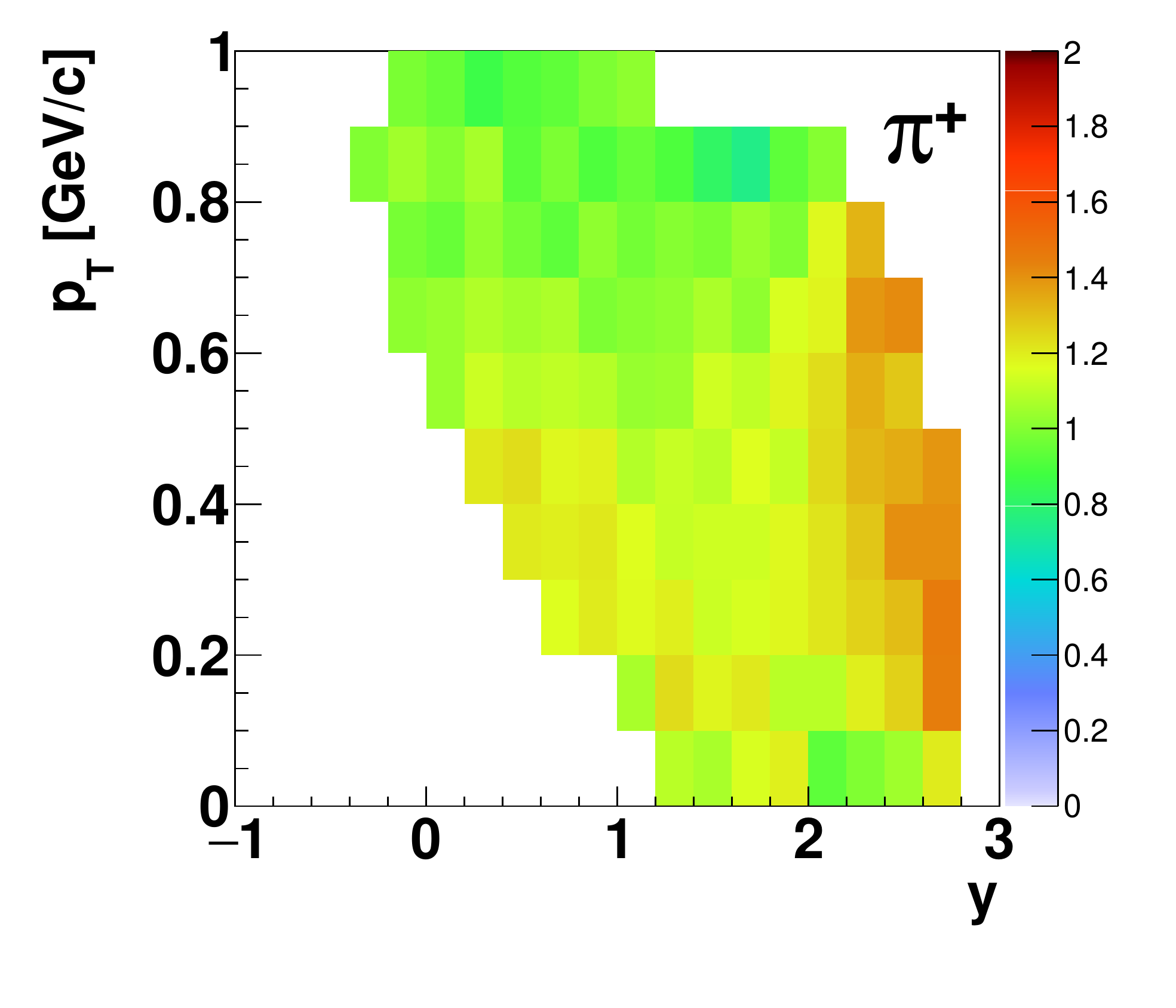}
\caption{(Color online) Two dimensional distributions $d^{2}n/(dp_{T}dy)$ of $\pi^{-}$, $\pi^{+}$, K$^{-}$, K$^{+}$, p and $\bar{\textrm{p}}$ produced in inelastic p+p interactions at 158~\GeVc divided by the \Urqmd 3.4 model~\cite{Bass:1998ca,Bleicher:1999xi} calculations.}
		\label{fig:2DUR158}
\end{figure}

Two dimensional distributions $d^{2}n/(dp_{T}dy)$ of $\pi^{-}$, $\pi^{+}$, K$^{-}$, K$^{+}$, p and $\bar{\textrm{p}}$ produced in inelastic p+p interactions divided by the \Epos model prediction are presented in Fig.~\ref{fig:2DEPOS20} at 20~\GeVc and in Fig.~\ref{fig:2DEPOS158} at 158~\GeVc and divided by the \Urqmd model calculations in Fig.~\ref{fig:2DUR20} at 20~\GeVc and in Fig.~\ref{fig:2DUR158} at 158~\GeV/c.

As demonstrated by Figs.~\ref{fig:2DEPOS20} and~\ref{fig:2DEPOS158} the \Epos model provides a good description of the measurements in most regions of phase space. Only at larger tansverse momenta, where particle yields are low, some underprediction occurs. This conclusion is supported by Figs.~\ref{fig:epos_comps20} and~\ref{fig:epos_comps158} which show projected $p_{T}$ distributions in selected rapidity intervals.

\begin{figure*}
\begin{center}
\newcolumntype{S}{>{\centering} m{0.03\textwidth} }
\newcolumntype{A}{>{\centering} m{0.19\textwidth} }
\hspace{-10mm}
\begin{tabular}{S A A A A A A}
& \tiny20~\GeVc & \hspace{-15mm}\tiny31~\GeVc & \hspace{-30mm}\tiny40~\GeVc & \hspace{-45mm}\tiny80~\GeVc  & \hspace{-60mm} \tiny158~\GeVc 
\tabularnewline
\vspace{-5mm}$\pi^{-}$ &
\includegraphics[width=0.18\textwidth]{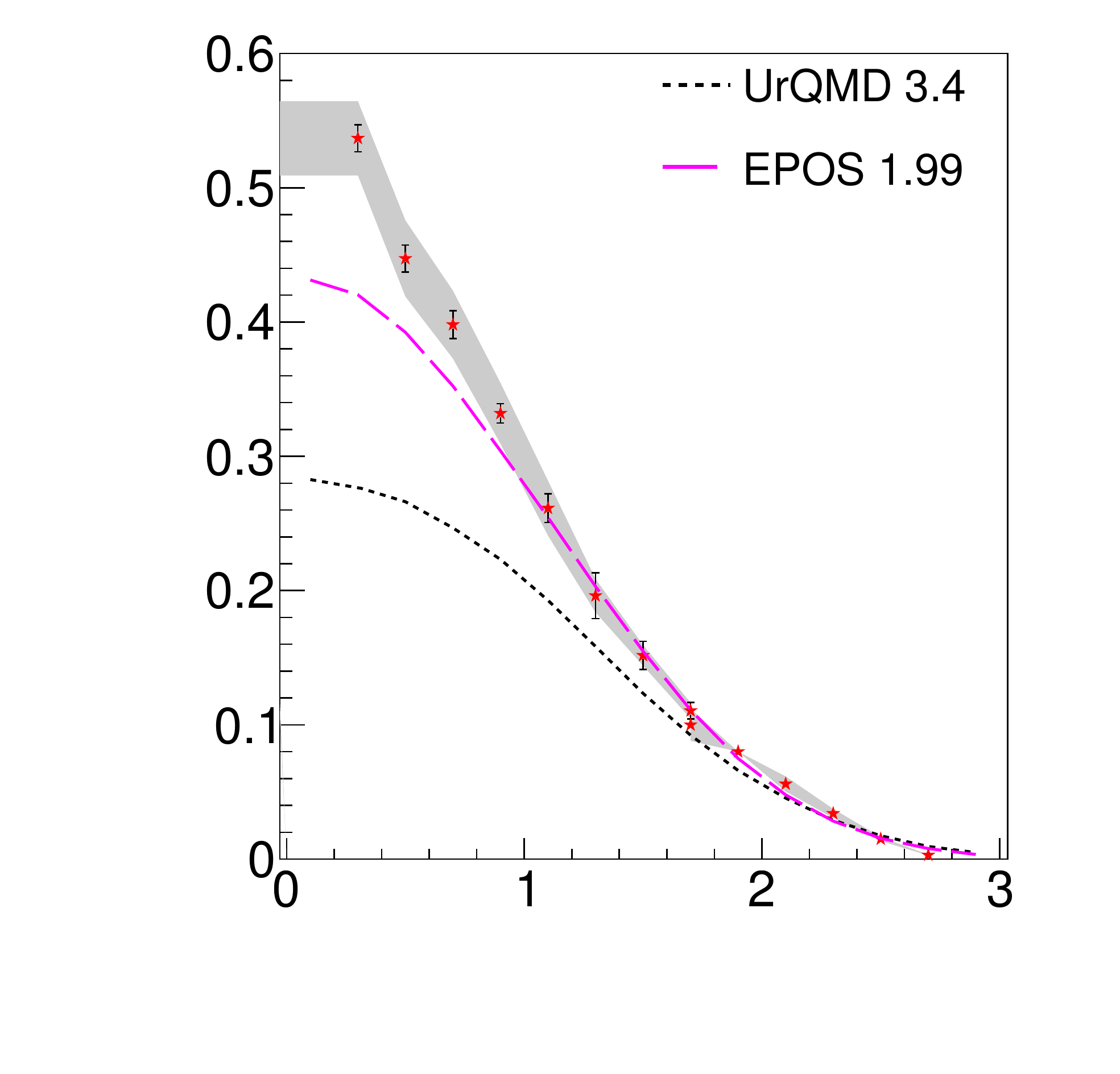} &
\hspace{-15mm}
\includegraphics[width=0.18\textwidth]{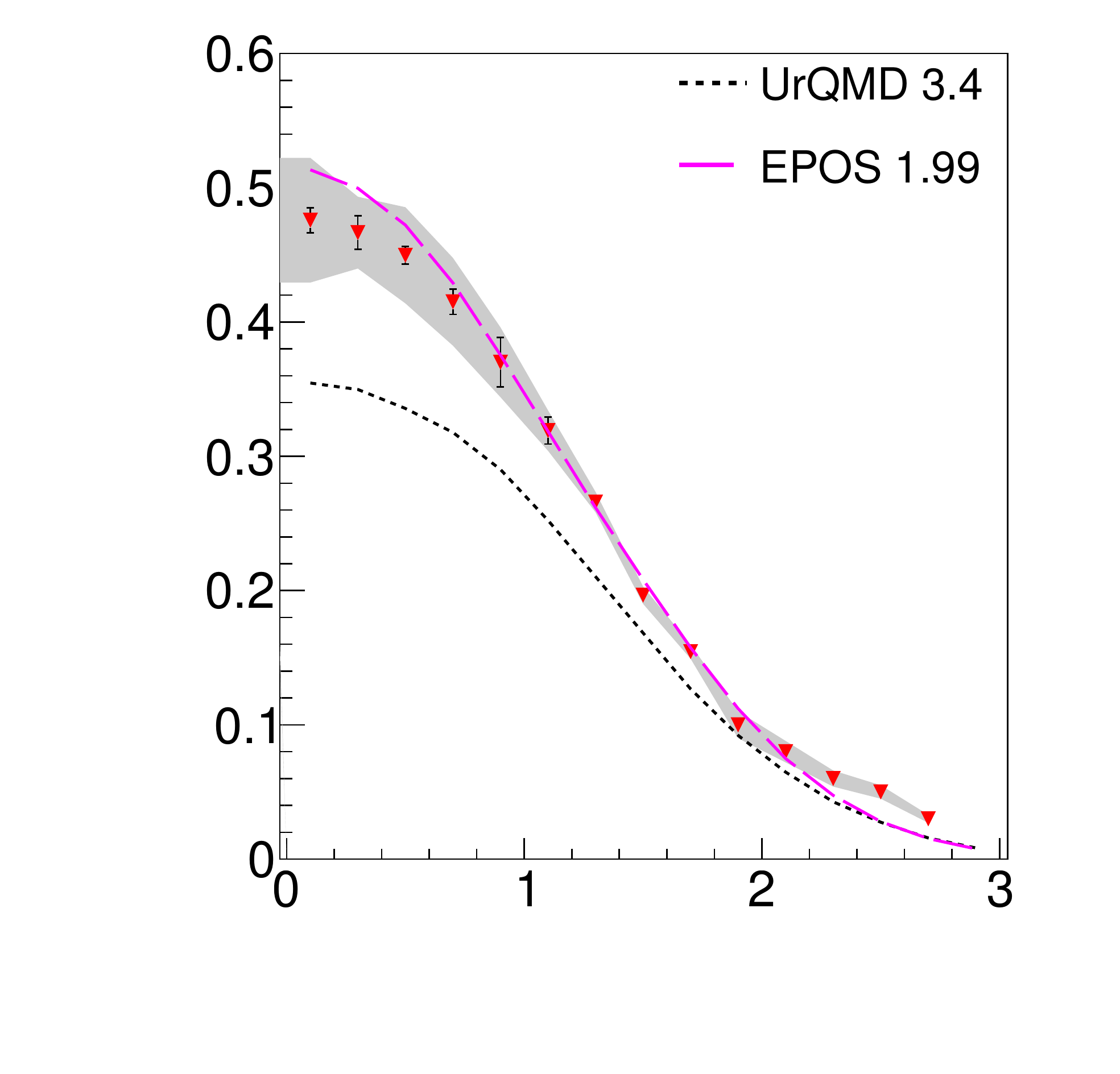} &
\hspace{-30mm}
\includegraphics[width=0.18\textwidth]{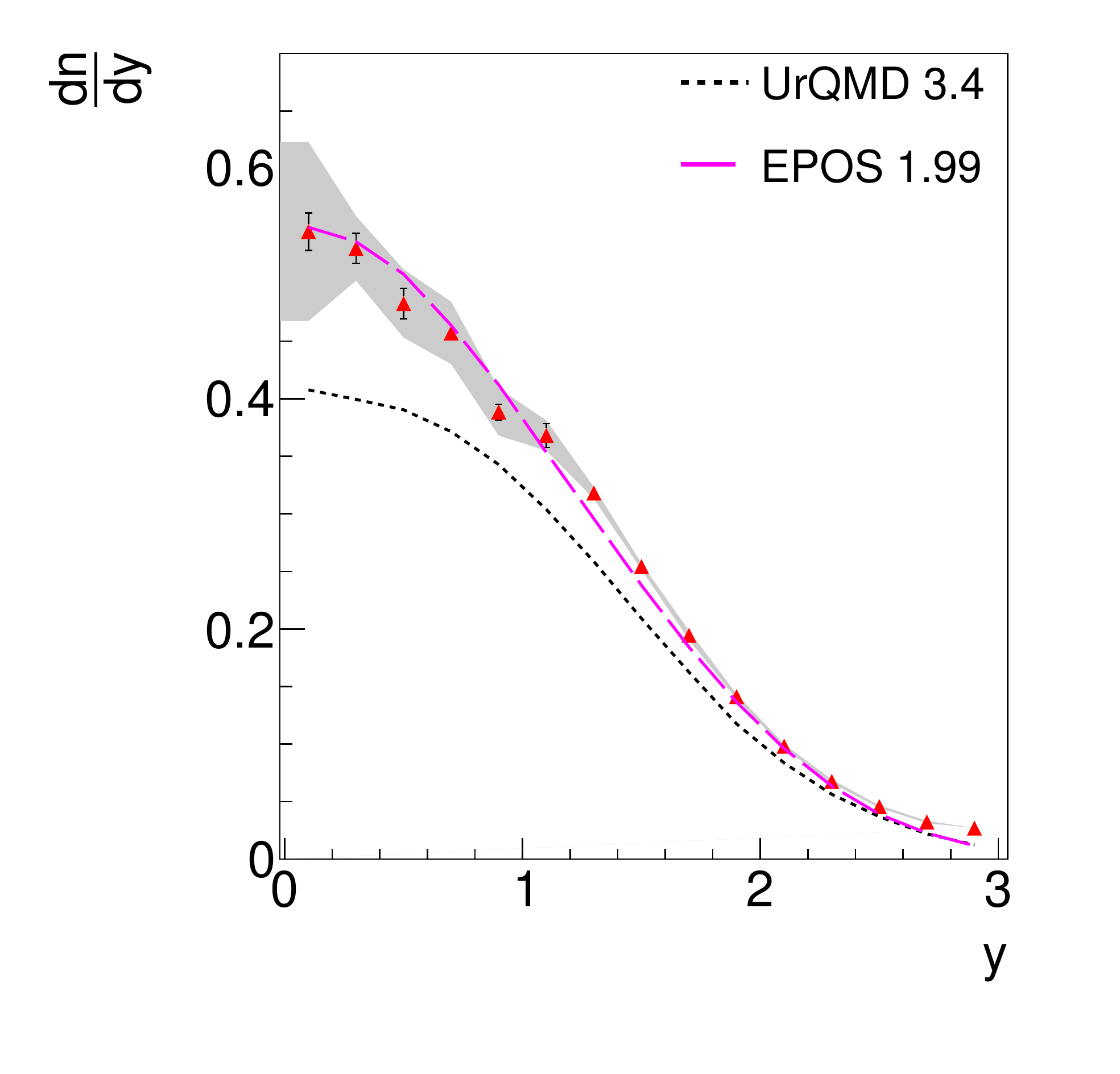} &
\hspace{-45mm}
\includegraphics[width=0.18\textwidth]{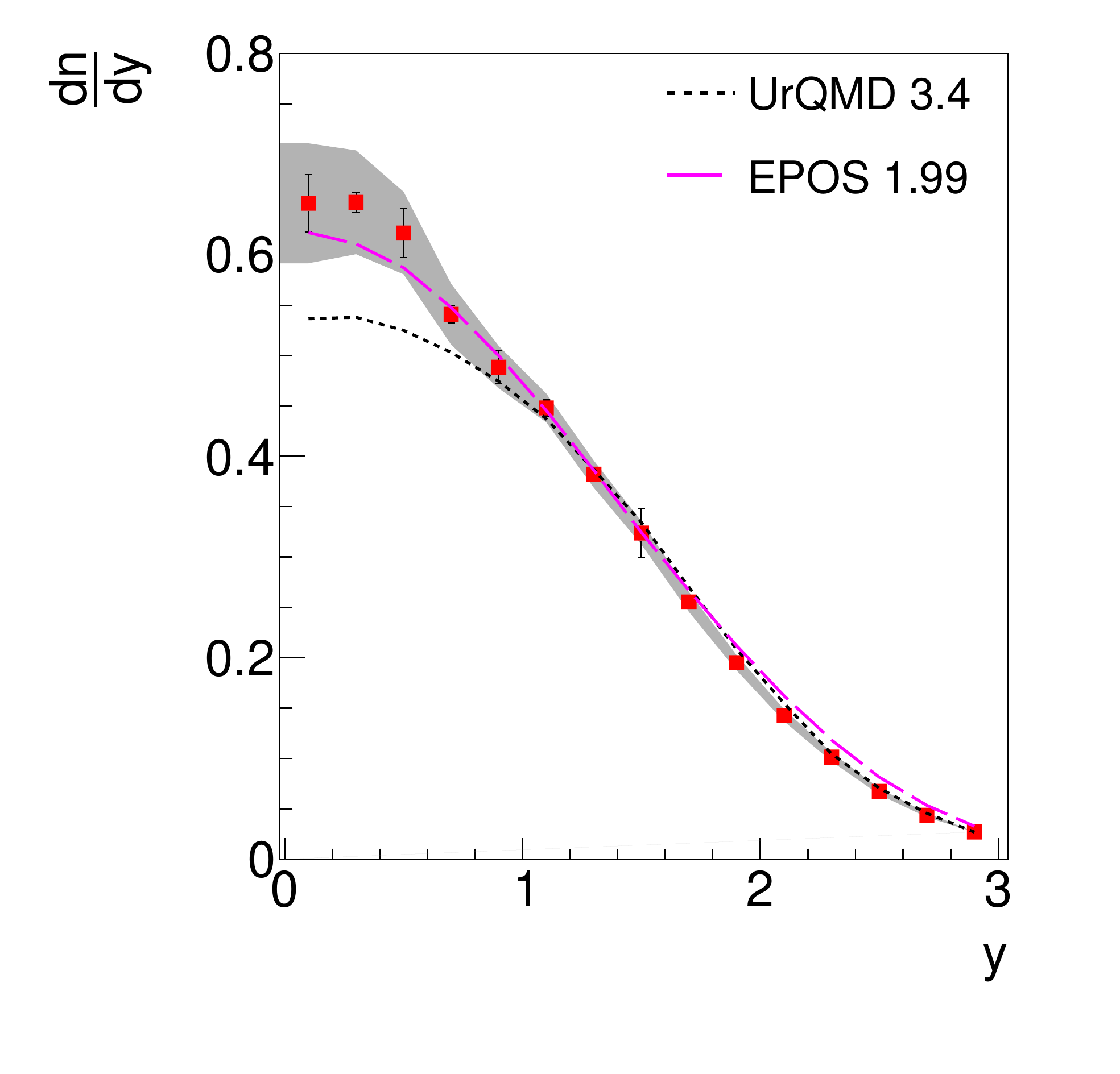} &
\hspace{-60mm}
\includegraphics[width=0.18\textwidth]{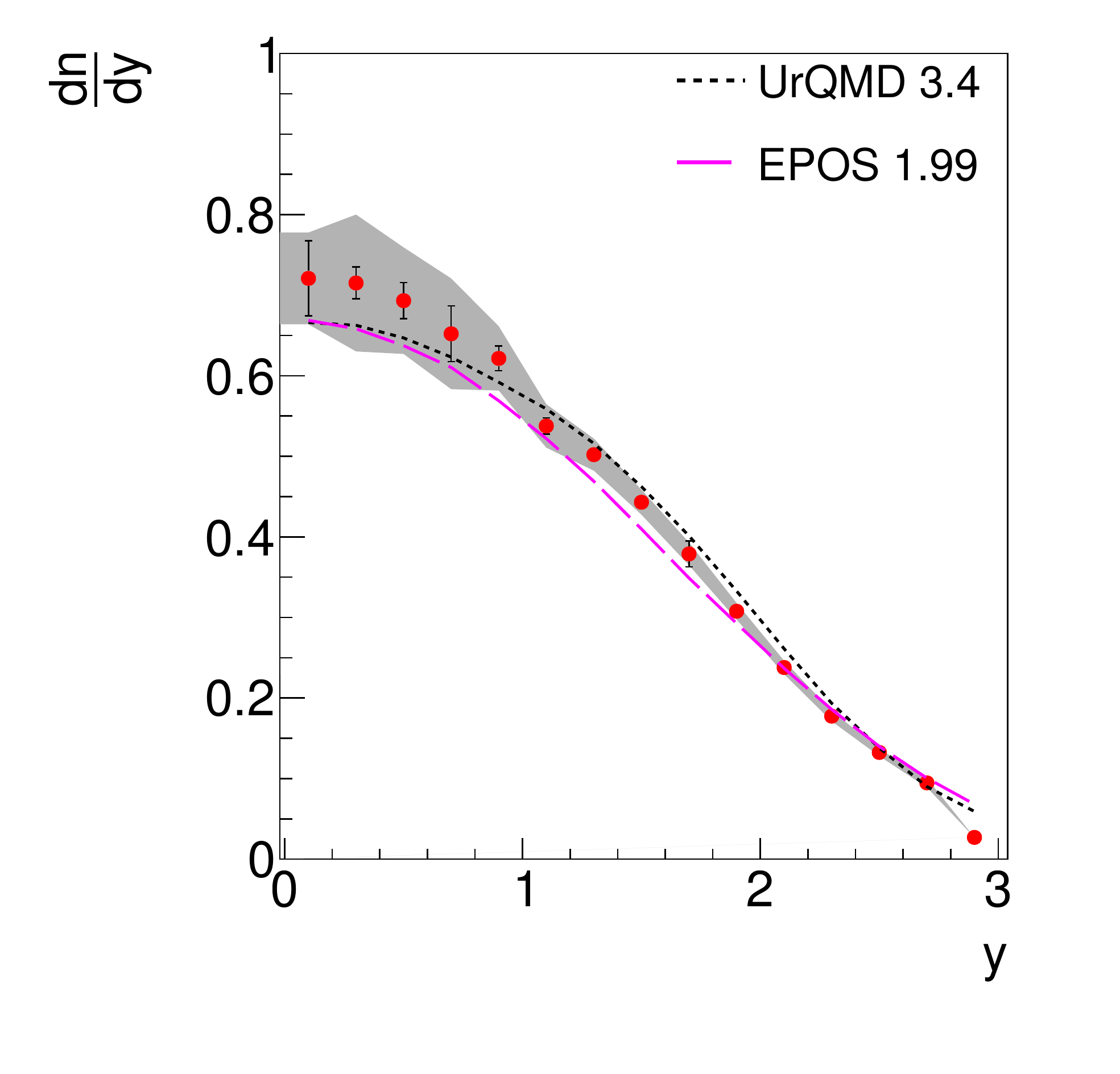} 
\tabularnewline
\vspace{-5mm}$\pi^{+}$ &
\includegraphics[width=0.18\textwidth]{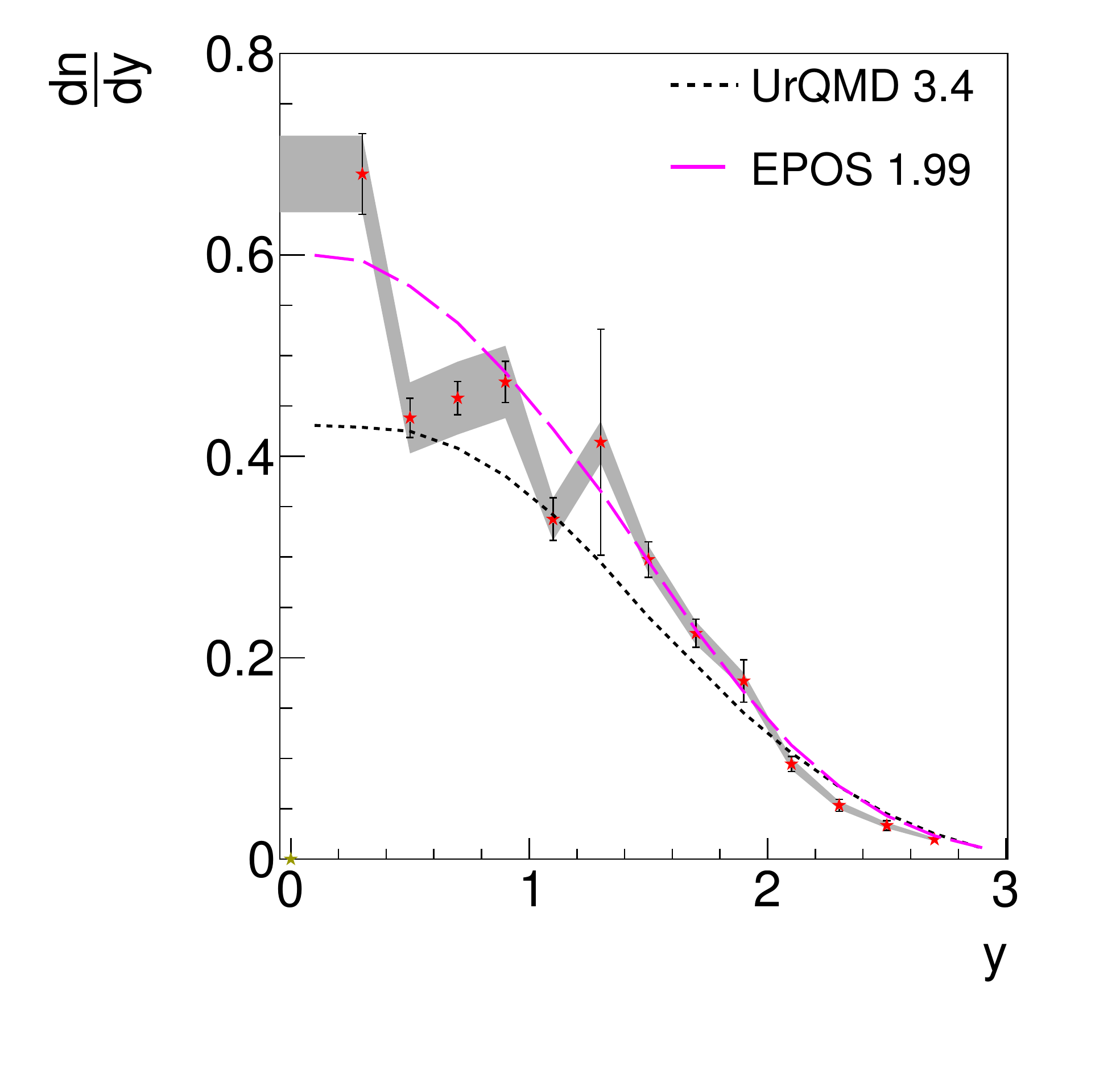} &
\hspace{-15mm}
\includegraphics[width=0.18\textwidth]{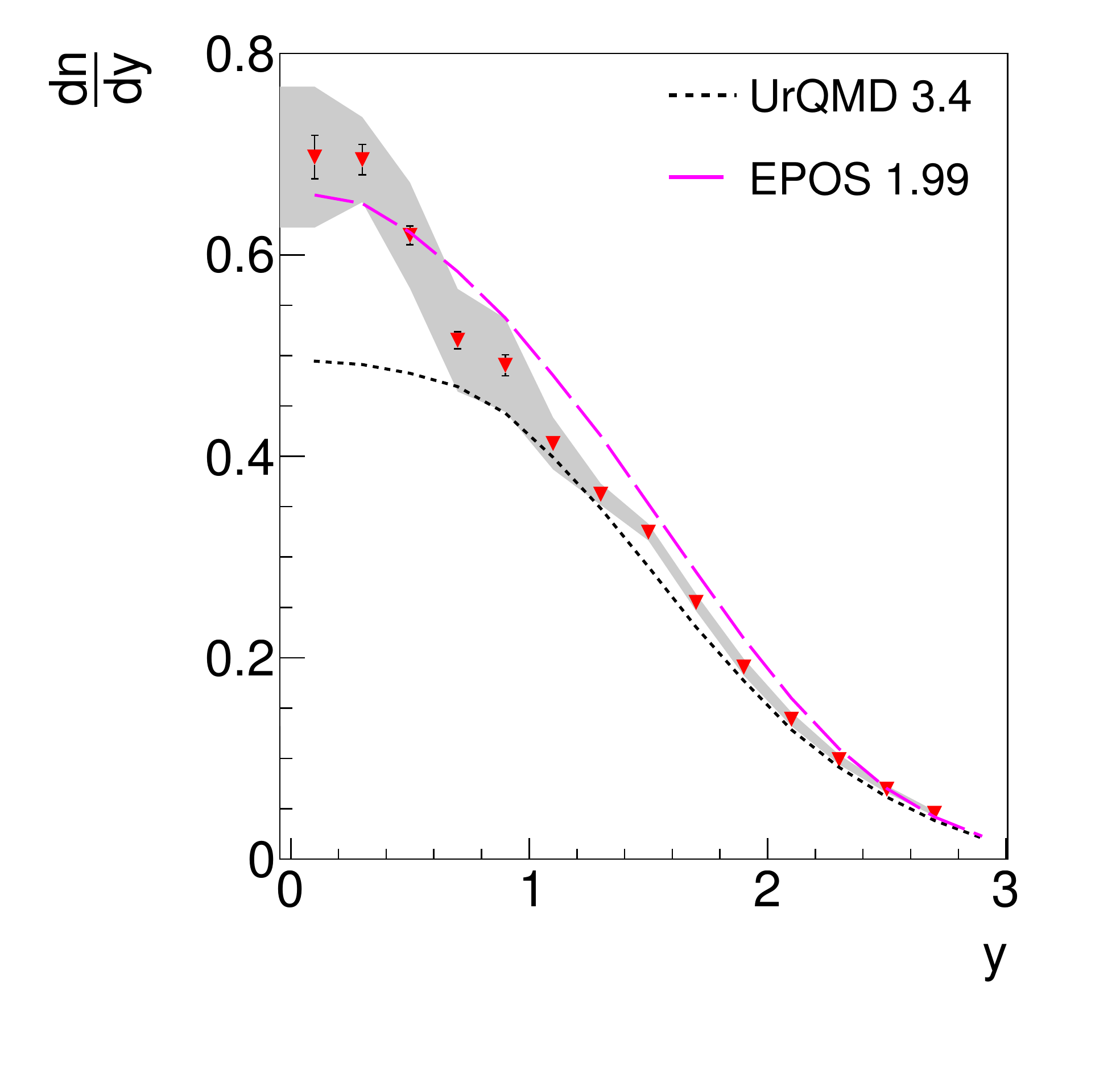} &
\hspace{-30mm}
\includegraphics[width=0.18\textwidth]{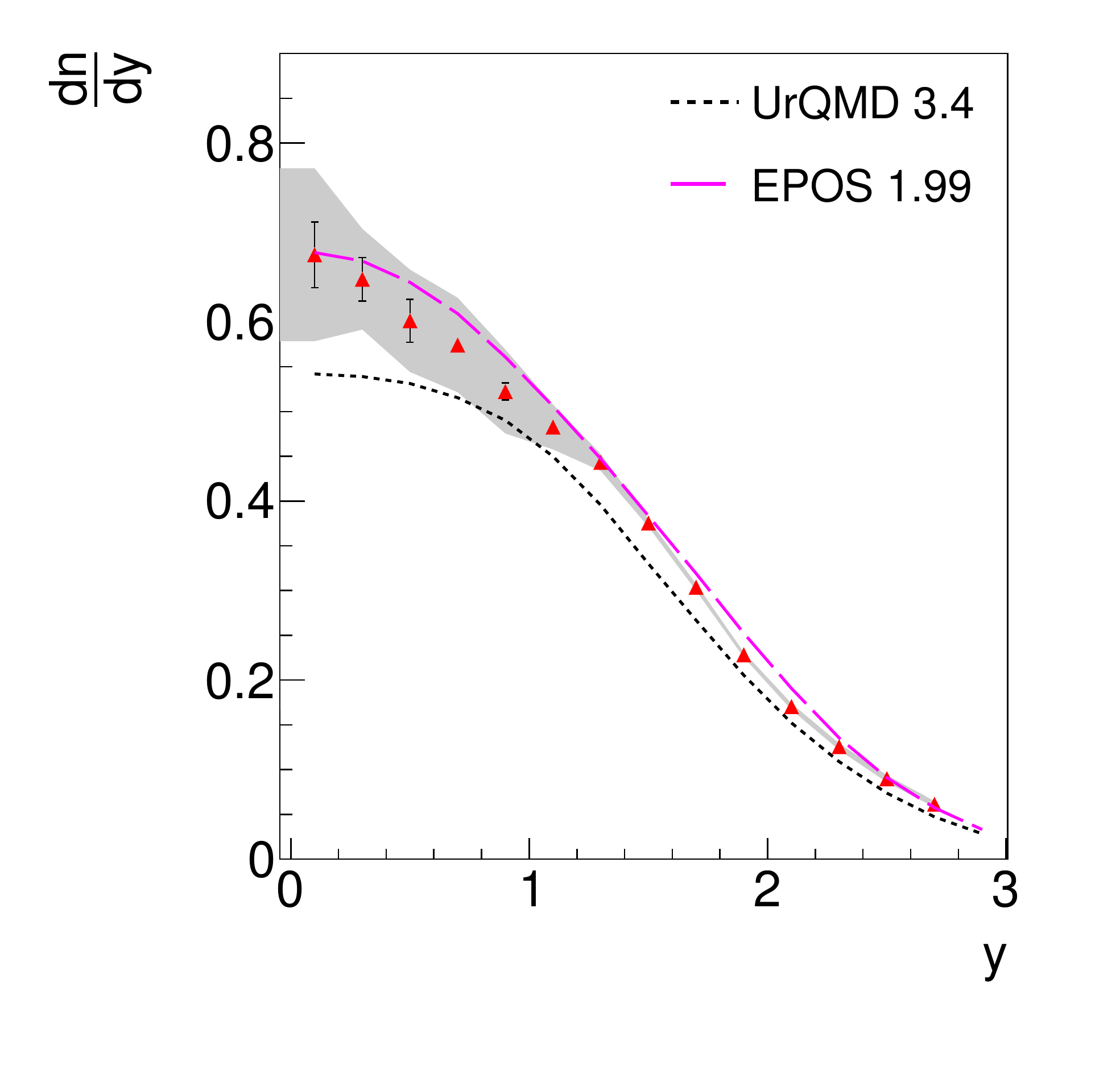} &
\hspace{-45mm}
\includegraphics[width=0.18\textwidth]{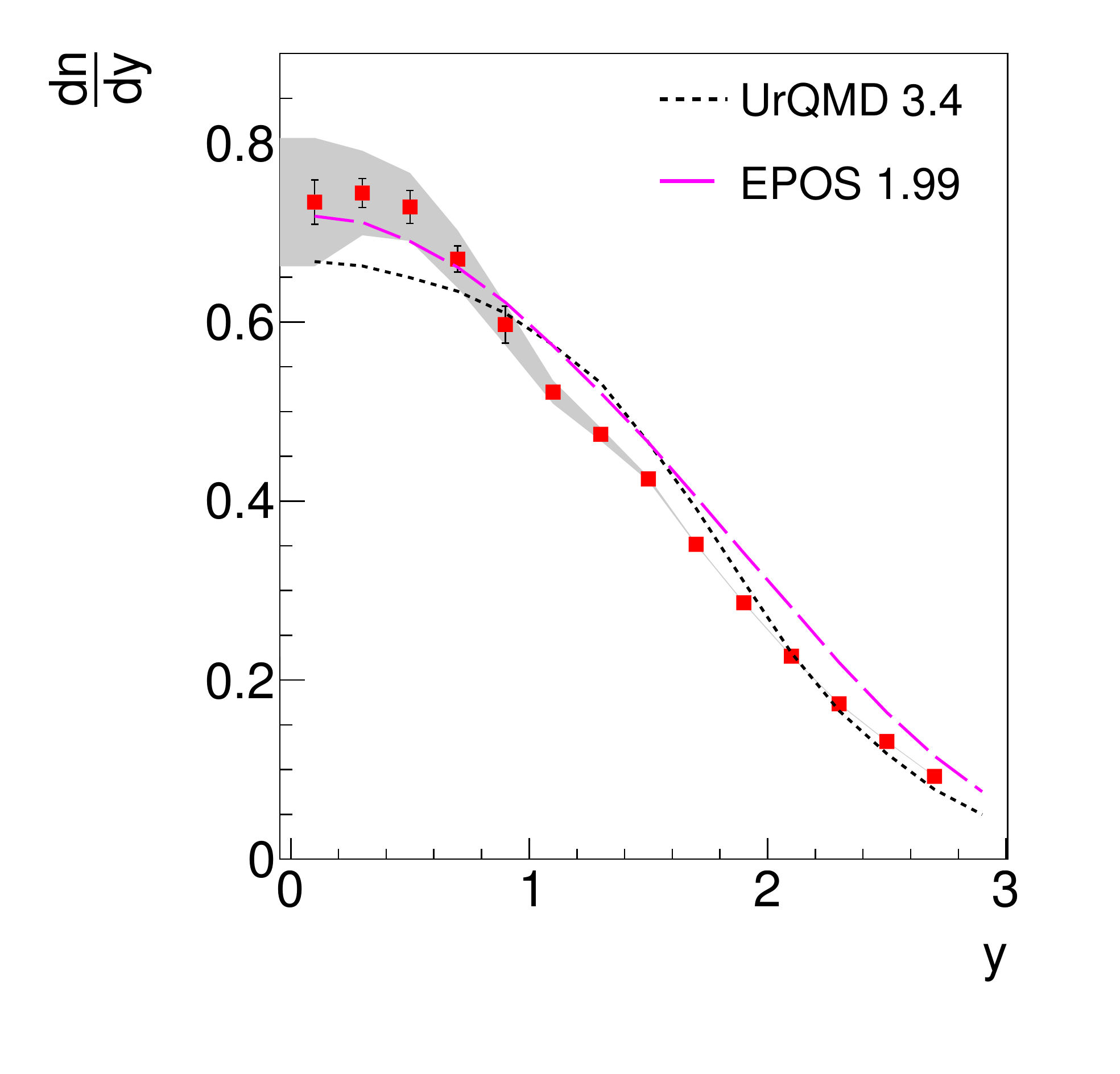} &
\hspace{-60mm}
\includegraphics[width=0.18\textwidth]{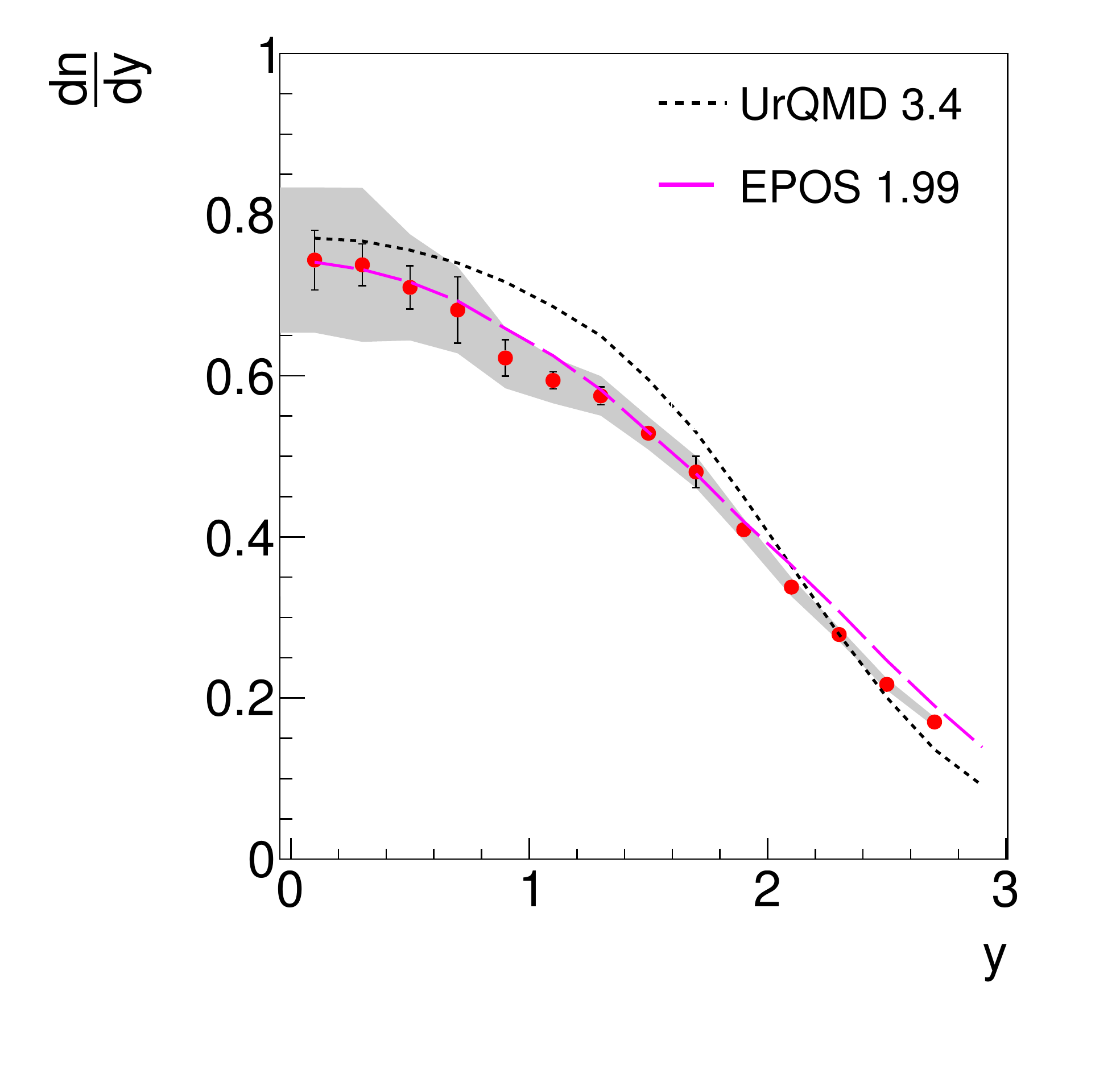} 
\tabularnewline
\vspace{-5mm}K$^{-}$ &
\includegraphics[width=0.18\textwidth]{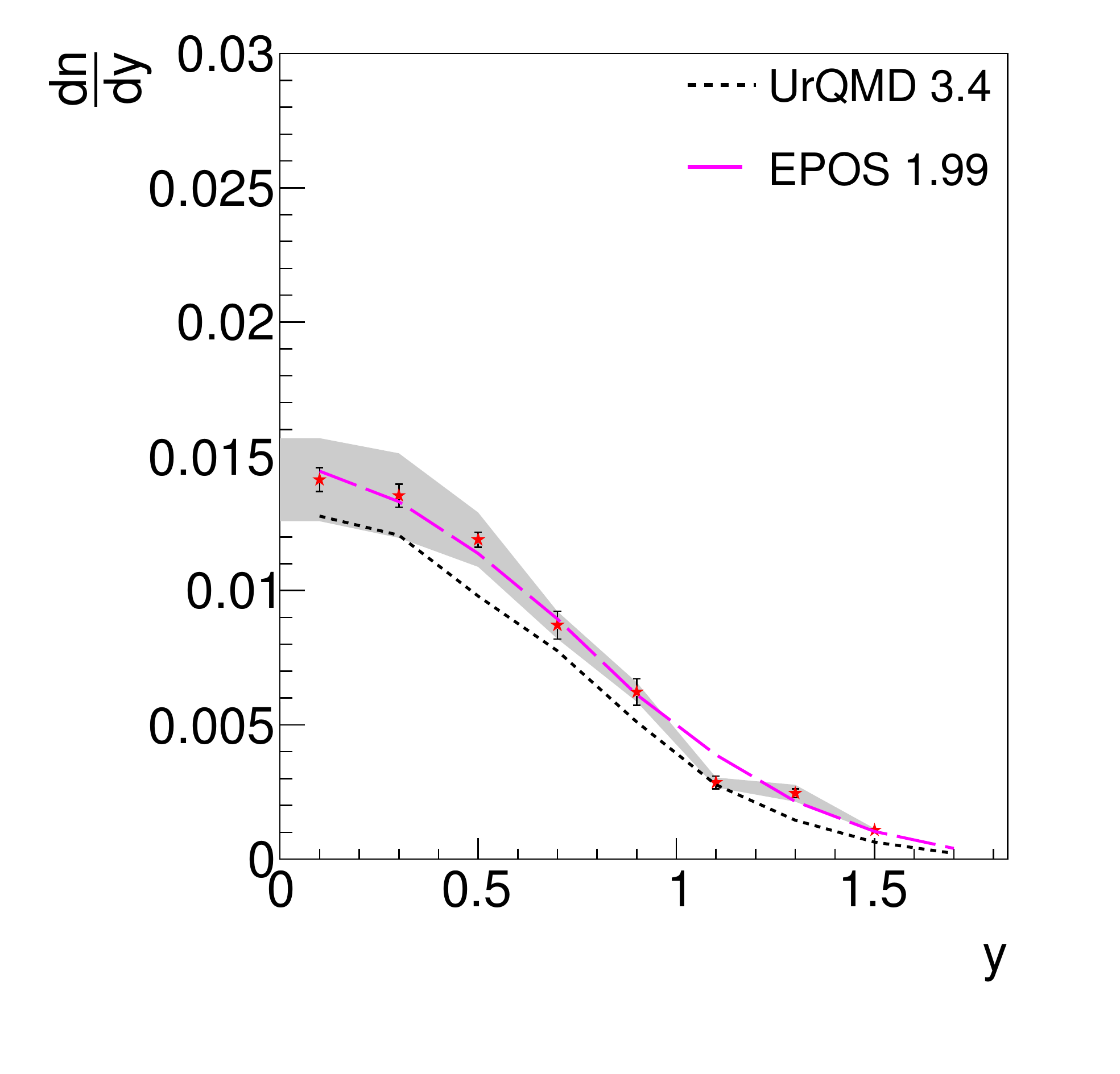} &
\hspace{-15mm}
\includegraphics[width=0.18\textwidth]{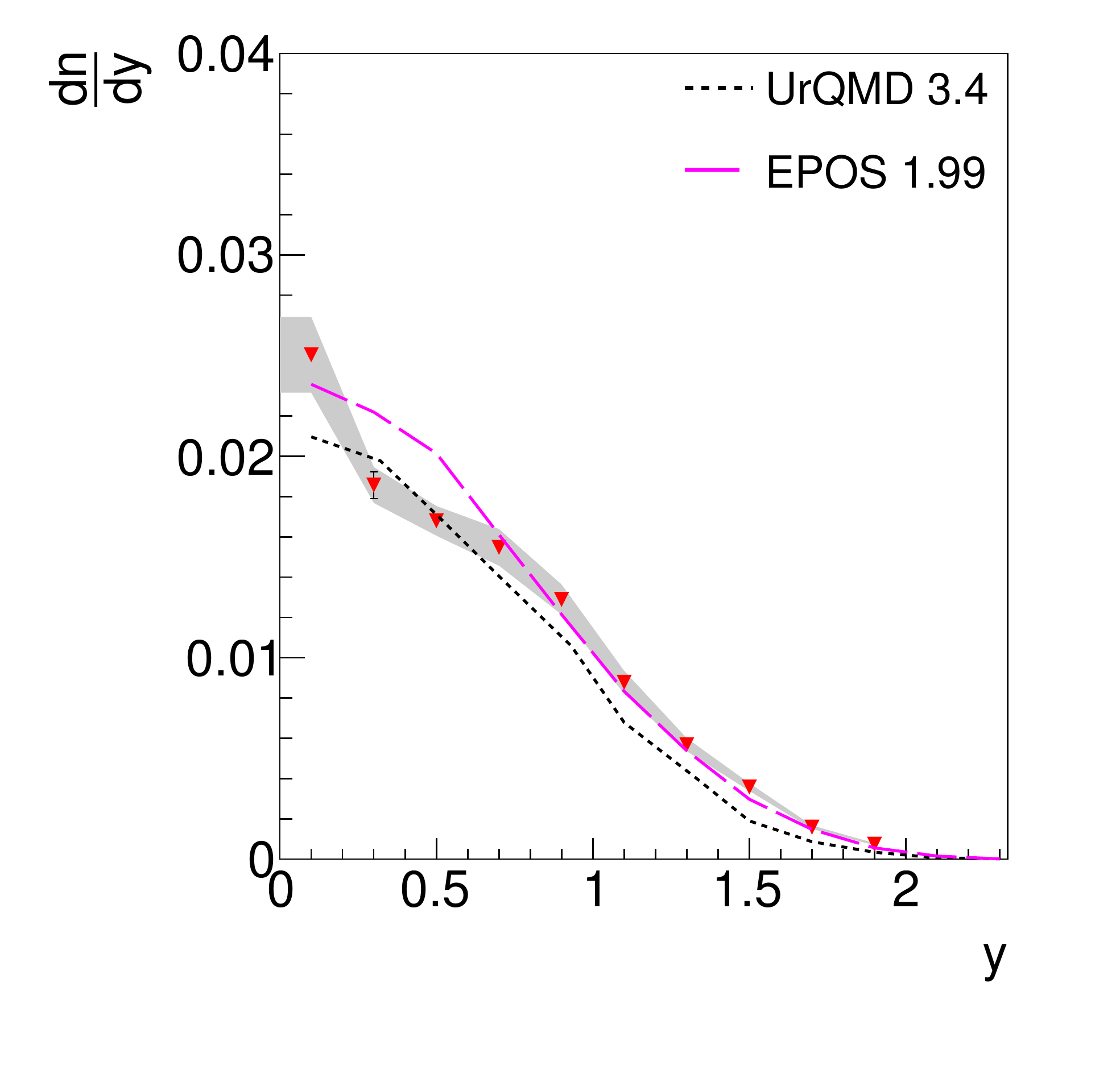} &
\hspace{-30mm}
\includegraphics[width=0.18\textwidth]{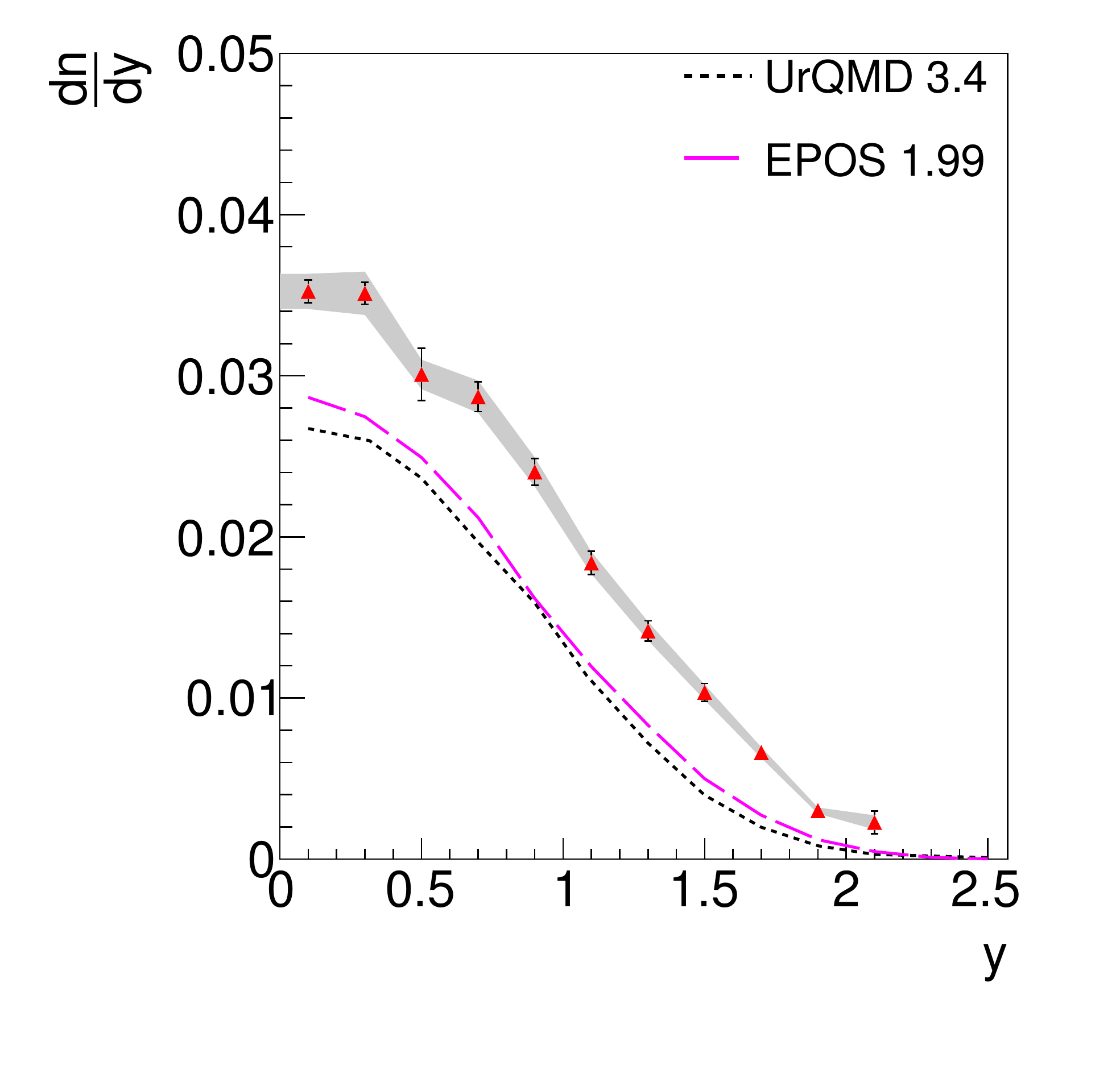} &
\hspace{-45mm}
\includegraphics[width=0.18\textwidth]{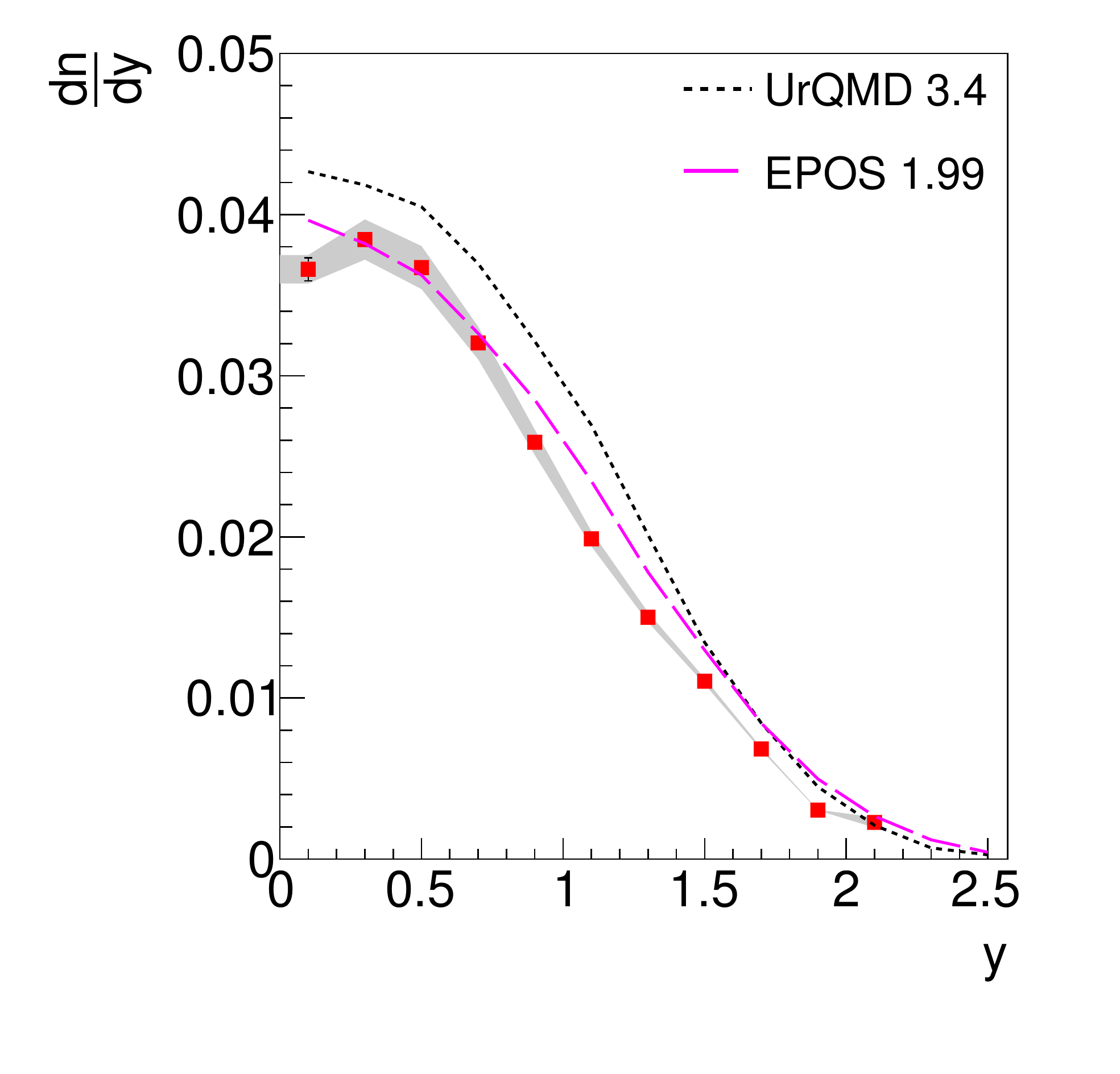} &
\hspace{-60mm}
\includegraphics[width=0.18\textwidth]{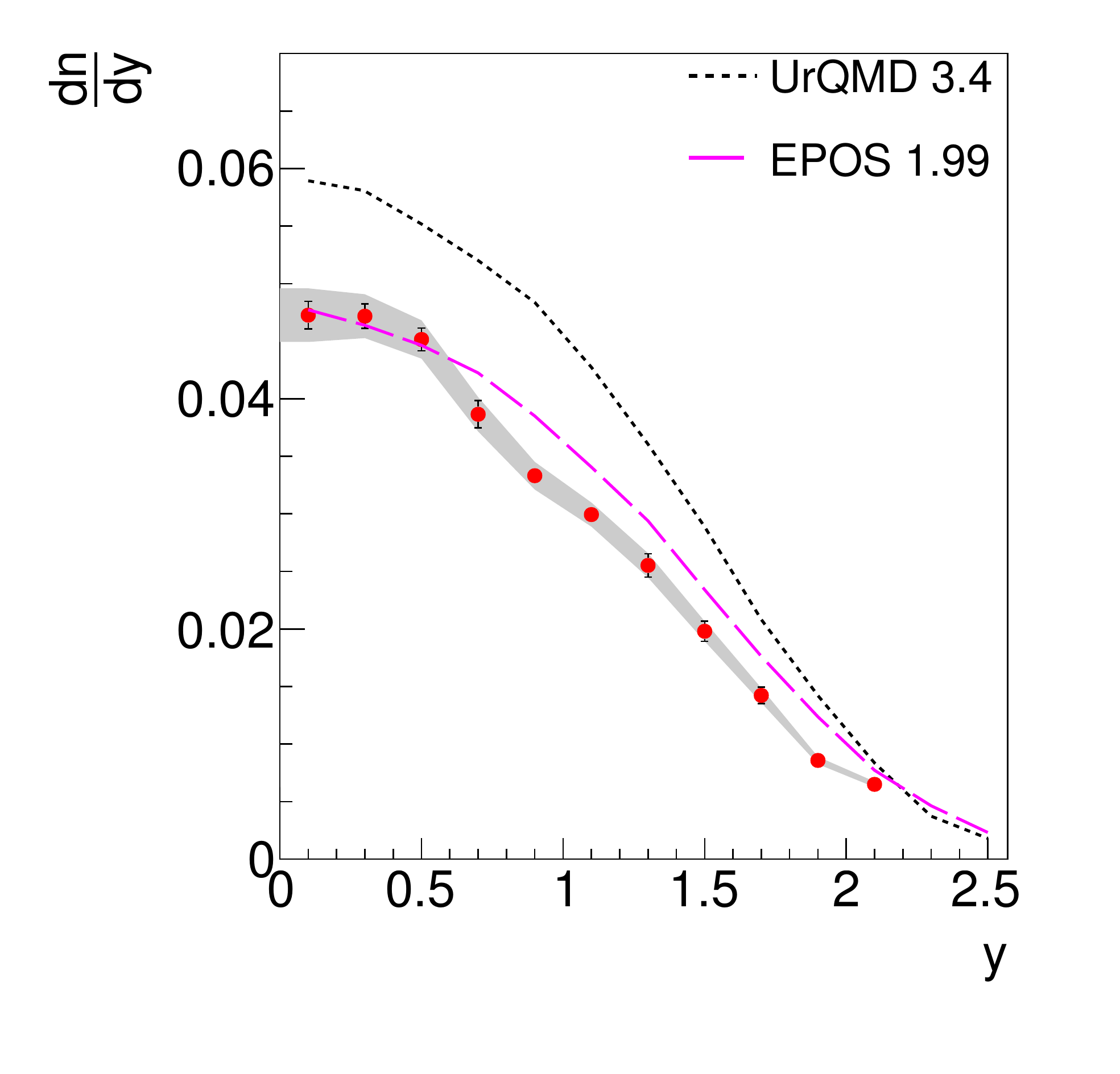} 
\tabularnewline
\vspace{-5mm}K$^{+}$ &
\includegraphics[width=0.18\textwidth]{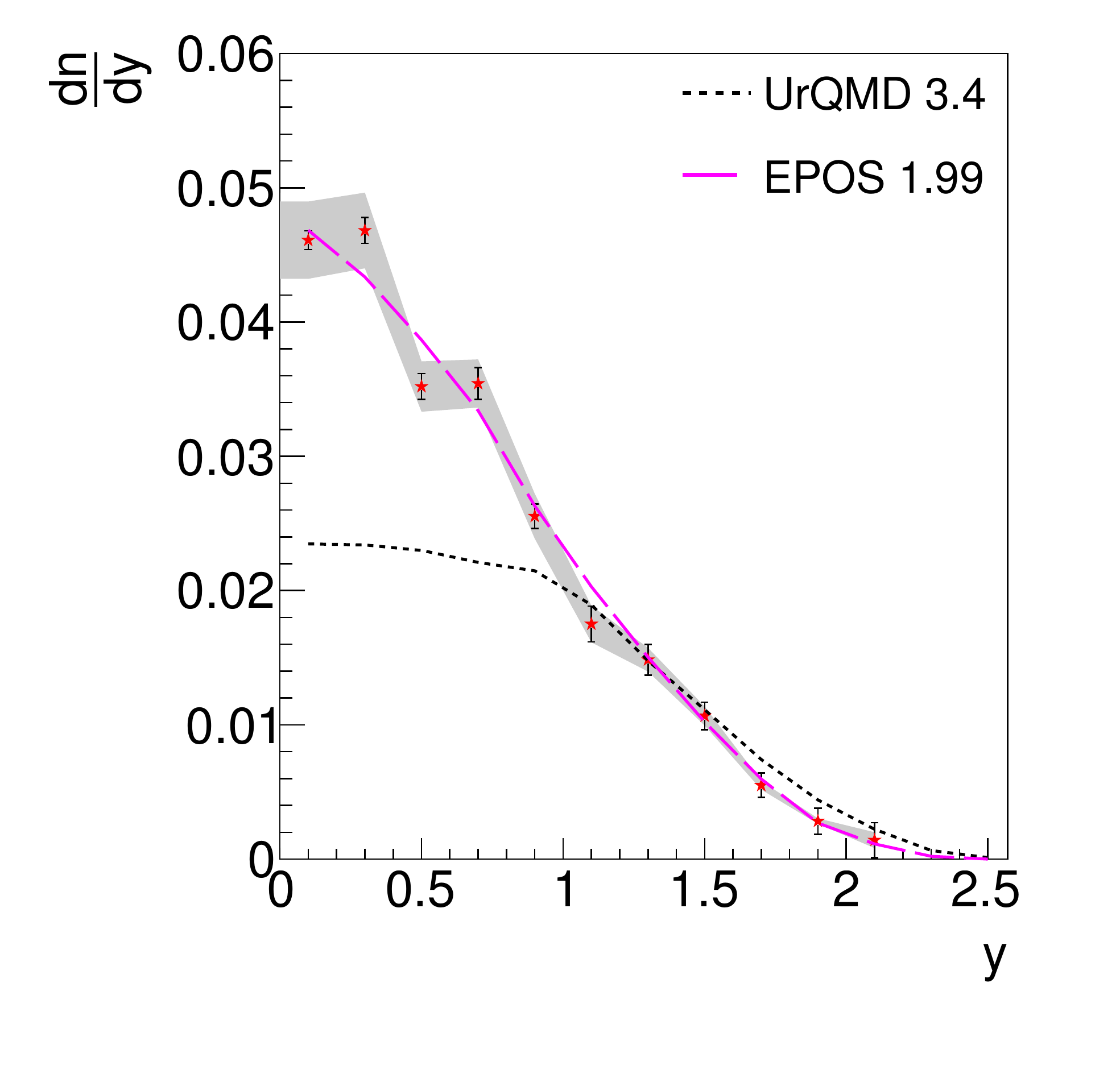} &
\hspace{-15mm}
\includegraphics[width=0.18\textwidth]{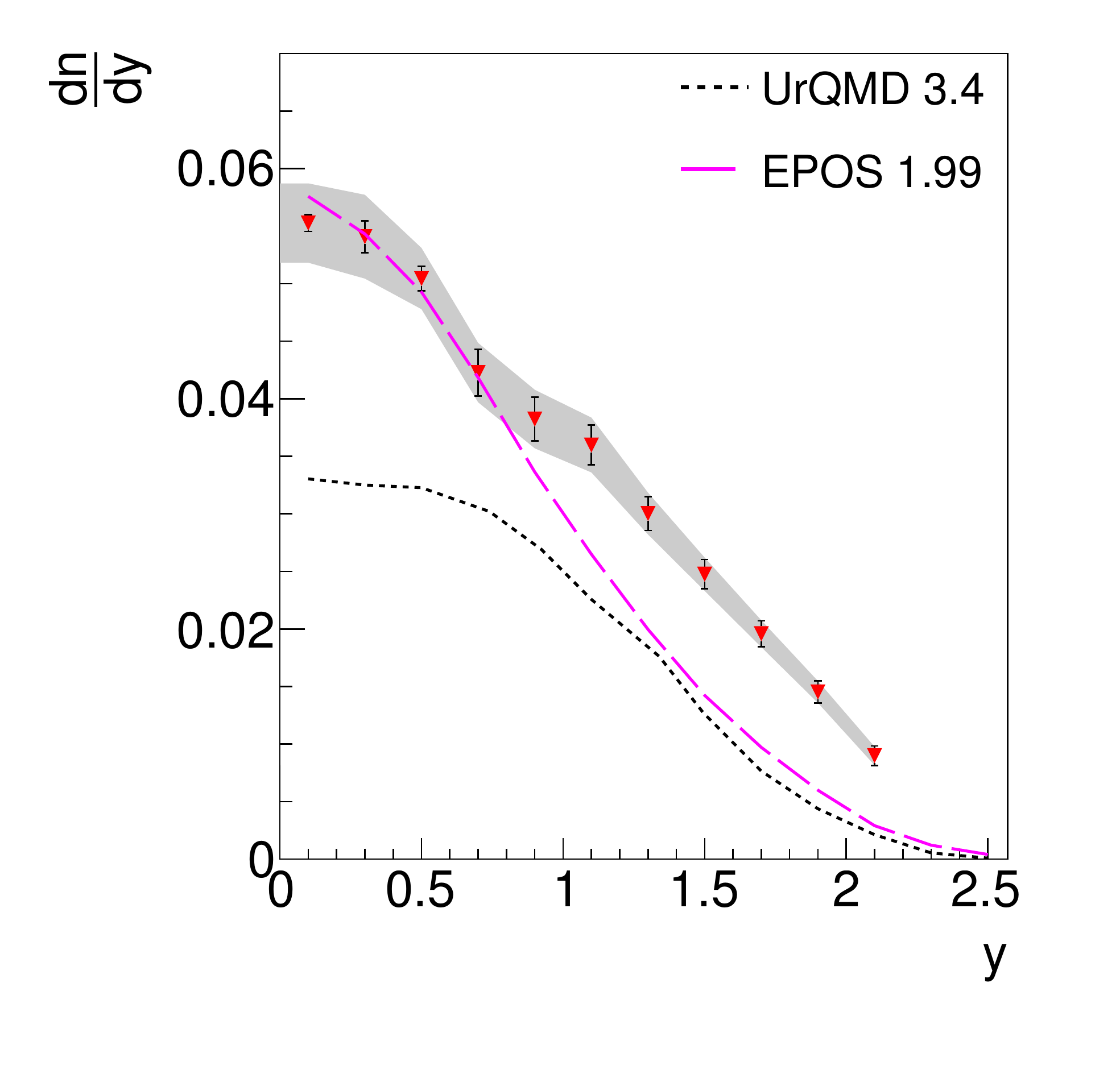} &
\hspace{-30mm}
\includegraphics[width=0.18\textwidth]{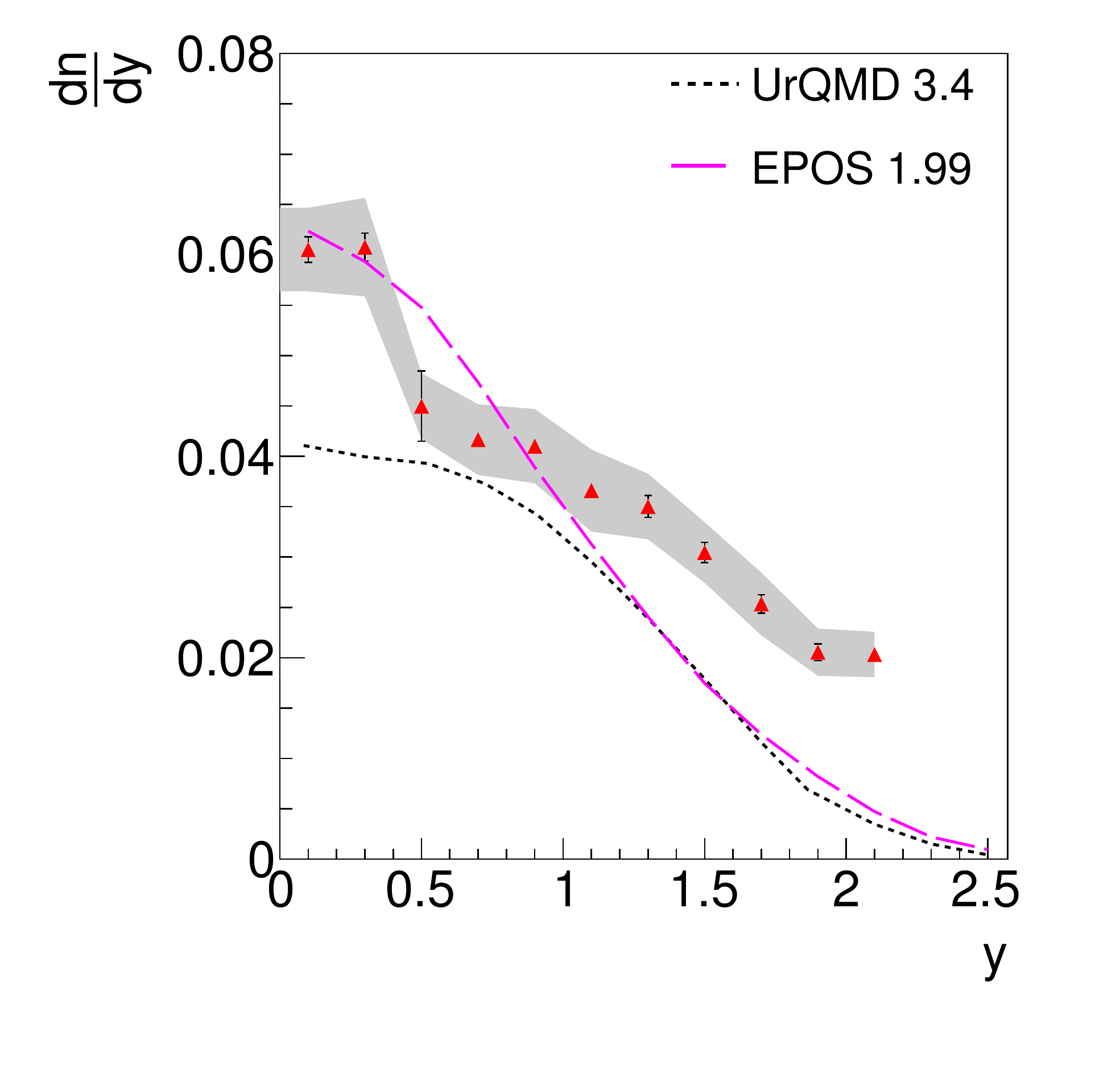} &
\hspace{-45mm}
\includegraphics[width=0.18\textwidth]{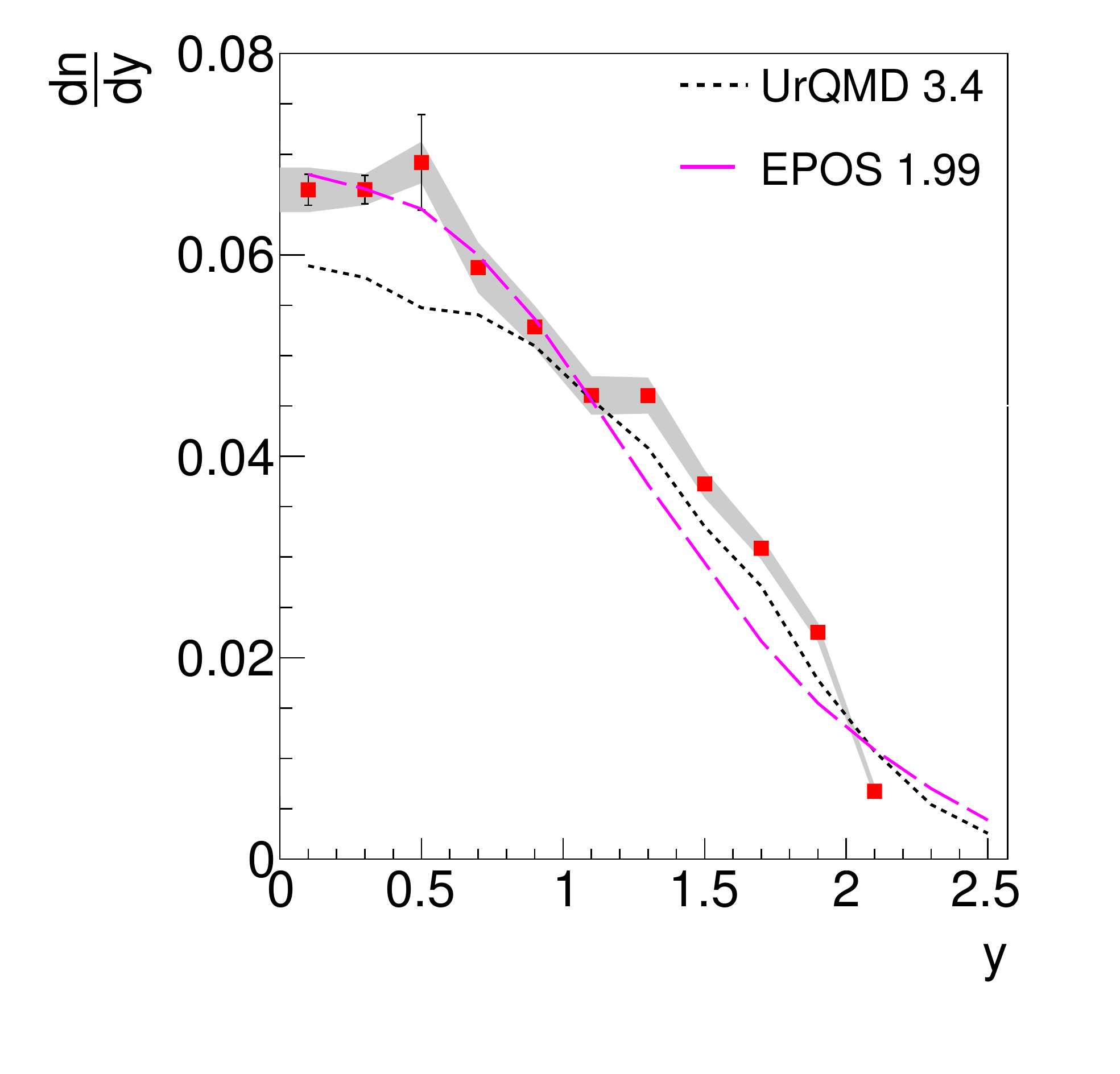} &
\hspace{-60mm}
\includegraphics[width=0.18\textwidth]{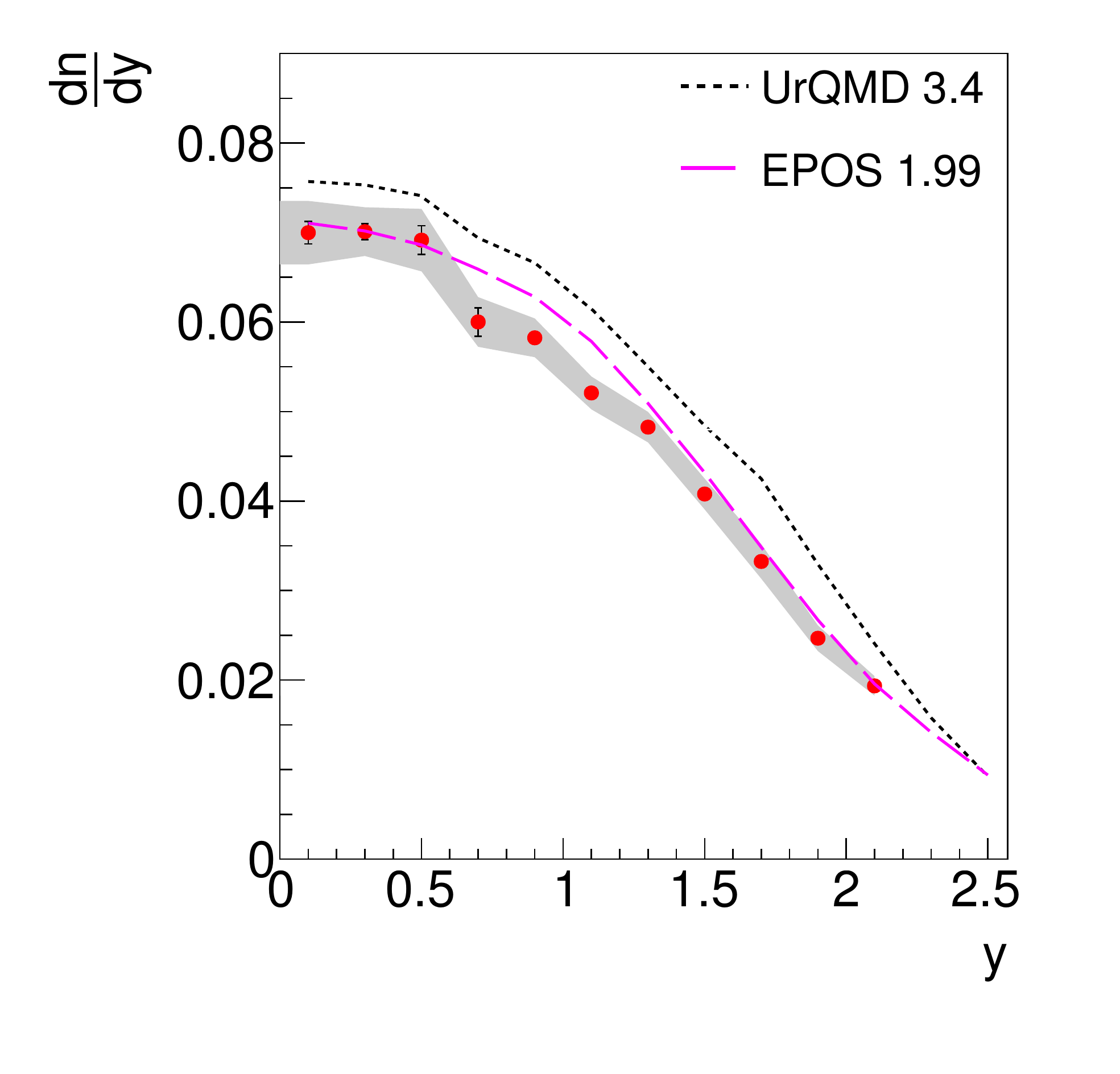} 
\tabularnewline
\vspace{-5mm}p &
\includegraphics[width=0.18\textwidth]{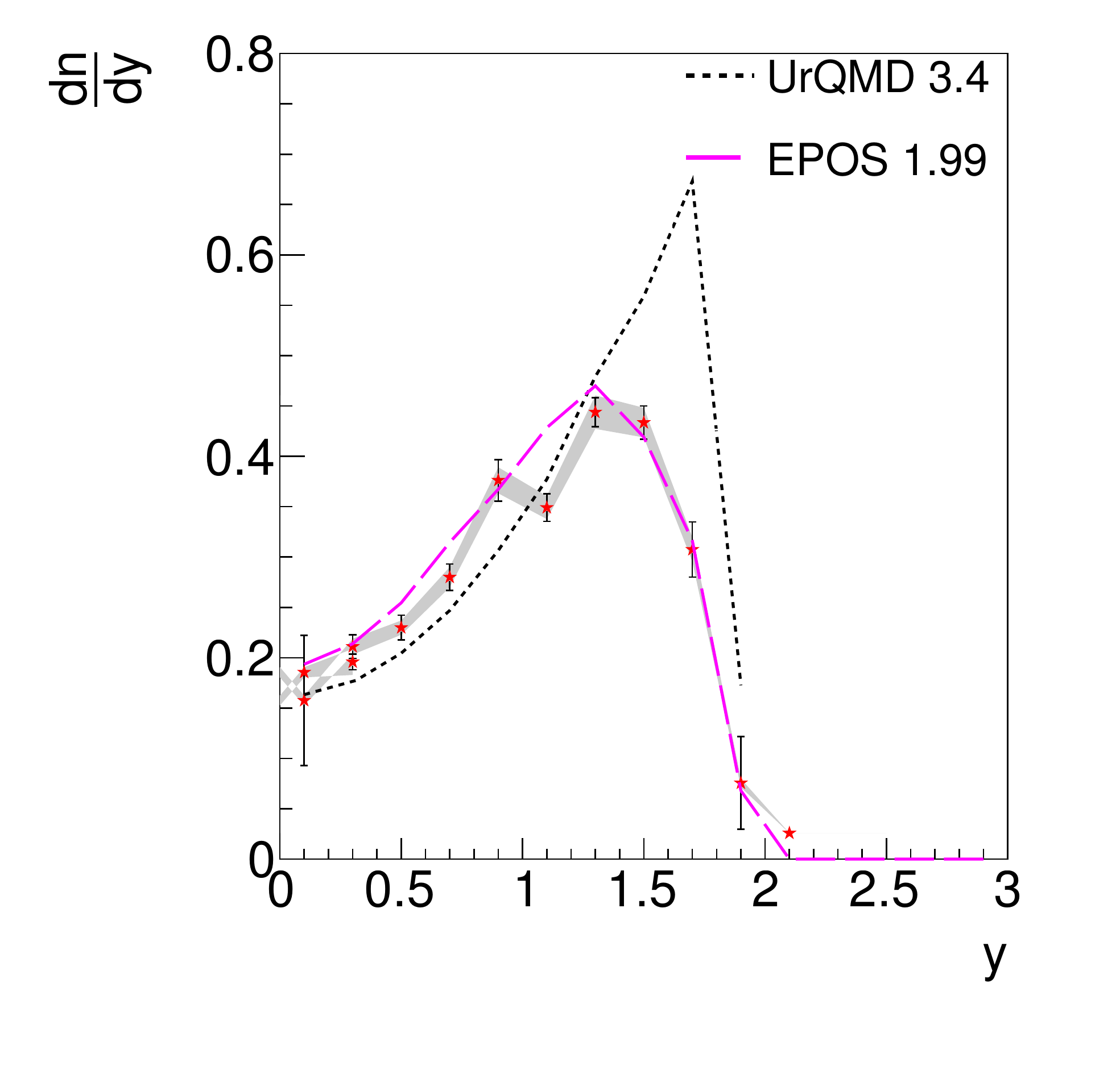} &
\hspace{-15mm}
\includegraphics[width=0.18\textwidth]{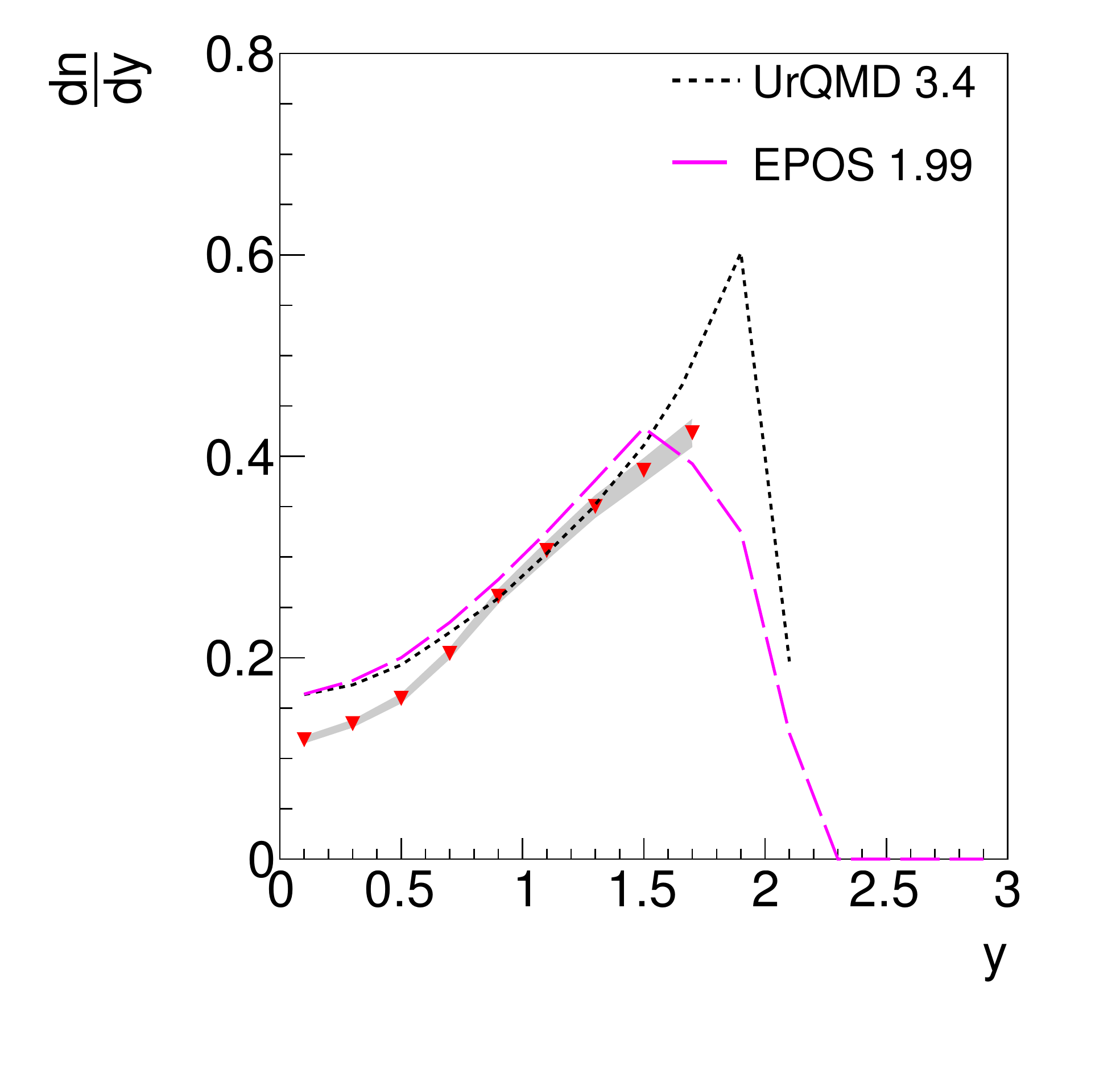} &
\hspace{-30mm}
\includegraphics[width=0.18\textwidth]{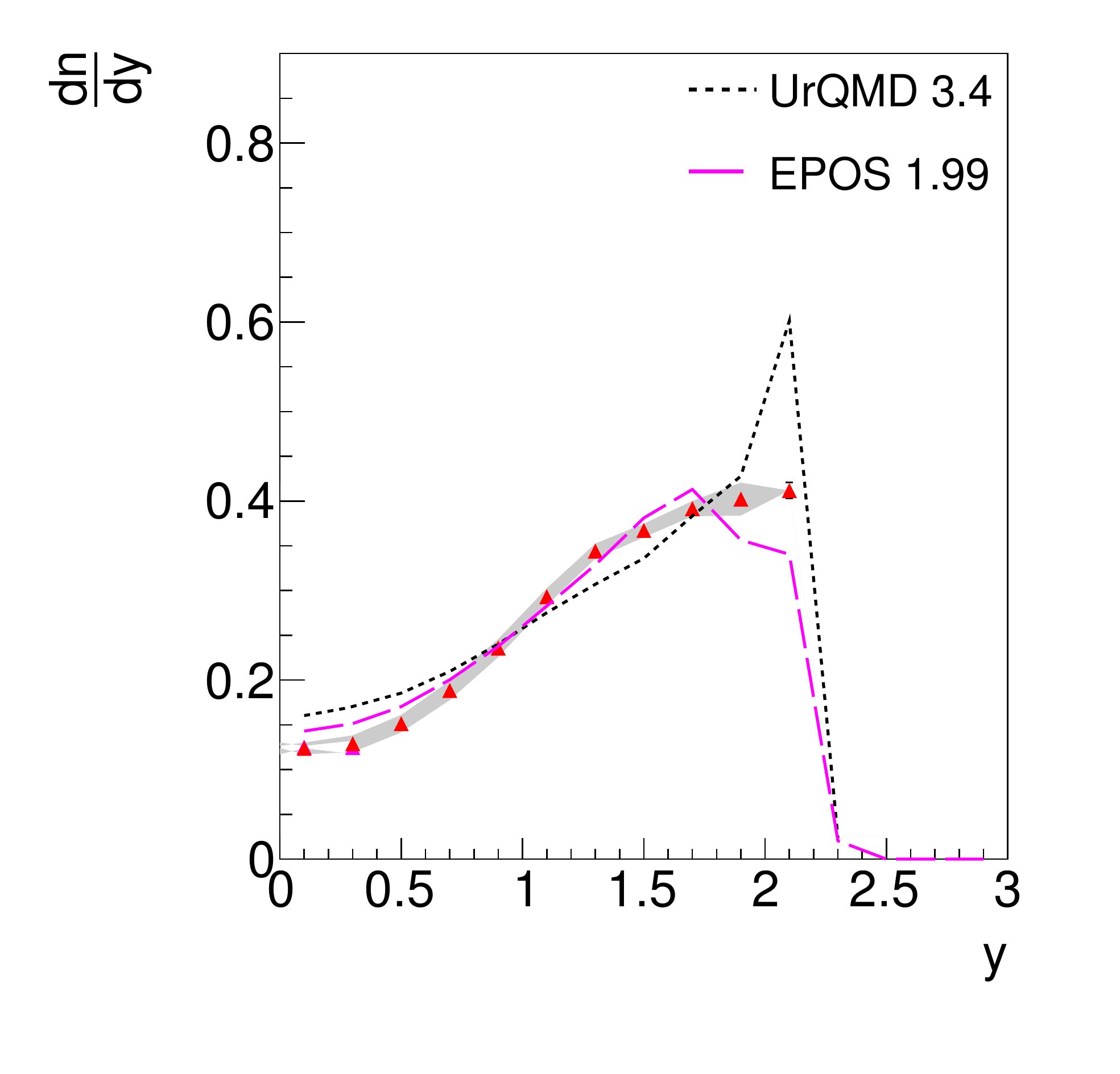} &
\hspace{-45mm}
\includegraphics[width=0.18\textwidth]{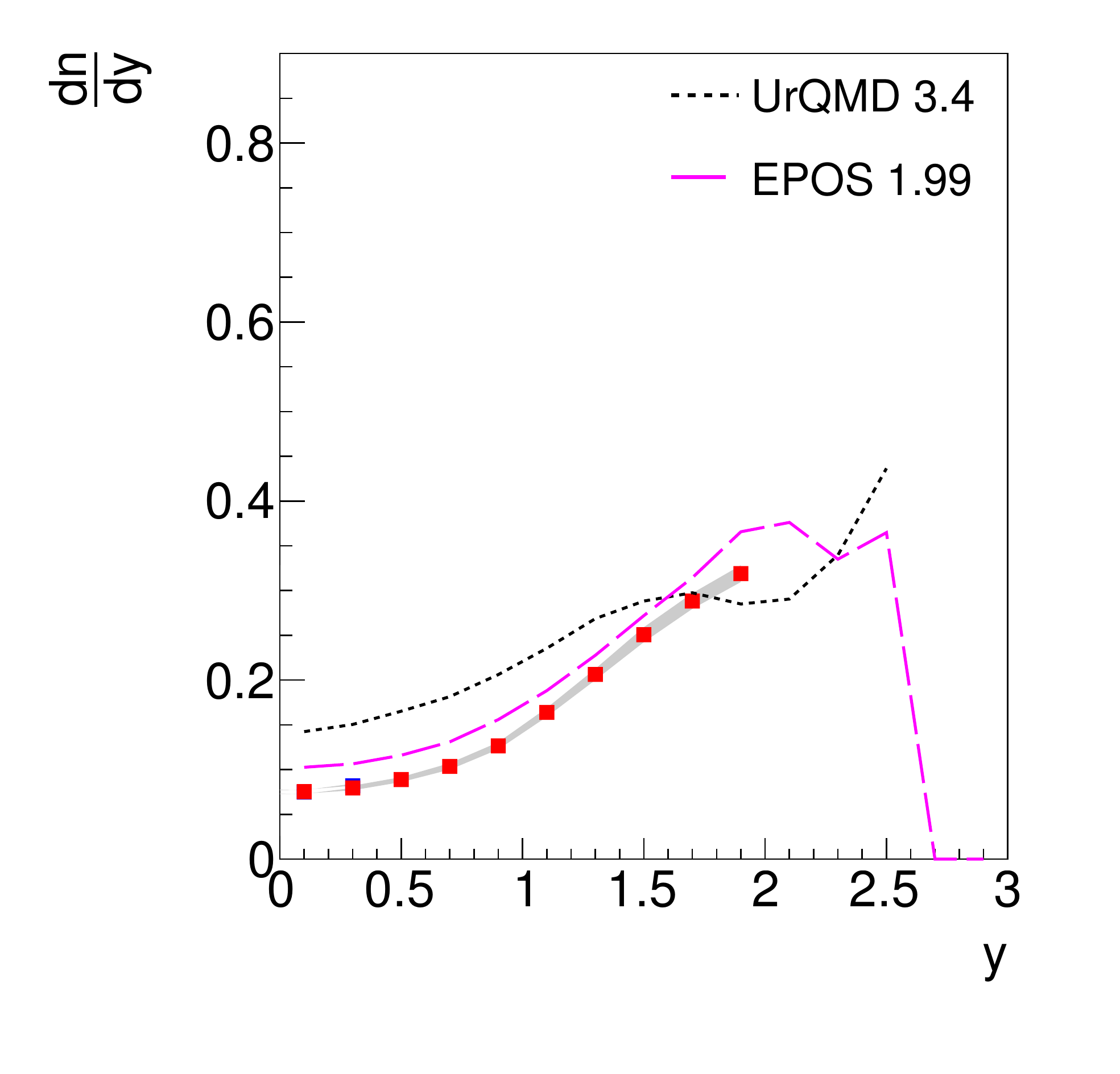} &
\hspace{-60mm}
\includegraphics[width=0.18\textwidth]{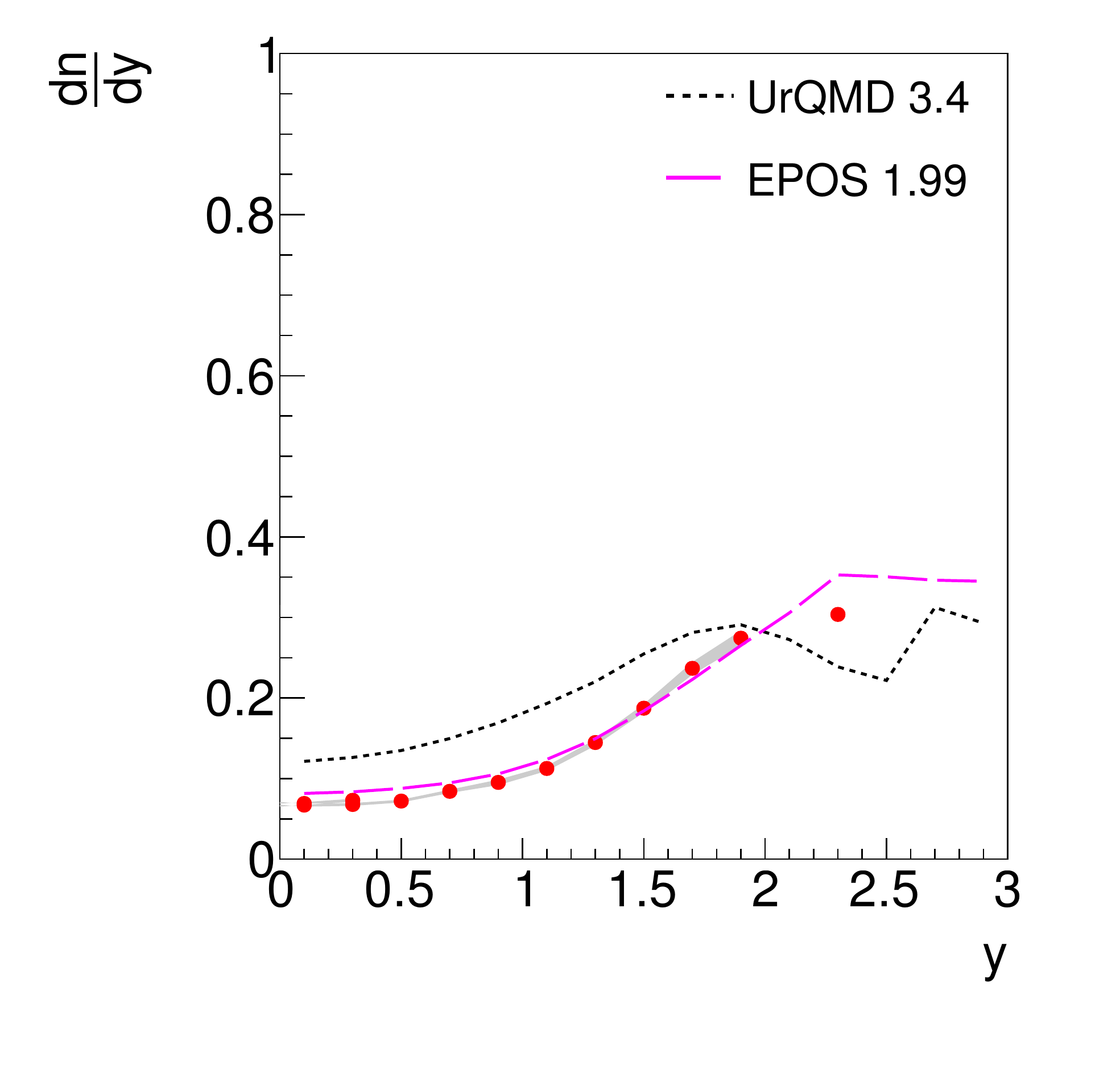} 
\tabularnewline
\vspace{-5mm}$\bar{\textrm{p}}$ &
&
\hspace{-15mm}
\includegraphics[width=0.18\textwidth]{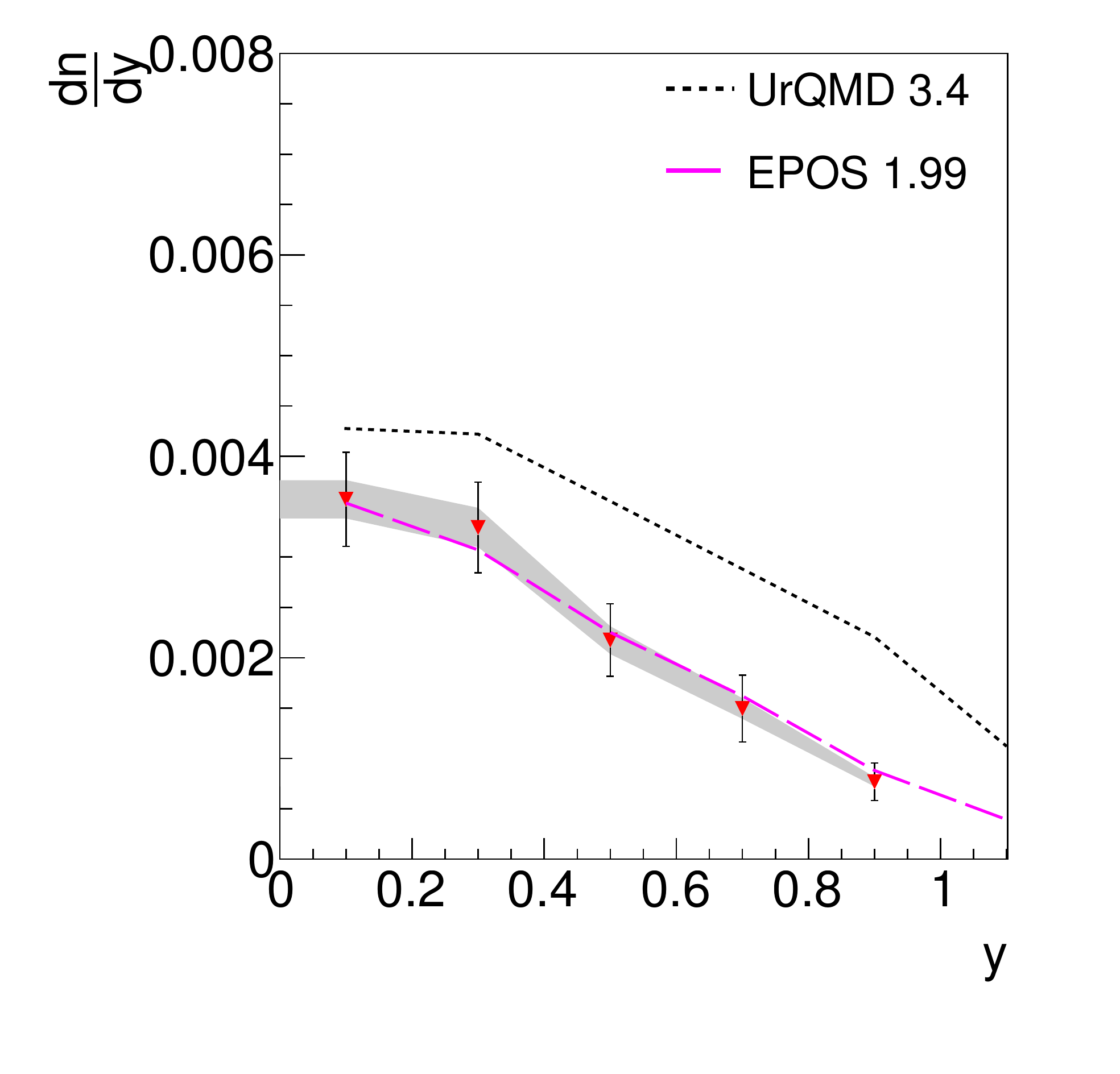} &
\hspace{-30mm}
\includegraphics[width=0.18\textwidth]{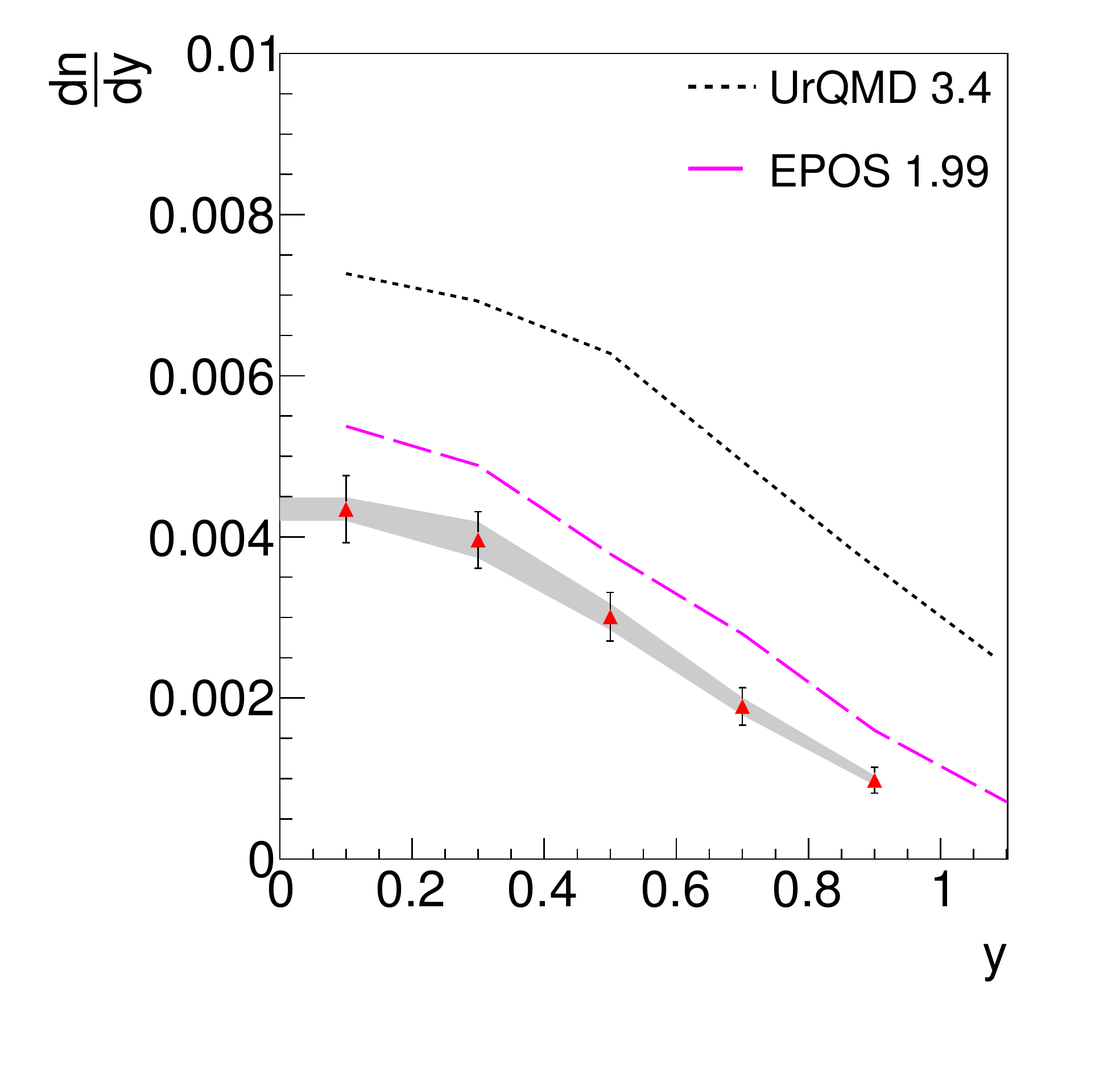} &
\hspace{-45mm}
\includegraphics[width=0.18\textwidth]{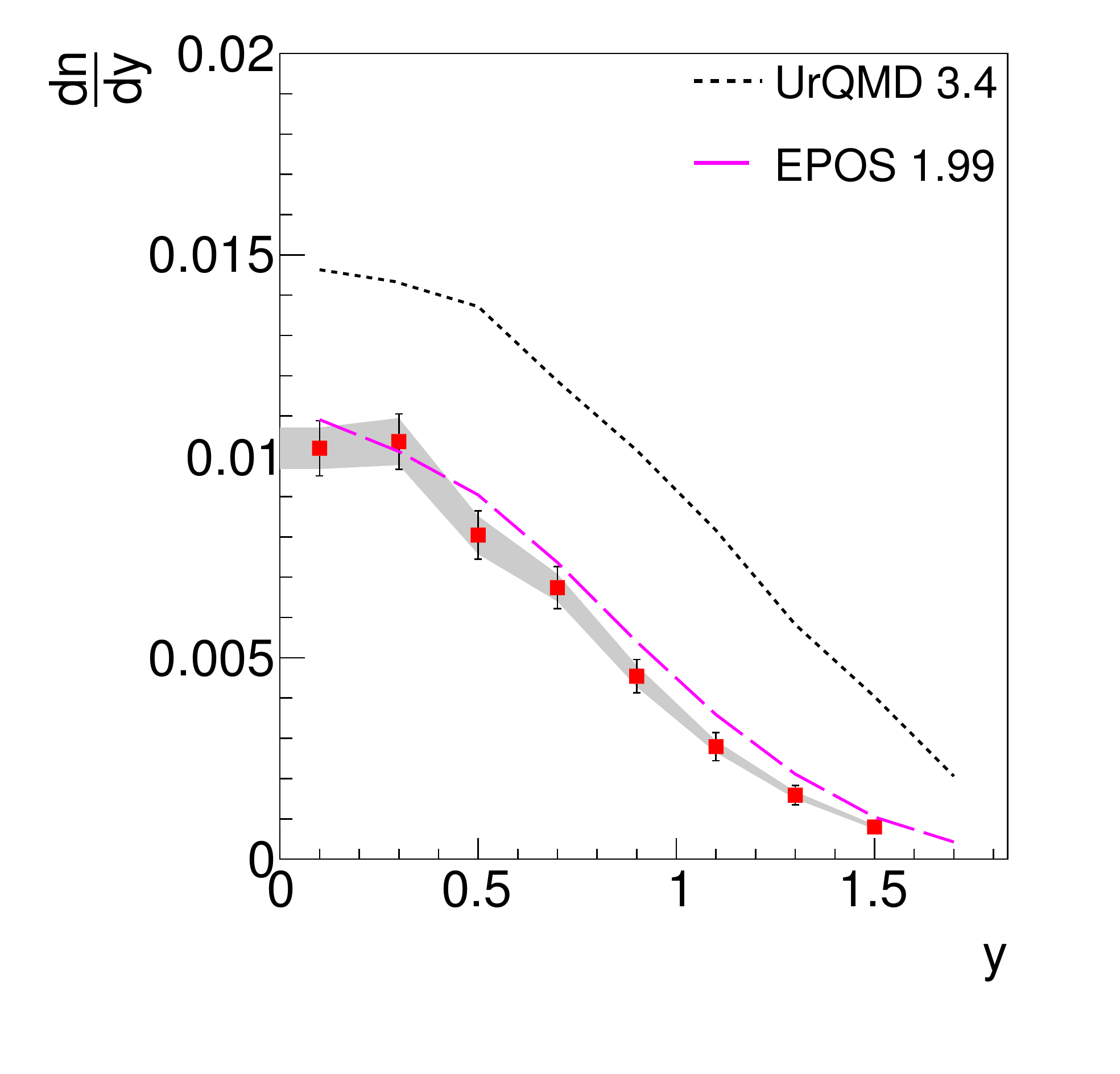} &
\hspace{-60mm}
\includegraphics[width=0.18\textwidth]{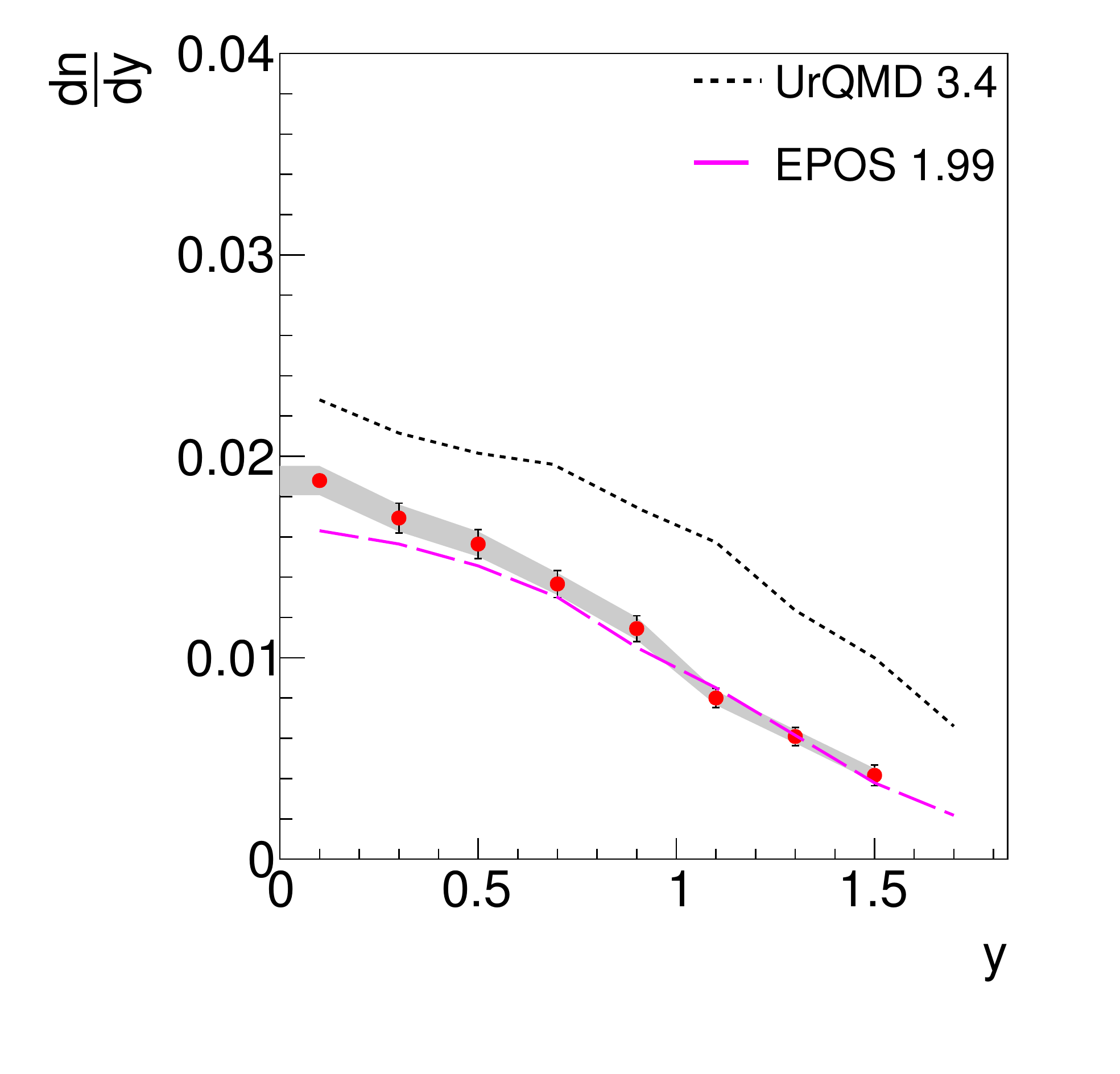} 
\tabularnewline
\end{tabular}
		\end{center}
		\caption{(Color online) Rapidity distribution of $\pi^{-}$, $\pi^{+}$, K$^{-}$, K$^{+}$, p and $\bar{\textrm{p}}$ produced in inelastic p+p interactions at 20, 31, 40, 80 and 158~\GeVc compared with predictions of the \Epos~\cite{Werner:2008zza} (dashed lines) and \Urqmd~\cite{Bass:1998ca,Bleicher:1999xi} (dotted lines) models. }
		\label{fig:modeldndy}
	\end{figure*}

Comparison of the  \Urqmd 3.4 calculations with the \NASixtyOne measurements show larger discrepancies as seen in Figs.~\ref{fig:2DUR20}, \ref{fig:2DUR158} and \ref{fig:modeldndy}. There are regions of underprediction for K$^{+}$ at lower energies and overprediction for K$^{-}$ at higher energies and especially $\bar{\textrm{p}}$ at all collision energies.

\begin{figure}[!ht]
	\begin{center}
	\includegraphics[width=0.42\textwidth]{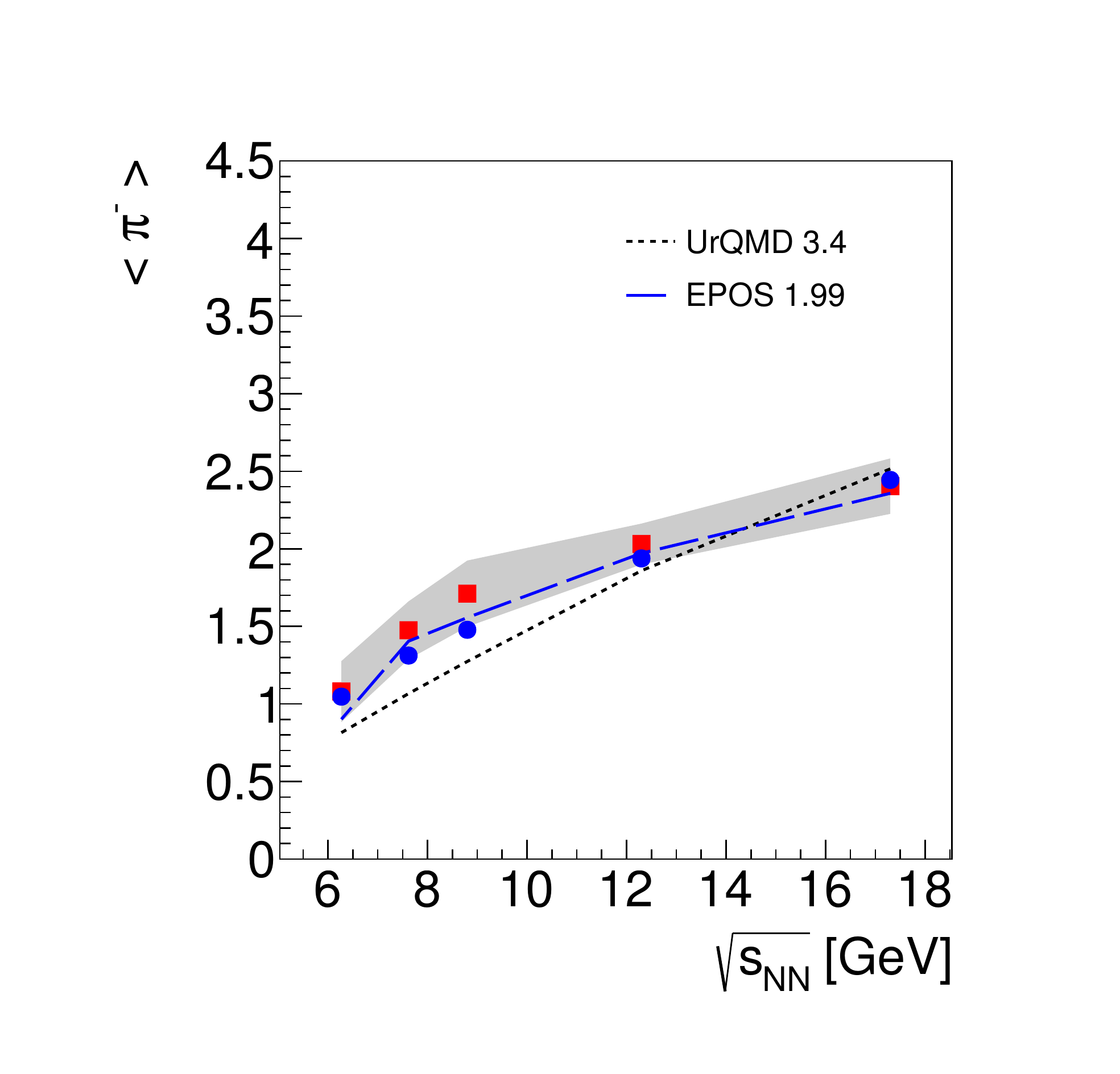}
	\includegraphics[width=0.42\textwidth]{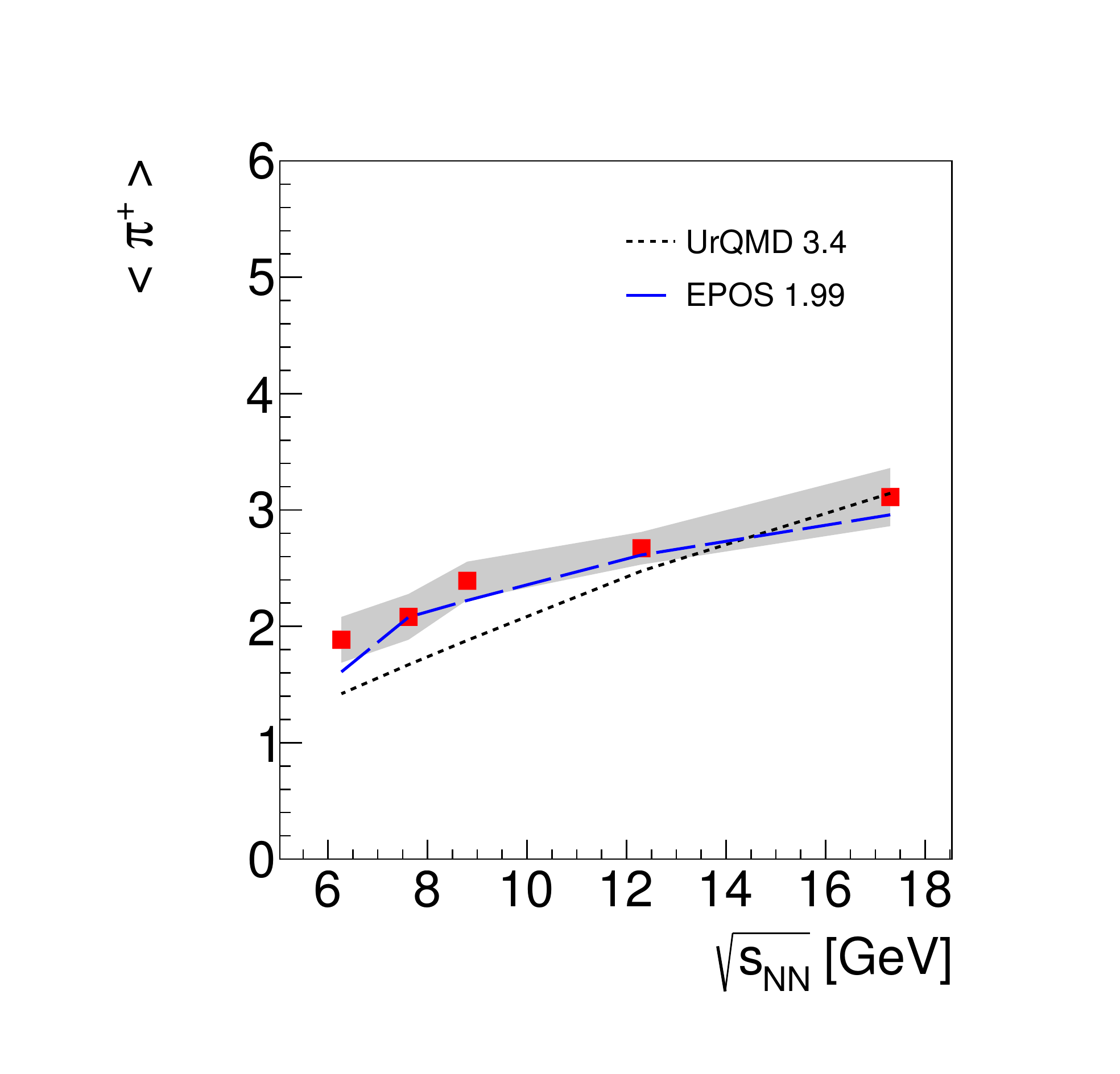}\\
	\includegraphics[width=0.42\textwidth]{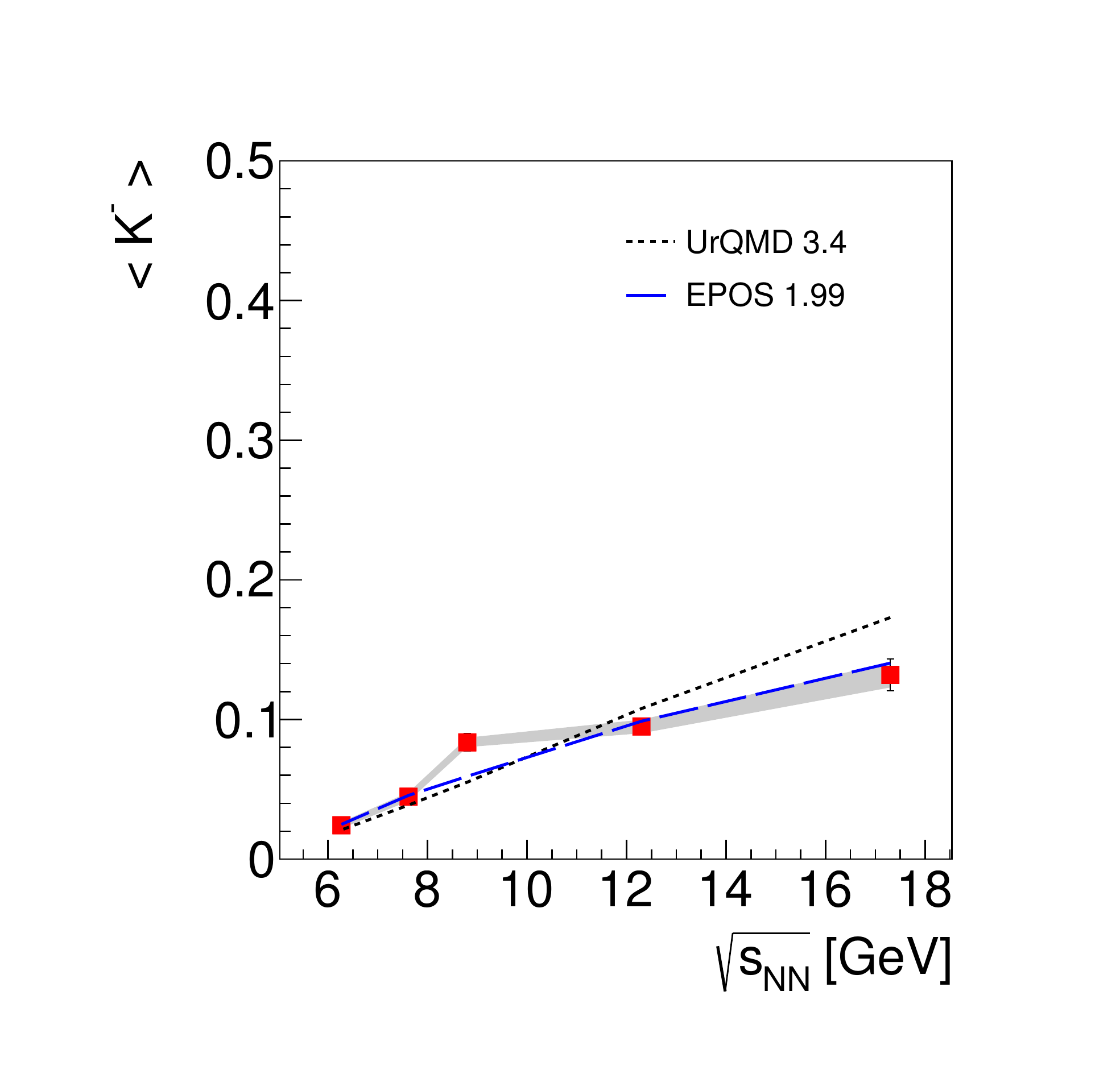}
	\includegraphics[width=0.42\textwidth]{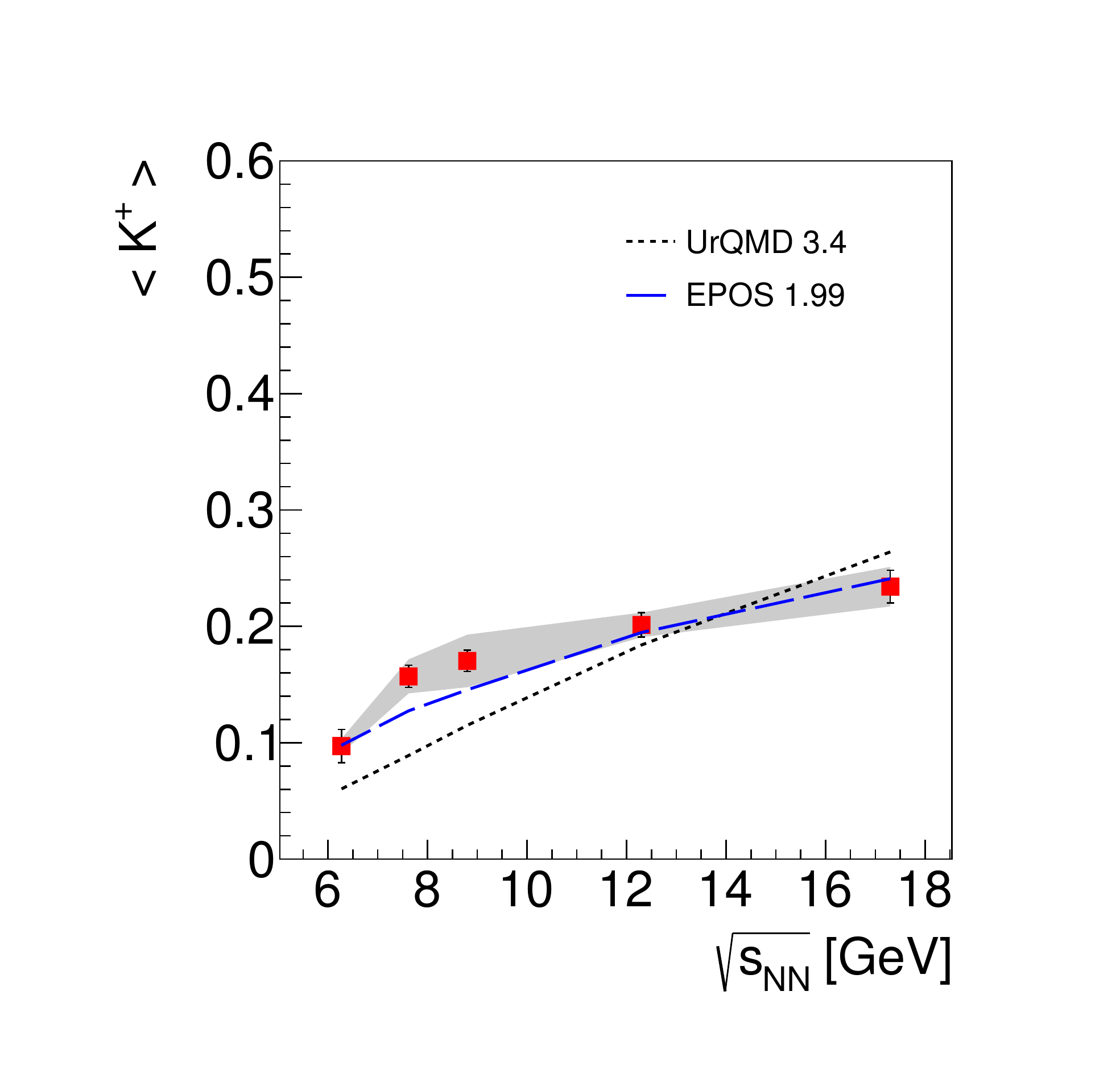}\\
	\includegraphics[width=0.42\textwidth]{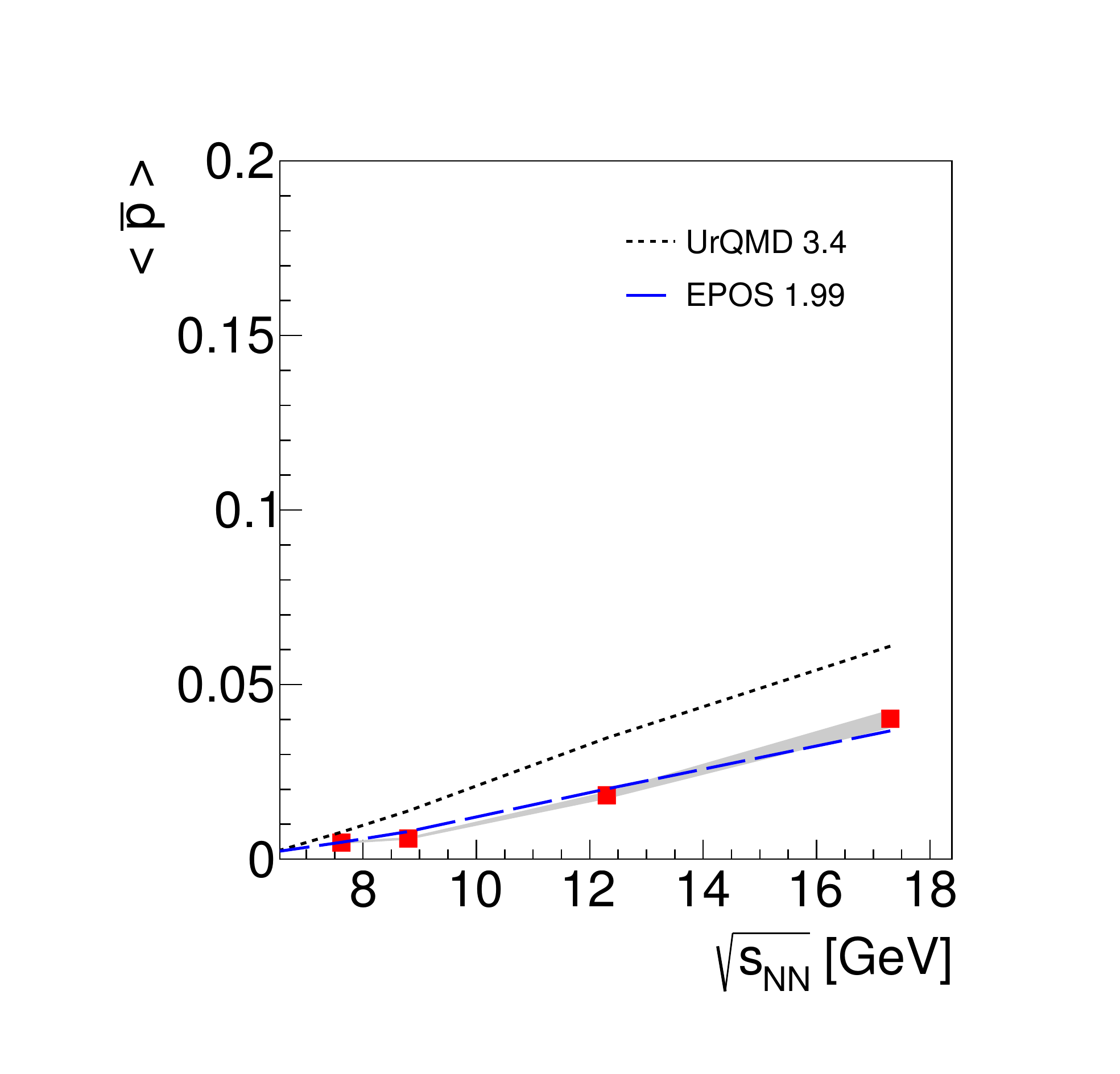}
	\end{center}
	\caption{(Color online) Mean multiplicity of $\pi^{-}$, $\pi^{+}$, K$^{-}$, K$^{+}$ and $\bar{\textrm{p}}$  multiplicity produced in inelastic p+p interactions as a function of collision energy. Measurements of \NASixtyOne are plotted as filled red squares, \Epos~\cite{Werner:2008zza} and \Urqmd~\cite{Bass:1998ca,Bleicher:1999xi} models predictions are shown as dashed and dotted lines, respectively. Additionally, blue filled circles correspond to previously published results obtained by \NASixtyOne via the $h^{-}$ method~\cite{Abgrall:2013pp_pim}. Statistical uncertainties are indicated by bars, systematic uncertainties of the \NASixtyOne measurements by shaded bands.}
	\label{fig:modelmean}
\end{figure}

Comparison of the measured rapidity distributions with predictions of both models are shown in Fig.~\ref{fig:modeldndy}.
The total multiplicities $\pi^{-}$, $\pi^{+}$, K$^{-}$, K$^{+}$ and $\bar{\textrm{p}}$ as function of collision energy 
are presented and compared in Fig.~\ref{fig:modelmean}. The \Epos 1.99 model provides a reasonable description
of the \NASixtyOne measurements, while significant discrepancies are evident for the \Urqmd 3.4
calculations.  


\begin{figure}[!ht]
	\begin{center}
	\includegraphics[width=0.46\textwidth]{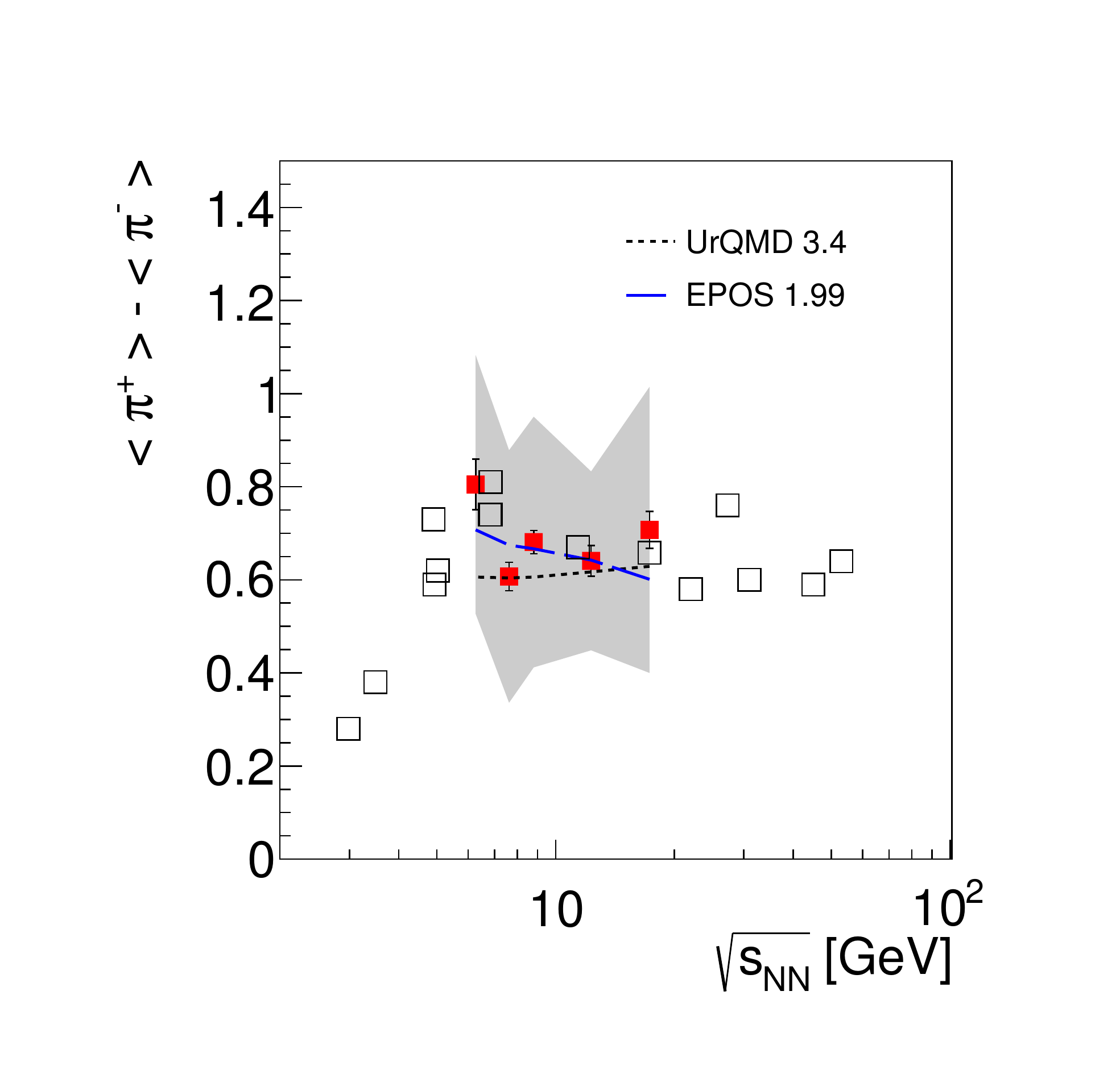}
	\end{center}
	\caption{(Color online) Difference $\left\langle \pi^{+} \right\rangle - \left\langle \pi^{-} \right\rangle$ in inelastic p+p interactions as function of collision energy. Solid red squares depict \NASixtyOne results from this analysis, open squares show previously published measurements~\cite{Alt:2005zq,Blobel:1975ka,AguilarBenitez:1991yy,Ammosov:1976zk}. Statistical uncertainties of the \NASixtyOne results are indicated by bars, systematic uncertainties by the shaded band. The \Epos~\cite{Werner:2008zza} and \Urqmd~\cite{Bass:1998ca,Bleicher:1999xi} model predictions are shown as dashed and dotted lines, respectively.}
	\label{fig:modeldiffpi}
\end{figure}

Additionally the difference $\left\langle \pi^{+} \right\rangle - \left\langle \pi^{-} \right\rangle$ is presented as function of collision energy in Fig.~\ref{fig:modeldiffpi}. It can be observed that this value is almost independent of collision energy.
The \NASixtyOne results agree with the measurements of previous experiments. Both \Epos and \Urqmd model predictions are consistent with the measurements.

\FloatBarrier
\section{Summary and outlook}
\label{sec:summary}

Spectra and multiplicities of $\pi^{+}$, $\pi^{-}$, K$^{+}$, K$^{-}$, p and $\bar{\textrm{p}}$ produced in inelastic p+p interactions were measured with the \NASixtyOne spectrometer at beam momenta of 20, 31, 40, 80, 158~\GeVc at the CERN SPS. New probabilistic analysis techniques based on the energy loss \dedx in the TPCs and the combination of the measurement of time of flight and energy loss $tof$-\dedx were employed. Statistical and systematic uncertainties were carefully evaluated. The \NASixtyOne results significantly improve the  world data both in precision and momentum coverage. The \Epos~1.99 model~\cite{Werner:2008zza} provides a good description of the measurements in the SPS energy range, while  predictions of the \Urqmd 3.4 model~\cite{Bass:1998ca,Bleicher:1999xi} significantly differ from the data.
The new \NASixtyOne measurements of particle production in inelastic p+p collisions provide the baseline for the systematic study of the system size dependence of the onset of deconfinement observed by the NA49 experiment in central Pb+Pb collisions in the SPS energy range. In the nearest future the \NASixtyOne collaboration will extend the p+p energy scan by results from p+p interactions at 400~\GeVc beam momentum.

\FloatBarrier

\section*{Acknowledgments}
We would like to thank the CERN EP, BE and EN Departments for the
strong support of NA61/SHINE.

This work was supported by the Hungarian Scientific Research Fund
(grants OTKA 68506 and 71989), the J\'anos Bolyai Research Scholarship
of the Hungarian Academy of Sciences, the Polish Ministry of Science
and Higher Education (grants 667\slash N-CERN\slash2010\slash0,
NN\,202\,48\,4339 and NN\,202\,23\,1837), the Polish National Center
for Science (grants~2011\slash03\slash N\slash ST2\slash03691,
2013\slash11\slash N\slash ST2\slash03879, 2014\slash13\slash N\slash
ST2\slash02565, 2014\slash14\slash E\slash ST2\slash00018, 2014\slash 12\slash T\slash ST2\slash00692,
2015\slash18\slash M\slash ST2\slash00125, 2015\slash 19\slash N\slash ST2 \slash01689 and 2014\slash15\slash B\slash ST2\slash02537) , the Foundation for Polish
Science --- MPD program, co-financed by the European Union within the
European Regional Development Fund, the Federal Agency of Education of
the Ministry of Education and Science of the Russian Federation (SPbSU
research grant 11.38.242.2015), the Russian Academy of Science and the
Russian Foundation for Basic Research (grants 08-02-00018, 09-02-00664
and 12-02-91503-CERN), the Ministry of Education, Culture, Sports,
Science and Tech\-no\-lo\-gy, Japan, Grant-in-Aid for Sci\-en\-ti\-fic
Research (grants 18071005, 19034011, 19740162, 20740160 and 20039012),
the German Research Foundation (grant GA\,1480/2-2), the EU-funded
Marie Curie Outgoing Fellowship, Grant PIOF-GA-2013-624803, the
Bulgarian Nuclear Regulatory Agency and the Joint Institute for
Nuclear Research, Dubna (bilateral contract No. 4418-1-15\slash 17),
Bulgarian National Science Fund (grant DN08/11), Ministry of Education
and Science of the Republic of Serbia (grant OI171002), Swiss
Nationalfonds Foundation (grant 200020\-117913/1), ETH Research Grant
TH-01\,07-3 and the U.S.\ Department of Energy, the National Research Nuclear
University MEPhI in the framework of the Russian Academic Excellence
Project (contract No.\ 02.a03.21.0005,\ 27.08.2013).

\bibliographystyle{na61Utphys}
\bibliography{na61References}
\newpage 
{\Large The \NASixtyOne Collaboration}
\bigskip

\noindent
A.~Aduszkiewicz$^{\,16}$,
Y.~Ali$^{\,13}$,
E.~Andronov$^{\,22}$,
T.~Anti\'ci\'c$^{\,3}$,
B.~Baatar$^{\,20}$,
M.~Baszczyk$^{\,14}$,
S.~Bhosale$^{\,11}$,
A.~Blondel$^{\,25}$,
M.~Bogomilov$^{\,2}$,
A.~Brandin$^{\,21}$
% ...


\noindent
% ...
$^{2}$~Faculty of Physics, University of Sofia, Sofia, Bulgaria\\
$^{3}$~Ru{\dj}er Bo\v{s}kovi\'c Institute, Zagreb, Croatia\\
% ...
$^{11}$~H. Niewodnicza\'nski Institute of Nuclear Physics of the
      Polish Academy of Sciences, Krak\'ow, Poland\\
% ...
$^{13}$~Jagiellonian University, Cracow, Poland\\
$^{14}$~University of Science and Technology, Cracow, Poland\\
% ...
$^{16}$~University of Warsaw, Warsaw, Poland\\
% ...
$^{20}$~Joint Institute for Nuclear Research, Dubna, Russia\\
$^{21}$~National Research Nuclear University (Moscow Engineering Physics Institute), Moscow, Russia\\
$^{22}$~St. Petersburg State University, St. Petersburg, Russia\\
% ...
$^{25}$~University of Geneva, Geneva, Switzerland\\
% ...

\clearpage
  \clearpage
\appendix
\counterwithin{figure}{section}
\section{ToF correction factors}
\label{apendB}
\begin{figure}[!ht]
	\begin{center}
	\includegraphics[width=0.2\textwidth]{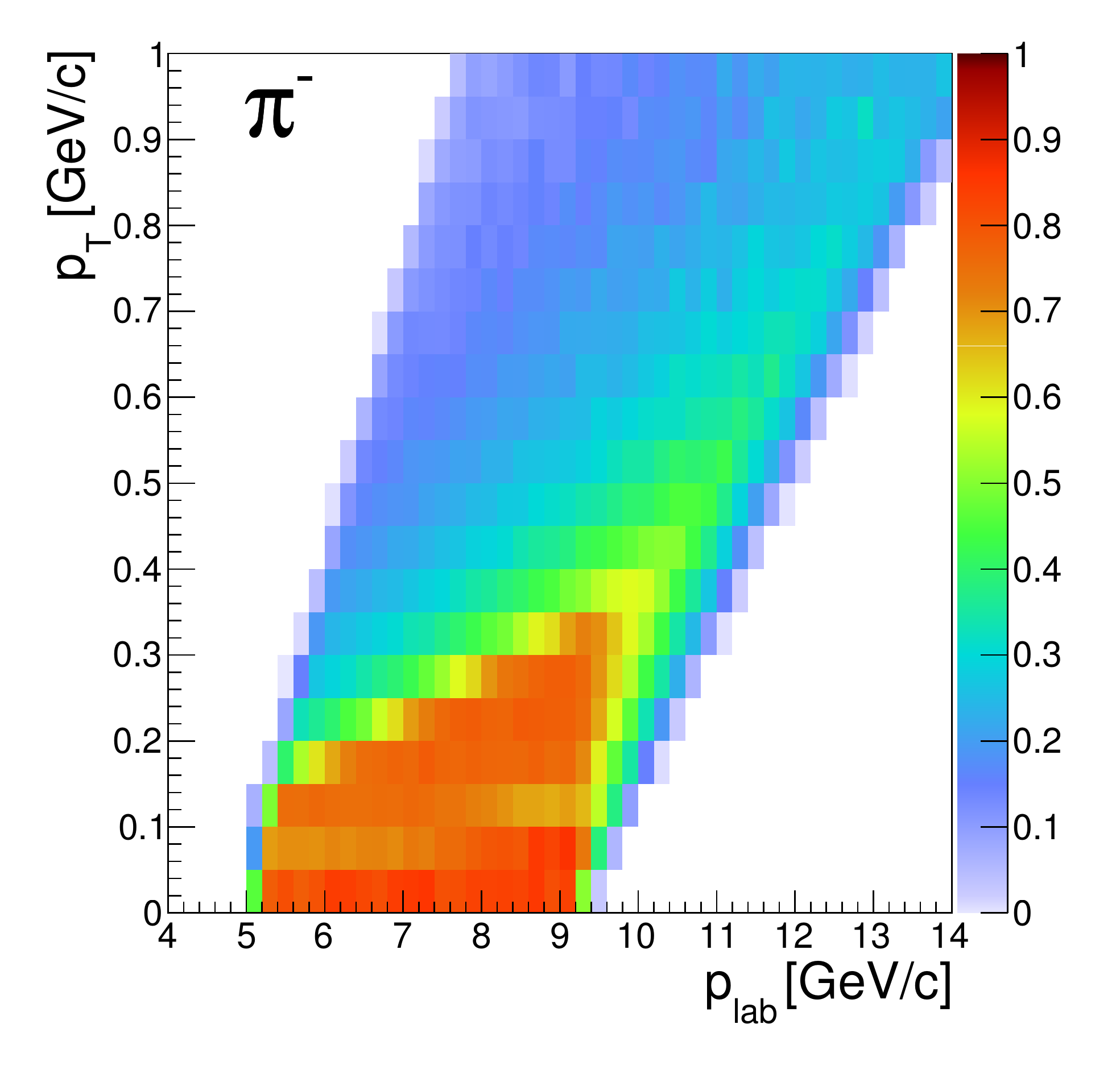}
	\includegraphics[width=0.2\textwidth]{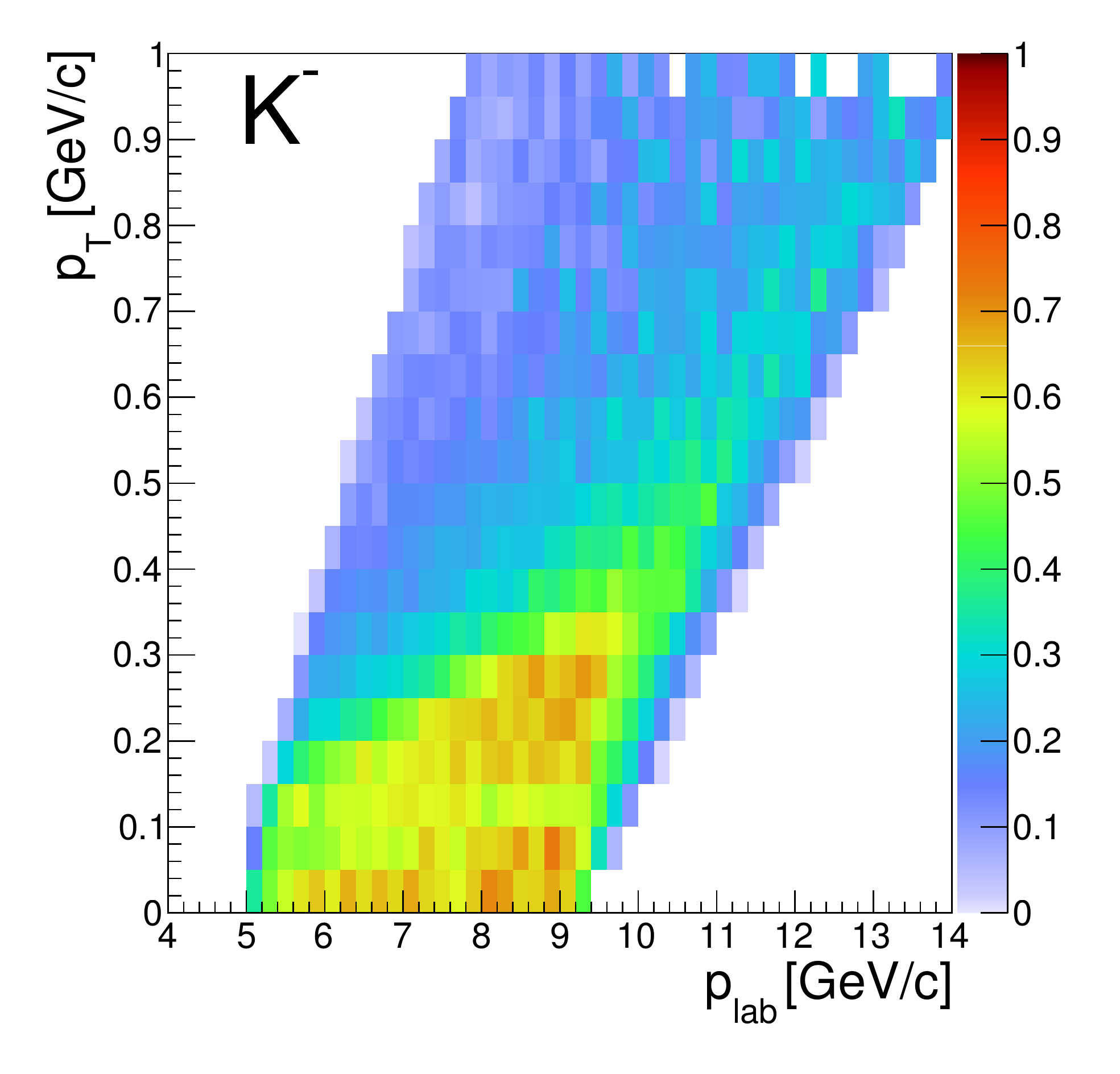}
	\includegraphics[width=0.2\textwidth]{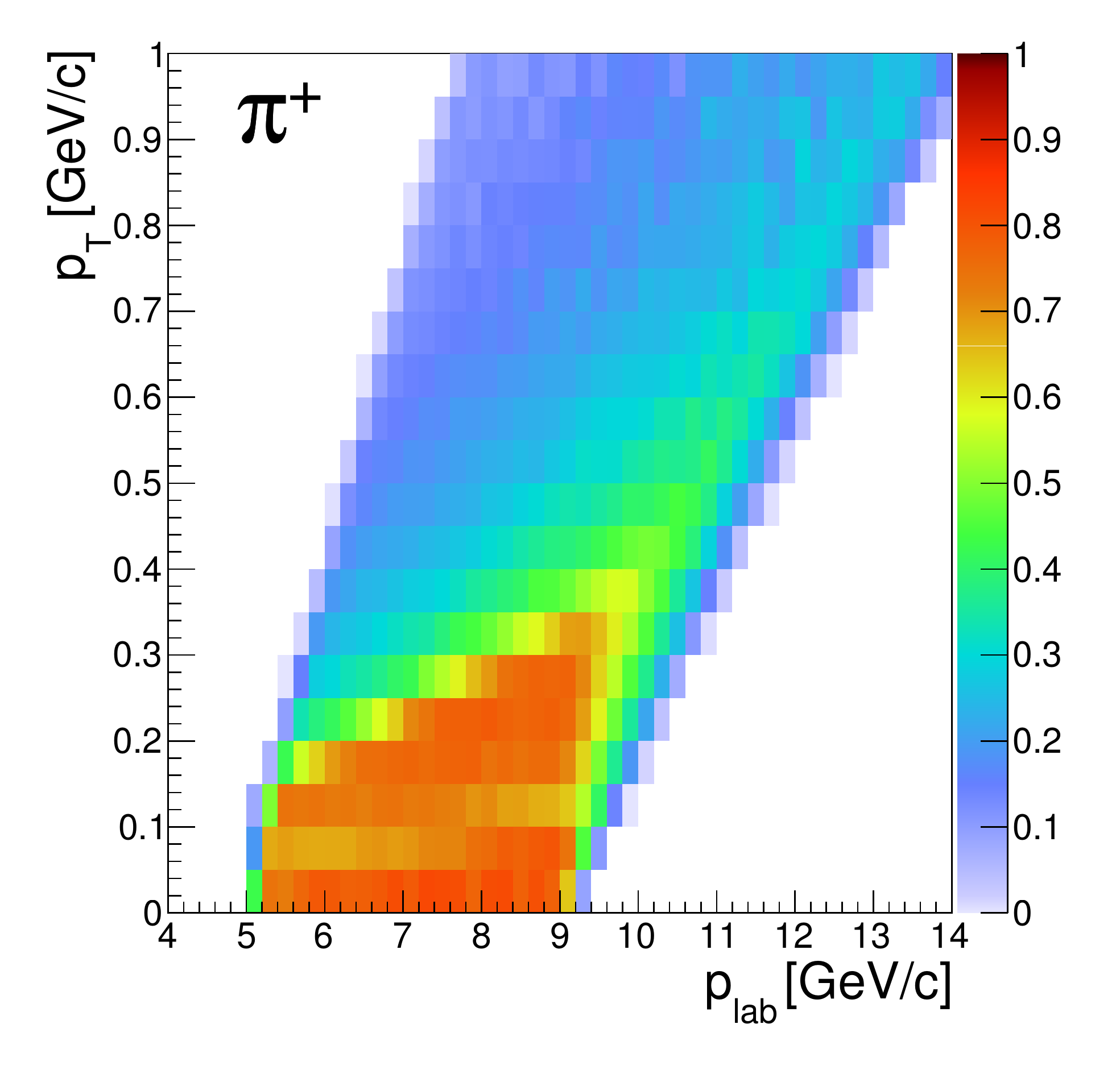}\\
	\includegraphics[width=0.2\textwidth]{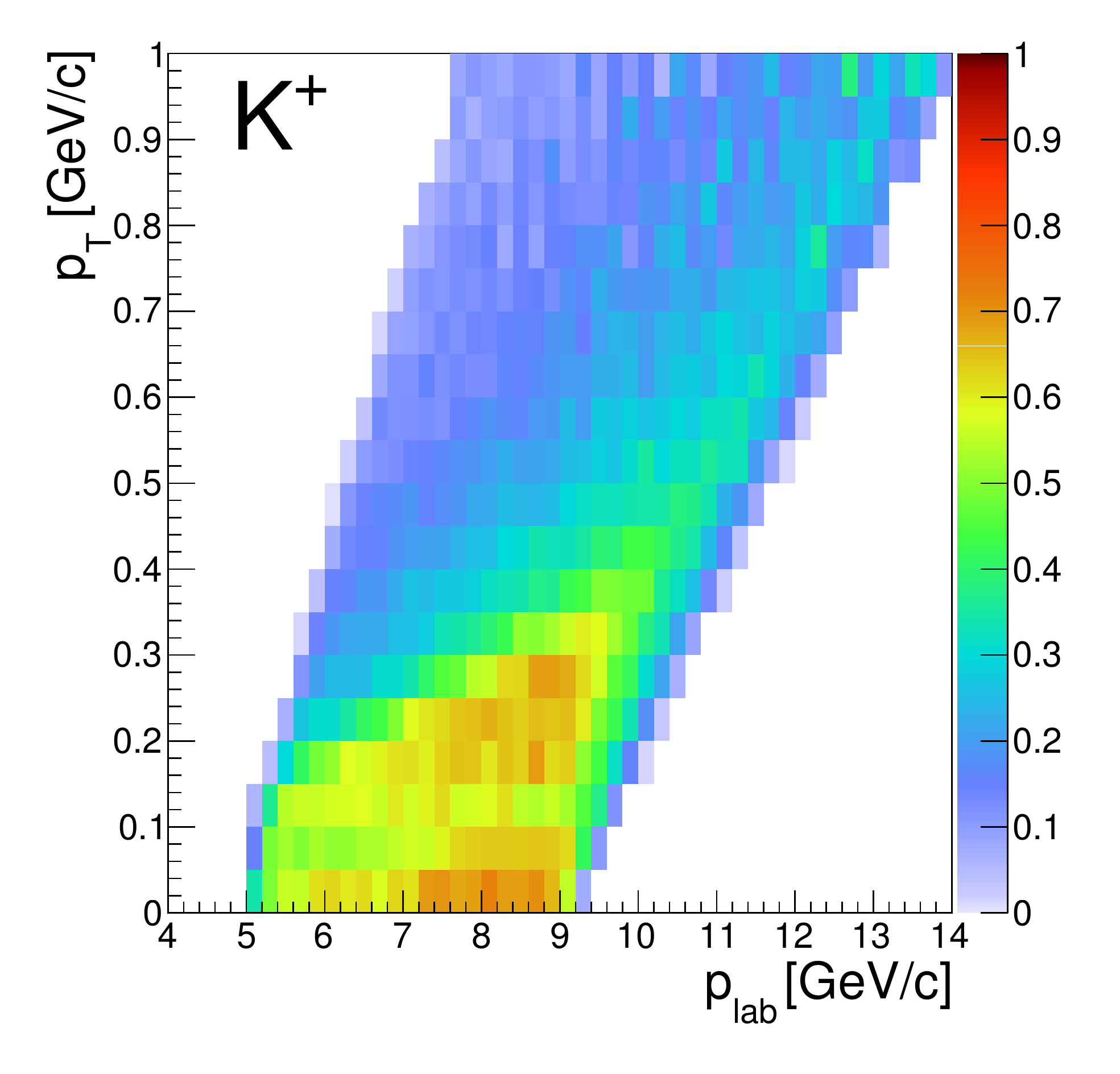}
	\includegraphics[width=0.2\textwidth]{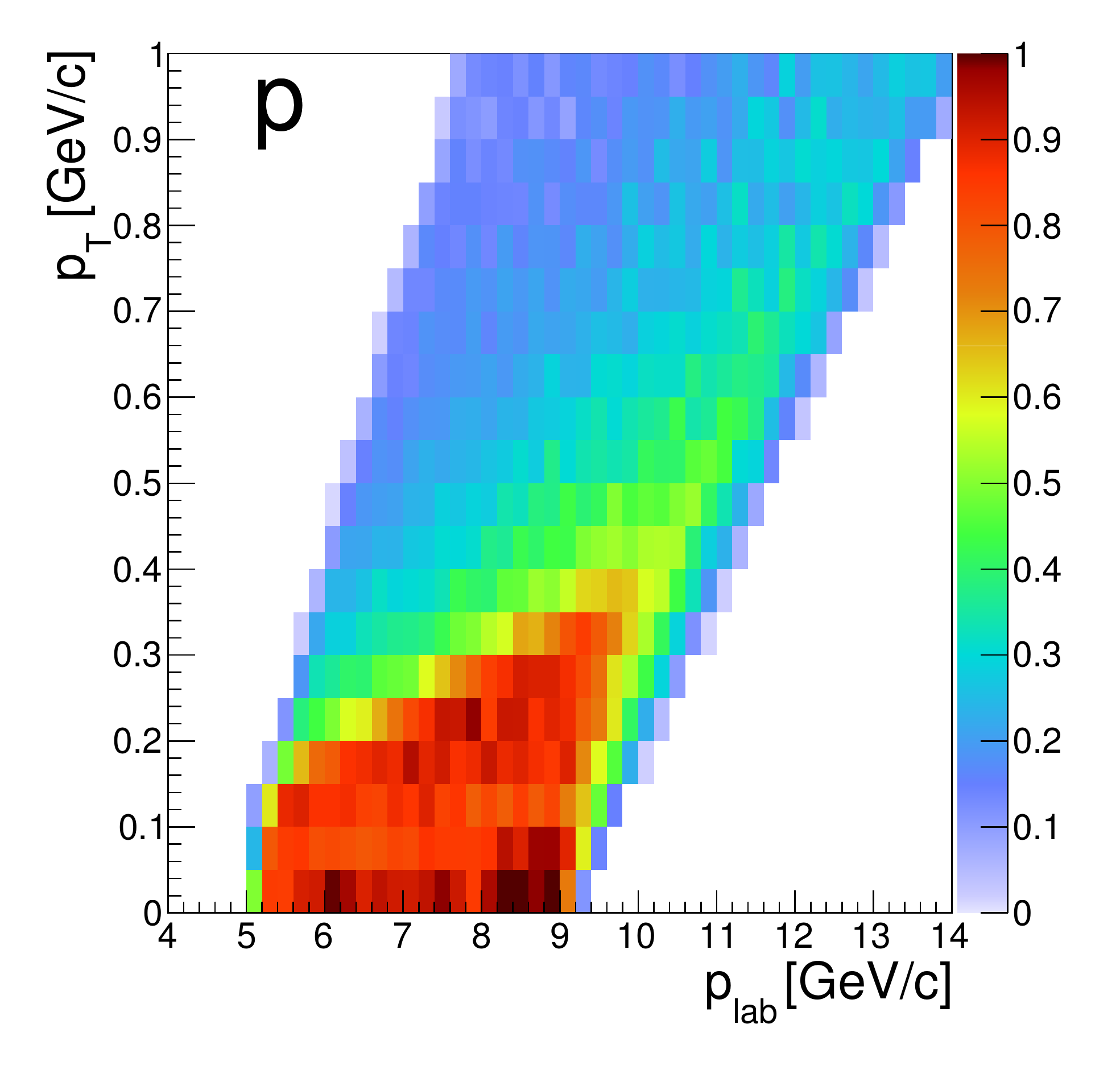}
	\end{center}
	\caption{(Color online) Acceptance for tracks extrapolating to a pixel of the ToF-L or ToF-R walls and having a last measured point in the two last padrows of the MTPC for p+p interactions at 158~\GeVc.}
	\label{fig:tofgeo}
\end{figure}

\begin{figure}[!ht]
	\begin{center}
	\includegraphics[width=0.2\textwidth]{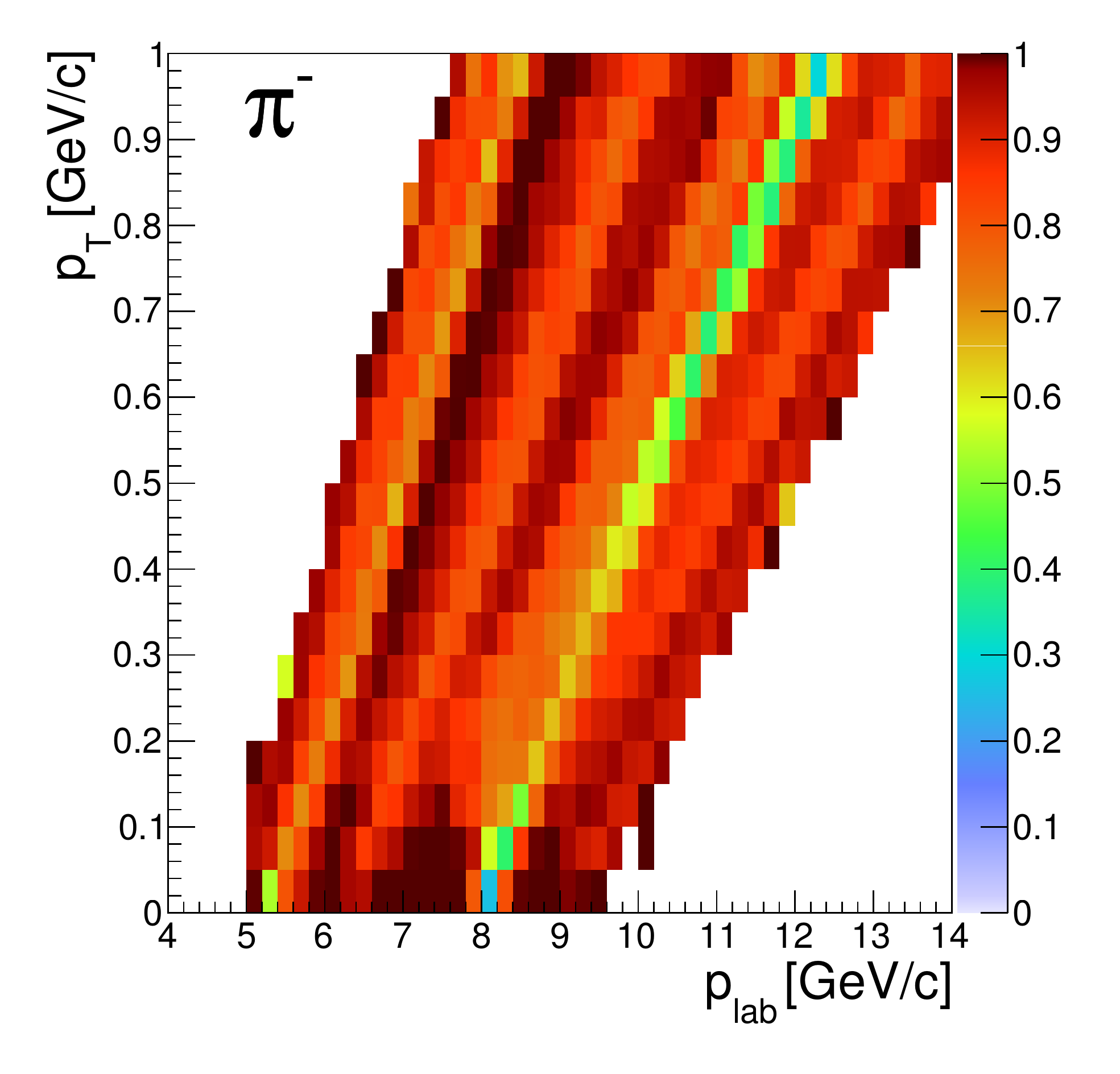}
	\includegraphics[width=0.2\textwidth]{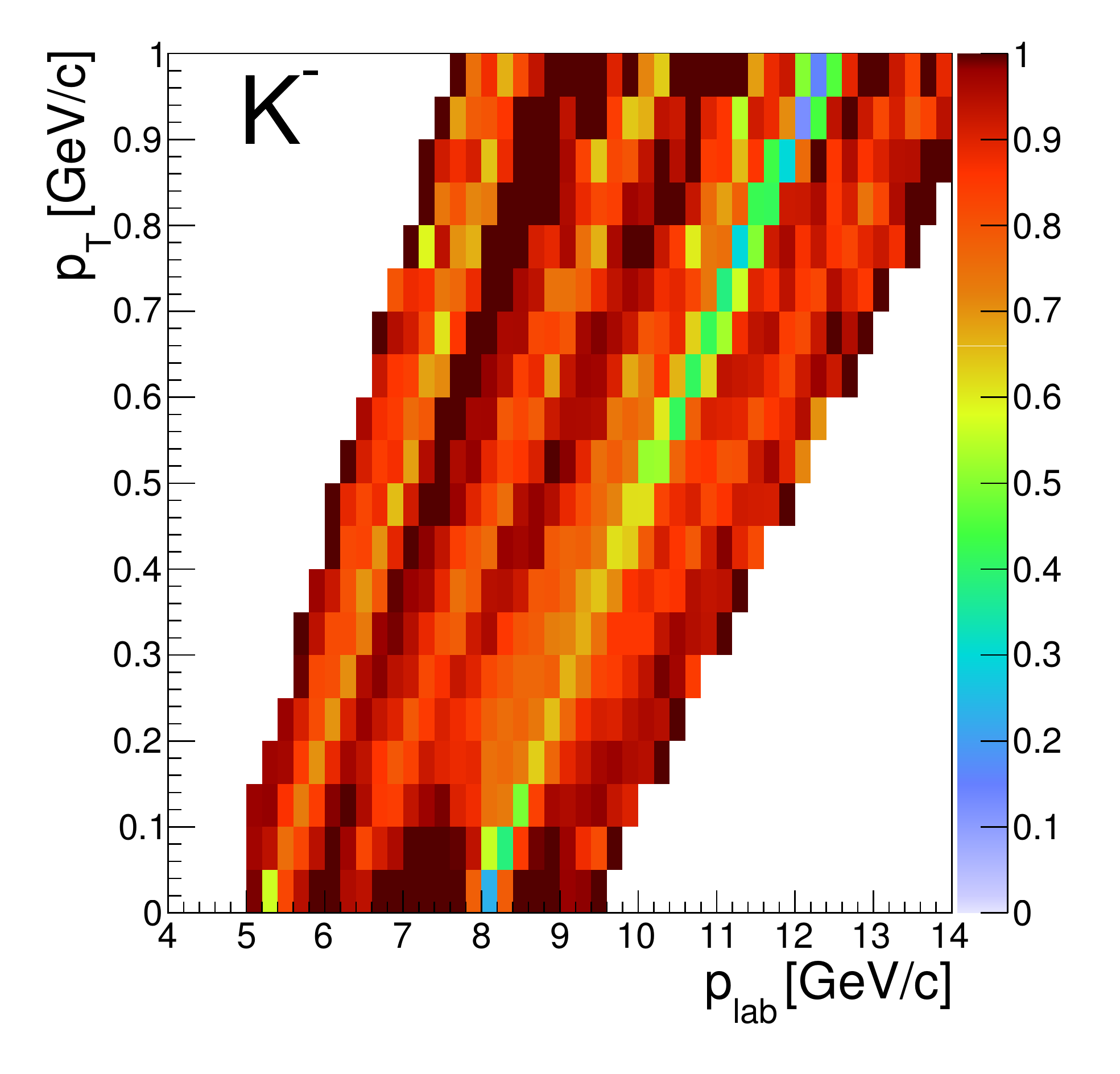}\\
	\includegraphics[width=0.2\textwidth]{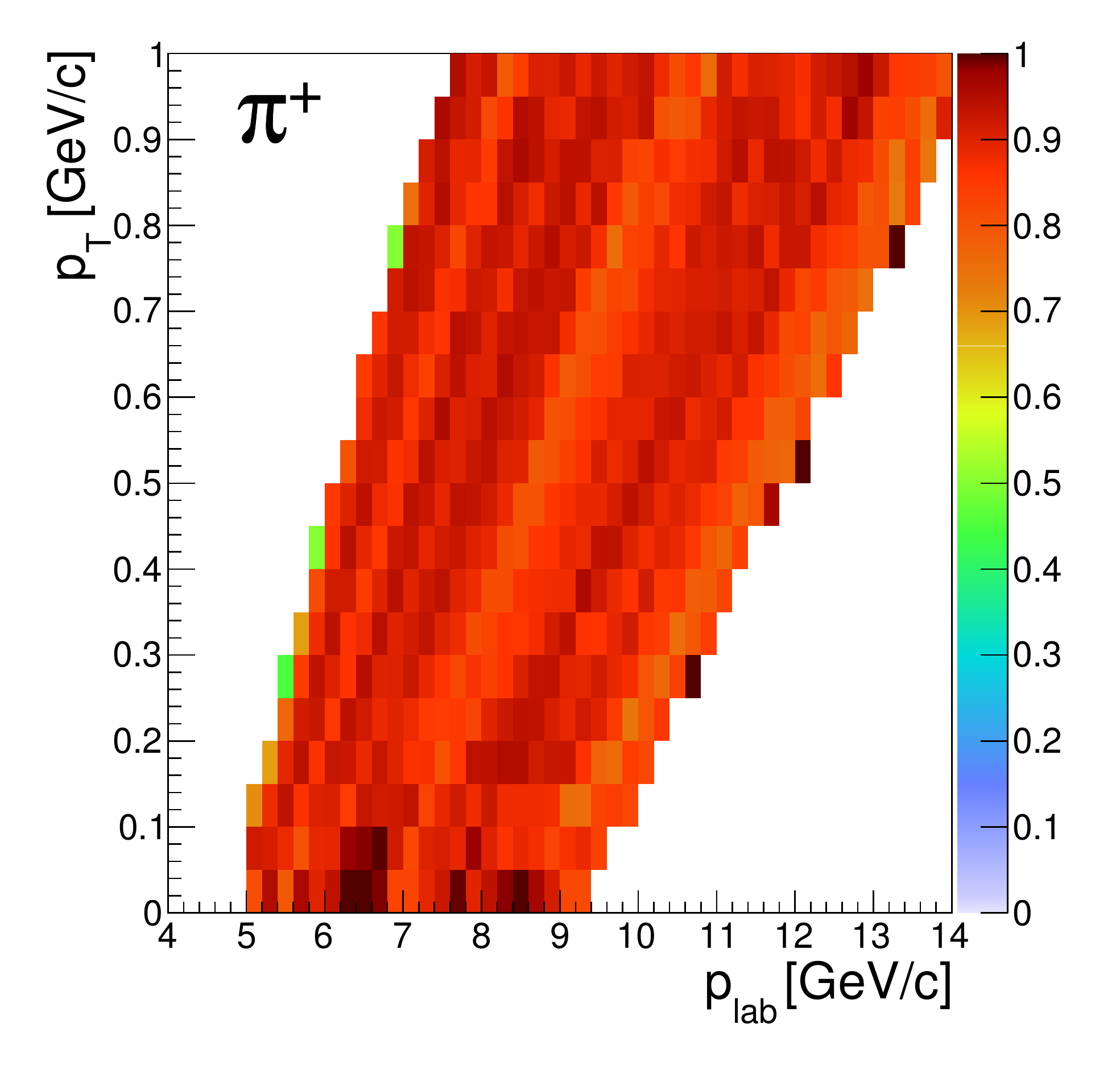}
	\includegraphics[width=0.2\textwidth]{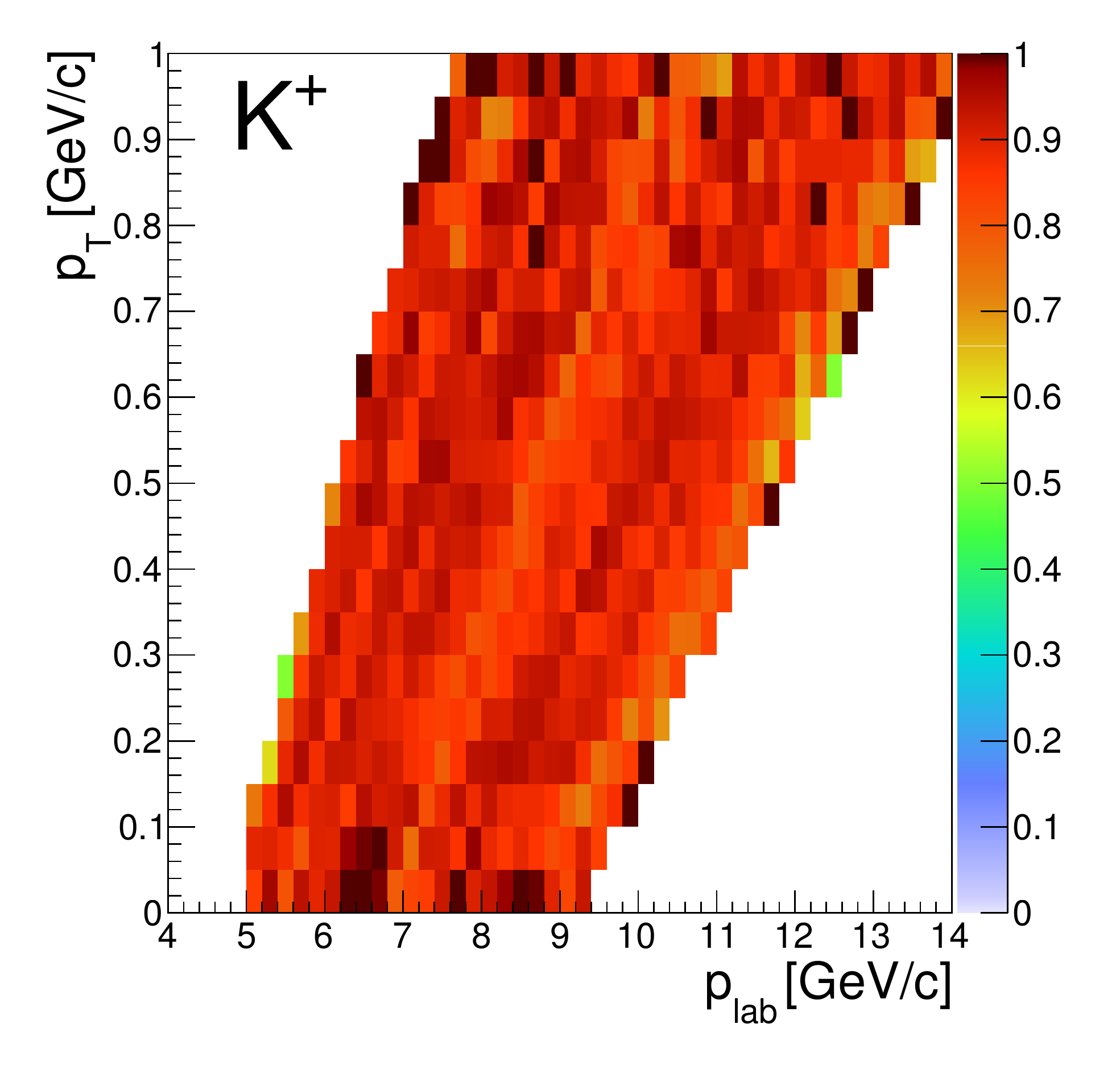}
	\includegraphics[width=0.2\textwidth]{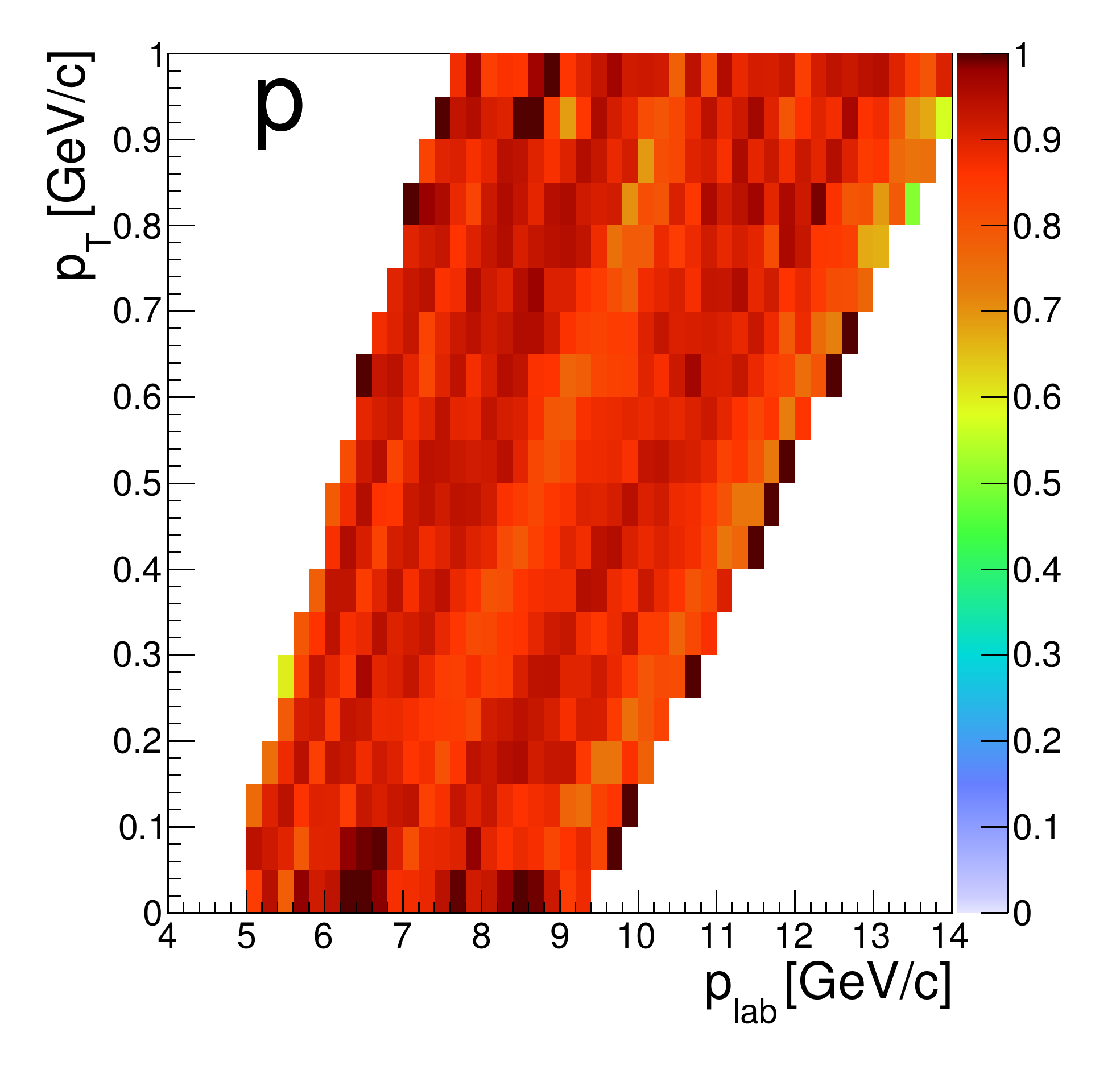}
	\end{center}
	\caption{(Color online) Fraction of pixels providing a valid ToF measurement for p+p interactions at 158~\GeVc. Visible strips with lower value are due to groups of not working pixels.}
	\label{fig:tofdead}
\end{figure}

\begin{figure}[!ht]
	\begin{center}
	\includegraphics[width=0.2\textwidth]{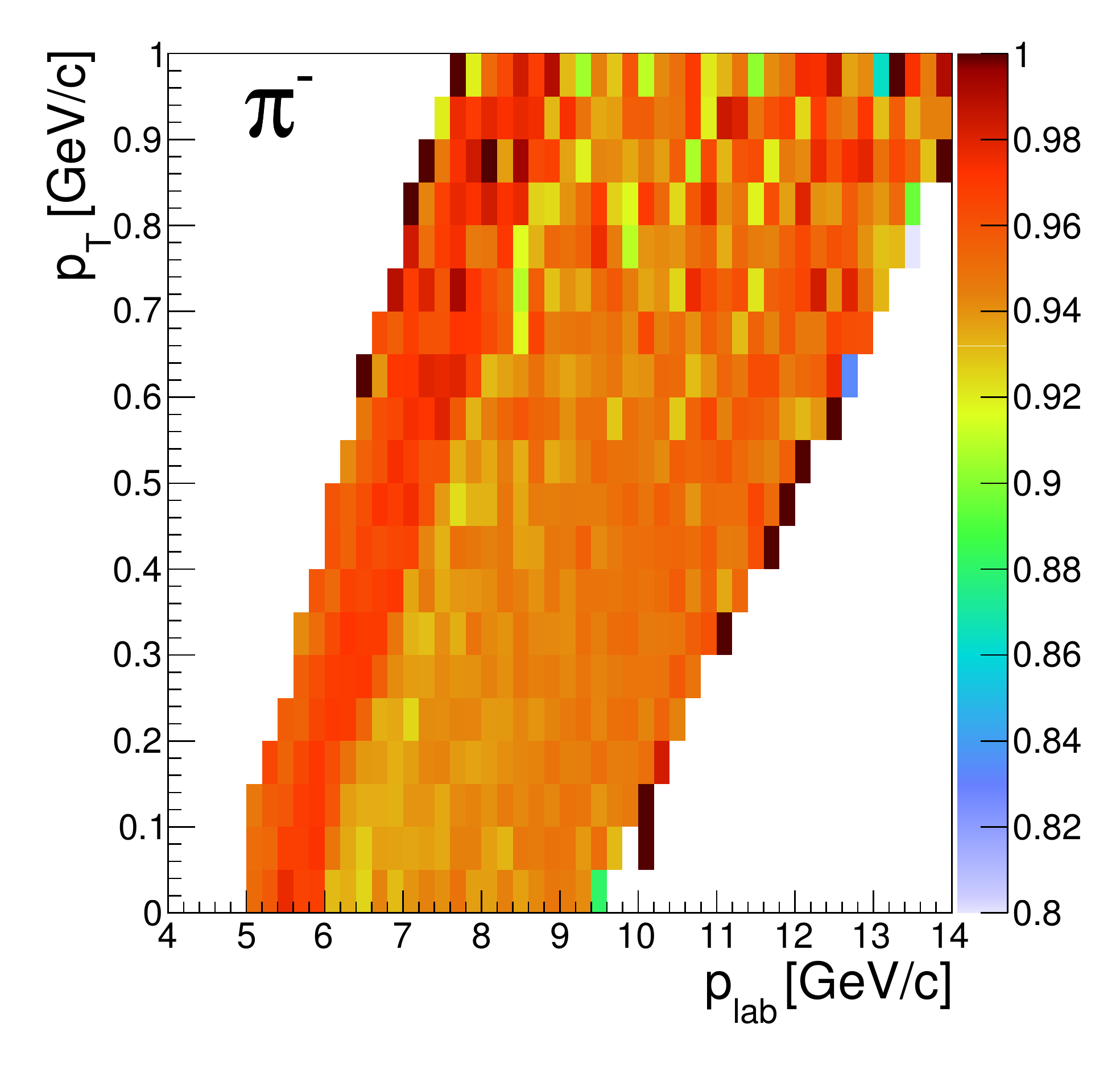}
	\includegraphics[width=0.2\textwidth]{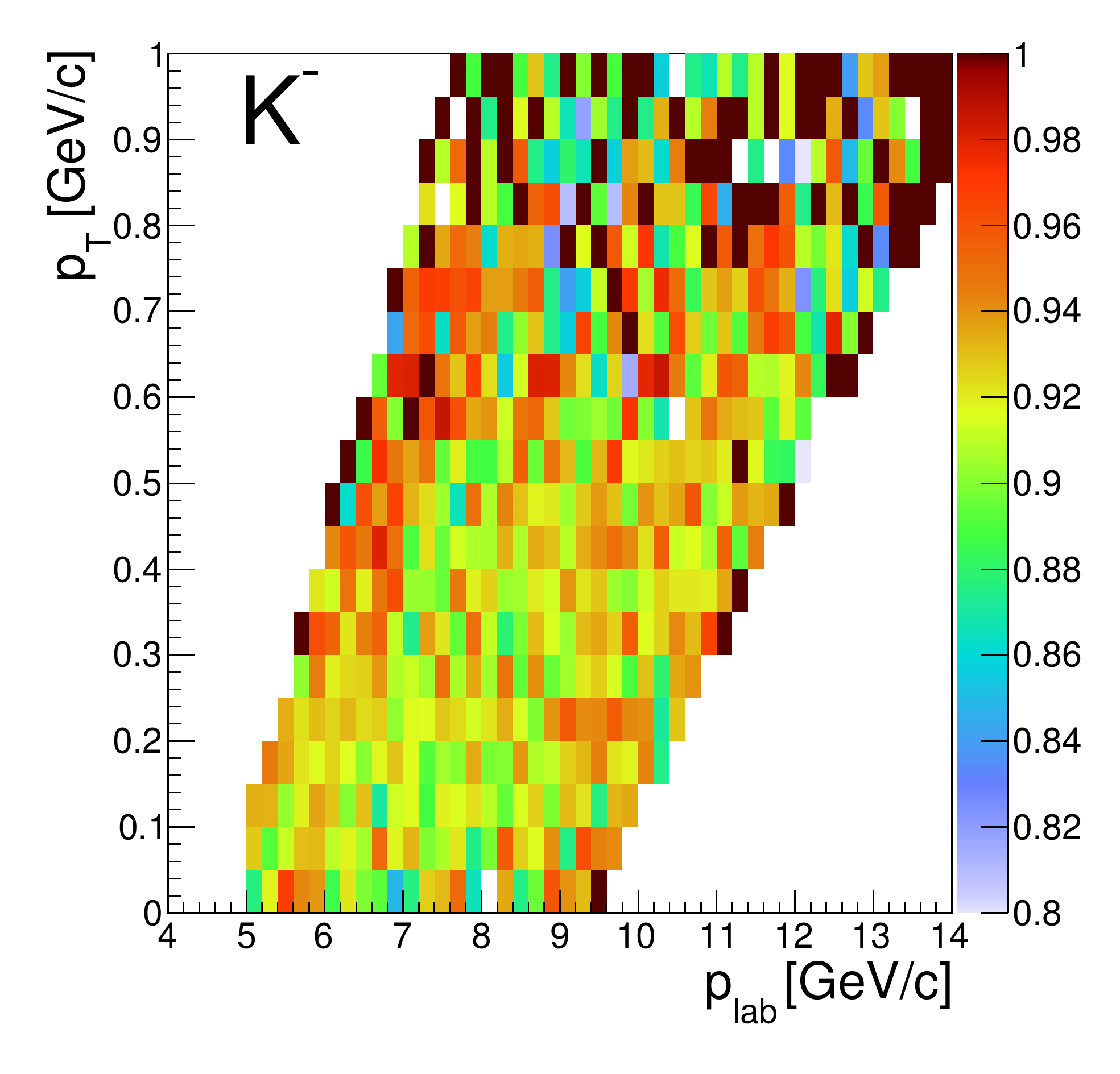}\\
	\includegraphics[width=0.2\textwidth]{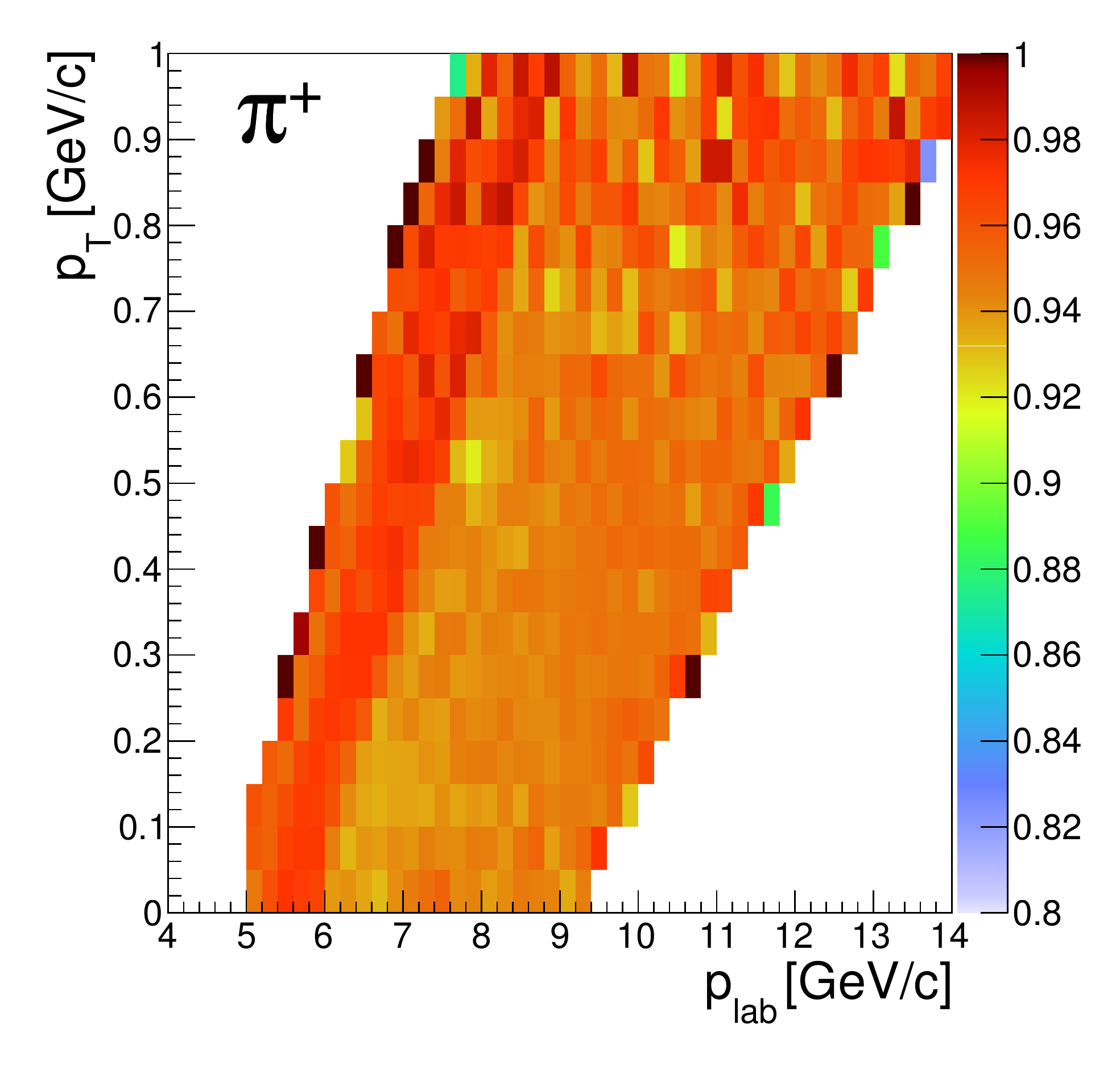}
	\includegraphics[width=0.2\textwidth]{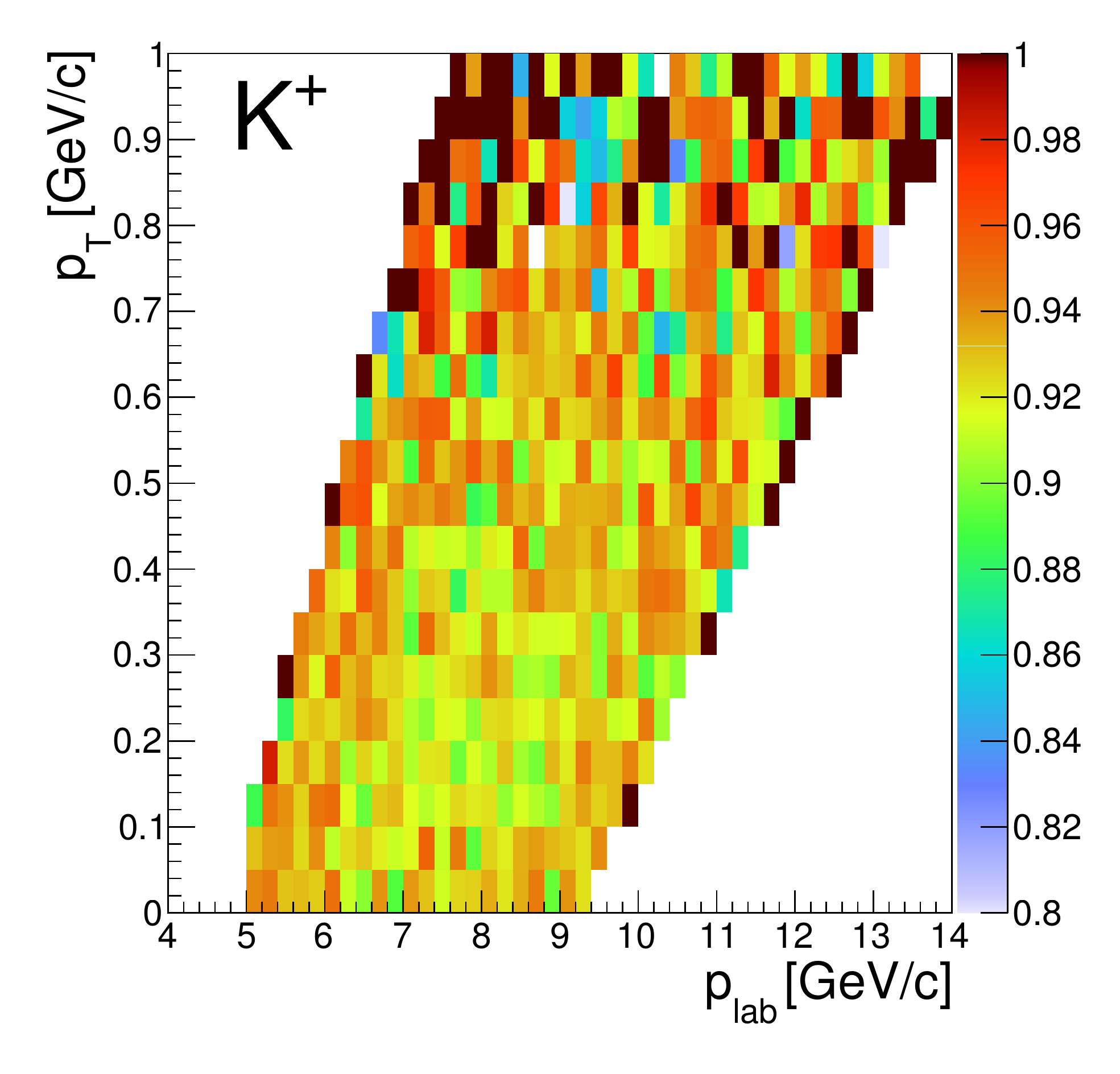}
	\includegraphics[width=0.2\textwidth]{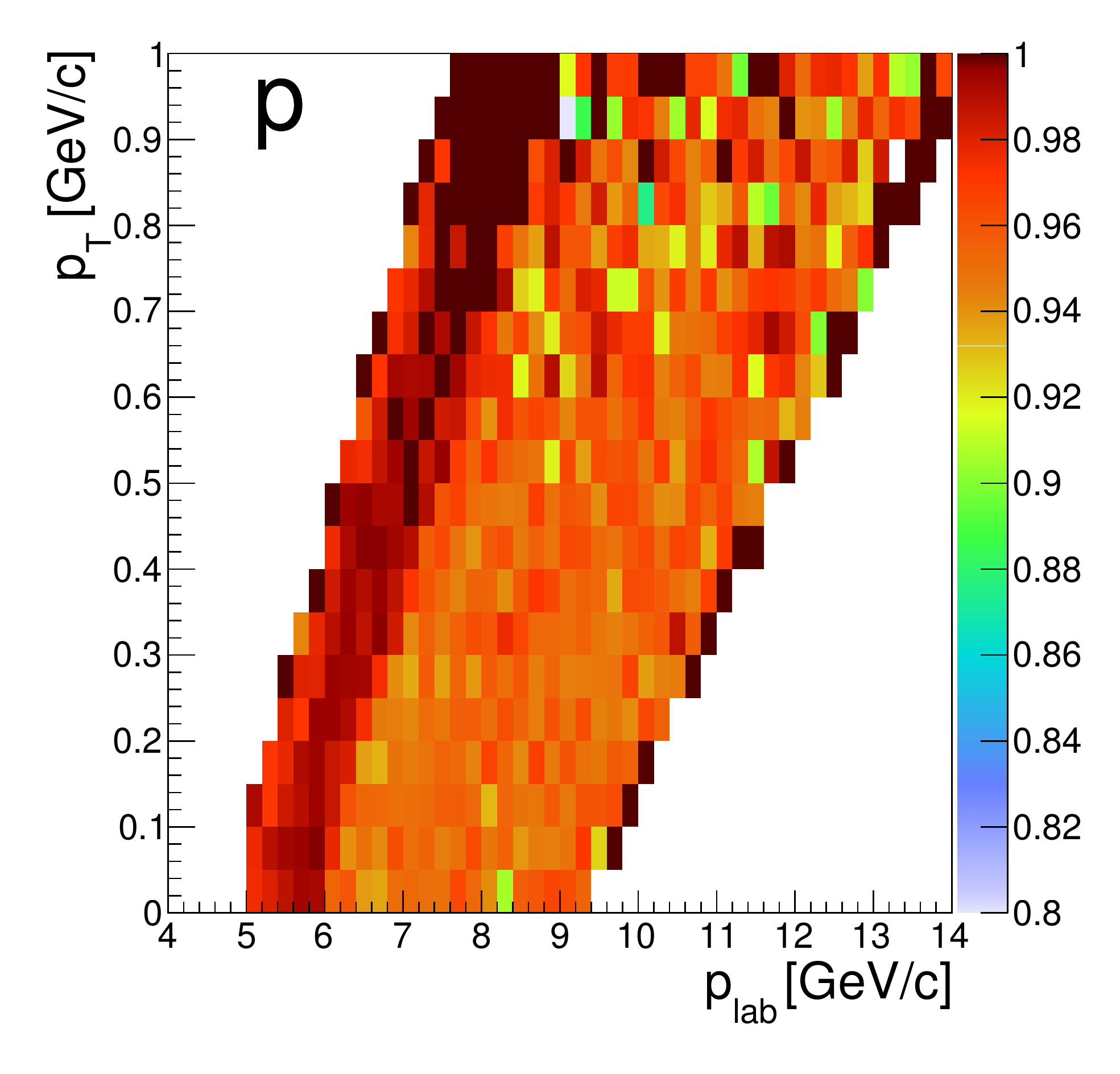}
	\end{center}
	\caption{(Color online) Survival probability of tracks between the last measured point in the MTPC 
and the ToF-L or ToF-R walls for p+p interactions at 158~\GeV/c.}
	\label{fig:tofdecay}
\end{figure}

\end{document}